\documentclass[11pt,twoside,a4]{book}
\def \1{\'{\i}}

\usepackage[english,activeacute]{babel}
\usepackage{epsfig}
\usepackage{amsmath}
\usepackage{epic}
\usepackage{amssymb}
\usepackage{fancybox}
\usepackage{wrapfig}
\usepackage[font=footnotesize,labelsep=period]{caption}
\usepackage{amsthm}
\usepackage{float}
 \usepackage[capitalise]{cleveref}
\usepackage{xcolor}
\definecolor{dark-red}{rgb}{0.6,0.15,0.15}
\definecolor{dark-blue}{rgb}{0.15,0.15,0.8}
\definecolor{medium-blue}{rgb}{0,0,0.6}
\allowdisplaybreaks
\usepackage{booktabs}

\usepackage{wrapfig}
\usepackage{cleveref}
\newcommand*{\KeepStyleUnderBrace}[1]{%
  \mathop{%
    \mathchoice
    {\underbrace{\displaystyle#1}}%
    {\underbrace{\textstyle#1}}%
    {\underbrace{\scriptstyle#1}}%
    {\underbrace{\scriptscriptstyle#1}}%
  }\limits
}

\newcommand{\dashedrightarrow}[2][]{\ext@arrow 0359\rightarrowfill@@{#1}{#2}}

\usepackage[all]{xy}
\usepackage{xypic, color, graphics}

\theoremstyle{plain}
\newtheorem{theorem}{Theorem}
\newtheorem{corollary}[theorem]{Corollary}
\newtheorem{proposition}[theorem]{Proposition}
\newtheorem{lemma}[theorem]{Lemma}
\newtheorem{example}{Example}

\theoremstyle{definition}
\newtheorem{definition}[theorem]{Definition}
\newtheorem{note}[theorem]{Note}

 \def\sp #1{{{\cal #1}}}

 \def\v #1{\vert #1\vert}             
 
 \def\m #1 #2{(-1)^{{\v #1} {\v #2}}} 

  \def\ota{j}

 \let \m=\medskip
 \let \b=\bigskip

 \def\degr{{\rm deg}}
 \def\ham{{\cal {H}}_\Lambda}

 \def\evo{{\cal {E}}_0}
 
 \def\mm{{r}}
 \def\JJ{{I}}
 \def\KK{{J}}
 \def\LL{{K}}
 \def\casU{{\cal C}} 
 \def\casS{{C}} 
 \def\ve{{v}}
 \def\FF{{F}}
 \def\aA{{A}}
 \def\bB{{B}}
 \def\cC{{\Upsilon}}
 \def\MM{{\mu}}
 \def\dd{{\rm d}}
 
 \def\ham{{\cal {H}}_\Lambda}
 \def\tX{ {X}_1}
 \def\ff{g} 
 \def\fff{g}
 \def\mm{\mu}
 \def\hh{\gamma} 
\def\ham{{\cal {H}}_\Lambda}
 \def\dd{{\rm d}}
\def\bea{\begin{eqnarray}}
 \def\eea{\end{eqnarray}}
\newcommand{\beq}{\begin{eqnarray}}
\newcommand{\eeq}{\end{eqnarray}}
\newcommand{\ba}{\begin{array}}
\newcommand{\ea}{\end{array}}
\newcommand{\be}{\begin {equation}}
\newcommand{\ee}{\end{equation}}
\def\picture #1 by #2 (#3){
  \vbox to #2{
    \hrule width #1 height 0pt depth 0pt
    \vfill
    \special{picture #3} 
    }
  }
\def\scaledpicture #1 by #2 (#3 scaled #4){{
  \dimen0=#1 \dimen1=#2
  \divide\dimen0 by 1000 \multiply\dimen0 by #4
  \divide\dimen1 by 1000 \multiply\dimen1 by #4
  \picture \dimen0 by \dimen1 (#3 scaled #4)}
  }
  
\usepackage{fancyhdr}
\fancypagestyle{plain}{ \fancyhead{}   }

\usepackage[Bjornstrup]{fncychap}
\usepackage{fancyhdr}
\fancypagestyle{plain}{ \fancyhead{}   }

\usepackage{appendix}
\usepackage[nottoc,numbib]{tocbibind}
\begin{document} 

\include{command}

 \thispagestyle{empty}
\begin{center}
%
%
%
 \vspace{1.5cm}
\centering{
\huge{\textsf{LIE SYSTEMS, LIE SYMMETRIES \\  \vspace{0.6cm} AND RECIPROCAL TRANSFORMATIONS.}}}

\vspace{1.2cm}
\centering{
\vspace{0.4cm}\Large{\textsf{Departamento de F\'isica Fundamental\\  \'Area F\'isica Te\'orica \\ \vspace{0.2cm} Universidad de Salamanca}}}

\vspace{1.2cm}
\centering{\includegraphics[scale=0.7]{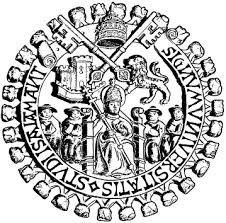}}

\vspace{1.5cm}{MEMORIA PARA OPTAR AL T\'ITULO DE DOCTOR}

\vspace{2.0cm}

\vspace{0.7cm}\Large{\textbf{Cristina Sard\'on Mu{\~n}oz}}\\
\vspace{0.2cm}{PhD Thesis, 2015}

\end{center}
%
%
%
%
%
%
%
%
%
%
\newpage

 \thispagestyle{empty} 
 \vspace{1.5cm}

{\bf D{\~n}a. Pilar Garc\'ia Est\'evez}, Catedr\'atica de la Universidad de Salamanca y {\bf D. Javier de Lucas Ara\'ujo}, ayudante Doctor de
la Universidad de Varsovia,
\vspace{1.5cm}
\begin{center}
\centering{CERTIFICAN:}
\end{center}
\vspace{1.5cm}
 
Que el trabajo de investigaci\'on que se recoge en la siguiente memoria titulado ``Sistemas de Lie, simetr\'ias de Lie y transformaciones rec\'iprocas'',
presentada por {\bf D{\~n}a. Cristina Sard\'on Mu{\~n}oz} para optar al t\'itulo de doctor y la Menci\'on de ``Doctorado Internacional'', 
ha sido realizada en su totalidad bajo su direcci\'on y autorizan su presentaci\'on.
\vspace{1.5cm}
\begin{center}
\centering{Abril, 2015}
\end{center}
\vspace{1.5cm}
\vspace{1.0cm}
 
\begin{flushleft}
D{\~n}a. PILAR GARC\'IA EST\'EVEZ\\
Catedr\'atica \\
{Universidad de Salamanca}
\end{flushleft}

 \vspace{1.0cm}
\vspace{0.5cm}

\begin{flushleft}
D. JAVIER DE LUCAS ARA\'UJO\\
Ayudante Doctor \\
{Universidad de Varsovia}
\end{flushleft}

\newpage

\thispagestyle{empty}
\phantom{hola!}
\vspace{3cm}
\begin{flushright}
My spine just squinted and my eye is weak\\
I can't find peace in a form of speech\\
My views change color red hot, ice cold\\
Black ain't a color, happiness ain't gold\\
It's hard for me to feel normal easy to feel free\\
It's hard for you to understand if you can't feel me\\
I'm a municipal and the mundane\\
I try to keep weird keep away from the same\\
I haven't read an outline on how to pass youth\\
Rather than get passed, I pass doobs\\
I'm out, I'm in, it's hard to live in this given culture\\
All the hers and hymns\\
All of the rats, snakes, and youth vultures around\\
My heart's out of shape and my head's in a cast\\
I am tired, I am weary\\
I could sleep for a thousand years\\
A thousand dreams that would awake me\\
different colors, made of tears.\\
\medskip
            {\it -The Growlers and The Velvet Underground-}
\end{flushright}

\newpage

\thispagestyle{empty}

\frontmatter

  \pagenumbering{roman} 
\pagestyle{myheadings}



\chapter*{\bf La esencia de todo arte bello, es la gratitud
\newline
{\small \it -Friederich Nietzsche}}

Cuando expresamos nuestra gratitud, no debemos olvidar que la mayor apreciaci\'on no va en las palabras, sino en las acciones, parafraseando a  J.F. Kennedy.
Las personas han sido muchas, las acciones fueron m\'as. 
Intentar hacer un resumen de cuatro a{\~n}os de trabajo y apoyo en un simple folio, son incompatibles. Sin embargo, en un breve y futil intento, siguiendo un orden cronol\'ogico,
me dispongo a agradecer debidamente:

A la Universidad de Salamanca, por cubrir econ\'onicamente este trabajo y a Pilar Garc\'ia Est\'evez por aceptarme como alumna.
Al Departamento de F\'isica Fundamental y todos sus integrantes, por la grata acogida y hacerme sentir parte de ellos. En especial,
a Jose Mar\'ia Cerver\'o por sus sabias aportaciones en la investigaci\'on asociada a esta tesis.

A mis dos directores de tesis, Pilar Garc\'ia Est\'evez y Javier de Lucas Ara\'ujo por depositar su confianza en m{\'i} como futura investigadora y 
tener la paciencia suficiente para formar a una novata, muy despistada. La discusi\'on cient\'ifica, las
explicaciones, las demostraciones y consejos en temas distendidos y/o personales, han sido de lo m\'as necesarios y bienvenidos.
Pero sobre todo, es de digno agradecimiento final, el esfuerzo final. La elaboraci\'on de este manuscrito y correcciones.

A la C\'atedra de M\'etodos Matem\'aticos en F\'isica de la Universidad de Varsovia, al IMPAN (Instituto de Matem\'aticas de la Academia 
Polaca de las Ciencias) y al Polish National Science Center, por todo tipo de acondicionamiento y apoyo econ\'omico para realizar mi estancia
internacional. En particular, querr\'ia agradecer la financiaci\'on del proyecto HARMONIA\footnote{DEC-2012/04/M/ST1/00523} y al Profesor Grabowski por la aceptaci\'on de mi colaboraci\'on y resolver la burocracia pertinente.

A la Universidad Roma Tr\'e en Roma, por aceptarme en una estancia breve y gratificante para la discusi\'on con los profesores de mi m\'as suma
admiraci\'on, Decio Levi, Orlando Ragnisco y Pavel Winternitz, con quienes espero mantener un estrecho contacto cient\'ifico de ahora en adelante. 
A Pep\'in Cari{\~n}ena, por transmitirme sus conocimientos y contribuir a mi formaci\'on en mis estancias en la Universidad de Zaragoza.
A la red de Geometr\'ia, Mec\'anica y Control por permitirme ser miembro de ella y sufragar mis gastos en viajes nacionales, que han contribu\'ido
muy provechosamente durante mi doctorado, permiti\'endome conocer a un gran n\'umero de cient\'ificos de renombre internacional.

A \'Angel Ballesteros por invitarme a la Universidad de Burgos y a Alfonso Blasco por trabajar conmigo. A mi colega Francisco Jos\'e Herranz, colega en sentido cient\'ifico, porque
en el plano personal le considero un gran amigo, que me ha servido de consejero, apoyo incondicional y el mejor compa{\~n}ero de viajes internacionales.
Adem\'as de lo obvio, por transmitirme sus conocimientos y ense{\~n}arme el perfeccionismo que pone en la investigaci\'on.
A Juan Domingo Lejarreta, por su metodismo y meticulosidad en sus c\'alculos, ayud\'andome en la realizaci\'on y comprobaci\'on de los m\'ios.
A Silvia Vilari{\~n}o, porque es admirable en su trabajo y ofrece verdadera confianza y profesionalidad y me ha ense{\~n}ado
a dar una buena charla y en la organizaci\'on de mi desorden matem\'atico.

A Cuchi, que ha contribuido en la estructuraci\'on y bonita presentaci\'on de esta tesis. Adem\'as de nuestra amistad.
A los marsopas \'Alvaro, Cuchi, Teresa, Alberto y Edu, porque no s\'olo de ciencia vive el hombre,
nos pegamos nuestras juergas.
A Lorena, estimada amiga y compa{\~n}ena de promoci\'on, de alegr\'ias y penas, de exploraci\'on internacional, cuya amistad y su profesionalidad investigadora
no han sido para m\'i sino el acicate para intentar conseguir parecerme y mantenerme junto a ella. A Bea, estimada amiga y compa{\~n}era de joven adolescencia,
de instituto, de promoci\'on, de alegr\'ias y penas, de exploraci\'on internacional, cuya amistad, profesionalidad y buen humor son el incentivo de j\'ubilo
y buenos ratos en mi vida, si me encuentro con ella.
A Rafa, Aser y Luis, porque son mis mejores amigos. Encuentro en ellos la mayor afinidad de cualquier \'indole.
Porque me han ense{\~n}ado que la vida puede ser simple pero insisto en hacerla complicada. Un diez por ciento es lo que nos ocurre,
el noventa es c\'omo actuamos. La despreocupaci\'on ha contribuido muy positivamente estos \'ultimos cuatro a{\~n}os. Y que lo pasamos de muerte.

Finalmente, me gustar\'ia agradecer a mis queridos padres el haberme acompa{\~n}ado siempre. Hoy me siento dichosa mujer, que gracias a ellos, no he sentido
nunca la soledad en este mundo. Que su ejemplo de sacrificio me ha servido en esforzarme en la finalizaci\'on de un doctorado y que me recuerden mi val\'ia,
en per\'iodos de privaci\'on de mi ocio y su comprensi\'on en la ansiedad, han sido ineludibles en esta etapa.

A todos vosotros, gracias de todo coraz\'on.


        \tableofcontents

\mainmatter
\pagestyle{fancy}

\chapter{Introduction}\markboth{Introduction}{Chapter 1}

\section{The notion of integrability}
\setcounter{equation}{0}
\setcounter{theorem}{0}
\setcounter{example}{0}

On a fairly imprecise, first approximation, the universal definition of integrability is understood as {\it the exact solvability or regular behavior
of solutions} of a system.
Nevertheless, in Mathematics and Physics, there are various distinct notions of integrable systems \cite{AbloClark,Abra,A89,Hoppe,Landau,Olver}.
The characterization and unified definition of integrable systems are two nontrivial matters. The study of integrability
from several different perspectives has led to an apparent fragmented notion of them.
The aim of this introduction is to give a comprehensive account of the variety of approaches to such an important and difficult concept as that of {\it integrability}.
In particular, we focus on those approaches applicable to mathematical and physical models described through ordinary differential equations (ODEs)
and partial differential equations (PDEs).

The field of integrable systems appeared with Classical Mechanics with a quest for exact solutions to Newton's equations of motion \cite{Babelon}.
Integrable systems usually present regular behavior of solutions or {\it conserved quantities} in time, such as energy, angular 
momentum, etc. Indeed, some of such systems present an infinite number of conserved quantities \cite{Babelon}.
But having a big amount of conserved quantities is more of an exception rather than a rule. 

Then, we are in need of a mathematically rigorous, constructive approach based on the study of algebraic or analytic
structures related to integrable systems. {The requirement of existence of such structure could be taken as the notion of integrability} \cite{Mikhailov}.


In the general theory of differential equations, we can endow different definitions and interpretations of integrability.

\subsection{Integrability of dynamical systems}

From a very geometrical point of view, dynamical systems are formulated based on their underlying geometric or algebraic structure.
Differential equations can be interpreted in terms of a system of differential one-forms, or a {\it Pfaffian system}. 
The idea is to see how these differential one-forms restrinct to a submanifold and how this
restriction is compatible with the exterior derivative.

Given a collection of one-forms, let us call them $\omega_i$ over a general manifold $N$, an {\it integral submanifold} is a submanifold $M$ whose tangent space
at every point is annihilated by each one-form at every point $p\in M$. We call a {\it maximal integral manifold} a submanifold $M$ 
such that the kernel of the restriction $\iota^{*}:\Omega_p^{(1)}(N)\rightarrow \Omega_p^{(1)}(M)$, with $\iota:M\hookrightarrow N$, is spanned by the one-forms $\omega_i$ of the Pfaffian system in 
every point of $M$. Additionally, the one-forms are linearly independent. 
A Pfaffian system is said to be {\it completely integrable}
if $N$ admits a foliation by maximal integral manifolds (the foliation need not be regular, the leaves of the foliation might not
be embedded submanifolds). The integrability condition is a condition on these one-forms that guarantees that there will be integral
submanifolds of sufficiently high dimension. 
In these terms, the {\it Fr\"obenius theorem} states the integrability condition for such systems:
if the ideal algebraically generated by the collection of one-forms is closed under exterior differentiation, then, the system admits
a foliation by maximal integral manifolds \cite{Abra,Fro}.

In the context of differentiable dynamical systems, the notion of integrability refers to {\it Liouville integrability} \cite{Liouville}, or the existence of invariant, regular foliations, i.e., ones whose leaves are embedded submanifolds of the smallest possible dimension that are invariant under the flow.

\subsection{Integrability of Hamiltonian systems}

In the case of integrable Hamiltonian systems with an associated Hamiltonian function, there exists a regular foliation of the phase space by invariant manifolds, such that
the Hamiltonian vector fields associated with the invariant manifolds of the foliation, span the tangent distribution.
Similarly, we can say that there exists a maximal set of invariants that commute with the Hamiltonian function $H$ under the Poisson bracket $\{\cdot,\cdot\}$, that
is a Lie bracket satisfying the Leibnitz rule
\begin{equation}
\{H,I\}=0,
\end{equation}
for a certain number of first-integrals $I$.
\begin{itemize}
\item If the space is symplectic, it has an even dimension $2n$ and the maximal number of independent Poisson commuting invariants (including H) is $n$.
\item When the number of independent Poisson commuting invariants is less than $n$, we say that the system is {\it partially integrable} \cite{Abra,A89}.
\item When there exist further
functionally independent invariants, beyond the maximal number that can be Poisson commuting, it is {\it superintegrable} \cite{Abra,A89}. This is equivalent to saying that
the dimension of the leaves of the invariant manifold is less than $n$. If there is a regular foliation with one-dimensional leaves (curves), this is called {\it maximally superintegrable} \cite{Abra,Fasso,Goldstein}.
\end{itemize}
When the Hamiltonian system is completely integrable in the Liouville sense and the level sets are compact, the flows are complete and the leaves of the invariant
foliation are tori. Then, there exist local canonical coordinates on the phase space known as {\it action-angle} variables \cite{Landau}, such that the invariant tori are the joint level set of the action variables.
The Lioviulle integrability provides a complete set of invariants of the Hamiltonian that are constants of motion. The angle variables are periodic and the motion on the tori is linear in these coordinates.

\section{Methods to guarantee integrability}

\setcounter{equation}{0}
\setcounter{theorem}{0}
\setcounter{example}{0}

In 1960, the discovery of {\it solitons} as strongly stable, localized solutions of certain PDEs could be understood by viewing these equations as infinite-dimensional
Hamiltonian systems. The study of soliton solutions led to multiple methods for solving PDEs. A fundamental one is the {\it inverse scattering method/transform} (IST) \cite{AbloClark,ARS2,AS}, which appeared as a result of the derivation
of Lax pairs \cite{Lax,Lax2}.

\subsection{Lax pairs and the IST}

A {\it Lax pair} (LP) or {\it spectral problem} is a pair of linear operators $L(t)$ and $P(t)$,
acting on a fixed Hilbert space $\mathcal{H}$, that
satisfy a corresponding differential equation, the so called {\it Lax equation}
\begin{equation}\label{Laxeq}
\frac{dL}{dt}=[P,L],
\end{equation}
where $[P,L]=PL-LP.$
The operator $L(t)$ is said to be {\it isospectral} if its spectrum of eigenvalues is independent of the evolution variable.
We call {\it eigenvalue problem} the relation
\begin{equation}\label{eigenpr}
L\psi=\lambda\psi,
\end{equation}
where $\psi\in \mathcal{H}$, henceforth called a {\it spectral function or eigenfunction}, and $\lambda$ is a {\it spectral value or eigenvalue}.

\noindent
The core observation is that there exists a unitary operator $u(s,t)$ such that
\begin{equation}
L(t)=u(s,t)L(s)u^{-1}(s,t),
\end{equation}
where $u(s,t)$ is the general solution of the Cauchy problem
\begin{equation}
\frac{du(s,t)}{dt}=P(t)u(s,t),\quad u(s,s)=1.
\end{equation}
Notice that $L(t)$ is self-adjoint and $P(t)$ is skew-adjoint.
In other words, to solve the eigenvalue problem \eqref{eigenpr} we can use the value $L(0)$
and for obtaining the solution at time $t$, we use the following condition and ODE
\begin{align}\label{eigenpr2}
\lambda(t)&=\lambda(0),\nonumber\\
\frac{\partial \psi}{\partial t}&=P\psi.
\end{align}

Lax pairs are interesting because they guarantee the integrability of certain differential equations \cite{AbloClark,AS}. 
Generally, nonlinear PDEs that are integrable, can equivalently be rewritten as the compatilibity condition \eqref{Laxeq} of a spectral problem \cite{EstGordoa,EstPrada}.
Sometimes, it is easier to solve the equations provided by the LP rather than the nonlinear problem. 
In this way, the IST is based on the aforementioned properties \cite{AbloClark}.

The {\it inverse scattering transform} is a method for solving certain nonlinear PDEs with an associated LP.
The IST guarantees the existence of analytical solutions of the PDE (when it can be applied).
The name inverse transform comes from the idea of recovering the time evolution of the potential $u(x,t)$ from the time evolution of its scattering data,
opposed to the direct scattering which finds the scattering matrix from the evolution of the potential.
This method has been applied to many exactly solvable models or completely integrable infinite-dimensional systems as: the Korteweg de Vries
equation \cite{KdV}, to be soon extended to the nonlinear Schr\"odinger equation, Sine-Gordon equation, etc., \cite{EstPrada1,ZakharovManakov}. It was first introduced by Gardner, Greene, Krustal, Miura {\it et al}
in 1967 \cite{GGKM1,GGKM2} and it can be summarized in a number of steps, in the case of $1+1$ dimensions. This process was later generalized
to higher dimensions (see \cite{Kono})
\begin{itemize}
\item Consider $L$ and $P$ acting
on $\mathcal{H}$, where $L$ depends on an unknown function $u(x,t)$ and $P$ is independent of it in the scattering
region.
\item We can compute the spectrum of eigenvalues $\lambda$ for $L(0)$ and obtain $\psi(x,0)$.
\item If $P$ is known, we can propagate the eigenfunction with the equation $\frac{\partial \psi}{\partial t}(x,t)=P\psi(x,t)$ with initial condition $\psi(x,0)$.
\item Knowing $\psi(x,t)$ in the scattering region, we construct $L(t)$ and reconstruct $u(x,t)$ by means of the {\it Gelfand--Levitan--Marchenko equation} (GLM) \cite{Mar}.
\end{itemize} 

\newpage

The following diagram shows the procedure


\bigskip

\bigskip

\bigskip

$
\xymatrix{*+<1cm>[F-,]{\text{Initial potential},\ u(x,t=0)} \ar[rrr]^{\textit{scattering data}} \ar@2{<->}[d]^{\textit{time difference}} &  &  &*+<1cm>[F-,]{\text{Spectrum}\ L(0),\ \psi(x,t=0)}\ar[d]^{\textit{scattering data}\ t>0}\\  *+<1cm>[F-,]{\text{Potential at time t},\ u(x,t)} &  &  &*+<1cm>[F-,]{ d\psi/dt=P\psi,\ \textit{i.c.}\ \psi(x,t=0) }\ar[lll]_(0.5){\textit{Inverse scattering data}}^{\textit{Reconstruct}\ L(t), t>0}}
$

\bigskip

\bigskip

\bigskip

\medskip

\subsection{The Hirota bilinear method}

Another method guaranteeing the integrability of a nonlinear PDE, together with the IST, is the {\it Hirota's bilinear method} (HBM) \cite{hirota}. The major
advantage of the HBM over the IST is the obtainance of possible {\it multi-soliton solutions} \cite{hirota,hirota2,Meng} by imposing Ans\"atze.
Additionally, the HBM can be applied to a greater number of equations and
it is solved algebraically instead of relying on  nontrivial analysis, as the IST \cite{Hietarinta1,Hietarinta2}.
Most PDEs are not initially expressed in a tractable format for subsequent mathematical analysis. Hirota noticed that the best dependent variables for constructing
soliton solutions are those in which the soliton appears as a finite number of exponentials. To apply this method it is necessary that the equation is quadratic and that the
derivatives can be expressed using Hirota's {\it D-operator} \cite{Hirota3} defined by
\begin{equation*}
D_x^n f\cdot g=(\partial_{x_1}-\partial_{x_2})^n f(x_1)g(x_2)\Big|_{x_2=x_1=x}.
\end{equation*}
Unfortunately, the process of bilinearization is far from being algorithmic, and it is hard to know how many variables are needed for bilinearization.

\subsection{The Painlev\'e test}

To discern whether our equation is integrable or not, involves a deep inspection of its geometrical properties and aforementioned methods. 
Now, a question arises: Is there any algorithmical method to check the integrability of a differential equation?
The answer is affirmative.

The Painlev\'e test establishes an algorithmical integrability criteria for differential equations \cite{Conte1,EstPrada,WeiTaborCarn}. It focuses on the {\it singularity analysis} of the differential
equation, attending to a fundamental property: being a {\it fixed} or a {\it movable singularity}, or a singularity
not depending or depending on the initial conditions, respectively.

These concepts were initiated by Fuchs and Sophia Kovalevskaya \cite{RogerCooke}. The latter centered herself in the study of equations for solid rigid
dynamics: singularities and properties of single-valuedness of poles of PDEs on the complex plane, etc. Eventually, she
expanded her results to other physical systems \cite{McLeodOlver}.

Let us illustrate the concept of fixed and movable singularity. Consider a manifold $N$ locally diffeomorphic to $\mathbb{R}\times {\rm T}\mathbb{R}$,
with local coordinates $\{t,u(t),u_t\}$. Consider the differential equation
\begin{equation}
(t-c)u_t=bu
\end{equation}
and $c,b\in \mathbb{R}$.
Its general solution reads 
\begin{equation}
u(t)=k_0(t-c)^b,
\end{equation}
where $k_0$ is a constant of integration.
Depending on the value of the exponent $b$, we have different types of singularities
\begin{itemize}
\item  If $b$ is a positive integer, then, $u(t)$ is a {\it holomorphic function}. 
\item If $b$ is a negative integer, then $c$ is a {\it pole singularity}.
\item In case of $b$ rational, $c$ is a {\it branch point}. 
\end{itemize}
Nevertheless, the singularity $t=c$ does not
depend on initial conditions.
We say that the singularity is {\it fixed}.

Let us now consider an ODE on $\mathbb{R}\times {\rm T}^2\mathbb{R}$ with local coordinates $\{t,u,u_t,u_{tt}\}$, which reads
\begin{equation}
buu_{tt}+(1-b)u_t^2=0,
\end{equation}
with $b\in \mathbb{R}$. The general solution to this equation is
\begin{equation}
u(t)=k_0(t-t_0)^b.
\end{equation}
If $b$ is a negative integer, the singularity $t=t_0$ is a
singularity that depends on the initial conditions through the constant of integration $t_0$.
In this case, we say that the singulary is {\it movable}.

Painlev\'e, Gambier {\it et al} oriented their study towards second-order differential equations. 
In particular, Painlev\'e focused on differential equations on $\mathbb{R}\times {\rm T}^2\mathbb{R}$ with local coordinates $\{t,u,u_t,u_{tt}\}$, of the type
\begin{equation}\label{pain2or}
u_{tt}=F(t,u,u_t),
\end{equation}
where $F$ is a rational function in $u,u_t$ and analytic in $t$.
He found that there were 50 different equations of type \eqref{pain2or} whose unique movable singularities were poles \cite{Pain1,Pain2}. Out of the 50 types, 44 were integrated in
terms of known functions as Riccati, elliptic, linear, etc., and the 6 remaining, although having meromorphic solutions, 
they do not possess algebraic integrals that permit us to reduce them by quadratures.
These 6 functions are the nonlinear ues of special functions and their solutions are called {\it Painlev\'e transcendents} $(P_{I}-P_{VI})$, because
they cannot be expressed in terms of elementary or rational functions or solutions expressible in terms of special functions \cite{CHJ}.

From here, we can establish the {\it Painlev\'e Property} (PP) for an algebraic ODE. We say that an {ODE has the PP if all the movable singularities of its solution are poles}. In other words,
if the solution is single-valued anywhere but in the fixed singularities of the coefficients.
The PP does not apply to other type of singularities as branch points, essential singularities, etc.

To establish the nature of the singularities and a criteria of integrability, we can make use of the {\it Painlev\'e test} (PT) for ODEs \cite{WeiTaborCarn}.
Given a general ODE on $\mathbb{R}\times {\rm T}^p\mathbb{R}$ with local coordinates $\{t,u,u_t,\dots,u_{\!\!\!\KeepStyleUnderBrace{ t,\dots,t}_{p-\text{times}}}\}$,
\begin{equation}\label{overtpr}
F=F(t,u(t),\dots,u_{\!\!\!\KeepStyleUnderBrace{ t,\dots,t}_{p-\text{times}}}),
\end{equation}
the PT analyzes local properties by proposing solutions in the form
\begin{equation}\label{painleveconj}
u(t)=\sum_{j=0}^{\infty}{a_j(t-t_0)^{(j-\alpha)}},
\end{equation}
where $t_0$ is the singularity, $a_j, \forall j$ are constants and $\alpha$ is necessarily a positive integer. If \eqref{painleveconj} is a solution of an ODE, then, the ODE is conjectured
integrable. To prove this, we have to follow a number of steps
\begin{enumerate}
\item We first need to determine the value of $\alpha$ by balance of dominant terms \cite{Steeb}, which will permit us to
obtain $a_0$, simultaneously. The values of $\alpha$ and $a_0$ are not necessarily unique, and $\alpha$ must be a positive integer. If there is more than one possibility, we say that there are branches
of expansion and the forthcoming steps need to be satisfied for every possible value of $\alpha$.
\item Having introduced \eqref{painleveconj} into the differential equation \eqref{overtpr}, we obtain a relation of recurrence for the
rest of coefficients $a_j$ that can be written as
\begin{equation}\label{rescond}
(j-\beta_1)\cdot \dots \cdot (j-\beta_n)a_j=F_j(t,\dots,u_k,(u_{k})_t,\dots),\quad k<j,
\end{equation}
which arises from setting equal to zero different orders in $(t-t_0)$. This gives us $a_j$ in terms of $a_k$ for $k<j$.
Observe that when $j=\beta_l$ with $l=1,\dots,n,$ the left-hand side of the equation is null and the associated $a_{\beta_l}$
is arbitrary.
Those values of $j$, are called {\it resonances} and the equation \eqref{rescond} turns into a relation for $a_k$ for $k<\beta_{l}$
which is known as the {\it resonance condition}. Considering that the number of arbitrary constants of motion that an ODE admits is equal to
its order, we must find the order of the equation minus one, as $t_0$ is already one of the arbitrary constants.
\item Make sure that the resonance conditions are satisfied identically, $F_j=0$ for every $j=\beta_{l}$. If so, we say that the ODE
posesses the PP and that it is integrable in the Painlev\'e sense. The resonances have to be positive except $j=-1$,
which is associated with the arbitrariness of $t_0$.
\item For systems of ODEs with $k$ dependent variables $u_1,\dots,u_k$, this procedure needs the addition of expansions for every $u_i$,
\begin{equation}
u_i(t)=\sum_{j=0}^{\infty}{a^{i}_j(t-t_0)^{(j-\gamma)}},\quad \forall i=1,\dots,k.
\end{equation}
\end{enumerate}

\noindent
In the case of PDEs, Weiss, Tabor and Carnevale carried out the generalization of the Painleve method, the so called {\it WTC method} \cite{WeiTaborCarn}.

Consider a system of $q$ partial differential equations of $p$-order, defined over a general manifold with local coordinates 
\begin{equation}
\left(x_i,u_j,(u_j)_{x_i},(u_j)_{x_{i_1}^{j_1},x_{i_2}^{j_2}},\dots, (u_j)_{x_{i_1}^{j_1},x_{i_2}^{j_2},\dots,x_{i_n}^{j_n}}\right)
\end{equation}
which reads
\begin{equation}\label{genpdeintro}
\Psi^l=\Psi^l\left(x_i,u_j,(u_j)_{x_i},(u_j)_{x_{i_1}^{j_1},x_{i_2}^{j_2}},\dots,(u_j)_{x_{i_1}^{j_1},x_{i_2}^{j_2},x_{i_3}^{j_3},\dots,x_{i_n}^{j_n}}\right),\qquad l=1,\ldots q,
\end{equation}
such that $j_1+\dots+j_n=p$, being $p$ the highest order of the equation and each $j_k$ indicates the number of appearances of each variable $x_{i_k}$.
The indices run such that $1\leq i_1,\dots,i_n\leq n$ and $x_1\leq x_i\leq x_n$, $j=1,\dots,k$. 

{\it The Ablowitz-Ramani-Segur conjecture} (ARS) says
that a PDE is integrable in the Painlev\'e sense \cite{Pain2}, if all of its reductions have the Painlev\'e property \cite{Weiss1}.
This is equivalent to saying that if a PDE is reducible
to an ODE that is not one of the {Painlev\'e transcendents} \cite{Fokas}, then, the PDE is not completely integrable \cite{ARS2}.
In other words, a PDE enjoys the PP if $u(t)$ is single-valued in the movable singularity manifold $\phi$ and $\alpha>0$ and integer.
McLeod and Olver \cite{McLeodOlver} tested a weaker version of this conjecture. No rigorous proof was given, but the extended study of its reduced versions suggests the
validity as a method to prove integrability.

Similarly, we can extend the Painlev\'e test to PDEs by substituting the function $(t-t_0)$ in \eqref{painleveconj} by
an arbitrary function $\phi(x_i)$ for all $i=1,\dots,n$, which receives the name of {\it movable singularity manifold}.
We propose a Laurent expansion series
\begin{equation}\label{Arsconj}
u_l(x_i)=\sum_{j=1}^n u^{(l)}_j(x_i)\phi(x_i)^{j-\alpha},\quad \forall x_i,\quad i=1,\dots,n,\quad \forall l=1,\dots,k,
\end{equation}
which incorporates $u_{l}(x_i)$ as functions of the coordinates $x_i$, instead of being constants as $a_j$ in \eqref{painleveconj}.
Steps 1--3  for ODEs can be reenacted in order to prove the integrability of a PDE. It is significant to notice that the resonance condition
and recursion equation \eqref{rescond} have now more general expressions as
\begin{equation}\label{rescondpde}
(j-\beta_1)\cdot \dots \cdot (j-\beta_n)u_j(x_i)=F_j(\phi_{x_i},\dots,u_k,(u_{k})_{x_i},\dots),
\end{equation}
for $k<j,\quad \forall i=1,\dots,n.$
It is important to mention that the PT is not invariant under changes of coordinates. This means that an equation
can be integrable in the Painlev\'e sense in certain variables, but not when expressed in others, i.e., the PT is not intrisecally a geometrical property.
There exist certain changes of coordinates to check the Painlev\'e integrability when the equation does not possess the PP
in its initial ones. In particular, we deal with a change of coordinates, the so called {\it reciprocal transformations} \cite{CourantFried,Est1,rogers3}.

\subsection{The singular manifold method}

The {\it singular manifold method} (SMM) focuses on solutions which arise from truncated series of the generalized PP method \eqref{Arsconj}. We require the solutions
of the PDE written in the form \eqref{Arsconj} to select the truncated terms \cite{EstGordoa}
\begin{equation}\label{smm}
u_l(x_i)\simeq u^{(l)}_0(x_i)\phi(x_i)^{-\alpha}+u^{(l)}_1(x_i)\phi(x_i)^{1-\alpha}+\dots+u^{(l)}_{\alpha}(x_i),
\end{equation}
for every $l$. In the case of several branches of expansion, this truncation needs to be formulated for every value of $\alpha$.
Here, the function $\phi(x_i)$ is no longer arbitrary, but a {\it singular manifold} whose expression arises from \eqref{smm} being a truncated
solution of the PDE under study.

The SMM is interesting because it contributes substantially in the derivation of a Lax pair. This procedure can be devised in \cite{EstPrada1,EstPrada,EstPrada2}.

\medskip

Some of the described methods, apart from guaranteeing the integrability of a differential equation, provide us with possible solutions. In the following chapters, we shall focus
on three primordial ones. Lie systems, Lie symmetries and Lax pairs.

\subsection{Lie systems}

\setcounter{equation}{0}
\setcounter{theorem}{0}
\setcounter{example}{0}

In short, a Lie system is a system of ODEs describing the integral curves of a $t$-dependent vector field taking values in a finite-dimensional 
Lie algebra $V$ of vector fields: a so called {\it Vessiot--Guldberg Lie algebra} (VG) for the Lie system \cite{CGM00,CGM07,Dissertationes,CLS122,PW}.

Equivalently, a Lie system is a system of ODEs admitting a {\it superposition principle} or {\it superposition rule}, i.e., a map
allowing us to express the general solution of the system of ODEs in terms of a family of particular solutions and a set of constants 
related to initial conditions \cite{CGM00,CGM07,Dissertationes,LS,PW}. These superposition principles are, in general, nonlinear. Among other reasons, superposition rules are interesting because they allow us to integrate systems of ODEs from the knowledge of a small set of particular solutions \cite{AHW81,AHW,Dissertationes}. 

Lie systems enjoy a plethora of mathematical properties. Geometrically, the {\it Lie--Scheffers theorem} states that a Lie system amounts to a curve within a Lie algebra of vector fields. Additionally, Cari\~nena, Grabowski and Marmo proven that superposition rules can be understood as a certain type of projective foliation on an appropriate bundle \cite{CGM07}. 
This idea gave rise to a geometric method for determining superposition rules that has been employed, along with the new techniques described in the thesis, to obtain superposition rules in an algebraic and geometric manner for certain types of Lie systems admitting compatible geometric structures \cite{BCHLS,cgls}.

From the point of view of their applications, Lie systems play a relevant role in Physics, Mathematics, and other fields of research (see \cite{Dissertationes} which provides more than 400 references on Lie systems and related topics). Some of the most representative Lie systems are the Riccati equations and their multiple derived versions (matrix Riccati equations, 
projective Riccati equations, etc.) \cite{HWA83,Lafortune,ORW87}. These latter differential equations frequently appear in Cosmology, Financial Mathematics, Control Theory and other disciplines \cite{cal,Dissertationes,Levin,SoriWinter}. It is also worth noting that Lie systems appear in the study of Wei--Norman equations \cite{CMN98}, Quantum Mechanical problems \cite{CLQuantum,Dissertationes} and Biology \cite{BBHLS}.

Additionally, many other differential equations can be studied through the theory of Lie systems, even though they are not Lie systems.
This is the case of the Kummer--Schwarz \cite{LS12} and  Milne--Pinney equations \cite{RSW97}, the Ermakov system \cite{SIGMA}, $t$-dependent frequency Winternitz--Smorodinsky oscillators \cite{BCHLS}, Buchdahl equations \cite{HLS}, among others \cite{CLS12,CLS122,LS12}. 

Many of the previously mentioned examples have been discovered by the author of this thesis and her collaborators. We have also added the Hamiltonian version of the second-order Riccati equations \cite{CLS12}, systems of ODEs that appear in the study of diffusion equations \cite{HLS}, new types of Riccati equations on different types of composition algebras, as the complex, quaternions, Study numbers \cite{EstLucasSar}, and many others \cite{HLS}. Other new Lie systems concern viral models \cite{BBHLS}, certain reductions of Yang--Mills equations \cite{EstLucasSar} and complex Bernoulli equations \cite{HLS}.

As a consequence of the new applications of Lie systems, we have significantly enlarged the field of potential applications of such systems. Moreover, most applications of Lie systems on the study of PDEs, e.g., in analyzing flat $\mathfrak{g}$-valued connections and partial Riccati equations \cite{EstLucasSar}, are described in this thesis \cite{cgls,EstLucasSar}.
Due to the special geometric structure of Lie systems, our findings open a new research approach to the study of the above mentioned differential equations and their related problems.

Very surprisingly, it was proven that the found new Lie systems admitted Vessiot--Guldberg algebras of Hamiltonian vector fields
with respect to some symplectic or Poisson structure \cite{CLS122}. This led to the study of an important particular case of
Lie systems, the so called {\it Lie--Hamilton systems}.

Lie--Hamilton systems are Lie systems that admit Vessiot--Guldberg Lie algebras of Hamiltonian vector fields with respect to a Poisson structure \cite{BCHLS,CIMM14,CLS122}. The author and collaborators proven that Lie--Hamilton systems posses a time-dependent Hamiltonian given by a curve
 in a finite-dimensional Lie algebra of functions with respect to a Poisson bracket related with the Poisson structure, a {\it Lie--Hamilton algebra} \cite{BCHLS}. 
For Lie systems, this structure plays the ue role of Hamiltonians in Hamiltonian Mechanics. From this, the writer has found numerous properties relative to the description
of constants of motion, momentum maps and Lie symmetries of Lie--Hamilton systems \cite{CLS122}. 

Among the developed methods for Lie--Hamilton systems, we consider of upmost importance the computation of
superposition principles through Poisson coalgebras \cite{BBHMR09,BBR06,BCR96,BHR08,BMO02}. The traditional method for the
computation of superposition principles for Lie systems relies in the integration of systems of ordinary or partial differential equations \cite{AHW81,CGM00,CGM07,PW}.
Nevertheless, we can obtain such principles for Lie--Hamilton systems by methods of algebraic and geometric nature \cite{BCHLS}.

Let us explain the last statement more precisely. Every Lie--Hamilton system induces a Lie--Hamilton algebra.
We construct a Poisson algebra of polynomials in the elements of a basis of the Lie--Hamilton algebra. To obtain superposition rules for a Lie--Hamilton system, we have to define the so called diagonal prolongations of the Lie--Hamilton system, which are again Lie--Hamilton systems \cite{BCHLS}. Hence, the diagonal prolongations are endowed with Poisson algebras in the same way. The point is that there exists a primitive coproduct passing from the Poisson algebra of the initial Lie--Hamilton system to the Poisson algebras of the successive diagonal prolongations, giving rise to a Poisson coalgebra structure \cite{BBHMR09,BBR06,BCR96,BHR08}.
The coproduct  and the Casimir elements of the Lie--Hamilton algebra of a Lie--Hamilton system enable us to obtain conserved quantities and Lie symmetries of the Lie--Hamilton system and their diagonal prolongations. 
Such constants of motion are employed to obtain superposition principles \cite{BCHLS}.
From now on, we call this procedure the {\it coalgebra method} for obtaining superposition principles. 

The coalgebra method has been developed by the author of this thesis and her collaborators in an original way, 
and they have shown its efficiency \cite{BCHLS}, by avoiding the long integration of systems of ODEs or PDEs
arising from traditional methods \cite{CGM07,Dissertationes,PW}. Moreover, it allows us to provide geometrical and algebraic interpretations of superposition rules.
A very relevant outcome has been the achievement of the superposition principle for the Riccati equations
 describing that the constant of motion that leads to such  a principle is basically given by the image, via a coproduct, of a Casimir function for $\mathfrak{sl}(2,\mathbb{R})$ \cite{BCHLS}.

Due to the interest shown in Lie--Hamilton systems, we have classified all the Vessiot--Guldberg 
Lie algebras of Hamiltonian vector fields on the plane with respect to a Poisson structure and have analyzed their properties \cite{BBHLS}.
We obtained twelve different nondiffeomorphic classes of Lie algebras of Hamiltonian vector fields, which completed several details of the previous study carried out by Olver, Kamran and Gonz\'alez \cite{GKP92}.

Our classification permitted the identification of physical and mathematical properties of the Lie--Hamilton systems on the plane \cite{HLS}.
For example, it helped in the understanding of trigonometric Lie systems appearing in the study of integrable systems \cite{ADR12},
diffusion equations \cite{BBHLS,HLS,SSVessiot--Guldberg11,SSVessiot--Guldberg14},  Smorodinsky--Winternitz oscillators \cite{MSVW67}, systems with medical applications \cite{EK05}, Lotka--Volterra systems \cite{BBHLS}, systems with periodic trajectories, etc.

But not every Lie system is a Lie--Hamilton system. The so called {\it no-go theorem} \cite{cgls} shows a very general condition that helps us identify 
when a Lie system is not a Lie--Hamilton one. We have found that numerous Lie systems admit a Vessiot--Guldberg Lie algebra which consists of
 Hamiltonian vector fields with respect to a certain geometric structure different from Poisson structures.
If the vector fields of the Vessiot--Guldberg Lie algebra are Hamiltonian with respect to a Dirac or Jacobi structure,
we say that they are Dirac--Lie \cite{cgls} or Jacobi--Lie \cite{HLS2} systems, correspondingly.
Along this thesis, we will give numerous new examples of Lie systems over different geometries, which do not only have great importance
from the mathematical point of view, but actually modelize the Physics of nature \cite{BCHLS,cgls,HLS2,LS12}.

A notorious example of Lie system compatible with another geometrical structure is that of Dirac--Lie systems \cite{cgls}.
These are Lie systems that possess a Vessiot--Guldberg Lie algebra of Hamiltonian vector fields with respect to a Dirac structure.
As Dirac structures describe Poisson structures as particular cases, Dirac--Lie systems cover Lie--Hamilton systems. 

In this thesis we show how Dirac--Lie systems describe, through similar techniques to those of the Poisson geometry,
 systems which cannot be described through Lie--Hamilton systems. For example, the third-order Kummer--Schwarz equations and certain systems appearing in the
 study of diffusion equations \cite{cgls,SSVessiot--Guldberg11}.
Also, some generalizations of the results for Lie--Hamilton systems were applied to the realm of Dirac--Lie systems. Moreover, it is interesting that
Dirac structures give rise to a particular type of Lie algebroid structures \cite{Marle08}. As a consequence, Lie algebroid structures can also be employed to study Dirac--Lie systems \cite{cgls}. Given the recurrent use of Lie algebroid structures in Physics is much discussed, we expect to continue this line of research in the future to analyze its relevance.  

The last type of geometry that has been employed in this thesis is the Jacobi geometry \cite{HLS2}. 
A Jacobi manifold is another generalization of a Poisson manifold.  We have defined and studied the denominated {\it Jacobi--Lie systems}, namely Lie systems admitting a Vessiot--Guldberg Lie algebra of Hamiltonian vector fields with respect to a Jacobi manifold. 
We generalize the theory of Lie--Hamilton systems to the realm of Jacobi--Lie systems and we detail a classification of Jacobi--Lie systems on the line and on the plane (see Table \ref{table9} in Appendix 1).
This entails the study of the structure of the space of constants of motion for such systems.

\subsection{Lie symmetries}

%

One of the most common techniques for the study of differential equations is the Lie symmetry method \cite{Nucci3}, iniciated by Sophus Lie in the XIX century  \cite{Lie81,Lie90} 
and all its variant versions developed along the last decades.
The Lie symmetry method can be summarized in the determination of a transformation that leaves invariant a group of equations \cite{Olver,PS,Stephani}.
Invariance under a transformation implies the possibility of reducing the number of independent variables by one, for each symmetry.

The classical Lie method for symmetry computation was generalized along the XX century. The development of computers has greatly helped in
its generalization and application to more complicated equations.

New software orientated towards symbolic calculus, for example {\it Maple}, has become very useful
in the treatment of cumbersome intermediate calculation.

An important generalization of the classical Lie method is the nonclassical approach by Bluman and Cole in 1969 \cite{BC1}, Olver and Rosenau \cite{OlverRose,OlverRose2}.
In this case, we look for a particular transformation that leaves invariant a subset of all the possible solutions of the equation.
The proposal of this type of symmetry started with the search of solutions for the heat equation \cite{Evans},
which were not deducible through the classical Lie method. Since then, the nonclassical method was popularized \cite{ArriBeck,ClarkWinter,Nucci1}.
A notorious difference between the classical Lie and nonclassical method is that the later provides us with no longer linear
systems of differential equations for the resolution of the symmetries.
In the last few years, the classical and nonclassical analysis have shown their efficiency in the treatment of equations of hydrodynamic origin,
in the context of Plasma Physics, cosmological models, Fluid Mechanics, etc., \cite{AbloClark,CH1,EstPrada,LNRW,Nucci2,NucciAmes}.

In our research, we find of particular interest the application of the Lie symmetry method (classical and nonclassical) to Lax pairs associated
with the nonlinear differential equations.
The inspection of symmetries of equations has been a very treated topic by lots of authors. Notwithstanding, the symmetries of their associated Lax pairs have been less investigated.

Our aim is to see how the Lax pairs reduce under the symmetry and in the case of nonisospectral problems,
to see if their nonisospectrality condition is propagated to lower dimensions \cite{EstLejaSar,EstLejaSar1,EstLejaSar2}.

\subsection{Reciprocal transformations}

Reciprocal transformations can be suitably used in the field of PDEs. These transformations consist on the mixing of the role of the dependent and independent
variables \cite{h00}, to achieve simpler versions or even linearized versions of the initial, nonlinear PDE \cite{Est1,EstPrada2}.
Reciprocal transformations, by experience of many worked examples, help in the identification of the plethora of PDEs available in the Physics and Mathematics
literature. Two different equations, although seemingly unrelated at
first sight, happen to be equivalent versions of a same equation, after a reciprocal transformation \cite{EstSar2}.
In this way, the big number of integrable equations in the literature, could
be greatly diminished by establishing a method to discern which equations are disguised versions of common problem. 
Then, the next question comes out: {Is there a way to identify different versions of a common nonlinear problem?}

In principle, the only way to ascertain is by proposing different transformations and obtain results by recurrent trial and error. It is desirable to derive
a canonical form by using the explained SMM, but it is still a conjecture to be proven.

%

\newpage 

The content of this thesis has either been published or it is on its way for soon publication. Here we show a list of our results.
\bigskip

{\bf PUBLICATIONS}
\bigskip

\begin{enumerate}

\item{Similarity reductions arising from nonisospectral Camassa Holm hierarchy in 2+1 dimensions,}\\
P.G. Est\'evez, J.D. Lejarreta, C. Sard\'on,\\
{\it J. Nonl. Math. Phys.} \textbf{18}, 9--28 (2011).

\item{A new Lie systems approach to second-order Riccati equations,}\\
J.F. Cari\~nena, J. de Lucas, C. Sard\'on,\\
{\it Int. J. Geom. Methods Mod. Phys.} \textbf{9}, 1260007 (2012).

\item{Miura reciprocal Transformations for hierarchies in 2+1 dimensions,}\\
P.G. Est\'evez, C. Sard\'on, \\
{\it J. Nonl. Math. Phys.} {\bf 20}, 552--564 (2013).

\item{Integrable 1+1 dimensional hierarchies arising from reduction of a non-isospectral problem in 2+1 dimensions,}\\
P.G. Est\'evez, J.D. Lejarreta, C. Sard\'on,\\
{\it Appl. Math. Comput.} {\bf 224}, 311--324 (2013).

\item{On Lie systems and Kummer--Schwarz equations,}\\
J. de Lucas, C. Sard\'on, \\
{\it J. Math. Phys.} {\bf 54}, 033505 (2013).

\item{From constants of motion to superposition rules of Lie--Hamilton systems,}\\
A. Ballesteros, J.F. Cari\~nena, F.J. Herranz,  J. de Lucas, C. Sard\'on,\\
{\it J. Phys. A: Math. Theor.} {\bf 46}, 285203 (2013).

\item{Lie--Hamilton systems: theory and applications,}\\
J.F. Cari\~nena, J. de Lucas, C. Sard\'on,\\
{\it Int. Geom. Methods Mod. Phys.} {\bf 10}, 0912982 (2013).

\item{Dirac--Lie systems and Schwarzian equations,}\\
J.F. Cari\~nena, J. Grabowski, J. de Lucas, C. Sard\'on,\\
{\it J. Differential Equations} {\bf 257}, 2303--2340 (2014).

\item{Lie--Hamilton systems on the plane: theory, classification and applications,}\\
A. Ballesteros, A. Blasco, F.J. Herranz, J. de Lucas, C. Sard\'on,\\
{\it J. Differential Equations} {\bf  258}, 2873--2907 (2015).

\item{Lie symmetries for Lie systems: applications to systems of ODEs and PDEs,}\\
P.G. Est\'evez,  F.J. Herranz, J. de Lucas, C. Sard\'on,\\
submitted to {\it Appl. Math. Comput.}\\
arXiv:1404.2740

\item{Lie--Hamilton systems on the plane: applications and superposition rules,}\\
A. Blasco, F.J. Herranz, J. de Lucas, C. Sard\'on,\\
submitted to {\it J. Phys. A: Math. Theor.}\\
arXiv:1410.7336

\end{enumerate}
\bigskip

{\bf PROCEEDINGS}
\bigskip

\begin{enumerate}
\item {Miura reciprocal transformations for two integrable hierarchies in 1+1 dimensions,}\\
P.G. Estevez, C. Sard\'on,\\
{\it Proceedings GADEIS}, Protaras, Cyprus (2012), \\
arXiv:1301.3636

\item{Jacobi--Lie systems: fundamentals and low-dimensional classification,}\\
F.J. Herranz, J. de Lucas, C. Sard\'on,\\
{\it Proceedings AIMS},  Madrid, Spain (2014),\\
arXiv:1412.0300
\end{enumerate}
\bigskip

{\bf PUBLICATIONS IN PREPARATION}
\bigskip

\begin{enumerate}
\item{Classical and nonclassical approach for a wave model in 2+1 dimensions,}\\
P.G. Est\'evez, J.D. Lejarreta, C. Sard\'on,\\
In preparation.

%

\end{enumerate}

\chapter{Geometric Fundamentals}\markboth{Geometric Fundamentals}{Chapter 2}
\label{Chap:GeomFund}

In short, Differential Geometry is the mathematical discipline generalizing the standard differential and integral calculus on vector spaces to more general spaces, the manifolds, in order to study the geometric properties of the structures defined on them. 

During the XVIII and XIX century, Differential Geometry was mostly concerned with the study of the intrinsic properties of curves and surfaces in $\mathbb{R}^n$, e.g., the {\it Gauss' Egregium Theorem}, and the notion of parallelism \cite{Gauss3,Gauss1,Gauss2}.

The first mature notion of manifold appears in Riemann's habilitation thesis in 1854 \cite{Riemann2}. Riemann defined the term {\it manifold}, coming from the german {\it Mannigfaltigkeit},  as a topological space that resembles the Euclidean space near each point, but not necessarily globally. Since then, manifolds became a central concept for Geometry and modern Mathematical Physics. Posteriorly,
manifolds arose as solution sets of systems of equations, graphs of functions, surfaces, etc. They allow us
to describe locally very complicated structures in terms of an Euclidean space and techniques from linear algebra. 

Riemannian settled metrics on manifolds as a way to allow us measure distances and angles. This led, in a natural way, to the existence of non-Euclidean geometries. It is worth noting that by Riemann's time, it was fundamental to understand the existence of such non-Euclidean geometries that had just appeared \cite{CHL,Benn}. 

Moreover, the Gauss' Egregium Theorem is based on the notion of a metric: {\it ``the curvature of a surface can be determined entirely by measuring angles and distances on the surface itself''}, without
further reference on how the surface is embedded in the 3-dimensional Euclidean space.
In more modern geometrical terms, the Gaussian curvature of a surface is invariant under local isometry. These were ideas fully established by Riemann: {\it relevant properties of geometric structures are independent of the spaces in which they might be embedded. }

Since the XIX century, Differential Geometry began involved with geometric aspects of the
theory of differential equations and geometric structures on manifolds. Indeed, Lie pioneered this line of research by developing a theory of invariance of
differential equations under groups of continuous transformations, as a way to achieve conserved quantities \cite{Lie1880,LSPartial,Lie93,LSII,LSIII,FunLS}.

Historically, the formulation of Differential Geometry and Symmetry Groups remained still
from Euler's theory description of the plane sound waves in fluids until Birkhoff retook the subject during World War II,
working on Fluid Dynamics including bazooka changes and problems of air-launched missiles entering water \cite{Birkhoff2,Birkhoff1}. 
He emphasized Group Theory for handling symmetries in Hydrodynamics and urged innovative numerical methods relying more heavily on computing. 
Indeed, he started working on the solution of partial differential equations by successive approximations.

By the end of the XX century, a big explosion in research of geometrical aspects of differential equations had taken place. In order to study
the Hamiltonian formalism of Classical Mechanics, the introduction of symplectic manifolds served as the mathematical description of the phase space.
More relevant manifolds were henceforth introduced, as the Lorentzian manifold model for space-time in General Relativity \cite{Benn}, etc.

Properties of nonlinear systems of partial differential equations with geometrical origin and natural description in the language of
infinite-dimensional Differential Geometry were settled. This helped the appearance of their Lagrangian and Hamiltonian formalisms, and
the obtention of low-dimensional conservation laws \cite{Zharinov}.
Many problems in Physics, Medicine, Finance and industry required descriptions by means of nonlinear partial differential equations. So, their
investigation through Differential Geometry  began an independent field of research, with lots of different directions: optimal transport problems,
free boundary problems, nonlinear diffusive systems, singular perturbations, stochastic partial differential equations, regularity issues and so on \cite{Dissertationes,CR,Hermann1}.

For all the above reasons, we find of importance to formulate the content of this thesis in geometrical terms. 
This introductory chapter provides an overview of the geometrical and standard notation we shall use along the forthcoming chapters. 
We will consider differential equations as submanifolds in jet bundles. Investigation of singularities is a very subtle theme which
we shall skip by assuming every geometric structure to be real, smooth and well defined globally. Minor technical details will be specified when strictly needed.

Additionally, this section provides a survey in other many differential geometric structures that will appear during this thesis. In particular, we will describe symplectic, presymplectic, Poisson, Dirac and Jacobi manifolds. In following chapters we will illustrate how these structures will allow us to describe constants of motion, symmetries, superposition rules and other properties of differential equations.
 
\section{Tangent spaces, vector fields and integral curves}
\setcounter{equation}{0}
\setcounter{theorem}{0}
\setcounter{example}{0}
Let $N$ be an $n$-dimensional differentiable manifold and let $\{y_i\,|\,i=1,\ldots,n\}$ be a coordinate system on $N$. This allows us to 
 denote each point $y\in N$ by an $n$-tuple $(y_1,\dots,y_n)\in\mathbb{R}^n$. We assume $\epsilon$ to be a coordinate  system on $\mathbb{R}$. 

We write curves in $N$ in the form $\gamma:\epsilon \in \mathbb{R}\mapsto \gamma(\epsilon)=(\gamma_1(\epsilon),\ldots,\gamma_n(\epsilon))\in N$. The space of curves in $N$ can be endowed with an equivalence relation $R$ given by
\begin{equation}\label{RelTan}
\gamma\,\,R\,\,\bar{\gamma}\,\,\,\Longleftrightarrow \,\,\,\,\gamma(0)=\bar{\gamma}(0)\,\,\, \,\,{\rm and}\,\,\,\, \frac{d\gamma_i}{d\epsilon}(0)=\frac{d\bar{\gamma}_i}{d\epsilon}(0),\qquad i=1,\ldots,n.
\end{equation}
In other words, two curves in $N$ are related if their first-order Taylor polynomials around $\epsilon=0$ coincide.
Observe that the above equivalence relation is independent of the chosen coordinate system. That is, if $\gamma \,R\,\bar{\gamma}$ for the coordinate systems $\epsilon$ on $\mathbb{R}$ and $\{y_1,\ldots,y_n\}$ on $N$, then $\gamma \,R\,\bar{\gamma}$ also for any other coordinate systems on $\mathbb{R}$ and $N$.
Hence, the equivalence relation (\ref{RelTan}) is independent of the chosen coordinate system and it therefore has a geometrical meaning. 

Every {\it equivalence class} of $R$ is called a {\it tangent vector}. We write $v_y$ for an equivalence class of curves passing through $y\in N$ for $\epsilon=0$. We call {\it tangent space at $y\in N$} the space ${\rm T}_yN$ of equivalence classes of the form $v_y$. In view of (\ref{RelTan}) each tangent vector $v_y\in T_yN$ can be identified with a 
set of numbers $(v_1,\ldots,v_n)$ and viceversa. Hence, the tangent space to each point $y\in N$ can be written as
\begin{equation}
T_{y}N=\{(v_1,\ldots,v_n)|(v_1,\ldots,v_n)\in \mathbb{R}^n\}.
\end{equation}
The space ${\rm T}_yN$ is naturally endowed with an $n$-dimensional vectorial space structure and there exists an isomorphism ${\rm T}_yN\simeq \mathbb{R}^n$. 
Consequently, given a local coordinate system $\{y_1,\ldots, y_n\}$ on $N$ around $y\in N$, we can identify each vector $(v_1,\ldots,v_n)\in\mathbb{R}^n$ with the tangent vector $v_y$ associated to those curves $\gamma:\epsilon\in\mathbb{R}\mapsto (\gamma_1(\epsilon),\ldots,\gamma_n(\epsilon))\in N$ satisfying that that $d\gamma_i/d\epsilon(0)=v_i$ for $i=1,\ldots,n$. It is worth noting that the above identification depends on the chosen coordinate systems due to $\gamma_i(\epsilon)=y_i(\gamma(\epsilon))$ for $i=1,\ldots,n$ and $\epsilon\in\mathbb{R}$.

Alternatively, tangent vectors can be understood as derivations on a certain $\mathbb{R}$-algebra, i.e., a real vector space along with an $\mathbb{R}$-bilinear commutative multiplication. More specifically, 
let $C^\infty(U)$ be the space of differentiable functions defined on an open subset $U\subset N$ containing $y$. 
The space $C^\infty(U)$ admits an equivalence relation $R_g$ given by 
$$
f\, R_g\, g\quad \Longleftrightarrow \quad f,g \,\, \text{coincide on an open subset containing}\,\,\, y.
$$ 
The equivalence classes of this relation, let us say $[f]$ with $f\in C^\infty(U)$, are called {\it germs}.   It is easy to verify that the quotient space, $C_g^\infty(U)$, related to the equivalence relation $R_g$  becomes a vectorial space with the sum $[f]+[g]=[f+g]$ and multiplication by scalars $\lambda\in\mathbb{R}$ given by $\lambda[f]=[\lambda f]$. Additionally, $C^\infty(U)$ forms an $\mathbb{R}$-algebra with the $\mathbb{R}$-bilinear commutative multiplication $[f][g]=[fg]$.

Each $v_y$ induces a mapping $D_{v_y}:[f]\in C_g^\infty(U)\mapsto D_{v_y}[f]\in \mathbb{R}$ given by
\begin{equation}\label{Der}
D_{v_y}[f]=\frac{d}{d\epsilon}\bigg|_{\epsilon=0}f\circ \gamma,\qquad \gamma\in v_y.
\end{equation}
Since the first-order Taylor polynomials around $0$ of each $\gamma\in v_y$ coincide by definition of $v_y$, the above mapping is well defined and it does not depend on the chosen $\gamma\in v_y$. It can be proven that each $D_{v_y}$ is a derivation on $C_g^\infty(U)$, i.e., $D_{v_y}$ is linear and satisfies the Leibniz rule $[D_{v_y}([f][g])](y)=D_{v_y}([f])g(y)+f(y)D_{v_y}([g])$ for each $[f],[g]\in C_g^\infty(U)$. Conversely, it can be proven that every derivation on $C_g^\infty(U)$ is of the form (\ref{Der}) for a certain $v_y\in {\rm T}_yN$. This motivates to identify each $v_y$ with a derivation $D_{v_y}$ and viceversa. In particular, the equivalence class $v_{v_y}$ related to the vector $(0,\ldots,1\,(i-{\rm position}),\ldots,0)\in\mathbb{R}^n$ induces the derivation $(\partial/\partial y_i)_y:[f]\in C^\infty_g(U)\mapsto \frac{\partial f}{\partial y_i}(y)\in \mathbb{R}$. 

We call {\it tangent space} to $N$ the space ${\rm T}N\equiv \bigcup_{y\in N}{\rm T}_yN$. We can endow ${\rm T}N$ with a differentiable structure. To do so, we related each coordinate system $\{y_1,\ldots,y_n\}$ on $N$ to a coordinate system on ${\rm T}N$ of the form
\begin{equation}
y_i(v_y)=y_i(y),\qquad v_i(v_y)=\frac{d\gamma_i}{d\epsilon}(0),\qquad i=1,\ldots,n,
\end{equation}
where $\epsilon \mapsto (\gamma_1(\epsilon),\ldots,\gamma_n(\epsilon))$ is a curve belonging to the class $v_y$. This gives rise to a coordinate system $\{y_1,\ldots,y_n,v_1,\ldots,v_n\}$ on ${\rm T}N$.  In consequence, if $N$ is an $n$-dimensional manifold, then ${\rm T}N$ becomes a $2n$-dimensional manifold. We call {\it tangent bundle} the bundle $({\rm T}N,N,\pi_N)$, where $\pi_N:v_y\in {\rm T}N\mapsto y\in N$.

A {\it vector field} on $N$ is a section of the tangent bundle ${\rm T}N$, i.e., a mapping $X:N\rightarrow {\rm T}N$ such that $\pi_N\circ X={\rm Id}_N$, with ${\rm Id}_N$ being the identity on $N$. In other words, a vector field is a mapping assigning to every $y\in N$
a tangent vector $v_y\in {\rm T}_yN$. In the coordinate system $\{y_1,\ldots,y_n\}$, a vector field can be brought into the form
\begin{equation}\label{vectfield} X(y)=\sum_{i=1}^n\eta_i(y) \frac{\partial}{\partial y_i},\end{equation}
for certain functions $\eta_i\in C^{\infty}(N)$ with $i=1,\dots,n.$ 

An {\it integral curve} $\gamma:\epsilon \in \mathbb{R} \mapsto \gamma(\epsilon)\in N$ for $X$ with initial condition $y_0\in N$ 
is the curve $\gamma:\epsilon \in \mathbb{R}\mapsto \gamma(\epsilon)\in N$ 
satisfying that
\begin{equation}\label{Sys1}
\left\{\begin{aligned}
&\gamma(0)=y_0,\\
&\frac{d\gamma_i}{d\epsilon}(\epsilon)=\eta_i(\gamma(\epsilon)),
\end{aligned}\right.\qquad \forall \epsilon \in \mathbb{R},\qquad i=1,\ldots,n.
\end{equation}
In view of the theorem of existence and uniqueness of solutions of systems of first-order ordinary differential equations, the solution to the above Cauchy problem exists and is unique. So, there exists a unique integral curve of a vector field at each point. 

Observe that if $\gamma:\epsilon \in \mathbb{R}\mapsto (\gamma_1(\epsilon),\ldots,\gamma_n(\epsilon))\in N$ is a particular solution to system (\ref{Sys1}), then  the coordinate expression for $\gamma$ in another coordinate system is a particular solution for (\ref{Sys1}) in the new coordinate system. This justifies to write system (\ref{Sys1}) in an intrinsic form as
\begin{equation}
\left\{\begin{aligned}
&\gamma(0)=y_0,\\
&\frac{d\gamma}{d\epsilon}(\epsilon)=X(\gamma(\epsilon)),
\end{aligned}\right.\end{equation} 
where $d\gamma/d\epsilon(\epsilon_0)$ is understood as the tangent vector associated with $\gamma$ at $\gamma(\epsilon_0)$, i.e., the equivalence class related to the curve $\epsilon\mapsto \bar \gamma(\epsilon)\equiv\gamma(\epsilon+\epsilon_0)$. Therefore, an integral curve $\gamma$ of a vector field $X$ on $N$ is a curve in $N$ whose tangent vector at each point $y\in {\rm Im}\,\gamma$ coincides with $X(y)$. We write $\gamma_{y_0}$ for the integral curve of $X$ with initial condition $\gamma_{y_0}(0)=y_0\in N$. 

By solving system (\ref{Sys1}) for each $y_0\in N$,
we obtain a family of integral curves $\{\gamma_{y_0}\}_{y_0\in N}$.  This gives rise to the so called {\it flow} of $X$, namely the mapping $\Phi:(\epsilon,y)\in\mathbb{R}\times N\mapsto \gamma_y(\epsilon)\in N$. Since $X(y)={d\gamma_y}/{d\epsilon}(0)$ for each $y\in N$, the flow uniquely determines the vector field $X$.

The flow allows us to define an $\epsilon$-parametric set of diffeomorphisms  
$\Phi_{\epsilon}: y\in N\mapsto \Phi(\epsilon,y)\equiv\gamma_y(\epsilon)\in N$. Observe that
 $\gamma_{\gamma_y(\bar{\epsilon})}(\epsilon)=\gamma_y(\epsilon+\bar{\epsilon}), \forall \epsilon,\bar{\epsilon}\in \mathbb{R},\forall y\in N$.
Indeed, fixed $\bar{\epsilon}$, the curve $\epsilon \mapsto \gamma_y(\epsilon+\bar{\epsilon})$  takes the value $\gamma_y(\bar{\epsilon})$ for $\epsilon=0$ and, defining $\tilde{\epsilon}\equiv \epsilon+\bar{\epsilon}$, we obtain
\begin{equation}
\frac{d\gamma_y(\epsilon+\bar{\epsilon})}{d\epsilon}=\frac{d\gamma_y(\tilde{\epsilon})}{d\tilde{\epsilon}}=X(\gamma_y(\tilde{\epsilon}))=X(\gamma_y(\epsilon+\bar{\epsilon})),\qquad \gamma_y(0+\bar{\epsilon})=\gamma_y(\bar{\epsilon}).
\end{equation}
Therefore, the curve $\epsilon \mapsto \gamma_y(\epsilon+\bar{\epsilon})$ is an integral curve of $X$  taking the value $\gamma_y(\bar{\epsilon})$ for $\epsilon=0$. So, it must coincide with  $\gamma_{\gamma_y(\bar{\epsilon})}(\epsilon)$ and $\gamma_{\gamma_y(\bar{\epsilon})}(\epsilon)=\gamma_y(\epsilon+\bar{\epsilon}).$
In consequence, we have
\begin{equation}
\Phi_{\epsilon}(\Phi_{\bar{\epsilon}}(y))=\Phi_{\epsilon}(\gamma_y(\bar{\epsilon}))=\gamma_{\gamma_y(\bar{\epsilon})}(\epsilon)=\gamma_y(\epsilon+\bar{\epsilon})=\Phi_{\epsilon+\bar{\epsilon}}(y),\quad\forall y\in N,\quad \epsilon,\bar \epsilon \in \mathbb{R}.
\end{equation}
On the other hand, $\Phi_0(y)=\gamma_y(0)=y$ for all $y\in N$. Thus,
\begin{equation}\label{action1}
\Phi_{\epsilon}\circ \Phi_{\bar{\epsilon}}=\Phi_{\epsilon+\bar{\epsilon}},\quad \Phi_{0}={\rm Id}_N,\qquad \forall \epsilon,\bar \epsilon \in\mathbb{R},
\end{equation}
where ${\rm Id}_N$ is the identity on $N$. The properties (\ref{action1}) tell us that $\{\Phi_{\epsilon}\}_{\epsilon \in\mathbb{R}}$ is an {\it $\epsilon$-parametric group of diffeomorphisms} on $N$. Indeed, the composition of elements of that family belongs to the family in virtue of (\ref{action1}). The element $\Phi_0={\rm Id}_N$ is the neutral element and $\Phi_{-\epsilon}\circ \Phi_{\epsilon}={\rm Id}_N$. In this way, $\Phi_{\epsilon}$ has inverse $\Phi_{-\epsilon}$, which makes each $\Phi_{\epsilon}$, with $\epsilon \in\mathbb{R}$, into a diffeomorphism. 
We can also understand the uniparametric group of diffeomorphisms $\{\Phi_\epsilon\}_{\epsilon\in\mathbb{R}}$ as a {\it Lie group action} $\Phi:(\epsilon,y)\in\mathbb{R}\times N\mapsto \Phi_\epsilon(y)\in N.$ 

Alternatively, every $\epsilon$-parametric group of diffeomorphisms $\{\Phi_\epsilon\}_{\epsilon \in\mathbb{R}}$ on $N$ enables us to define a vector field on ${N}$  of the form
\begin{equation}\label{induced}
{X}_{\Phi}(y)\equiv \frac{d}{d\epsilon}\Big|_{\epsilon=0}\Phi_{\epsilon}(y),\qquad \forall  y\in N,
\end{equation}
whose integral curves are $\gamma_y(\epsilon)=\Phi_\epsilon(y)$, with $y\in N$. If $\{\Phi_\epsilon\}_{\epsilon \in\mathbb{R}}$ is coming from the flow of a vector field $X$, then $X_\Phi=X$. This shows that each vector field is equivalent to an $\epsilon$-parametric group of diffeomorphisms.

Assuming $X$ to be given by (\ref{vectfield}), the corresponding flow gives rise to a set of $\epsilon$-parametric diffeomorphisms which, for infinitesimal $\epsilon$, takes the form
\begin{equation}
y_i\mapsto  y_i+\epsilon \, \eta_{i}(y)+O(\epsilon^2),\qquad i=1,\ldots,n,
\end{equation}
where $O(\epsilon^2)$ stands, as usual, for a function of $\epsilon$ satisfying that $\lim_{\epsilon\rightarrow 0}O(\epsilon^2)/\epsilon=0$.
This transformation is invertible for each fixed value of the parameter $\epsilon$. The set of all transformations for different values of $\epsilon$ forms a {\it group}.

\section{Time-dependent vector fields}
\setcounter{equation}{0}
\setcounter{theorem}{0}
\setcounter{example}{0}
In order to study systems of non-autonomous first-order ordinary differential equations from a geometrical viewpoint, we introduce $t$-dependent vector fields.  In this section we present some of their basic properties.

Let $\pi_2:(t,y)\in\mathbb{R}\times N\mapsto y\in N$ be the projection onto the second factor with $t$ being  the natural coordinate system on $\mathbb{R}$. We now write $t$ for the variable in $\mathbb{R}$ because this variable mostly stands for the physical time in the applications described in this thesis.
A {\it $t$-dependent vector field} $X$ on $N$ is a map $X:(t,y)\in
\mathbb{R}\times N\mapsto X(t,y)\in {\rm T}N$ satisfying the commutative diagram
\vskip 0.1cm
\centerline{
\xymatrix{&{\rm T}N\ar[d]^{\pi_N}\\
\mathbb{R}\times N\ar[ur]^X\ar[r]^{\pi_2}&N}}

\noindent that is, $\pi_N\circ X=\pi_2$.  Hence, $X(t,y)\in \pi_N^{-1}(y)={\rm
T}_yN$ and $X_t:y\in N\mapsto X_t(y)\equiv
X(t,y)\in{\rm T}_yN\subset {\rm T}N$ is a vector field on $N$ for every
$t\in\mathbb{R}$. Conversely, every $t$-parametric family $\{X_t\}_{t\in\mathbb{R}}$ of vector fields on $N$ gives rise to a unique $t$-dependent vector field $X:\mathbb{R}\times N \mapsto X_t\in {\rm T}N$. Thus, each $t$-dependent vector
field $X$ is equivalent to a family $\{X_t\}_{t\in\mathbb{R}}$ of vector fields
on $N$. 

The $t$-dependent vector fields enable us to describe, as a particular instance, 
standard vector fields. Indeed, every vector field $X:y\in N\mapsto X(y)\in TN$
can naturally be regarded as a $t$-dependent vector field $X:(t,y)\in \mathbb{R}\times N\mapsto X(y)\in {\rm T}N$. Conversely, a `constant' $t$-dependent
vector field $X$ on $N$, i.e., $X_t=X_{t'}$ for every
$t,t'\in\mathbb{R}$, can be understood as a vector field $X=X_0$ on $N$.

Each $t$-dependent vector field $X$ is equivalent to a linear morphism $X:f\in C^\infty(N)\mapsto (Xf)\in C^\infty (\mathbb{R}\times N)$, with $(Xf)(t,y)\equiv (X_tf)(y)$ for every $(t,y)\in \mathbb{R}\times N$, satisfying a Leibniz rule, namely $[X(fg)](t,y)=(X_tf)(y)g(y)+f(y)(X_tg)(y)$ for arbitrary $f,g\in C^\infty(N)$ and $(t,y)\in \mathbb{R}\times N$.

Similarly to vector fields, $t$-dependent vector fields also admit local integral curves. To define them, we make use of the so called
autonomization of $X$. The {\it autonomization} of a vector field $X$ on $N$ is the only vector field $\bar X$ on $\mathbb{R}\times N$ satisfying that $\iota_{\bar X}dt=1$ and $(\bar X\pi_2^*f)(t,y)=(Xf)(t,y)$ for an arbitrary function $f\in C^\infty(N)$ and $(t,y)\in \mathbb{R}\times N$.   In coordinates, if $X=\sum_{i=1}^nX_i(t,y)\partial/\partial y_i$, then
\begin{equation}
\bar X(t,y)=\frac{\partial}{\partial t}+\sum_{i=1}^nX_i(t,y)\frac{\partial}{\partial y_i}.
\end{equation}

Now we say that $\gamma:\mathbb{R}\rightarrow \mathbb{R}\times N$ is an integral curve of the $t$-dependent vector field $X$ if $\gamma$ is 
an integral curve of $\bar X$. If $\gamma:s\in\mathbb{R}\mapsto (t(s),y(s))\in \mathbb{R}\times N$ is an integral curve of $X$, then
\begin{equation}
\left\{\begin{aligned}
\frac{dy_i}{ds}&=X_i(t,y),\\
\frac{dt}{ds}&=1,
\end{aligned}\right.\qquad i=1,\ldots,n.\end{equation} 
By using the reparametrization $t=t(s)$, we obtain
\begin{equation}\label{nonauto}
\frac{dy_i}{dt}=X_i(t,y),\qquad i=1,\ldots,n.
\end{equation}
The particular solutions to this system can be considered as integral curves of $X$ with a preferred parametrization $t\rightarrow (t,y(t))$. The above system is called the {\it associated system} with $X$. Conversely, given a first-order system in normal form (\ref{nonauto}), we can define a $t$-dependent vector field of the form $X(t,y)=\sum_{i=1}^nX_i(t,y)\partial/\partial y_i$ whose integral curves $t\rightarrow (t,y(t))$ are the particular solutions to (\ref{nonauto}).  This motivates to employ $X$ to represent a nonautonomous system of ordinary differential equations and its related $t$-dependent vector field.

Every $t$-dependent vector field $X$ on $N$ gives rise to its {\it generalized flow} $g^X$, i.e., the map
$g^X:\mathbb{R}\times N\rightarrow N$ such that
$g^X(t,y)\equiv g^X_t(y)=\gamma_y(t)$ with $\gamma_y$ being the particular solution to $X$ such that $\gamma_y(0)=y$. We already showed that
for $X$ being a standard vector field the particular solutions satisfy that $\gamma_{\gamma_y(s)}(t)=\gamma_y(t+s)$ for each $t,s\in\mathbb{R}$ and $y\in N$. As a consequence, the  generalized flow for the $t$-dependent vector field $X$ associated with an autonomous vector field leads to a uni-parametric group of diffeomorphisms $\{g^X_t\}_{t\in\mathbb{R}}$. If $X$ is not related to an autonomous vector field, this property is not longer valid.

Given a fiber vector bundle ${\rm pr}:P \rightarrow N$, we denote by $\Gamma({\rm pr})$
the $C^\infty(N)$-module of 
its smooth sections.  So, if $\tau_N:{\rm T}N\rightarrow N$ and $\pi_N:
{\rm T}^*N\rightarrow N$ are the canonical projections
 associated with the tangent and cotangent bundle to $N$, respectively, then
$\Gamma(\tau_N)$ and $\Gamma(\pi_N)$
  designate the $C^\infty(N)$-modules of vector fields and one-forms on
$N$, correspondingly.

We call {\it generalized distribution} $\mathcal{D}$ on $N$, a correspondence relating each $y\in N$ to  a linear
 subspace $\mathcal{D}_y\subset {\rm T}_yN$. A generalized distribution is said to be
{\it regular} at $y'\in N$ when the function 
  ${\rm r}:y\in N\mapsto \dim\mathcal{D}_y\in\mathbb{N}\cup \{0\}$  is locally constant
around $y'$. Similarly, $\mathcal{D}$ is regular on an open $U\subset N$ when ${\rm r}$ is constant on $U$. Finally,
a vector field  $Y\in\Gamma(\tau_N)$ is said to
   take values in $\mathcal{D}$, in short $Y\in\mathcal{D}$, when
$Y_y\in\mathcal{D}_y$ for all $y\in N$. Likewise, similar 
   notions can be defined for a {\it generalized codistribution}, namely a
correspondence mapping relating every $y\in N$ to a linear subspace of ${\rm T}_y^*N$.

It will be very important to our purposes to relate $t$-dependent vector fields to the so called Lie algebras as follows. 
A {\it Lie algebra} is a pair $(V,[\cdot,\cdot])$, where $V$ stands for a
real linear space endowed with a Lie
 bracket $[\cdot\,,\cdot]:V\times V\rightarrow V$, namely an $\mathbb{R}$-bilinear antisymmetric mapping satisfying the Jacobi identity. Given two subsets
$\mathcal{A}, \mathcal{B}\subset V$, we write $[\mathcal{A},\mathcal{B}]$ for the real vector space spanned by the Lie brackets between elements 
 of $\mathcal{A}$ and $\mathcal{B}$, and we define ${\rm
Lie}(\mathcal{B},V,[\cdot,\cdot])$ to be  the smallest Lie subalgebra
  of $V$ containing $\mathcal{B}$. 
Note that ${\rm Lie}(\mathcal{B},V,[\cdot,\cdot])$ is expanded by
\begin{equation}
\mathcal{B},[\mathcal{B},\mathcal{B}],[\mathcal{B},[\mathcal{B},\mathcal{B}]],[
\mathcal{B},[\mathcal{B},[\mathcal{B},\mathcal{B}]]],[[\mathcal{B},\mathcal{B}],[\mathcal{B},\mathcal{B}]],\ldots
\end{equation}
From now on, we use ${\rm Lie}(\mathcal{B})$ and $V$ to represent ${\rm
Lie}(\mathcal{B},V,[\cdot,\cdot])$ and $(V,[\cdot,\cdot])$, 
correspondingly, when their meaning is clear
 from the context.
 
 Given a $t$-dependent vector field $X$, we call {\it minimal algebra} of $X$ the smallest Lie algebra, $V^X$, of vector fields (relative to the Lie bracket of vector fields) containing all the vector fields $\{X_t\}_{t\in\mathbb{R}}$, namely $V^X={\rm
Lie}(\{X_t\}_{t\in\mathbb{R}})$. 

Given a $t$-dependent vector field $X$ on $N$, its {\it
associated distribution}, 
$\mathcal{D}^X,$ is the generalized distribution on $N$ spanned by the vector
fields of $V^X$, i.e.,
\begin{equation}
\mathcal{D}^X_x=\{Y_x\mid Y\in V^X\}\subset T_xN,\end{equation} 
and its {\it associated co-distribution}, $\mathcal{V}^X$, is the generalized
co-distribution on $N$ of the form
\begin{equation}
\mathcal{V}^X_x=\{\vartheta\in T_x^*N\mid \vartheta(Z_x)=0,\forall
\,\,Z_x\in \mathcal{D}_x^X\}=(\mathcal{D}^X_x)^\circ\subset T_x^*N,\end{equation} 
where $(\mathcal{D}^X_x)^\circ$ is the {\it annihilator} of $\mathcal{D}_x^X$. 
\begin{proposition}\label{NuX} A function $f:U\rightarrow \mathbb{R}$
is a local $t$-independent constant of  
motion for a system $X$ if and only if $df\in \mathcal V^X|_U$. 
\end{proposition}
\begin{proof} If $f$ is assumed to be a $t$-independent constant of motion, then $X_tf|_U=df(X_t)|_U=0$ for all
$t\in\mathbb{R}$. Consequently, $df$ also vanishes on the successive Lie
brackets of elements from $\{X_t\}_{t\in\mathbb{R}}$ and  hence
\begin{equation}
df(Y)|_U=Yf|_U=0,\qquad \forall  Y\in {\rm Lie}(\{X_t\}_{t\in\mathbb{R}}).\end{equation} 
Since the elements of $V^X$ span the generalised distribution
  $\mathcal{D}^X$, then $df_x(Z_x)=0$ for all $x\in U$ and $Z_x\in
\mathcal{D}^X_x$, i.e., $df\in \mathcal{V}^X|_U$. The converse directly follows
from the above considerations. 
\end{proof}

The following lemma can easily be proven \cite{CLS122}.

\begin{lemma}\label{basisVX} Given a system $X$, its associated co-distribution
$\mathcal{V}^X$ admits a local basis around
 every $x\in U^X$ of the form $df_1,\ldots,df_{p(x)}$, with $p(x)= {\rm r}^X(x)$
and $f_1,\ldots,f_{p(x)}:U\subset U^X\rightarrow\mathbb{R}$ 
 being a family of (local) $t$-independent constants of motion for $X$.
Furthermore, the $\mathbb{R}$-linear space $\mathcal{I}^X|_U$ of $t$-independent
constants of motion of $X$ on $U$ can be written as
\begin{equation}
\mathcal{I}^X|_U=\{g\in C^\infty(U)\mid \exists F:U\subset
\mathbb{R}^{p(x)}\rightarrow\mathbb{R}, \ g=F(f_1,\ldots, f_{p(x)})\}.\end{equation} 
\end{lemma}

\section{Higher-order tangent spaces and related notions}
\setcounter{equation}{0}
\setcounter{theorem}{0}
\setcounter{example}{0}
Given two curves $\rho,\sigma:\mathbb{R}\mapsto N$ such that
$\rho(0)=\sigma(0)=y\in N$, we say that they have a {\it contact of order $p$ at $y$}, with
$p\in\mathbb{N}\cup \{0\}$,  if they satisfy
\begin{equation}
\frac{d^p(f\circ\rho)}{dt^p}(0)=\frac{d^p(f\circ\sigma)}{dt^p}(0),
\end{equation}
for every function $f\in C^{\infty}(N)$. The relation `to have a contact of order $p$ at $y$' is
an equivalence relation. Observe that this relation is purely geometrical, i.e., if two curves are related with respect to a certain coordinate system, then they are so in any other coordinate system. Note also that the above relation amounts to saying that the two curves have the same Taylor expansion around $0$ up to order $p$. 

Each equivalence class of the previous equivalence relation, let us say ${\bf t}^p_{y}$, is called a {\it $p$-tangent vector}
at $y$. More specifically, ${\bf t}^p_{y}$ stands for an equivalence class of contacts of order $p$ with a curve $\sigma(t)$ with $\sigma(0)=y$.  We write ${\rm T}^p_{y}N$ for the space of all $p$-tangent vectors at $p$ and we
define
$$
{\rm T}^pN=\bigcup_{y\in N}{\rm T}^p_{y}N.
$$
It can be proven that ${\rm T}^pN$ can be endowed with a differential structure turning it into a differential manifold. Additionally,
 $({\rm T}^pN,\pi,N)$, with $\pi:{\bf t}^p_{y}\in{\rm
T}^pN\mapsto y\in N$, is a fiber bundle. Let
us briefly analyze these facts.

Every coordinate system $t$ on $\mathbb{R}$ and $\{y_1,\ldots,y_n\}$ on $N$ induces a natural coordinate system on
the space ${\rm T}^p N$. Indeed, consider again a curve $\rho:t\in \mathbb{R}\mapsto \rho(t)\in N$ with coordinates $\rho_1(t),\ldots, \rho_n(t)$. The $p$-tangent vector, ${\bf
t}^p_{\rho(0)}$, associated with this curve admits a representative
\begin{equation}
\rho_i(0)+\frac{t}{1!}\frac{d\rho_i}{dt}(0)+\ldots+\frac{t^p}{p!}\frac{d^p\rho_i}{dt^p}(0),\qquad
i=1,\ldots,n,
\end{equation}
which can be characterized by the coefficients
\begin{equation}
y_i(0)=\rho_i(0),\quad y^{1)}_i(0)=\frac{d\rho_i}{dt}(0),\quad \ldots,\quad
y^{p)}_i(0)=\frac{d^p\rho_i}{dt^p}(0),\qquad i=1,\ldots,n.
\end{equation}
In consequence, the mapping $\varphi:{\bf t}^p_{y(0)}\in{\rm T}^pN\mapsto
(y_i(0),y^{1)}_i(0),\ldots,y^{p)}_i(0))\in \mathbb{R}^{n(p+1)}$ gives a coordinate system for ${\rm
T}^pN$. Obviously, the map $\pi$ becomes a smooth submersion which makes ${\rm T}^pN$
into a fiber bundle with base $N$. We hereby denote each element of ${\rm T}^pN$ by
${\bf t}^p_y=(y,y^{1)},\ldots,y^{p)})$.

Now, given a curve $c:t\in \mathbb{R}\mapsto c(t)\in N$, we call {\it prolongation to ${\rm
T}^pN$ of $c$} the curve ${\bf t}^pc:t\in \mathbb{R} \mapsto {\bf t}^pc(t)\in {\rm
T}^pN$, associating with every $t_0$ the corresponding equivalence class of $c(t+t_0)$ given
in coordinates by
\begin{equation}
{\bf t}^pc(t_0)=\left(c(t_0),c^{1)}(t_0),\ldots,c^{p)}(t_0)\right).
\end{equation}

\section{Jet bundles}
 \setcounter{equation}{0}
\setcounter{theorem}{0}
\setcounter{example}{0}
Jet bundles are certain types of bundles constructed out of the sections of a fiber bundle. Our interest in them is due to the geometrical description of systems of higher-order ordinary and partial differential equations and their Lie symmetries as structures on an appropriate jet bundle. 

For simplicity, consider a projection $\pi:(x,u)\in \mathbb{R}^n\times N\equiv N_{\mathbb{R}^n}\mapsto x\in \mathbb{R}^n$ giving rise to a trivial bundle $(N_{\mathbb{R}^n}, \mathbb{R}^n,\pi)$.
Let us hereafter assume $N$ to be a $k$-dimensional manifold and let $\{x_1,\ldots,x_n\}$ be a global coordinate system on $\mathbb{R}^n$.

We say that two sections $\sigma_1,\sigma_2:\mathbb{R}^n\rightarrow N_{\mathbb{R}^n}$  are {\it p-equivalent at a point $x\in \mathbb{R}^n$} or they have a {\it contact of order $p$ at $x$}  if they have the same Taylor expansion of order $p$ at $x\in \mathbb{R}^n$. Equivalently,
\begin{equation}
\sigma_1(x)=\sigma_2(x),\qquad \frac{\partial^{|J|} (\sigma_1)_i}{\partial x_1^{j_1}\ldots\partial x_n^{j_n}}(x)=\frac{\partial^{|J|} (\sigma_2)_i}{\partial x_1^{j_1}\ldots\partial x_n^{j_n}}(x),
\end{equation}
for every multi-index $J=(j_1,\ldots,j_n)$ such that $0<|J|\equiv j_1+\ldots+j_n\leq p$ and $i=1,\dots,n$. Being $p$-equivalent induces an equivalence relation in the space $\Gamma(\pi)$ of sections of the bundle $(N_{\mathbb{R}^n},\mathbb{R}^n,\pi)$. Observe that if two sections have a contact of order $p$ at a point $x$, then they do have a contact at that point of the same type for any other coordinate systems on $\mathbb{R}^n$ and $N$, i.e., this equivalence relation is geometric.

We write $j_{x}^p\sigma$ for the equivalence class of sections that have a {\it contact of $p$-order} at $x\in \mathbb{R}^n$ with a section $\sigma$. Every such an equivalence
class is called a {\it p-jet}. We write ${\rm J}^{p}_x\pi$ for the space of all jets of order $p$ of sections at $x$. We will denote by ${\rm J}^p\pi$ the space of all jets of order $p$. Alternatively, we will write ${\rm J}^p(\mathbb{R}^n,\mathbb{R}^k)$ for the jet bundle of sections of the bundle $\pi:(x,u)\in\mathbb{R}^n\times\mathbb{R}^k\mapsto  x\in\mathbb{R}^n$.

Given a section $\sigma:\mathbb{R}^n\rightarrow {\rm J}^p\pi$, we can define the functions
\begin{equation}
(u_j)_J(j^p_x\sigma)=\frac{\partial^{|J|} \sigma_j}{\partial x_1^{j_1}\ldots\partial x_n^{j_n}}(x),\quad \forall j, \quad |J|\leq p.\end{equation} 
For $|J|=0$, we define $u_J(x)\equiv u(x)$. Coordinate systems on $\mathbb{R}^n$ and $N$ along with the previous functions give rise to a local coordinate system on ${\rm J}^p\pi$. We will also hereafter denote
the $n$-tuple and $k$-tuple, respectively, by $x=(x_1,\ldots,x_n),\,\, u=(u_1,\ldots,u_k)$, then
\begin{equation}\label{nose1}
(u_j)_J=u_{x_{i_1}^{j_1}\dots x_{i_n}^{j_n}}=\frac{\partial^{|J|} u_j}{\partial x_{i_1}^{j_1}\ldots \partial x_{i_n}^{j_n}},\quad \forall j,\quad |J|\leq 0.
\end{equation}

All such local coordinate systems give rise to a manifold structure on ${\rm J}^p\pi$. In this way, every point of ${\rm J}^p\pi$ can be written as
\begin{equation}
\left(x_i,u_j,(u_j)_{x_i},(u_j)_{x_{i_1}^{j_1}x_{i_2}^{2-j_1}},(u_j)_{x_{i_1}^{j_1}x_{i_2}^{j_2}x_{i_3}^{3-j_1-j_2}},\dots,(u_j)_{x_{i_1}^{j_1}x_{i_2}^{j_2}\dots x_{i_n}^{p-\sum_{i=1}^{n-1}j_i}}\right),
\end{equation} 
where the numb indices run $i_1,\ldots,i_p=1,\dots,n$, $j=1,\dots,k,$ $j_1+\dots+j_n\leq p$. 

For small values of $p$, jet bundles have simple descriptions: ${\rm J}^{0}\pi=N_{\mathbb{R}^n}$ and ${\rm J}^1\pi\simeq \mathbb{R}^n\times {\rm T}N$.

The projections $\pi_{p,l}:j^p_x\sigma\in {\rm J}^p\pi\mapsto j^l_x\sigma\in {\rm J}^l\pi$ with $l<p$ lead to define the smooth bundles $({\rm J}^p\pi,{\rm J}^l\pi,\pi_{p,l})$.  
Conversely, for each section $\sigma: \mathbb{R}^n\rightarrow N_{\mathbb{R}^n}$, we have a natural embedding $j^p\sigma:\mathbb{R}^n\ni x\mapsto j^{p}_x\sigma \in {\rm J}^p\pi$. 

Consider now a vector field on $N_{\mathbb{R}^n}$ of the form
\begin{equation}
X(x,u)=\sum_{i=1}^n\xi_i(x,u)\frac{\partial}{\partial x_i}+\sum_{j=1}^k\eta_j(x,u)\frac{\partial}{\partial u_j}.
\end{equation}
This gives rise to an $\epsilon$-parametric group of transformations

\begin{equation}
\left\{\begin{aligned}
{x}_i&\mapsto x_i+\epsilon \xi_i(x,u) +O\left(\epsilon^2\right),\\
{u}_j&\mapsto u_j+\epsilon \eta_j(x,u) +O\left(\epsilon^2\right),\\
\end{aligned}\right.\qquad i=1,\ldots,n,\qquad\,j=1,\ldots,k.\end{equation} 

%
We can extend this infinitesimal transformation up to $p$-order derivatives
\begin{equation}
\left\{\begin{aligned}\label{pointtransep2}
\bar{x}_i&=x_i+\epsilon \xi_i(x,u) +O\left(\epsilon^2\right),\\
\bar{u}_j&=u_j+\epsilon \eta_j(x,u)+O\left(\epsilon^2\right),\\
(\bar{u}_j)_J&=(u_j)_J+\epsilon (\eta_j)_J(x,u) +O\left(\epsilon^2\right),\\
\end{aligned}\right.\end{equation} 
for $i=1,\ldots,n,\,\,j=1,\ldots,k,$ $0<|J|\leq p$ and $(u_j)_J$ represents \eqref{nose1} and the $(\eta_j)_J$ are given in the forthcoming steps. 

In coordinates, the above $\epsilon$-parametric group of diffeomorphisms has the associated vector field

\begin{equation}
X^p=\sum_{i=1}^n\xi_i\frac{\partial}{\partial x_i}+\sum_{j=1}^k\eta_j\frac{\partial}{\partial u_j}+\sum_{0<|J|\leq p}\sum_{j=1}^k \mathfrak{Pr}_J \eta_j\frac{\partial}{\partial (u_j)_{J}},
\end{equation}
where $\mathfrak{Pr}_J\eta_j$ denotes the prolongations of $\eta_j$ for the multi-index $J$.

\begin{lemma} Let   $\{\Phi_\epsilon: (x,u)\in N_{\mathbb{R}^n}\mapsto (\bar x,\bar u)\in N_{\mathbb{R}^n}\}_{\epsilon\in\mathbb{R}}$ be the one-parametric group of transformations induced by the vector field $X=\sum_{i=1}^n\xi_i(t,u)\partial/\partial x_i+\sum_{j=1}^k\eta_j(t,u)\partial/\partial u_j$ and let $\sigma:x\in\mathbb{R}^n\mapsto (x,u(x))\in N_{\mathbb{R}^n}$ be a section of the bundle $\pi:N_{\mathbb{R}^n}\rightarrow \mathbb{R}^n$.  The section
$\bar{\sigma}_\epsilon=\Phi_\epsilon \circ \sigma$ has slopes
$\partial \bar u_j/\partial \bar x_i=\partial  u_j/\partial x_i+\epsilon (\eta_{j})_{x_{ i}} +O(\epsilon^2)$, for any two fixed values of $1\leq i\leq n$ and $1\leq j\leq k$, where
\begin{equation}\label{prolong2}
\mathfrak{Pr}_{J\equiv x_i} \eta_j=(\eta_{j})_{x_{ i}}=\frac{D\eta_{{j}}}{Dx_{i}}-\sum_{q=1}^n(u_{j})_{x_q}\frac{D\xi_{q}}{Dx_{i}},
\end{equation}
and ${D}/{Dx_i}$ is the total derivative with respect to $x_i$, namely
\begin{equation}\label{opsymi}
\frac{D}{Dx_{i}}=\frac{\partial}{\partial x_{i}}+\sum_{l=1}^k(u_{l})_{x_i}\frac{\partial}{\partial u_l}.
\end{equation}

\end{lemma}

\begin{proof}
Observe that 
\begin{equation}
\delta \bar{u}_{ j}=\delta u_{ j}+\epsilon \delta \eta_{j}+O\left(\epsilon^2\right)=\sum_{i=1}^n\left[\frac{\partial u_{ j}}{\partial x_i}+\epsilon\left(\frac{\partial \eta_{j}}{\partial x_i}+\sum_{l=1}^k\frac{\partial \eta_{j}}{\partial u_l}\frac{\partial u_l}{\partial x_i}\right)\right]\delta x_i+O\left(\epsilon^2\right)
\end{equation}
and 
\begin{equation}
\delta \bar{x}_{i}=\delta x_{i}+\epsilon \delta \xi_{i}+O\left(\epsilon^2\right)=\sum_{m=1}^n\left[\delta_{m}^{ i}+\epsilon\left(\frac{\partial \xi_{ i}}{\partial x_m}+\sum_{l=1}^k\frac{\partial \xi_{ i}}{\partial u_l}\frac{\partial u_l}{\partial x_m}\right)\right]\delta x_m+O\left(\epsilon^2\right).
\end{equation}
Using the operator defined in \eqref{opsymi}, we can rewrite these two expressions as
%
\begin{equation}
\delta\bar{u}_{j}=\sum_{l=1}^k\left((u_{ j})_{x_l}+\epsilon\frac{D\eta_{ j}}{Dx_l}\right)\delta x_l+O\left(\epsilon^2\right),\quad
\delta\bar{x}_{ i}=\sum_{m=1}^n\left(\delta^{ i}_{m}+\epsilon\frac{D\xi_{i}}{Dx_m}\right)\delta x_m+O\left(\epsilon^2\right).
\end{equation}
Hence, 
\begin{align}
(\bar{u}_{j})_{\bar x_{i}}&=\frac{\sum_{l=1}^n\left((u_{j})_{x_l}+\epsilon\frac{D\eta_{j}}{Dx_l}\right)}{\sum_{m=1}^n\left(\delta^{i}_m+\epsilon\frac{D\xi_{i}}{Dx_m}\right)}\delta_m^l+O\left(\epsilon^2\right)\nonumber\\
&=\sum_{l=1}^n\left((u_{j})_{x_l}+\epsilon\frac{D\eta_{j}}{Dx_l}+O\left(\epsilon^2\right)\right)\cdot \left(\delta^{l}_{i}-\epsilon\frac{D\xi_{l}}{Dx_{i}}+O\left(\epsilon^2\right)\right)\nonumber\\
&=(u_{j})_{x_{i}}+\epsilon\left(\frac{D\eta_{j}}{Dx_{i}}-\sum_{q=1}^n(u_{j})_{x_{q}}\frac{D\xi_{q}}{Dx_{i}}\right)+O\left(\epsilon^2\right),
\end{align}
which finishes the proof.
\end{proof}
 
In a similar way as we deduced the expression for the first-order prolongation, we can deduce it for higher-order prolongations.
We will deduce the expression by generalization of results in lower dimensional cases.

\begin{lemma} Let $\sigma:x\in\mathbb{R}^n\mapsto (x,u(x))\in N_{\mathbb{R}^n}$  and let $\Phi_\epsilon:(x,u)\in N_{\mathbb{R}^n}\mapsto (\bar x,\bar u)\in N_{\mathbb{R}^n}$ be a one-parametric group of transformations induced by the vector field $X=\sum_{i=1}^n\xi_i(t,u)\partial/\partial x_i+\sum_{j=1}^k\eta_j(t,u)\partial/\partial u_j$. We obtain that the section
$\bar{\sigma}_\epsilon=\Phi_\epsilon \circ \sigma$ of $\pi:N_{\mathbb{R}^n}\rightarrow \mathbb{R}^n$ has slope $\partial^2 \bar{u}_{j}/\partial \bar{x}_{i_1}\partial \bar{x}_{i_2}=\partial^2 u_j/\partial x_{i_1}\partial x_{i_2}+\epsilon (\eta_j)_{x_{i_1}x_{i_2}}+O(\epsilon^2)$, where 
\begin{align}\label{prolong3}
(\eta_{j})_{x_{i_1}x_{i_2}}=\frac{D^2 \eta_{j}}{Dx_{i_1}Dx_{i_2}}-&\sum_{l=1}^n (u_{j})_{x_{i_1},x_l}\frac{D\xi_{l}}{Dx_{i_2}}\nonumber\\
&-\sum_{l=1}^n (u_{j})_{x_{i_2},x_l}\frac{D\xi_{l}}{Dx_{i_1}}-\sum_{l=1}^n (u_{j})_{x_l}\frac{D\xi_{l}}{Dx_{i_1}Dx_{i_2}}
\end{align}
and $D/Dx_i$, when acting on functions of $J^1\pi$, stands for
\begin{equation}\label{totalderiv}
\frac{Df}{Dx_i}=\frac{\partial f}{\partial x_i}+\sum_{j=1}^k\left[(u_j)_{x_i}\frac{\partial f}{\partial u_j}+\sum_{l=1}^n(u_j)_{x_lx_i}\frac{\partial f}{\partial (u_j)_{x_l}}\right].
\end{equation}
\end{lemma}

\begin{proof}
We have $(\bar{u}_j)_{\bar{x}_{i_1}}=(u_{j})_{x_{i_1}}+\epsilon (\eta_{j})_{x_{i_1}}$. Therefore,
\begin{equation}
\delta[(\bar{u}_{j})_{\bar{x}_{i_1}}]=\delta[(u_{j})_{x_{i_1}}]+\epsilon \delta[(\eta_{j})_{x_{i_1}}]=\sum_{l=1}^n\left(\frac{D [(u_{j})_{x_{i_1}}]}{Dx_{l}}+\epsilon\frac{D[(\eta_{j})_{x_{i_1}}]}{D x_{l}}\right)\delta x_{l}
\end{equation} 
and
\begin{equation}
\delta \bar{x}_{i_2}=\delta x_{i_2}+\epsilon\delta \xi_{i_2}=\sum_{q=1}^n\left(\delta_{q}^{i_2}+\epsilon\frac{D\xi_{i_2}}{D x_{q}}\right)\delta x_q.
\end{equation} 
Therefore, proceeding in similar fashion as for the first-order prolongation,
\begin{align}\label{nose2}
(\bar{u}_{j})_{\bar{x}_{i_1}\bar{x}_{i_2}}&=\frac{\delta[(\bar{u}_{j})_{\bar{x}_{i_1}}]}{\delta\bar{x}_{i_2}}=\frac{\sum_{l=1}^n\left(\frac{D[(u_{j})_{x_{i_1}}]}{D x_{l}}+\epsilon\frac{D[(\eta_{j})_{x_{i_1}}]}{Dx_{l}}\right)}{\sum_{q=1}^n\left(\delta_{q}^{i_2}+\epsilon \frac{D\xi_{i_2}}{Dx_q}\right)}\delta^{l}_q+O\left(\epsilon^2\right) \nonumber\\
&=  \sum_{l=1}^n \left(\frac{D [(u_{j})_{x_{i_1}}]}{D x_{l}}+\epsilon\frac{D[(\eta_{j})_{x_{i_1}}]}{Dx_{l}}\right)\cdot \left(\delta^{i_2}_{l}-\epsilon\frac{D\xi_{l}}{Dx_{i_2}}\right)+O\left(\epsilon^2\right)\nonumber\\
&= (u_{j})_{x_{i_1}x_{i_2}}+\epsilon \left(\frac{D[(\eta_{j})_{x_{i_1}}]}{Dx_{i_2}}-\sum_{l=1}^n(u_{j})_{x_{i_1}x_{l}}\frac{D\xi_{l}}{Dx_{i_2}}\right)+O\left(\epsilon^2\right).
\end{align}
Introducing the value of $(\eta_{j})_{x_{i_1}}$ in \eqref{nose2}, using \eqref{prolong2}, we arrive at \eqref{prolong3}.

\end{proof}

Applying this process recursively, we obtain the {\it $p$-order prolongations}
\begin{align}\label{prolongcomplete}
&(\eta_{j})_{x_{i_1}^{j_1}x_{i_2}^{j_2}\dots x_{i_n}^{j_n}}=\frac{D\left((\eta_{j})_{x_{i_1}^{j_1}\ldots x_{i_{n-1}}^{j_{n-1}} x_{i_{n}}^{(j_{n})-1}}\right)}{Dx_{i_n}}-\sum_{l=1}^n(u_{j})_{x_{i_1}^{j_1}x_{i_2}^{j_2}\dots x_{i_{n-1}}^{j_{n-1}} x_{i_{n}}^{(j_{n})-1} x_{l}}\frac{D\xi_{l}}{Dx_{i_n}},
\end{align}
or equivalently,

\begin{align}
\mathfrak{Pr}_J \eta_j& =(\eta_{j})_{x_{i_1}^{j_1}x_{i_2}^{j_2}\dots x_{i_n}^{j_n}}=\frac{D^{|J|} \eta_j}{Dx_{i_1}^{j_1} Dx_{i_2}^{j_2} \dots Dx_{i_n}^{j_n}}-\sum_{l=1}^n (u_j)_{x_l} \frac{D^{|J|} \eta_j}{Dx_{i_1}^{j_1} Dx_{i_2}^{j_2} \dots Dx_{i_n}^{j_n}}-\dots \nonumber \\
&-\sum_{i,i'=1}^n \sum_{l=1}^n (u_j)_{x_l \dots x_{i_{n-2}}^{j_{n-2}}} \frac{D^2 \xi_l}{Dx_{i}Dx_{i'}}-\sum_{i=1}^n \sum_{l=1}^n (u_j)_{x_l \dots  x_{i_{n-1}}^{j_{n}-1}} \frac{D \xi_l}{Dx_{i}},
\end{align}
where $x_1\leq x_{i_1},\ldots,x_{i_n}\leq x_n$ and $j_1+\dots+j_n\leq p.$
The associated vector field for this transformation reads 
\begin{equation}
X^p=\xi\frac{\partial}{\partial t}+\sum_{j=1}^n\eta_j\frac{\partial}{\partial u_j}+\sum_{j=1}^n\sum_{|J|=p} \mathfrak{Pr}_J \eta_j\frac{\partial}{\partial (u_j)_{J}},
\end{equation}

\section{Poisson algebras}
\setcounter{equation}{0}
\setcounter{theorem}{0}
\setcounter{example}{0}
A {\it Poisson algebra} is a triple $(A,\star,\{\cdot,\cdot\})$, where $A$ is a real vector space endowed with two bilinear maps, namely $`\star$' and $\{\cdot,\cdot\}$, such that
$`\star$' is a commutative and associative real algebra and $(A,\{\cdot,\cdot\})$ is a real Lie algebra whose Lie bracket, the
{\it Poisson bracket}, satisfies the {\it Leibnitz rule} with respect to $`\star$'
$$
\{f\star g,h\}=f\star \{g,h\}+\{f,h\}\star g,\qquad \forall f,g,h \in A.
$$
In other words, $\{\cdot,h\}$ is a derivation.

A {\it Poisson algebra morphism} is a morphism $\mathcal{T}:(A,\star_A,\{\cdot,\cdot\}_A)\rightarrow (B,\star_B,\{\cdot,\cdot\}_B)$ of $\mathbb{R}$-algebras
$\mathcal{T}:(A,\star_A)\rightarrow (B,\star_B)$ that satisfies that $\mathcal{T}(\{a,b\}_A)=\{\mathcal{T}(a),\mathcal{T}(b)\}_B$ for every $a,b\in A$. 

Two of the types of Poisson algebras to be used, are: the symmetric and universal Poisson algebras. Let us describe their
main characteristics. Given a finite-dimensional real Lie algebra $(\mathfrak{g},[\cdot,\cdot]_\mathfrak{g})$, its {\it universal algebra}, $U_\mathfrak{g}$, is obtained from the quotient
$T_\mathfrak{g}/\mathcal{R}$ of the tensor algebra $(T_\mathfrak{g},\otimes)$ of $\mathfrak{g}$
by the bilateral ideal  $\mathcal{R}$ spanned by the elements
$v\otimes w-w\otimes v-[v,w]$, with $v,w\in \mathfrak{g}$. Given the quotient map $\pi:T_\mathfrak{g}\rightarrow U_\mathfrak{g}$, the space $U_\mathfrak{g}$ becomes an $\mathbb{R}$-algebra  $(U_\mathfrak{g},\widetilde \otimes)$ when endowed with the product $\widetilde \otimes: U_\mathfrak{g}\times U_\mathfrak{g}\rightarrow U_\mathfrak{g}$  given by
$\pi(P)\,\widetilde\otimes\,\pi(Q)\equiv \pi(P\otimes Q)$, for every $P,Q\in T_\mathfrak{g}$. The Lie
bracket on $\mathfrak{g}$ can be extended to a Lie bracket $\{\cdot,\cdot\}_{U_\mathfrak{g}}$ on
$U_\mathfrak{g}$ by imposing it to be a derivation of
$(U_\mathfrak{g},\widetilde\otimes)$ on each factor.  This turns
$U_\mathfrak{g}$ into a Poisson algebra $(U_\mathfrak{g},\widetilde\otimes,\{\cdot,\cdot\}_{U_\mathfrak{g}})$ \cite{CL99}. The elements
of its Casimir subalgebra are henceforth dubbed as {\it Casimir elements} of $\mathfrak{g}$ \cite{Be81}.

If we set $\mathcal{R}$ to be the bilateral ideal spanned by the elements $v\otimes w-w\otimes v$  in the above procedure, we obtain a new commutative Poisson algebra $S_\mathfrak{g}$ called
{\it symmetric algebra} of $(\mathfrak{g},[\cdot,\cdot]_\mathfrak{g})$. The elements of $S_\mathfrak{g}$ are polynomials on the elements of $\mathfrak{g}$. Via the isomorphism 
$\mathfrak{g}\simeq (\mathfrak{g}^*)^*$,
they can naturally be understood as polynomial functions on $\mathfrak{g}^*$ \cite{AMA75,CL99}. The Casimir elements of
this Poisson algebra are called {\it Casimir functions} of $\mathfrak{g}$.

The Poisson algebras $U_\mathfrak{g}$ and $S_\mathfrak{g}$ are related by the
{\it symmetrizer map} \cite{AMA75,Be81,Var98}, i.e. the linear
isomorphism $\lambda:S_\mathfrak{g}\rightarrow 
U_\mathfrak{g}$ of the form
\begin{equation}
\label{symmap}
\lambda(v_{i_1})=\pi(v_{i_1}),\quad
\lambda(v_{i_1}v_{i_2}\ldots v_{i_l})=\frac{1}{l!}\sum_{s\in
\Pi_l}\lambda(v_{s(i_1)})\, \widetilde\otimes\ldots\widetilde\otimes\, \lambda (v_{s(i_l)}),
\end{equation}
for all $v_{i_1},\ldots,v_{i_l}\in\mathfrak{g}$ and with $\Pi_l$ being the set of permutations of $l$ elements. Moreover, 
\begin{equation}\label{UgSg}
\lambda^{-1} \left(\{ v,P\}_{U_\mathfrak{g}} \right)=\{v,\lambda^{-1}(P)\}_{S_\mathfrak{g}}, \qquad \forall P\in
U_\mathfrak{g},\quad\forall v\in\mathfrak{g}.
\end{equation}
So, $\lambda^{-1}$ maps the Casimir elements of $\mathfrak{g}$ into Casimir elements of $S_\mathfrak{g}$.
If $(A,\star_A,\{\cdot,\cdot\}_A)$ and $(B,\star_B,\{\cdot,\cdot\}_B)$ are Poisson algebras and operations: $\star_A$ and $\star_B$ are commutative, then
$A\otimes B$ becomes a Poisson algebra $(A\otimes B,\star_{A\otimes B},\{\cdot,\cdot\}_{A\otimes
B})$ by defining
\begin{equation}
\begin{aligned}
  (a\otimes b)\star_{A\otimes B} (c\otimes d)&=(a\star_A c)\otimes (b\star_B d),\\
  \{a\otimes b, c\otimes d\}_{A\otimes B}&=\{a, c\}_A\otimes b\star_B d+a\star_A
c\otimes\{b,d\}_B .
\end{aligned}\end{equation} 
for all $a,c\in A,\quad \forall b,d\in B.$ 

Similarly, a Poisson structure on $A^{(m)}\equiv\   \stackrel{m-{\rm times}}{\overbrace{  A\otimes \ldots\otimes A}}$ can be constructed by induction.

We say that $(A,\star_A,\{\cdot,\cdot\}_A,\Delta)$ is a {\it Poisson coalgebra} if $(A,\star_A,\{\cdot,\cdot\}_A)$ is a Poisson
algebra and $\Delta:(A,\star_A,\{\cdot,\cdot\}_A)\rightarrow (A\otimes
A,\star_{A\otimes A},\{\cdot,\cdot\}_{A\otimes A})$, the so called {\it coproduct}, is a
Poisson algebra
homomorphism
which is  {\it coassociative} \cite{CP}, i.e. $
(\Delta \otimes {\rm Id}) \circ \Delta=({\rm Id} \otimes \Delta) \circ \Delta$.
Then, the {$m$-th coproduct} map
$\Delta ^{(m)}:  A\rightarrow  A^{(m)}$
can be defined recursively as follows
\begin{equation}\label{copr}
{\Delta}^{(m)}= ({\stackrel{(m-2)-{\rm times}}{\overbrace{{\rm
Id}\otimes\ldots\otimes{\rm Id}}}}\otimes {\Delta^{(2)}})\circ \Delta^{(m-1)},\qquad m>2,
\end{equation}
and $\Delta\equiv \Delta^{(2)}$. Such an induction ensures that $\Delta^{(m)}$ is also a Poisson map.

In particular, $S_\mathfrak{g}$ is a Poisson coalgebra with {\em primitive coproduct map} given by
$\Delta(v)= v\otimes 1+1\otimes v,$ for all $v\in\mathfrak{g}\subset S_\mathfrak{g}$.
The coassociativity of $\Delta$ is straightforward, and its $m$-th generalization reads
\begin{equation}
  \Delta^{(m)}(v)=v\otimes{ \stackrel{(m-1)-{\rm times}}{\overbrace{{\rm
1}\otimes\ldots\otimes{\rm 1}}}}
 +\, {\rm 1}\otimes v   \otimes{ \stackrel{(m-2)-{\rm times}}{\overbrace{{\rm
1}\otimes\ldots\otimes{\rm 1}}}}  +
\ldots
  +  { \stackrel{(m-1)-{\rm times}}{\overbrace{{\rm
1}\otimes\ldots\otimes{\rm 1}}}}  \otimes v.\end{equation} 

for all $v\in\mathfrak{g}\subset S_\mathfrak{g}$.

\section{Symplectic, presymplectic and Poisson manifolds}
\setcounter{equation}{0}
\setcounter{theorem}{0}
\setcounter{example}{0}

 A {\it symplectic manifold} is a pair $\left( N, \omega \right) $,
where $N$ stands for a manifold and $\omega$ is a non-degenerate closed two-form
on $N$. We say that a
vector field $X$ on $N$ is Hamiltonian with respect to $(N,\omega)$ if there
exists a function $f\in C^\infty(N)$
such that\footnote{In Geometric Mechanics, the convention $\iota_X\omega=df$ is used.}
\begin{equation}\label{conHam}
\iota_X\omega=-df.
\end{equation}
In this case, we say that $f$ is a {\it Hamiltonian function} for $X$.
Conversely, given a
function $f$, there exists a unique vector field $X_f$ on $N$, the
so called {\it Hamiltonian vector field} of $f$, satisfying (\ref{conHam}). This allows us to define a bracket $\{\cdot,\cdot \}:C^{\infty}
\left( N \right)\times C^\infty(N)\rightarrow C^\infty(N)$ given by\footnote{In Geometric Mechanics $\{f,g\}=X_gf$. So, the mapping $f\rightarrow X_f$ becomes a Lie algebra anti-homomorfism.}
\begin{equation}\label{PB}
\left\{f,g\right\}=\omega(X_f,X_g)=X_f(g).
\end{equation}
This bracket turns $C^\infty(N)$ into a {\it Poisson algebra} $(C^\infty(N),\cdot,\{\cdot,\cdot\})$, i.e., $\{
\cdot, \cdot \}$ is a Lie bracket on $C^\infty(N)$ which additionally holds the
{\it Leibniz rule} with respect to
the standard product `${\bf \cdot}$' of functions
\begin{equation}
 \{\{ f,g \}, h\} = \{\{ f, h\}, g\} + \{g, \{ f, h\}\} \qquad
   \forall f, g, h \in C^\infty(N). \end{equation} 
 The Leibniz rule can be rephrased by saying that $\{f,\cdot\}$ is a derivation of the
associative algebra $(C^\infty(N),\cdot)$ for each $f \in C^\infty(N)$. Actually, this derivation is represented by the Hamiltonian vector field $X_f$.
The bracket
$\{\cdot,\cdot\}$ is called the
{\it Poisson bracket} of $(C^\infty(N),\cdot,\{\cdot,\cdot\})$. Note that if $(N,\omega)$ is a symplectic manifold, the non-degeneracy condition for
$\omega$ implies that $N$ is even dimensional \cite{Abra}.

The above observations lead to the concept of a Poisson manifold which is a natural generalization of the symplectic one. A {\it
Poisson manifold} is a pair $\left(N,\{\cdot,\cdot\}\right)$, where
$\{\cdot,\cdot\}:C^\infty(N)\times
C^\infty(N)\rightarrow C^\infty(N)$ is the Poisson bracket of $(C^{\infty}( N),\cdot,\{\cdot,\cdot\})$ which is also referred to as a {\it Poisson structure} on $N$. In
view of this and (\ref{PB}), every symplectic manifold is a particular type of
Poisson manifold. Moreover, by noting that $\{f,\cdot\}$ is a derivation on
$(C^\infty(N),\cdot)$ for every $f\in C^\infty(N)$, we can associate with every function $f$  a
single vector field $X_f$, called the {\it Hamiltonian vector field} of $f$, such that $\{f,g\}=X_fg$ for all $g\in C^\infty(N)$, like in the symplectic case.

As the Poisson
structure is a derivation in each entry, it gives rise to a bivector field $\Lambda$, i.e., an element of $
\Gamma(\bigwedge^2{\rm T}N)$, the referred to as {\it Poisson bivector}, such that
$\{f,g\}=\Lambda(df,dg)$. It is known that the Jacobi identity for
$\{\cdot,\cdot\}$ amounts to
$[\Lambda,\Lambda]_{SN}=0$, with $[\cdot,\cdot]_{SN}$ being the {\it
Schouten--Nijenhuis bracket} \cite{IV}. Conversely, a
bivector $\Lambda$ satisfying $[\Lambda,\Lambda]_{SN}=0$ gives rise to a Poisson bracket on $C^\infty(N)$ by
setting $\{f,g\}=\Lambda(df,dg)$. Hence, a Poisson manifold can be
considered, equivalently, as $(N,\{\cdot,\cdot\})$ or $(N,\Lambda)$. It is remarkable that $\Lambda$ induces
a bundle morphism $\widehat\Lambda:\alpha_x\in {\rm T}^*N\rightarrow \widehat\Lambda(\alpha_x)\in {\rm T}N$, where
$\bar\alpha_x(\widehat\Lambda(\alpha_x))=\Lambda_x(\alpha_x,\bar\alpha_x)$ for all $\bar\alpha_x\in T_x^*N$
, which enables us to write $X_f=\widehat \Lambda(df)$ for every $f\in C^\infty(N)$.

Another way of generalizing a symplectic structure is to consider a two-form $\omega$ which is merely closed (not necessarily of constant rank), forgetting the non-degeneracy assumption. In this case,
$\omega$ is said to be a {\it presymplectic form}
and the pair $\left( N, \omega \right)$ is called a
{\it presymplectic manifold} \cite{LM87}. Like in the symplectic case, we
call a vector field $X$ on $N$ a {\it Hamiltonian vector field} if there exists a function $f\in C^\infty(N)$, a {\it Hamiltonian function} for
$X$, such that (\ref{conHam}) holds for the presymplectic form $\omega$.

The possible degeneracy of $\omega$ introduces several differences
with respect to the symplectic setting.
For example, given an $f\in C^\infty(N)$,
we cannot ensure neither the existence nor the uniqueness of a vector field $X_f$
satisfying $\iota_{X_f}\omega=-df$. If it exists, we say that $f$ is an {\it
admissible function}
with respect to $\left( N, \omega \right)$. Since the linear combinations and multiplications of admissible functions are also admissible functions, the space ${\rm Adm}(N,\omega)$ of admissible functions of $(N,\omega)$ is a real associative algebra. It is canonically also a Poisson algebra. Indeed, observe that every
$f\in {\rm Adm}(N,\omega)$ is associated to a family of Hamiltonian vector fields of the form
$X_f+Z$, with $Z$ being a vector field taking values in $\ker \omega$. Hence,  (\ref{PB}) does not depend on the representatives $X_f$ and $X_g$ and becomes a Poisson bracket on the space ${\rm Adm}(N,\omega)$, making the latter into a Poisson algebra. It is also remarkable that
\begin{equation}
\iota_{[X_f,X_g]}\omega=\mathcal{L}_{X_f}\iota_{X_g}\omega-
\iota_{X_g}\mathcal{L}_{X_f}\omega=-\mathcal{L}_{X_f}dg=-d\{f,g\}\,.
\end{equation} 
In consequence, $[X_f,X_g]$ is a Hamiltonian vector field with a Hamiltonian function $\{f,g\}$.

\begin{proposition} The Casimir co-distribution of a Poisson manifold is
involutive. 
\end{proposition}
\begin{proof} Given two sections $\omega,\omega'\in\mathcal{C}^\Lambda$, we have
that
 $\widehat \Lambda(\omega)=\widehat\Lambda(\omega')=0$ and then
$\Lambda(\omega,\omega')=0$. In consequence, 
\begin{equation}
[\omega,\omega']_\Lambda=\mathcal{L}_{\widehat\Lambda(\omega)}\omega'-
\mathcal{L}_{\widehat\Lambda(\omega')}\omega-d\Lambda(\omega,\omega')=0.\end{equation} 
\end{proof}

\section{Dirac manifolds}
\setcounter{equation}{0}
\setcounter{theorem}{0}
\setcounter{example}{0}
The concept of {\it Dirac structure} was proposed by Dorfman \cite{Do87} in the Hamiltonian framework of integrable evolution equations and defined in \cite{Co87} as a subbundle of the {\it  Whitney sum}, $T N\oplus_N T^\ast N$, called the {\it extended tangent} or {\it Pontryagin bundle}.
It was thought-out as a common generalization of Poisson and presymplectic structures. It was also designed to deal with constrained systems, including constraints induced by degenerate Lagrangians investigated by Dirac \cite{Di50}, hence their name.
See \cite{Bu11,Co87,Co90,CW88,JR11,IV,NTZ12} for details.

\begin{definition}
We call a {\it Pontryagin bundle} $\mathcal{P}N$  a vector bundle ${\rm T}N
\oplus_N T^{\ast} N$ on $N$.
\end{definition}
\begin{definition} An {\it almost-Dirac manifold} is a pair $(N, L)$, where $L$ is a  maximally isotropic  subbundle of
$\mathcal{P}N$ with respect to the pairing
\begin{equation}
 \langle X_x + \alpha_x, \bar{X}_x + \bar{\alpha}_x\rangle _+ \equiv
\frac{1}{2}  (
   \bar{\alpha}_x(X_x) + \alpha_x(\bar{X}_x)),\end{equation} 
where $ X_x + \alpha_x, \bar{X}_x
+
   \bar{\alpha}_x \in T_xN \oplus
T_x^{\ast} N=\mathcal{P}_xN.$
In other words, $L$ is isotropic and has rank $n=\dim N$.
\end{definition}
\begin{definition}
A {\it Dirac manifold} is an almost-Dirac manifold $(N,L)$  whose subbundle $L$,
its {\it Dirac structure}, is
involutive relative to the {\it Courant--Dorfman bracket} \cite{Co87,Do87,GG11,JR11}, namely
\begin{equation}
 [[X + \alpha, \bar{X} + \bar{\alpha}]]_C \equiv [X, \bar{X}] + \mathcal{L}_X
   \bar{\alpha} - \iota_{\bar{X}}d\alpha\,,\end{equation} 
where
$X+\alpha, \bar X+\bar{\alpha}\in \Gamma({\rm T}N\oplus_N{\rm T}^*N)$.

\end{definition}
Note that the Courant--Dorfman bracket satisfies the Jacobi identity in the form
\begin{equation}\label{Jacobi} [[\,[[e_1,e_2]]_C,e_3]]_C
\!=\![[e_1,[[e_2,e_3]]_C]]_C-\![[e_2,[[e_1,e_3]]_C]]_C,\,\, \forall e_1,e_2,e_3\!\in\! \Gamma(\mathcal{P}N),
\end{equation}
but is not skew-symmetric. It is, however, skew-symmetric on sections of the Dirac subbundle $L$, defining a {\it Lie algebroid} structure
$(L,[[\cdot,\cdot]]_C,\rho)$, where $\rho
: L\ni X_x + \alpha_x \mapsto X_x\in  {\rm T}N$. This means that $(\Gamma(L),[[\cdot,\cdot]]_C)$ is a Lie algebra and the vector bundle morphism  $\rho:L\rightarrow {\rm T}N$, the {\it anchor}, satisfies
\begin{equation}\label{anchor}[[e_1,fe_2]]_C=(\rho(e_1)f)e_2+f[[e_1,e_2]]_C
\end{equation}
for all $e_1,e_2\in \Gamma(L)$ and $f\in C^\infty(N)$ \cite{Co87}.
One can prove that, automatically, $\rho$ induces a Lie algebra morphism of $(\Gamma(L),[[\cdot,\cdot]]_C)$ into the Lie algebra of vector fields on $N$.
The generalized distribution $\rho(L)$, called the {\it characteristic distribution} of the Dirac structure, is therefore integrable in the sense of Stefan--Sussmann \cite{Su73}.

\begin{definition}
A vector field $X$ on $N$ is said to be an $L$-{\it Hamiltonian vector
field} (or simply a {\it Hamiltonian vector field} if $L$ is fixed) if there exists a
$f \in C^{\infty} (N)$ such that $X + df \in \Gamma(L)$. In this case, $f$
is an $L$-{\it Hamiltonian function} for $X$ and an {\it admissible function} of
$(N,L)$. Let us denote by
Ham$(N,L)$ and Adm$(N, L)$ the spaces of Hamiltonian vector fields and admissible functions of $(N,L)$, respectively.
\end{definition}

The space ${\rm Adm}(N,L)$ becomes a Poisson algebra $({\rm Adm}(N, L), \cdot,\{ \cdot, \cdot \}_L)$ relative to the standard product of functions  and the Lie bracket
 given by $\{f,\bar f\}_L =X \bar f $
where $X$ is an $L$-Hamiltonian vector field for $f$.
 Since $L$ is isotropic, $\{f,\bar f\}_L$ is
 well defined, i.e., its value is independent on the choice of the $L$-Hamiltonian vector field associated to $f$. The elements $f\in {\rm Adm}(N,L)$
 possessing trivial Hamiltonian vector fields are called the {\it Casimir functions} of
 $(N,L)$ \cite{NTZ12}. We write ${\rm Cas}(N,L)$ for the set of Casimir functions of $(N,L)$. We can also distinguish the space $G(N,L)$ of $L$-Hamiltonian vector fields which admit
 zero (or, equivalently, any constant) as an $L$-Hamiltonian function. We call them {\it gauge vector fields} of the Dirac structure.

Note that, if $X$ and $\bar X$ are $L$-Hamiltonian vector
fields with Hamiltonian functions $f$ and $\bar f$, then $\{f, \bar f \}_L$ is a Hamiltonian for $[X, \bar X]$
\begin{equation}
[[X+df,\bar X+d\bar f]]_C=[X,\bar X]+\mathcal{L}_{X}d\bar f-\iota_{\bar
X}d^2f=[X,\bar X]+d\{f,\bar f\}_L.
\end{equation}
This implies that
$({\rm Ham}(N,L),[\cdot,\cdot])$ is a Lie algebra in which $G(N,L)$ is a Lie ideal.
Denote the quotient Lie algebra ${\rm Ham}(N,L)/G(N,L)$ by $\widehat{\rm Ham}(N,L)$.

\begin{proposition}\label{CasCon}
If  $(N, L)$ is a Dirac manifold, then  $\{{\rm Cas} (N, L), {\rm Adm}(N,L)\}_L=0$, i.e.,  {\rm Cas}$(N, L)$ is an ideal
  of the Lie algebra $( {\rm Adm} (N, L), \{ \cdot, \cdot \}_L)$. Moreover, we have the following exact
sequence of Lie algebra homomorphisms
\begin{equation}\label{PreSymSeqII}
0\hookrightarrow {\rm Cas}(N,L)\hookrightarrow {\rm Adm}(N,L)\stackrel{B_L}{\longrightarrow} \widehat{\rm Ham}(N,L)\rightarrow 0\,,
\end{equation}
with $B_L(f)=\pi(X_f)$, where the vector field $X_f$ is an $L$-Hamiltonian vector field of $f$, and $\pi$ is the canonical projection $\pi:{\rm Ham}
(N,L)\rightarrow \widehat{\rm Ham}(N,L)$.
\end{proposition}

For every Dirac manifold $(N,L)$, we have a canonical linear map
$\Omega^L_x : \rho (L)_x\subset T_xN \rightarrow \rho (L)_x^*\subset T^*_xN$ given by
\begin{equation}\label{twoform}
[\Omega^L_x (X_x)](\bar X_x) = -{\alpha}_x (\bar{X}_x), \qquad X_x,\bar X_x\in
\rho(L),
\end{equation}
where $\alpha_x\in T^*_xN$ is such that $X_x+\alpha_x\in L$.
Note that, as  $L$ is isotropic, $\Omega_x^L$ is well defined, i.e., the
value of
\begin{equation}
\Omega_x^L(X_x,\bar X_x)=[\Omega^L_x(X_x)](\bar X_x)\end{equation} 
 is independent of the particular
${\alpha}_x$ and defines a skew-symmetric bilinear form $\Omega^L$ on the (generalized) distribution $\rho(L)$. Indeed, given
$X_x+\bar{\bar{\alpha}}_x\in L$, we have that ${\alpha}_x-\bar{\bar{\alpha}}_x\in
L$. Since $L$ is isotropic, $\langle {\alpha}_x-\bar{\bar{\alpha}}_x,
\bar{X}_x+\bar{\alpha}_x\rangle_+=({\alpha}_x-\bar{\bar{\alpha}}_x)\bar{X}_x/2=0$ for all
$ \bar{X}_x+\bar{\alpha}_x\in L$. Then,  $[\Omega_x^L(X_x)](\bar X_x)=-\bar{\bar
{\alpha}}_x( \bar{X}_x)=-{\alpha}_x(\bar{X}_x)$ for all $\bar{X}_x\in \rho(L)$ and
$\Omega^L$ is well defined.

It is easy to see that gauge vector fields generate the {\it gauge distribution} $\ker \Omega^L$. Moreover, the involutivity of $L$ ensures that $\rho(L)$ is an integrable
generalized distribution in the sense of Stefan--Sussmann \cite{Su73}.
Therefore, it induces a (generalized) foliation $\mathfrak{F}^L=\{\mathfrak{F}^L_x: x\in N\}$ on $N$.

Since $\rho(L_x)=T_x\mathfrak{F}_x^L$, if the elements $X_x+\alpha_x$ and $X_x+\bar{\alpha}_x$, with $X_x\in T_x\mathfrak{F}_x^L$, are in $L_x\subset \mathcal{P}_xN=T_xN\oplus T^*_xN$, then $\alpha_x-\bar{\alpha}_x$ is in the annihilator of $T_x\mathfrak{F}_x^L$, so the image of $\alpha_x$ under the canonical restriction
$\sigma:\alpha_x\in T^*_xN\mapsto \alpha_x|_{T_x\mathfrak{F}_x^L}\in T^*_x\mathfrak{F}_x^L$ is uniquely determined. One can verify that $\sigma(\alpha_x)=-\Omega^L_x(X_x)$. The two-form $\Omega^L$ restricted to $\mathfrak{F}_x^L$ turns out to be closed, so that $\mathfrak{F}_x^L$ is canonically a presymplectic manifold, and the canonical restriction of $L$ to $\mathfrak{F}_x^L$ is the graph of this form \cite{Co87}.

As particular instances,
Poisson and presymplectic
manifolds are particular cases of Dirac
manifolds. On one hand, consider a presymplectic manifold $(N,\omega)$ and
define $L^\omega$ to be the graph of minus the fiber bundle morphism
$\widehat\omega:X_x\in {\rm T}N\mapsto \omega_x(X_x,\cdot) \in {\rm T}^*N$. The generalized
distribution $L^\omega$ is isotropic, as
\begin{equation}
\langle
X_x-\widehat {\omega}(X_x),\bar{X}_x-\widehat {\omega}(\bar{X}_x)\rangle_+=-({\omega}_x(X_x,\bar{X}_x)+\omega_x(\bar{X}_x,X_x))/2=0\,.                                                                                                                                      \end{equation} 
As $L^\omega$ is the graph of $-\widehat{\omega}$, then $\dim L^\omega_x=\dim N$ and $L^\omega$ is a maximally
isotropic subbundle of $\mathcal{P}N$. In addition, its integrability relative to
the Courant--Dorfman bracket comes from the fact that $d\omega=0$. Indeed, for arbitrary $X,X'\in \Gamma({\rm T}N)$, we have
\begin{equation}
[[X-\iota_X\omega,X'-\iota_{X'}\omega]]_C=[X,X']-\mathcal{L}_X \iota_{X'}\omega+\iota_{X'}d\iota_{X}\omega=
[X,X']-\iota_{[X,X']}\omega\,,\end{equation} 
since
\begin{equation}
\mathcal{L}_X \iota_{X'}\omega-\iota_{X'}d\iota_{X}\omega=\mathcal{L}_X \iota_{X'}\omega-\iota_{X'}\mathcal{L}_{X}\omega=\iota_{[X,X']}\omega\,.\end{equation} 
In this case, $\rho:L^\omega\rightarrow {\rm T}N$ is a bundle isomorphism.
Conversely, given a
Dirac manifold whose $\rho:L\rightarrow {\rm T}N$ is a bundle isomorphism, its characteristic distribution satisfies $\rho(L)={\rm T}N$ and it admits a unique integral leaf, namely
$N$, on which $\Omega^L$ is a closed two-form, i.e., $(N,\Omega^L)$ is a presymplectic manifold.

On the other hand, every Poisson manifold $(N,\Lambda)$ induces a subbundle
 $L^\Lambda$ given by the graph of $\widehat\Lambda$. It is isotropic,
\begin{equation}
\langle \widehat{\Lambda}(\alpha_x)+ \alpha_x,\widehat{\Lambda}(\bar{\alpha}_x)+
\bar{\alpha}_x\rangle_+=(\Lambda_x(\bar{\alpha}_x,\alpha_x)+\Lambda_x(\alpha_x,
\bar{\alpha}_x))/2=0,\end{equation} 
for all $ \alpha_x,\bar{\alpha}_x\in T^*_xN$  and $x\in N$,
and of rank $\dim N$ as the graph of $\widehat \Lambda$ is a map from ${\rm T}^*N$. Additionally, $L^\Lambda$ is integrable. Indeed, as $\widehat \Lambda (d\{f,g\})=[\widehat \Lambda(df),\widehat \Lambda(dg)]$ for every $f,g\in C^\infty(N)$ \cite{IV}, we have
\begin{align}
\![[\widehat \Lambda(df)\!+\!df,\!\widehat \Lambda(dg)+dg]]_C\!\!&=\!\![\widehat \Lambda(df),\!\widehat \Lambda(dg)]+\mathcal{L}_{\widehat \Lambda(df)}dg-\iota_{\widehat \Lambda(dg)}d^2\!f\!\!=\!\! \nonumber \\
& \widehat \Lambda(d\{f,g\})+d\{f,g\}\end{align} 
and the involutivity follows from the fact that the module of 1-forms is generated locally by exact 1-forms.

Conversely, every Dirac manifold $(N,L)$ such
that $\rho^*:L\rightarrow {\rm T}^*N$ is a bundle isomorphism is the graph of $\widehat\Lambda$ of a Poisson bivector.

Let us motivate our terminology. We call  $\rho(L)$ the
 characteristic distribution of $(N,L)$, which follows the terminology of \cite{NTZ12}
instead of the original one by Courant \cite{Co87}. This is done because when
$L$ comes from a Poisson manifold, $\rho(L)$ coincides with the characteristic
distribution of the Poisson structure \cite{IV}. Meanwhile, the vector fields taking values in $\ker \Omega^L$ are called {\it gauge vector fields}. In this way, when $L$ is the graph of a presymplectic
structure, such vector fields are its gauge vector fields
\cite{EMR99}.

From here, we see that the Dirac structure incorporates the  presymplectic and Poisson structures as particular cases. Courant \cite{Co87,Co90} provided the theory to this
statement.

Recall that every presymplectic manifold $(N,
\omega)$ gives rise to a Dirac manifold $(N,L^\omega)$ whose distribution
$L^\omega$ is spanned by elements
of $\Gamma({\rm T}N\oplus_N{\rm T}^*N)$ of the form $X - \iota_{X} \omega$ with $X\in\Gamma({\rm T}N)$. Obviously, this shows
that the Hamiltonian vector fields for $(N,\omega)$ are $L$-Hamiltonian vector
fields relative to $(N,L)$.

\section{Jacobi manifolds}
\setcounter{equation}{0}
\setcounter{theorem}{0}
\setcounter{example}{0}

Jacobi manifolds were independently introduced by Kirillov and Lichnerowicz \cite{Kiri,Li77}. We now briefly survey their most fundamental properties. Several known results will be illustrated during the proof of the main results of the paper.

\begin{definition}A {\it Jacobi manifold} is a triple $(N,\Lambda,R)$, where $\Lambda$ is a bivector field on $N$ and $R$ is a vector field, the referred to as {\it Reeb vector field}, satisfying:
\begin{equation}
[\Lambda,\Lambda]_{SN}=2R\wedge \Lambda,\qquad [R,\Lambda]_{SN}=0.\end{equation} 
\end{definition}

\begin{example} Every Poisson manifold $(N,\Lambda)$ can be considered as a Jacobi manifold $(N,\Lambda,R=0)$.
\end{example}

\begin{example}\label{BivectorRH}
The {\it continuous Heisenberg group} \cite{We00} can
be described as the space of matrices
\begin{equation}\label{Hei}
\mathbb{H}=\left\{\left(\begin{array}{ccc}1&x&z\\0&1&y\\0&0&1\end{array}
\right)\bigg|\,x,y,z\in\mathbb{R}\right\},
\end{equation}
endowed with the standard matrix multiplication,
where $\{x,y,z\}$ is the natural coordinate system on $\mathbb{H}$ induced by (\ref{Hei}). Consider the bivector field on $\mathbb{H}$ given by
\begin{equation}\label{BivectorH}
\Lambda_\mathbb{H}\equiv -y\frac{\partial}{\partial y}\wedge\frac{\partial}{\partial z}+\frac{\partial}{\partial x}\wedge\frac{\partial}{\partial y}
\end{equation}
and the vector field $R_\mathbb{H}\equiv \partial/\partial z$. After a simple calculation, we obtain that
\begin{equation}
[\Lambda_\mathbb{H},\Lambda_\mathbb{H}]_{SN}=2\frac{\partial}{\partial x}\wedge\frac{\partial}{\partial y}\wedge\frac{\partial}{\partial z}= 2R_\mathbb{H}\wedge \Lambda_\mathbb{H},\qquad [R_\mathbb{H},\Lambda_\mathbb{H}]_{SN}=0.\end{equation} 
So, $(\mathbb{H},\Lambda_\mathbb{H},R_\mathbb{H})$ is a Jacobi manifold.
\end{example}
\begin{definition} We say that $X$ is a {\it Hamiltonian
vector field} with respect to the Jacobi manifold $(N,\Lambda,R)$ if there exists a function $f\in C^\infty(N)$ such
that
\begin{equation}
X=[\Lambda,f]_{SN}+fR=\widehat\Lambda({\rm d}f)+fR.\end{equation} 
In this case, $f$ is said to be a {\it Hamiltonian function} of $X$ and we write $X=X_f$. If $f$ is also a first-integral of $R$, we say that $f$ is
a {\it good Hamiltonian function} and we call $X_f$ a {\it good Hamiltonian vector field}.
\end{definition}
\begin{example} Given the Jacobi manifold $(\mathbb{H},\Lambda_\mathbb{H},R_\mathbb{H})$ and the vector field $X_1^L\equiv \partial/\partial x$, we have that
\begin{equation}
X_1^L=[\Lambda_\mathbb{H},-y]_{SN}-yR_\mathbb{H}=\widehat{\Lambda}_\mathbb{H}(-{\rm d}y)-yR_\mathbb{H}.\end{equation} 
Hence, $X_1^L$ is a Hamiltonian vector field with Hamiltonian function $h_1^{L}=-y$ with respect to $(\mathbb{H},\Lambda_\mathbb{H},R_\mathbb{H})$.
\end{example}
Each function gives rise to a unique Hamiltonian vector field. Nevertheless, each vector field may admit several Hamiltonian functions. This last result will be illustrated afterwards within relevant results concerning Lie--Jacobi systems. We write ${\rm Ham}(\Lambda,R)$ for the space of Hamiltonian vector fields relative to $(N,\Lambda,R)$. It can be proven that ${\rm Ham}(\Lambda,R)$ is a Lie algebra with respect to the standard Lie bracket of vector fields. Additionally,
a Jacobi manifold allow us to define a Lie bracket on $C^\infty(N)$ given by
\begin{equation}
\{f,g\}_{\Lambda,R}=\Lambda({\rm d}f,{\rm d}g)+f Rg-gRf.\end{equation} 
This Lie bracket becomes a Poisson bracket if and only if $R=0$. Moreover, the morphism $\phi_{\Lambda,R}:f\in C^\infty(N)\mapsto X_f\in {\rm Ham}(\Lambda,R)$ is a Lie algebra morphism. It is important to emphasize that it may not be injective.

\chapter{Lie systems}\markboth{Lie systems}{Chapter 3}
\label{Chap:LieSystems}
\section{The interest of Lie systems}
\setcounter{equation}{0}
\setcounter{theorem}{0}
\setcounter{example}{0}

Lie systems constitute a very specific class of nonautonomous systems of first-order ordinary differential equations 
admitting a {\it superposition rule}, i.e., a $t$-independent function allowing us to write its general solution 
in terms of a generic family of its particular solutions and a set of constants to be related to initial conditions \cite{Dissertationes,LSPartial,LSII,LSIII,LS,Ve93,Ve93II,Ve94}.

The analysis of Lie systems dates back to the end of the XIX century, when Vessiot, Guldberg and Lie \cite{Gu93,LS,Ve93} found the fundamental properties of such systems.
In that time, Lie proven the nowadays called {\it Lie--Scheffers Theorem} \cite{LS}, which states that a nonautonomous system of first-order ordinary differential equations
in normal form admits a superposition rule if and only if it describes the integral curves of a $t$-dependent vector field taking values in a
finite-dimensional real Lie algebra $V$ of vector fields \cite{CGM07}. We call $V$ a {\it Vessiot--Guldberg Lie algebra} \cite{Dissertationes} of the Lie system.

The theory of Lie systems was deeply analyzed in its origins only to be forgotten soon after.
In the XX century, Winternitz and coworkers revived the theory by deriving superposition rules for many relevant systems of first-order ordinary differential and superdifferential equations and studied the existence of particular types of Lie systems on $\mathbb{R}^n$ \cite{AHW81,BecHusWin86,BecHusWin86b,BecGagHusWin90,PW}. 

Posteriorly, there has been strong interest in using modern geometrical techniques \cite{BCHLS,Blaz,BM09II,Ramos,PW} to investigate Lie systems \cite{CLMP,CLR11,CLS12,HLS,Leach1,Leach2,LSKummer}.
For instance, Cari{\~n}ena, Grabowski and Marmo interpreted superposition rules as a particular type of projective foliations \cite{CGM07}. Also, the outlook of Lie systems as curves within finite-dimensional real Lie algebras of vector fields \cite{LS,ShniderWinter} suggests the inspection of generalized distributions of vector fields and Lie group actions to study their properties \cite{CGM00,CGM07,Dissertationes}. 

The Lie--Scheffers Theorem shows that there exists a quite restricted number of Lie systems in the literature: most differential equations cannot be described as a curve in a  
finite-dimensional Lie algebras of vector fields \cite{Dissertationes,In72}. Nevertheless, this fact does not downplay their great importance. For instance,
Lie systems appear in the study of relevant physical models, Mathematics and Control theory \cite{CR,CR05,JP09,SoriWinter}. More specifically, Lie systems have also been employed to study the integrability of Riccati and matrix Riccati equations \cite{CarRam,HWA83,Lafortune}, Floquet theory \cite{Ru08,Ru10,FLV,EV11}, etc. They can also be employed in Quantum Mechanics \cite{CLQuantum,CR} or in the computation of geometric phases in \cite{FLV}.
Furthermore, they have served as a way for the geometric investigation of stochastic equations \cite{JP09}, superequations
\cite{BecGagHusWin87,BecGagHusWin90}, and other topics \cite{CGL11,CR}.

In recent years, the concepts of Lie systems and superposition rules have been extrapolated to higher-order systems of ordinary differential equations \cite{CGL11}. This was employed to study systems of higher-order ordinary differential equations appearing in Classical Mechanics and Mathematics. For instance, the higher-order Riccati equation \cite{CLS12,GL12}, the second- and third-order Kummer--Schwarz \cite{LSKummer}, Milne--Pinney
and dissipative Milne--Pinney equations \cite{CLMP,Leach2}, etcetera \cite{CLR10}.

The theory of Lie systems has also been expanded to the realm of PDE Lie systems by Odzijewicz and Grundland in \cite{OG99} and Cari\~nena, Grabowski and Marmo in \cite{CGM07}.
Additionally, other extensions of Lie systems to other realms have been carried out during latter decades, e.g., to stochastic equations \cite{JP09}.

Many new examples of Lie systems have been found along the research for this thesis. Concerning PDEs Lie systems \cite{CGM07}, the findings of this thesis are almost the only applications of this theory existing in the actual literature \cite{EstLucasSar}. Additionally, we unveil new applications of Lie systems. Among other systems, we analyze through our methods the Hamilton equations describing certain second-order Riccati equations, Cayley--Klein Riccati equations \cite{EstLucasSar}, planar Riccati equations, coupled Riccati
equations, Smorodinsky--Winternitz oscillators \cite{BCHLS}, systems of trigonometric nonlinearities \cite{BCHLS}, among many others \cite{CLS122}.
A plethora of other Lie systems and applications can be consulted in \cite{BCHLS,CGL11,CLS122,HLS}.	

Although the Lie--Scheffers theorem determines whether a system is a Lie system, it can be difficult to determine when a system is a Lie system using this theorem. This is due to the fact that it is complicated to distinguish whether a system possesses a Vessiot--Guldberg Lie algebra of very high dimension or it has no Vessiot--Guldberg Lie algebra at all. In such cases, it is helpful to know all the possible Vessiot--Guldberg Lie algebras on the manifold where the system is defined as this delimites the dimension and structures of the possible Vessiot--Guldberg algebras for the system (see \cite{GL12}). 

Considering the definition of a Lie system, a classification of finite-dimensional Lie algebras of vector fields on a manifold $N$ amounts to a classification
of Lie systems on $N$ \cite{GKP92}. There exists a classification of finite-dimensional Lie algebras on the real plane by Lie \cite{Lie93,FunLS}. This classification
was retaken and refurbished by Gonzalez, Kamran and Olver giving rise to the henceforth called {\it GKO classification} \cite{GKP92}.

As a byproduct of the big number of new applications of Lie systems found in recent years by the author of this thesis and her collaborators, it was found that many of them possess a Vessiot--Guldberg Lie algebra
of Hamiltonian vector fields \cite{CLS122,BBHLS,BCHLS}. This led us to focus on studying geometric structures compatible with Lie systems.
The idea consists in identifying Lie systems admitting Vessiot--Guldberg Lie algebras of Hamiltonian vector fields with respect to a certain geometric structure \cite{cgls,CLS122}.
The geometric structures that we used simplified very much the study of Lie systems.
For example, the case of Hamiltonian vector fields with respect to a Poisson structure led to the derivation of a superposition rule for Riccati equations by using the Casimir function of $\mathfrak{sl}(2,\mathbb{R})$ \cite{BCHLS}.

Lie systems admitting a Vessiot--Guldberg Lie algebra of Hamiltonian vector fields with respect to a Poisson structure, are known
as {\it Lie--Hamilton systems.} Some of the abovementioned systems, as some second-order Riccati equations or the second-order Kummer--Schwarz equations, are Lie--Hamilton systems.
On the other hand, the third-order Kummer--Schwarz equation is not a Lie--Hamilton system. Our {\it no-go theorem} explains how a Lie--Hamilton
system is always a particular type of Lie system but the converse is not true \cite{cgls}.

It is also possible to classify Lie--Hamilton systems attending to their underlying Vessiot--Guldberg Lie algebras on the plane. We give such a classification
in this part of this thesis. Based on the previous GKO classification, we classify Vessiot--Guldberg Lie algebras of Hamiltonian planar vector fields relative to a symplectic structure. Their associated Poisson or Hamiltonian functions shall be obtained \cite{BBHLS}. Table \ref{table3} in Appendix 1 summarizes our results.

One of the main advantages of Lie--Hamilton systems is that they allow us to use a Poisson coalgebra approach to derive their superposition rules algebraically. An ue of the structure of a coalgebra symmetry, first noticed by Ragnisco, Ballesteros and coworkers for the study of integrable systems \cite{BBHMR09,BBR06,BCR96,BR98}, naturally appears in the study of Lie--Hamilton systems. Each Lie--Hamilton system is endowed with a finite-dimensional Lie algebra of Hamiltonian functions. The polynomials on the elements of a basis of this Lie algebra form a Poisson algebra.  Next, the diagonal prolongations of the Lie--Hamilton system, which are employed to obtain its superposition rules \cite{CGM07}, naturally induces another Poisson algebra for each diagonal prolongation. There also exists a primitive coproduct passing from the initial Poisson algebra to the prolongated ones. Using this structure, the Casimir operators of the Vessiot--Guldberg Lie algebra for the Lie--Hamilton system induce naturally, via the coproduct, constants of motion for the diagonal prolongations of the Lie--Hamilton system. These constants of motion allow us to derive superposition rules. 
This method is much easier and geometrically more clarifying than previous methods  to obtain superposition rules based on integrating vector fields that involved solving ODEs or PDEs \cite{CGM00,CGM07,PW}.

Meanwhile, there are certain systems which do not admit a symplectic nor a Poisson structure turning any of their Vessiot--Guldberg Lie algebras into Hamiltonian vector fields. Instead, they do admit a {\it presymplectic form},
a closed two form, turning one of their Vessiot--Guldberg Lie algebras into Hamiltonian vector fields \cite{cgls}.
Indeed, the third-order Kummer--Schwarz, which is not a Lie--Hamilton system, admits such a property.

A class of systems including all the previous types of Lie systems with respect to the mentioned structures are the so called {\it Dirac--Lie systems}. This type of Lie systems
admits Vessiot--Guldberg Lie algebras of Hamiltonian vector fields of a very general kind. Most properties of other Hamiltonian vector
fields with respect to the other structures can naturally  be extended to these ``generalized" vector fields \cite{cgls}.
{\it Generalized Hamiltonian vector fields} imply {\it generalized admissible functions} \cite{cgls} with respect to a Dirac structure and will enjoy
a plethora of properties that will be explained in this thesis. Examples of physical interest will be accompanying as well.

To conclude, it is important to mention {\it Jacobi--Lie systems}. In a similar fashion, Jacobi--Lie systems are
Lie systems admitting Hamiltonian vector fields with respect to a {\it Jacobi structure}.  Furthermore, we can find {\it $k$-symplectic Lie systems}, which admit a Vessiot--Guldberg Lie algebra of Hamiltonian vector fields with respect to the presymplectic forms 
of a $k$-symplectic structure \cite{LVT,LucasVil}. 

\medskip

The content of this third Chapter by sections corresponds with

\begin{itemize}

%

\item {\bf Section 1: Lie systems}:  We define Lie systems, superposition rules and Vessiot--Guldberg Lie algebras. We describe the Lie--Scheffers theorem.
Subsequently, four important examples of Lie systems are contemplated: the Riccati equation, a planar Riccati equation with $t$-dependent real coefficients,
a system of coupled Riccati equations and a matrix Riccati equation. Next, we review Lie systems on Lie groups whose superposition rules are given through the action of the Lie group on itself.
The theorem of reduction of Lie systems is briefly commented to solve Lie systems defined on Lie groups.
We show how to derive superposition rules by using the so called diagonal prolongations of the vector fields of a Vessiot--Guldberg Lie algebra.
As a new application, we study three Lie systems: the second-order Riccati equation and the Kummer--Schwarz equations of second- and third-order.
We obtain a solution in terms of a Lie group action for all these Lie systems, which have Vessiot--Guldberg Lie algebras isomorphic to $\mathfrak{sl}(2,\mathbb{R})$. 
To conclude, we show the link between the Schwarzian derivative and the third-order Kummer--Schwarz equation and we find a superposition rule for the latter.

\newpage

\item {\bf Section 2: Lie--Hamilton systems}: We introduce Lie--Hamilton systems.
We show that planar Riccati equations are Lie--Hamilton systems and we study their properties. Some new examples of Lie--Hamilton systems are given: the coupled Riccati equations, the Hamilton equations of the second-order Riccati equations, the Hamilton equations of the second-order Kummer--Schwarz equations,
and the Hamilton equations of the Smorodinsky--Winternitz oscillator.

We study $t$-dependent Hamiltonians for Lie--Hamilton systems.
We analyze the properties of constants of motion for Lie--Hamilton systems. We use Poisson coalgebras to divise an algebraic and geometric method for the
construction of superposition rules. 
We will explicitly construct superposition rules for certain systems of physical and mathematical relevance: the classical Ermakov system, four coupled Riccati equations, the Hamilton equations of the
second-order Kummer--Schwarz equations, the Hamilton equations of the Smorodinsky--Winternitz oscillator and a system with trigonometric nonlinearities.

\item {\bf Section 3: Classification of Lie--Hamilton systems on the plane}: First, we introduce the notion generic points of Lie algebras, primitive and imprimitive Lie algebras, integrating factors and modular generating systems. From these concepts, 
a number of new results arises and classifies Vessiot--Guldberg Lie algebras of Hamiltonian vector fields on the plane (with respect to a Poisson structure). To do so, we rely on the GKO classification for finite-dimensional Lie algebras on the plane. 
We obtain that, out of the initial nondiffeomorphic 28 classes of finite-dimensional Lie algebras of vector fields on $\mathbb{R}^2$ given by the GKO classification, 
only 12 of them consist of Hamiltonian vector fields. We will inspect all of them and describe their properties.

\item {\bf Section 4: Applications of Lie--Hamilton systems on the plane}:
The classification of Lie--Hamilton systems on the plane enables us to classify Lie--Hamilton systems on the plane with physical applications. 
 In this section we aim to give examples of Lie--Hamilton systems on the plane of physical and mathematical interest. 

We start by Lie--Hamilton systems with Lie algebras isomorphic to $\mathfrak{sl}(2,\mathbb{R})$. Some examples are: the Milne--Pinney equation, the second-order Kummer--Schwarz equations,
and the complex Riccati equation with $t$-dependent real coefficients. 

Then, we study Lie--Hamilton systems with biological and physical applications: the generalized Buchdahl equations, time-dependent Lotka-Volterra systems, qua\-dra\-tic polynomial models and viral infection models.
More Lie--Hamilton systems on the plane with relevance are displayed: the Cayley--Klein Riccati equations which is a compendium of Double-Clifford or split-complex Riccati equations and the Dual-Study Riccati equation.
Next, we present other  Lie--Hamilton systems with Vessiot--Guldberg Lie algebras isomorphic to  $\mathfrak{sl}(2,\mathbb{R})$ and we study the equivalence between them. More specifically, we analyze coupled Riccati equations, the Milne--Pinney equations,
the second-order Kummer--Schwarz equation and a planar diffusion Riccati system.
Later, we present Lie--Hamilton systems of two-photon type, which encompasses: the dissipative harmonic oscillator and the second-order Riccati equation.
To conclude, we study Lie systems with Vessiot--Guldberg Lie algebras isomorphic to $\mathfrak{h}_2$. This covers: the complex Bernoulli equation with $t$-dependent real coefficients, generalized Buchdahl equations
$t$-dependent Lotka--Volterra systems, etc.

\item {\bf Section 6: Dirac--Lie systems}: We motivate the study of Dirac--Lie systems by showing that third-order Kummer--Schwarz equations
and a Riccati system cannot be studied through Lie--Hamilton systems. Nevertheless, they admit a Vessiot--Guldberg Lie algebra of Hamiltonian vector fields relative to a presymplectic structure. This has led us to make use of Dirac structures and its corresponding geometry.
The non Lie--Hamiltonian nature of the mentioned two examples also suggests to enunciate the so called ``no-go'' theorem, which tells us when a Lie system cannot be a Lie--Hamilton one.


We state how a Dirac--Lie system covers Lie systems with Hamiltonian Vessiot--Guldberg Lie algebras with respect to presymplectic and Poisson structures as particular cases.
 In this way, we generalize Lie--Hamilton systems to a more general class of Lie systems  covering the aforementioned structures.

In similar fashion as did for Lie--Hamilton systems, we derive Dirac--Lie Hamiltonians and all their corresponding concepts are reformulated: the concept of diagonal prolongation is redefined
to work out their superposition rules. Certain properties for their constants of motion are reconsidered.
To conclude, we show a superposition rule for third-order Kummer--Schwarz equations treated with Dirac--Lie systems theory. It is also obvious
that the previously studied coalgebra method to derive superposition rules can also be extrapolated to Dirac--Lie systems. To conclude, we introduce {\it bi-Lie-Dirac systems}, or Lie systems admitting
two compatible Dirac structures. A method for generating bi-Lie-Dirac systems is given. We find out that our techniques can be applied to Schwarzian Korteweg-de Vries (SKdV) equations \cite{As10,Mar11}.
 This provides a new approach to the study of these equations. 

Furthermore, we derive soliton-type solutions for Schwarzian--KdV equations, namely shape-preserving traveling wave solutions. 
We show how Lie systems and our methods can be applied to provide B\"acklund transformations for certain solutions of these equations.
This can be considered as one of the first applications of Dirac--Lie systems in the study of PDEs of physical and mathematical interest from the point of view of the theory of Lie systems.

\item {\bf Section 7: Jacobi--Lie systems}: In this section we search for Vessiot--Guldberg Lie algebras of Hamiltonian vector fields relative to a Jacobi structure on the plane. 

As in past sections, we introduce the concept of Hamiltonian vector field with respect to a Jacobi structure and the consequent definition
of Jacobi--Lie systems.
Also, it is possible to define Jacobi--Lie Hamiltonians and to find a Lie algebra of Hamiltonian functions with respect to a defined Jacobi--Lie bracket.

It is our purpose to classify all Jacobi--Lie systems on the plane (up to diffeomorphisms). For this, we have relied on our previous classification of
Lie--Hamilton systems on the plane and the GKO classification. 
The classification of Jacobi--Lie systems on the plane is displayed in Table \ref{table9} in Appendix 1.
An explicit example is given: the continuous Heisenberg group is understood as a Jacobi manifold and a Jacobi--Lie system is constructed out of its left-invariant
vector fields.
\end{itemize}

\newpage

\section{Lie systems}

\setcounter{equation}{0}
\setcounter{theorem}{0}
\setcounter{example}{0}
On a first approximation, a {\it Lie system} is a first-order system of ODEs that admits a superposition rule. In this section we will describe Lie systems, their main properties and their most relevant results.

\subsection{Lie systems and superposition rules}
Let us now turn to describe some fundamental notions appearing in the theory of Lie
systems.

\begin{definition} A {\it superposition rule} depending on $m$ particular
solutions for a system $X$ on $N$ 
 is a function $\Phi:N^{m}\times N\rightarrow
N$, $x=\Phi(x_{(1)}, \ldots,x_{(m)};\lambda)$, such that the general
 solution $x(t)$ of $X$ can be brought into the form
  $x(t)=\Phi(x_{(1)}(t), \ldots,x_{(m)}(t);\lambda),$
where $x_{(1)}(t),\ldots,x_{(m)}(t)$ is any generic family of
particular solutions and $\lambda$ is a point of $N$ to be related to initial
conditions. 
 \end{definition}

\begin{example}\normalfont
The Riccati equation on the real line \cite{PW}
\begin{equation}\label{Ric}
\frac{dx}{dt}=a_0(t)+a_1(t)x+a_2(t)x^2,
\end{equation}
where $a_0(t),a_1(t),a_2(t)$ are arbitrary $t$-dependent functions, admits the {superposition rule}
$\Phi:(x_{(1)},x_{(2)},x_{(3)};k)\in \mathbb{R}^3\times \mathbb{R}\mapsto x\in \mathbb{R}$ given by
\begin{equation}\label{SupRiccati}
x=\frac{x_{(1)}(x_{(3)}-x_{(2)})+kx_{(2)}(x_{(1)}-x_{(3)})}{(x_{(3)}-x_{(2)})+k(x_{(1)}-x_{(3)})}.
\end{equation}
In other words, the general solution, $x(t)$, to (\ref{Ric}) can be brought into the form
\begin{equation}
x(t)=\frac{x_{(1)}(t)(x_{(3)}(t)-x_{(2)}(t))+kx_{(2)}(t)(x_{(1)}(t)-x_{(3)}(t))}{(x_{(3)}(t)-x_{(2)}(t))+k(x_{(1)}(t)-x_{(3)}(t))}.
\end{equation}
where $x_{(1)}(t),x_{(2)}(t),x_{(3)}(t)$ are three different particular
solutions and $k$ is an arbitrary constant.
\end{example}

The conditions ensuring that a system $X$ possesses a superposition rule are
stated
 by the {\it Lie--Scheffers Theorem} \cite[Theorem 44]{LS}. A modern statement
of this relevant result is described
 next (for a modern geometric description see \cite[Theorem 1]{CGM07}). 

\begin{theorem}{\bf (The Lie--Scheffers Theorem)} A first-order system
\begin{equation}\label{genLiesys1}
 \frac{dx}{dt}=F(t,x),\qquad x\in N,
\end{equation}
admits a superposition rule if and only if $X$ can be written as
 \begin{equation}\label{genLiesys}
X_t={{\sum_{\alpha=1}^r}}b_\alpha(t)X_\alpha
\end{equation}
for a certain family $b_1(t),\ldots,b_r(t)$  of $t$-dependent functions and a
family  $X_1,\ldots,X_r$ of vector fields  on N spanning 
an $r$-dimensional real Lie algebra of vector fields.
\end{theorem}

The Lie--Scheffers 
Theorem yields that every Lie system $X$ is related to (at least) one
finite-dimensional real Lie algebra of vector fields 
$V$, a so called {\it Vessiot--Guldberg Lie algebra}, satisfying that
$\{X_t\}_{t\in\mathbb{R}}\subset V$. This implies 
that $V^X$ must be finite-dimensional. Conversely, if $V^X$ is
finite-dimensional, 
this Lie algebra can be chosen as a Vessiot--Guldberg Lie algebra for $X$. This
proves the following
 theorem, which motivates, among other reasons, the definition of $V^X$ 
\cite{Dissertationes}. 

\begin{theorem}\label{ALST}{\bf (The abbreviated Lie--Scheffers Theorem)} A
system $X$ admits a superposition rule if and only if $V^X$ is
finite-dimensional.
\end{theorem}
\begin{example}\normalfont
In order to illustrate the above concepts, let us consider the {Riccati equation} \eqref{Ric}. 
 Observe that (\ref{Ric}) is  the system associated with
the $t$-dependent vector field $X=a_0(t)X_1+a_1(t)X_2+a_2(t)X_3$, where
\begin{equation}
X_1=\frac{\partial}{\partial x},\qquad X_2=x\frac{\partial}{\partial x},\qquad X_3=x^2\frac{\partial}{\partial x}
\end{equation}
span a Vessiot--Guldberg Lie algebra $V$ for (\ref{Ric}) isomorphic to $\mathfrak{sl}(2,\mathbb{R})$. The Lie--Scheffers
Theorem shows that Riccati equations must admit a superposition rule. Indeed,
the general solution of Riccati equation (\ref{Ric}) can be brought into the form
$\Phi:(x_{(1)},x_{(2)},x_{(3)};k)\in\mathbb{R}^3\times\mathbb{R}\mapsto
x\in\mathbb{R}$ given by (\ref{SupRiccati})
enabling us to write their general solutions as $x(t)=\Phi(x_{(1)}(t),x_{(2)}(t),x_{(3)}(t);k)$.

\end{example}

\begin{example} \normalfont
Consider the system of differential equations
\begin{equation}\label{Riccati21}
\frac{{\rm d} x}{{\rm d} t}=a_0(t)+a_1(t)x+a_2(t)(x^2-y^2),\qquad \frac{{\rm d} y}{{\rm d} t}=a_1(t)y+a_2(t)2xy,
\end{equation}
with $a_0(t),a_1(t),a_2(t)$ being arbitrary $t$-dependent real functions. This system is a particular type of planar Riccati equation briefly studied in \cite{Eg07}. By writing $z=x+{\rm i}y$, we find that (\ref{Riccati21}) is equivalent to
\begin{equation}
\frac{{\rm d} z}{{\rm d} t}=a_0(t)+a_1(t)z+a_2(t)z^2,\qquad z\in\mathbb{C},
\end{equation}
which is a  particular type of complex Riccati equations, whose study has attracted some attention.  Particular solutions of periodic equations of this type have been investigated in \cite{Ju97,Or12}.

Every particular solution $(x(t),y(t))$ of (\ref{Riccati21}) obeying that $y(t_0)=0$ at $t_0\in\mathbb{R}$ satisfies that $y(t)=0$ for every $t\in\mathbb{R}$. In such a case, $x(t)$ is a particular solution of a {\em real} Riccati equation \cite{PW}. This suggests us to restrict ourselves to studying (\ref{Riccati21}) on
$
\mathbb{R}^2_{y\neq 0}\equiv\{(x,y)\,|\, y \neq 0\}\subset \mathbb{R}^2
$.

Let us show that (\ref{Riccati21}) on $\mathbb{R}^2_{y\neq 0}$ is a Lie system. This is related to the $t$-dependent vector field 
\begin{equation}\label{PR}
X_t=a_0(t)X_1+a_1(t)X_2+a_2(t)X_3,
\end{equation}
where
\begin{equation}
X_1= \frac{\partial}{\partial x},\qquad X_2= x\frac{\partial}{\partial x}+y\frac{\partial}{\partial y} ,\qquad X_3= (x^2-y^2)\frac{\partial}{\partial x}+2xy\frac{\partial}{\partial y}
\label{vectRiccati21}
\end{equation}
span a Vessiot--Guldberg real Lie algebra $V\simeq \mathfrak{sl}(2)$
with commutation relations
\begin{equation}\label{aa}
[X_1,X_2]=X_1,\qquad [X_1,X_3]=2X_2,\qquad [X_2,X_3]=X_3 .
\end{equation}

 Hence, $\{X_t\}_{t\in\mathbb{R}}\subset V^X\subset V$ and $V^X$ is finite-dimensional, which makes $X$ into a Lie system. Hence, it admits a superposition rule.
\end{example}

\begin{example}\normalfont
Consider the following system of Riccati equations \cite{Dissertationes}
\begin{equation}\label{coupRiceq}
\frac{{\rm d}x_i}{{\rm d}t}=a_0(t)+a_1(t)x_i+a_2(t)x_i^2,\qquad i=1,\ldots,n,
\end{equation}
with $a_0(t),a_1(t),a_2(t)$ being arbitrary $t$-dependent functions.
System \eqref{coupRiceq} is associated with the $t$-dependent vector field $X_{R}=a_0(t)X_1+a_1(t)X_2+a_2(t)X_3,$
where
\begin{equation}\label{VF1}
X_1=\sum_{i=1}^n\frac{\partial}{\partial x_i},\qquad X_2=\sum_{i=1}^nx_i\frac{\partial}{\partial x_i},\qquad X_3=\sum_{i=1}^nx_i^2\frac{\partial}{\partial x_i}.
\end{equation}
Hence, $X_{R}$ takes values in a Vessiot--Guldberg Lie algebra $\langle X_1,X_2,X_3\rangle \simeq \mathfrak{sl}(2,\mathbb{R})$. 
Lie proven that each Lie system on the real line is, up to a local change of variables, a particular case of system (\ref{coupRiceq}) for $n=1$ \cite{Dissertationes,Lie1880}.
\end{example}

\begin{example}\normalfont
Consider the matrix Riccati equations on $\mathbb{R}^2$~\cite{Abou,HLS,Levin}, namely
 \begin{equation}\label{MatRic3}
\left\{
\begin{aligned}
\frac{{\rm d}x}{{\rm d}t}&=a_0(t)+a_{11}(t)x+a_{12}(t)y+b_0(t) x^2+ b_1(t) xy,\\
\frac{{\rm d}y}{{\rm d}t}&=a_1(t)+a_{21}(t)x+a_{22}(t)y+b_0(t)xy+b_1(t)y^2,
\end{aligned}\right.
\end{equation}
with $a_0(t),\ a_1(t),\ a_{11}(t),\ a_{12}(t),\ a_{21}(t),\ a_{22}(t),\ b_0(t),\ b_1(t)$ being any $t$-dependent functions. 

\newpage

The associated $t$-dependent vector field reads

\begin{align}
X=a_0(t)X_0+a_1(t)X_1+a_{11}(t)X_2+&a_{12}(t)X_3+a_{21}(t)X_4+\nonumber \\
&a_{22}(t)X_5+b_0(t)X_6+b_1(t)X_7,
\end{align}
where the vector fields
\begin{equation}
\begin{gathered}
X_0=\frac{\partial}{\partial x},\qquad X_1=\frac{\partial}{\partial y},\qquad X_2=x\frac{\partial}{\partial x},\qquad X_3=y\frac{\partial}{\partial x},\qquad X_4=x\frac{\partial}{\partial y}, \\[2pt]
 X_5=y\frac{\partial}{\partial y},\qquad X_6=x^2\frac{\partial}{\partial x}+xy\frac{\partial}{\partial y},\qquad X_7=xy\frac{\partial}{\partial x}+y^2\frac{\partial}{\partial y}\label{vv}
\end{gathered}
\end{equation}
 span an eight-dimensional Lie algebra $V^{\rm MR}$  with non-vanishing commutation relations
\begin{equation}\label{ConRel3}
\begin{array}{lllll}
[X_0,X_2]=X_0,    &\,\, [X_0,X_4]=X_1,&\,\, [X_0,X_6]=2X_2+X_5, &[X_0,X_7]=X_3,\\[2pt]
[X_1,X_3]=X_0,    &\,\, [X_1,X_5]=X_1,  &\, [X_1,X_6]=X_4,& [X_1,X_7]=X_2+2X_5,\\[2pt]
[X_2,X_3]=-X_3,  &\,\,[X_2,X_4]=X_4,&\,\, [X_2,X_6]=X_6,&[X_3,X_4]=X_5-X_2,\\[2pt]
[X_3,X_5]=-X_3,&\,\,[X_3,X_6]=X_7,&\,\,[X_4,X_5]=X_4, & [X_4,X_7]=X_6.   \\[2pt]
  [X_5,X_7]=X_7.& & &
\end{array}
\end{equation}
Since $X$ takes value in the Lie algebra $V^{\rm MR}$, it becomes a Lie system \cite{HLS}. Indeed, it can be proven that $V^{\rm MR}$ is isomorphic to $\mathfrak{sl}(3,\mathbb{R})$
(see Table \ref{table1}  in Appendix 1 \cite{GKP92}).

\end{example}

\subsection{Lie systems on Lie groups}

Every Lie system $X$ associated with a Vessiot--Guldberg Lie algebra $V$  gives
rise by integrating $V$
to a (generally local) Lie group action $\varphi:G\times
N\rightarrow N$ whose fundamental vector fields are the elements of $V$ and such that $T_eG\simeq V$ with $e$ being the neutral element of $G$ \cite{Palais}. If we assume $X$ to be of the form (\ref{genLiesys}), its associated Lie group action
allows us to write the general solution of $X$ as
\begin{equation}\label{mix}
x(t)=\varphi(g_1(t),x_0), \qquad x_0\in\mathbb{R}^n,
\end{equation}
with $g_1(t)$ being a particular solution of
\begin{equation}\label{eqLie}
\frac{dg}{dt}=-\sum_{\alpha=1}^rb_\alpha(t)X_\alpha^R(g),
\end{equation}
where $X^R_1,\ldots,X_r^R$ is a certain basis of right-invariant vector fields on $G$
such that $X^R_\alpha(e)={\rm a_\alpha}\in T_eG$, with $\alpha=1,\ldots,r$, and each ${\rm a_\alpha}$ is the element of $T_eG$ associated with the fundamental vector field $X_\alpha$ (see \cite{CGM00} for
details). 

Since $X^R_1,\ldots,X_r^R$ span a finite-dimensional real Lie algebra, the Lie--Scheffers Theorem guarantees that (\ref{eqLie}) admits a superposition rule and becomes a Lie system. Indeed, as the right-hand
side of (\ref{eqLie}) is invariant under every $R_{g_0*}$ with $R_{g_0}:g'\in G\mapsto g'\cdot g_0\in G$ being the multiplication on the right on $G$ by $g_0\in G$, the general solution to \eqref{eqLie}
can be brought into the form
\begin{equation}\label{SupGroup}
g(t)=R_{g_0}g_1(t),\qquad g_0\in G,
\end{equation}
where $g_1(t)$ is a particular solution of (\ref{eqLie}) \cite{CGM00}. 
In other words, (\ref{eqLie}) admits a superposition rule.

Several methods can now be applied to solve (\ref{eqLie}). If $\mathfrak{g}\simeq T_eG$ is solvable, the theory of reduction of Lie systems allows us to integrate (\ref{eqLie}) by quadratures. More generally, we can use a  {\it Levi decomposition} of $\mathfrak{g}$ to write $\mathfrak{g}\simeq \mathfrak{r}\oplus_s \left(\mathfrak{s}_1\oplus\cdots\oplus\mathfrak{s}_l\right)$, 
where $\mathfrak{r}$ is the radical of $\mathfrak{g}$, the Lie algebra $\mathfrak{s}_1\oplus\cdots\oplus\mathfrak{s}_l$ is the direct sum of a family 
of simple Lie subalgebras of $\mathfrak{g}$, and $\oplus_s$ denotes a semi-direct sum of $\mathfrak{r}$ and $\mathfrak{s_1}\oplus\cdots\oplus\mathfrak{s}_l$. 
The {\it theorem of reduction of Lie systems} \cite[Theorem 2]{CarRamGra} yields, with the aid of the previous decomposition, that the solution of a Lie system (\ref{eqLie}) defined on a Lie group with 
Lie algebra $\mathfrak{s}_1\oplus\dots\oplus\mathfrak{s}_l$ enables us to construct a $t$-dependent change of variables that transforms (\ref{eqLie}) into a Lie system on a Lie group with 
Lie algebra $\mathfrak{r}$, which is integrable by quadratures  (see \cite{CarRamGra} for details). Summarizing, the explicit integration of (\ref{eqLie}) reduces to providing a particular solution of a Lie system related to $\mathfrak{s}_1\oplus\cdots\oplus\mathfrak{s}_l$.

\subsection{An algorithm to derive superposition rules}

General solutions of Lie systems can also be investigated through superposition rules. There exist different procedures to derive them \cite{AHW81,CGM07,PW}, but we hereafter use the method
devised in \cite{CGM07}, which is based on the {\it diagonal prolongation} notion \cite{CGM07,Dissertationes}.

\begin{definition} Given a $t$-dependent vector field $X$ on $N$, its  {\it diagonal prolongation} $\widetilde X$ to $N^{(m+1)}$ is the unique $t$-dependent vector field on $N^{(m+1)}$ such that
\begin{itemize}
\item Given ${\rm pr}:(x_{(0)},\ldots,x_{(m)})\in N^{(m+1)}\mapsto x_{(0)}\in N$, we have that  ${\rm pr}_*\widetilde X_t=X_t$  $\forall t\in\mathbb{R}$. 
\item $\widetilde X$ is invariant under the permutations $x_{(i)}\leftrightarrow x_{(j)}$, with $i,j=0,\ldots,m$.
\end{itemize}
\end{definition}
In coordinates, we have that
\begin{equation}
X_j=\sum_{i=1}^n X^i(t,x)\frac{\partial}{\partial x_i}\Rightarrow \widetilde{X}=\sum_{j=0}^{m+1} X_{j}=\sum_{j=0}^{m+1}\sum_{i=1}^n X^i(t,x)\frac{\partial}{\partial x_i}.
\end{equation}

\newpage

The procedure to determine superposition rules described in \cite{CGM07} goes as follows

\begin{itemize}
 \item Take a basis $X_1,\ldots,X_r$ 
of a Vessiot--Guldberg Lie algebra $V$ associated with the Lie system.
\item  Choose the minimum integer $m$ so that 
the diagonal prolongations to $N^{m}$ of $X_1,\dots,X_r$ are linearly independent at a generic point. 
\item Obtain $n$ functionally independent first-integrals $F_1,\ldots,F_n$ common to all the diagonal prolongations, $\widetilde X_1,\ldots,\widetilde X_r,$ to $N^{(m+1)}$, 
for instance, by the {\it method of characteristics}. We require such functions to hold that
\begin{equation}
\frac{\partial (F_1,\dots,F_n)}{\partial \left((x_1)_{(0)},\dots,(x_n)_{(0)}\right)}\neq 0.
\end{equation}
\item Assume that these integrals take certain constant values, i.e., $F_i=k_i$ with $i=1,\ldots,n$, and
employ these equalities to express the variables $(x_1)_{(0)},\dots,(x_n)_{(0)}$ in terms of the variables of the other copies of $N$ within $N^{(m+1)}$ and the constants $k_1,\ldots,k_n$. The obtained 
 expressions constitute a superposition rule in terms of any generic family of $m$ particular solutions and $n$ constants.
\end{itemize}

There exists another method to obtain superposition rules frequently used in applications \cite{PW}. This second method goes as follows. The general solution $x(t)$ of $X$ can be written as $x(t)=\varphi(g_1(t),x(0))$, where $g_1(t)$ is the solution of the associated Lie system (\ref{eqLie}) with $g_1(0)=e$ and $\varphi$ is the Lie group action obtained by integrating  a Vessiot--Guldberg Lie algebra of $X$. The curve $g_1(t)$ can
be characterized as the unique solution to the algebraic system $x_p(t)=\varphi(g_1(t),x_p(0))$, where $x_p(t)$ ranges over a ``sufficient large set'' of
particular solutions of $X$ \cite{PW}. Obtaining $g_1(t)$ from these equations and substituting in $x(t)=\varphi(g_1(t),x(0))$, we can write  $x(t)$ in terms of the particular solutions $x_p(t)$ and the parameter $x(0)$ giving rise to a superposition rule for $X$. In general, this and previous method are convenient to derive superposition rules for different types of systems.

\subsection{New applications of Lie systems}

Here we show a few new applications of Lie systems through examples of physical relevance. These applications were obtained by the author
of this thesis and her coworkers for the first time.

\subsubsection{The second order Riccati equation}

The most general class of second-order Riccati equations is given by the family of second-order 
differential equations of the form
\begin{equation}\label{NLe2}
\frac{d^2x}{dt^2}+(f_0(t)+f_1(t)x)\frac{dx}{dt}+c_0(t)+c_1(t)x+c_2(t)x^2+c_3(t)x^3=0,
\end{equation}
with
\begin{equation}
f_1(t)=3\sqrt{c_3(t)},\qquad f_0(t)=\frac{c_2(t)}{\sqrt{c_3(t)}}-\frac{1}{2c_3(t)}\frac{dc_3}{dt}(t), \qquad c_3(t)\neq 0
\end{equation}
and $c_0(t),c_1(t),c_2(t)$ being arbitrary $t$-dependent functions with $c_3(t)>0$ for every $t\in\mathbb{R}$. 
These equations arise by reducing third-order linear differential equations through a dilation symmetry and a $t$-reparametrization \cite{CRS05}. Their interest is due to their use in the study of several physical and mathematical problems
 \cite{CL11Sec,CRS05,CC87,GGL08,GL99}.

It was recently discovered that a quite general family of second-order Riccati equations (\ref{NLe2}) admits a $t$-dependent non-natural regular Lagrangian of the form
\begin{equation}
L(t,x,v)=\frac{1}{v+U(t,x)},
\end{equation}
with $U(t,x)=a_0(t)+a_1(t)x+a_2(t)x^2$ and $a_0(t),a_1(t),a_2(t)$ being certain functions related to the
 $t$-dependent coefficients of (\ref{NLe2}), see \cite{CRS05}. Therefore,
\begin{equation}\label{Legtrricsec}
p=\frac{\partial L}{\partial v}=\frac{-1}{(v+U(t,x))^2},
\end{equation}
and the image of the Legendre transform $\mathbb{F}L:(t,x,v)\in \mathcal{W}\subset \mathbb{R}\times {\rm T}\mathbb{R}\mapsto (t,x,p)\in \mathbb{R}\times {\rm T}^*\mathbb{R}$, where $\mathcal{W}=\{(t,x,v)\in \mathbb{R}\times {\rm T}\mathbb{R}\mid v+U(t,x)\neq 0\}$, is the open submanifold $\mathbb{R}\times\mathcal{O}$ where $\mathcal{O}=\{(x,p)\in {\rm T}^*\mathbb{R}\mid p< 0\}$. The Legendre transform is not injective, as $(t,x,p)=\mathbb{F}L(t,x,v)$ for $v={\pm 1}/{\sqrt{-p}}-U(t,x)$. Nevertheless, it can become so by restricting it to the open $\mathcal{W}_+=\{(t,x,v)\in \mathbb{R}\times {\rm T}\mathbb{R}\mid v+U(t,x)>0\}$. In such a case, $v=1/{\sqrt{-p}}-U(t,x)$ and we can define over $\mathbb{R}\times\mathcal{O}$ the $t$-dependent Hamiltonian
\begin{equation}
 h(t,x,p)=p\left(\frac 1{\sqrt{-p}}-U(t,x)\right)-\sqrt{-p}=-2\sqrt{-p}- p\, U(t,x).
\end{equation}
Its Hamilton equations read
\begin{equation}
\left\{
\begin{aligned}\label{Hamil12}
\frac{dx}{dt}&=\frac{\partial h}{\partial p}=\frac{1}{\sqrt{-p}}-U(t,x)=\frac{1}{\sqrt{-p}}-a_0(t)-a_1(t)x-a_2(t)x^2,\\
\frac{dp}{dt}&=-\frac{\partial h}{\partial x}= p\frac{\partial U}{\partial x}(t,x)= p(a_1(t)+2a_2(t)x).
\end{aligned}\right.
\end{equation}

Since the general solution $x(t)$ of a second-order Riccati equation (\ref{NLe2}) can be recovered from the general solution $(x(t),p(t))$ of its corresponding system (\ref{Hamil12}), the analysis of the latter provides information about general solutions of such second-order Riccati equations.

The important point now is that system (\ref{Hamil12}) is a Lie system. Indeed, consider the vector fields on $\mathcal{O}$ of the form
\begin{equation}
\begin{gathered}\label{commrelsecordric}
X_1=\frac{1}{\sqrt{-p}}\frac{\partial}{\partial x},\quad
X_2=\frac{\partial}{\partial x},\quad
X_3=x\frac{\partial}{\partial x}-p\frac{\partial}{\partial p},\quad
X_4=x^2\frac{\partial}{\partial x}-2xp\frac{\partial}{\partial p},\\
X_5=\frac{x}{\sqrt{-p}}\frac{\partial}{\partial x}+2\sqrt{-p}\frac{\partial}{\partial p}.\\
\end{gathered}
\end{equation}

Their non-vanishing commutation relations read
\begin{equation}\label{ComRel21}
\begin{gathered}
\left[X_1,X_3\right]=\frac 12X_1,\qquad [X_1,X_4]=X_5,\qquad
[X_2,X_3]=X_2,\qquad [X_2,X_4]=2X_3,\\  \left[X_2,X_5\right]=X_1,
\qquad \left[X_3,X_4\right]=X_4,\qquad [X_3,X_5]=\frac 12 X_5,\\
\end{gathered}
\end{equation}
and therefore span a five-dimensional Lie algebra $V$ of vector fields. Additionally, the $t$-dependent 
vector field $X_t$ associated with (\ref{Hamil12}) holds
\begin{equation}
X_t=X_1-a_0(t)X_2-a_1(t)X_3-a_2(t)X_4.\label{F22}
\end{equation}
In view of (\ref{F22}), system (\ref{Hamil12}) is a Lie system. Note also that a similar result would have been obtained by restricting the Legendre transform over the open $\mathcal{W}_-=\{(t,x,v)\in\mathbb{R}\times{\rm T}\mathbb{R}\mid v+U(t,x)<0\}$.

Let us use the theory of Lie systems to reduce the integration of (\ref{Hamil12}) to solving a Lie system on a Lie group. Using a Levi decomposition of $V$, we get $V\simeq V_1\oplus_sV_2$, with $V_2=\langle X_2,X_3,X_4\rangle$ being a semisimple Lie algebra isomorphic to 
   $\mathfrak{sl}(2,\mathbb{R})$ and $V_1=\langle X_1,X_5\rangle$ being the radical of $V$. Hence, 
   $V$ is isomorphic to the Lie algebra of a Lie group $G\equiv\mathbb{R}^2\rtimes SL(2,\mathbb{R})$, where
    $\rtimes$ denotes a semidirect product of $\mathbb{R}^2$ by $SL(2,\mathbb{R})$, and there exists a local 
    action $\Phi:G\times \mathcal{O}\rightarrow \mathcal{O}$ whose fundamental vector fields are 
     those of $V$. It is a long, but simple, computation to show that 
\begin{equation}
\Phi\left(\left(
(\lambda_1,\lambda_5),
\left(\begin{array}{cc}
 \alpha&\beta\\
\gamma&\delta
\end{array}\right)\right),(x,p)\right)=\left(\frac{\sqrt{-\bar p}\bar x-\lambda_1}{\sqrt{-\bar p}+\lambda_5
},-(\sqrt{-\bar p}+\lambda_5)^2\right),
\end{equation}
where $\bar x=(\alpha x+ \beta)/(\gamma x+\delta)$, $\bar p={p}\left(\gamma x+\delta\right)^2$ and $\alpha\delta-\beta\gamma=1$, is one of such actions (for a detailed example of how to derive these actions see \cite[Ch. 2]{Dissertationes}).

The above action enables us to write the general solution $\xi(t)$ of system (\ref{Hamil12}) in the form $\xi(t)=\Phi(g(t),\xi_0)$, 
where $\xi_0\in N$ and $g(t)$ is a particular solution of
\begin{equation}\label{HamGrup}
\frac{dg}{dt}=-\left(X^R_1(g)-a_0(t)X^R_2(g)-a_1(t)X^R_3(g)-a_2(t)X^R_4(g)\right),\qquad g(0)=e,
\end{equation}
on $G$, with the $X^R_\alpha$ being a family of right-invariant vector fields on $G$ whose vectors 
$X^R_\alpha(e)\in T_eG$ close on the opposite commutation relations of the vector fields $X_\alpha$ (cf. \cite{Dissertationes}). 

We now turn to apply the theory of reduction for Lie systems. Since $T_eG\simeq \mathbb{R}^2\oplus_s
 \mathfrak{sl}(2,\mathbb{R})$, a particular solution of a Lie system of the form (\ref{HamGrup}) but on $SL(2,\mathbb{R})$, which amounts to integrating (first-order) Riccati equations (cf. \cite{AHW81,Dissertationes}), provides us with a transformation which maps system (\ref{HamGrup}) into an easily integrable Lie system on $\mathbb{R}^2$. In short, the explicit determination of the general solution of a second-order Riccati equation reduces to solving Riccati equations. 

Another way of analyzing the solutions of (\ref{Hamil12}) is based on the determination of a superposition rule. 
According to  the method sketched in Subsection 2.3, a superposition rule for a Lie system (\ref{Hamil12}), which 
admits a decomposition of the form (\ref{F22}), can be obtained through two common functionally independent first-integrals $I_1,I_2$, for the 
diagonal prolongations $\widetilde{X}_{1},\widetilde{X}_{2},\widetilde{X}_{3},\widetilde{X}_{4},\widetilde{X}_{5}$ 
to a certain ${\rm T}^*N^{(m+1)}$ provided their prolongations to ${\rm T}^*\mathbb{R}^{m}$ are linearly independent at a generic point and
\begin{equation}
\frac{\partial (I_1,I_2)}{\partial (x_{(0)},p_{(0)})}\neq 0.
\end{equation}
 In our case, it can easily be verified that $m=3$. The resulting first-integrals, derived through a long but easy calculation (see \cite{CL11Sec} for a similar procedure), read
\begin{equation}
\begin{aligned}
F_0=(x_{(2)}-x_{(3)})\sqrt{p_{(2)}p_{(3)}}+(x_{(3)}-x_{(1)})\sqrt{p_{(3)}p_{(1)}}+(x_{(1)}-x_{(2)})\sqrt{p_{(1)}p_{(2)}},\\
F_1=(x_{(1)}-x_{(2)})\sqrt{p_{(1)}p_{(2)}}+(x_{(2)}-x_{(0)})\sqrt{p_{(2)}p_{(0)}}+(x_{(0)}-x_{(1)})\sqrt{p_{(0)}p_{(1)}},\\
F_2=(x_{(1)}-x_{(3)})\sqrt{p_{(1)}p_{(3)}}+(x_{(3)}-x_{(0)})\sqrt{p_{(3)}p_{(0)}}+(x_{(0)}-x_{(1)})\sqrt{p_{(0)}p_{(1)}}.
\end{aligned}
\end{equation}
Note that given a family of solutions $(x_{(i)}(t),p_{(i)}(t))$, with $i=0,\ldots,3$, of (\ref{Hamil12}), then $d\bar F_j/dt=\widetilde X_tF_j=0$ for $j=0,1,2$ and $\bar F_j=F_j(x_{(0)}(t),p_{(0)}(t),\ldots,x_{(3)}(t),p_{(3)}(t))$.

In order to derive a superposition rule, we just need to obtain the value of $p_{(0)}$ from the equation $k_1=F_1$, where $k_1$ is a real constant, to get
\begin{equation}
\sqrt{-p_{(0)}}=\frac{k_1+(x_{(2)}-x_{(1)})\sqrt{p_{(1)}p_{(2)}}}{(x_{(2)}-x_{(0)})\sqrt{-p_{(2)}}+(x_{(0)}-x_{(1)})\sqrt{-p_{(1)}}},
\end{equation}
and then plug this value into the equation $k_2=F_2$ to have
\begin{equation}
\begin{aligned}
\hspace*{-.7em}x_{(0)}&=\frac{k_1\Gamma(x_{(1)},p_{(1)},x_{(3)},p_{(3)})+k_2\Gamma(x_{(2)},p_{(2)},x_{(1)},p_{(1)})-F_0 x_{(1)} \sqrt{-p_{(1)}}}
{k_1(\sqrt{-p_{(1)}}-\sqrt{-p_{(3)}})+k_2(\sqrt{-p_{(2)}}-\sqrt{-p_{(1)}})-\sqrt{-p_{(1)}}F_0},\\
\hspace*{-.7em}p_{(0)}&=-\left[{k_1/F_0(\sqrt{-p_{(3)}}-\sqrt{-p_{(1)}})+k_2/F_0(\sqrt{-p_{(1)}}-\sqrt{-p_{(2)}})+\sqrt{-p_{(1)}}}\right]^2,\\
\end{aligned}
\end{equation}
where $\Gamma(x_{(i)},p_{(i)},x_{(j)},p_{(j)})=\sqrt{-p_{(i)}}x_{(i)}-\sqrt{-p_{(j)}}x_{(j)}$. The above expressions give us a superposition rule $\Phi:(x_{(1)},p_{(1)},x_{(2)},p_{(2)},x_{(3)},p_{(3)};k_1,k_2)\in{\rm T}^*\mathbb{R}^3\times\mathbb{R}^2\mapsto (x_{(0)},p_{(0)})\in {\rm T}^*\mathbb{R}$ for system (\ref{Hamil12}). Finally, since every $x_{(i)}(t)$ is a particular solution for (\ref{NLe2}), the map $\Upsilon=\tau\circ \Phi$ gives the general solution of second-order Riccati equations in terms of three generic particular solutions
 $x_{(1)}(t),x_{(2)}(t),x_{(3)}(t)$ of (\ref{NLe2}), the corresponding $p_{(1)}(t),p_{(2)}(t),p_{(3)}(t)$ and two real constants
  $k_1,k_2$.

\subsubsection{Kummer--Schwarz equations}

Here we analyze second- and third-order Kummer--Schwarz equations (KS-2 and KS-3, respectively), \cite{Be82,Be88,GCG,AL08}, 
 with the aid of Lie systems. 

The KS-2 equations take the form
\begin{equation}\label{KS22}
\frac{d^2x}{dt^2}=\frac 3{2x}\left(\frac{dx}{dt}\right)^2-2c_0x^3+2b_1(t)x,
\end{equation}
with $c_0$ being a real constant and $b_1(t)$ a $t$-dependent function. 
KS-2
equations are a particular case of second-order Gambier equations
\cite{CGL11,GCG} and appear in the study of cosmological models \cite{NR02}. 
In addition, their relations to other differential equations like Milne--Pinney
equations \cite{GCG}, make them an
alternative approach to the analysis of many physical problems
\cite{CGL11,IK03,AL08}. 

In the case of KS-3 equations, i.e., we have
\begin{equation}\label{KS33}
\frac{d^3x}{dt^3}=\frac 32\left(
\frac{dx}{dt}\right)^{-1}\!\!\left(\frac{d^2x}{dt^2}\right)^{2}\!\!-2c_0(x)\left(\frac{
dx}{dt}\right)^3\!\!+2b_1(t)\frac{dx}{dt},
\end{equation}
where $c_0=c_0(x)$ and $b_1=b_1(t)$ are arbitrary. 

The relevance of KS-3 equations resides in
their relation to
the Kummer's problem \cite{Be82,Be88,Be07}, Milne--Pinney \cite{AL08}
and Riccati
equations \cite{Co94,EEL07,AL08}. Such  relations can be useful in the
interpretation of physical systems through KS-3 equations, e.g., the case of
quantum
non-equilibrium dynamics of many body systems \cite{GBD10}. Furthermore, KS-3
equations with $c_0 = 0$ can be rewritten as $\{x, t\} = 2b_1(t)$, where $\{x,
t\}$ is the
{\it Schwarzian derivative} \cite{LG99} of the function $x(t)$ relative to $t$.

KS-2 and KS-3 appear in other related themes \cite{As10,Be82,Be88,BR97,Ma94,Ta97,Ta89}.
Moreover, there is some
interest in studying the particular solutions of KS-2 and KS-3 equations, which have been analyzed in several manners in the
literature, e.g., through
non-local transformations or in terms of solutions to other differential
equations \cite{Be82,Be07,BR97,GCG,AL08}. From a physical viewpoint, KS-2 and KS-3
equations occur in the study of Milne--Pinney equations, Riccati
equations, and time-dependent frequency harmonic oscillators \cite{CGL11,Co94,GCG,AL08}, which are of certain relevancy in
two-body problems \cite{Be89,Be81}, Quantum Mechanics \cite{IK03,Kr09}, 
Classical Mechanics \cite{NR02}, etcetera \cite{Pe05}. 

We show that KS-2 equations can be
studied through two $\mathfrak{sl}(2,\mathbb{R})$-Lie systems \cite{Pi12}, i.e., Lie systems that describe the integral curves of a
$t$-dependent vector field taking values in {Vessiot--Guldberg Lie algebra} isomorphic to $\mathfrak{sl}(2,\mathbb{R})$. This new result slightly generalizes previous findings about these equations \cite{CGL11}. 
 
Afterwards, we obtain two Lie group actions whose
fundamental vector fields correspond with those of the abovementioned Vessiot--Guldberg Lie algebras. These actions 
allow us to prove that the explicit integration of KS-2 and KS-3
equations is equivalent to working out a particular solution of a Lie system on
$SL(2,\mathbb{R})$. Further, we  will see that 
Riccati and Milne--Pinney equations exhibit similar features. 

We show that
the knowledge of the general solution of any of the reviewed equations enables us to solve simultaneously any other related to the same equation on $SL(2,\mathbb{R})$. This fact provides a new powerful and general way of linking solutions of these equations, which were previously known to be related through {ad hoc} expressions in certain cases \cite{Co94,AL08}.

\subsubsection*{Second-order Kummer Schwarz equations}

Consider the first-order system associated with \eqref{KS22}
\begin{equation}\label{FirstOrderKummer}
\left\{\begin{aligned}
\frac{dx}{dt}&=v,\\
\frac{dv}{dt}&=\frac 32 \frac{v^2}x-2c_0x^3+2b_1(t)x,
\end{aligned}\right.
\end{equation}
on ${\rm T}\mathbb{R}_0$, with
$\mathbb{R}_0=\mathbb{R}-\{0\}$, obtained by adding the new variable
$v\equiv dx/dt$ to the KS-2 equation (\ref{KS22}). This
system describes the integral curves of the $t$-dependent vector field
\begin{equation}\label{Dec0}
\!X_t=v\frac{\partial }{\partial x}+\left(\frac 32
\frac{v^2}x\!-\!2c_0x^3+2b_1(t)x\right)\frac{\partial }{\partial
v}\!=\!M_3+b_1(t)M_1,
\end{equation}
where 
\begin{equation}\label{VFKS2}
\begin{gathered}
M_1=2x\frac{\partial}{\partial v},\qquad M_2=x\frac{\partial}{\partial
x}+2v\frac{\partial }{\partial v},\qquad M_3=v\frac{\partial}{\partial
x}+\left(\frac 32\frac{v^2}x-2c_0x^3\right)\frac{\partial}{\partial v}
\end{gathered}
\end{equation}
satisfy the commutation relations
\begin{equation}
[M_1,M_3]=2M_2,\quad [M_1,M_2]=M_1,\quad [M_2,M_3]=M_3.
\end{equation}
These vector fields span a three-dimensional real Lie algebra $V$ of vector fields
isomorphic to $\mathfrak{sl}(2,\mathbb{R})$ \cite{CGL11,Dissertationes}.  Hence, in view of (\ref{Dec0}) and the 
Lie--Scheffers Theorem, $X$ admits a superposition rule and becomes a Lie system
associated with a Vessiot--Guldberg Lie algebra isomorphic to $\mathfrak{sl}(2,\mathbb{R})$, i.e.,
an $\mathfrak{sl}(2,\mathbb{R})$-Lie system. 

The knowledge of 
a Lie group action $\varphi_{2KS}:G\times {\rm T}\mathbb{R}_0\rightarrow {\rm T}\mathbb{R}_0$ whose fundamental vector fields are $V$ and $T_eG\simeq V$ allows us to express
 the general solution of $X$ in
the form (\ref{mix}), in terms of a particular solution of a Lie system (\ref{eqLie}) on
$G$. Let us determine $\varphi_{2KS}$ in such a
way that our procedure can easily be extended to third-order
Kummer--Schwarz equations. 

Consider the basis of matrices of $\mathfrak{sl}(2,\mathbb{R})$ given by
\begin{equation}\label{Base}
{\rm a}_1=\left(
\begin{array}{cc}
0&1\\
0&0 
\end{array}
\right),\quad {\rm a}_2=\frac 12\left(
\begin{array}{cc}
-1&0\\
0&1
\end{array}
\right)
,\quad {\rm a}_3=\left(
\begin{array}{cc}
0&0\\
-1&0 
\end{array}
\right)
\end{equation}
satisfying the commutation relations
\begin{equation}
[{\rm a}_1,{\rm a}_3]=2{\rm a}_2,\qquad [{\rm a}_1,{\rm a}_2]={\rm a}_1,\qquad
[{\rm a}_2,{\rm a}_3]={\rm a}_3,
\end{equation}
which match those satisfied by $M_1,M_2$ and $M_3$. So, the linear
function $\rho:\mathfrak{sl}(2,\mathbb{R})\rightarrow V$ mapping ${\rm
a}_\alpha$ into
$M_\alpha$, with $\alpha=1,2,3$, is a Lie algebra isomorphism. If we consider it as
an infinitesimal Lie group action, we can then ensure that there
exists a local Lie group action $\varphi_{2KS}:SL(2,\mathbb{R})\times {\rm
T}\mathbb{R}_0\rightarrow{\rm T}\mathbb{R}_0$ obeying the required properties. In particular, 
\begin{equation}
\frac{d}{ds}\varphi_{2KS}(\exp(-s{\rm a}_\alpha),{\bf
t}_{x})=M_\alpha(\varphi_{2KS}(\exp(-s{\rm a}_\alpha),{\bf t}_{x})), 
\end{equation}
where ${\bf t}_{x}\equiv (x,v)\in {\rm T}_{x}\mathbb{R}_0\subset {\rm
T}\mathbb{R}_0$, $\alpha=1,2,3$, and $s\in\mathbb{R}$. This condition determines the action
on ${\rm T}\mathbb{R}_0$ of the
elements of $SL(2,\mathbb{R})$ of the form
$\exp(-s{\rm a}_\alpha)$, with $\alpha=1,2,3$ and $s\in\mathbb{R}$.  By integrating $M_1$ and $M_2$, we obtain
\begin{equation}\label{FunEl}
\begin{aligned}
\varphi(\exp_{2KS}(-\lambda_1{\rm a}_1),{\bf
t}_{x})&=\left(x,v+2x\lambda_1\right),\\
\varphi(\exp_{2KS}(-\lambda_2{\rm a}_2),{\bf
t}_{x})&=\left(xe^{\lambda_2},ve^{2\lambda_2}
\right).\\
\end{aligned}
\end{equation}
Observe that $M_3$ is not defined on ${\rm T}\mathbb{R}_0$. Hence, its integral curves, let us
say $(x(\lambda_3),v(\lambda_3))$, must be fully contained in either ${\rm
T}\mathbb{R}^+$ or ${\rm T}\mathbb{R}^-$. These integral curves are
determined by the system
\begin{equation}\label{LC}
\frac{dx}{d\lambda_3}=v,\qquad \frac{dv}{d\lambda_3}=\frac 32\frac {v^2}x-2c_0x^3.
\end{equation}
When $v\neq 0$, we obtain
\begin{equation}
\frac{dv^2}{dx}=\frac{3v^2}{x}-4c_0x^3\Longrightarrow
v^2(\lambda_3)=x^3(\lambda_3) \Gamma-4c_0x^4(\lambda_3),
\end{equation}
for a real constant $\Gamma$. Hence, for each integral curve $(x(\lambda_3),v(\lambda_3))$, we have 
\begin{equation}
\Gamma=\frac{v^2(\lambda_3)+4c_0x^4(\lambda_3)}{x^3(\lambda_3)}.
\end{equation}
Moreover, $d\Gamma/d\lambda_3=0$ not only for solutions of (\ref{LC}) with $v(\lambda_3)\neq 0$ for every $\lambda_3$, but for any solution of (\ref{LC}). 
Using the above results and (\ref{LC}), we
see that
\begin{equation}\label{cur}
\frac{dx}{d\lambda_3}={\rm sg}(v)\sqrt{\Gamma x^3-4c_0x^4 }\Rightarrow
x(\lambda_3)=
\frac{x(0)}{F_{\lambda_3}(x(0),v(0))},
\end{equation}
where ${\rm sg}$ is the well-known {\it sign function} and
\begin{equation}
F_{\lambda_3}({\bf t}_{x})=\left(1-\frac{v\lambda_3}{2x}\right)^2+
c_0x^2\lambda_3^2.
\end{equation}
Now, from (\ref{cur}) and taking into account the first equation within (\ref{LC}),
it immediately follows that 
\begin{equation}\label{FunEl2}
\begin{aligned}
\varphi_{2KS}(\exp(-\lambda_3{\rm a}_3),{\bf
t}_{x})&=\left(\frac{x}{F_{\lambda_3}({\bf
t}_{x})},\frac{v-\frac{
v^2+4c_0x^4}{2x}\lambda_3}{F_{\lambda_3}^{2}({\bf t}_{x})}\right)\!.
\end{aligned}
\end{equation}

Let us employ previous results to determine the action  on $\mathfrak{sl}(2,\mathbb{R})$
of those
elements $g$ close to the neutral element $e\in SL(2,\mathbb{R})$. 
Using the so called {\it canonical coordinates of the second kind} \cite{It87},
we can write $g$ within an open neighborhood $U$ of $e$ in a unique form as
\begin{equation}\label{decomposition}
g=\exp(-\lambda_3{\rm a}_3)\exp(-\lambda_2{\rm a}_2)\exp(-\lambda_1{\rm a}_1),
\end{equation}
for real constants $\lambda_1,\lambda_2$ and $\lambda_3$. This allows us to
obtain the action of every
$g\in U$  on ${\rm T}\mathbb{R}_0$ through the composition of the actions of
elements $\exp(-\lambda_\alpha
{\rm a}_\alpha)$, with $\lambda_\alpha\in\mathbb{R}$ for $\alpha=1,2,3$. 
To do so, we determine the
constants
$\lambda_1,\lambda_2$ and $\lambda_3$ associated with each $g\in U$ in (\ref{decomposition}).

Considering the standard matrix representation of $SL(2,\mathbb{R})$, we can
express every
$g\in SL(2,\mathbb{R})$ as
\begin{equation}\label{rep}
g=\left(\begin{array}{cc}
\alpha &\beta\\
\gamma &\delta 
\end{array}\right),\qquad \alpha\delta-\beta\gamma=1,\qquad
\alpha,\beta,\gamma,\delta\in\mathbb{R}.
\end{equation}
In view of (\ref{Base}), and comparing (\ref{decomposition}) and (\ref{rep}), we obtain
\begin{equation}
\alpha=e^{\lambda_2/2},\quad \beta=-e^{\lambda_2/2}\lambda_1,\qquad
\gamma=e^{\lambda_2/2}\lambda_3. 
\end{equation}
Consequently, 
\begin{equation}
\lambda_1=-\beta/\alpha,\qquad \lambda_2=2\log \alpha,\qquad
\lambda_3=\gamma/\alpha, 
\end{equation}
and, from the basis (\ref{Base}), the decomposition (\ref{decomposition}) and expressions (\ref{FunEl}) and (\ref{FunEl2}), the action reads
\begin{equation}
\varphi_{2KS}\left(g,
{\rm \bf t}_{x}\right)=\left(\frac{x}{F_g({\bf
t}_{x})},\frac{1}{F_g^{2}({\bf t}_{x})}\left[
(v\alpha-2x\beta)\left(\delta-\frac{\gamma v}{2x}\right)-2c_0x^3\alpha\gamma\right]\right),
\end{equation}
where 
\begin{equation}
F_g({\bf
t}_{x})=\left(\delta-\frac{\gamma v}{2x}\right)^2
+c_0x^2\gamma^2.
\end{equation}
Although this expression has been derived for $g$ being close to $e$, it can be proven
that the action is properly defined at points $(g,{\bf t}_x)$ such that $F_g({\bf t}_x)\neq 0$. If $c_0>0$, then $F_{g}({\bf t}_x)>0$ for all $g\in SL(2,\mathbb{R})$ and ${\bf t}_x\in{\rm T}\mathbb{R}_0$. So, $\varphi_{2KS}$ becomes globally defined. Otherwise,
 $F_{g}({\bf t}_x)>0$ for $g$ close enough to $e$. Then, $\varphi_{2KS}$ is only defined on a neighborhood of $e$.

The action $\varphi_{2KS}$ also permits us to write the general solution of system
(\ref{FirstOrderKummer}) in the form $(x(t),v(t))=\varphi_{2KS}(g(t),{\bf
t}_{x})$,
with
 $g(t)$ being a particular solution of 
\begin{equation}\label{reduced}
\frac{dg}{dt}=-X^R_3(g)-b_1(t)X^R_1(g),
\end{equation}
where $X^R_\alpha$, with $\alpha=1,2,3$, are the right-invariant
vector fields on $SL(2,\mathbb{R})$ satisfying $X_\alpha^R(e)={\rm a}_\alpha$ \cite{CGM00,Dissertationes}. 
Additionally, as $x(t)$ is the general solution of KS-2 equation
(\ref{KS22}), we readily see that
\begin{equation}\label{action}
x(t)=\tau\circ \varphi_{2KS}(g(t),{\bf t}_{x}),
\end{equation}
with $\tau:(x,v)\in{\rm
T}\mathbb{R}\mapsto x\in \mathbb{R}$, the natural tangent bundle projection, provides us with the general solution of (\ref{KS22})  in terms
of a particular
solution of (\ref{reduced}).

Conversely, we prove that
we can recover a particular solution to (\ref{reduced}) from the knowledge of the general solution of (\ref{KS22}). For simplicity, we will determine 
the particular solution $g_1(t)$ with $g_1(0)=e$. Given two particular solutions $x_1(t)$ and $x_2(t)$ of (\ref{KS22}) with $dx_1/dt(t)=dx_2/dt(t)=0$, the expression (\ref{action}) implies that
\begin{equation}
(x_i(t),v_i(t))=\varphi_{2KS}(g_1(t),(x_i(0),0)),\qquad i=1,2.
\end{equation}
Writing the above expression explicitly, we get
\begin{equation}\label{sys11}
\begin{aligned}
-\frac{x_i(0)v_i(t)}{2x_i^2(t)}&=\beta(t)\delta(t)+c_0 x_i^2(0)\alpha(t)\gamma(t),\\
\frac{x_i(0)}{x_i(t)}&=\delta^2(t)+c_0x_i^2(0)\gamma^2(t),
\end{aligned}
\end{equation}
for $i=1,2$. 
The first two equations allow us to determine the value of $\beta(t)\delta(t)$ and $\alpha(t)\gamma(t)$. Meanwhile, we can obtain the value of $\delta^2(t)$ and $\gamma^2(t)$ from the other two ones. As $\delta(0)=1$, we know that $\delta(t)$ is positive when close to $t=0$. Taking into account that we have already worked out $\delta^2(t)$, we can determine $\delta(t)$ for small values of $t$. Since we have already obtained $\beta(t)\delta(t)$, we can also derive $\beta(t)$ for small values of $t$ by using $\delta(t)$. Note that $\alpha(0)=1$. So, $\alpha(t)$ is positive for small values of $t$, and the sign of $\alpha(t)\gamma(t)$ determines
the sign of $\gamma(t)$ around $t=0$. In view of this, the value of $\gamma(t)$ can be determined from $\gamma^2(t)$ in the interval around $t=0$. Summing up, we can obtain algebraically a particular solution of (\ref{sys11}) with $g_1(0)=e$ from the general solution of (\ref{KS22}). 

\subsubsection*{Third-order Kummer--Schwarz equation}

Let us write KS-3 equations \eqref{KS33} as a first-order system
\begin{equation}\label{firstKS33}
\left\{\begin{aligned}
\frac{dx}{dt}&=v,\\
\frac{dv}{dt}&=a,\\
\frac{da}{dt}&=\frac 32 \frac{a^2}v-2c_0(x)v^3+2b_1(t)v,
\end{aligned}\right.
\end{equation}
in the open submanifold $\mathcal{O}_2=\{(x,v,a)\in {\rm T}^2\mathbb{R}\mid
v\neq 0\}$ of ${\rm T}^2\mathbb{R}\simeq \mathbb{R}^3$, the referred to as {\it
second-order tangent bundle} of $\mathbb{R}$ \cite{Abra}.
 
Consider now the set of vector fields on $\mathcal{O}_2$ given by
\begin{equation}\label{VFKS1}
\begin{aligned}
N_1&=2v\frac{\partial}{\partial a},\\
N_2&=v\frac{\partial}{\partial v}+2a\frac{\partial}{\partial a},\\
N_3&=v\frac{\partial}{\partial x}+a\frac{\partial}{\partial v}+\left(\frac 32
\frac{a^2}v-2c_0(x)v^3\right)\frac{\partial}{\partial a},
\end{aligned}
\end{equation}
 which satisfy
the commutation relations
\begin{equation}\label{cond1}
[N_1,N_3]=2N_2,\quad [N_1,N_2]=N_1,\quad [N_2,N_3]=N_3.
\end{equation}
Thus, they span a three-dimensional Lie algebra of vector fields $V$ isomorphic
to $\mathfrak{sl}(2,\mathbb{R})$. Since (\ref{firstKS33})  is determined by the
$t$-dependent vector field
\begin{equation}
X_t=v\frac{\partial}{\partial x}+a\frac{\partial}{\partial v}+\left(\frac 32
\frac{a^2}v-2c_0(x)v^3+2b_1(t)v\right)\frac{\partial}{\partial
a},
\end{equation}
we can write $X_t=N_3+b_1(t)N_1.$ Consequently, $X$ takes values in the 
finite-dimensional Vessiot--Guldberg Lie algebra $V$ and becomes an $\mathfrak{sl}(2,\mathbb{R})$-Lie system. 
This generalizes the result provided in
\cite{GL12} for $c_0(x)=const.$

We shall now reduce the integration of (\ref{firstKS33}) with $c_0(x)=const.$, and in consequence the integration of the related (\ref{KS33}),
to working out a
particular solution of the Lie system (\ref{reduced}). To do so, we employ the Lie group action
$\varphi_{3KS}:SL(2,\mathbb{R})\times \mathcal{O}_2\rightarrow  \mathcal{O}_2$
whose
infinitesimal action is given by the Lie algebra isomorphism
$\rho:\mathfrak{sl}(2,\mathbb{R})\rightarrow V$ satisfying that
$\rho({\rm a}_\alpha)=N_\alpha$, with $\alpha=1,2,3$. This Lie group action holds that 
\begin{equation}
\frac{d}{ds}\varphi_{3KS}(\exp(-s{\rm a}_\alpha),{\bf
t}^2_x)=N_\alpha(\varphi_{3KS}(\exp (-s{\rm a}_\alpha),{\bf t}^2_x)),
\end{equation}
with ${\bf t}^2_x\equiv (x,v,a) \in\mathcal{O}_2$ and $\alpha=1,2,3$.
Integrating $N_1$ and $N_2$,
we obtain that 
\begin{equation}
\varphi_{3KS}\left(\exp(-\lambda_1{\rm a}_1),{\bf
t}_{x}^2\right)=\left(\begin{array}{c}
x\\
v\\
a+2v\lambda_1\\ 
\end{array}\right)
\end{equation}
and
\begin{equation}
\varphi_{3KS}\left(\exp(-\lambda_2{\rm a}_2),{\bf t}_{x}^2\right)=
\left(\begin{array}{c}
x\\
ve^{\lambda_2}\\
ae^{2\lambda_2}\\ 
\end{array}\right).
\end{equation}
To integrate $N_3$, we need to obtain the solutions of
\begin{equation}\label{Sys3}
\frac{dx}{d\lambda_3}=v,\qquad \frac{dv}{d\lambda_3}=a,\qquad \frac{da}{d\lambda_3}=\frac 32\frac {a^2}v-2c_0v^3.
\end{equation}
Proceeding mutatis mutandis, as in the analysis of system (\ref{LC}), we obtain
\begin{equation}
v(\lambda_3)\!=
\!\frac{v(0)}{F_{\lambda_3}(x(0),v(0),a(0))},
\end{equation}
with 
\begin{equation}
F_{\lambda_3}({\bf
t}^2_{x})=\left(1-\frac{a\lambda_3}{2v}\right)^2+c_0v^2\lambda_3^2.
\end{equation}
Taking into account this and the first two equations within (\ref{Sys3}), we see that 
\begin{equation}
\varphi_{3KS}\left(e^{-\lambda_3{\rm a}_3},
{\bf t}_{x}^2\right)=\left(
\begin{array}{c}
x+v\int^{\lambda_3}_0F^{-1}_{\lambda'_3}({\bf t}_{x}^2)d\lambda'_3\\[1ex]
F^{-1}_{\lambda_3}({\bf t}_{x}^2)v\\[1ex]
v\partial (F^{-1}_{\lambda_3}({\bf t}_{x}^2))/\partial \lambda_3
\end{array}\right).
\end{equation}
Using decomposition (\ref{decomposition}), we can reconstruct the new action 
\begin{equation}
\varphi_{3KS}\left(g,
{\bf t}_{x}^2\right)=\left(
\begin{array}{c}
x+v\int^{\gamma/\alpha}_0\bar F^{-1}_{\lambda_3,g}({\bf
t}^2_x)d\lambda_3\\[1ex]
\bar{F}^{-1}_{\gamma/\alpha,g}({\bf t}^2_x)v\\[1ex]
v\frac{\partial (\bar{F}^{-1}_{\lambda_3,g}({\bf t}^2_x))}{\partial
\lambda_3}\big|_{\lambda_3=\gamma/\alpha}
\end{array}\right),
\end{equation}
with $\bar F_{\lambda_3,g}({\bf t}_x^2)=\alpha^{-2}F_{\lambda_3}(x,v\alpha^2,(a\alpha-2v\beta)\alpha^3\lambda_3)$, i.e., 
\begin{equation}
\bar{F}_{\lambda_3,g}({\bf
t}^2_{x})=\left(\frac{1}{\alpha}-\frac{a\alpha -2v\beta}{2 v}\lambda_3\right)^2
+c_0v^2\alpha^2\lambda_3^2.
\end{equation}
This action enables us to write the general solution of
(\ref{firstKS33}) as
\begin{equation}
(x(t),v(t),a(t))=\varphi_{3KS}(g(t),{\bf t}^2_{x}),
\end{equation}
where ${\bf t}^2_{x}\in
\mathcal{O}_2$ and $g(t)$ is a particular solution of the equation on 
$SL(2,\mathbb{R})$ given by (\ref{reduced}). Hence, if ${\tau^{2)}}:(x,v,a)\in
{\rm T}^2\mathbb{R}\mapsto x\in
\mathbb{R}$ is the fiber bundle projection corresponding to the second-order
tangent
bundle on $\mathbb{R}$, we can write the general solution of
(\ref{KS33}) in the form
\begin{equation}
x(t)=\tau^{2)}\circ \varphi_{3KS}(g(t),{\bf t}^2_{x}),
\end{equation}
where $g(t)$ is any particular solution of (\ref{reduced}).

Conversely, given the general solution of (\ref{KS33}), we can obtain a particular solution of (\ref{reduced}). As before, we focus on obtaining the particular solution $g_1(t),$ with $g_1(0)=e$. In this case, 
given two particular solution $x_1(t),x_2(t)$ of (\ref{KS33}) with $d^2 x_1/dt^2(0)=d^2x_2/dt^2(0)=0$, we obtain that the $t$-dependent coefficients $\alpha(t)$, $\beta(t)$, $\gamma(t)$ 
and $\delta(t)$ corresponding to the matrix expression of $g_1(t)$ obey a system similar to (\ref{sys11}) where $v$ and $x$ have been replaced by $a$ and $v$, respectively.

\subsubsection*{The Schwarzian derivative and Kummer--Schwarz}
The Schwarzian derivative of a real function $f=f(t)$ is
defined by
\begin{equation}\label{schderi}
\{f,t\}=\frac{d^3f}{dt^3}\left(\frac{df}{dt}\right)^{-1}-\frac
32\left[\frac{d^2f}{dt^2}\left(\frac{df}{dt}\right)^{-1}\right]^{2}.
\end{equation}
This derivative is clearly related to KS-3 equations (\ref{KS33}) with $c_0=0$, which can be written as $\{x,t\}=2b_1(t)$. 

Although a superposition rule for studying KS-3 equations was developed in \cite{CGL11}, the result provided in there was not
valid
when $c_0=0$, which retrieves the relevant equation $\{x,t\}=2b_1(t)$. This is why we aim to reconsider this case and its important connection to the Schwarzian derivative.

The vector fields $N_1,N_2,N_3$ are linearly independent at a generic point of
$\mathcal{O}_2\subset{\rm T}^2\mathbb{R}_0$.
Therefore, obtaining a superposition rule for (\ref{firstKS33}) amounts to
obtaining three functionally independent first-integrals $F_1,F_2,F_3$ common to all diagonal
prolongations $\widetilde N_1,\widetilde N_2,\widetilde N_3$ in
$(\mathcal{O}_2)^2$, such that
\begin{equation}
\frac{\partial (F_1,F_2,F_3)}{\partial (x_0,v_0,a_0)}\neq 0.
\end{equation}
As $[\widetilde N_1,\widetilde N_3]=2\widetilde N_2$, it
suffices to obtain common first-integrals for $\widetilde N_1,\widetilde
N_3$ to describe first-integrals common to the integrable distribution $\mathcal{D}$
spanned by $\widetilde N_1,\widetilde N_2,\widetilde N_3$.

Let us start by solving $\widetilde N_1F=0$, with
$F:\mathcal{O}_2\rightarrow\mathbb{R}$, i.e.,
\begin{equation}
v_0\frac{\partial F}{\partial a_0}+v_1\frac{\partial F}{\partial a_1}=0.
\end{equation}
The method of characteristics shows that $F$ must be constant along the
solutions of the associated {\it Lagrange--Charpit equations} \cite{De97}, namely
\begin{equation}
\frac{da_0}{v_0}=\frac{da_1}{v_1},\qquad dx_0=dx_1=dv_0=dv_1=0.
\end{equation}
Such solutions are the curves $(x_0(\lambda),v_0(\lambda),a_0(\lambda),x_1(\lambda),v_1(\lambda),a_1(\lambda))$ within $\mathcal{O}_2$ with $\Delta=v_1(\lambda)a_0(\lambda)-a_1(\lambda)v_0(\lambda)$, for a real constant $\Delta\in\mathbb{R},$ and constant
$x_i(\lambda)$ and $v_i(\lambda)$, with $i=0,1$. In other words, there exists a function
$F_2:\mathbb{R}^5\rightarrow\mathbb{R}$ such that
$F(x_0,v_0,a_0,x_1,v_1,a_1)=F_2(\Delta,x_0,x_1,v_0,v_1)$.

If we now impose $\widetilde N_3F=0$, we obtain
\begin{multline}
\widetilde N_3F=\widetilde N_3F_2=\frac{\Delta+a_1v_0}{v_1}\frac{\partial
F_2}{\partial v_0}+a_1\frac{\partial F_2}{\partial v_1}
\\
+v_0\frac{\partial
F_2}{\partial x_0}
+v_1\frac{\partial F_2}{\partial x_1}+\frac{3\Delta^2+6\Delta
a_1v_0}{2v_1v_0}\frac{\partial F_2}{\partial\Delta}=0.
\end{multline}
We can then write that $\widetilde N_3F_2=(a_1/v_1)\Xi_1F_2+\Xi_2F_2=0$, where

\begin{equation}
\begin{aligned}
\Xi_1&=v_0\frac{\partial}{\partial v_0}+v_1\frac{\partial}{\partial
v_1}+3\Delta\frac{\partial}{
\partial \Delta}
,\\
 \Xi_2&=v_0\frac{\partial}{\partial x_0}+v_1\frac{\partial}{\partial
x_1}+\frac{\Delta}{v_1}\frac{\partial}{\partial
v_0}+\frac{3\Delta^2}{2v_0v_1}\frac{\partial}{\partial \Delta}.\end{aligned}\end{equation}

As $F_2$ does not depend on $a_1$ in the chosen coordinate system, it follows
$\Xi_1F_2=\Xi_2F_2=0$. Using the characteristics method again, we obtain that
$\Xi_1F_2=0$ implies the existence of a new function
$F_3:\mathbb{R}^4\rightarrow\mathbb{R}$ such that
$F_2(\Delta,x_0,x_1,v_0,v_1)=F_3(K_1\equiv v_1/v_0,K_2\equiv v_0^3/\Delta,
x_0,x_1)$. 

The only condition remaining is $\Xi_2F_3=0$. In the local coordinate system
$\{K_1,K_2,$ $x_0,x_1\}$, this equation reads
\begin{equation}
v_0\left(\frac 3{2K_1}\frac{\partial F_3}{\partial
K_2}-\frac{1}{K_2}\frac{\partial F_3}{\partial K_1}+\frac{\partial F_3}{\partial
x_0}+K_1\frac{\partial F_3}{\partial x_1}\right)=0,
\end{equation} 
and its Lagrange--Charpit equations become
\begin{equation}
-K_2dK_1=\frac{2K_1dK_2}{3}={dx_0}=\frac{dx_1}{K_1}.
\end{equation}
From the first equality, we obtain that $K_1^3K_2^2=\Upsilon_1$ for a
certain real constant $\Upsilon_1$. In view of this and with the aid of the
above system, it turns out
\begin{equation}
\frac 23 K_1^2dK_2=dx_1\longrightarrow \frac
23\Upsilon_1^{2/3}K_2^{-4/3}dK_2=dx_1.
\end{equation}
Integrating, we see that $-2K_2K_1^2-x_1=\Upsilon_2$ for a certain real
constant
$\Upsilon_2$. Finally, these previous results are used to solve the last part of the
Lagrange--Charpit system, i.e.,
\begin{equation}
dx_0=\frac{dx_1}{K_1}=\frac{4\Upsilon_1dx_1}{(x_1+\Upsilon_2)^2}\longrightarrow
\Upsilon_3=x_0+\frac{4\Upsilon_1}{x_1+\Upsilon_2}.
\end{equation}
Note that $\partial(\Upsilon_1,\Upsilon_2,\Upsilon_3)/\partial(x_0,v_0,a_0)\neq 0$. Therefore,
considering $\Upsilon_1=k_1$, $\Upsilon_2=k_2$ and $\Upsilon_3=k_3$, we can 
obtain a mixed superposition rule. From these equations, we easily obtain 
\begin{equation}\label{sup0}
x_0=\frac{x_1k_3+k_2k_3-4k_1}{x_1+k_2}. 
\end{equation}
Multiplying numerator and denominator of the right-hand side by a non-null constant $\Upsilon_4$, the above expression 
can be rewritten as
\begin{equation}\label{sup1}
x_0=\frac{\alpha x_1+\beta}{\gamma x_1+\delta},
\end{equation}
with $\alpha=\Upsilon_4k_3,\beta=\Upsilon_4(k_2k_3-4k_1), \gamma=\Upsilon_4,\delta=k_2\Upsilon_4$. Observe that
\begin{equation}
\alpha\delta-\gamma\beta=4\Upsilon_4^2\Upsilon_1=\frac{4\Upsilon_4^2v^3_0v_1^3}{(v_1a_0-a_1v_0)^2}\neq 0.
\end{equation}
Then, choosing an appropriate $\Upsilon_4$, we obtain that (\ref{sup0}) can be rewritten as (\ref{sup1}) for a family of constants $\alpha,\beta,\gamma,\delta$ such that $\alpha\delta-\gamma\beta=\pm 1$. It is important to recall that the matrices
\begin{equation}
\left(\begin{array}{cc}
\alpha&\beta\\
\gamma&\delta\\
\end{array}\right),\qquad I=\alpha\delta-\beta\gamma=\pm 1,
\end{equation}
are the matrix description of the Lie group $PGL(2,\mathbb{R})$. Recall that we understand each matrix as the equivalence class related to it.

Operating, we also obtain that
\begin{equation}
v_0=\frac{Iv_1}{(\gamma x_1+\delta)},\qquad a_0=I\left[\frac{a_1}{(\gamma x_1+\delta)^2}-\frac{2v_1^2\gamma}{(\gamma x_1+\delta)^3}\right].
\end{equation}

The above expression together with (\ref{sup1}) become a superposition rule for KS-3 equations with $c_0=0$ (written as a first-order system). In other words, the general solution $(x(t),v(t),a(t))$ of (\ref{KS33}) with $c_0=0$ can be written as
\begin{equation}
(x(t),v(t),a(t))=\Phi(A,x_1(t),v_1(t),a_1(t)),
\end{equation}
with $(x_1(t),v_1(t),a_1(t))$ being a particular solution, $A\in PGL(2,\mathbb{R})$ and
\begin{equation}
\Phi(A,x_1,v_1,a_1)=\left(\frac{\alpha x_1+\beta }{\gamma x_1+\delta},\frac{Iv_1}{(\gamma x_1+\delta)^2},I\left[\frac{a_1(\gamma x_1+\delta)-2v_1^2\gamma}{(\gamma x_1+\delta)^3}\right]\right).
\end{equation}
Moreover, $x(t)$, which is the general solution of a KS-3 equation with $c_0=0$, can be determined out of a particular solution $x_1(t)$ and three constants through
\begin{equation}\label{Basic}
x(t)=\tau^{2)}\circ\Phi\left(A,x_1(t),\frac{dx_1}{dt}(t),\frac{d^2x_1}{dt^2}(t)\right),
\end{equation}
where we see that the right-hand part does merely depend on $A$ and $x_1(t)$. This constitutes a {\it basic superposition rule} \cite{CGL11} for equations $\{x(t),t\}=2b_1(t)$, i.e., it is an expression that allows us to
describe the general solution of any of these equations in terms of a particular solution (without involving its derivatives) and some constants to be related to initial conditions. 
We shall now employ this superposition rule to describe some properties of the
Schwarzian derivative.

From the equation above, we analyze the relation between two particular solutions $x_1(t)$ and $x_2(t)$ of the same equation $\{x,t\}=2b_1(t)$, i.e., $\{x_1(t),t\}=\{x_2(t),t\}$. Our basic superposition rule (\ref{Basic}) tells us that from  $x_1(t)$ we can generate every other solution of the equation. In particular, there must exist certain real constants $c_1,c_2,c_3,c_4$ such that 
\begin{equation}
x_2(t)=\frac{c_1 x_1(t)+c_2}{c_3 x_1(t)+c_4},\qquad c_1c_4-c_2c_3\neq 0.
\end{equation} 
In this way, we recover a relevant property of this type of equations \cite{OT09}. 

Our superposition rule (\ref{Basic}) also provides us with information about the Lie symmetries of 
$\{x(t),t\}=2b_1(t)$. Indeed, note that (\ref{Basic}) implies that the local Lie group action
$\varphi:PGL(2,\mathbb{R})\times \mathbb{R}\rightarrow \mathbb{R}$ 
\begin{equation}
\varphi(A,x)=\frac{\alpha x+\beta}{\gamma x+\delta},
\end{equation}
transforms solutions of $\{x(t),t\}=2b_1(t)$ into solutions of the same equation. The prolongation \cite{Bo01,Ol93} $\widehat\varphi: PGL(2,\mathbb{R})\times {\rm T}^2\mathbb{R}_0\rightarrow {\rm T}^2\mathbb{R}_0$ of $\varphi$ to ${\rm T}^2\mathbb{R}_0$, i.e.,
\begin{equation}
\widehat\varphi(A,{\bf t}^2_x)\!=\!\left(\frac{\alpha x+\beta}{\gamma x+\delta},\frac{Iv}{(\gamma x+\delta)^2},I\frac{a(\gamma x+\delta)-2\gamma v^2}{(\gamma x+\delta)^3}\right),
\end{equation}
gives rise to a group of symmetries $\varphi(A,\cdot)$ of (\ref{firstKS33}) when $c_0=0$. The fundamental vector fields of this action are spanned
by

\begin{equation}
\begin{gathered}
Z_1=-\frac{\partial}{\partial x},\qquad Z_2=x\frac{\partial}{\partial x}+v\frac{\partial}{\partial v}+a\frac{\partial}{\partial a},\\
Z_3=-\left[x^2\frac{\partial}{\partial x}+2vx\frac{\partial}{\partial v}+2(ax+v^2)\frac{\partial}{\partial a}\right],
\end{gathered}\end{equation} 

which close on a Lie algebra of vector fields isomorphic to $\mathfrak{sl}(2,\mathbb{R})$ and commute with $X_t$ for every $t\in\mathbb{R}$. In addition, their projections onto $\mathbb{R}$ must be Lie symmetries of $\{x(t),t\}=2b_1(t)$. Indeed, they read
\begin{equation}
S_1=-\frac{\partial}{\partial x},\qquad S_2=x\frac{\partial}{\partial x},\qquad S_3=-x^2\frac{\partial}{\partial x},
\end{equation}
which are the known Lie symmetries for these equations \cite{OT05}.

Consider now the equation $\{x(t),t\}=0$. Obviously, this equation admits the
particular solution $x(t)=t$. This, together with our basic superposition rule, show
that the general solution of this equation is
\begin{equation}\label{symsol}
x(t)=\frac{\alpha t+\beta}{\gamma t+\delta},\qquad \alpha\delta-\gamma\beta\neq 0,
\end{equation}
recovering another relevant known solution of these equations.

\section{Lie--Hamilton systems}
\setcounter{equation}{0}
\setcounter{theorem}{0}
\setcounter{example}{0}

We have found that many instances of relevant Lie systems possess Vessiot--Guldberg Lie algebras of Hamiltonian vector fields with respect to a Poisson structure.
Such Lie systems are hereafter called {\it Lie--Hamilton systems}.

Lie--Hamilton systems admit a plethora of geometric properties. For instance, we have proven that every Lie--Hamilton system admits a $t$-dependent Hamiltonian which can be understood as a curve in finite-dimensional
Lie algebra of functions (with respect to a certain Poisson structure). These $t$-dependent Hamiltonians, the so called {\it Lie--Hamiltonian structures},
are the key to understand the properties of Lie--Hamilton systems \cite{ADR12,CGM00,CLS122,Ru10}. Additionally,
Lie--Hamilton systems appear in the analysis of relevant physical and mathematical problems, like second-order Kummer--Schwarz equations and $t$-dependent Winternitz--Smorodinsky oscillators \cite{CLS122}. 

In this section we study Lie--Hamilton systems and some of their applications. In particular, our achievements are employed to study superposition rules, Lie symmetries and constants of motion for these systems.

\subsection{On the necessity of Lie--Hamilton systems}

In this section we start by showing how several relevant Lie systems admit a Vessiot--Guldberg Lie algebra of Hamilton vector fields with respect to a Poisson or symplectic structure. This will allow us to justify the interest of defining a new particular type of Lie systems enjoying such a property: the Lie--Hamilton systems.

\begin{example}\normalfont
Let us show that {\bf planar Riccati equations} (\ref{Riccati21}), described geometrically by the $t$-dependent vector field $X=a_0(t)X_1+a_1(t)X_2+a_2(t)X_3$ given by (\ref{PR}), admit a Vessiot--Guldberg Lie algebra consisting of Hamiltonian vector fields with respect to a symplectic structure. To do so, we search for a symplectic form, let us say $\omega=f(x,y){\rm d}x\wedge \dd y$, turning the Vessiot--Guldberg Lie algebra $V=\langle X_1,X_2,X_3\rangle$ into a Lie algebra of Hamiltonian vector fields with respect to it. To ensure that the basis for $V$ given by the vector fields $X_1,X_2$ and $X_3$, (see (\ref{vectRiccati21})), are locally Hamiltonian relative to $\omega$, we impose $\mathcal{L}_{X_i}\omega=0$ ($i=1,2,3$), where $\mathcal{L}_{X_i}\omega$ stands for the Lie derivative of $\omega$ with respect to $X_i$. In coordinates, these conditions read
\begin{equation}
\frac{\partial f}{\partial x}=0,\qquad x\frac{\partial f}{\partial x}+y\frac{\partial f}{\partial y}+2 f=0,\qquad
(x^2-y^2)\frac{\partial f}{\partial x}+2xy\frac{\partial f}{\partial y}+4xf=0.
\end{equation}
From the first equation, we obtain $f=f(y)$. Using this in the second equation, we find that $f=y^{-2}$ is a particular solution of both equations (the third one is therefore automatically fulfilled). This leads to a closed and non-degenerate two-form on $\mathbb{R}^2_{y\neq 0}\equiv\{(x,y)\in\mathbb{R}^2\,|\,y\neq 0\}$, namely
\begin{equation}
\omega=\frac{{\rm d} x \wedge {\rm d} y}{y^2}  .
\label{aa1}
\end{equation}
Using the relation $\iota_{X}\omega={\rm d}h$ between a Hamiltonian vector field $X$ and one of its corresponding Hamiltonian functions $h$, we observe that $X_1,X_2$ and $X_3$ are Hamiltonian vector fields with Hamiltonian functions
 \begin{equation}
h_1=-\frac 1y,\qquad h_2=-\frac xy,\qquad h_3=-\frac{x^2+y^2}{y},
\label{ab}
\end{equation}
 respectively. Obviously, the remaining vector fields of $V$ become also Hamiltonian. Thus, planar Riccati equations (\ref{Riccati21}) admit a Vessiot--Guldberg Lie algebra of Hamiltonian vector fields relative to (\ref{aa1}).
\end{example}

\begin{example}\normalfont
Let us now focus on analyzing the Hamilton equations for a $n$-dimensional
{\bf Smorodinsky--Winternitz oscillator} \cite{WSUF67} on ${\rm T}^*\mathbb{R}^n_0$,  where $\mathbb{R}^n_0=\mathbb{R}-{0}$, of the form
\begin{equation}\label{LieS}
\left\{\begin{aligned}
\frac{dx_i}{dt}&=p_i,\\ 
\frac{dp_i}{dt}&=-\omega^2(t)x_i+\frac{k}{x_i^3},
\end{aligned}\right.\qquad i=1,\ldots,n,
\end{equation}
with $\omega(t)$ being any $t$-dependent function, $\mathbb{R}^n_0\equiv \{(x_1,\ldots,x_n)\in \mathbb{R}_0^n\,|\,x_1, \ldots, x_n\neq 0\}$ and $k\in \mathbb{R}$. These oscillators have
attracted quite much attention in Classical and Quantum Mechanics for their
special properties \cite{GPS06,HBS05,YNHJ11}. In addition, observe that Winternitz--Smorodinsky oscillators reduce to $t$-dependent isotropic harmonic oscillators when $k=0$.

System (\ref{LieS}) describes the integral curves of the $t$-dependent vector
field
\begin{equation}
X_t=\sum_{i=1}^n\left[p_i\frac{\partial}{\partial
x_i}+\left(-\omega^2(t)x_i+\frac{k}{x_i^3}\right)\frac{\partial}{\partial
p_i}\right]
\end{equation}
on ${\rm T}^*\mathbb{R}^{n}_0$. This cotangent bundle admits a natural Poisson
bivector $\Lambda$ related to the restriction to ${\rm T}^*\mathbb{R}^n_0$ of
the canonical symplectic structure on ${\rm T}^*\mathbb{R}^n$, namely $\Lambda=\sum_{i=1}^n\partial/\partial x_i\wedge \partial/\partial p_i$. If we consider
the vector fields
\begin{equation}\label{Vessiot--GuldbergSec11}
\begin{gathered}
X_1=-\sum_{i=1}^nx_i\frac{\partial}{\partial p_i},\qquad\qquad
X_2=\sum_{i=1}^n\frac{1}{2}\left(p_i\frac{\partial}{\partial
p_i}-x_i\frac{\partial}{\partial x_i}\right),\\
X_3=\sum_{i=1}^n\left(p_i\frac{\partial}{\partial
x_i}+\frac{k}{x_i^3}\frac{\partial}{\partial p_i}\right),
\end{gathered}
\end{equation}
we can write $X_t=X_3+\omega^2(t)X_1$. Additionally, since
\begin{equation}\label{relWS}
[X_1,X_3]=2X_2,\qquad [X_1,X_2]=X_1,\qquad [X_2,X_3]=X_3,
\end{equation}
it follows that (\ref{LieS}) is a Lie system related to a Vessiot--Guldberg Lie
algebra isomorphic to $\mathfrak{sl}(2,\mathbb{R})$. In addition, this Lie
algebra is again made of Hamiltonian vector fields. In fact, it is easy to check
that $X_\alpha=-\widehat\Lambda(dh_\alpha)$, with $\alpha=1,2,3$ and 
\begin{equation}\label{HamWS}
h_1=\frac{1}{2}\sum_{i=1}^{n}{x_i^2},\qquad h_2=-\frac
12\sum_{i=1}^{n}{x_ip_i},\qquad
h_3=\frac{1}{2}\sum_{i=1}^{n}{\left(p_i^2+\frac{k}{x_i^2}\right)}.
\end{equation}
Consequently, system (\ref{LieS}) admits a Vessiot--Guldberg Lie algebra of Hamiltonian vector fields relative to the Poisson structure $\Lambda$.
\end{example}

\begin{example}\normalfont

Finally, let us turn to the system of Riccati equations (\ref{coupRiceq}) for $n=4$. The relevance of this system is due to the fact that it appears in the calculation of superposition rules for Riccati equations \cite{Dissertationes}. 

We already know that {\bf coupled Riccati equations} system (\ref{coupRiceq}) is a Lie system possessing a Vessiot--Guldberg Lie algebra isomorphic to $\mathfrak{sl}(2,\mathbb{R})$ spanned by the vector fields $X_1,X_2,X_3$ given by (\ref{VF1}). Let us prove that this Vessiot--Guldberg Lie algebra consists of Hamiltonian vector fields. Indeed, we can consider the Poisson bivector
\begin{equation}
\Lambda_R=(x_1-x_2)^2\frac{\partial}{\partial x_1}\wedge \frac{\partial}{\partial x_2}+(x_3-x_4)^2\frac{\partial}{\partial x_3}\wedge \frac{\partial}{\partial x_4},
\end{equation}
on  $\mathcal{O}=\{(x_1,x_2,x_3,x_4)|(x_1-x_2)(x_2-x_3)(x_3-x_4)\neq 0\}\subset \mathbb{R}^4$.
We have that $X_i=-\widehat\Lambda({\rm d}h_i)$ on $(\mathcal{O},\omega_R)$ for

\begin{equation}
\begin{gathered}
{h_1}=\frac{1}{x_1-x_2}+\frac{1}{x_3-x_4},\quad
h_2=\frac 12\left(\frac{x_1+x_2}{x_1-x_2}+\frac{x_3+x_4}{x_3-x_4}\right),\\
h_3=\frac{x_1 x_2}{x_1-x_2}+\frac{x_3 x_4}{x_3-x_4}.
\end{gathered}\end{equation} 

Hence, the vector fields $X_1,X_2,X_3$ given in (\ref{VF1}) are Hamiltonian with respect to $(\mathcal{O},\Lambda_{R})$ and our system of Riccati equations admits a Vessiot--Guldberg Lie algebra of Hamiltonian vector fields. The relevance of
this result is due to the fact that it allowed us to obtain the superposition rule for Riccati equations through a Casimir of $\mathfrak{sl}(2,\mathbb{R})$ \cite{BCHLS}.
\end{example}

All previous examples lead to propose the following definition. 

\begin{definition} A system $X$ on $N$ is said to be a {\it Lie--Hamilton system} if
$N$ can be endowed with a Poisson bivector $\Lambda$ in such a way that $V^X$ becomes a finite-dimensional real
Lie algebra of Hamiltonian vector fields relative to $\Lambda$.
\end{definition}

Although Lie--Hamilton systems form a particular class of Lie systems, it turns out that it is easier to find applications of Lie--Hamilton systems than applications of other types of Lie systems. Moreover, Lie--Hamilton systems admit much more geometrical properties than general Lie systems.

\subsection{Lie--Hamilton structures}

Standard Hamiltonian systems can be described through a Hamiltonian. Likewise, Lie--Hamilton systems admit a similar structure playing an ous role: the Lie--Hamiltonian structures.
 
\begin{definition} A {\it Lie--Hamiltonian structure} is a triple
$(N,\Lambda,h)$, where $(N,\Lambda)$ stands for  a Poisson manifold and $h$
represents a $t$-parametrized family of functions $h_t:N\rightarrow \mathbb{R}$
such that ${\rm Lie}(\{h_t\}_{t\in\mathbb{R}},\{\cdot,\cdot\}_\Lambda)$ is a
finite-dimensional real Lie algebra.  
\end{definition}

\begin{definition} A $t$-dependent vector field $X$ is said to admit, or
to possess, a Lie--Hamiltonian structure $(N,\Lambda,h)$ if $X_t$ is the Hamiltonian vector field corresponding to $h_t$ for each $t\in\mathbb{R}$. The Lie algebra ${\rm Lie}(\{h_t\}_{t\in\mathbb{R}},\{\cdot,\cdot\}_\Lambda)$ is called a {\it Lie--Hamilton algebra} for $X$.
\end{definition}

\begin{example}\normalfont
Recall that the vector fields $X_1$, $X_2$ and $X_3$ for planar Riccati equations (\ref{Riccati21}) are Hamiltonian with respect to the symplectic form $\omega$ given by (\ref{aa1}) and form a basis for a Vessiot--Guldberg Lie algebra $V\simeq \mathfrak{sl}(2,\mathbb{R})$. Assume that the minimal Lie algebra for a planar Riccati equation, $X$, is $V$, i.e., $V^X=V$.
If $\{\cdot,\cdot\}_\omega:C^\infty(\mathbb{R}^2_{y\neq 0})\times C^\infty(\mathbb{R}^2_{y\neq 0})\rightarrow C^\infty(\mathbb{R}^2_{
 y\neq 0})$ stands for the Poisson bracket induced by $\omega$ (see \cite{IV}), then
\begin{equation}
\label{sl2Rh}
\{h_1,h_2\}_\omega=-h_1,\qquad \{h_1,h_3\}_\omega=-2h_2,\qquad \{h_2,h_3\}_\omega=-h_3.
 \end{equation}
 Hence, the planar Riccati equation $X$ possesses a Lie--Hamiltonian structure of the form $\left(\mathbb{R}^2_{y\neq 0},\omega,h=a_0(t)h_1+a_1(t)h_2+a_2(t)h_3\right)$
 and, as $V^X\simeq \mathfrak{sl}(2)$, we have that $(\mathcal{H}_\Lambda,\{\cdot,\cdot\}_\omega)\equiv (\langle h_1,h_2,h_3\rangle,\{\cdot,\cdot\}_\omega)$ is a Lie--Hamilton algebra for $X$ isomorphic to $\mathfrak{sl}(2)$.
 \end{example}

\begin{proposition}\label{First} If a system $X$ admits a Lie--Hamiltonian
structure, then $X$ is a Lie--Hamilton system. 
\end{proposition}
\begin{proof} Let $(N,\Lambda,h)$ be a Lie--Hamiltonian structure for $X$. Thus,
${\rm Lie}(\{h_t\}_{t\in\mathbb{R}})$ is a finite-dimensional
 Lie algebra. Moreover, $\{X_t\}_{t\in\mathbb{R}}\subset \widehat\Lambda\circ d
[{\rm Lie}(\{h_t\}_{t\in\mathbb{R}})]$, and  as 
 $\widehat \Lambda\circ d $ is a Lie algebra morphism, it follows that
$V=\widehat\Lambda\circ d [{\rm Lie}(\{h_t\}_{t\in\mathbb{R}})]$
  is a finite-dimensional Lie algebra of Hamiltonian vector fields containing
$\{X_t\}_{t\in\mathbb{R}}$. Therefore, $V^X\subset V$ and $X$ is
   a Lie--Hamilton system.
\end{proof}

\begin{example}\normalfont
Consider again the second-order Riccati equation \eqref{NLe2} in Hamiltonian form \eqref{Hamil12} and the vector fields \eqref{commrelsecordric}
which span a five-dimensional Vessiot--Guldberg Lie algebra for such equations.  All the vector fields of this Lie
algebra are Hamiltonian vector fields with respect to the Poisson
bivector $\Lambda=\partial/\partial x\wedge\partial/\partial p$ on
$\mathcal{O}$. Indeed, note that $X_\alpha=-\widehat \Lambda(d h_\alpha)$, with
$\alpha=1,\ldots,5$ and 
\begin{equation}\label{equFun1}
\begin{gathered}
h_1(x,p)=-2\sqrt{-p},\quad\quad
h_2(x,p)=p,\quad\quad h_3(x,p)=xp,\quad\quad
h_4(x,p)=x^2p,\\
h_5(x,p)=-2x\sqrt{-p}.\\
\end{gathered}
\end{equation}
This system is therefore a Lie--Hamilton system. Moreover, the previous Hamiltonian functions span along with $\tilde h_0=1$ a   Lie algebra of functions  isomorphic to the {\it two-photon Lie algebra} $\mathfrak{h}_6$~\cite{BCHLS,CLS122} with non-vanishing Lie brackets given by
\be
\begin{aligned}
  \{h_1,h_3\}_\omega &=-\frac  12 \tilde h_1 , & \{h_1,h_4\}_\omega &=-h_5, &\{h_1,h_5\}_\omega &=2h_0, \\
\{h_2,h_3\}_\omega &=-h_2, &
  \{h_2,h_4\}_\omega &=-2 h_3,& \{h_2,h_5\}_\omega &=-h_1, \\ \{h_3,h_4\}_\omega &=-h_4, &\{h_3,h_5\}_\omega &=-\frac 12 h_5  .&\end{aligned}
\label{ComRel2}
\ee
In consequence, system \eqref{NLe2} admits a $t$-dependent Hamiltonian structure given by $({\rm T}^*\mathbb{R},\Lambda, h=h_1-a_0(t)h_2-a_1(t)h_3-a_2(t)h_4).$

\end{example}

Consider $X$ to be a Lie--Hamilton system admitting a Lie--Hamiltonian structure $(N,\Lambda,h)$ leading to a Lie--Hamilton algebra ${\rm Lie}(\{h_t\}_{t\in\mathbb{R}},\{\cdot,\cdot\}_\Lambda)$. Let us now analyze the relations between all the previous geometric structures.

\begin{lemma}\label{IsoRule} Given a system $X$ on $N$ possessing a 
Lie--Hamiltonian structure $(N,\Lambda,h)$, we have that
\begin{equation}\label{exacseq}
0\hookrightarrow {\rm Cas}(N,\Lambda)\cap {\rm
Lie}(\{h_t\}_{t\in\mathbb{R}})\hookrightarrow {\rm
Lie}(\{h_t\}_{t\in\mathbb{R}})\stackrel{\mathcal{J}_\Lambda}{\longrightarrow}
V^X\rightarrow 0,
\end{equation}
where $\mathcal{J}_\Lambda: f\in {\rm Lie}(\{h_t\}_{t\in\mathbb{R}})\mapsto
\widehat\Lambda\circ df\in V^X$, is an exact sequence of Lie algebras.

\end{lemma}
\begin{proof} Consider the exact sequence of (generally) infinite-dimensional real Lie
algebras
\begin{equation}
0\hookrightarrow {\rm Cas}(N,\Lambda)\hookrightarrow
C^\infty(N)\stackrel{\widehat\Lambda\circ d}{\longrightarrow}{\rm
Ham}(N,\Lambda)\rightarrow 0.
\end{equation}
Since $X_t=-\widehat\Lambda\circ d h_t$, we see that $V^X={\rm Lie}(\widehat
\Lambda \circ d(\{h_t\}_{t\in\mathbb{R}}))$. 
Using that $\widehat \Lambda\circ d$ is a Lie algebra morphism, 
we have $V^X=\widehat \Lambda \circ d[{\rm
Lie}(\{h_t\}_{t\in\mathbb{R}})]=\mathcal{J}_\Lambda({\rm
Lie}(\{h_t\}_{t\in\mathbb{R}}))$. 
Additionally, as $\mathcal{J}_\Lambda$ is the restriction to ${\rm
Lie}(\{h_t\}_{t\in\mathbb{R}})$ of $\widehat\Lambda\circ d$, 
we obtain that its kernel consists of Casimir functions belonging  to
 ${\rm Lie}(\{h_t\}_{t\in\mathbb{R}})$, i.e. $\ker \mathcal{J}_\Lambda={\rm
Lie}(\{h_t\}_{t\in\mathbb{R}})\cap {\rm Cas}(N,\Lambda)$. 
 The exactness of sequence (\ref{exacseq})  easily follows from these results. 
\end{proof}

The above proposition entails that every $X$ that possesses a Lie--Hamiltonian
structure $(N,\Lambda,h)$ is such 
that ${\rm Lie}(\{h_t\}_{t\in\mathbb{R}})$ is a Lie algebra extension of $V^X$
by ${\rm Cas}(N,\Lambda)\cap {\rm Lie}(\{h_t\}_{t\in\mathbb{R}})$, i.e., the
sequence of Lie algebras (\ref{exacseq}) is exact. Note that if ${\rm Lie}(\{h_t\}_{t\in\mathbb{R}})$ is finite-dimensional by assumption, all the Lie
algebras appearing in such a sequence are finite-dimensional.
 For instance, the first-order system (\ref{Hamil12}) associated with
second-order Riccati equations in Hamiltonian form admits a Lie--Hamiltonian structure 
\begin{equation}
\left(\mathcal{O},\frac{\partial}{\partial x}\wedge \frac{\partial}{\partial
p},h_1-a_0(t)h_2-a_1(t)h_3-a_2(t)h_4\right),
\end{equation}
where $h_1,h_2,h_3,h_4$ are given by (\ref{equFun1}). Note that ${\rm
Lie}(\{h_t\}_{t\in\mathbb{R}})$, for generic functions $a_0(t)$,$a_1(t)$,$a_2(t)$,
is a six-dimensional Lie algebra of functions $\mathfrak{W}\simeq V^X
\oplus\mathbb{R}$. Hence, we see that $\mathfrak{W}$ is an extension of $V^X$. 

It is worth noting that every $t$-dependent vector field admitting a
Lie--Hamiltonian structure necessarily possesses many other Lie--Hamiltonian
structures. For instance, if a system $X$ admits $(N,\Lambda,h)$, then it also admits a Lie--Hamiltonian structure
$(N,\Lambda,h')$, with $h':(t,x)\in \mathbb{R}\times N\mapsto
h(t,x)+f_\mathcal{C}(x)\in \mathbb{R}$, where $f_\mathcal{C}$ is any Casimir
function with respect to $\Lambda$. Indeed, it is easy to see that if
$h_1,\ldots,h_r$ is a basis for ${\rm Lie}(\{h_t\}_{t\in\mathbb{R}})$, then
$h_1,\ldots,h_r,f_\mathcal{C}$ span ${\rm Lie}(\{h'_t\}_{t\in\mathbb{R}})$,
which also becomes a finite-dimensial real Lie algebra. As shown later, this has relevant implications
for the linearization of Lie--Hamilton systems.

We have already proven that every system $X$ admitting a Lie--Hamiltonian
structure must possess several ones. Nevertheless, we have not yet studied 
the conditions ensuring that a Lie--Hamilton system $X$ possesses a Lie--Hamiltonian structure. Let
us answer this question.

\begin{proposition} Every Lie--Hamilton system admits a Lie--Hamiltonian
structure. 
\end{proposition} 
\begin{proof}
Assume $X$ to be a Lie--Hamilton system on a manifold $N$ with respect to a
Poisson bivector $\Lambda$. 
Since $V^X\subset {\rm Ham}(N,\Lambda)$ is finite-dimensional, there exists a
finite-dimensional linear space $\mathfrak{W}_0\subset C^\infty(N)$
 isomorphic to $V^X$ and such that $\widehat\Lambda\circ d(\mathfrak{W}_0)=V^X$.
Consequently, there exists a curve $h_t$ in $\mathfrak{W}_0$ 
 such that
$X_t=-\widehat{\Lambda}\circ d (h_t)$. To ensure that $h_t$ gives rise to a
Lie--Hamiltonian structure, we need to demonstrate that 
${\rm Lie}(\{h_t\}_{t\in\mathbb{R}},\{\cdot,\cdot\}_\Lambda)$ is
finite-dimensional. This will be done by constructing a finite-dimensional Lie algebra of functions containing the curve $h_t$.

Define the linear isomorphism $T:X_f\in V^X\mapsto -f\in \mathfrak{W}_0\subset
C^\infty(N)$ associating each vector field in $V^X$ 
with minus its unique Hamiltonian function within $\mathfrak{W}_0$.  This can be
done by choosing a representative
 for each element of a basis of $V^X$ and extending the map by linearity.

Note that this mapping needs not be a Lie algebra morphism and
 hence ${\rm Im}\, T=\mathfrak{W}_0$ does not need to be a Lie algebra. Indeed, we can
define a bilinear map 
 $\Upsilon: V^X\times V^X\rightarrow C^\infty(N)$ of the form
\begin{equation}\label{formula2}
\Upsilon(X_f, X_g)=\{f,g\}_\Lambda -T[X_f,X_g],
\end{equation}
measuring the obstruction for  $T$ to be a Lie algebra morphism,
 i.e. $\Upsilon$ is identically null if and only if $T$ is a Lie algebra
morphism. In fact, 
 if $\mathfrak{W}_0$ were a Lie algebra, then $\{f,g\}_\Lambda$ would be the
only element of $\mathfrak{W}_0$ with Hamiltonian 
 vector field $-[X_f,X_g]$, i.e. $T[X_f,X_g]$, and $\Upsilon$ would be a zero function. 

Note that $\Upsilon(X_f,X_g)$ is the difference between two functions, namely
$\{f,g\}_\Lambda$ and $T[X_f,X_g]$, sharing the same Hamiltonian vector field.
Consequently, ${\rm Im}\,\Upsilon\subset {\rm Cas}(N,\Lambda)$ and it can be injected
into a finite-dimensional Lie algebra of Casimir functions of the form
\begin{equation}
\mathfrak{W}_\mathcal{C}\equiv \langle \Upsilon(X_i,X_j)\rangle,\qquad
i,j=1,\ldots,r,
\end{equation}
where $X_1,\ldots,X_r$ is a basis for $V^X$. From here, it follows that  
\begin{equation}
\{\mathfrak{W}_\mathcal{C},\mathfrak{W}_\mathcal{C}\}_\Lambda=0,\quad
\{\mathfrak{W}_\mathcal{C},\mathfrak{W}_0\}_\Lambda=0,\quad
\{\mathfrak{W}_0,\mathfrak{W}_0\}_\Lambda\subset
\mathfrak{W}_\mathcal{C}+\mathfrak{W}_0.
\end{equation}
Hence, $\mathfrak{W}\equiv \mathfrak{W}_0+\mathfrak{W}_\mathcal{C}$ is a
finite-dimensional Lie algebra of functions containing the curve $h_t$. 
From here, it readily follows that $X$ admits a Lie--Hamiltonian structure
$(N,\Lambda,-TX_t)$. 
\end{proof}

Since every Lie--Hamilton system possesses a Lie--Hamiltonian
structure and every Lie--Hamiltonian structure determines a Lie--Hamilton
system, we obtain  the following theorem.

\begin{theorem}\label{HamLieSys}$\!$ A system $X\!$  admits a Lie--Hamiltonian
structure if and only if it is a Lie--Hamilton system.
\end{theorem}

\begin{example}\normalfont
Let us consider the Hamilton equations for the second-order Kummer--Schwarz equations \eqref{KS22} in Hamiltonian form \cite{Be07,CGL11}. They read
\begin{equation}\label{Hamil2}
\left\{
\begin{aligned}
\frac{dx}{dt}&=\frac{px^3}{2},\\
\frac{dp}{dt}&=-\frac{3p^2x^2}{4}-4c_0+\frac{4b_1(t)}{x^2},
\end{aligned}\right.
\end{equation}
on ${\rm T}^*\mathbb{R}_0$, where $\mathbb{R}_0=\mathbb{R}-\{0\}$. Once again,
the above system is a Lie system as it describes the integral curves of the
$t$-dependent vector field  $X_t=X_3+b_1(t)X_1$, where
\begin{equation}\label{2KSVecFiel}
X_1=\frac{4}{x^2}\frac{\partial}{\partial p},\quad
X_2=x\frac{\partial}{\partial x}-p\frac{\partial}{\partial p},\quad
X_3=\frac{px^3}{2}\frac{\partial}{\partial
x}-\left(\frac{3p^2x^2}{4}+4c_0\right) \frac{\partial}{\partial p},
\end{equation}
span a three-dimensional Lie algebra $V^{2KS}$ isomorphic to
$\mathfrak{sl}(2,\mathbb{R})$. Indeed, 
\begin{equation}
[X_1,X_3]=2X_2,\qquad [X_1,X_2]=X_1,\qquad [X_2,X_3]=X_3.
\end{equation}
Apart from providing a new approach to Kummer--Schwarz equations (see
\cite{CGL11} for a related method), our new description gives an additional
relevant property: $V^{2KS}$ consists of Hamiltonian vector fields with respect
to the Poisson bivector $\Lambda=\partial/\partial x\wedge\partial/\partial p$
on ${\rm T}^*\mathbb{R}_0$. In fact, $X_\alpha=-\widehat\Lambda(dh_\alpha)$ with
$\alpha=1,2,3$ and
\begin{equation}\label{FunKS}
h_1=\frac 4 x,\qquad h_2=xp,\qquad h_3=\frac 14p^2x^3+4c_0x. 
\end{equation}
Therefore (\ref{Hamil2}) is a Lie--Hamilton system. Moreover, we have that
\begin{equation}
\{h_1,h_2\}=-h_1,\quad \{h_1,h_3\}=-2h_2,\quad \{h_2,h_3\}=-h_3.
\end{equation}
Therefore, \eqref{Hamil2} admits a Lie--Hamiltonian structure $({\rm T}^*\mathbb{R}_0,\Lambda,h=h_3+b_1(t)h_1)$ as ensured by Theorem \ref{HamLieSys}.
\end{example}

\subsection{$t$-independent constants of motion}
Let us now study the structure of the space of $t$-independent constants of motion for Lie--Hamilton systems. In particular, we are interested in investigating the use of Poisson structures to study such constants of motion.

\begin{proposition}\label{Cas}
Given a system $X$ admitting a Lie--Hamiltonian structure $(N,\Lambda,h)$,
then $\mathcal{C}^\Lambda\subset \mathcal{V}^X$, where we recall that $\mathcal{C}^\Lambda$ is the Casimir distribution relative to $\Lambda$.  
\end{proposition}

\begin{proof}
Consider a $\theta_x\in\mathcal{C}_x^\Lambda$, with $x\in N$. As $X$ is a Lie--Hamilton system, 
for every $Y\in V^X$ there exists a function $f\in C^\infty(N)$ such that
$Y=-\widehat\Lambda (df)$. Then,
\begin{equation}
\theta_x(Y_x)=-\theta_x(\widehat{\Lambda}_x(df_x))=-\Lambda_x(df_x,\theta_x)=0,
\end{equation}
where $\widehat{\Lambda}_x$ is the restriction of $\widehat \Lambda$ to
$T^*_xN$. As the vectors $Y_x$, with $Y\in V^X$, span $\mathcal{D}^X_x$, then
$\theta_x\in \mathcal{V}_x^X$ and $\mathcal{C}^\Lambda\subset \mathcal{V}^X$. 
\end{proof}

Observe  that different Lie--Hamiltonian structures for a Lie--Hamilton system
$X$ may lead to different families of Casimir functions, which may determine
different constants of motion for $X$.

\begin{theorem}\label{IntLie} Let $X$\! be a system admitting a Lie--Hamiltonian
structure $(N,\!\Lambda,h)$, the space $\mathcal{I}^X|_U\!$ of $t$-independent
constants of motion of $X\!$\! on an open $U\!\subset\! U^X$ is a Poisson algebra.
Additionally, the codistribution $\mathcal{V}^X|_{U_X}$ is involutive with respect to
the Lie bracket $[\cdot,\cdot ]_\Lambda$ induced by $\Lambda$ on the space
$\Gamma(\pi_N)$ of smooth one-forms on $N$. 
\end{theorem}
\begin{proof}
Let  $f_1,f_2:U\rightarrow\mathbb{R}$ be two $t$-independent functions constants of motion
for $X$, i.e., $X_tf_i=0$, for $i=1,2$ and $t\in \mathbb{R}$. As $X$ is a
Lie--Hamilton system, all the elements of $V^X$ are Hamiltonian vector fields
and we can write $Y\{f,g\}_\Lambda=\{Yf,g\}_\Lambda+\{f,Yg\}_\Lambda$ for every
$f,g\in C^\infty(N)$ and $Y\in V^X$. In particular,
$X_t(\{f_1,f_2\}_\Lambda)=\{X_tf_1,f_2\}_\Lambda+\{f_1,X_tf_2\}_\Lambda=0$. Hence, the
Poisson bracket of $t$-independent constants of motion is a new one.
As $\lambda f_1+\mu f_2$ and $f_1\cdot f_2$ are also $t$-independent constants
of motion for every $\lambda,\mu\in\mathbb{R}$, it easily follows that
$\mathcal{I}^X|_U$ is a Poisson algebra. 

In view of 
Lemma \ref{basisVX} in Chapter \ref{Chap:GeomFund}, the co-distribution $\mathcal{V}^X$ admits a local basis of
exact forms $df_1,\ldots,df_{p(x)}$ for every point $x\in U^X$, where
$\mathcal{V}^X$ has local constant rank $p(x)\equiv\dim\, N-\dim\,\mathcal{D}^X_x$.  Now, 
$[df_i,df_j]_\Lambda=d(\{f_i,f_j\}_\Lambda)$ for $i,j=1,\ldots,p(x)$. We already proven that the
function $\{f_i,f_j\}_\Lambda$ is another first-integral. Therefore, from Lemma
\ref{basisVX} in Chapter \ref{Chap:GeomFund}, it easily follows that $\{f_i,f_j\}_\Lambda=G(f_1,\ldots,f_{p(x)})$.
Thus, $[df_i,df_j]_\Lambda \in\mathcal{V}^X|_{U_X}$. Using the latter and the properties
of the Lie bracket $[\cdot,\cdot]_\Lambda$, it directly turns out that the Lie bracket of two one-forms
taking values in $\mathcal{V}^X|_{U^X}$ belongs to $\mathcal{V}^X|_{U^X}$. Hence, 
$\mathcal{V}^X|_{U^X}$ is involutive. 
\end{proof}
\begin{corollary} Given a Lie--Hamilton system $X$, the space
$\mathcal{I}^X|_U$, where $U\subset U^X$ is such that $\mathcal{V}^X$ admits a local basis of exact forms, is a function
group, that is
\begin{enumerate}
 \item The space $\mathcal{I}^X|_U$ is a Poisson algebra.
\item There exists a family of functions $f_1,\ldots,f_s\in\mathcal{I}^X|_U$ such that 
every element $f$ of $\mathcal{I}^X|_U$ can be put in the form
$f=F(f_1,\ldots,f_s)$ for a certain function
$F:\mathbb{R}^s\rightarrow\mathbb{R}$.
\end{enumerate}
\end{corollary}
\begin{proof}
In view of the previous theorem, $\mathcal{I}^X|_U$ is a Poisson algebra with
respect to a certain Poisson bracket. Taking into account Proposition \ref{NuX} in Chapter \ref{Chap:GeomFund} and the form of
$\mathcal{I}^X|_U$ given by Lemma \ref{basisVX} in Chapter \ref{Chap:GeomFund}, we obtain that this space becomes
a function group.
\end{proof}

The above properties do not necessarily hold for systems other than
Lie--Hamilton systems, as they do not need to admit any {\it a priori} relation
among a Poisson bracket of functions and the $t$-dependent vector field
describing the system. Let us exemplify this. 
\begin{example}\normalfont

Consider the Poisson manifold
$(\mathbb{R}^3,\Lambda_{GM})$, where 
\begin{equation}
\Lambda_{GM}=\sigma_3\frac{\partial}{\partial
\sigma_2}\wedge\frac{\partial}{\partial
\sigma_1}-\sigma_1\frac{\partial}{\partial
\sigma_2}\wedge\frac{\partial}{\partial
\sigma_3}+\sigma_2\frac{\partial}{\partial
\sigma_3}\wedge\frac{\partial}{\partial \sigma_1}
\end{equation}
and $(\sigma_1,\sigma_2,\sigma_3)$ is a coordinate basis for $\mathbb{R}^3$, 
appearing in the study of Classical XYZ Gaudin Magnets \cite{BR98}. The system
$X=\partial/\partial \sigma_3$ is not a Lie--Hamilton system with respect to
this Poisson structure as $X$ is not Hamiltonian, namely
$\mathcal{L}_X\Lambda_{GM}\neq 0$. In addition, this system admits two
first-integrals $\sigma_1$ and $\sigma_2$. Nevertheless, their Lie bracket reads
$\{\sigma_1,\sigma_2\}=-\sigma_3$, which is not a first-integral for $X$. On the
other hand, consider the system
\begin{equation}
Y=\sigma_3\frac{\partial}{\partial \sigma_2}+\sigma_2\frac{\partial}{\partial
\sigma_3}.
\end{equation}
This system is a Lie--Hamilton system, as it can be written in the form
$Y=-\widehat\Lambda_{GM}(d\sigma_1)$, and it possesses two first-integrals given
by $\sigma_1$ and $\sigma_2^2-\sigma_3^2$. Unsurprisingly,
$Y\{\sigma_1,\sigma^2_2-\sigma_3^2\}=0$, i.e., the Lie bracket of two
$t$-independent constants of motion is also a constant of motion. 
\end{example}

Let us prove some final interesting results about the $t$-independent constants
of
motion for Lie--Hamilton systems.
\begin{proposition} Let $X$ be a Lie--Hamilton system that admits a
Lie--Hamiltonian structure $(N,\Lambda,h)$. The function $f:N\rightarrow
\mathbb{R}$ is a constant of motion for $X$ if and only if $f$ Poisson commutes
with all elements of ${\rm
Lie}(\{h_t\}_{t\in\mathbb{R}},\{\cdot,\cdot\}_{\Lambda})$.
\end{proposition}
\begin{proof}
The function $f$ is a $t$-independent constant of motion for $X$ if and only if 
\begin{equation}\label{con21}
0=X_tf=\{f,h_t\}_{\Lambda},\qquad \forall t\in\mathbb{R}.
\end{equation}
From here,
\begin{equation}
\{f,\{h_t,h_{t'}\}_{\Lambda}\}_{\Lambda}=\{\{f,h_t\}_{\Lambda},h_{t'}\}_{\Lambda
}+\{h_t,\{f,h_{t'}\}_{\Lambda}\}_{\Lambda}=0,\qquad \forall t,t'\in\mathbb{R},
\end{equation}
and inductively follows that $f$ Poisson commutes with all  successive Poisson
brackets of elements of $\{h_t\}_{t\in\mathbb{R}}$ and their linear combinations. 
As these elements span ${\rm
Lie}(\{h_t\}_{t\in\mathbb{R}})$, we get that $f$ Poisson commutes with ${\rm
Lie}(\{h_t\}_{t\in\mathbb{R}})$. 

Conversely, if $f$ Poisson commutes with ${\rm Lie}(\{h_t\}_{t\in\mathbb{R}})$,
it Poisson commutes with the elements $\{h_t\}_{t\in\mathbb{R}}$, and, in view
of (\ref{con21}), it becomes a constant of motion for $X$.
\end{proof}

In order to illustrate the above proposition, let us show an example.

\begin{example}\normalfont
 Consider a Smorodinsky--Winternitz system 
(\ref{LieS}) with $n=2$. Recall that this system admits a Lie--Hamiltonian structure $({\rm T}^*\mathbb{R}_0^2,\Lambda,h=h_3+\omega^2(t)h_1)$, 
where $\Lambda=\sum_{i=1}^2\partial/\partial x_i\wedge\partial/\partial p_i$ is a  Poisson bivector 
on ${\rm T}^*\mathbb{R}_0^2$ and the functions $h_1,h_3$ are given within (\ref{HamWS}). 
For non-constant $\omega(t)$, we obtain that 
${\rm Lie}(\{h_t\}_{t\in\mathbb{R}},\{\cdot,\cdot\}_\Lambda)$ is a real Lie algebra of functions isomorphic
to $\mathfrak{sl}(2,\mathbb{R})$ generated by the functions $h_1,h_2$ and $h_3$ detailed in (\ref{HamWS}).
When $\omega(t)=\omega_0\in\mathbb{R}$, the Lie algebra ${\rm Lie}(\{h_t\}_{t\in\mathbb{R}},\{\cdot,\cdot\}_\Lambda)$ becomes a one-dimensional
Lie subalgebra of the previous one. In any case, it is known that
\begin{equation}\label{Inte}
I=(x_1p_2-p_1x_2)^2+k\left[\left(\frac{x_1}{x_2}\right)^2+\left(\frac{x_2}{x_1}\right)^2\right]
\end{equation}
is a $t$-independent constant of motion (cf. \cite{CLR08}). A simple calculation shows that
\begin{equation}
\{I,h_\alpha\}_\Lambda=0,\qquad \alpha=1,2,3.
\end{equation}
Then, the function $I$ always Poisson commutes with the whole
${\rm Lie}(\{h_t\}_{t\in\mathbb{R}},\{\cdot,\cdot\}_\Lambda),$
 as expected. 
\end{example}

Obviously, every autonomous Hamiltonian system is a Lie--Hamilton system possessing a Lie--Hamiltonian structure
$(N,\Lambda,h)$, with $h$ being a $t$-independent Hamiltonian. Consequently, Proposition 11 shows that the $t$-independent 
first-integrals for a Hamiltonian system are those functions that Poisson commute with its Hamiltonian, recovering as a particular
case this wide known result. 

Moreover, the above proposition suggests that the role played by autonomous Hamiltonians for Hamiltonian systems is performed by the finite-dimensional Lie algebras of functions associated with Lie--Hamiltonian structures in the case of Lie--Hamilton systems. This can be employed to study $t$-independent first-integrals of Lie--Hamilton systems or analyze the maximal number of such first-integrals in involution, which would lead to the interesting analysis of integrability/superintegrability of Lie--Hamilton systems. 

\subsection{Symmetries, linearization and comomentum maps}

\begin{definition}
We say that a Lie system $X$ admitting a Lie--Hamilton structure $(N,\Lambda,h)$ possesses a compatible
 {\it strong comomentum map} with respect to this Lie--Hamilton structure if there
exists a Lie algebra morphism
$\lambda:V^X\rightarrow {\rm Lie}(\{h_t\}_{t\in\mathbb{R}},\{\cdot,\cdot \}_\Lambda)$ such that the following diagram
$$
\xymatrix{&&V^X\ar[d]^\iota\ar[lld]_\lambda\\
{\rm Lie}(\{h_t\}_{t\in\mathbb{R}},\{\cdot,\cdot \}_\Lambda)\ar[rr]^{\widehat\Lambda\circ d}&&{\rm Ham}(N,\Lambda)
}
$$
where $\iota:V^X\hookrightarrow {\rm Ham}(N,\Lambda)$ is the natural injection
of $V^X$ into ${\rm Ham}(N,\Lambda)$, is commutative.
\end{definition}

\begin{proposition}\label{linearization} Let $X$ be a Lie system possessing a Lie--Hamiltonian structure $(N,\Lambda,h)$
compatible with a strong comomentum map $\lambda$  such that $\dim\,\mathcal{D}^X_x=\dim
N=\dim V^X$ at a generic $x\in N$. Then, there exists a local coordinate system defined on a neighborhood of each $x$
such that $X$ and $\Lambda$ are
simultaneously linearizable and where  $X$ possesses a linear superposition rule. 
\end{proposition}
\begin{proof}
As it is assumed that $n\equiv\dim N=\dim V^X=\dim \mathcal{D}^X_x$ at a generic $x$, every basis $X_{1},\ldots,X_{n}$
of $V^X$ gives rise to a basis for the tangent bundle $TN$ on a neighborhood of $x$. 
Since $X$ admits a strong comomentum map compatible with $(N,\Lambda,h)$, 
we have $(V^X,[\cdot,\cdot])\simeq (\lambda(V^X),\{\cdot,\cdot\}_\Lambda)$ and the family of 
functions, $h_\alpha=\lambda(X_\alpha)$, with $\alpha=1,\ldots,n,$ form a basis for 
the Lie subalgebra $\lambda(V^X)$. Moreover, since $\widehat\Lambda\circ d\circ \lambda (V^X)=V^X$ and 
$\dim\,V^X=\dim \mathcal{D}^X_{x'}$  for $x'$ 
 in a neighborhood of $x$,
 then $\widehat \Lambda_{x'}\circ d (\lambda(V^X))\simeq T_{x'}N$
 and $dh_1\wedge \ldots \wedge dh_n\neq 0$ at a generic point. Hence, the set $(h_1,\ldots,h_n)$ is a
coordinate system on an open dense subset of $N$. Now, using again that $(\lambda(V^X), \{\cdot,\cdot\}_\Lambda)$ 
is a real Lie algebra, the Poisson bivector
$\Lambda$ can be put in the form
\begin{equation}\label{linear}
\Lambda=\frac 12\sum_{i,j=1}^n\{h_i,h_j\}_\Lambda\frac{\partial}{\partial
h_i}\wedge\frac{\partial}{\partial h_j}=\frac
12\sum_{i,j,k=1}^nc_{ijk}h_k\frac{\partial}{\partial
h_i}\wedge\frac{\partial}{\partial h_j},
\end{equation}
for certain real $n^3$ constants $c_{ijk}$. In other words, the Poisson bivector
$\Lambda$ becomes linear in the chosen coordinate system.

Since we can write $X_t=-\widehat\Lambda (d\bar h_t)$, with $\bar{h}_t=-\lambda (X_t)$ being a curve in the Lie
 algebra $\lambda(V^X)\subset {\rm
Lie}(\{h_t\}_{t\in\mathbb{R}})$, expression (\ref{linear}) yields
\begin{equation}
\begin{aligned}
X_t&=-\widehat\Lambda(d\bar h_t)=-\widehat\Lambda\circ
d\left(\sum_{l=1}^nb_l(t)h_l\right)\\&=-\sum_{l=1}^nb_l(t)(\widehat\Lambda\circ
dh_l)=-\sum_{l,j,k=1}^nb_l(t)c_{ljk}h_k\frac{\partial}{\partial h_j},
\end{aligned}
\end{equation}
and $X_t$  is  linear in this  coordinate system. Consequently, as every linear system, $X$ admits a linear superposition rule in the coordinate system
$(h_1,\ldots,h_n)$.
\end{proof}

Let us turn to describing some features of $t$-independent Lie symmetries for
Lie--Hamilton systems. Our exposition will be based upon the properties of the
hereafter called {\it symmetry distribution}.

\begin{definition} Given a Lie--Hamilton system $X$ that possesses a Lie--Hamiltonian structure $(N,\Lambda,h)$, we
define its {\it symmetry distribution}, $\mathcal{S}^X_\Lambda$, by
\begin{equation}
(\mathcal{S}^X_\Lambda)_x=\widehat{\Lambda}_x(\mathcal{V}_x^X)\in T_xN,\qquad
x\in N.
\end{equation}
\end{definition}

As its name indicates, the symmetry distribution can be employed to investigate the $t$-independent Lie symmetries of a Lie--Hamilton system.
Let us give some basic examples of how this can be done.

\begin{proposition} Given a Lie--Hamilton system $X$ with a Lie--Hamiltonian
structure $(N,\Lambda,h)$, then
\begin{enumerate}
 \item The symmetry distribution $\mathcal{S}^X_\Lambda$ associated with $X$ and $\Lambda$ is involutive on an open subset of $U^X$, where $U^X$ is the open dense subset of $N$ where $\mathcal{V}^X$ is differentiable.
\item  If $f$ is a $t$-independent constant of motion for $X$, then $\widehat
\Lambda(df)$ is a $t$-independent Lie symmetry of $X$.
\item The distribution $\mathcal{S}^X_\Lambda$ admits a local basis of $t$-independent Lie
symmetries of $X$ defined around a generic point of $N$. The elements of such a basis
are Hamiltonian vector fields of $t$-independent constants of motion of $X$.
\end{enumerate}
\end{proposition}
\begin{proof}
By definition of $\mathcal{S}^X_\Lambda$ and using that $\mathcal{V}^X$ has constant rank on the 
connected components of an open $U\subset U_X$, where $\Lambda$ has locally constant rank, we can ensure that 
given two vector fields in $Y_1,Y_2\in \mathcal{S}^X_\Lambda|_{U}$, there exist two forms
$\omega,\omega'\in\mathcal{V}^X|_{U}$ such that $Y_1=\widehat\Lambda(\omega)$,
$Y_2=\widehat\Lambda(\omega')$. Since $X$ is a Lie--Hamilton system, 
$\mathcal{V}^X|_{U}$ is involutive and $\widehat \Lambda$ is an anchor, i.e.,a Lie algebra 
morphism from $(\Gamma(\pi_N),[\cdot,\cdot]_\Lambda)$ to $(\Gamma(\tau_N),[\cdot,\cdot])$, then 
\begin{equation}
[Y_1,Y_2]=[\widehat\Lambda (w),\widehat\Lambda (w')]=\widehat\Lambda
([w,w']_\Lambda)\in\mathcal{S}^X_\Lambda.
\end{equation}
In other words, since $\mathcal{V}^X$ is involutive on $U$, then
$\mathcal{S}^X_\Lambda$ is so, which proves $(1)$.

To prove $(2)$, note that
\begin{equation}
[X_t,\widehat \Lambda(df)]=-[\widehat\Lambda (dh_t),\widehat
\Lambda(df)]=-\widehat\Lambda (d\{h_t,f\}_\Lambda)=\widehat\Lambda [d(X_tf)]=0.
\end{equation}
Finally, the proof of $(3)$ is based upon the fact that $\mathcal{V}^X$ admits,
around a point $x\in U^X\subset N$, a local basis of one-forms 
$df_1,\ldots, df_{p(x)}$, with $f_1,\ldots,f_{p(x)}$ being a family of
$t$-independent constants of motion for $X$ and
$p(x)=\dim\,N-\dim\mathcal{D}_x^X$. From $(2)$, the vector fields $X_{f_1},\ldots, X_{f_{p(x)}}$
form a family of Lie symmetries of $X$ locally spanning $\mathcal{S}^X_\Lambda$.
Hence, we can easily choose among them a local basis for
$\mathcal{S}^X_\Lambda$. 
\end{proof}

\begin{example}\normalfont

As a particular example of the usefulness of the above result, 
let us turn to a two-dimensional
Smorodinsky--Winternitz oscillator $X$ given by (\ref{LieS}) 
and its known constant of motion (\ref{Inte}). In view of the previous proposition, 
$Y=\widehat\Lambda(dI)$ must be
a Lie symmetry for these systems. A little calculation leads to	
\begin{multline}
Y=2(x_1p_2-p_1x_2)\left(x_2\frac{\partial}{\partial x_1}-x_1\frac{\partial}{\partial x_2}\right)
+2\left[(x_1p_2-p_1x_2)p_2
\vphantom{\frac{x_1^4-x_2^4}{x_1^3x_2^2}}\right.\\\left.
+k\frac{x_1^4-x_2^4}{x_1^3x_2^2}\right]\frac{\partial}{\partial p_1}
-2\left[(x_1p_2-p_1x_2)p_1+k\frac{x_1^4-x_2^4}{x_2^3x_1^2}\right]\frac{\partial}{\partial p_2},
\end{multline}
and it is straightforward to verify that $Y$ commutes with $X_1,X_2,X_3$, given by (\ref{Vessiot--GuldbergSec11}), and therefore with every $X_t$, with $t\in\mathbb{R}$, i.e., $Y$ is a Lie symmetry for $X$.

\end{example}

\begin{proposition} Let $X$ be a Lie--Hamilton system with a Lie--Hamiltonian
structure $(N,\Lambda,h)$. If $[V^X,V^X]=V^X$ and $Y\in {\rm Ham}(N,\Lambda)$ is
a Lie symmetry of $X$, then $Y\in \mathcal{S}_\Lambda^X$.
\end{proposition}
\begin{proof}
As $Y$ is a $t$-independent Lie symmetry, then  $[Y,X_t]=0$ for every $t\in\mathbb{R}$. Since $Y$ is a
Hamiltonian vector field, then $Y=-\widehat\Lambda\circ df$ for a certain $f\in
C^\infty(N)$. Using that $X_t=-\widehat \Lambda(dh_t)$, we obtain
\begin{equation}
0=[Y,X_t]=[\widehat\Lambda (df),\widehat \Lambda (dh_t)]=\widehat \Lambda
(d\{f,h_t\}_\Lambda)=\widehat \Lambda [d (X_tf)].
\end{equation}
Hence, $X_tf$ is a Casimir function. Therefore, as every $X_{t'}$ is a Hamiltonian vector field for all $t'\in\mathbb{R}$, it turns out that $X_{t'}X_tf=0$ for every
$t,t'\in \mathbb{R}$ and, in consequence, $Z_1f$ is a Casimir function for every
$Z_1\in V^X$. Moreover,  as every $Z_2\in V^X$ is Hamiltonian, we have
\begin{equation}
Z_2Z_1f=Z_1Z_2f=0\Longrightarrow (Z_2Z_1-Z_1Z_2)f=[Z_2,Z_1]f=0.
\end{equation}
As $[V^X,V^X]=V^X$, every element $Z$ of $V^X$ can be written as the commutator of two elements of $V^X$ and, in view of the above expression,  $Zf=0$ which shows that $f$ is a $t$-independent
constant of motion for $X$. Finally, as $Y=-\widehat \Lambda(df)$, then $Y\in
\mathcal{S}^X_\Lambda$.
\end{proof}
Note that, roughly speaking, the above proposition ensures that, when $V^X$ is
{\it perfect}, i.e., $[V^X,V^X]=V^X$ (see \cite{Ca03}), then $\mathcal{S}_\Lambda^X$ contains all Hamiltonian Lie symmetries
of $X$. This is the case for Smorodinsky--Winternitz systems (\ref{LieS}) with a non-constant $\omega(t)$, whose
$V^X$ was already shown to be isomorphic to $\mathfrak{sl}(2,\mathbb{R})$.

\subsection{On $t$-dependent constants of motion}
The aim of this section is to present an analysis of the algebraic properties
 of the $t$-dependent constants of motion for Lie--Hamilton systems.
More specifically,
we prove that a Lie--Hamiltonian structure $(N,\Lambda,h)$ for a
Lie--Hamilton system $X$ induces a Poisson
bivector on $\mathbb{R}\times N$. This allows us
to endow the space of constants of motion for $X$ with a Poisson algebra structure, which can be used to produce new constants of motion from known ones.  

\begin{definition}
 We call autonomization of a vector field $X$, to the explicit $t$-dependent vector field $\bar{X}=X+d/dt.$
\end{definition}

Given a system $X$ on $N$, a constant of motion for $X$ is a first-integral $f\in C^\infty(\mathbb{R}\times N)$ of the autonomization $\bar X$.
\begin{equation}\label{EqIn}
\frac{\partial f}{\partial t}+Xf=\bar Xf=0,
\end{equation}
where $X$ is understood as a vector field on $\mathbb{R}\times N$. Using this, we can straightforwardly prove the following proposition.

\begin{proposition}\label{Prop:Ralg} The space $\bar{\mathcal{I}}^X$ of $t$-dependent constants of motion for a
system $X$ forms an $\mathbb{R}$-algebra $(\bar{\mathcal{I}}^X,\cdot)$.
\end{proposition}

 
\begin{lemma} Every Poisson manifold $(N,\Lambda)$ induces a Poisson manifold $(\mathbb{R}\times N,\bar\Lambda)$ with Poisson structure
\begin{equation}\label{newBrack}
\{f,g\}_{\bar\Lambda}(t,x)\equiv\{f_t,g_t\}_{\Lambda}(x),\qquad (t,x)\in
\mathbb{R}\times N.
\end{equation}
\end{lemma}
\begin{definition} Given a Poisson manifold $(N,\Lambda)$,
the associated Poisson manifold $(\mathbb{R}\times N,\bar \Lambda)$ is called the {\it autonomization} of $(N,\Lambda)$. Likewise,
 the Poisson bivector $\bar {\Lambda}$ is called the {\it autonomization} of $\Lambda$.
\end{definition}

The following lemma allows us to prove that $(\bar{\mathcal{I}}^X,\cdot,\{\cdot,\cdot\}_{\bar\Lambda})$
is a Poisson algebra.

\begin{lemma}\label{AutLieHam} Let $(N,\Lambda)$ be a Poisson manifold and
$X$ be a Hamiltonian vector field on $N$ relative to $\Lambda$. Then,
$\mathcal {L}_{\bar{X}}\bar{\Lambda}=0$.
\end{lemma}
\begin{proof}
Given a coordinate system $\{x_1,\ldots,x_n\}$ for $N$ and $x_0 \equiv t$ in $\mathbb{R}$, we can naturally define a coordinate system $\{x_0,x_1,\ldots,x_n\}$ on $\mathbb{R}\times N$.
Since $(x_0)_t=t$ is constant as a function on $N$, then $\bar\Lambda(dx_0,df)=\{(x_0)_{t},f_t\}_\Lambda=0$ for every $f\in C^\infty(\mathbb{R}\times N)$. Additionally, 
$(x_i)_t=x_i$ for $i=1,\ldots,n$. Hence, we have that $\bar\Lambda({\rm d} x_i,{\rm d} x_j)=\{x_i,x_j\}_{\bar\Lambda}$ is a $x_0$-independent function for $i,j=0,\ldots,n$. So,	
\begin{equation}
(\mathcal{L}_{\bar{X}}\bar{\Lambda})(t,x)=\left[\mathcal{L}_{\frac{\partial}{\partial x_0}+X}\left(\sum_{i<j=1}^n\{x_i,x_j\}_\Lambda\frac{\partial}{\partial x_i}\wedge\frac{\partial}{\partial x_j}\right)\right](x)=(\mathcal{L}_{X}\Lambda)(x).
\end{equation}
Since $X$ is Hamiltonian, we obtain $0=(\mathcal{L}_{X}\Lambda)(x)=(\mathcal{L}_{\bar{X}}\bar\Lambda)(t,x)=0$.
\end{proof}

Now, we can establish the main result of this section.

\begin{proposition}\label{AutLieHam2}  Let $X$ be a Lie--Hamilton system on $N$
with a Lie--Hamiltonian structure $(N,
\Lambda,h)$, then the space $(\overline{\mathcal{I}}^X,\cdot,\{\cdot,\cdot\}_{\bar\Lambda})$
is a Poisson algebra.
\end{proposition}
\begin{proof} From Proposition \ref{Prop:Ralg}, we see that
$(\bar{\mathcal{I}}^X,\cdot)$ is an $\mathbb{R}$-algebra. To demonstrate that $(\bar{\mathcal{I}}^X,\cdot,\{\cdot,\cdot\}_{\bar\Lambda})$ is a Poisson algebra, it remains to prove that the Poisson bracket of any two elements $f,g$ of $\bar{\mathcal{I}}^X$ remains in it, i.e.,~$X\{f,g\}_{\bar{\Lambda}}=0$.
By taking into account that the vector fields $\{X_t\}_{t\in\mathbb{R}}$ are Hamiltonian relative to $(N,\Lambda)$ and
Lemma \ref{AutLieHam}, we find that $\bar\Lambda$ is invariant under the autonomization of each vector field $X_{t'}$ with $t'\in\mathbb{R}$, i.e., $
\mathcal{L}_{\overline{X_{t'}}}\bar\Lambda=0$. Therefore,
\begin{align}
  \bar X\{f,g\}_{\bar{\Lambda}}(t',x)&=\overline {X_{t'}}\{f,g\}_{\bar{\Lambda}}(t',x)= \{\overline{X_{t'}}f,g\}_{\bar{\Lambda}}(t',x)+
\{f,\overline{ X_{t'}}g\}_{\bar{\Lambda}}(t',x)
\nonumber\\
 &=\{\bar{X}f,g\}_{\bar{\Lambda}}(t',x)+
\{f,\bar{X}g\}_{\bar{\Lambda}}(t',x)=0.
\end{align}
That is, $\{f,g\}_{\bar{\Lambda}}$ is a $t$-dependent constant of
motion for $X$.
\end{proof}

\subsection{Lie integrals}

\begin{definition} Given a Lie--Hamilton system $X$ on $N$ possessing a Lie--Hamiltonian structure
$(N,\Lambda,h)$, a {\it Lie integral} of $X$ with respect to
$(N,\Lambda,h)$ is a (generally $t$-dependent) constant of motion $f$ of $X$ such that $\{f_t\}_{t\in\mathbb{R}}\subset \ham$. In other words, given a basis $h_1,\ldots,h_r$
of the Lie algebra $(\ham,\{\cdot,\cdot\}_\Lambda)$, we have that $\bar Xf=0$ and
$f_t=\sum_{\alpha=1}^r f_\alpha(t)h_\alpha$ for every $t\in\mathbb{R}$ and certain $t$-dependent functions $f_1,\ldots,f_r$.
\end{definition}

The Lie integrals of a Lie--Hamilton system $X$ relative to
 a Lie--Hamiltonian structure $(N,\Lambda,h)$ are the solutions of  the equation
\begin{equation}
0=\bar{X}f=\frac{\partial f}{\partial t}+Xf=\frac{\partial f}{\partial t}+\{f,h\}_{\bar\Lambda}\Longrightarrow \frac{\partial f}{\partial t}=\{h,f\}_{\bar\Lambda}.
\end{equation}
Since $f$ and $h$ can be understood as curves $t\mapsto f_t$ and $t\mapsto g_t$ within $\ham$, the above equation can be rewritten as
\begin{equation}\label{EqIn2}
\frac{{\rm d}f_t}{{\rm d} t}=\{h_t,f_t\}_{\Lambda},
\end{equation}
which can be thought of as an
Euler equation
on   the Lie algebra $(\mathcal{H}_\Lambda,\{\cdot,\cdot\}_\Lambda)$. Equations of this type quite
frequently appear in the literature such as in the Lewis--Riesenfeld method and
works concerning Lie--Hamilton systems \cite{Ru10,LT05,Ma95}.

\begin{proposition}
Given a  Lie--Hamilton system $X$ with a Lie--Hamiltonian structure $(N,\Lambda,h)$, the
space $\mathfrak{L}^\Lambda_h$ of Lie integrals relative to $(N,\Lambda,h)$
gives rise to a Lie algebra $(\mathfrak{L}^\Lambda_h,\{\cdot ,\cdot
\}_{\bar{\Lambda}})$ isomorphic to $(\ham,\{\cdot,\cdot\}_{\Lambda})$.
\end{proposition}
\begin{proof}
Since the Lie integrals of $X$ are the solutions of the system of ODEs (\ref{EqIn2}) on $\ham$, they span an $\mathbb{R}$-linear space of dimension $\dim\,\ham$. In
view of Proposition \ref{AutLieHam2}, the Poisson bracket $\{f,g\}_{\bar\Lambda}$ of two constants of motion $f,g$ for $X$ is another
constant of motion. If $f$ and $g$ are Lie integrals, the function $\{f,g\}_{\bar \Lambda}$ is then a new constant of motion that can additionally be considered as a curve $t\mapsto \{f_t,g_t\}_{\Lambda}$ taking values in $\ham$, i.e.,~a new Lie integral.

Consider now the linear morphism   $\evo:f\in \mathfrak{L}^\Lambda_h\mapsto f_0\in \ham$ relating every Lie integral to its value in $\ham$ at $t=0$. As every initial condition in $\ham$ is related to a single solution of (\ref{EqIn2}), we can relate every $v\in \ham$ to a unique Lie integral $f$ of $X$ such that $f_0=v$. Therefore, $\evo$ is an isomorphism. Indeed, it is a Lie algebra isomorphism since $(\{f,g\}_{\bar\Lambda})_0=\{f_0,g_0\}_\Lambda$ for every $f,g\in\mathfrak{L}^\Lambda_h$.
\end{proof}

\begin{proposition} Given a Lie--Hamilton system $X$ on $N$ which possesses a Lie--Hamiltonian structure $(N,\Lambda,h)$,
then $\mathfrak{L}^\Lambda_h$ consists of $t$-independent constants of motion if and only if $\ham$ is abelian.
\end{proposition}
 \begin{proof}
If $(\ham,\{\cdot,\cdot\}_\Lambda)$ is abelian, then $\{f_t,h_t\}_{\Lambda}=0$ and the system (\ref{EqIn2}) reduces to
${{\rm d}f_t}/{{\rm d} t}=0$, whose solutions are of the form $f_t=g\in\ham$, i.e.,~$\mathfrak{L}^\Lambda_h=\ham$.
  Conversely, if $\mathfrak{L}^\Lambda_h=\ham$, then every $g\in \ham$ is a solution of (\ref{EqIn2}) and $\{g,h_t\}_\Lambda=0$   $\forall t\in\mathbb{R}$. 
  Hence, every $g\in \ham$ commutes with the whole $\ham$, which becomes Abelian.
\end{proof}

\subsection{Polynomial Lie integrals}

Let us formally define and investigate a remarkable class of constants
of motion for Lie--Hamilton systems appearing in the literature  \cite{Ru10,Ma95}, the hereafter called {\it Lie integrals}, and a relevant generalization of them, 
the {\it polynomial Lie integrals}. We first prove that Lie integrals can be characterized by an Euler equation on a finite-dimensional real Lie algebra of functions, 
retrieving as a particular case a result given in \cite{Ru10}. 
Then, we show that Lie integrals form a finite-dimensional real Lie algebra and we devise several methods to determine them. Our
results can easily be extended to investigate certain quantum mechanical systems \cite{LT05}. Finally, we investigate polynomial Lie integrals
and the relevance of Casimir functions to derive them.

\begin{definition} Given a Lie--Hamilton system $X$ on $N$ possessing a Lie--Hamiltonian structure
$(N,\Lambda,h)$, a {\it Lie integral} of $X$ with respect to
$(N,\Lambda,h)$ is a constant of motion $f$ of $X$ such that $\{f_t\}_{t\in\mathbb{R}}\subset \ham$. In other words, given a basis $h_1,\ldots,h_r$
of the Lie algebra $(\ham,\{\cdot,\cdot\}_\Lambda)$, we have that $\bar Xf=0$ and
$f_t=\sum_{\alpha=1}^r f_\alpha(t)h_\alpha$ for every $t\in\mathbb{R}$ and certain $t$-dependent functions $f_1,\ldots,f_r$.
\end{definition}

The Lie integrals of a Lie--Hamilton system $X$ relative to
 a Lie--Hamiltonian structure $(N,\Lambda,h)$ are the solutions of  the equation
\begin{equation}
0=\bar{X}f=\frac{\partial f}{\partial t}+Xf=\frac{\partial f}{\partial t}+\{f,h\}_{\bar\Lambda}\Longrightarrow \frac{\partial f}{\partial t}=\{h,f\}_{\bar\Lambda}.
\end{equation}
Since $f$ and $h$ can be understood as curves $t\mapsto f_t$ and $t\mapsto g_t$ within $\ham$, the above equation can be rewritten as
\begin{equation}\label{EqIn21}
\frac{{\rm d}f_t}{{\rm d} t}=\{h_t,f_t\}_{\Lambda},
\end{equation}
which can be thought of as an
Euler equation
on   the Lie algebra $(\mathcal{H}_\Lambda,\{\cdot,\cdot\}_\Lambda)$ \cite{Ru10}. Equations of this type quite
frequently appear in the literature such as in the Lewis--Riesenfeld method and
works concerning Lie--Hamilton systems \cite{Ru10,LT05,Ma95}.

\subsubsection*{Polynomial Lie integrals and Casimir functions}

We here investigate a generalization of Lie integrals: the hereafter called {\it polynomial Lie integrals}.
Although we prove that these constants of motion can be determined by Lie integrals, we also show that 
their determination can be simpler in some cases. In particular, we can obtain polynomial Lie integrals algebraically by means of the Casimir functions related to the Lie algebra of Lie integrals.

\begin{definition} Let $X$ be a Lie--Hamilton system admitting a compatible Lie--Hamiltonian
structure $(N,\Lambda,h)$. A {\it polynomial Lie integral} for $X$ with respect to $(N,\Lambda,h)$
is a constant of motion $f$ for $X$ of the form $f_t=\sum_{\JJ\in M}\lambda_\JJ(t)h^\JJ$,
where the $\JJ$'s are $r$-multi-indexes, i.e.,~sets $(i_1,\ldots,i_r)$ of nonnegative integers, the set $M$ is a finite family of multi-indexes, the $\lambda_\JJ(t)$ are certain $t$-dependent functions, and $h^\JJ=h_1^{i_1}\cdot\ldots\cdot h_r^{i_r}$ for a fixed basis $\{h_1,\ldots,h_r\}$ for $\ham$.
\end{definition}

The study of polynomial Lie integrals can be approached through the symmetric Lie algebra $S_\mathfrak{g}$,
where $\mathfrak{g}\simeq \ham$.

\begin{lemma}
\label{62} Every Lie algebra isomorphism $\phi:(\mathfrak{g},[\cdot,\cdot]_{\mathfrak{g}})\rightarrow
(\ham,\{\cdot,\cdot\}_\Lambda)$
can be extended to a Poisson algebra morphism $D:(S_\mathfrak{g},\cdot,\{\cdot,\cdot\}_{S_\mathfrak{g}})\rightarrow (C^\infty(N),\cdot,\{\cdot,\cdot\}_\Lambda)$ in a unique way.
Indeed, if $\{v_1,\ldots,v_r\}$ is a basis for $\mathfrak{g}$, then
$
D(P(v_1,\ldots,v_r))=P(\phi(v_1),\ldots,\phi(v_r)),
$
for every polynomial $P\in S_\mathfrak{g}$.
\end{lemma}

\begin{proof}
The elements of $S_\mathfrak{g}$ given by $v^\JJ\equiv v_1^{i_1}\cdot\ldots\cdot v_r^{i_r}$,
where the $\JJ$'s are $r$-multi-indexes, ``$\cdot$" denotes the product of elements of $\mathfrak{g}$ as functions on $\mathfrak{g}^*$ and $\{v_1,\ldots,v_r\}$ is a basis for $\mathfrak{g}$, form a basis of $S_\mathfrak{g}$.
 Then, every $P\in S_\mathfrak{g}$ can be written in a unique way as $P=\sum_{\JJ\in M}\lambda_\JJ v^\JJ$, where $M$ is a finite family of multi-indexes and each $\lambda_\JJ\in\mathbb{R}$. 
 Hence, the $\mathbb{R}$-algebra morphism  $D:(S_\mathfrak{g},\cdot)
\rightarrow
(C^\infty(N),\cdot)$ extending $\phi : \mathfrak{g}\rightarrow \ham$ is determined by the image of the elements of a basis for $\mathfrak{g}$. Indeed,
\begin{equation}\label{dec2}
D(P)=\sum_\JJ\lambda_\JJ D(v^\JJ)=\sum_\JJ\lambda_\JJ\phi(v^{i_1}_1)\cdot\ldots\cdot \phi(v^{i_r}_r).
\end{equation}
Let us prove that $D$ is also a $\mathbb{R}$-algebra morphism. From (\ref{dec2}),
we see that $D$ is linear. Moreover, $D(PQ)=D(P)D(Q)$
for every $P,Q\in S_\mathfrak{g}$. In fact, if we write $Q=\sum_{\KK\in M}\lambda_\KK v^\KK$, we obtain
\begin{align}
  D(PQ)&=D\left(\sum_\JJ\lambda_\JJ v^\JJ\sum_\KK\lambda_\KK v^\KK\right)=\sum_\LL
\sum_{\JJ+\KK=\LL}\lambda_\JJ\lambda_\KK D(v^\LL)
\nonumber\\
&=\sum_\JJ\lambda_\JJ D(v^\JJ)\sum_\KK\lambda_\KK D(v^\KK)=D(P)D(Q),
\end{align}
where $\JJ+\KK=(i_1+j_1,\ldots,i_r+j_r)$ with $\JJ=(i_1,\ldots,i_r)$ and $\KK=(j_1,\ldots, j_r)$.

Let us show that  $D$ is also a Poisson morphism. By linearity, this reduces to proving that
$D\left ( \{v^\JJ,v^\KK\}_{S_\mathfrak{g}}\right)=\{D(v^\JJ),D (v^\KK)\}_\Lambda$ for arbitrary $\JJ$ and $\KK$. Define $|\JJ|=i_1+\ldots+i_r$.
If $|\KK|=0$ or $|\JJ|=0$ this is satisfied, as a Poisson bracket vanishes when any entry is a constant. We now prove by induction 
the remaining cases. For $|\JJ|+|\KK|=2$, we have 
\begin{equation}
  {D}\left( \{v_\alpha,v_\beta\}_{S_\mathfrak{g}} \right)={\phi}([v_\alpha,v_\beta]_{\mathfrak{g}})=\{ \phi (v_\alpha),\phi (v_\beta)\}_\Lambda=\{{D} (v_\alpha),{D}
(v_\beta)\}_\Lambda ,\quad \end{equation} 
for all $\alpha,\beta=1,\ldots,r.$ If $D$ is a Poisson morphism for $|\JJ|+|\KK|=m>2$,
then for $|\JJ|+|\KK|=m+1$ we can set $v^\JJ=v^{\bar \JJ}v^{i_{\gamma}}_\gamma$ for $i_\gamma\neq 0$ and some $\gamma$
to obtain
\begin{align}
  D\left(\{v^\JJ,v^\KK\}_{S_\mathfrak{g}}\right)&=D\left(\{v^{\bar \JJ}v_\gamma^{i_\gamma},v^\KK\}_{S_\mathfrak{g}}\right)=D\left(\{v^{\bar \JJ},v^\KK\}_{S_\mathfrak{g}}
v_\gamma^{i_\gamma}+v^{\bar \JJ}\{v_\gamma^{i_\gamma},v^\KK\}_{S_\mathfrak{g}}\right)
\nonumber\\
 &=
\{D(v^{\bar \JJ}),D(v^\KK)\}_\Lambda D (v_\gamma^{i_\gamma})+
D( v^{\bar \JJ})\{D(v_\gamma^{i_\gamma}),D (v^\KK)\}_\Lambda
\nonumber\\
&=\{D( v^{\bar \JJ})D (v_\gamma^{i_\gamma}),D (v^\KK)\}_\Lambda=
\{D(v^\JJ),D( v^\KK)\}_\Lambda.
\end{align}
By induction, $D\left ( \{v^\JJ,v^\KK\}_{S_\mathfrak{g}}\right)=\{D(v^\JJ),D (v^\KK)\}_\Lambda$ for any $I$ and $J$.

\end{proof}

Recall that ``$\cdot$'' denotes the standard product of elements
of $S_\mathfrak{g}$ understood as polynomial functions on $S_\mathfrak{g}$. 
It is remarkable that $D$ does not need to be injective, which causes that $S_\mathfrak{g}$ 
is not in general isomorphic to the space of polynomials on the elements of a basis of $\ham$. For instance, 
consider the Lie algebra isomorphism $\phi:(\mathfrak{sl}(2,\mathbb{R}),[\cdot,\cdot]_{\mathfrak{sl}(2,\mathbb{R})})\rightarrow (\ham,\{\cdot,\cdot\}_{\Lambda})$, 
with $\{v_1,v_2,v_3\}$ being a basis of $\mathfrak{sl}(2,\mathbb{R})$, of the form  $\phi(v_1)=p^2$, $\phi(v_2)=xp$ and $\phi(v_3)=x^2$ 
and $\{\cdot,\cdot\}_{\Lambda}$ being the standard Poisson structure on ${\rm T}^*\mathbb{R}$. Then, $D(v_1v_3-v_2^2)=\phi(v_1)\phi(v_3)-\phi^2(v_2)=0$.  

The following notion enables us to simplify the statement and proofs of our results.
\begin{definition} Given a curve $P_t$ in $S_\mathfrak{g}$, its {\it degree}, $\degr(P_t)$, is the highest degree of the polynomials $\{P_t\}_{t\in\mathbb{R}}$. If there exists no finite highest degree, we say that $\degr(P_t)=\infty$.
\end{definition}

\begin{proposition}\label{Th:Lift} A function $f$ is a polynomial Lie integral for a Lie--Hamilton system $X$
with respect to the Lie--Hamiltonian  structure $(N,\Lambda,h)$ if and only
if  for every $t\in\mathbb{R}$ we have $f_t=D(P_t)$, where $D$ is the Poisson algebra morphism
$D:(S_\mathfrak{g},\cdot,\{\cdot,\cdot\}_{{S}_\mathfrak{g}})\rightarrow (C^\infty(N),\cdot,\{\cdot,\cdot\}_\Lambda)$ induced by $\phi:(\mathfrak{g},[\cdot,\cdot]_{\mathfrak{g}})\rightarrow
(\ham,\{\cdot,\cdot\}_\Lambda)$, and the curve $P_t$ is a solution of finite degree for
\begin{equation}\label{PolLie}
\frac{{\rm d}P}{{\rm d} t}+\{P, w_t\}_{S_\mathfrak{g}}=0,\qquad P\in S_\mathfrak{g},
\end{equation}
where $w_t$ stands for a curve in $\mathfrak{g}$ such that $D(w_t)=h_t$ for every $t\in\mathbb{R}$.
\end{proposition}
\begin{proof}
Let $P_t$ be a particular solution of (\ref{PolLie}). Since $D$ is a Poisson algebra morphism and  $h_t=D(w_t)$ for every $t\in\mathbb{R}$, we obtain by applying $D$ to (\ref{PolLie}) that
\begin{equation}\label{PolLie3}
\frac{\partial D(P_t)}{\partial t}+\{D(P_t), D(w_t)\}_{\bar\Lambda}=\frac{\partial D(P_t)}{\partial t}+\{D(P_t), h_t\}_{\bar\Lambda}=0.
\end{equation}
So, $D(P_t)$ is a Lie integral for $X$. Note that this does not depend on the chosen curve $w_t$ satisfying $D(w_t)=h_t$.

Conversely, given a polynomial Lie integral $f$ for $X$, there exists a curve $P_t$ of finite degree
such that $D(P_t)=f_t$ for every $t\in\mathbb{R}$. Hence, we see that
\begin{gather}\label{PolLie4}
  D\left(\frac{{\rm d}P_t}{{\rm d} t}+\{P_t, w_t\}_{S_\mathfrak{g}}\right)=\frac{\partial D(P_t)}{\partial t}+\{D(P_t), D(w_t)\}_{\bar\Lambda}=0
\nonumber\\
\Longrightarrow \frac{{\rm d}P_t}{{\rm d} t}+\{P_t, w_t\}_{S_\mathfrak{g}}=\xi_t,
\end{gather}
where $\xi_t$ is a curve in $\ker D$. As $\degr(dP_t/{\rm d} t)$ and $\degr(\{P_t,w_t\}_{{S_\mathfrak{g}}})$ are at most $\degr(P_t)$, then $\degr(\xi_t)\leq \degr(P_t)$.
Next, consider the equation
\begin{equation}
\label{PolLie5}
\frac{{\rm d}\eta}{{\rm d} t}+\{\eta, w_t\}_{S_\mathfrak{g}}=\xi_t,\qquad \degr(\eta)\leq \degr(P)\quad\,\, {\rm and} \,\,\quad\eta\subset \ker D.\end{equation} 
Note that this equation is well defined. Indeed, since $\degr(\eta)\leq \degr(P)$  and $\degr(w_t)\leq 1$ for every $t\in\mathbb{R}$, then $\degr (\{\eta, w_t\}_{S_\mathfrak{g}})\leq\degr(P)$ for all $t\in\mathbb{R}$. In addition, as $D(\eta_t)\subset \ker D$, then $\{\eta, w_t\}_{S_\mathfrak{g}}\in \ker D$. Then, the above equation can be restricted to the finite-dimensional space of elements of $\ker D$ with degree at most $\degr(P_t)$. Given a particular
solution $\eta_t$ of this equation, which exists for the Theorem of existence and uniqueness, we have that $P_t-\eta_t$ is a solution of (\ref{PolLie}) projecting
into $f_t$.
\end{proof}

\begin{proposition}\label{Pr:De}
Every polynomial Lie integral $f$ of a Lie--Hamilton system
$X$ admitting a Lie--Hamiltonian structure $(N,\Lambda,h)$ can be brought into the form
$f=\sum_{\JJ\in M}c_\JJ l^\JJ,$
where $M$ is a finite set of multi-indexes, the $c_\JJ$'s are certain real constants, and
$l^\JJ=f_1^{i_1}\cdot\ldots\cdot f_r^{i_r}$, with $f_1,\ldots,f_r$ being a basis of the space
$\mathfrak{L}^\Lambda_h$.
\end{proposition}
\begin{proof}
From Proposition \ref{Th:Lift}, we have that $f_t=D(P_t)$ for a solution $P_t$ of finite degree $p$ for (\ref{PolLie}). So, it is a solution of the restriction of this system
to $\mathbb{P}(p,\mathfrak{g})$, i.e.,~the elements of $S_{\mathfrak{g}}$ of degree at most $p$. Given the isomorphism $\phi:\mathfrak{g}\rightarrow \ham$, define $\phi^{-1}(f_j)$, with $j=1,\ldots,r$, to be the curve in $\mathfrak{g}$ of the form $t\mapsto \phi^{-1}(f_j)_t$. Note that $v_1\equiv\phi^{-1}(f_1)_0,\ldots, v_r\equiv\phi^{-1}(f_r)_0$ form a basis of $\mathfrak{g}$. Hence, their polynomials up to order $p$ span a basis for $\mathbb{P}(p,\mathfrak{g})$ and we can write $P_0=\sum_{\JJ\in M}c_\JJ v^\JJ$, where $v^\JJ=v_1^{i_1}\cdot\ldots\cdot v_r^{i_r}$. As $P'_t=\sum_{I\in M}c_I[\phi^{-1}(f_1)]^{i_1}_t\cdots [\phi^{-1}(f_r)]^{i_r}_t$ and $P_t$ are solutions with the same initial condition of the restriction of (\ref{PolLie}) to $\mathbb{P}(p,\mathfrak{g})$, they must be the same in virtue of the Theorem of existence and uniqueness of systems of differential equations. Applying $D$, we obtain that $f_t=D(P_t)=D(\sum_{\JJ\in M}c_\JJ [\phi^{-1}(f_1)_t]^{i_1}\cdots [\phi^{-1}(f_r)_t]^{i_r})=\sum_{\JJ\in 
M}c_\JJ l^\JJ_t$.
\end{proof}

\begin{corollary}\label{ST}
Let $X$ be a Lie--Hamilton system that possesses a Lie--Hamiltonian structure $(N,\Lambda,h)$ inducing a Lie algebra isomorphism $\phi:\mathfrak{g}\rightarrow\ham$ and a Poisson algebra morphism $D:S_\mathfrak{g}\rightarrow C^\infty(N)$. The function $\FF=D(\casS)$, where $\casS$ is a Casimir element of $S_\mathfrak{g}$, is a
$t$-independent constant of motion of $X$. If $\casU$ is
a Casimir element of $U_\mathfrak{g}$, then $F=D(\lambda^{-1}(\casU))$ is
$t$-independent constant of motion for $X$.
\end{corollary}

Note that if $C$ is a constant of motion for $X$, it is also for any other $X'$ whose $V^{X'}\subset V^X$.    
From  Proposition \ref{Pr:De} and Corollary \ref{ST}, it follows that $F=D(\casS)=\sum_{\JJ\in M}c_\JJ l^\JJ$. Therefore,
the knowledge of Casimir elements provides not only constants of motion for Lie--Hamilton systems, but also information
about the Lie integrals of the system.  

As Casimir functions are
known for many Lie
algebras, we can use them to derive constants of motion for the corresponding Lie--Hamilton systems algebraically instead of applying the usual procedure, i.e.,~by solving a system of  PDEs or ODEs.

In particular, Casimir functions for (semi)simple Lie algebras of arbitrary dimension are known \cite{GO64,PP68}. The same is true for the so called ``quasi-simple" Lie algebras, which can be obtained from simple Lie algebras through contraction techniques \cite{Vulpi}. Moreover, the Casimir invariants (Casimir elements of the Poisson algebra $(C^\infty(\mathfrak{g}^*),\{\cdot,\cdot\}_{\mathfrak{g}^*}$), being $\{\cdot,\cdot\}_{\mathfrak{g}^*}$
the Poisson structure induced by the Lie bracket for all real Lie algebras with dimension $d\leq 5$ were given in \cite{invariantsWint} (recall that the Casimir invariants for some of the solvable cases are not polynomial, i.e., they cannot be considered as elements of $S_\mathfrak{g}$), and the literature dealing with Casimir invariants for solvable and nilpotent Lie algebras is not scarce (see, e.g., \cite{campoamor1,campoamor,campoamor2}).

\subsection{The coalgebra method, constants of motion and superposition rules}

We here prove that each Lie--Hamiltonian structure $(N,\Lambda,h)$ for a Lie--Hamilton system $X$ gives rise  in a natural way to a Poisson coalgebra  $(S_\mathfrak{g},\cdot,\{\cdot,\cdot\}_{S_\mathfrak{g}},\Delta)$ where $\mathfrak{g}
\simeq(\ham,\{\cdot,\cdot\}_\Lambda)$.
This allows us to use the coproduct of this coalgebra
to construct new Lie--Hamiltonian structures
for all the diagonal prolongations of $X$ and to derive algebraically $t$-independent constants of  motion for 
such diagonal prolongations. In turn, these constants can further be employed
to obtain a superposition rule for the initial system. Our findings, which are only applicable to 
``primitive" Poisson coalgebras,
give rigorous proof and generalizations of previous achievements established in \cite{BBHMR09,BCR96,BR98}.


\begin{lemma}\label{PA} If $X$ is a Lie--Hamilton system with a Lie--Hamiltonian structure $(N,\Lambda,h)$, then
the space $(S_\mathfrak{g},\cdot,\{\cdot,\cdot\}_{S_\mathfrak{g}},\Delta)$, with $\mathfrak{g}
\simeq(\ham,\{\cdot,\cdot\}_\Lambda)$, is a Poisson coalgebra with a coproduct ${\Delta}:
S_\mathfrak{g}\rightarrow
S_\mathfrak{g}\otimes S_\mathfrak{g}$ satisfying
\begin{equation}\label{Con}
{\Delta}(v)=v\otimes 1+1\otimes v,\qquad \forall  v\in\mathfrak {g}\subset S_\mathfrak{g}.
\end{equation}
\end{lemma}

\begin{proof}
We know that $(S_\mathfrak{g},\cdot,\{\cdot,\cdot\}_{S_\mathfrak{g}})$ and $(S_\mathfrak{g}\otimes S_\mathfrak{g},\cdot_{S_\mathfrak{g}\otimes S_\mathfrak{g} },\{\cdot,\cdot\}_{S_\mathfrak{g}\otimes S_\mathfrak{g} })$ are Poisson algebras. The coassociativity property for the coproduct map (\ref{Con}) is straightforward. Therefore, let us prove
that there exists a Poisson algebra morphism $\Delta:(S_\mathfrak{g},\cdot,\{\cdot,\cdot\}_{S_\mathfrak{g}})
\rightarrow (S_\mathfrak{g}\otimes S_\mathfrak{g},\cdot_{S_\mathfrak{g}\otimes S_\mathfrak{g}},\{\cdot,\cdot\}_{S_\mathfrak{g}\otimes S_\mathfrak{g}})$
satisfying (\ref{Con}), which turns $(S_\mathfrak{g},\cdot,\{\cdot,\cdot\}_{S_\mathfrak{g}},\Delta)$ into a Poisson coalgebra.

The elements of $S_\mathfrak{g}$ of the form $v^\JJ\equiv v_1^{i_1}\cdot\ldots\cdot v_r^{i_r}$, where the $\JJ$'s are $r$-multi-index with $r=\dim\,\mathfrak{g}$, form a basis for $S_\mathfrak{g}$ (considered as a linear space).
 Then, every $P\in S_\mathfrak{g}$ can be written in a unique way as $
P=\sum_{\JJ\in M}\lambda_\JJ v^\JJ$, 
where the $\lambda_\JJ$ are real constants and $\JJ$ runs all the elements of a finite set $M$. In view of this, 
an $\mathbb{R}$-algebra morphism  $\Delta:S_\mathfrak{g}\rightarrow
S_\mathfrak{g}\otimes S_\mathfrak{g}$ is determined by the image of the elements of a basis for $\mathfrak{g}$, i.e.,
\begin{equation}\label{dec}
\Delta(P)=\sum_\JJ\lambda_\JJ\Delta(v^\JJ)=\sum_\JJ\lambda_\JJ\Delta(v^{i_1}_1)\cdot\ldots\cdot\Delta(v^{i_r}_r).
\end{equation}
Therefore, two $\mathbb{R}$-algebra morphisms that coincide on the elements on $\mathfrak{g}$ are necessarily the same.
 Hence, if
there exists such  a morphism, it is unique. Let us prove that there exists an $\mathbb{R}$-algebra morphism
$\Delta$ satisfying (\ref{Con}).

From (\ref{dec}), we easily see that $\Delta$ is $\mathbb{R}$-linear. Let us also prove that $\Delta(PQ)=\Delta(P)\Delta(Q)$
for every $P,Q\in S_\mathfrak{g}$, which shows that $\Delta$ is an $\mathbb{R}$-algebra morphism. If we write $Q=\sum_{\KK\in M}\lambda_\KK v^\KK$, we obtain that
\begin{align}
  \Delta(PQ)&=\sum_\LL\left(
\sum_{\JJ+\KK=\LL}\lambda_\JJ\lambda_\KK\right)\Delta(v^\LL)=\sum_\JJ\lambda_\JJ\Delta(v^\JJ)\sum_\KK\lambda_\KK\Delta(v^\KK )
\nonumber\\
&=\Delta(P)\Delta(Q).\end{align}

Finally, we show that  $\Delta$ is also a Poisson morphism. By linearity, this reduces to proving that $\Delta(\{v^\JJ,v^\KK\}_{S_\mathfrak{g}})=\{\Delta (v^\JJ),\Delta( v^\KK)\}_{S_\mathfrak{g}\otimes S_\mathfrak{g}}$.  If $|\JJ|=0$ or $|\KK|=0$, this result is immediate as the Poisson bracket involving a constant is zero. For the remaining cases and starting by $|\JJ|+|\KK|=2$, we have that ${\Delta}(\{v_\alpha,v_\beta\}_{S_\mathfrak{g}})=\{{\Delta}( v_\alpha),{\Delta}(
v_\beta)\}_{S_\mathfrak{g}\otimes S_\mathfrak{g}}$, $\forall \alpha,\beta=1,\ldots,r.$ Proceeding by induction, we prove that this holds for every value of $|\JJ|+|\KK|$; by writing $v^\JJ= v^{\bar \JJ}v_\gamma^{i_\gamma}$ and using induction hypothesis, we get
\begin{align}
  \Delta\left(\{v^\JJ,v^\KK\}_{S_\mathfrak{g}}\right)&=\Delta\left (\{v^{\bar \JJ}v_\gamma^{i_\gamma},v^\KK\}_{S_\mathfrak{g}}\right)=\Delta\left(\{v^{\bar \JJ},v^\KK\}_{S_\mathfrak{g}}v_\gamma^{i_\gamma}+v^{\bar \JJ}\{v_\gamma^{i_\gamma},v^\KK\}_{S_\mathfrak{g}}\right)
\nonumber\\
  &= \left\{\Delta (v^{\bar \JJ}),\Delta(v^\KK)\right\}_{S_\mathfrak{g}\otimes S_\mathfrak{g}}\!\! \Delta( v_\gamma^{i_\gamma})+\Delta (v^{\bar \JJ})\left\{\Delta (v_\gamma^{i_\gamma}),\Delta (v^\KK)\right\}_{S_\mathfrak{g}\otimes S_\mathfrak{g}}
 \nonumber\\
  &=\left\{\Delta (v^{\bar \JJ})\Delta (v_\gamma^{i_\gamma}),\Delta (v^\KK)\right\}_{S_\mathfrak{g}\otimes S_\mathfrak{g}}=\left\{\Delta (v^\JJ),\Delta (v^\KK)\right\}_{S_\mathfrak{g}\otimes S_\mathfrak{g}}.
 \end{align}\end{proof}

The coproduct defined in the previous lemma gives rise to a new Poisson algebra morphism as stated in the
following immediate lemma.
\begin{lemma}
\label{L72}
The map $\Delta^{(m)}:(S_\mathfrak{g},\cdot,\{\cdot,\cdot \}_{S_\mathfrak{g}})\rightarrow (S_\mathfrak{g}^{(m)},\cdot_{S_\mathfrak{g}^{(m)}},\{\cdot,\cdot\}_{S_\mathfrak{g}^{(m)}})$ , with $m>1$ and $\Delta^{(2)}=\Delta$ given by (\ref{Con}), is a Poisson algebra morphism. 

\end{lemma}

The injection $\iota:\mathfrak{g}\rightarrow  \ham\subset
C^\infty(N)$ is a Lie algebra morphism that can be extended to a Poisson algebra morphism $D:S_\mathfrak{g}\rightarrow C^\infty(N)$
given by
$D(P(v_1,\ldots,v_r))=P(\iota(v_1),\ldots,\iota(v_r))$. Recall that this map need not to be
injective.

\begin{lemma}
\label{TuTu}
The Lie algebra morphism $\mathfrak{g}\hookrightarrow C^\infty(N)$ gives rise to a family of
Poisson algebra morphisms $D^{(m)}:S^{(m)}_\mathfrak{g} \hookrightarrow C^\infty(N)^{(m)} \subset
C^\infty(\, {N^m}\,)$ satisfying, for all $v_1,\ldots,v_m\in \mathfrak{g}\subset S_\mathfrak{g}$,  that
\begin{equation}\label{induced}
  \left[D^{(m)}(v_{1}\otimes\ldots\otimes
v_{m})\right]\!(x_{(1)},\ldots,x_{(m)})\!=\![D(v_{1})](x_{(1)})\cdot\ldots\cdot
[D(v_{m})](x_{(m)}),
\end{equation}
where $x_{(i)}$ is a point of the manifold $N$ placed in the $i$-position within the product $N\times\ldots\times N\,\equiv N^m$.
\end{lemma}

From the above results, we can easily demonstrate the following statement which
shows that the diagonal prolongations of a Lie--Hamilton system $X$ are also
Lie--Hamilton ones admitting a structure induced by that of $X$.

\begin{proposition}\label{MT1} If $X$ is a Lie--Hamilton system on $N$ with a Lie--Hamiltonian structure $(N,\Lambda,h)$,
then the diagonal
prolongation $\widetilde X$ to each $N^{m+1}$ is also a Lie--Hamilton system endowed with a Lie--Hamiltonian structure $(N^{m+1},
\Lambda^{m+1},\widetilde h)$ given by
$\Lambda^{m+1}(x_{(0)},\ldots,x_{(m)})=\sum_{a=0}^m\Lambda(x_{(a)})$, where we make use of the vector bundle
isomorphism ${\rm T}N^{m+1}\simeq {\rm T}N\oplus\cdots\oplus{\rm T}N$,
and $\widetilde h_t=D^{(m+1)}({\Delta}^{(m+1)}(h_t))$, where $D^{(m+1)}$ is the Poisson algebra morphism (\ref{induced}) induced
by the Lie algebra morphism $\mathfrak{g}\hookrightarrow \ham\subset C^\infty(N)$.
\end{proposition}

The above results enable us to prove the following theorem that provides  a method
to obtain $t$-independent constants of motion for the diagonal
prolongations of a Lie--Hamilton system. From this theorem, one
may obtain superposition rules for Lie--Hamilton
systems in an algebraic way.  Additionally,  this theorem is a generalization,  only valid in the case of primitive coproduct maps, of the integrability theorem for coalgebra symmetric systems given in \cite{BR98}.

\begin{theorem}\label{MT2}
 If $X$ is a Lie--Hamilton system with a
Lie--Hamiltonian structure $(N,\Lambda,h)$ and $\casS$ is a Casimir element of $(S_\mathfrak{g}
,\cdot,\{,\}_{S_\mathfrak{g}})$, where $\mathfrak{g}\simeq \ham$, then\\
(i)
The functions defined as
\begin{equation}\label{invA}
\FF^{(k)} = D^{(k)}(\Delta^{(k)}({\casS})) ,   \qquad k=2,\ldots,m,
\end{equation}
are  $t$-independent constants of motion  for the diagonal prolongation
$\widetilde X$ to $N^m$. Furthermore, if all the $\FF^{(k)}$ are non-constant functions,  they form a set of  $(m-1)$      functionally independent functions  in involution.
\\
(ii) The functions given by
\begin{equation}\label{invB}
\FF_{ij}^{(k)}=S_{ij} ( \FF^{(k)}   ) , \qquad 1\le  i<j\le  k,\qquad k=2,\ldots,m,
\end{equation}
where $S_{ij}$ is the permutation of variables $x_{(i)}\leftrightarrow
x_{(j)}$, are $t$-independent constants of motion  for the diagonal prolongation
$\widetilde X$ to $N^m$.
\end{theorem}
  \begin{proof}
Every $P\in S^{(j)}_\mathfrak{g}$ can naturally be considered
as an element
$P\otimes \stackrel{(k-j)-{\rm times}}{\overbrace{ 1\otimes\ldots\otimes
1}}\,\, \in
S^{(k)}_\mathfrak{g}.
$
Since $j\le k$, we have that $\{\Delta^{(j)}(\bar v),\Delta^{(k)}( v)\}_{S^{(k)}_\mathfrak{g}}=
\{\Delta^{(j)}(\bar v),\Delta^{(j)}( v)\}_{S^{(j)}_\mathfrak{g}}$, $\forall \bar v, v\in\mathfrak{g}$. So,
\begin{equation}
\{\Delta^{(j)}({\casS}),\Delta^{(k)}(v)\}_{S^{(k)}_\mathfrak{g}}=
\{\Delta^{(j)}({\casS}),\Delta^{(j)}(v)\}_{S^{(j)}_\mathfrak{g}}=\Delta^{(j)}\left(\{{\casS},v\}_{S^{(j)}_{ {\mathfrak{g}}  }}\right)=0.\end{equation} 
Hence, by using that every function $f\in C^\infty(N^j)$ can be understood as a function $\pi^*f\in C^\infty(N^k)$, being
$\pi:N^j\times N^{k-j}\rightarrow N^j$ the projection onto the first factor, and by applying the Poisson algebra morphisms   introduced in Lemma~\ref{TuTu} we get
\begin{equation}
\left \{D^{(j)}(\Delta^{(j)}({\casS})),D^{(k)}(\Delta^{(k)}(v))\right\}_{\Lambda^k}
=\left \{\FF^{(j)} ,D^{(k)}(\Delta^{(k)}(v))\right\}_{\Lambda^k}=0 , \,\, \forall v\in\mathfrak{g},\end{equation} 
 which leads to $ \left\{\FF^{(j)},\FF^{(k)} \right\}_{\Lambda^k}=0$, that is,
the functions  (\ref{invA}) are in involution. By construction (see Lemma~\ref{L72}), if these are non-constant, then they are functionally independent functions since $\FF^{(j)}$ lives in  $N^{j}$, meanwhile  $\FF^{(k)}$ is defined on  $N^{k}$.

Let us prove now that all the functions (\ref{invA}) and (\ref{invB}) are   $t$-independent constants of motion for $\widetilde X$.
Using that $\ham\simeq \mathfrak{g}$ and
$X_t=-\widehat\Lambda\circ d h_t$, we see that $X$ can be brought into the form $X_t=-\widehat\Lambda\circ d\circ
D(v_t)$ for a unique curve $t\rightarrow v_t$ in $\mathfrak{g}$. From this and Proposition \ref{MT1}, it follows
\begin{align}
  \widetilde X_t=-\Lambda^m\circ & d
D^{(m)}(\Delta^{(m)}(v_t) )\Longrightarrow \nonumber\\
&\widetilde X_t
( \FF^{(k)})=\left\{D^{(k)}(\Delta^{(k)}({\casS})),D^{(m)}(\Delta^{(m)}(v_t))\right\}_{\Lambda^m}=0.\end{align} 
Then,  $\FF^{(k)}$ is a common
first-integral for every $\widetilde X_t$. Finally, consider the permutation operators $S_{ij}$, with $1\le i <j \le k$ for $k=2,\dots,m$. Note that
\begin{align}
0=S_{ij} \left\{\FF^{(k)},D^{(m)}(\Delta^{(m)}(v_t))\right\}_{\Lambda^m} &=  \left\{S_{ij}  ( \FF^{(k)} ),S_{ij} \left(D^{(m)}(\Delta^{(m)}(v_t)) \right)\right\}_{\Lambda^m} 
\nonumber\\
&=\left\{ F_{ij}^{(k) }, D^{(m)}(\Delta^{(m)}(v_t))  \right\}_{\Lambda^m} =\widetilde X_t   (F_{ij}^{(k) }  ) .
\end{align}
Consequently, the functions $F_{ij}^{(k) }$ are $t$-independent constants of motion for $\widetilde X$.
  \end{proof}

Note that the ``omitted" case with $k=1$ in the set of constants (\ref{invA}) is, precisely, the one provided by Corollary \ref{ST} as $\FF^{(1)}\equiv \FF=D({\casS})$. Depending on the system $X$, or more specifically, on the associated $\ham$, the function $\FF$ can be 
either a useless trivial constant or a relevant function. It is also worth noting that  constants (\ref{invB}) need not be functionally independent, but we can always  choose those fulfilling such a property. Finally, observe that if $X'$ is such that $V^{X'}\subset V^X$, then the functions (\ref{invA}) and (\ref{invB}) are also constants of motion for the diagonal prolongation of $X'$ to $N^m$.

\section{Applications of the geometric theory of Lie--Hamilton systems}
\setcounter{equation}{0}
\setcounter{theorem}{0}
\setcounter{example}{0}
In this section we aim to achieve superposition rules and constants of motion through the coalgebra formalism. We find of
special interest our achievement of finding a superposition rule for Riccati equations with the aid of the coalgebra method.

\subsection{The Ermakov system} 
Let us consider the classical Ermakov system \cite{Dissertationes}
\begin{equation}
\left\{\begin{aligned}
\frac{{\rm d}^2x}{{\rm d} t^2}&=-\omega^2(t)x+\frac{b}{x^3},\\
\frac{{\rm d}^2y}{{\rm d} t^2}&=-\omega^2(t)y,
\end{aligned}\right.
\end{equation}
with a non-constant $t$-dependent frequency $\omega(t)$, being $b$  a real constant. This system appears in a number of
applications related to problems in Quantum and Classical Mechanics \cite{AL08}.
By writting this system as a first-order one
\begin{equation}\label{ClasErm}
\left\{\begin{aligned}
\frac{{\rm d} x}{{\rm d} t}&=v_x, &\frac{{\rm d}v_x}{{\rm d} t}&=-\omega^2(t)x+\frac{b}{x^3},\\
\frac{{\rm d}y}{{\rm d} t}&=v_y, &\frac{{\rm d}v_y}{{\rm d} t}&=-\omega^2(t)y,
\end{aligned}
\right.
\end{equation}
we can apply the theory of Lie systems. Indeed, this is a Lie system related to a Vessiot--Guldberg Lie
algebra $V$ isomorphic to $\mathfrak{sl}(2,\mathbb{R})$  \cite{SIGMA}. In fact, system
(\ref{ClasErm}) describes the integral curves of the $t$-dependent vector field $X=X_3+\omega^2(t)X_1$,
where the vector fields
\begin{equation}
\begin{gathered}
  X_1=-x\frac{\partial}{\partial v_x}-y\frac{\partial}{\partial v_y}, \qquad
X_2=\frac12 \left({v_x}\frac{\partial}{\partial v_x}+{v_y}\frac{\partial}{\partial v_y}-x\frac{\partial}{\partial x}-y\frac{\partial}{\partial y}\!\right),  
\\
X_3=v_x\frac{\partial}{\partial x}+
v_y\frac{\partial}{\partial y}+\frac{b}{x^3}\frac{\partial}{\partial v_x},
\end{gathered}\end{equation} 
satisfy the commutation relations
\begin{equation}
\label{sl2Rtb}
[X_1,X_2]=X_1,\qquad [X_1,X_3]= 2X_2,\qquad [X_2,X_3]=X_3.
\end{equation}

As a first new result we show that this is a Lie--Hamilton system.  The vector fields are Hamiltonian
with respect to the  Poisson bivector $\Lambda=\partial/\partial x\wedge \partial/\partial v_x+
\partial/\partial y\wedge \partial/\partial v_y$ provided that  $X_\alpha=- \widehat\Lambda(d h_\alpha)$ for $\alpha=1,2,3$. Thus,
 we find the  following Hamiltonian functions which form a basis for $(\ham,\{\cdot,\cdot\}_\Lambda)\simeq (\mathfrak{sl}(2,\mathbb{R}),[\cdot,\cdot])$
\begin{equation}
h_1=\frac 12(x^2+y^2),\qquad h_2=-\frac 12 (xv_x+yv_y),\qquad h_3=\frac 12\left(v_x^2+v_y^2+\frac{b}{x^2}\right),
\end{equation}
 as they fulfil
\begin{equation}
\label{sl2Rh1}
\{h_1,h_2\}=-h_1,\qquad \{h_1,h_3\}=-2h_2,\qquad \{h_2,h_3\}=-h_3.
\end{equation}

Since $X=X_3+\omega^2(t)X_1$ and $\omega(t)$ is not a constant, every $t$-independent constant of motion $f$ for $X$ is a common first-integral for $X_1,X_2,X_3$. 
Instead of searching an $f$ by solving the system of
PDEs given by $X_1f=X_2f=X_3f=0$,
we use Corollary~\ref{ST}. This easily provides such a first integral  through the Casimir element  of the symmetric algebra of
${\mathfrak{sl}(2,\mathbb{R})}$. Explicitly,
 given a basis $\{\ve_1,\ve_2,\ve_3\}$
for $\mathfrak{sl}(2,\mathbb{R})$ satisfying
\begin{equation}
\label{sl2Rve}
[\ve_1,\ve_2]=-\ve_1,\qquad  [\ve_1,\ve_3]=-2\ve_2,\qquad [\ve_2,\ve_3]=-\ve_3,
\end{equation}
the Casimir element
 of $\mathfrak{sl}(2,\mathbb{R})$ reads $\casU=\frac 12 (\ve_1\widetilde\otimes \ve_3+\ve_3\widetilde\otimes \ve_1)-\ve_2\widetilde\otimes \ve_2\in  U_{\mathfrak{sl}(2,\mathbb{R})}$. Then, the inverse of symmetrizer
morphism (\ref{symmap}), $\lambda^{-1}: U_{\mathfrak{sl}(2,\mathbb{R})} \rightarrow S_{\mathfrak{sl}(2,\mathbb{R})}
$, gives rise to  the Casimir element of $S_{\mathfrak{sl}(2,\mathbb{R})}$
\begin{equation}
\label{casSL2}
\casS=\lambda^{-1}(\casU)=\ve_1\ve_3-\ve^2_2 .
 \end{equation}
 According to Lemma \ref{62} we consider the  Poisson algebra morphism $D$  induced by the isomorphism $\phi:  {\mathfrak{sl}(2,\mathbb{R})} \to \ham$ defined by $\phi(\ve_\alpha)= h_\alpha$ for $\alpha=1,2,3$. 
 Subsequently, via Corollary~\ref{ST}, we obtain
\begin{align}
  F=D(\casS)= \phi(\ve_1)\phi(\ve_3)-\phi^2(\ve_2)=&h_1h_3-h_2^2=\nonumber\\
&=(v_y x-v_x y)^2+b\left(1+\frac{y^2}{x^2}\right) .
\end{align}
 In this way, we recover,
 up to an additive and multiplicative constant, the well-known Lewis--Riesenfeld
invariant \cite{AL08}. Note that when $\omega(t)$ is a constant, then $V^X\subset V$ and the function $F$ is also a constant of motion
for $X$ (\ref{ClasErm}).

\subsection{A superposition rule for Riccati equations}

Let us turn to the system of Riccati equations  on $\mathcal{O}=\{(x_1,x_2,x_3,x_4)\,|\, x_i\neq x_j,i\neq j=1,\ldots,4\}\subset \mathbb{R}^4,$ given by
\begin{equation}\label{CoupledRic}
\frac{{\rm d} x_i}{{\rm d} t}=a_0(t)+a_1(t)x_i+a_2(t)x_i^2,\qquad i=1,\ldots,4,
\end{equation}
where $a_0(t),a_1(t),a_2(t)$ are arbitrary $t$-dependent functions.
The knowledge of a non-constant $t$-independent constant of motion for any system of this type
leads to obtaining a superposition rule for Riccati equations \cite{CGM07}.
Usually, this requires the integration of a system of PDEs~\cite{CGM07} or ODEs~\cite{PW}.
We here obtain such a $t$-independent constant of motion through algebraic methods by showing that (\ref{CoupledRic})
is a Lie--Hamilton system with a given Lie--Hamiltonian structure and obtaining an associated polynomial Lie integral.

Observe that (\ref{CoupledRic}) is  a Lie system related to
a $t$-dependent vector field $X_R=a_0(t)X_1+a_1(t)X_2+a_2(t)X_3$, where
\begin{equation}
X_1=\sum_{i=1}^4\frac{\partial}{\partial x_i},\qquad X_2=\sum_{i=1}^4x_i\frac{\partial}{\partial x_i},\qquad X_3=\sum_{i=1}^4x_i^2\frac{\partial}{\partial x_i}
\end{equation}
span a Vessiot--Guldberg Lie algebra $V$ for (\ref{CoupledRic}) isomorphic to $\mathfrak{sl}(2,\mathbb{R})$ satisfying the same commutation relations (\ref{sl2Rtb}). For simplicity, we assume $V^X=V$. Nevertheless, our final results are valid for any other case.

To show that (\ref{CoupledRic}) is a Lie--Hamilton system for arbitrary functions $a_0(t)$, $a_1(t)$, $a_2(t)$, we need to search for a symplectic form $\omega$ such that $V$ consists of Hamiltonian vector fields. By impossing $\mathcal{L}_{X_\alpha}\omega=0$, for $\alpha=1,2,3$, we obtain the 2-form
\begin{equation}\label{wR}
\omega_R=\frac{{\rm d} x_1\wedge {\rm d} x_2}{(x_1-x_2)^2}+\frac{{\rm d} x_3\wedge {\rm d} x_4}{(x_3-x_4)^2},
\end{equation}
which is closed and non-degenerate on $\mathcal{O}$. Now, observe that
$\iota_{X_\alpha}\omega={\rm d}h_\alpha$, with $\alpha=1,2,3$ and
\begin{equation}
\begin{gathered}
  h_1=\frac{1}{x_1-x_2}+\frac{1}{x_3-x_4}, \quad
h_2=\frac 12\left(\frac{x_1+x_2}{x_1-x_2}+\frac{x_3+x_4}{x_3-x_4}\right), \nonumber\\
h_3=\frac{x_1 x_2}{x_1-x_2}+\frac{x_3 x_4}{x_3-x_4}.
\end{gathered}
\end{equation}
So, $h_1,h_2$ and $h_3$ are Hamiltonian functions for $X_1$, $X_2$ and $X_3$, correspondingly. Using the Poisson bracket $\{\cdot,\cdot\}_\omega$ induced by $\omega$, we obtain that $h_1,h_2$ and $h_3$ satisfy the commutation relations (\ref{sl2Rh1}), 
and $(\langle h_1,h_2,h_3\rangle,\{\cdot,\cdot\}_\omega) \simeq \mathfrak{sl}(2,\mathbb{R})$.
Next, we again express  $\mathfrak{sl}(2,\mathbb{R})$ in the   basis $\{\ve_1,\ve_2,\ve_3\}$  with Lie brackets   (\ref{sl2Rve}) and Casimir function (\ref{casSL2}), 
and we consider the Poisson algebra morphism $D: S_{{\mathfrak{sl}(2,\mathbb{R})}}\to   C^\infty ({\cal O})$  given by the isomorphism $\phi(\ve_\alpha)=\,h_\alpha$ for $\alpha=1,2,3$.
As $(\mathcal{O},\{\cdot,\cdot\}_\omega,h_t=a_0(t)h_1+a_1(t)h_2+a_2(t)h_3)$ is a Lie--Hamiltonian structure for $X$ and applying Corollary~\ref{ST}, we obtain  
the $t$-independent constant of motion for $X$
\begin{equation}
F=D(\casS)= h_1h_3-h_2^2=\frac{(x_1-x_4)(x_2-x_3)}{(x_1-x_2)(x_3-x_4)}.
\end{equation}
As in the previous example, if $V^X\subset V$, then $F$ is also a constant of motion for $X$. It is worth noting that $F$ is the
known constant of motion obtained for deriving a superposition rule for Riccati equations
\cite{CGM07,PW}, which is here deduced through a simple algebraic calculation.

It is also interesting that $V$ also becomes  a Lie algebra of Hamiltonian vector fields with respect to a
 second symplectic structure given by $\omega=\sum_{1\le i<j}^4\frac{{\rm d} x_i\wedge {\rm d} x_j}{(x_i-x_j)^2}\,$.
Consequently, the system (\ref{CoupledRic}) can be considered, in fact, as a {\it bi--Lie--Hamilton system}.

\subsection{Kummer--Schwarz equations in Hamilton form}
 It was proven  in  \cite{CLS122}  that the   second-order Kummer--Schwarz equations \cite{BB95,CGL11}
admit a $t$-dependent Hamiltonian which can be used to work out their Hamilton's equations, namely
\begin{equation}\label{Hamilsi}
\left\{
\begin{aligned}
\frac{{\rm d} x}{{\rm d} t}&=\frac{px^3}{2},\\
\frac{{\rm d} p}{{\rm d} t}&=-\frac{3p^2x^2}{4}- \frac{b_0}{4}+\frac{4b_1(t)}{x^2},
\end{aligned}\right.
\end{equation}
where $b_1(t)$ is assumed to be a non-constant $t$-dependent function, $(x,p)\in {\rm T}^*\mathbb{R}_0$ with  $\mathbb{R}_0\equiv \mathbb{R}-\{0\}$, and $b_0$ is a real constant.
 This is  a  Lie system  associated with the $t$-dependent vector field
$X=X_3+b_1(t)X_1$~\cite{CLS122}, where the vector fields
\begin{equation}
\begin{gathered}
  X_1=\frac{4}{x^2}\frac{\partial}{\partial p},\qquad
X_2=x\frac{\partial}{\partial x}-p\frac{\partial}{\partial p},
\\
X_3=\frac{px^3}{2}\frac{\partial}{\partial
x}-\frac 14\left({3p^2x^2}+ b_0\right) \frac{\partial}{\partial p}
\end{gathered}\end{equation} 
span a Vessiot--Guldberg Lie algebra $V$   isomorphic to $\mathfrak{sl}(2,\mathbb{R})$   fulfilling   (\ref{sl2Rtb}).
 Moreover, $X$ is a Lie--Hamilton system as $V$ consists of
Hamiltonian vector fields with respect to the Poisson bivector
$\Lambda={\partial}/{\partial x}\wedge\partial/ \partial p$ on ${\rm T}^* \mathbb{R}_0$. Indeed,
$X_\alpha=-\widehat\Lambda(dh_\alpha)$, with $\alpha=1,2,3$ and
\begin{equation}\label{FunKS}
h_1=\frac 4 x,\qquad h_2= xp,\qquad h_3=\frac 14\left(p^2x^3+ b_0x \right)
\end{equation}
are a basis of    a Lie algebra isomorphic to $\mathfrak{sl}(2,\mathbb{R})$ satisfying the commutation relations (\ref{sl2Rh1}).
Therefore, (\ref{Hamilsi}) is a Lie--Hamilton system possessing a Lie--Hamiltonian structure
$({\rm T}^*\mathbb{R}_0,\Lambda,h)$, where $h_t=h_3+b_1(t)h_1$. 

To obtain a superposition rule for $X$ we need to determine an integer $m$ so
that the diagonal prolongations  of $X_\alpha$ to ${\rm T}^*\mathbb{R}^m_0$ $(\alpha=1,2,3)$
become linearly independent at a generic point (see \cite{CGM07}). This happens for $m=2$. We consider a coordinate system in ${\rm T}\mathbb{R}^3_0$, namely $\{ x_{(1)}, p_{(1)},
x_{(2)}, p_{(2)},x_{(3)}, p_{(3)} \}$. A superposition rule for $X$ can be obtained
by determining two common first integrals for the diagonal prolongations
$\widetilde X_\alpha$ to ${\rm T}^*\mathbb{R}^3_0$
 satisfying
\begin{equation}\label{F1F2}
\frac{\partial(F_1,F_2)}{\partial(x_{(1)},p_{(1)})}\neq 0.
\end{equation}
Instead of searching $F_1,F_2$ in the standard way, i.e.,~by solving the system of PDEs given by $\widetilde X_\alpha f =0$,
we make use of Theorem \ref{MT2}. This provides such first integrals through the Casimir element $\casS$  (\ref{casSL2}) of the symmetric algebra of
$\ham\simeq {\mathfrak{sl}(2,\mathbb{R})}$. Indeed, the coproduct (\ref{Con}) enables
us to define the elements
\begin{equation}
\begin{aligned}
  \Delta (C)&=\Delta(\ve_1)\Delta(\ve_3)-\Delta(\ve_2)^2
\\
&=(\ve_1 \otimes  1 + 1 \otimes  \ve_1)(\ve_3 \otimes  1 + 1 \otimes  \ve_3)\!-\!(\ve_2 \otimes
 1 + 1 \otimes  \ve_2)^2,
\\
  \Delta^{(3)} (C)&=\Delta^{(3)}(\ve_1)\Delta^{(3)}(\ve_3)-\Delta^{(3)}(\ve_2)^2
\\
&=
(\ve_1 \otimes  1\otimes  1+1 \otimes  \ve_1 \otimes  1 + 1 \otimes  1\otimes
  \ve_1)
 (\ve_3 \otimes  1 \otimes  1 + 1 \otimes  \ve_3 \otimes  1 
\\
&\hspace{1em}+ 1 \otimes 1 \otimes
  \ve_3)-(\ve_2 \otimes  1 \otimes
 1 + 1 \otimes  \ve_2 \otimes  1 + 1 \otimes 1 \otimes  \ve_2)^2,
\end{aligned}\end{equation} 
for $S_{\mathfrak{sl}(2,\mathbb{R})}^{(2)}$ and $S_{\mathfrak{sl}(2,\mathbb{R})}^{(3)}$, respectively. By applying $D$, $D^{(2)}$ and $D^{(3)}$ coming from
   the isomorphism $\phi(\ve_\alpha)=\,h_\alpha$ for the Hamiltonian functions (\ref{FunKS}),     we obtain,  via Theorem \ref{MT2},
 the following constants of motion of the type (\ref{invA})
 
\begin{equation}
\begin{aligned}
   F&=D(\casS) = h_1(x_1,p_1) h_3(x_1,p_1)-h_2^2(x_1,p_1)=  b_0, \\
  \FF^{(2)} &= D^{(2) } (\Delta(\casS) )
\\
&=  \left[h_1(x_1,p_1)+h_1(x_2,p_2)\right]\left[h_3(x_1,p_1)+h_3(x_2,p_2)\right] 
\\
   &
\hspace{1em}-\left[h_2(x_1,p_1)+h_2(x_2,p_2)\right]^2
\\
&
=\frac{b_0\left(x_1 +x_2\right)^2+\left(p_1x_1^2-p_2x_2^2\right)^2}{x_1x_2}=\frac{b_0(x_1^2+x_2^2)+\left(p_1x_1^2-p_2x_2^2\right)^2}{x_1x_2}+2b_0, 
\\
  \FF^{(3)} & =D^{(3) } (\Delta(\casS) )
= \sum_{i=1}^3 h_1(x_i,p_i) \sum_{j=1}^3h_3(x_j,p_j)-\left( \sum_{i=1}^3 h_2(x_i,p_i)   \right)^2
\\
  &= \sum_{1\le i<j}^3 \frac{b_0(x_i +x_j)^2+(p_ix_i^2-p_jx_j^2)^2}{x_ix_j} - 3 b_0,
  \label{intKS}
\end{aligned}\end{equation} 
    where, for the sake of simplicity, hereafter we   denote $(x_i,p_i)$ the coordinates $(x_{(i)},p_{(i)})$. Thus $F$ simply  gives rise to the constant $b_0$, while $\FF^{(2)}$ and $\FF^{(3)}$ are, by construction,   two functionally independent constants of motion for $\widetilde X$ fulfilling (\ref{F1F2}) which, in  turn, allows us to   to derive a superposition rule for $X$. Furthermore, the function $\FF^{(2)}\equiv \FF^{(2)}_{12}$ provides two other constants of the type (\ref{invB}) given by  $\FF_{13}^{(2)}=S_{13}(\FF^{(2)})$ and  $\FF_{23}^{(2)}=S_{23}(\FF^{(2)})$ that verify $  \FF^{(3)}=  \FF^{(2)}+\FF_{13}^{(2)}+\FF_{23}^{(2)}-3b_0 .$
Since it is simpler to work with $\FF_{23}^{(2)}$ than with $  \FF^{(3)}$, we choose the pair $ \FF^{(2)}$, $\FF_{23}^{(2)}$ as the two  functionally independent  first integrals   to obtain a superposition rule. We set
\begin{equation}
\label{xa}
\FF^{(2)}=k_1+2b_0,\qquad \FF_{23}^{(2)}=k_2+2b_0,
\end{equation}
and   compute $x_1,p_1$ in terms of the other variables and $k_1,k_2$.
From (\ref{xa}), we have
\begin{equation}
\label{p1}
p_1=p_1(x_1,x_2,p_2,x_3,p_3,k_1)=\frac{p_2x_2^2\pm\sqrt{k_1 x_1 x_2-b_0(x_1^2+x_2^2)}}{x_1^2}.
\end{equation}
Substituying in the second relation within (\ref{xa}), we obtain
 \begin{equation}
\label{x1}
x_1=x_1(x_2,p_2,x_3,p_3,k_1,k_2)= \frac{\aA^2\bB_++b_0 \bB_- (x_2^2-x_3^2) \pm 2 \aA\sqrt{\cC}}{\bB_-^2+4b_0 \aA^2} ,
\end{equation}
provided that the functions $\aA,\bB_\pm,\cC$ are defined by
\begin{equation}
 \begin{gathered}
\aA=p_2x_2^2-p_3 x_3^2,\qquad \bB_\pm=k_1 x_2\pm k_2 x_3,\\
\begin{aligned}
\cC=\aA^2\left[ k_1k_2x_2x_3\right.&-2b_0^2\left.(x_2^2+x_3^2)-
b_0\aA^2\right]
\\
&+b_0x_2x_3B_-(k_2x_2-k_1x_3)-b_0^3\left(x_2^2-x_3^2\right)^2.                                                                            \end{aligned}
\end{gathered}
\end{equation}
By introducing this result  into (\ref{p1}), we obtain $p_1=p_1(x_2,p_2,x_3,p_3,k_1,k_2)$ which, along with $x_1=x_1(x_2,p_2,x_3,p_3,k_1,k_2)$, provides  a superposition rule for $X$.

In particular for (\ref{Hamilsi})  with  $b_0=0$, it results
\begin{equation}
\begin{aligned}
  x_1&= \frac{ \aA^2\left(\bB_+   \pm 2  \sqrt{k_1k_2x_2x_3} \,\right)}{\bB_-^2} ,
\\
 p_1&= \bB_-^3\frac{\left[ \bB_- p_2 x_2^2\pm \aA\sqrt{k_1 x_2\left(\bB_+ \pm 2\sqrt{k_1k_2 x_2 x_3} \,\right)} \,\right] }  {\aA^4\left( \bB_+ \pm 2\sqrt{k_1k_2 x_2 x_3}\, \right)^2} ,\end{aligned}\end{equation} 
where the functions $\aA,\bB_\pm$ remain in the above same  form. As the constants of motion were derived for non-constant $b_1(t)$, when $b_1(t)$ is constant
we have $V^{X}\subset V$. As a consequence, the functions $F$, $F^{(2)}$, $F^{(3)}$  and so on are still constants of motion for the diagonal prolongation $\widetilde X$ and the superposition rules are still valid for any system  (\ref{Hamilsi}).

\subsection{The $n$-dimensional Smorodinsky--Winternitz systems}
Let us focus on the  $n$-dimensional  Smorodinsky--Winternitz systems \cite{WSUF65,WSUF67} with unit mass and  
a non-constant $t$-dependent frequency $\omega(t)$ whose  Hamiltonian is given by
\begin{equation}
h=\frac 12  \sum_{i=1}^n p_i^2+ \frac 12 \omega^2(t) \sum_{i=1}^n x_i^2+\frac 12 \sum_{i=1}^n \frac{b_i}{x_i^2},
\end{equation}
where the $b_i$'s are $n$ real constants.
  The corresponding Hamilton's equations read
\begin{equation}\label{ClasErm2}
\left\{\begin{aligned}
\frac{{\rm d} x_i}{{\rm d} t}&=p_i,\\
\frac{{\rm d} p_i}{{\rm d} t}&=-\omega^2(t)x_i+\frac{b_i}{x_i^3},\\
\end{aligned}
\right.\qquad i=1,\ldots,n.
\end{equation}
 These systems have been recently attracting quite much attention in Classical and
Quantum Mechanics for their special properties \cite{CLS122,GPS06,SIGMAvulpi,YNHJ11}. Observe that Ermakov systems (\ref{ClasErm}) arise  
as the particular case   of  (\ref{ClasErm2}) for  $n=2$ and $b_2=0$. For $n=1$
the above system maps into the Milne--Pinney equations, which are of interest
in the study of several cosmological models \cite{Pi50,Dissertationes}, through the diffeomorphism $(x,p)\in {\rm T}^*\mathbb{R}_0\mapsto (x,v=p)\in {\rm T}\mathbb{R}_0$.

Let us show that the
system (\ref{ClasErm2}) can be endowed with a Lie--Hamiltonian structure.
This system
 describes the integral curves of the $t$-dependent vector
field on ${\rm T}^*\mathbb{R}^{n}_0$ given by $X=X_3+\omega^2(t)X_1$, where the vector fields
\begin{equation}
\begin{gathered}
  X_1=-\sum_{i=1}^nx_i\frac{\partial}{\partial p_i},\qquad
X_2=\frac{1}{2}\sum_{i=1}^n\left(p_i\frac{\partial}{\partial p_i}-x_i\frac{\partial}{\partial x_i}\right),
 \\
 X_3=\sum_{i=1}^n
\left(p_i\frac{\partial}{\partial x_i}+
\frac{b_i}{x_i^3}\frac{\partial}{\partial p_i}\right),
\label{Vessiot--GuldbergSec111}
\end{gathered}\end{equation} 
fulfil the commutation rules (\ref{sl2Rtb}). Hence, (\ref{ClasErm2}) is a Lie system. The space ${\rm T}^*\mathbb{R}^n_0$ admits a natural Poisson bivector $\Lambda=\sum_{i=1}^n\partial/\partial x_i\wedge \partial/\partial p_i$ related to the
restriction to this space of the canonical symplectic structure
on ${\rm T}^*\mathbb{R}^n$.  Moreover, the preceding vector
fields are Hamiltonian
vector fields with Hamiltonian functions
\begin{equation}
\label{SWh}
h_1=\frac 12\sum_{i=1}^nx_i^2,\qquad h_2=-\frac 12\sum_{i=1}^nx_ip_i,\qquad h_3=\frac 12\sum_{i=1}^n\left(p_i^2+\frac{b_i}{x_i^2}\right)
\end{equation}
which satisfy the commutation relations (\ref{sl2Rh1}), so that $\ham\simeq \mathfrak{sl}(2,\mathbb{R})$.
Consequently, every curve $h_t$ that takes values in the Lie algebra spanned by
$h_1,h_2$ and
$h_3$ gives rise to a Lie--Hamiltonian structure $({\rm
T}^*\mathbb{R}_0^n,\Lambda,h)$. Then,
the system (\ref{ClasErm2}), described by the $t$-dependent vector field
$X=X_3+\omega^2(t)X_1=-\widehat\Lambda(dh_3+\omega^2(t)dh_1),
$ is a Lie--Hamilton system with a
Lie--Hamiltonian structure $({\rm T}^*\mathbb{R}_0^n,\Lambda,
h_t=h_3+\omega^2(t)h_1)$.

Subsequently, we derive an explicit superposition rule for   the  simplest case of the system (\ref{ClasErm2}) corresponding to $n=1$,
and  proceed as in the previous subsection. The prolongations of $X_\alpha$ $(\alpha=1,2,3)$ again become linearly independent for $m=2$ and we need to obtain two first integrals for the diagonal prolongations $\widetilde X_\alpha$ of ${\rm T}^*\mathbb{R}^3_0$ fulfilling (\ref{F1F2}) for the coordinate system $\{ x_{(1)}, p_{(1)},
x_{(2)}, p_{(2)},x_{(3)}, p_{(3)} \}$ of ${\rm T}^*\mathbb{R}^3_0$.
Similarly to the previous example, we have an injection $D:\mathfrak{sl}(2,\mathbb{R})\rightarrow C^\infty({\rm T}^*\mathbb{R}_0)$ which leads to the morphisms   $D^{(2)}$ and $D^{(3)}$.

\newpage

 Then, by taking into account the Casimir function (\ref{casSL2}) and the Hamiltonians (\ref{SWh}), we apply Theorem \ref{MT2} obtaining the following first integrals
 
\begin{equation}
\begin{aligned}
 \FF^{(2)} &= D^{(2) } (\Delta(\casS) )
= \frac 14
(x_1p_2-x_2p_1)^2+\frac{b(x_1^2+x_2^2)^2}{4x_1^2x_2^2}, 
\\[1ex]
 \FF^{(3)}  &=D^{(3) } (\Delta(\casS) )
= \smash{ \frac 14 \sum_{1\le i<j}^3 \left[
(x_ip_j-x_jp_i)^2+\frac{b(x_i^2+x_j^2)^2}{x_i^2x_j^2} \right]- \frac 34 b},
\\
    \FF_{13}^{(2)}&=S_{13}(\FF^{(2)}),
\\  \FF_{23}^{(2)}&=S_{23}(\FF^{(2)}),
\\
   \FF^{(3)}&=  \FF^{(2)}+\FF_{13}^{(2)}+\FF_{23}^{(2)}-3b/4 ,
   \label{intSW}
\end{aligned}\end{equation} 
    where  $(x_i,p_i)$ denote the coordinates $(x_{(i)},p_{(i)})$; notice that   $F=D(\casS) =    { b}/4$.
 We choose $\FF^{(2)}$ and $ \FF_{23}^{(2)}$ as the two functionally independent constants of motion and we shall use $ \FF_{13}^{(2)}$ in order to simplify the results. Recall that
 these functions are exactly the first integrals obtained in other works, e.g.,~\cite{SIGMA}, for describing superposition rules of dissipative Milne--Pinney
equations (up to the diffeomorphism
$\varphi:(x,p)\in {\rm T}^*\mathbb{R}_0\mapsto (x,v)=(x,p)\in {\rm
T}\mathbb{R}_0$ to system (\ref{ClasErm2}) with $n=1$), and lead straightforwardly to
deriving a superposition rule for these equations \cite{CL08b}.

 Indeed, we set
 \begin{equation}
\label{ff1}
 \FF^{(2)}=\frac{k_1}4+\frac{b}2 ,\qquad   \FF_{23}^{(2)}=\frac{k_2}4+\frac{b}2 ,\qquad   \FF_{13}^{(2)}=\frac{k_3}4+\frac{b}2 ,\qquad
\end{equation}
 and from the first equation we
    obtain
 $p_1$ in terms of the remaining variables and $k_1$
\begin{equation}
\label{pp1}
p_1=p_1(x_1,x_2,p_2,x_3,p_3,k_1)=\frac{p_2x_1^2x_2\pm\sqrt{k_1 x_1^2 x_2^2-b(x_1^4+x_2^4)}}{x_1 x_2^2}.
\end{equation}
By  introducing this value in the second expression  of (\ref{ff1}),
 one can determine the expression of $x_1$ as a function of $x_2,p_2,x_3,p_3$ and the constants $k_1,k_2$. Such a result is rather simplified when the third constant of  (\ref{ff1}) enters, 
 yielding
 \begin{equation}
\begin{aligned}
x_1&=x_1(x_2,p_2,x_3,p_3,k_1,k_2)=x_1(x_2,x_3,k_1,k_2,k_3)\\
&=\left\{ {\MM_1 x_2^2+\MM_2 x_3^2\pm \sqrt{\MM \left[k_3 x_2^2 x_3^2 -b (x_2^4+x_3^4) \right]}}    \right\}^{1/2} ,
 \end{aligned}
 \label{xx1}
\end{equation}
where the constants $\MM_1,\MM_2,\MM$ are defined in terms of $k_1,k_2,k_3$ and $b$  as
\begin{equation}
  \begin{gathered}
\MM_1=\frac{2b k_1-k_2k_3}{4b^2-k_3^2},\qquad \MM_2=\frac{2b k_2-k_1k_3}{4b^2-k_3^2},
\\
\MM=\frac{4\left[4b^3 +k_1 k_2 k_3 - b(k_1^2+k_2^2+k_3^2) \right]}{(4b^2-k_3^2)^2}.                                                                                   \end{gathered}
\end{equation}
 And by introducing (\ref{xx1}) into (\ref{pp1}), we   obtain
\begin{equation}
 p_1=p_1(x_2,p_2,x_3,p_3,k_1,k_2)=p_1(x_2,p_2,x_3,k_1,k_2,k_3),\end{equation} 
  which together with   (\ref{xx1})  provide a superposition rule for (\ref{ClasErm2}) with $n=1$.
  These expressions constitute the known superposition rule for Milne--Pinney equations \cite{CL08b}.
Observe that,
instead of solving systems of PDEs for obtaining  the first integrals   as
in \cite{SIGMA,CL08b}, we have obtained them algebraically in a simpler way. When $b=0$ we recover, as expected, 
the superposition rule for the harmonic oscillator with a $t$--dependent frequency. Similarly to previous examples, the above
superposition rule is also valid when $\omega(t)$ is constant.

\subsection{A trigonometric system}
 Let us study a final example appearing in the
study of integrability of classical systems \cite{ADR12,AW05}. Consider the
system
\begin{equation}
\left\{\begin{aligned}
\frac{{\rm d} x}{{\rm d} t}&=\sqrt{1-x^2}\left(B_x(t)\sin p-B_y(t)\cos p\right),\\
\frac{{\rm d} p}{{\rm d} t}&=-(B_x(t) \cos p+B_y (t)\sin p)\frac{x}{\sqrt{1-x^2}}-B_z(t),
\end{aligned}\right.
\end{equation}
where $B_x(t),B_y(t), B_z(t)$ are arbitrary $t$-dependent
functions and $(x,p)\in {\rm T}^*{\rm I}$, with ${\rm I}=(-1,1)$. This system describes the integral curves of the $t$-dependent vector
field
\begin{multline}
  X=\sqrt{1-x^2}(B_x(t)\sin\,p-B_y(t)\cos p)\frac{\partial}{\partial x}
\\
-\left[\frac{(B_x(t) \cos
p+B_y (t)\sin p)x}{\sqrt{1-x^2}}+B_z(t)\right]\frac{\partial}{\partial p},
\end{multline}
which can be brought into the form $X=B_x(t)X_1+B_y(t)X_2+B_z(t)X_3$, where
\begin{equation}
\begin{aligned}
  X_1&=\sqrt{1-x^2}\sin\, p\frac{\partial}{\partial x}\!-\!\frac{x}{\sqrt{1-x^2}}\cos
p\frac{\partial}{\partial p},
\\
X_2&=-\sqrt{1-x^2}\cos\,p\frac{\partial}{\partial x}\!-\!\frac{x}{\sqrt{1-x^2}}\sin
p\frac{\partial}{\partial p},
\end{aligned}\end{equation} 
and $X_3=-\partial/\partial p$ satisfy the commutation relations
\begin{equation}
 [X_1,X_2]=X_3, \qquad [X_3,X_1]=X_2, \qquad [X_2,X_3]=X_1.\end{equation} 
In other words, $X$ describes a Lie system associated with a Vessiot--Guldberg
Lie algebra isomorphic to $\mathfrak{su}(2)$. As in the previous examples, we assume $V^X=V$. Now, the vector fields
 $X_\alpha$ $(\alpha=1,2,3)$ are Hamiltonian ones with Hamiltonian functions given by
\begin{equation}
\label{so3}
h_1=-\sqrt{1-x^2}\cos p,\qquad h_2=-\sqrt{1-x^2}\sin p,\qquad h_3=x,
\end{equation}
thus spanning a real Lie algebra isomorphic to $\mathfrak{su}(2)$. Indeed,
\begin{equation}
 \{h_1,h_2\}=-h_3 ,\qquad \{h_3,h_1\}=-h_2 ,\qquad \{h_2,h_3\}=-h_1 .
\end{equation}
Next we consider a basis $\{\ve_1,\ve_2,\ve_3\}$
for $\mathfrak{su}(2)$ satisfying
\begin{equation}
[\ve_1,\ve_2]=-\ve_3,\qquad  [\ve_3,\ve_1]=-\ve_2,\qquad [\ve_2,\ve_3]=-\ve_1,
\end{equation}
so that  $\mathfrak{su}(2)$ admits the Casimir $\casU= \ve_1\widetilde\otimes \ve_1+\ve_2\widetilde\otimes \ve_2+\ve_3\widetilde\otimes \ve_3\in  U_{\mathfrak{su}(2)}$. Then,   the Casimir element of $S_{\mathfrak{su}(2)}$ reads 
$
\casS=\lambda^{-1}(\casU)=\ve_1^2+ \ve^2_2 + \ve^2_3 .
$

The diagonal prolongations of $X_1,X_2,X_3$ are linearly independent at a generic point for $m=2$ and we have to derive
two first integrals for the diagonal prolongations $\widetilde X_1,\widetilde X_2,\widetilde X_3$ on ${\rm T}^*{\rm I}^3$ satisfying (\ref{F1F2}) 
working with the coordinate system $\{ x_{(1)}, p_{(1)}, x_{(2)}, p_{(2)} , x_{(3)}, p_{(3)} \}$ of ${\rm T}^*{\rm I}^3$. Then, by taking into account the Casimir function $C$, 
the Hamiltonians (\ref{so3}), the isomorphism $\phi(\ve_\alpha)=\,h_\alpha$ and the injection $D:\mathfrak{sl}(2,\mathbb{R})\rightarrow C^\infty({\rm T}^*{\rm I})$, we apply Theorem \ref{MT2} obtaining the following first integrals:
 
\begin{equation}
\begin{aligned}
  \FF^{(2)}
&= 2\left[\sqrt{1-x^2_1}\sqrt{1-x^2_2}\cos (p_1-p_2)+x_1x_2+1\right] ,\\
  \FF^{(3)}
&=  2  \sum_{1\le i<j}^3 \left[
\sqrt{1-x^2_i}\sqrt{1-x^2_j}\cos (p_i-p_j)+x_ix_j \right]+3,
\\
    \FF_{13}^{(2)}&=S_{13}(\FF^{(2)}),  
 \\
 \FF_{23}^{(2)}&=S_{23}(\FF^{(2)}),
\\
   \FF^{(3)}&=  \FF^{(2)}+\FF_{13}^{(2)}+\FF_{23}^{(2)}-3  ,
 \end{aligned}\end{equation} 
and  $F=D(\casS) =   1$.
 We again choose $\FF^{(2)}$ and $ \FF_{23}^{(2)}$ as the two functionally independent constants of motion,
  which provide us, after   cumbersome but straightforward computations, with a superposition rule
for these systems.
This leads to a quartic equation, whose solution can be obtained
through known methods. All our results are also valid for the case when $V^X\subset V$.

\section{Classification of Lie--Hamilton systems on the plane}
\setcounter{equation}{0}
\setcounter{theorem}{0}
\setcounter{example}{0}
In this section we classify finite-dimensional real Lie algebras of Hamiltonian vector fields on the plane. In view of the definition
of a Lie--Hamilton system, our classification of finite-dimensional Lie algebras
on the plane implies a classification of Lie--Hamilton systems on the plane. This classification
will be of primordial importance, given the number of physical system which can be interpreted in terms of Lie--Hamilton systems. 
This will be illustrated in the following sections.

\subsection{On general definitions and properties}

A natural problem in the theory of Lie systems is the classification of Lie systems on a fixed manifold, which amounts to classifying finite-dimensional Lie algebras of vector fields on it.
Lie accomplished the local classification of finite-dimensional real Lie algebras of  vector fields on the real line. More precisely, he showed that each such a Lie algebra is locally diffeomorphic to a Lie subalgebra of $\langle \partial_x,x\partial_x,x^2\partial_x\rangle \simeq \mathfrak{sl}(2)$ on a neighborhood of each {\it generic point} $x_0$ of the Lie algebra~\cite{GKP92}. He also performed the local classification of finite-dimensional real Lie algebras of planar vector fields and started the study of the ous problem on $\mathbb{C}^3$~\cite{HA75}.

Lie's local classification on the plane presented some unclear points which were misunderstood by several authors during the following decades. To solve this, {Gonz{\'a}lez-L{\'o}pez, Kamran and Olver} retook the problem and provided a clearer insight in \cite{GKP92}. They proven that every non-zero Lie algebra of vector fields on the plane is locally diffeomorphic around each generic point to one of the finite-dimensional real Lie algebras (the GKO classification) given in Table \ref{table1} in Appendix 1.

As every Vessiot--Guldberg Lie algebra on the plane has to be locally diffeomorphic to one of the Lie algebras of the GKO classification,  every Lie system on the plane is locally diffeomorphic to a Lie system taking values in a Vessiot--Guldberg Lie algebra within the GKO classification. So, the local properties of all Lie systems on the plane can be studied through the Lie systems related to the GKO classification. As a consequence, we say that the GKO classification gives the local classification of Lie systems on the plane.
Their classification divides finite-dimensional Lie algebras of vector fields on $\mathbb{R}^2$ into $28$ nondiffeomorphic classes, which, in fact, can be regarded as a local classification of Lie systems on the plane \cite{BBHLS,GKP92}.

For instance, we see from the matrix Riccati equations (\ref{ConRel3}) that system (\ref{MatRic3}) is related to a non-solvable eight-dimensional Vessiot--Guldberg Lie algebra. In view of the GKO classification in Table \ref{table1} (cf. \cite{BBHLS,GKP92}), all Lie algebras of this type are locally diffeomorphic to P$_8\simeq\mathfrak{sl}(3)$. That is why we say that the system (\ref{MatRic3}) is {\it a Lie system of class {\rm P}$_8$}. 


The GKO classification of finite-dimensional Lie algebras of vector fields $X_i$ $(i=1,\dots,n)$ on the plane covers two subclasses called {\em primitive} (8 cases P$_x$)  and {\em imprimitive}  (20 cases  I$_x$)~\cite{GKP92}.

\begin{definition}
A finite-dimensional real Lie algebra $V$  of vector fields on an open subset $U\subset \mathbb{R}^2$ is {\it imprimitive} when there exists a one-dimensional distribution $\mathcal{D}$ on $\mathbb{R}^2$ invariant under the action of $V$ by Lie brackets, i.e., for every  $X\in V$ and every vector field $Y$ taking values in $\mathcal{D}$, we have that $[X,Y]$ takes values in $\mathcal{D}$. Otherwise, $V$ is called   {\em primitive}.
\end{definition}

\begin{example}\normalfont
Consider the Lie algebra ${\rm I}_4$ which is spanned by the vector fields $X_1,X_2$ and $X_3$ given in Table \ref{table1} in Appendix 1. If we define $\mathcal{D}$ to be the distribution  on $\mathbb{R}^2$ generated
by $Y=\partial_x$, we see that
\begin{equation}
[X_1,Y]=0,\qquad [X_2,Y]=-Y,\qquad [X_3,Y]=-2xY.
\end{equation}
We infer from this that $\mathcal{D}$ is a one-dimensional distribution invariant under the action of I$_4$. Hence, I$_4$ is an imprimitive Lie algebra of vector fields.
\end{example}

 To determine which of the 28 classes can be considered as Vessiot--Guldberg Lie algebras of Hamiltonian vector fields,  a symplectic form
\begin{equation}
\omega=f(x,y){\rm d}x\wedge \dd y
\end{equation}
 must be found so that each $X_i$, belonging to the basis of the considered Lie algebra, becomes Hamiltonian (see \cite{BBHLS} for details). In particular, 
$X_i$ are Hamiltonian vector fields with respect to $\omega$ whenever   the Lie derivative of $\omega$ relative to $X_i$ vanishes, that is, $\mathcal{L}_{X_i}\omega=0$. If $\omega$ exists, then the vector fields $X_i$ become Hamiltonian and their
 corresponding Hamiltonian functions $h_i$ are obtained by using the relation $\iota_{X_i}\omega={\rm d}h_i$. The functions $\langle h_1,\ldots,h_n\rangle$ and their successive brackets  with respect to the Lie bracket 
\begin{equation}
 \{\cdot,\cdot\}_\omega\ :\ C^\infty\left(\mathbb{R}^2\right)\times C^\infty\left(\mathbb{R}^2\right)\rightarrow C^\infty\left(\mathbb{R}^2\right)\end{equation} 
   induced by $\omega$ span a finite-dimensional Lie algebra of functions $(\mathfrak{W},  \{\cdot,\cdot\}_\omega)$. We will call
 this Lie algebra a {\it Lie--Hamilton algebra} (LH algebra, in short).

In the case of the Vessiot--Guldberg Lie algebra $V^{MR}$ for matrix Riccati equations, we see that there exists no such a structure. Indeed, if $X_0=\partial_x$ and $X_2=x\partial_x$ which belong to $V^{MR}$
were Hamiltonian relative to a symplectic structure, let us say $\omega=f(x,y)\dd x\wedge \dd y$, we would have $\mathcal{L}_{X_0}\omega=\mathcal{L}_{X_2}\omega=0$. This implies that $\partial_x f=x\partial_xf+f=0$. Hence, $f=0$ and $\omega$ cannot be symplectic.  In other words, $V^{MR}$ is not a Lie algebra of Hamiltonian vector fields.

This example illustrates that not every Lie system admits a Vessiot--Guldberg Lie algebra of Hamiltonian vector fields with respect to a Poisson structure \cite{IV}. When a Lie system does, we call it a {\it Lie--Hamilton system} (LH system) \cite{CLS122}. 

The problem of which of the Lie algebras within the GKO classification consist of Hamiltonian vector fields with respect to some Poisson structure has recently been solved by the present writer and collaborators in~\cite{BBHLS}, showing that, among the $28$ classes of the GKO classification,  only  $12$ remain as Lie algebras of Hamiltonian vector fields. Furthermore, examples of LH systems having applications in the above mentioned fields have also been worked out in~\cite{BBHLS}. 
 We recall that one advantage of  LH systems is that their superposition rules might be obtained straightforwardly by  applying a coalgebra approach as it has been formulated in the past section ~\cite{BCHLS}.

 It has recently been proven that the initial $8+20$ cases of the GKO classification  are reduced to $4+8$ classes of finite-dimensional Lie algebras of Hamiltonian vector fields~\cite{BBHLS}.
  The final result  is summarized in Table~\ref{table1} in Appendix 1,  where we  indicate the Lie algebra $ \mathfrak{g}$ of Hamiltonian vector fields $X_i$, a family of their corresponding Hamiltonian functions $h_i$ and an associated symplectic form $\omega$.

\subsubsection*{Minimal Lie algebras of Lie--Hamilton systems on the plane}

In this section we study the local structure of the minimal Lie algebras of Lie--Hamilton systems on the plane around their generic points.

Our main result, Theorem \ref{Char}, and the remaining findings of this section enable us  to give the local classification of Lie--Hamilton systems on the plane in the following subsections.  To simplify the notation, $U$ will hereafter stand for a contractible open subset of $\mathbb{R}^2$.

\begin{definition}\label{GenPon} Given a finite-dimensional Lie algebra $V$ of vector fields on a manifold $N$, we say that $\xi_0\in N$ is a {\it generic point} of $V$ when the rank of the generalized distribution
\begin{equation}
\mathcal{D}^V_\xi=\{X(\xi)\mid X\in V\}\subset T_\xi N,\qquad \xi\in N,
\end{equation}
i.e., the function $r^V(\xi)=\dim \mathcal{D}^V_\xi$, is locally constant around $\xi_0$. We call {\it generic domain} or simply {\it domain} of $V$ the set of generic points of $V$.
\end{definition}

\begin{example}\normalfont
 Consider the Lie algebra ${\rm I}_4=\langle X_1,X_2,X_3\rangle$ of vector fields on $\mathbb{R}^2$ detailed in Table \ref{table1}. By using the  expressions of $X_1$, $X_2$ and $X_3$ in coordinates, we see that $r^{{\rm I}_4}(x,y)$ equals the rank of the matrix
\begin{equation}
\left(\begin{array}{ccc}
1&x&x^2\\
1&y&y^2\\
\end{array}
\right),\end{equation} 
which is two for every $(x,y)\in\mathbb{R}^2$ except for points with $y-x=0$, where the rank is one. So, the domain of I$_4$ is $
\mathbb{R}^2_{x\neq y}=\{(x,y)\,|\, x \neq y\}\subset \mathbb{R}^2
$.
\end{example}

\begin{lemma}\label{lem:local_sym} Let $V$ be a finite-dimensional real Lie algebra of Hamiltonian vector fields on $\mathbb{R}^2$ with respect to a Poisson structure and let $\xi_0\in\mathbb{R}^2$ be a generic point of $V$. There exists a $U\ni\xi_0$ such that $V|_{U}$ consists of Hamiltonian vector fields relative to a symplectic structure.
\end{lemma}
\begin{proof}

If $\dim \mathcal{D}^V_{\xi_0}=0$, then $\dim \mathcal{D}_\xi^V=0$ for every $\xi$ in a $U\ni \xi_0$ because the rank of $\mathcal{D}^V$ is locally constant around generic points. Consequently, $V|_U=0$  and its unique element become Hamiltonian relative to the restriction of
$\omega=\dd x\wedge \dd y$ to $U$.
Let us assume now $\dim \mathcal{D}^V_{\xi_0}\neq 0$. By assumption, the elements of $V$ are Hamiltonian vector fields with respect to a Poisson bivector $\Lambda\in \Gamma(\Lambda^2{\rm T}\mathbb{R}^2)$.
Hence, $\mathcal{D}^V_\xi\subset \mathcal{D}^\Lambda_\xi$ for every $\xi\in\mathbb{R}^2$, with $\mathcal{D}^\Lambda$ being the {\it characteristic distribution} of $\Lambda$ \cite{IV}. Since $\dim \mathcal{D}_{\xi_0}^V\neq 0$ and $r^V$ is locally constant at $\xi_0$, then $\dim\,\mathcal{D}^V_\xi>0$ for every $\xi$ in a $U\ni \xi_0$. Since the rank of $\mathcal{D}^\Lambda$ is even at every point of $\mathbb{R}^2$ and $\mathcal{D}^V_\xi\subset \mathcal{D}^\Lambda_\xi$ for every $\xi\in U$, the rank of $\mathcal{D}^\Lambda$ is two on $U$. So, $\Lambda$ comes from a symplectic structure on $U$ and $V|_U$ is a Lie algebra of Hamiltonian vector fields relative to it.
\end{proof}

Roughly speaking, the previous lemma establishes that any Lie--Hamilton system $X$ on $\mathbb{R}^2$ can be considered around each generic point of ${V^X}$ as a Lie--Hamilton system admitting a minimal Lie algebra of Hamiltonian vector fields with respect to a symplectic structure. As our study of such systems is local, we hereafter focus on analysing  minimal Lie algebras of this type.

A {\it volume form} $\Omega$ on an $n$-dimensional manifold $N$ is a non-vanishing $n$-form on $N$. The divergence of a vector field $X$ on $N$ with respect to $\Omega$ is the unique function ${\rm div} X:N\rightarrow\mathbb{R}$ satisfying $\mathcal{L}_X\Omega=({\rm div}\, X)\, \Omega$. An {\it integrating factor} for $X$ on $U\subset N$ is a function $f:U\rightarrow \mathbb{R}$ such that $\mathcal{L}_{fX}\Omega=0$ on $U$. Next we have the following result~\cite{LP12}.

\begin{lemma}\label{Lem:IntFac}  Consider the volume form $\Omega=\dd x\wedge \dd y$ on a $U\subset \mathbb{R}^2$ and a vector field $X$ on $U$. Then, $X$ is
Hamiltonian with respect to a symplectic form $\omega=f \Omega$ on $U$  if and only if $f:U\rightarrow \mathbb{R}$ is a non-vanishing integrating factor of $X$ with respect to $\Omega$, i.e., $Xf=-f{\rm div} X$ on $U$.
\end{lemma}
\begin{proof} Since $\omega$ is a symplectic form on $U$, then $f$ must be non-vanishing. As
\begin{equation}
\mathcal{L}_X\omega=\mathcal{L}_X(f\Omega)=(Xf)\Omega+f\mathcal{L}_X\Omega=(Xf+f{\rm div} X)\Omega=\mathcal{L}_{fX}\Omega,
\end{equation}
then $X$ is locally Hamiltonian with respect to $\omega$, i.e., $\mathcal{L}_X\omega=0$, if and only if $f$ is a non-vanishing integrating factor for $X$ on $U$. As $U$ is a contractible open {\ subset}, the Poincar\'e Lemma ensures that $X$ is a local Hamiltonian vector field if and only if it is a Hamiltonian vector field. Consequently, the lemma follows.
\end{proof}

\begin{definition} Given a vector space $V$ of vector fields on $U$, we say that $V$ admits a {\it modular generating system} $(U_1,X_1,\ldots,X_p)$ if $U_1$ is a dense open subset of $U$ such that every $X\in V|_{U_1}$ can be brought into the form
$
X|_{U_1}=\sum_{i=1}^p\ff_i X_i|_{U_1}
$
for certain functions $\ff_1,\ldots,\ff_p\in C^\infty(U_1)$ and vector fields $X_1,\ldots,X_p\in V$.
\end{definition}

\begin{example}\normalfont
Given the Lie algebra P$_3\simeq \mathfrak {so}(3)$ on $\mathbb{R}^2$ of Table~\ref{table1} of Appendix 1, the vector fields
\begin{equation}
X_1=y\frac{\partial}{\partial x}-x\frac{\partial}{\partial y}  ,\qquad X_2=(1+x^2-y^2)\frac{\partial}{\partial x}+2xy\frac{\partial }{\partial y}
\end{equation}
of P$_3$ satisfy that $X_3=\ff_1X_1+\ff_2X_2$ on $U_1=\{(x,y)\in \mathbb{R}^2\mid x\neq 0\}$ for the functions $\ff_1,\ff_2\in C^\infty(U_1)$
\begin{equation}
\ff_1=\frac{x^2+y^2-1}{x},\qquad \ff_2=\frac y {x}.
\end{equation}
 Obviously, $U_1$ is an open dense subset of $\mathbb{R}^2$. As every element of $V$ is a linear combination of $X_1,X_2$ and $X_3=\ff_1X_1+\ff_2X_2$, then  every $X\in V|_{U_1}$ can be written as a linear combination with smooth functions on $U_1$ of $X_1$ and $X_2$. So, $(U_1,X_1,X_2)$ form a generating modular system for P$_3$.
\end{example}

In Table \ref{table1} in Appendix 1 we detail a modular generating system, which is indicated by the first one or two vector fields written between brackets in the list of the $X_i$'s, for every finite-dimensional Lie algebra of vector fields of the GKO classification.

\begin{theorem}\label{Char} Let $V$ be a Lie algebra of vector fields on $U\subset \mathbb{R}^2$ admitting a modular generating system $(U_1,X_1,\ldots,X_p)$.  We have that
\medskip

\noindent
1) The space $V$ consists of Hamiltonian vector fields relative to a symplectic form 
on $U$ if and only if
\begin{itemize}
 \item{i) Let $\ff_1,\ldots,\ff_p$ be certain smooth functions on $U_1\subset U$. Then,
 \begin{equation}\label{DivCon}
 X|_{U_1}=\sum_{i=1}^p \ff_i X_i|_{U_1}\in V|_{U_1}\Longrightarrow {\rm div} X|_{U_1}=\sum_{i=1}^p \ff_i {\rm div}X_i|_{U_1}.
 \end{equation}}
\item{ii) The elements $X_1,\ldots,X_p$ admit a common non-vanishing integrating factor on $U$.}
\end{itemize}

\noindent
2) If the rank of $\mathcal{D}^V$ is two on $U$, the symplectic form is unique up to a multiplicative non-zero constant.
\end{theorem}

\begin{proof} Let us prove the direct part of 1). Since $(U_1,X_1,\ldots,X_p)$ form a modular generating system for $V$, we have that every $X|_{U_1}\in V|_{U_1}$ can be brought into  the form $X|_{U_1}=\sum_{i=1}^p\ff_i X_i|_{U_1}$ for certain $\ff_1,\ldots,\ff_p\in C^\infty(U_1)$. As $V$ is a Lie algebra of Hamiltonian vector fields with respect to a symplectic structure on $U$, let us say
 \begin{equation}\label{ww}
\omega=f(x,y)\dd x\wedge \dd y,
 \end{equation}
then Lemma \ref{Lem:IntFac} ensures that  $Yf=-f{\rm div} Y$ for every $Y\in V$.  Then,
\begin{equation}
\begin{gathered}
f{\rm div} X=-X f=-\sum_{i=1}^p \ff_i X_i f=f \sum_{i=1}^p \ff_i {\rm div}X_i 
\\
\Longleftrightarrow
f\left({\rm div} X-\sum_{i=1}^p \ff_i {\rm div}X_i\right)=0
\end{gathered}\end{equation} 
on $U_1$.  
As $\omega$ is non-degenerate, then $f$ is non-vanishing and $i)$ follows. Since all the vector fields of $V$ are Hamiltonian with respect to $\omega$, they share a common non-vanishing integrating factor, namely $f$. From this, $ii)$ holds.

Conversely, if $ii)$ is satisfied, then Lemma \ref{Lem:IntFac} ensures that $X_1,\ldots,X_p$ are Hamiltonian with respect to  (\ref{ww}) on $U$, with $f$ being a non-vanishing integrating factor. As $(U_1,X_1,\ldots,X_p)$ form a generating modular system for $V$, every $X\in V$ can be written as $\sum_{i=1}^p\ff_i X_i$ on $U_1$ for certain functions $\ff_1,\ldots,\ff_p\in C^\infty(U_1)$. From $i)$ we obtain
$
{\rm div}\, X=\sum_{i=1}^p\ff_i{\rm div}X_i
$
on $U_1$. Then,
\begin{equation}
Xf=\sum_{i=1}^p\ff_i X_i f=-f\sum_{i=1}^p\ff_i{\rm div}X_i=-f{\rm div X}
\end{equation}
on $U_1$ and, since the elements of $V$ are smooth and $U_1$ is dense on $U$, the above expresion also holds  on $U$. Hence, $f$ is a non-vanishing integrating factor for $X$, which becomes a Hamiltonian vector field with respect to $\omega$ on $U$ in virtue of Lemma \ref{Lem:IntFac}. Hence, part 1) is proven.

As far as part 2) of the theorem is concerned, if the vector fields of $V$ are Hamiltonian with respect to two different symplectic structures on $U$, they admit two different non-vanishing integrating factors $f_1$ and $f_2$. Hence,
\begin{equation}
X(f_1/f_2)=(f_2Xf_1-f_1Xf_2)/f_2^2=(f_2f_1{\rm div} X-f_1f_2{\rm div}X)/f_2^2=0
\end{equation}
and $f_1/f_2$ is a common constant of motion for all the elements of $V$. Hence, it is a constant of motion for every vector field taking values in the distribution $\mathcal{D}^V$. Then rank of $\mathcal{D}^V$ on $U$ is two by assumption. So, $\mathcal{D}^V$ is generated by the vector fields $\partial_x$ and $\partial_y$ on $U$. Thus, the only constants of motion on $U$
common to all the vector fields taking values in $\mathcal{D}^V$, and consequently common to the elements of $V$, are constants. Since $f_1$ and $f_2$ are non-vanishing, then $f_1=\lambda f_2$ for a $\lambda\in\mathbb{R}\backslash \{0\}$ and the associated symplectic structures are the same up to an irrelevant non-zero  proportionality constant.
\end{proof}

Using Theorem \ref{Char}, we can immediately prove the following result.

\begin{corollary}\label{Cor:NonDiv}
If a Lie algebra of vector fields $V$ on a $U\subset\mathbb{R}^2$ consists of Hamiltonian vector fields with respect to a symplectic form and admits a modular generating system whose elements are divergence free, then every element of $V$ is divergence free.
\end{corollary}

\subsection{Lie--Hamilton algebras}

It is known that given a Lie--Hamilton system $X$, its Lie--Hamilton algebras are not uniquely defined in general. Moreover, the existence of different types of Lie--Hamilton algebras for the same Lie--Hamilton system is important in its linearization and the use of certain methods~\cite{CLS122}. For instance, if a Lie--Hamilton system $X$ on $N$ admits a Lie--Hamilton algebra isomorphic to $V^X$ and $\dim V^X=\dim N$, then $X$ can be linearized together with its associated Poisson structure \cite{CLS122}.

\begin{example}\normalfont
Consider again the Lie--Hamilton system $X$ given by the complex Riccati equations with $t$-dependent real coefficients (\ref{Riccati21}) and assume $V^X\simeq \mathfrak{sl}(2)$. Recall that $X$ admits a Lie--Hamilton algebra $(\mathcal{H}_\Lambda,\{\cdot,\cdot\}_\omega)\simeq \mathfrak{sl}(2)$ spanned by the Hamiltonian functions $h_1,h_2,h_3$ given by (\ref{ab}) relative to the symplectic structure $\omega$ detailed in (\ref{aa1}). We can also construct a second (non-isomorphic) Lie--Hamilton algebra for $X$ with respect to $\omega$. The vector fields $X_i$, with $i=1,2,3$, spanning $V^X$ (see (\ref{vectRiccati21})) have also Hamiltonian functions $\bar h_i=h_i+1$, for $i=1,2,3$, respectively. Hence, $(\mathbb{R}^2_{y\neq 0},\omega,h=a_0(t)\bar{h}_1+a_1(t)\bar{h}_2+a_2(t)\bar{h}_3)$ is a Lie--Hamiltonian structure for $X$
giving rise to a Lie--Hamilton algebra  $(\overline{\mathcal{H}}_\Lambda,\{\cdot,\cdot\}_\omega)\equiv (\langle \bar{h}_1,\bar{h}_2,\bar{h}_3,1\rangle,\{\cdot,\cdot\}_\omega) \simeq \mathfrak{sl}(2)\oplus \mathbb{R}$ for $X$.
\end{example}

\begin{proposition}\label{LieHam1} A Lie--Hamilton system $X$ on a symplectic connected manifold $(N,\omega)$ possesses an associated Lie--Hamilton algebra $(\mathcal{H}_\Lambda,\{\cdot,\cdot\}_\omega)$ isomorphic to $V^X$ if and only if every Lie--Hamilton algebra non-isomorphic to $V^X$ is isomorphic to $V^X\oplus \mathbb{R}$.
\end{proposition}
\begin{proof} Let $(\overline{\mathcal{H}}_\Lambda,\{\cdot,\cdot\}_\omega)$ be an arbitrary Lie--Hamilton algebra for $X$.
As $X$ is defined on a connected manifold, the sequence of Lie algebras
\begin{equation}\label{ExacSeq}
0\hookrightarrow (\overline{\mathcal{H}}_\Lambda,\{\cdot,\cdot\}_\omega)\cap \langle 1\rangle \hookrightarrow (\overline{\mathcal{H}}_\Lambda,\{\cdot,\cdot\}_\omega)\stackrel{\varphi}{\longrightarrow} V^X\rightarrow 0,
\end{equation}
where $\varphi:\overline{\mathcal{H}}_\Lambda\rightarrow V^X$ maps every function of $\overline{\mathcal{H}}_\Lambda$ to minus its Hamiltonian vector field, is always exact (cf.~\cite{CLS122}). Hence, $(\overline{\mathcal{H}}_\Lambda,\{\cdot,\cdot\}_\omega)$ can be isomorphic either to $V^X$ or to   a Lie algebra extension of $V^X$ of  dimension $\dim V^X+1$. 

If $(\mathcal{H}_\Lambda,\{\cdot,\cdot\}_\omega)$ is isomorphic to $V^X$ and there exists a second Lie--Hamilton algebra $(\overline{\mathcal{H}}_\Lambda,\{\cdot,\cdot\}_\omega)$ for $X$ non-isomorphic to $V^X$, we see from (\ref{ExacSeq}) that $1\in \overline{\mathcal{H}}_\Lambda$ and $1\notin \mathcal{H}_\Lambda$. Given a basis $X_1,\ldots,X_r$ of $V^X$, each element $X_i$, with $i=1,\ldots,r$, has a Hamiltonian function $\overline{h}_i\in\overline{\mathcal{H}}_\Lambda$ and another $h_i\in{\mathcal{H}}_\Lambda$. As $V^X$ is defined on a connected manifold, then $h_i=\overline{h}_i-\lambda_i\in \overline{\mathcal{H}}_\Lambda$ with $\lambda_i\in\mathbb{R}$ for every $i=1,\ldots,r$. From this and using again that $1\in \overline{\mathcal{H}}_\Lambda\backslash \mathcal{H}_\Lambda$, we obtain that $\{h_1,\ldots,h_r,1\}$ is a basis for $\overline{\mathcal{H}}_\Lambda$ and $(\overline{\mathcal{H}}_\Lambda,\{\cdot,\cdot\}_\omega)\simeq (\mathcal{H}_\Lambda \oplus\mathbb{R},\{ \cdot,\cdot\}_\omega)$.

Let us assume now that every Lie--Hamilton algebra $(\overline{\mathcal{H}}_\Lambda,\{\cdot,\cdot\}_\omega)$ non-isomorphic to $V^X$ is isomorphic to $V^X\oplus\mathbb{R}$. We can define a Lie algebra anti-isomorphism $\mu :V^X\rightarrow \overline{\mathcal{H}}_\Lambda$ mapping each element of $V^X$ to a Hamiltonian function belonging to a Lie subalgebra of $(\overline{\mathcal{H}}_\Lambda,\{\cdot,\cdot\}_\omega)$ isomorphic to $V^X$. Hence, $(N,\omega,h=\mu(X))$, where $h_t=\mu(X_t)$ for each $t\in\mathbb{R}$, is a Lie--Hamiltonian structure for $X$ and $(\mu(V^X),\{\cdot,\cdot\}_\omega)$ is a Lie--Hamilton algebra for $X$ isomorphic to $V^X$.
\end{proof}

\begin{proposition}\label{LieHam2} If a Lie--Hamilton system $X$ on a symplectic connected manifold $(N,\omega)$ admits an associated Lie--Hamilton algebra $(\mathcal{H}_\Lambda,\{\cdot,\cdot\}_\omega)$ isomorphic to $V^X$, then it admits a Lie--Hamilton algebra isomorphic to $V^X\oplus \mathbb{R}$.
\end{proposition}

\begin{proof}
Let $(N,\omega,h)$ be a Lie--Hamiltonian structure for $X$ giving rise to the Lie--Hamilton algebra $(\mathcal{H}_\Lambda,\{\cdot,\cdot\}_\omega)$. Consider the linear space $L_h$ spanned by linear combinations of the functions $\{h_t\}_{t\in\mathbb{R}}$. Since we assume $\mathcal{H}_\Lambda\simeq V^X$, the exact sequence (\ref{ExacSeq}) involves that $1\notin L_h$. Moreover, we can write $h=\sum_{i=1}^pb_i(t)h_{t_i}$, where $h_{t_i}$ are the values of $h$ at certain times $t_1,\ldots,t_p$ such that $\{h_{t_1},\ldots,h_{t_p}\}$ are linearly independent and $b_1,\ldots,b_p$ are certain $t$-dependent functions. Observe that the vector fields $(b_1(t),\ldots,b_p(t))$, with $t\in\mathbb{R}$, span a $p$-dimensional linear space. If we choose a $t$-dependent Hamiltonian $\bar{h}=\sum_{i=1}^{p}b_i(t)h_{t_i}+b_{p+1}(t)$, where $b_{p+1}(t)$ is not a linear combination of $b_1(t),\ldots,b_p(t)$, and we recall that $1,h_{t_1},\ldots,h_{t_p}$ are linearly independent over $\mathbb{R}$,
we obtain that the linear hull of the functions $\{\bar h_t\}_{t\in\mathbb{R}}$ has dimension $\dim L_h+1$. Moreover, $(N,\{\cdot,\cdot\}_\omega,\bar h)$ is a Lie--Hamiltonian structure for $X$. Hence, they span, along with their successive Lie brackets, a Lie--Hamilton algebra isomorphic to $\mathcal{H}_\Lambda\oplus \mathbb{R}$.
\end{proof}

\begin{corollary}\label{NoGo2} If $X$ is a Lie--Hamilton system with respect to a symplectic connected manifold $(N,\omega)$ admitting a Lie--Hamilton algebra $(\mathcal{H}_\Lambda,\{\cdot,\cdot\}_\omega)$ satisfying that $1\in \{\mathcal{H}_\Lambda,\mathcal{H}_\Lambda\}_\omega$, then $X$ does not possess any Lie--Hamilton algebra isomorphic to $V^X$.
\end{corollary}
\begin{proof} If $1\in \{\mathcal{H}_\Lambda,\mathcal{H}_\Lambda\}_\omega$, then $\mathcal{H}_\Lambda$ cannot be isomorphic to $V^X\oplus\mathbb{R}$ because the derived Lie algebra of $\mathcal{H}_\Lambda$, i.e. $\{\mathcal{H}_\Lambda,\mathcal{H}_\Lambda\}_\omega$, contains the constant function $1$ and the derived Lie algebra of a $\mathcal{H}_\Lambda$ isomorphic to $V^X\oplus \mathbb{R}$ does not. In view of Proposition \ref{LieHam1}, system $X$ does not admit any Lie--Hamilton algebra isomorphic to $V^X$.
\end{proof}

\begin{proposition}\label{UniExt} If $X$ is a Lie--Hamilton system on a connected manifold $N$ whose $V^X$ consists of Hamiltonian vector fields with respect to a symplectic structure $\omega$ that does not possess any Lie--Hamilton algebra $(\mathcal{H}_\Lambda,\{\cdot,\cdot\}_\omega)$ isomorphic to $V^X$, then all its Lie--Hamilton algebras (with respect to  $\{\cdot,\cdot\}_\omega$) are {\b isomorphic}.
\end{proposition}
\begin{proof} Let $(\mathcal{H}_\Lambda,\{\cdot,\cdot\}_\omega)$ and $(\overline{\mathcal{H}}_\Lambda,\{\cdot,\cdot\}_\omega)$ be two Lie--Hamilton algebras for $X$. Since they are not isomorphic to $V^X$ and in view of the exact sequence (\ref{ExacSeq}), then $1\in \mathcal{H}_\Lambda\cap \overline{\mathcal{H}}_\Lambda$ . Let $X_1,\ldots,X_r$ be a basis of $V^X$. Every vector field $X_i$ admits a Hamiltonian function $h_i\in\mathcal{H}_\Lambda$ and another $\bar{h}_i\in\overline{\mathcal{H}}_\Lambda$. The functions $h_1,\ldots,h_r$ are linearly independent and the same applies to $\bar h_1,\ldots,\bar h_r$. Then,  $\{h_1,\ldots,h_r,1\}$ is a basis for $\mathcal{H}_\Lambda$ and  $\{\bar{h}_1,\ldots,\bar{h}_r,1\}$   is a basis for $\overline{\mathcal{H}}_\Lambda$. As $N$ is connected, then $h_i=\bar{h}_i -\lambda_i$ with $\lambda_i\in\mathbb{R}$ for each $i\in\mathbb{R}$. Hence, the functions $h_i$ belong to $\overline{\mathcal{H}}_\Lambda$ and the functions $\bar{h}_i$ belong to ${\mathcal{H}}_\Lambda$. Thus $\mathcal{H}_\Lambda=\overline{\mathcal{H}}_\Lambda$.
\end{proof}

For instance, the case ${\rm P}_1$ from Table \ref{table1} in Appendix 1 corresponds to the two-dimensional Euclidean algebra
$ {\mathfrak{iso}}(2)\simeq\langle X_1,X_2,X_3\rangle$, but the Hamiltonian functions $\langle h_1,h_2,h_3,h_0=1\rangle$
span the centrally extended  Euclidean algebra $\overline {\mathfrak{iso}}(2)$.  A similar fact arises in classes ${\rm P}_3\simeq \mathfrak{so}(3)$, ${\rm P}_5\simeq \mathfrak{sl}(2 ) \ltimes\mathbb{R}^2$,
${\rm I}_8\simeq  { {\mathfrak{iso}}}(1,1)$ (the $(1+1)$-dimensional Poincar\'e algebra), ${\rm I}_{14B}\simeq \mathbb{R} \ltimes \mathbb{R}^{r}$ and ${\rm I}_{16}\simeq {\mathfrak{h}_2 \! \ltimes \! \mathbb{R}^{r+1}}$. Among them, only the family ${\rm P}_3\simeq \mathfrak{so}(3)$ is a simple Lie algebra in such a manner that $h_0=1$ gives rise to a trivial central extension which means that the LH  algebra is isomorphic to
  $\mathfrak{so}(3)\oplus\mathbb{R}$; otherwise the central extension is a non-trivial one and it cannot be `removed'.

  In this respect, notice    that the appearance of a non-trivial central extension is the  difference between  the family  ${\rm I}_{14B}$ with respect to  ${\rm I}_{14A}$.   We also recall that the LH  algebra corresponding  to the class P$_5$, that is  $\overline{\mathfrak{sl}(2 )\ltimes \mathbb{R}^2}$,  is isomorphic to the two-photon Lie algebra   $\mathfrak{h}_6$ (see~\cite{BBF09,Gilmore} and references therein)
  and, therefore,  also to   the  $(1+1)$-dimensional centrally extended Schr\"odinger  Lie algebra~\cite{Schrod}.

We stress that   the Lie algebra  $\mathfrak{sl}(2 )$ appears four times (classes ${\rm P}_2$,  ${\rm I}_3$, ${\rm I}_4$ and ${\rm I}_5$ in the GKO classification) which means that there may be
{different} LH systems sharing isomorphic Vessiot--Guldberg Lie algebras that are nondiffeomorphic, namely there exists no diffeomorphism mapping the elements of one into the other. In other words, only LH  systems  belonging to each class can be related through a change of variables.   We shall explicitly apply this property throughout the paper.

\subsection{Local classification}\label{Classification}

In this section we describe the local structure of Lie--Hamilton systems on the plane, i.e., given the minimal Lie algebra of a Lie--Hamilton system $X$ on the plane, we prove that $V^X$ is locally diffeomorphic around a generic point of $V^X$ to one of the Lie algebras given in Table \ref{table3} in Appendix 1. We also prove that, around a generic point of $V^X$, the Lie--Hamilton algebras of $X$ must have one of the algebraic structures described in Table \ref{table3}, Appendix 1.

If $X$ is a Lie--Hamilton system, its minimal Lie algebra must be locally diffeomorphic to one of the Lie algebras of the GKO classification that consists of Hamiltonian vector fields with respect to a Poisson structure. As we are concerned with generic points of minimal Lie algebras, Lemma \ref{lem:local_sym} ensures that $V^X$ is locally diffeomorphic around generic points to a Lie algebra of Hamiltonian vector fields with respect to a symplectic structure. So, its minimal Lie algebra is locally diffeomorphic to one of the Lie algebras of the GKO classification consisting of Hamiltonian vector fields with respect to a symplectic structure on a certain open contractible subset of its domain. By determining which of the Lie algebras of the GKO classification admit such a property, we can classify the local structure of all Lie--Hamilton systems on the plane. 

\begin{proposition}\label{NoGo} The primitive Lie algebras ${\rm  P}^{\alpha\ne 0}_1$, {\rm P}$_4$, {\rm P}$_6$--{\rm P}$_8$ and the imprimitive ones {\rm I}$_2$, {\rm I}$_3$, {\rm I}$_6$, {\rm I}$_7$, {\rm I}$^{(\alpha\neq -1)}_8$, {\rm I}$_9$--{\rm I}$_{11}$, {\rm I}$_{13}$, {\rm I}$_{15}$, ${\rm I}^{(\alpha\neq -1)}_{16}$, {\rm I}$_{17}$--{\rm I}$_{20}$ are not Lie algebras of Hamiltonian vector fields on any $U\subset \mathbb{R}^2$.
\end{proposition}

\begin{proof}
Apart from I$_{15}$, the remaining Lie algebras detailed in this statement  admit a modular generating system whose elements are divergence free on the whole $\mathbb{R}^2$ (see the elements written between brackets in Table \ref{table1}). At the same time, we also observe in Table \ref{table1} that these Lie algebras admit a vector field with non-zero divergence on any $U$. In view of Corollary \ref{Cor:NonDiv}, they cannot be Lie algebras of Hamiltonian vector fields with respect to any symplectic structure on any $U\subset\mathbb{R}^2$.

In the case of the Lie algebra I$_{15}$, we have that $(\mathbb{R}^2_{y\neq 0},X_1=\partial_x,X_2=y\partial_y)$ form a generating modular system of I$_{15}$. Observe that $X_2=y\partial_y$ and $X_3=\eta_1(x)\partial_y$, where $\eta_1$ is a non-null function ---it forms with $\eta_2(x),\ldots,\eta_r(x)$ a basis of solutions of a system of $r$ first-order linear homogeneous differential equations in normal form with constant coefficients (cf. \cite{GKP92,Lie1880})--- satisfy ${\rm div} X_2=1$ and ${\rm div}X_3=0$. Obviously, ${\rm div} X_3\neq \eta_1{\rm div}X_2/y$ on any $U$. So,
I$_{15}$ does not satisfy Theorem \ref{Char} on any $U$ and it is not a Lie algebra of Hamiltonian vector fields on any  $U\subset \mathbb{R}^2$.
\end{proof}

To simplify the notation, we assume in this Section that all objects are defined on a contractible $U\subset \mathbb{R}^2$ of the domain of the Lie algebra under study. Additionally, $U_1$ stands for a dense open subset of $U$. In the following two subsections, we explicitly show that {\em all} of  the Lie algebras of the GKO classification {\em not} listed in Proposition \ref{NoGo} consist of Hamiltonian vector fields on any $U$ of their domains. For each Lie algebra, we compute the integrating factor $f$ of $\omega$ given by (\ref{ww}) turning the elements of a basis of the Lie algebra into Hamiltonian vector fields and we work out their Hamiltonian functions.
Additionally, we obtain the algebraic structure of all the Lie--Hamilton algebras of the Lie--Hamilton systems admitting such minimal Lie algebras. 

We stress that the main results covering the resulting Hamiltonian functions $h_i$, the symplectic form $\omega$ and the Lie--Hamilton algebra are summarized in Table~\ref{table3} accordingly to the GKO classification of Table~\ref{table1}.
We  point out that  the Lie algebras of the class I$_{14}$   give rise to two non-isomorphic Lie--Hamilton algebras: I$_{14A}$ whenever $1\notin \langle \eta_1,\ldots,\eta_r\rangle$ and I$_{14B}$ otherwise. Consequently, we  obtain twelve finite-dimensional real  Lie algebras of Hamiltonian vector fields on the plane.

In order to shorten the presentation of the following results, we remark that for some of such Lie--Hamilton algebras their symplectic structure is just the standard one.

\begin{proposition}\label{GoSym}
The Lie algebras ${\rm P}^{(\alpha=0)}_1$, {\rm P}$_5$, {\rm I}$_8^{(\alpha=-1)}$,  {\rm I}$_{14B}$ and {\rm I}$_{16}^{(\alpha=-1)}$   are Lie algebras of Hamiltonian vector fields  with respect to the standard symplectic form $\omega=\dd x\wedge \dd y$, that is, $f\equiv  1$.
\end{proposition}
\begin{proof}
We see in Table \ref{table1} that all the aforementioned Lie algebras admit a modular generating system $(U,X_1=\partial_x,X_2=\partial_y)$ and all their elements have zero divergence. So, they satisfy condition (\ref{DivCon}). The vector fields $X_1,X_2$ are Hamiltonian with respect to the symplectic structure $\omega=\dd x\wedge \dd y$. In view of Theorem \ref{Char}, the whole Lie algebra consists of Hamiltonian vector fields relative to $\omega$.
\end{proof}

\subsubsection{Primitive Lie algebras}

\subsubsection*{Lie algebra P$^{(\alpha=0)}_{1}$: $A_0\simeq  {\mathfrak{iso}}(2) $}

Proposition \ref{GoSym} states that $A_0$ is a Lie algebra of Hamiltonian vector fields with respect to the symplectic form $\omega=\dd x\wedge \dd y$. The basis of vector fields $X_1, X_2, X_3$ of  $A_0$ given in Table \ref{table1} in Appendix 1 satisfy the commutation relations
\begin{equation}
 [X_1,X_2]=0,\qquad [X_1,X_3]= -X_2,\qquad [X_2,X_3]=X_1.
\end{equation}
So, $A_0$ is isomorphic to the  two-dimensional Euclidean algebra $ {\mathfrak{iso}}(2) $.
Using the relation $\iota_{X}\omega={\rm d}h$ between a Hamiltonian vector field and one of its Hamiltonian functions, we get that the Hamiltonian functions for $X_1,X_2,X_3$ read
\begin{equation}
h_1=y,\qquad h_2=-x,\qquad h_3=\tfrac 12(x^2+y^2),
\end{equation}
correspondingly. Along with $h_0=1$, these functions  fulfil 
\begin{equation}
\{h_1,h_2\}_\omega=h_0,\quad \{h_1,h_3\}_\omega=h_2,\quad \{h_2,h_3\}_\omega=-h_1,\quad  \{h_0,\cdot\}_\omega=0 .
\end{equation}
Consequently, if $X$ is a Lie--Hamilton system admitting a minimal Lie algebra $A_0$, i.e. $X=\sum_{i=1}^3b_i(t)X_i$ for certain $t$-dependent functions $b_1(t),b_2(t),b_3(t)$ such that $V^X\simeq A_0$, then it admits a Lie--Hamiltonian structure $(U,\omega,h=\sum_{i=1}^3b_i(t)h_i$) and a Lie--Hamilton algebra $(\ham,\{\cdot,\cdot\}_\omega)$ generated by the functions $\langle h_1,h_2,h_3,h_0\rangle$. Hence, $(\ham,\{\cdot,\cdot\}_\omega)$ is a finite-dimensional real Lie algebra of Hamiltonian functions isomorphic to the {\em centrally extended}  Euclidean algebra $\overline{\mathfrak{iso}}(2) $~\cite{azca}.
Indeed, note that $1\in \{\mathcal{H}_\Lambda,\mathcal{H}_\Lambda\}_\omega$. In virtue of Corollary \ref{NoGo2}, system $X$ does not admit any Lie--Hamilton algebra isomorphic to $V^X$. Moreover, Proposition~\ref{UniExt} ensures that all Lie--Hamilton algebras for $X$ are isomorphic to $\overline{\mathfrak{iso}}(2)$.

\subsubsection*{Lie algebra P$_2$: $\mathfrak{sl}(2)$}

We have already proven that the Lie algebra of vector fields P$_2$, which is spanned by the vector fields (\ref{vectRiccati21}), is a Lie algebra of Hamiltonian vector fields with respect to the symplectic structure (\ref{aa1}).
The   Hamiltonian functions $h_1,h_2,h_3$ for $X_1$, $X_2$ and $X_3$  were found to be (\ref{ab}), correspondingly. Then, a Lie system $X$ with minimal Lie algebra P$_2$, i.e. a system of the form $X=\sum_{i=1}^3b_i(t)X_i$ for certain $t$-dependent functions $b_1(t),b_2(t),b_3(t)$ such that $V^X={\rm P}_2$, is a Lie--Hamilton system
with respect to the Poisson bracket induced by (\ref{aa1}). Then, $X$ admits a Lie--Hamiltonian structure $(U,\omega,h=\sum_{i=1}^3b_i(t)h_i)$ and  a Lie--Hamilton algebra isomorphic to $\mathfrak{sl}(2)$ with commutation relations (\ref{ab}). In view of Proposition \ref{LieHam2}, any Lie--Hamilton system associated with P$_2$ also admits a Lie--Hamilton algebra isomorphic to $\mathfrak{sl}(2)\oplus\mathbb{R}$. In view of Proposition \ref{LieHam1}, these are the only algebraic structures of the Lie--Hamilton algebras for such Lie--Hamilton systems.

\subsubsection*{Lie algebra P$_3$: $\mathfrak{so}(3)$}

In this case, we must determine a symplectic structure $\omega$ turning the elements of the modular generating system $(U_1,X_1,X_2)$ of P$_3$ into locally Hamiltonian vector fields with respect to a symplectic structure $\omega$ (\ref{ww}). In view of Theorem \ref{Char}, this ensures that every element of P$_3$ is Hamiltonian with respect to $\omega$. The condition $\mathcal{L}_{X_1}\omega=0$ gives
\begin{equation}
 y\frac{\partial f}{\partial x}-x\frac{\partial f}{\partial y} =0.
\end{equation}
Applying the characteristics method, we find that $f$ must be constant along the integral curves of the system
$x\, \dd x+ y\, \dd y =0$,  namely curves with $x^2+y^2=k\in\mathbb{R}$. So, $f=f(x^2+y^2)$. If we now require  $\mathcal{L}_{X_2}\omega=0$, we obtain that
\begin{equation}
 (1+x^2-y^2)\frac{\partial f}{\partial x}+2xy\frac{\partial f}{\partial y}+4xf=0.
\end{equation}
Using that $f=f(x^2+y^2)$, we have
\begin{equation}
\frac{f'}{f}=-\frac{2}{1+x^2+y^2} \Rightarrow f(x^2+y^2)=(1+x^2+y^2)^{-2}.
\end{equation}
  Then,
\begin{equation}
\omega=\frac{\dd x\wedge \dd y}{(1+x^2+y^2)^2}.
\end{equation}
So, P$_3$ becomes a Lie algebra of Hamiltonian vector fields relative to $\omega$. The vector fields $X_1$, $X_2$ and $X_3$ admit the Hamiltonian functions
\begin{equation}
h_1=-\frac{1}{2(1+x^2+y^2)},\qquad h_2=\frac{y}{1+x^2+y^2},\qquad h_3=-\frac{x}{1+x^2+y^2},
\end{equation}
which along $h_0=1$ satisfy the commutation relations
\begin{align}
&\{h_1,h_2\}_\omega=-h_3,\qquad \{h_1,h_3\}_\omega= h_2 ,\nonumber \\
& \{h_2,h_3\}_\omega=-4h_1-h_0,\qquad   \{h_0,\cdot\}_\omega=0,
\end{align}
with respect to the Poisson bracket induced by $\omega$. Then, $\langle h_1,h_2,h_3, h_0\rangle$ span a Lie algebra of Hamiltonian functions isomorphic to a {\it central extension} of  ${\mathfrak{so}}(3)$, denoted $\overline{\mathfrak{so}}(3)$. It is well-known \cite{azca} that the central extension associated with $h_0$ is a trivial one; if we define $\bar h_1= h_1+h_0/4$, then $\langle \bar h_1,h_2,h_3\rangle$ span a Lie algebra  isomorphic to ${\mathfrak{so}}(3)$ and $\overline{\mathfrak{so}}(3)\simeq \mathfrak{so}(3)\oplus\mathbb{R}$. In view of this and using Propositions \ref{LieHam1} and \ref{LieHam2}, a Lie system admitting a minimal Lie algebra P$_3$ admits Lie--Hamilton structures isomorphic to $\mathfrak{so}(3)\oplus\mathbb{R}$ and $\mathfrak{so}(3)$.

\subsubsection*{Lie algebra P$_5$: $\mathfrak{sl}(2 )\ltimes \mathbb{R}^2$}
From Proposition \ref{GoSym}, this Lie algebra consists of Hamiltonian vector fields with respect to the symplectic form $\omega=\dd x\wedge \dd y$. The vector fields of the basis
$X_1,\ldots,X_5$ for P$_5$ given in Table~\ref{table1} in Appendix 1 are Hamiltonian vector fields relative to $\omega$ with  Hamiltonian functions
\begin{equation}
h_1=y,\qquad h_2=-x,\qquad h_3=xy,\qquad h_4=\tfrac 12 {y^2},\qquad h_5=-\tfrac
 12 x^2,
\end{equation}
correspondingly. These functions together with $h_0=1$ satisfy the relations
\begin{equation}
\begin{aligned}
&\{h_1,h_2\}_\omega=h_0, &&
 \{h_1,h_3\}_\omega=-h_1,&&
 \{h_1,h_4\}_\omega=0,
\\& \{h_1,h_5\}_\omega=-h_2,
&&
\{h_2,h_3\}_\omega=h_2,&&
\{h_2,h_4\}_\omega=-h_1,
\\&
\{h_2,h_5\}_\omega=0,&&
\{h_3,h_4\}_\omega=2h_4, &&
\{h_3,h_5\}_\omega=-2h_5,
\\&
\{h_4,h_5\}_\omega=h_3, &&
  \{h_0,\cdot\}_\omega=0 . &&
\end{aligned}\end{equation} 
Hence $\langle h_1,\ldots,h_5  , h_0\rangle$ span a Lie algebra $\overline{\mathfrak{sl}(2 )\ltimes \mathbb{R}^2}$ which is a {\it non-trivial} central extension of P$_5$, i.e.,~it is not isomorphic to P$_5\oplus\mathbb{R}$. In fact, it is isomorphic to the so called two-photon Lie algebra   $\mathfrak{h}_6$ (see~\cite{BBF09} and references therein); this  can be proven to be $\mathfrak{h}_6\simeq \mathfrak{sl}(2)
\oplus_s \mathfrak{h}_3$,
where $\mathfrak{sl}(2 ) \simeq \langle h_3,h_4,h_5\rangle $,
$\mathfrak{h}_3\simeq  \langle h_1,h_2,h_0\rangle$ is the {\it Heisenberg--Weyl Lie algebra}, and
$\oplus_s$ stands for a semidirect sum. Furthermore, $\mathfrak{h}_6$ is also  isomorphic to the  $(1+1)$-dimensional centrally {\it extended Schr\"odinger  Lie algebra}~\cite{Schrod}.

In view of Corollary \ref{NoGo2}, Proposition \ref{UniExt} and following the same line of reasoning than in previous cases, a Lie system admitting a minimal Lie algebra P$_5$ only possesses Lie--Hamilton algebras isomorphic 	to $\mathfrak{h}_6$.

\subsubsection{Imprimitive Lie algebras}

\subsubsection*{Lie algebra I$_1$: $\mathbb{R}$}
Note that $X_1=\partial_x$ is a modular generating basis of I$_1$. By solving the PDE $\mathcal{L}_{X_1}\omega=0$ with $\omega$ written in the form (\ref{ww}), we obtain that $\omega=f(y)\dd x\wedge \dd y$ with $f(y)$ being any non-vanishing function of $y$. In view of Theorem \ref{Char}, the Lie algebra I$_1$ becomes a Lie algebra of Hamiltonian vector fields with respect to $\omega$. Observe that $X_1$, a basis of I$_1$, has a Hamiltonian function, $h_1=\int^y f(y')\dd y'$. As $h_1$ spans a Lie algebra isomorphic to $\mathbb{R}$, it is obvious that a system $X$ with $V^X\simeq I_1$ admits a Lie--Hamilton algebra isomorphic to I$_1$. Proposition \ref{LieHam2} yields that $X$ possesses a Lie--Hamilton algebra isomorphic to $\mathbb{R}^2$. In view of Proposition \ref{LieHam1}, these are the only algebraic structures for the Lie--Hamilton algebras for $X_1$.

\subsubsection*{Lie algebra I$_4$: $\mathfrak{sl}(2)$ of type II}
This Lie algebra admits a modular generating system $(\mathbb{R}^2_{x\neq y},X_1= {\partial}_x+  {\partial}_y, X_2= x{\partial}_x +y{\partial}_y)$.
Let us search for a symplectic form $\omega$ (\ref{ww})  turning $X_1$ and $X_2$ into local Hamiltonian vector fields with respect to it. In rigour of Theorem \ref{Char},
if $X_1,X_2$ become locally Hamiltonian, then we can ensure that every element of I$_4$ is Hamiltonian with respect to $\omega$.  By imposing $\mathcal{L}_{X_i}\omega=0$ ($i=1,2$), we find that
\begin{equation}
\frac{\partial f}{\partial x}+\frac{\partial f}{\partial y}=0,\qquad x\frac{\partial f}{\partial x}+y\frac{\partial f}{\partial y}+2 f=0.
\end{equation}
 Applying the method of characteristics to the first equation, we have that
 $\dd x=\dd y $.  Then $f=f(x-y)$. Using this in the second equation, we obtain  a particular solution $f= (x-y)^{-2}$ which  gives rise to a  closed and non-degenerate two-form, namely
\begin{equation}
\omega=\frac{{\rm d} x \wedge {\rm d} y}{(x -y)^2}  .
\label{aa2}
\end{equation}
Hence,
 \begin{equation}
 h_1=\frac{1}{x-y} ,\qquad
h_2=  \frac{x +y}{2(x -y)}  ,\qquad
h_3=\frac{xy}{x -y}
\end{equation}
are Hamiltonian functions of the vector fields $X_1,X_2,X_3$ of the basis for I$_4$ given in Table \ref{table1} in Appendix 1, respectively. Using the Poisson bracket $\{\cdot,\cdot\}_\omega$ induced by (\ref{aa2}), we obtain that $h_1,h_2$ and $h_3$ satisfy
\begin{equation}
\{h_1,h_2\}_\omega=-h_1,\qquad \{h_1,h_3\}_\omega=-2h_2,\qquad \{h_2,h_3\}_\omega=-h_3.
\end{equation}
Then, $(\langle h_1,h_2,h_3\rangle,\{\cdot,\cdot\}_\omega) \simeq \mathfrak{sl}(2)$.
Consequently, if $X$ is a Lie--Hamilton system admitting a minimal Lie algebra I$_4$, it admits a Lie--Hamilton algebra that is isomorphic to  $\mathfrak{sl}(2)$ or, from Proposition \ref{LieHam2}, to $\mathfrak{sl}(2)\oplus\mathbb{R}$. From Proposition \ref{LieHam1}, these are the only algebraic structures for its Lie--Hamilton algebras.

\subsubsection*{Lie algebra I$_5$: $\mathfrak{sl}(2)$ of type III}
Observe that $(U,X_1=\partial_x,X_2=2x\partial_x+y\partial_y)$ form a modular generating system of I$_5$. The conditions $\mathcal{L}_{X_1}\omega=\mathcal{L}_{X_2}\omega=0$ ensuring that $X_1$ and $X_2$ are locally Hamiltonian with respect to $\omega$ give rise to the equations
\begin{equation}
 \frac{\partial f}{\partial x}=0,\qquad 2x\frac{\partial f}{\partial x}+y\frac{\partial f}{\partial y}+3f=0,
\end{equation}
so that  $f(x,y)=y^{-3}$ is a particular solution of the second equation right above, and $X_1,X_2$ become locally Hamiltonian vector fields relative to the symplectic form
\begin{equation}
\omega=\frac{\dd x\wedge \dd y}{y^3} .
\end{equation}
In view of Theorem \ref{Char}, this implies that every element of I$_5$ is Hamiltonian with respect to $\omega$.
  Hamiltonian functions for the elements of the basis $X_1,X_2,X_3$ for I$_5$ given in Table \ref{table1} in Appendix 1, read
\begin{equation}
h_1=-\frac{1}{2y^{2}},\qquad  h_2= -\frac{x}{y^{2}},\qquad  h_3=-\frac{x^2}{2 y^{2}  }.
\end{equation}
They span a Lie algebra isomorphic to $\mathfrak{sl}(2)$
\begin{equation}
\{h_1,h_2\}_\omega=-2h_1,\qquad \{h_1,h_3\}_\omega=-h_2,\qquad \{h_2,h_3\}_\omega=-2h_3.
\end{equation}
Therefore, a Lie system possessing a minimal Lie algebra I$_5$ possesses a Lie--Hamilton algebra isomorphic to $\mathfrak{sl}(2)$ and, in view of Proposition \ref{LieHam2}, to $\mathfrak{sl}(2)\oplus\mathbb{R}$. In view of Proposition \ref{LieHam1}, these are the only possible algebraic structures for the Lie--Hamilton algebras for $X$.

\subsubsection*{Lie algebra I$_8^{(\alpha=-1)}$: $B_{-1}\simeq{\mathfrak{iso}}(1,1) $}

In view of Proposition \ref{GoSym}, this Lie algebra consists of Hamiltonian vector fields with respect to the standard symplectic structure $\omega=\dd x\wedge \dd y$.
The elements of the basis for $B_{-1}$ detailed in Table \ref{table1} in Appendix 1 satisfy the commutation relations
\begin{equation}
 [X_1,X_2]=0,\qquad [X_1,X_3]= X_1,\qquad [X_2,X_3]=-X_2 .
\end{equation}
Hence, these vector fields
 span a Lie  algebra isomorphic to the  (1+1)-dimensional Poincar\'e algebra $ {\mathfrak{iso}}(1,1) $. Their corresponding  Hamiltonian functions turn out to be
\begin{equation}
h_1=y,\qquad h_2=-x,\qquad h_3=xy ,
\end{equation}
which together with a central generator $h_0=1$ fulfil the commutation relations
\begin{equation}
\{h_1,h_2\}_\omega=h_0,\qquad \{h_1,h_3\}_\omega=-h_1,\qquad \{h_2,h_3\}_\omega=h_2,\qquad \{h_0,\cdot\}_\omega=0 .
\end{equation}
Thus, a Lie  system $X$ admitting a minimal Lie algebra $B_{-1}$ possesses a Lie--Hamilton algebra isomorphic to the centrally extended Poincar\'e algebra $ {\overline{\mathfrak{iso}}}(1,1) $ which, in turn, is also isomorphic to the harmonic oscillator  algebra $\mathfrak{h}_4$. As is well  known~\cite{azca}, this Lie algebra is not of the form $\mathfrak{iso}(1,1)\oplus\mathbb{R}$, then Proposition \ref{LieHam1} ensures that $X$ does not admit any Lie--Hamilton algebra isomorphic to $\mathfrak{iso}(1,1)$. Moreover, Proposition \ref{UniExt} states that all Lie--Hamilton algebras of $X$ must be isomorphic to $ {\overline{\mathfrak{iso}}}(1,1)$.

\subsubsection*{Lie algebra I$_{12}$: $\mathbb{R}^{r+1}$}

The vector field $X_1=\partial_y$ is a modular generating system for I$_{12}$ and all the elements of this Lie algebra have zero divergence. By solving the PDE $\mathcal{L}_{X_1}\omega=0$, where we recall that $\omega$ has the form (\ref{ww}),  we see that $f=f(x)$ and $X_1$ becomes Hamiltonian for any non-vanishing function $f(x)$. In view of Theorem \ref{Char}, the remaining elements of I$_{12}$ become automatically Hamiltonian with respect to $\omega$. Then, we obtain that $X_1,
\ldots,X_{r+1}$ are Hamiltonian vector fields relative to the symplectic structure $\omega=f(x)\dd x\wedge \dd y$ with   Hamiltonian functions
\begin{equation}
h_1=- \int^{x}{f(x')\dd x'} ,\qquad  h_{j+1} =-  \int^{x} f(x')\xi_j(x')\dd x' ,
\end{equation}
with $j=1,\ldots,r ,$ and $r\ge 1$, which span the Abelian Lie algebra $\mathbb{R}^{r+1}$. In consequence, a Lie--Hamilton system $X$ related to a minimal Lie algebra I$_{12}$ possesses a Lie--Hamilton algebra isomorphic to $\mathbb{R}^{r+1}$. From Propositions \ref{LieHam1} and \ref{LieHam2}, it only admits an additional Lie--Hamilton algebra isomorphic to $\mathbb{R}^{r+2}$.

\subsubsection*{Lie algebra I$_{14}$: $\mathbb{R}\ltimes\mathbb{R}^{r}$}
The functions $\eta_1(x),\ldots,\eta_r(x)$ form a fundamental set of solutions of a system of $r$ differential equations with constant coefficients \cite{HA75,GKP92}. Hence, none of them vanishes in an open interval of $\mathbb{R}$ and I$_{14}$ is such that
$(U_1,X_1,X_2)$, where $X_1$ and $X_2$ are given in Table \ref{table1} in Appendix 1,  form a modular generating system. Since all the elements of I$_{14}$ have divergence zero and using Theorem \ref{Char}, we infer that I$_{14}$  consists of Hamiltonian vector fields relative to a symplectic structure if and only if $X_1$ and $X_2$ are locally Hamiltonian vector fields with respect to a symplectic structure. By requiring $\mathcal{L}_{X_i}\omega=0$, with $i=1,2$ and $\omega$ of the form (\ref{ww}), we obtain that
\begin{equation}
 \frac{\partial f}{\partial x}=0,\qquad \eta_j(x)  \frac{\partial f}{\partial y}=0,\qquad j=1,\ldots,r.
\end{equation}
So, I$_{14}$ is only compatible, up to nonzero multiplicative factor, with $\omega=\dd x\wedge \dd y$. The Hamiltonian functions corresponding to $X_1,\ldots,X_{r+1}$ turn out to be
\begin{equation}\label{hh}
h_1=y ,\qquad  h_{j+1} =-  \int^{x} \eta_j(x')\dd x' ,
\end{equation}
with $j=1,\ldots,r ,$ and $r\ge 1$. We remark that different Lie--Hamilton algebras, corresponding to the Lie algebras hereafter called I$_{14A}$ and I$_{14B}$,  are spanned by the above Hamiltonian functions:

\begin{itemize}

\item Case I$_{14A}$: If $1\notin \langle \eta_1,\ldots,\eta_{r}\rangle$,  then the functions (\ref{hh}) span a Lie algebra $\mathbb{R}\ltimes\mathbb{R}^{r}$ and, by considering Propositions~\ref{LieHam1} and \ref{LieHam2}, this case only admits an additional Lie--Hamilton algebra isomorphic to  $(\mathbb{R}\ltimes \mathbb{R}^r)\oplus\mathbb{R}$.

\item Case I$_{14B}$: If $1\in \langle \eta_1,\ldots,\eta_{r}\rangle$, we can choose a basis of I$_{14}$ in such a way that there exists a function, let us say $\eta_1$, equal to $1$. Then the Hamiltonian functions (\ref{hh}) turn out to be
\begin{equation}
h_1=y ,\qquad  h_2=-x,\qquad h_{j+1} =-  \int^{x} \eta_j(x')\dd x' ,\end{equation} 
with $j=2,\ldots,r ,$ and $r\ge 1$ which require a central generator $h_0=1$  in order to close   a centrally extended Lie algebra $(\ham,\{\cdot,\cdot\}_\omega)\simeq  \overline{(\mathbb{R} \ltimes \mathbb{R}^{r})}$.

\end{itemize}

In view of the above, a Lie system $X$ with a minimal Lie algebra I$_{14}$ is a Lie--Hamilton system. Its Lie--Hamilton algebras can be isomorphic to I$_{14}$ or I$_{14}\oplus \mathbb{R}$ when  $1\notin \langle \eta_1,\ldots,\eta_{r}\rangle$ (class I$_{14A}$). If $1\in \langle \eta_1,\ldots,\eta_{r}\rangle$ (class I$_{14B}$), a Lie--Hamilton algebra is isomorphic to $\overline{\mathbb{R} \ltimes \mathbb{R}^{r}}$ and since  $1\in \{\ham ,\ham\}_\omega$, we obtain from Corollary \ref{NoGo2} and Proposition \ref{UniExt} that every Lie--Hamilton algebra for $X$ is isomorphic to it.

\subsubsection{Lie algebra I$^{(\alpha=-1)}_{16}$: $C_{-1}^r \simeq \mathfrak{h}_2\ltimes\mathbb{R}^{r+1}$}

In view of Proposition \ref{GoSym}, this Lie algebra consists of Hamiltonian vector fields with respect to the standard symplectic structure. The resulting  Hamiltonian functions for $X_1,\ldots, X_{r+3}$ are given by
   \begin{equation}
h_1=y,\qquad h_2=-x,\qquad h_3=xy,\qquad \dots,\qquad h_{j+3}=-\frac{x^{j+1}}{j+1} ,
\end{equation} 
with $j=1,\ldots,r ,$ and $r\ge 1$, which again require an additional central generator $h_0=1$ to close on a finite-dimenisonal Lie algebra. The commutation relations for this Lie algebra are given by
\begin{equation}
\begin{aligned}
&\{h_1,h_2\}_\omega=h_0,&&
\{h_1,h_3\}_\omega=-h_1,
\\
 &\{h_2,h_3\}_\omega=h_2,&&
\{h_1,h_{4}\}_\omega=-h_{2},
\\
 &\{h_1,h_{k+4}\}_\omega=-( k +1)h_{k+3},&&
\{h_2,h_{j+3}\}_\omega=0  ,\\
&  \{h_3,h_{j+3} \}_\omega=-(j+1) h_{j+3},&&
\{h_{j+3} ,h_{k+4}\}_\omega= 0,  
\\
&\{h_0,\cdot\}_\omega=0 ,&&
\end{aligned}\end{equation} 
with $j=1,\dots, r$ and $k=1,\dots,r-1$, which define the centrally extended Lie algebra $\overline{\mathfrak{h}_2\ltimes\mathbb{R}^{r+1}}$. This Lie algebra is not a trivial extension of $\overline{\mathfrak{h}_2\ltimes\mathbb{R}^{r+1}}$.

Then, given a Lie system $X$ with a minimal Lie algebra $C_{-1}^r$, the system is a Lie--Hamilton one which admits a Lie--Hamilton algebra isomorphic to $\overline{\mathfrak{h}_2\ltimes \mathbb{R}^{r+1}}$. As $1\in \left\{\overline{\mathfrak{h}_2\ltimes\mathbb{R}^{r+1}},\overline{\mathfrak{h}_2\ltimes\mathbb{R}^{r+1}}\right\}_\omega$, Corollary \ref{NoGo2} and Proposition \ref{UniExt} ensure that  every Lie--Hamilton algebra for $X$ is isomorphic to $\overline{\mathfrak{h}_2\ltimes\mathbb{R}^{r+1}}$.

\section{Applications of Lie--Hamilton systems on the plane}
\setcounter{equation}{0}
\setcounter{theorem}{0}
\setcounter{example}{0}
Here we show Lie--Hamilton systems of relevance in Science.
Given the recurrent appearance of the $\mathfrak{sl}(2,\mathbb{R})$-Lie algebra, we show a few instances of Lie--Hamilton systems with a $\mathfrak{sl}(2,\mathbb{R})$ 
Vessiot--Guldberg Lie algebra. These are: the Milne--Pinney and second-order Kummer--Schwarz equations, the complex Riccati equations with t-dependent real coefficients,
the coupled Riccati equations, planar diffusion Riccati equations, etc. We will establish certain equivalences among some of the mentioned systems.

The importance of Lie--Hamilton systems in biological applications is undoubtable. Some examples will be displayed: the generalized Buchdahl equations,
certain Lotka--Volterra systems, quadratic polynomial systems and  models for viral infections.
Another two Lie algebras worth of mention in the realm of Lie--Hamilton systems are the two-photon and $\mathfrak{h_2}$-Lie algebras. Their respective
applications are: the dissipative harmonic oscillator and the second-order Riccati equation for the two photon Lie algebra, and the complex Bernoulli and generalized Buchdahl equations
and certain Lotka--Volterra systems for the $\mathfrak{h_2}$-Lie algebra.
To conclude, we will add other interesting Lie--Hamilton systems as the Cayley--Klein Riccati equation and double-Clifford or split complex Riccati equations.
The forthcoming subsections show the abovementioned systems in detail.

\subsection{$\mathfrak{sl}(2,\mathbb{R})$-Lie--Hamilton systems}

Let us employ our techniques to study $\mathfrak{sl}(2)$-Lie--Hamilton systems \cite{LSKummer,Pi12}. More specifically, we analyse 
Lie systems  used to describe Milne--Pinney equations \cite{SIGMA}, Kummer--Schwarz equations \cite{CGL11} and complex Riccati equations with real $t$-dependent coefficients  \cite{Eg07}. As a byproduct, our results also cover the 
$t$-dependent frequency harmonic oscillator.


\begin{example}\normalfont
The {\bf Milne--Pinney equation} is well-known for its multiple properties and applications in Physics (see~\cite{AL08} and references therein).
For example, it is useful to modelize the propagation of laser beams in nonlinear media, plasma Dynamics, Bose-Einstein condensates
through the so called Gross-Pitaevskii equation, etc.

It takes the form
\begin{equation}
\frac{\dd^2x}{\dd t^2}=-\omega^2(t)x+\frac{c}{x^3},\end{equation} 
where $\omega(t)$ is any $t$-dependent function and $c$ is a real constant.  It was first introduced by Ermakov
as a way to find first-integrals of its corresponding $t$-dependent harmonic oscillator when $c=0$.
By adding a new variable $y\equiv \dd x/\dd t$, we can study these equations through the first-order system
\begin{equation}\label{FirstLie1}
\left\{
\begin{aligned}
\frac{\dd x}{\dd t}&=y,\\
\frac{\dd y}{\dd t}&=-\omega^2(t)x+\frac{c}{x^3},
\end{aligned}\right.
\end{equation}
which is a Lie system \cite{SIGMA,PW}. We recall that (\ref{FirstLie1}) can be regarded as  the equations of motion of the one-dimensional Smorodinsky--Winternitz system~\cite{BCHLS, WSUF65}; moreover, when the parameter $c$ vanishes, this reduces to the harmonic oscillator (both with   a $t$-dependent frequency).
Explicitly, (\ref{FirstLie1}) is the associated system with the $t$-dependent vector field
$
X_t=X_3+\omega^2(t)X_1,
$
where
\begin{equation}\label{FirstLieA}
X_1=-x\frac{\partial}{\partial y},\qquad X_2=\frac 12 \left(y\frac{\partial}{\partial y}-x\frac{\partial}{\partial x}\right),\qquad X_3=y\frac{\partial}{\partial x}+\frac{c}{x^3}\frac{\partial}{\partial y},
\end{equation}
span a finite-dimensional real Lie algebra $V$ of vector fields isomorphic to $\mathfrak{sl}(2)$ with commutation relations given by
\begin{equation}\label{FirstLieB}
[X_1,X_2]=X_1,\qquad [X_1,X_3]=2 X_2,\qquad  [X_2,X_3]=X_3 .
\end{equation}

There are {\em  four} classes of finite-dimensional Lie algebras of vector fields isomorphic to $\mathfrak{sl}(2)$ in the GKO  classification: P$_2$ and I$_3$--I$_5$. To determine which one is locally diffeomorphic to $V$, we first  find out whether $V$ is imprimitive or not. In this respect,
 recall that $V$ is  {\em imprimitive} if there exists a one-dimensional distribution $\mathcal{D}$ invariant under the action (by Lie brackets) of the elements of $V$. Hence,  $\mathcal{D}$  is spanned by
 a non-vanishing vector field
\begin{equation}
Y=\fff_x(x,y)\frac{\partial}{\partial x}+\fff_y(x,y)\frac{\partial}{\partial y},
\end{equation}
 which must be invariant under the action of $X_1$, $X_2$ and $X_3$. As $\fff_x$ and $\fff_y$ cannot vanish simultaneously, $Y$ can be taken either of the following local forms
\begin{equation}\label{xxx}
Y=\frac{\partial}{\partial x}+\fff_y\frac{\partial}{\partial y},\qquad Y=\fff_x\frac{\partial}{\partial x}+\frac{\partial}{\partial y}.
\end{equation}
Let us assume that $\mathcal{D}$ is spanned by the first one and search for   $Y$. Now, if $\mathcal{D}$ is invariant under the Lie brackets of the elements of $V$, we have that
\begin{subequations}
\begin{align}
&\quad\qquad \qquad
 \mathcal{L}_{X_1}Y=\left(1-x\frac{\partial \fff_y}{\partial y}\right)\frac{\partial}{\partial y}=\hh_1 Y,\label{con2}\\
&\mathcal{L}_{X_2}Y=\frac 12\left[\frac{\partial}{\partial x}+\left(y\frac{\partial \fff_y}{\partial y}-x\frac{\partial \fff_y}{\partial x}- \fff_y\right)\frac{\partial}{\partial y}\right]=\hh_2 Y,\label{con1}\\
&\mathcal{L}_{X_3}Y=-\fff_y\frac{\partial}{\partial x}+\left(\frac{3c}{x^4}+y\frac{\partial \fff_y}{\partial x}+\frac c {x^3}\frac{\partial \fff_y}{\partial y}\right)\frac{\partial}{\partial y}=\hh_3 Y,\label{con3}
\end{align}
\end{subequations}
for certain functions $\hh_1,\hh_2,\hh_3$ locally defined on $\mathbb{R}^2$. The left-hand side of (\ref{con2}) has no term $\partial_x$ but the right-hand one has it provided $\hh_1\neq 0$. Therefore, $\hh_1=0$ and $\fff_y= { y}/{x}+G$ for a certain $G=G(x)$. Next by introducing this result in
  (\ref{con1}), we find that $\hh_2=1/2$ and $2G + x    G^\prime =0$, so that $G(x)=\mm/x^2$ for   $\mm\in\mathbb{R}$.
   Substituting this into (\ref{con3}), we obtain that $\hh_3=-(\mm+x y)/x^2$
and $ \mm^2=-4c$. Consequently, when  $c>0$  it does not exist any non-zero $Y$ spanning locally $\mathcal{D}$ satisfying (\ref{con2})--(\ref{con3})  and $V$ is therefore primitive, whilst   if $c\le 0$   there exists a vector field
\begin{equation}
Y= \frac{\partial}{\partial x}+\left(\frac{y}{x} +\frac{\mm}{x^2}\right) \frac{\partial}{\partial y},\qquad \mm^2=-4c,
\end{equation} 
   which spans  $\mathcal{D}$, so that $V$ is  imprimitive.   The case of $\mathcal{D}$ being   spanned by the second form of $Y$   (\ref{xxx}) can be analysed ously and drives to the same conclusion. 

Therefore the system     (\ref{FirstLie1})  belongs to different classes 
within the GKO classification according to  the value of the parameter $c$. The final result   is established in the following statement.

\begin{proposition}\label{prop71}  The Vessiot--Guldberg Lie algebra for system   (\ref{FirstLie1}), corresponding to the the Milne--Pinney equations,  is locally diffeomorphic to   {\rm P}$_2$ for $c>0$, {\rm I}$_4$ for $c<0$ and {\rm I}$_5$ for $c=0$.
\end{proposition}
\begin{proof}  Since $V$ is  primitive when $c>0$ and  this is  isomorphic
to $\mathfrak{sl}(2)$,  the GKO classification given in Table \ref{table1} in Appendix 1 implies  that $V$ is locally diffeomorphic to
the primitive class P$_2$.

 Let us now consider that $c<0$ and prove that the system is
 diffeomorphic to  the class  I$_4$. We do this by showing that there exists  a local diffeomorphism 
$\phi:(x,y)\in U\subset \mathbb{R}^2_{x\ne y}\mapsto \bar U\subset (u,v)\in\mathbb R^2_{u\neq 0}$, satisfying that $\phi_*$ maps the basis  for I$_4$  listed in Table \ref{table1} in Appendix 1 into
 (\ref{FirstLieA}). Due to the Lie bracket $[X_1,X_3]=2X_2$, verified in both bases,  it is only necessary to search    the map for the generators $X_1$ and $X_3$  (so for  $X_2$ this will be automatically fulfilled). 
  By writing  in coordinates
\begin{equation}
\phi_*(\partial_x+\partial_y)=-x\partial_y,\qquad  \phi_*(x^2\partial_x+y^2\partial_y)=y\partial_x+c/x^3\partial_y ,
\end{equation}
we obtain
$$
\left(
\begin{array}{cc}
\frac{\partial u}{\partial x}&\frac{\partial u}{\partial y}\\[4pt]
\frac{\partial v}{\partial x}&\frac{\partial v}{\partial y}\\
\end{array}
\right)\left(\begin{array}{c}
              1\\[4pt]
              1
             \end{array}\right)=
             \left(
             \begin{array}{c}
              0\\[4pt] -u
             \end{array}\right),\qquad
             \left(
\begin{array}{cc}
\frac{\partial u}{\partial x}&\frac{\partial u}{\partial y}\\[4pt]
\frac{\partial v}{\partial x}&\frac{\partial v}{\partial y}\\
\end{array}
\right)\left(\begin{array}{c}
              x^2\\[4pt]
              y^2
             \end{array}\right)=
             \left(
             \begin{array}{c}
              v\\ [4pt]  {c}/{u^3}
             \end{array}\right).
$$
Hence, ${\partial_x u}+{\partial_y u}=0\Rightarrow u=\ff(x-y)$ for a certain $\ff:z\in\mathbb{R}\mapsto g(z)\in \mathbb{R}$. Since $x^2\partial_x u +y^2\partial_y u  =v$, then $v=(x^2-y^2) \ff'$, where $\ff'$ is the derivative of $\ff(z)$ in terms of $z$.
Using now that ${\partial_x v}+{\partial_y v}=-u$ we get $2(x-y)\ff^\prime=-\ff$ so that $\ff=\lambda/ |x-y|^{1/2}$  where $\lambda\in\mathbb{R}\backslash \{0\}$. Substituting  this into the remaining equation $x^2\partial_x v +y^2\partial_y v  =c/u^3$,  we find that $\lambda^4=-4 c$. 
Since $c<0$, we consistently find that 
\begin{equation}
u=\frac{\lambda}{|x-y|^{1/2}},\qquad v=-\frac{\lambda(x+y )}{2|x-y|^{1/2}},\qquad \lambda^4=-4 c.
\end{equation}
Finally,  let us set $c=0$ and search for a   local diffeomorphism 
$\phi:(x,y)\in U\subset \mathbb{R}^2_{y\ne 0}\mapsto \bar U\subset (u,v)\in\mathbb R^2$ such that $\phi_*$ maps the basis corresponding to I$_5$ into (\ref{FirstLieA}); namely
\begin{equation}
\phi_*(\partial_x )=-x\partial_y,\qquad  \phi_*(x^2\partial_x+xy\partial_y)=y\partial_x,
\end{equation}
yielding
$$
\left(
\begin{array}{cc}
\frac{\partial u}{\partial x}&\frac{\partial u}{\partial y}\\[4pt]
\frac{\partial v}{\partial x}&\frac{\partial v}{\partial y}\\
\end{array}
\right)\left(\begin{array}{c}
              1\\[4pt]
              0
             \end{array}\right)=
             \left(
             \begin{array}{c}
              0\\[4pt] -u
             \end{array}\right),\qquad
             \left(
\begin{array}{cc}
\frac{\partial u}{\partial x}&\frac{\partial u}{\partial y}\\[4pt]
\frac{\partial v}{\partial x}&\frac{\partial v}{\partial y}\\
\end{array}
\right)\left(\begin{array}{c}
              x^2\\[4pt]
              x y
             \end{array}\right)=
             \left(
             \begin{array}{c}
              v\\ [4pt] 0
             \end{array}\right).
$$
Hence, ${\partial_x u}=0\Rightarrow u=\ff_1(y)$ for a certain $\ff_1:\mathbb{R}\rightarrow\mathbb{R}$. Since $\partial_x v =-u$, then $v=-\ff_1(y)x+\ff_2(y)$ for another   $\ff_2:\mathbb{R}\rightarrow \mathbb{R}$.
Using now the PDEs of the second matrix, we see that 
$xy {\partial_y u}  =xy  { \ff^\prime _1} =v=-\ff_1 x+\ff_2  $, so that  
  $\ff_2=0$ and $\ff_1=\lambda/y $, where $\lambda\in\mathbb{R}\backslash \{0\}$ and $\lambda\neq 0$ to avoid $\phi$ not to be a diffeomorphism.
 It can be checked that the remaining equation is so fulfilled. Therefore $u= \lambda/y $ and $v=-\lambda x/y$.
  \end{proof}

We remark that, since  the three classes P$_2$, I$_4$ and I$_5$ appear in Table \ref{table3} in Appendix 1, system (\ref{FirstLie1}) can always be associated
with a symplectic form turning their vector fields Hamiltonian. In this respect, 
  recall that  it was recently proven, that the  system (\ref{FirstLie1})  is a
 Lie--Hamilton one for any value of $c$~\cite{BCHLS}.
However,   we shall show that  identifying   it  to one of the classes of
 the GKO  classification will be useful  to study the relation of this system to other ones.
\end{example}

\begin{example}\normalfont

Let us turn now to reconsider the {\bf second-order Kummer--Schwarz equation} \eqref{KS22} written as a first-order system \eqref{FirstOrderKummer}

\begin{equation}\label{FirstOrderKummer2}
\left\{\begin{aligned}
\frac{dx}{dt}&=v,\\
\frac{dv}{dt}&=\frac 32 \frac{v^2}x-2c_0x^3+2b_1(t)x,
\end{aligned}\right.
\end{equation}
on ${\rm T}\mathbb{R}_0$, with
$\mathbb{R}_0=\mathbb{R}-\{0\}$, obtained by adding the new variable
$v\equiv dx/dt$ to the KS-2 equation (\ref{KS22}).

It is well-known that (\ref{FirstOrderKummer2}) is a Lie system \cite{CGL11,LSKummer}. In fact, it describes the integral curves of the $t$-dependent vector field
$X_t=X_3+b_1(t)X_1$ where the vector fields $X_1,X_2,X_3$ correspond to those in \eqref{VFKS2}. Since their corresponding commutation relations are the same as in \eqref{VFKS2}.
spanning a Vessiot--Guldberg Lie algebra $V$ isomorphic to $\mathfrak{sl}(2)$. Thus, $V$ can be isomorphic to one of the  four $\mathfrak{sl}(2)$-Lie algebras of vector fields in the GKO classification.  

As in the previous subsection, we analyse  if there exists a distribution $\mathcal{D}$ stable under $V$ and locally generated by a vector field $Y$ of the first form given in (\ref{xxx}) (the same results can be obtained by assuming the second one).
Notice that in this section we change the notation in the variable $v$ in \eqref{FirstOrderKummer2}, as $v\leftrightarrow y$. So, impossing  $\mathcal{D}$ to be stable under $V$ yields
\begin{subequations}
\begin{align}
&\qquad\qquad\,\,\,\,\mathcal{L}_{X_1}Y=2\left(x\frac{\partial \fff_y}{\partial y} -1\right)\frac{\partial}{\partial y}=\hh_1Y\label{con2p}
,\\
&\qquad\,\, \mathcal{L}_{X_2}Y=-\frac{\partial}{\partial x}+\left(x\frac{\partial \fff_y}{\partial x}+2y\frac{\partial \fff_y}{\partial y}-2\fff_y\right)\frac{\partial}{\partial y}=\hh_2Y,\label{con1p}\\
&\mathcal{L}_{X_3}Y=-\fff_y\frac{\partial}{\partial x}+\left[X_3\fff_y+\frac{3y^2}{2x^2}+6c\,x^2-\frac{3y}x\fff_y\right]\frac{\partial}{\partial y}=\hh_3 Y\label{con3p},
\end{align}
\end{subequations}
for certain functions $\hh_1,\hh_2,\hh_3$ locally defined on $\mathbb{R}^2$.
The left-hand side of (\ref{con2p}) has no term $\partial_x$ and the right-hand one does not have it provided $\hh_1=0$. Hence, $\hh_1=0$ and $\fff_y= y/x+F$ for a $F=F(x)$. In view of (\ref{con1p}), we then obtain  $\hh_2=-1$ and 
$F -x    F^\prime=0$, that is, $F(x) =   \mm x$ for   $\mm\in\mathbb{R}$. Substituting $g_y$ in (\ref{con3p}),  
we obtain that $\hh_3=-\mm x-y/x$
and $ \mm^2=-4c$. Hence, as in the Milne--Pinney equations, we find that 
if $c>0$  it does not exist any   $Y$ spanning locally $\mathcal{D}$ satisfying (\ref{con2p})--(\ref{con3p})  and $V$ 
is primitive, meanwhile  if $c\le 0$, then   $\mathcal{D}$ is spanned by the vector field
\begin{equation}
Y= \frac{\partial}{\partial x}+\left(\frac{y}{x} + {\mm}{x}\right) \frac{\partial}{\partial y},\qquad  \mm^2=-4c,
\end{equation}
and $V$ is  imprimitive.

The precise classes of the  GKO classification  corresponding to the system (\ref{FirstOrderKummer}) are summarized in the following proposition.

\begin{proposition}\label{prop72} The Vessiot--Guldberg Lie algebra for system   (\ref{FirstOrderKummer2}), associated with the second-order Kummer--Schwarz    equations,    is locally diffeomorphic to   P$_2$ for $c>0$, I$_4$ for $c<0$ and I$_5$ for $c=0$.  
\end{proposition}

\begin{proof}  The case with   $c>0$ provides the primitive class P$_2$  since $Y=0$. If   $c<0$ we look for a local diffeomorphism 
$\phi:(x,y)\in U\subset \mathbb{R}^2_{x\ne y}\mapsto \bar U\subset (u,v)\in\mathbb R^2_{u\neq 0}$, such that $\phi_*$ maps the basis of  I$_4$  into  (\ref{VFKS2}), that is,
\begin{equation}
\phi_*(\partial_x+\partial_y)=2x\partial_y,\qquad  \phi_*(x^2\partial_x+y^2\partial_y)=y\partial_x+ (\tfrac32 y^2/{x}-2c\,x^3 )\partial_y .
\end{equation}
Then
$$
\left(
\begin{array}{cc}
\frac{\partial u}{\partial x}&\frac{\partial u}{\partial y}\\[4pt]
\frac{\partial v}{\partial x}&\frac{\partial v}{\partial y}\\
\end{array}
\right)\left(\begin{array}{c}
              1\\[4pt]
              1
             \end{array}\right)=
             \left(
             \begin{array}{c}
              0\\[4pt] 2u
             \end{array}\right),\qquad
             \left(
\begin{array}{cc}
\frac{\partial u}{\partial x}&\frac{\partial u}{\partial y}\\[4pt]
\frac{\partial v}{\partial x}&\frac{\partial v}{\partial y}\\
\end{array}
\right)\left(\begin{array}{c}
              x^2\\[4pt]
              y^2
             \end{array}\right)=
             \left(
             \begin{array}{c}
              v\\ [4pt]  \tfrac 32 v^2/u -2{c}\,{u^3}
             \end{array}\right).
$$
Proceeding as in    the proof of Proposition~\ref{prop71}, we find that   $u=\ff(x-y)$ and  $v=(x^2-y^2) \ff^\prime$
 for   $\ff:\mathbb{R}\rightarrow\mathbb{R}$. 
  As  now   ${\partial_x v}+{\partial_y v}=2u$ we obtain $2(x-y)\ff^\prime=2\ff$, so that $\ff=\lambda (x-y) $  with $\lambda\in\mathbb{R}\backslash \{0\}$ and assume $\lambda\neq 0$ to avoid $\phi$ not being a diffeomorphism. The remaining equation $x^2\partial_x v +y^2\partial_y v  =\tfrac 32 v^2/u -2{c}\,{u^3}$ implies that 
   $4\lambda^2=-1/ c$, which is consistent with the value $c<0$. Then
\begin{equation}
u=  {\lambda } (x-y) ,\qquad v=  {\lambda }(x^2-y^2) ,\qquad 4\lambda^2=-1/ c.
\end{equation}

In the third possibility  with $c=0$  we require that $\phi_*$     maps the basis of  I$_5$  into  (\ref{VFKS2}) fulfilling
\begin{equation}
\phi_*(\partial_x )=2x\partial_y,\qquad  \phi_*(x^2\partial_x+xy\partial_y)=y\partial_x+ \tfrac32 y^2/{x} \partial_y,
\end{equation}
that is,
$$
\left(
\begin{array}{cc}
\frac{\partial u}{\partial x}&\frac{\partial u}{\partial y}\\[4pt]
\frac{\partial v}{\partial x}&\frac{\partial v}{\partial y}\\
\end{array}
\right)\left(\begin{array}{c}
              1\\[4pt]
              0
             \end{array}\right)=
             \left(
             \begin{array}{c}
              0\\[4pt] 2u
             \end{array}\right),\qquad
             \left(
\begin{array}{cc}
\frac{\partial u}{\partial x}&\frac{\partial u}{\partial y}\\[4pt]
\frac{\partial v}{\partial x}&\frac{\partial v}{\partial y}\\
\end{array}
\right)\left(\begin{array}{c}
              x^2\\[4pt]
              x y
             \end{array}\right)=
             \left(
             \begin{array}{c}
              v\\ [4pt]  \tfrac 32 v^2/u
             \end{array}\right).
$$
By taking into account the proof of Proposition~\ref{prop71}, it is straightforward  to check  that the four PDEs are satisfied for
$u= \lambda y^2$ and $v=2\lambda  x y^2$ with  $\lambda\in\mathbb{R}\backslash \{0\}$.    \end{proof}

 \end{example}

\begin{example}\normalfont

Let us return to {\bf complex Riccati equations with $t$-dependent real coefficients} in the form (\ref{Riccati21}). We already showed that this system has a Vessiot--Guldberg Lie algebra P$_2\simeq \mathfrak{sl}(2)$. Therefore, it is locally diffeomorphic to the Vessiot--Guldberg Lie algebra appearing in the above Milne--Pinney  and Kummer--Schwarz   equations whenever the parameter $c>0$.
 In view of the GKO classification, there exist local diffeomorphisms relating the three  first-order systems associated with these equations. For instance, we can search for a local diffeomorphism $\phi:(x,y)\in U\subset \mathbb{R}^2_{y\ne 0}\mapsto \bar U\subset (u,v)\in\mathbb R^2_{u\neq 0}$ mapping every system (\ref{Riccati21}) into one of the form (\ref{FirstLie1}), e.g.,~satisfying that $\phi_*$ maps the basis of P$_2$ in Table \ref{table1} in Appendix 1, related to the planar Riccati equation, into the basis (\ref{FirstLieA}) associated with the Milne--Pinney one. By writing  in coordinates
\begin{equation}
\phi_*(\partial_x)=-x\partial_y,\qquad  \phi_*[(x^2-y^2)\partial_x+2xy\partial_y]=y\partial_x+c/x^3\partial_y ,
\end{equation}
we obtain
$$
\left(
\begin{array}{cc}
\frac{\partial u}{\partial x}&\frac{\partial u}{\partial y}\\[4pt]
\frac{\partial v}{\partial x}&\frac{\partial v}{\partial y}\\
\end{array}
\right)\left(\begin{array}{c}
              1\\[4pt]
              0
             \end{array}\right)=
             \left(
             \begin{array}{c}
              0\\[4pt] -u
             \end{array}\right),\qquad
             \left(
\begin{array}{cc}
\frac{\partial u}{\partial x}&\frac{\partial u}{\partial y}\\[4pt]
\frac{\partial v}{\partial x}&\frac{\partial v}{\partial y}\\
\end{array}
\right)\left(\begin{array}{c}
              x^2-y^2\\[4pt]
              2xy
             \end{array}\right)=
             \left(
             \begin{array}{c}
              v\\ [4pt]  {c}/{u^3}
             \end{array}\right).
$$
Similar computations to those performed in the proof of Proposition~\ref{prop71} for the three PDEs 
${\partial_x u}=0$, $\partial_x v =-u$ and $(x^2-y^2) {\partial_x u}+  2xy {\partial_y u}=v$ gives
$u=\lambda/|y|^{1/2}$ and $v=-\lambda x/|y|^{1/2}$ with  $\lambda\in\mathbb{R}\backslash \{0\}$.
Substituting these results  into the remaining equation we find that $\lambda^4=c$ which is consistent with the positive value of $c$.   Consequently, this maps the system (\ref{Riccati21}) into (\ref{FirstLie1}) and the solution of the first one is locally equivalent to solutions of the second.

Summing up, we have explicitly proven that the three $\mathfrak{sl}(2)$ Lie algebras of the classes  P$_2$, I$_4$ and I$_5$ given in Table \ref{table3} in Appendix 1 cover the following $\mathfrak{sl}(2)$-Lie systems

\begin{itemize} 
 
\item P$_2$:  Milne--Pinney  and Kummer--Schwarz   equations for $c>0$ as well as complex Riccati equations with $t$-dependent coefficients,

\item I$_4$:  Milne--Pinney  and Kummer--Schwarz   equations for $c<0$,

\item I$_5$:  Milne--Pinney  and Kummer--Schwarz   equations for $c=0$ and
the harmonic oscillator with $t$-dependent frequency.

\end{itemize} 

This means that, only within each class,  they are locally diffeomorphic  and, therefore, there   exists a  
local change of variables mapping  one into another. Thus, for instance, there does not exist any diffeomorphism mapping the Milne--Pinney  and Kummer--Schwarz equations with $c\ne 0$  to the harmonic oscillator. These results also  explain from an algebraic point of view the existence of the known diffeomorphism mapping   Kummer--Schwarz equations to Milne--Pinney equations \cite{AL08} provided that the sign of $c$ is the same in both systems. 
\end{example}

\subsection{Lie--Hamilton biological models}

In this section  we focus on new applications of the Lie--Hamilton approach  to  Lotka--Volterra-type systems and to a viral infection model. We also consider here the analysis of Buchdahl equations which can be studied through a Lie--Hamilton system diffeomorphic to a precise $t$-dependent  Lotka--Volterra system.


\begin{example}\normalfont
 We call {\bf generalized Buchdahl equations}~\cite{Bu64, CSL05, CN10}  the  second-order differential equations
 \begin{equation}\label{Bucheq}
 \frac{\dd^2 x}{\dd t^2}=a(x)\left(\frac{\dd x}{\dd t}\right)^2+b(t)\frac{\dd x}{\dd t},
 \end{equation}
where $a(x)$ and $b(t)$ are arbitrary functions of their respective arguments. In order to analyse whether these equations can be studied through a Lie system,
 we add the variable $y\equiv \dd x/\dd t $ and consider the first-order   system  of differential equations
  \begin{equation}\label{Buchdahl1}
\left\{ \begin{aligned}
 \frac{\dd x}{\dd t}&=y,\\
 \frac{\dd y}{\dd t}&=a(x)y^2+b(t)y .
 \end{aligned}\right.
 \end{equation}
 Note that if $(x(t),y(t))$ is a particular solution of this system with $y(t_0)=0$ for a particular $t_0\in\mathbb{R}$, then $y(t)=0$ for every $t\in\mathbb{R}$ and $x(t)=\lambda \in\mathbb{R}$. Moreover, if $a(x)=0$ then the solution of the above system is also trivial. As a consequence, we can restrict ourselves to studying  particular solutions on $\mathbb R^2_{y\neq 0}$ with $a(x)\neq 0$.

 Next let us  prove that (\ref{Buchdahl1}) is a Lie system. Explicitly, (\ref{Buchdahl1}) is associated with the $t$-dependent vector field
 $X_t=X_1+b(t)X_2$, 
 where 
 \begin{equation}\label{VectBuch}
 X_1=y\frac{\partial }{\partial x}+a(x)y^2\frac{\partial}{\partial y},\qquad  X_{2}=y\dfrac{\partial}{\partial y}.
 \end{equation}
Since
\begin{equation}
 [X_1,X_2]=-X_1,\end{equation}  
these vector fields span a Lie algebra $V$ isomorphic to $\mathfrak{h}_2$ leaving invariant the distribution $\mathcal{D}$ spanned by $Y\equiv X_1$.  Since the rank of $\mathcal{D}^V$ is two,  $V$ is locally diffeomorphic to the imprimitive class ${\rm I}_{14A}$ with $r=1$ and $\eta_1(x)={\rm e}^x$  given in Table~\ref{table3} in Appendix 1. This proves for the first time that generalized Buchdahl equations written as the     system (\ref{Buchdahl1})  are, in fact, not only a Lie system \cite{CGL11} but a Lie--Hamilton one.

Next by determining a symplectic form obeying $\mathcal{L}_{X_i}\omega=0$, with $i=1,2$ for the vector fields   (\ref{VectBuch}) and  the  generic $\omega$ (\ref{ww}), it can be shown that this reads
\begin{equation}
\omega=\dfrac{\exp\left(-\int a(x)\dd x\right)}{y}\,\dd x\wedge \dd y,
\end{equation}
which turns   $\tX$ and $X_2$ into Hamiltonian vector fields with Hamiltonian functions
\begin{equation}
h_1=y\exp\left(-\int^x a(x')\dd x'\right),\qquad h_2=-\int^x\exp\left(-\int^{x'} a(\bar x)d\bar x\right)dx',
\end{equation}
respectively. Note that all the these structures are properly defined on $\mathbb R^2_{y\neq 0}$ and   hold
$\{h_{1},h_{2}\}=h_{1}.
$
Consequently, the system (\ref{Buchdahl1}) has a $t$-dependent Hamiltonian given by
$
h_t=h_{1}+b(t) h_{2}.
$
\end{example}

\begin{example}\normalfont
 Consider the specific {\bf $t$-dependent Lotka--Volterra system}~\cite{JHL05,Tr96} of the form
  \begin{equation}\label{LV1}
  \left\{
 \begin{aligned}
 \frac{\dd x}{\dd t}&=ax-g(t)(x-ay)x,\\
 \frac{\dd y}{\dd t}&=ay-g(t)(bx-y)y,
 \end{aligned}\right.
\end{equation}
where $g(t)$ is a $t$-dependent function  representing the variation of the seasons and $a,b$ are certain real parameters
describing the interactions among the species. We hereafter focus on the case $a\neq 0$, as otherwise the above equation becomes, up to a $t$-reparametrization,
an autonomous differential equation that can easily be integrated. We also assume $g(t)$ to be a non-constant function and we restrict ourselves to particular solutions on $\mathbb{R}_{x,y\ne 0}=\{(x,y)|x\neq 0, y\neq 0\}$ (the remaining ones can be trivially  obtained).

Let us prove that (\ref{LV1}) is a Lie system and that for some values of the real parameters $a\ne 0$ and $b$ this is   a Lie--Hamilton system as well. This system describes the integral curves of the $t$-dependent vector field $
X_t=X_1+g(t)X_2$
 where
\begin{equation}
 X_1=ax\frac{\partial}{\partial x}+ay\frac{\partial}{\partial y},\qquad X_2=-(x-ay)x\frac{\partial}{\partial x}-(bx-y)y\frac{\partial}{\partial y} ,
\end{equation}
satisfy 
\begin{equation}
[X_1,X_2]=a X_2,\qquad a\ne 0.
\end{equation}
 Hence, $X_1$ and $X_2$ are the generators of  a Lie algebra $V$ of vector fields isomorphic to $\mathfrak{h}_2$  leaving invariant the distribution $\mathcal{D}$  on $\mathbb{R}_{x,y\ne 0}$ spanned by $Y\equiv X_2$.
 According to the values of the parameters $a\ne 0$ and $b$ we find that
 
 \begin{itemize}
 \item When  $a=b=1$,  the rank of $\mathcal{D}^V$  on the domain of $V$ is one. In view of Table \ref{table1} in Appendix 1, the Lie algebra $V$ is thus isomorphic to I$_2$ and, by taking into account Table \ref{table3} in Appendix 1,  we conclude that $X$ is a Lie system, but not a Lie--Hamilton one.
 
 \item Otherwise,  the rank of $\mathcal{D}^V$ is two, so that this Lie algebra is locally diffeomorphic to I$_{14A}$ with $r=1$ and $\eta_1={\rm e}^{a x}$ given  in Table \ref{table3} in Appendix 1 and, consequently, $X$ is a Lie--Hamilton system.  As a straightforward consequence, when $a=1$ and $b\ne 1$ the system (\ref{LV1})  is locally diffeomorphic to the  
  generalized Buchdahl equations (\ref{Buchdahl1}).

 \end{itemize}

Let us now derive a symplectic structure (\ref{ww}) turning the elements of $V$ into Hamiltonian vector fields by solving the system of PDEs $\mathcal{L}_{X_1}\omega=\mathcal{L}_{X_2}\omega=0$. The first condition reads in local coordinates
\begin{equation}
\mathcal{L}_{X_1}\omega=(X_1f+2af)\dd x\wedge \dd y=0.
\end{equation}
So we obtain that $f=F(x/y)/y^2$ for any function $F:\mathbb{R}\rightarrow \mathbb{R}$.
By imposing that $\mathcal{L}_{X_2}\omega=0$, we find
\begin{equation}
\mathcal{L}_{X_2}\omega=\left[(b-1)x^2+(a-1)yx\right]\frac{\partial f}{\partial x}+f\left[(b-2)x+ay\right]=0.
\end{equation}
Notice that, as expected,   $f$ vanishes when $a=b=1$. We study separately the remaining cases: i) $a\ne 1$ and  $b\ne 1$; ii) $a=1$ and $b\ne 1$; and iii) $a\ne 1$ and $b= 1$.

When  both $a,b\ne 1$  we write $f=F(x/y)/y^2$, thus obtaining that $\omega$ reads, up to a non-zero multiplicative constant, as
\begin{equation}
\omega=\frac 1{y^2}\left(\frac{x}{y}\right)^{\frac a{1-a}}\left[1-a+(1-b)\frac xy\right]^{\frac 1{a-1}+\frac 1{b-1}}\dd x\wedge \dd y,\qquad a,b\neq 1.
\end{equation}
From this, we   obtain the following Hamiltonian functions for $X_1$ and $X_2$
\begin{equation}
\begin{aligned}
h_{1}&=a (1-b)^{1+\frac{1}{a-1}+\frac{1}{b-1}}\left(\dfrac{x}{y}\right)^{\frac{1}{b-1}}\times
\\
&\hspace{1em}\times\! \,_2F_1\left(
\dfrac{1}{1-b},\dfrac{1}{1-a}+\dfrac{1}{1-b};\dfrac{b-2}{b-1};\dfrac{y(1-a)}{x(b-1)}
\right),
\\
\qquad h_{2}&=-y \left(\dfrac{x}{y}\right)^{\frac{1}{1-a}}\left[
(1-a)+(1-b)\dfrac{x}{y}
\right]^{\frac{1}{a-1}+\frac{1}{b-1}+1},
\end{aligned}\end{equation} 
where $\,_2F_1(\alpha,\beta, \gamma,\zeta)$ stands for the hypergeometric function 
\begin{equation}
\,_2F_1(\alpha,\beta,\gamma,z)=\sum_{n=0}^\infty [(\alpha)_n(\beta)_n/(\gamma)_n]z^n/n!
\end{equation}
with $(\delta)_n=\Gamma(\delta+n)/\Gamma(\delta)$ being the rising {\it Pochhmaler symbol}. 
As expected, $\{h_1,h_2\}_\omega=-a h_2$.

When $a=1$ and $b\neq 1$, the symplectic form for $X$ becomes
\begin{equation}
\omega=\frac 1{y^2}\exp\left(\frac{y-(b-2)x \ln|x/y|}{(b-1)x}\right)\dd x\wedge \dd y, \qquad b \neq 1,
\end{equation}
and the Hamiltonian functions for $X_1$ and $X_2$ read
\begin{equation}
\begin{aligned}
h_{1}&=-\left(\dfrac{1}{1-b}\right)^{\frac{1}{b-1}}\Gamma\left(
\dfrac{1}{1-b},\dfrac{y}{x(1-b)} 
\right),
\\
h_{2}&=(b-1)x\left(\dfrac{x}{y}\right)^{\frac{1}{b-1}} \text{exp}\left(
\dfrac{y}{(b-1)x}
\right),
\end{aligned}\end{equation} 
with $\Gamma(u,v)$ being the {\it incomplete Gamma function},  which satisfy  $\{h_{1}, h_{2}\}_\omega=-h_{2}$.

Finally, when $b=1$ and $a\neq 1$, we have
\begin{equation}
\omega=\dfrac{1}{y^{2}}\left(
\dfrac{x}{y}
\right)^{\frac{a}{1-a}}\text{exp}\left(
\dfrac{x}{y (a-1)}
\right)\text{d} x \wedge \text{d}y ,\qquad a\neq 1.
\end{equation}
Then, the Hamiltonian functions for $X_1$ and $X_2$ are, in this order, 
\begin{equation}
\begin{aligned}
h_{1}&=a(1-a)^{\frac{1}{1-a}}\, \Gamma \left(\dfrac{1}{1-a},\dfrac{x}{y(1-a)}\right),
\\
h_{2}&=(a-1)y\,\exp\left(
\dfrac{x}{y(a-1)}
\right)\left(
\dfrac{x}{y}
\right)^{\frac{1}{1-a}} .
\end{aligned}\end{equation} 
Indeed,  $\{h_{1}, h_{2}\}_\omega=-a h_{2}$.

\end{example}

\begin{example}\normalfont
The system of differential equations \cite{LlibreValls2}, is known as {\bf a class of quadratic polynomial Lie systems}
\begin{equation} \label{SLV}
\left\{
\begin{aligned}
\frac{\dd x}{\dd t}&= b(t)x + c(t)y +d(t)x^{2}+e(t)x y + f(t)y^{2},\\
\frac{\dd y}{\dd t}&= y,\\
\end{aligned}\right.
\end{equation}
where $b(t), c(t), d(t), e(t)$ and $f(t)$ are arbitrary $t$-dependent functions, is an interacting species model of Lotka--Volterra type that belongs to the class of quadratic-linear polynomial systems with a unique singular point at the origin \cite{LlibreValls2}.

In general, this system is not a Lie system. For instance, consider the particular system associated with the $t$-dependent vector field
$
X_t=d(t) X_1+ e(t)X_2+X_3,$
\begin{equation}
X_1=x^2\frac{\partial}{\partial x}   ,\qquad X_2=xy\frac{\partial}{\partial x},\qquad X_3=y\frac{\partial}{\partial y},
\end{equation} 
where $d(t)$ and $e(t)$ are non-constant and non-proportional functions. Notice that $V^X$ contains $X_1,X_2$ and their successive  Lie brackets, i.e. the vector fields
\begin{equation}
\stackrel{n-{\rm times}}{\overbrace{[X_2,\ldots [X_2}},X_1]\ldots]= x^2 y^n \frac{\partial}{\partial x} \equiv Y_n .
\end{equation} 
Hence, $[X_2,Y_n]=Y_{n+1}$ and  the family of vector fields $X_1,X_2,X_3,Y_1,Y_2,\dots$
  span an infinite-dimensional family of linearly independent vector fields over $\mathbb{R}$. Then, $X$ is not a Lie system.

 Hereafter we analyse the cases of (\ref{SLV}) with $d(t)=e(t)=0$ which provides quadratic polynomial systems that are Lie systems. We call them {\it quadratic polynomial Lie systems}; these are    related to the system of differential equations.
  \cite{LlibreValls2}
\begin{equation} \label{SLV2}
\left\{
\begin{aligned}
\frac{\dd x}{\dd t}&=b(t)x+ c(t)y +f(t) y^{2},\\
\frac{\dd y}{\dd t}&=y.
\end{aligned}\right.
\end{equation}
Note that if a solution $(x(t),y(t))$ of the above system is such that $y(t_0)=0$ for a certain $t_0$, then $y(t)=0$ for all $t\in\mathbb{R}$ and the corresponding $x(t)$ can be then easily obtained. In view of this, we   focus on those particular solutions within $\mathbb R^2_{y\neq 0}$. The   system (\ref{SLV2})  is    associated with the $t$-dependent vector field on $\mathbb R^2_{y\neq 0}$  of the form
$X_{t}=X_1+b(t)X_{2}+c(t) X_{3}+f(t) X_{4}$, where
\begin{equation}
X_1=y\dfrac{\partial}{\partial y},\qquad X_2=x \dfrac{\partial}{\partial x},\qquad X_3=y\dfrac{\partial}{\partial x}, \qquad X_4=y^{2}\dfrac{\partial}{\partial x} ,
\end{equation}
satisfy the commutation rules
\begin{equation}
\begin{array}{lll}
[X_1,X_2]=0,  &\qquad [X_1,X_3]=X_3,&\qquad [X_1,X_4]=2 X_4,\\[2pt]
[X_2,X_3]=-X_3,&\qquad [X_2,X_4]=-X_4,&\qquad  [X_3,X_4]=0 .
\end{array}\end{equation} 
Note that $V\simeq V_1\ltimes V_2$ where $V_1=\langle X_1,X_2\rangle\simeq\mathbb{R}^2$ and $V_2=\langle X_3,X_4\rangle\simeq \mathbb{R}^2$. In addition, the distribution $\mathcal{D}$ spanned by $Y\equiv \partial_
x$ is invariant under the action of the above vector fields. So, $V$ is imprimitive. In view of Table \ref{table1} in Appendix 1, we find that (\ref{SLV2}) is a Lie system corresponding to the imprimitive class I$_{15}$ with $V\simeq \mathbb{R}^2\ltimes\mathbb{R}^2$. By taking into account our classification given in Table \ref{table3} in Appendix 1, we know that this is
 not a Lie algebra of vector fields with respect to any symplectic structure.
\end{example}


\begin{example}\normalfont
We now   consider a subcase of (\ref{SLV2}) that provides a Lie--Hamilton system. We will refer to such subcase as {\bf quadratic polynomial Lie--Hamilton systems}.
 In view of Table \ref{table3} in Appendix 1, the Lie subalgebra $\mathbb{R}\ltimes \mathbb{R}^2$ of $V$ is a Lie algebra of Hamiltonian vector fields with respect to a symplectic structure, that is, ${\rm I}_{14}\subset {\rm I}_{15}$. So, it is natural consider the restriction of (\ref{SLV2}) to
\begin{equation} \label{SLV3}
\left\{
\begin{aligned}
\frac{\dd x}{\dd t}&=b\, x+ c(t)y +f(t) y^{2},\\
\frac{\dd y}{\dd t}&=y,
\end{aligned}\right.
\end{equation}
where   $b\in \mathbb{R}\backslash\{1,2\}$ and $c(t)$, $f(t)$ are still   $t$-dependent functions.
The system
  (\ref{SLV3})  is    associated   to the $t$-dependent vector field
 $X_{t}=X_1+c(t) X_{2}+f(t) X_{3}$ on $\mathbb {R}^2_{y\ne 0}=\{(x,y)\in \mathbb{R}\mid y\neq 0\}$, where
\begin{equation}\label{vecsA}
X_1=b\,   x \dfrac{\partial}{\partial x}+y\dfrac{\partial}{\partial y},\qquad X_2=y\dfrac{\partial}{\partial x}, \qquad X_3=y^{2}\dfrac{\partial}{\partial x}
\end{equation}
satisfy the commutation relations
\begin{equation}\label{algG}
[\tX, X_{2}]=(1-b)X_{2}, \qquad  [\tX, X_{3}]=(2-b)X_{3}, \qquad [X_{2}, X_{3}]=0 .
\end{equation}
Therefore, the vector fields (\ref{vecsA})  generate a Lie algebra $V\simeq V_1\ltimes V_2$, where $V_1=\langle X_1 \rangle\simeq\mathbb{R}$ and $V_2=\langle X_2,X_3\rangle\simeq \mathbb{R}^2$. The domain of $V$ is $\mathbb {R}^2_{y\ne 0}$ and the rank of $\mathcal{D}^V$ is two. Moreover, the distribution $\mathcal{D}$ spanned by the vector field $Y\equiv \partial_x$ is stable under the action of the elements of $V$, which turns $V$ into an imprimitive Lie algebra. So, $V$ must be locally diffeomorphic to the imprimitive Lie algebra I$_{14}$ displayed in Table \ref{table1} in Appendix 1 for $r=2$. We already know that the class I$_{14}$ is a Lie algebra of Hamiltonian vector fields with respect to a   symplectic structure.

By imposing $\mathcal{L}_{X_i}\omega=0$ for  the vector fields (\ref{vecsA}) and  the  generic symplectic form (\ref{ww}), it can be shown that $\omega$ reads
\begin{equation}
\omega=\dfrac{\dd x\wedge \dd y}{y^{b+1}} ,
\end{equation}
which turns    (\ref{vecsA}) into Hamiltonian vector fields with Hamiltonian functions
\begin{equation}
h_1=-\frac{x}{y^b},\qquad h_2=\frac  {   y^{1-b}}{  1-b},\qquad h_3=\frac  {   y^{2-b}}{  2-b}, \qquad b\in \mathbb{R}\backslash\{1,2\}.
\end{equation} 
Note that all the above structures are properly defined on $\mathbb {R}^2_{y\ne 0}$.
The above Hamiltonian  functions span a three-dimensional Lie algebra with commutation relations
\begin{equation}
\{h_{1},h_{2}\}_\omega=(b-1)h_{2}  , \qquad  \{h_{1}, h_{3}\}_\omega=(b-2)h_{3} ,\qquad  \{h_{2},h_{3}\}_\omega=0 .\end{equation} 
Consequently, $V$ is locally diffeomorphic  to the imprimitive Lie algebra I$_{14A}$ of Table \ref{table3} such that the Lie--Hamilton algebra is $\mathbb{R}\ltimes \mathbb{R}^{2}$ (also  $(\mathbb{R}\ltimes \mathbb{R}^{2})\oplus\mathbb{R}$).
The system (\ref{SLV3}) has a $t$-dependent
Hamiltonian
\begin{equation}
h_t=b \,h_{1}+c(t) h_{2}+d(t)h_{3}=-b\dfrac{x}{y^b}+c(t) \frac  {   y^{1-b}}{  1-b} + d(t) \frac  {   y^{2-b}}{  2-b}.
\end{equation}

We point out that the cases of  (\ref{SLV3}) with either $b=1$ or $b=2$     also lead to Lie--Hamilton systems,  but  now belonging, both of them, to the   class I$_{14 B}$ of Table \ref{table3} as a central generator is  required. For instance if $b=1$, the commutation relations (\ref{algG}) reduce to
\begin{equation}\label{algG2}
[\tX, X_{2}]= 0 , \qquad  [\tX, X_{3}]=X_{3}, \qquad [X_{2}, X_{3}]=0 ,
\end{equation}
while    the symplectic form and the Hamiltonian functions are found to be
\begin{equation}
\omega=\dfrac{\dd x\wedge \dd y}{y^{2}} ,\qquad h_1=-\frac{x}{y},\qquad h_2=\ln y ,\qquad h_3= y ,
\end{equation} 
which together with $h_0=1$ close the   (centrally extended)  Lie--Hamilton algebra 
$\overline{\mathbb{R} \ltimes \mathbb{R}^2 }$, that is,
\begin{equation}\label{algG3}
\{h_{1},h_{2}\}_\omega=-h_0 , \qquad  \{h_{1}, h_{3}\}_\omega=-h_{3} ,\qquad  \{h_{2},h_{3}\}_\omega=0 ,\qquad  \{h_{0},\cdot\}_\omega=0 .
\end{equation}
A similar result can be found for $b=2$.
\end{example}


\begin{example}\normalfont 

Finally, let us consider a {\bf simple viral infection model} given by \cite{EK05}
\begin{equation}\left\{
\begin{aligned}
\dfrac{{\rm d}x}{{\rm d}t}&=(\alpha(t)-g(y))x,\\
\dfrac{{\rm d}y}{{\rm d}t}&=\beta(t) x y-\gamma(t) y,
\end{aligned}\label{VS1}\right.
\end{equation}
where $g(y)$ is a real positive function taking into account the power of the infection. Note that if a particular solution satisfies $x(t_0)=0$ or $y(t_0)=0$ for a $t_0\in\mathbb{R}$, then
$x(t)=0$ or $y(t)=0$, respectively, for all $t\in\mathbb{R}$. As these cases are trivial, we restrict
ourselves to studying particular solutions within $\mathbb R^2_{x,y\ne 0}=\{(x,y)\in \mathbb{R}^2\mid x\neq 0,  y\neq 0\}$.

The simplest possibility consists in setting  $g(y)=\delta$, where $\delta$ is a constant.
 Then, (\ref{VS1})    describes the integral curves of the $t$-dependent vector field
$X_t=(\alpha(t)-\delta)X_{1}+\gamma(t) X_{2} +\beta(t) X_{3}$, 
on $\mathbb R^2_{x,y\ne 0}$, where  the vector fields
\begin{equation}
X_1=x\dfrac{\partial}{\partial x}, \qquad X_2=- y \dfrac{\partial}{\partial y}, \qquad X_3=x y \dfrac{\partial}{\partial y} ,
\end{equation}
satisfy the   relations (\ref{algG2}).
 So, $X$ is a Lie system related to a Vessiot--Guldberg Lie algebra $V\simeq \mathbb{R}\ltimes\mathbb{R}^2$ where
  $ \langle X_1 \rangle\simeq\mathbb{R}$ and $ \langle X_2,X_3\rangle\simeq \mathbb{R}^2$. The distribution $\mathcal{D}^V$ has rank two on $\mathbb R^2_{x,y\ne 0}$. Moreover, $V$ is imprimitive, as the distribution $\mathcal{D}$ spanned by $Y\equiv \partial_y$ is invariant under the action of vector fields of $V$. Thus $V$ is locally diffeomorphic to the
     imprimitive Lie algebra I$_{14 B}$  for $r=2$ and, in view of Table \ref{table3}, the system $X$ is
       a Lie--Hamilton one.

 Next we   obtain that $V$ is a Lie algebra of Hamiltonian vector fields with respect to the symplectic form
 \begin{equation}
\omega=\dfrac{\dd x \wedge \dd y}{x y}.
\end{equation}
Then, the vector fields $X_1$, $X_2$ and $X_3$ have Hamiltonian functions:
$
h_1=\ln{|y|}, h_2=\ln{|x|}, h_3=-x,
$
which along $h_0=1$ close the   relations (\ref{algG3}).
If we assume $V^X=V$, the $t$-dependent Hamiltonian $
h_t=(\alpha(t) -\delta)h_1+\gamma(t) h_2+\beta(t)h_3
$
gives rise to a Lie--Hamiltonian structure $(\mathbb R^2_{x,y\ne 0},\omega, h)$ for $X$ defining the  Lie--Hamilton algebra 
$\overline{(\mathbb{R} \ltimes \mathbb{R}^2 ) }$. 

\end{example}

\subsection{Other Lie--Hamilton systems on the plane}

%
\begin{example}\normalfont

Let us reconsider the {\bf Cayley--Klein  Riccati equation} reviewed in \eqref{Riccati21},

\begin{equation}\label{Riccati213}
\frac{{\rm d} x}{{\rm d} t}=a_0(t)+a_1(t)x+a_2(t)(x^2-y^2),\qquad \frac{{\rm d} y}{{\rm d} t}=a_1(t)y+a_2(t)2xy,
\end{equation}
with $a_0(t),a_1(t),a_2(t)$ being arbitrary $t$-dependent real functions
with t-dependent vector field
\begin{equation}\label{tdepenRiccati21}
 X_t=a_0(t)X_1+a_1(t)X_2+a_2(t)X_3,
\end{equation}
where $X_1$, $X_2$, $X_3$ have commutation relations \eqref{aa}.
According to  Table~\ref{table1} in Appendix 1, system (\ref{Riccati21}) is a LH system possessing a Vessiot--Guldberg Lie algebra diffeomorphic to the primitive Lie algebra P$_2$ and the vector fields $X_1,X_2,X_3$ are Hamiltonian vector fields with respect to the symplectic form
 \be
\omega= \frac{ {\rm d}x\wedge {\rm  d}y }{ y^{2} } .
\label{ua}
\ee
Then, some Hamiltonian functions  $h_i$ of $X_i$ (\ref{vectRiccati21}) can be chosen to be
\be
h_1= - \frac 1y, \qquad h_2= -\frac xy, \qquad h_3= -\frac{x^2+y^2}{y} ,
\label{ub}
\ee
which satisfy
\begin{equation}
\label{sl2Rh}
\{h_1,h_2\}_\omega=-h_1,\qquad \{h_1,h_3\}_\omega=-2h_2,\qquad \{h_2,h_3\}_\omega=-h_3.
\end{equation}
Hence, $ (\langle h_1,h_2,h_3\rangle,\{\cdot,\cdot\}_\omega)$ is a Lie--Hamilton algebra for $X$ isomorphic to $\mathfrak{sl}(2)$
and
\begin{equation}
h=a_0(t)h_1+a_1(t)h_2+a_2(t)h_3
\end{equation}
 is a  $t$-dependent Hamiltonian function.

The above result can be generalized by making use of
  analytic continuation and contractions which can also be understood as a Cayley--Klein approach~\cite{CK2d,Gromova,Ro88,Yaglom}, which underlies the structure of the referred to as two-dimensional Cayley--Klein geometries.

Consider the real plane with coordinates $\{x,y\}$ and an `additional' unit $\ota$ such that
\begin{equation}
\label{ba}
\ota^2 \in\{ -1,+1,0\}    .
\end{equation}
Next,  we define
\begin{equation}
\label{bb4}
 z:=x+\ota y,\qquad \bar z:=x-\ota y,\qquad  (x,y)\in\mathbb R^2.
\end{equation}
Assuming that $\ota$ commutes with real numbers, we can write
\begin{equation}
\label{bc}
 |z|^2:= z\bar z= x^2-\ota^2 y^2,\qquad z^2=x^2+\ota^2 y^2 +2 \ota x y.
\end{equation}

 In this way we find that the number $z$ in (\ref{bb4})  comprises three possibilities

 \begin{itemize}

 \item $\ota^2=-1$. In this case we are dealing with the usual {\it complex numbers} $\ota:= i$ and $z\in\mathbb C$. Hence
\begin{equation}
 \label{bca}
   |z|^2= z\bar z= x^2+ y^2,\qquad z^2=x^2- y^2 +2 i x y,\qquad z\in\mathbb C.
\end{equation}

  \item $\ota^2=+1$. Now we  are dealing with the  so called {\it split-complex numbers} $z\in\mathbb C^\prime$. The additional unit  is usually known as  the {\em double} or {\em Clifford} unit $\ota:= e$~\cite{Yaglom}. Thus,
 \begin{equation}
 \label{bcb}
   |z|^2= z\bar z= x^2- y^2,\qquad z^2=x^2+y^2 +2 e x y,\qquad z\in \mathbb C^\prime.
\end{equation}

  \item $\ota^2=0$. In this last possibility   $z$ is known as a {\it dual} or {\it Study number}~\cite{Yaglom},  which can be regarded as a {\it contracted} case since
   \begin{equation}
 \label{bcc}
   |z|^2= z\bar z= x^2 ,\qquad z^2=x^2+ 2 \varepsilon x y,\qquad z\in \mathbb D.
\end{equation}

  \end{itemize}

 With these ingredients we    shall    call the {\em Cayley--Klein  Riccati equation}~\cite{EstLucasSar}  the generalization of the complex Riccati equation (\ref{Riccati21}) to   $z\in \{ \mathbb C,\mathbb C^\prime, \mathbb D\}$, that is,
\begin{equation}\label{CKRE}
\frac{{\rm d} z}{{\rm d} t}=a_0(t)+a_1(t)z+a_2(t)z^2,\qquad z:= x+\ota y ,
\end{equation}
which, for real $t$-dependent coefficients $a_0(t),a_1(t),a_2(t)$,  gives rise to a system of two differential equations
\begin{equation}\label{CKRE2}
\frac{{\rm d} x}{{\rm d} t}=a_0(t)+a_1(t)x+a_2(t)(x^2+\ota^2 y^2),\qquad \frac{{\rm d} y}{{\rm d} t}=a_1(t)y+a_2(t)2xy,
\end{equation}
that generalizes the planar Riccati equation (\ref{Riccati21}).

We now prove separately that the two remaining cases with $\ota^2=+1$ and $\ota^2=0$ are also Lie--Hamilton systems.

\end{example}


\begin{example}\normalfont

If we set $\ota^2:=+1$, the system (\ref{CKRE2}) is known as a {\bf Double-Clifford or split-complex Riccati equation}
\begin{equation}\label{CKRE3}
\frac{{\rm d} x}{{\rm d} t}=a_0(t)+a_1(t)x+a_2(t)(x^2+  y^2),\qquad \frac{{\rm d} y}{{\rm d} t}=a_1(t)y+a_2(t)2xy.
\end{equation}
The point is to analyze whether this is also a Lie system and, in affirmative case, whether it is furthermore a Lie--Hamilton one with a Vessiot--Guldberg Lie algebra diffeomorphic to a class given in  Table~\ref{table1} in Appendix 1.

Indeed, system (\ref{CKRE3}) is  related to the $t$-dependent vector field of the form  (\ref{tdepenRiccati21}) with
\begin{equation}
X_1= \frac{\partial}{\partial x},\qquad X_2= x\frac{\partial}{\partial x}+y\frac{\partial}{\partial y} ,\qquad X_3= (x^2+y^2)\frac{\partial}{\partial x}+2xy\frac{\partial}{\partial y} .
\label{ca}
\end{equation}
These again span a Vessiot--Guldberg real Lie algebra $V\simeq \mathfrak{sl}(2)$ with the same commutation rules given by (\ref{aa}). Hence (\ref{CKRE3}) is a Lie system.

Vector fields (\ref{ca}) do not arise exactly in the Lie algebras isomorphic to $ \mathfrak{sl}(2)$ in Table~\ref{table1} in Appendix 1, namely, ${\rm P}_2$,  ${\rm I}_4$ and ${\rm I}_5$.
Nevertheless, if  we introduce in (\ref{ca}) the new variables $\{u,v\}$ defined by
\begin{equation}\label{change}
\begin{aligned}
&u=x+y,\qquad v=x-y,\qquad 
&x=\tfrac 12(u+v),\qquad y=\tfrac 12 (u-v), 
\end{aligned}
\end{equation}
we find that
\begin{equation}
X_1= \frac{\partial}{\partial u}+ \frac{\partial}{\partial v},\qquad X_2= u\frac{\partial}{\partial u}+v\frac{\partial}{\partial v} ,\qquad X_3= u^2\frac{\partial}{\partial u}+v^2\frac{\partial}{\partial v} ,
\end{equation}
which are, exactly,
 the vector fields  appearing in the imprimitive Lie algebra I$_4\simeq \mathfrak{sl}(2)$ of Table~\ref{table1} and Table~\ref{table3} of Appendix 1. Thus  these are endowed with   a  closed and non-degenerate two-form; the latter  and  some associated  Hamiltonian functions are given by
\begin{equation}
\omega=\frac{{\rm d} u \wedge {\rm d} v}{(u -v)^2}  ,\qquad  h_1=\frac{1}{u-v} ,\qquad
h_2=  \frac{u +v}{2(u -v)}  ,\qquad
h_3=\frac{uv}{u -v} ,
\label{aa3}
\end{equation}
 which satisfy the same commutation relations (\ref{sl2Rh}).

In terms of the initial variables $\{x,y\}$, the expressions (\ref{aa3}) turn out to be
\begin{equation}
\omega=-\frac{{\rm d} x \wedge {\rm d} y}{ 2 y^2}  ,
\qquad  h_1=\frac{1}{2 y} ,\qquad
h_2=  \frac{x}{2 y}  ,\qquad
h_3=\frac{x^2-y^2}{2 y} ,
\end{equation}
to be compared with (\ref{ua}) and (\ref{ub}).

\end{example}

\begin{example}\normalfont
When $\ota^2=0$ the system (\ref{CKRE2}) reduces to a {\bf Dual-Study Riccati equation}
\begin{equation}\label{CKRE4}
\frac{{\rm d} x}{{\rm d} t}=a_0(t)+a_1(t)x+a_2(t) x^2 ,\qquad \frac{{\rm d} y}{{\rm d} t}=a_1(t)y+a_2(t)2xy ,
\end{equation}
which can be regarded as a `contracted' system  from either (\ref{Riccati21}) or (\ref{CKRE3}). We stress that this system appears as a part of the Riccati system employed in the resolution of diffusion-type equations~\cite{SSVessiot--Guldberg11,SSVessiot--Guldberg14}. Its relevance is due to the fact that its general solution allows us to solve the whole Riccati system and to map diffusion-type equations into an easily integrable PDE.

Let us prove that (\ref{CKRE4})  is  a Lie--Hamilton system.
It is clear that this is a Lie system as it is  related to the $t$-dependent vector field $X_t=a_0(t)X_1+a_1(t)X_2+a_2(t)X_3$,
 where
\begin{equation}
X_1= \frac{\partial}{\partial x},\qquad X_2= x\frac{\partial}{\partial x}+y\frac{\partial}{\partial y} ,\qquad X_3= x^2\frac{\partial}{\partial x}+2xy\frac{\partial}{\partial y} ,
\label{da}
\end{equation}
span a Vessiot--Guldberg real Lie algebra $V\simeq \mathfrak{sl}(2)$ with   commutation rules given by (\ref{aa}).  These vector fields do not appear again in  Table \ref{table1} in Appendix 1.
Nevertheless, we can map them into a basis given in Table \ref{table1}, by choosing an appropiate change of variables.

We assume $y>0$, while the case $y<0$ can be studied ously giving a similar result.
Consider the new variables $\{u,v\}$ defined by
\begin{equation}
\begin{aligned}
&u=x,\qquad v=\sqrt {y},\qquad  &x=u ,\qquad y= v^2. 
\end{aligned}
\end{equation}
By writting vector fields (\ref{da}) in the new variables, we obtain
\begin{equation}
X_1= \frac{\partial}{\partial u},\qquad X_2= u\frac{\partial}{\partial u}+\frac 12 v \frac{\partial}{\partial v} ,\qquad X_3= u^2\frac{\partial}{\partial u}+uv\frac{\partial}{\partial v} ,
\label{db} 
\end{equation}
which are those   appearing in the imprimitive Lie algebra I$_5\simeq \mathfrak{sl}(2)$ of  Table \ref{table1} and Table \ref{table3}  and so
we get\begin{equation}\label{dc}
\omega=\frac{\dd u\wedge \dd v}{v^3} ,\qquad h_1=-\frac{1}{2v^{2}},\qquad  h_2= -\frac{u}{2v^{2}},\qquad  h_3=-\frac{u^2}{2 v^{2}  }.
\end{equation}
These Hamiltonian functions satisfy the commutation relations (\ref{sl2Rh}). Consequently, 
\noindent

$h_1$,$h_2$,$h_3$ span a Lie algebra isomorphic to $\mathfrak{sl}(2)$.
\noindent

Finally we write  the expressions (\ref{dc})    in terms of the initial variables $\{x,y\}$
\begin{equation}
\omega=\frac{{\rm d} x \wedge {\rm d} y}{ 2 y^2}  ,\qquad  h_1=-\frac{1}{2 y} ,\qquad
h_2= - \frac{x}{2 y}  ,\qquad
h_3=-\frac{x^2}{2 y} ,
\label{dee}
\end{equation}
which fulfill the commutation relations (\ref{sl2Rh}).  Therefore, the system (\ref{CKRE4}) is, once more,  an $\mathfrak{sl}(2)$-Lie-Hamilton system possessing a Vessiot--Guldberg Lie algebra diffeomorphic to I$_5$ and a LH algebra $(\langle h_1,h_2,h_3\rangle,\{\cdot,\cdot\}_\omega) \simeq \mathfrak{sl}(2)$.

 We conclude that the Cayley--Klein  Riccati equation  (\ref{CKRE}) comprises in a unified way the {\em three} nondiffeomorphic  Vessiot--Guldberg Lie algebras of Hamiltonian vector fields isomorphic to  $\mathfrak{sl}(2)$: P$_2$, I$_4$ and I$_5$.

\end{example}

\subsubsection{Other $\mathfrak{sl}(2)$-Lie--Hamilton systems and `equivalence'}

In this section we present some $\mathfrak{sl}(2)$-Lie--Hamilton systems of mathematical and physical  interest. To the keep notation simple, we say that a second-order differential equation is a Lie system (resp.~LH system) when the first-order system obtained from it by adding a new variable $y:={\rm d}x/{\rm d}t$, is a Lie system (resp.~Lie--Hamilton system). 

We here study the coupled Riccati equation, Milne--Pinney and second-order Kummer--Schwarz equations, certain Lie systems appearing in the study of diffusion equations, the Smorodinsky--Winternitz system and the   harmonic oscillator, both  with     a $t$-dependent frequency. 
Next we establish, according to Table~\ref{table1} in Appendix 1, the equivalence among them and also among the three cases covered by the Cayley--Klein Riccati equation.
More precisely, we establish which of all of the above systems are locally diffeomorphic.

\begin{example}\normalfont

Consider the system of {\bf coupled Riccati equations} given in \eqref{CoupledRic} for $n=2$ \cite{Mariton}, with $x_1=x$ and $x_2=y$.
\begin{equation}
\frac{{\rm d}x}{{\rm d}t}=a_0(t)+a_1(t)x+a_2(t)x^2,\qquad \frac{{\rm d}y}{{\rm d}t}=a_0(t)+a_1(t)y+a_2(t)y^2,
\label{cR1}
\end{equation}
This system can be expressed as a $t$-dependent vector field (\ref{tdepenRiccati21})
where
\begin{equation}
X_1= \frac{\partial}{\partial x}+ \frac{\partial}{\partial y},  \qquad X_2= x\frac{\partial}{\partial x}+y\frac{\partial}{\partial y} ,\qquad X_3= x^2\frac{\partial}{\partial x}+y^2\frac{\partial}{\partial y} ,
\end{equation}
so that these  vector fields  exactly reproduce those given in   Table~\ref{table1}  for the   class I$_4\simeq \mathfrak{sl}(2)$ which, in turn, means that this system is
locally diffeomorphic to the   split-complex Riccati equation (\ref{CKRE3}).

\end{example}

\begin{example}\normalfont
The {\bf Milne--Pinney equation} \cite{AL08} was shown in \eqref{FirstLie1} as a first-order system
\begin{equation}\label{FirstLie1again}
\left\{
\begin{aligned}
\frac{\dd x}{\dd t}&=y,\\
\frac{\dd y}{\dd t}&=-\omega^2(t)x+\frac{c}{x^3},
\end{aligned}\right.
\end{equation}

It has an associated  $t$-dependent vector field
$
X=X_3+\omega^2(t)X_1,
$
with vector fields

\begin{equation}\label{FirstLieA2}
X_1=-x\frac{\partial}{\partial y},\qquad X_2=\frac 12 \left(y\frac{\partial}{\partial y}-x\frac{\partial}{\partial x}\right),\qquad X_3=y\frac{\partial}{\partial x}+\frac{c}{x^3}\frac{\partial}{\partial y},
\end{equation}

The vector fields $X_1,X_2,X_3$
span a Lie algebra isomorphic to $\mathfrak{sl}(2)$ with Lie brackets given by (\ref{FirstLieB}).



\begin{proposition} The system  (\ref{FirstLie1})  is a Lie--Hamilton system of class    {\rm P}$_2$ for $c>0$, {\rm I}$_4$ for $c<0$ and {\rm I}$_5$ for $c=0$ given in  Table~\ref{table1} in Appendix 1.
\end{proposition}

Therefore, as the Cayley--Klein Riccati equation (\ref{CKRE2}), the Milne--Pinney equations include the three possibilities of Vessiot--Guldberg Lie algebras isomorphic to $\mathfrak{sl}(2)$ of Hamiltonian vector fields.

\end{example}

\begin{example}\normalfont

Reconsider the {\bf second-order Kummer--Schwarz equation} studied in~\cite{CGL11} and it was earlier shown in \eqref{KS22}, which was rewritten in
terms of a first-order system \eqref{FirstOrderKummer},

\begin{equation}\label{FirstOrderKummer3}
\left\{\begin{aligned}
\frac{dx}{dt}&=v,\\
\frac{dv}{dt}&=\frac 32 \frac{v^2}x-2c_0x^3+2b_1(t)x,
\end{aligned}\right.
\end{equation}
on ${\rm T}\mathbb{R}_0$, with
$\mathbb{R}_0=\mathbb{R}-\{0\}$, obtained by adding the new variable
$v\equiv dx/dt$ to the KS-2 equation (\ref{KS22}).

This system has an associated $t$-dependent vector field $M=M_3+\eta(t)M_1,$ where the vector fields correspond with
\begin{equation}\label{VFKS22}
\begin{gathered}
M_1=2x\frac{\partial}{\partial v},\qquad M_2=x\frac{\partial}{\partial
x}+2v\frac{\partial }{\partial v},\qquad M_3=v\frac{\partial}{\partial
x}+\left(\frac 32\frac{v^2}x-2c_0x^3\right)\frac{\partial}{\partial v}
\end{gathered}
\end{equation}
satisfy the commutation relations
\begin{equation}
[M_1,M_3]=2M_2,\quad [M_1,M_2]=M_1,\quad [M_2,M_3]=M_3.
\end{equation}
that span a Lie algebra isomorphic to $\mathfrak{sl}(2)$.
It can be proven that (\ref{FirstOrderKummer3})  comprises, once more, the three Vessiot--Guldberg Lie algebras of Hamiltonian vector fields isomorphic to $\mathfrak{sl}(2)$ given in  Table~\ref{table1} in Appendix 1 according to  the value of the parameter $c$~\cite{BBHLS}.

\begin{proposition} The system  (\ref{FirstOrderKummer3}) is a LH system of class  {\rm P}$_2$ for $c>0$, {\rm I}$_4$ for $c<0$ and {\rm I}$_5$ for $c=0$ given in  Table~\ref{table1} in Appendix 1.
\end{proposition}

\end{example}

\begin{example}\normalfont

A diffusion equation can be transformed into a simpler PDE by solving a system of seven first-order ordinary differential equations (see \cite{SSVessiot--Guldberg11} and \cite[p.~104]{SSVessiot--Guldberg14} for details). This system can be easily solved by integrating its projection to $\mathbb{R}^2$,
known as the {\bf planar diffusion Riccati system} given by
\begin{equation}\label{diff}
\frac{{\rm d}x}{{\rm d}t}=-b(t)+2c(t)x+4a(t)x^2+a(t)c_0y^4,\qquad \frac{{\rm d}y}{{\rm d}t}=\bigl(c(t)+4a(t)x\bigr)y,
\end{equation}
where $a(t),b(t)$ and $c(t)$ are arbitrary $t$-dependent functions and $c_0\in \{0,1\}$. We call this system  planar diffusion Riccati system. This system is related to the $t$-dependent vector field
\begin{equation}
X=a(t)X_3-b(t)X_1+c(t)X_2,
\end{equation}
where
\begin{equation}\label{PlanarDif}
X_1=\frac{\partial}{\partial x},\qquad X_2=2x\frac{\partial}{\partial x}+y\frac{\partial}{\partial y}, \qquad X_3=(4x^2+c_0y^4)\frac{\partial}{\partial x}+4xy\frac{\partial}{\partial y},
\end{equation}
satisfy the commutation relations
\begin{equation}
[X_1,X_2]=2X_1,\qquad [X_1,X_3]=4X_2,\qquad [X_2,X_3]=2X_3.
\end{equation}
Consequently, they span a Lie algebra isomorphic to $\mathfrak{sl}(2)$. For $c_0=1$ the change of variables
\begin{equation}
u=2x+y^2,\qquad v=2x-y^2,\qquad x=\tfrac 14( u+v),\qquad y=\sqrt{(u-v)/2}
\end{equation}
maps these vector fields to a basis of I$_4$ (see Table \ref{table1}). Writing the symplectic structure and the Hamiltonian functions in the coordinate system $\{x,y\}$, we obtain
\begin{equation}
\omega=\frac{{\rm d}x\wedge {\rm d}y}{y^3},\qquad h_1=-\frac{1}{2y^2},\qquad h_2=-\frac{x}{y^2},\qquad h_3=\frac{-2x^2}{y^2}+\frac 12y^2,
\end{equation}
that satisfy
\begin{equation}
\{h_1,h_2\}_\omega=-2 h_1,\qquad \{h_1,h_3\}_\omega=-4 h_2,\qquad \{h_2,h_3\}_\omega=-2 h_3.
\end{equation}

For the case $c_0=0$, we have that the vector fields (\ref{PlanarDif}) form a  basis of I$_5$ (see (\ref{db})). Hence, their associated symplectic form and some corresponding Hamiltonian functions can easily be obtained from Table \ref{table1} in Appendix 1.
\end{example}

\subsubsection{Equivalence among  $\mathfrak{sl}(2)$-Lie-Hamilton systems}

Consequently, by taking into account all the above results, we are led to the following statement.

\begin{theorem}\label{Main} The $\mathfrak{sl}(2)$-Lie-Hamilton systems  (\ref{FirstOrderKummer}), (\ref{CoupledRic}), (\ref{FirstLie1}), (\ref{CKRE2}) and  (\ref{diff}) are equivalent through local diffeomorphisms whenever their Vessiot--Guldberg Lie algebras belong to the same class in  Table~\ref{table1}, that is,

\begin{itemize}

\item P$_2$:  Milne--Pinney  and Kummer--Schwarz   equations for $c>0$ as well as complex Riccati equations.

\item I$_4$:  Milne--Pinney  and Kummer--Schwarz   equations for $c<0$, coupled Riccati equations, split-complex Riccati equations and the planar diffusion Riccati system with $c_0=1$.

\item I$_5$:  Milne--Pinney  and Kummer--Schwarz   equations for $c=0$ as well as dual-Study  Riccati equations, planar diffusion Riccati system with $c_0=0$ and the harmonic oscillator with $t$-dependent frequency.

\end{itemize}

\noindent
All of the above systems are considered to have t-dependent coefficients.

\end{theorem}

Only within each class,  these  systems  are locally diffeomorphic  and, therefore,  there   exists a
local $t$-independent change of variables mapping  one into another.  For instance, there does not exist any diffeomorphism on $\mathbb{R}^2$ mapping the Milne--Pinney  and Kummer--Schwarz equations with $c\ne 0$  to the harmonic oscillator with a $t$-dependent frequency as the latter  correspond to set  $c=0$ and belong to class  I$_5$. Moreover, these results also  explain   the existence of the known diffeomorphism mapping   Kummer--Schwarz equations to Milne--Pinney equations, which from our approach, should be understood in a `unified map'~\cite{AL08}, that is, the value of  the parameter $c$ should be considered and everything works whenever  the same sign of  $c$ is preserved.

\subsection{Two-photon Lie--Hamilton systems}

In this section, we study two different Lie--Hamilton systems that belong to the same class  P$_5$  in  Table~\ref{table1} in Appendix 1; these are a dissipative harmonic oscillator and the second-order Riccati equation in Hamiltonian form. As a consequence, 
there exists a diffeomorphism mapping one into the other when they are written as first-order systems.

The five generators $X_1,\dots,X_5$ written in Table~\ref{table1} in Appendix 1 satisfy the Lie brackets
\begin{equation}
\begin{aligned}
&[X_1,X_2] = 0, &&
 [X_1,X_3] =X_1,&&
 [X_1,X_4] =0,&&
 [X_1,X_5] =X_2,
\\
&[X_2,X_3] =-X_2,&&
[X_2,X_4] =X_1,&&
 [X_2,X_5] =0,&&
[X_3,X_4] =-2X_4, 
\\
&[X_3,X_5] =2X_5,&&
[X_4,X_5] =-X_3, &&
   &&
\end{aligned}
\label{uf}\end{equation} 
in such a manner that they span a Lie algebra isomorphic to $\mathfrak{sl}(2 )\ltimes \mathbb{R}^2$ where $\mathbb{R}^2=\langle X_1,X_2 \rangle$ and $\mathfrak{sl}(2 )=\langle X_3,X_4,X_5 \rangle$.  Such vector fields are Hamiltonian with respect  to the canonical symplectic form $\omega=\dd x\wedge \dd y$. Nevertheless,  the corresponding Hamiltonian functions must be enlarged with a central generator $h_0=1$ giving rise to the centrally extended Lie algebra $\overline{\mathfrak{sl}(2 )\ltimes \mathbb{R}^2}$ which is, in fact, isomorphic to the two-photon Lie algebra $\mathfrak{h}_6=\langle h_1,h_2,h_3,h_4,h_5,h_0 \rangle$~\cite{BBF09,Gilmore}. That is why we shall call these systems {\it two-photon Lie--Hamilton systems}. The Lie brackets  between these functions read
\be
\begin{aligned}
&\{h_1,h_2\}_\omega=h_0, &&
 \{h_1,h_3\}_\omega=-h_1,&&
 \{h_1,h_4\}_\omega=0,&& \{h_1,h_5\}_\omega=-h_2,
\\
&\{h_2,h_3\}_\omega=h_2,&&
\{h_2,h_4\}_\omega=-h_1,&&
 \{h_2,h_5\}_\omega=0,&&
\{h_3,h_4\}_\omega=2h_4,
 \\
&\{h_3,h_5\}_\omega=-2h_5,&&
\{h_4,h_5\}_\omega=h_3, &&
 \{h_0,\cdot\}_\omega=0 . &&
\end{aligned}
\label{ug}
\ee
Notice that  $\mathfrak{h}_6\simeq \mathfrak{sl}(2)
\ltimes \mathfrak{h}_3$, where $\mathfrak{h}_3\simeq  \langle h_0,h_1,h_2\rangle$ is the Heisenberg--Weyl Lie algebra and   $\mathfrak{sl}(2 ) \simeq \langle h_3,h_4,h_5\rangle $. Since  $\mathfrak{h}_4\simeq  \langle h_0,h_1,h_2,h_3\rangle$ is the harmonic oscillator Lie algebra (isomorphic to 
  $  {\overline {\mathfrak{iso}}}(1,1)$ in the class I$_8$), we have the embeddings
$\mathfrak{h}_3\subset \mathfrak{h}_4\subset \mathfrak{h}_6$.

\begin{example}\normalfont
The $t$-dependent Hamiltonian for the {\bf dissipative harmonic oscillator} studied in~\cite{CR}  is given by
\begin{equation}\label{disharoscsys}
h(t,q,p)=\alpha(t)\, \frac{p^2}{2}+\beta(t)\, \frac{pq}{2}+\gamma(t)\, \frac{q^2}{2}+\delta(t)p+\epsilon(t)q+\phi(t),
\end{equation}
where $\alpha(t),\beta(t),\gamma(t),\delta(t),\epsilon(t),\phi(t)$ are real $t$-dependent functions.
The corresponding  Hamilton equations read
\begin{equation}
\begin{aligned}\label{disharoscsys2}
\frac{{\rm d}q}{{\rm d}t}&=\frac{\partial h}{\partial p}=\alpha(t)\, p+\frac{\beta(t)}{2}\, q+\delta(t),\\
\frac{{\rm d}p}{{\rm d}t}&=-\frac{\partial h}{\partial q}=-\left( \beta(t)\, \frac{p}{2}+\gamma(t)q+\epsilon(t)\right) .
\end{aligned}\end{equation} 
This system has an associated $t$-dependent vector field
\begin{equation}
X=\delta(t)X_1-\epsilon(t)X_2+ \frac{\beta(t)}{2}X_3+\alpha(t)X_4-\gamma(t)X_5
\end{equation}
where
\begin{equation}
\begin{gathered}
 X_1=\frac{\partial}{\partial q},\qquad X_2=\frac{\partial}{\partial p},\qquad
 X_3=q\frac{\partial}{\partial q}-p\frac{\partial}{\partial p},
\\
X_4=p\frac{\partial}{\partial q}, \qquad X_5=q\frac{\partial}{\partial p},
\end{gathered}\end{equation} 
are, up to a trivial change of variables $x=q$ and $y=p$, the vector fields of the basis of P$_5$ given in Table \ref{table1} in Appendix 1. Hence, their Hamiltonian functions with respect to the symplectic structure $\omega=\dd q\wedge \dd p$ are indicated in Table \ref{table1} in Appendix 1. 
\end{example}

\begin{example}\normalfont
We reconsider the family of {\bf second-order Riccati equations} \eqref{NLe2} which arose by reducing third-order linear differential equations through a dilation symmetry and a $t$-reparametrization \cite{CRS05}.
The crucial point is that a quite general family of second-order Riccati equations (\ref{NLe2}) admits  a   $t$-dependent Hamiltonian (see~\cite{CRS05,LSKummer} for details) given by
\begin{equation}
 h(t,x,p)= -2\sqrt{-p}- p\left(a_0(t)+a_1(t)x+a_2(t)x^2 \right) ,\qquad p<0,
\end{equation}
where  $a_0(t),a_1(t),a_2(t)$ are certain functions related to the
$t$-dependent coefficients of (\ref{NLe2}). The corresponding Hamilton equations are \eqref{Hamil12}
and the associated $t$-dependent vector field has the expression
\begin{equation}
X=X_1-a_0(t)X_2-a_1(t)X_3-a_2(t)X_4, \label{F2}
\end{equation}
such these vector fields correspond with \eqref{commrelsecordric} and close the commutation relations \eqref{ComRel21}.

If we set
\begin{equation}
\begin{gathered}
Y_1= -\tfrac 1{\sqrt{2} } X_5,\qquad  Y_2=\tfrac 1{\sqrt{2} }  X_1,\qquad  Y_3=-2 X_3,\\  Y_4=-X_4,
\qquad
  Y_5= X_2,
\end{gathered}
\end{equation}
we find that the commutation relations for $Y_1,\ldots,Y_5$ coincide with (\ref{uf})
and therefore span a  Lie algebra isomorphic to $ \mathfrak{sl}(2 )\ltimes \mathbb{R}^2$. Indeed, it is a Lie--Hamilton system endowed with a canonical  symplectic form $\omega=\dd x\wedge \dd p$. The   Hamiltonian functions corresponding to the vector fields \eqref{commrelsecordric}
 turn out to be
\begin{equation}
\begin{gathered}\label{equFun}
\tilde h_1=-2\sqrt{-p}, \qquad  \tilde h_2=p,\qquad \tilde h_3=xp,
\\
\tilde h_4=x^2p,\qquad \tilde h_5=-2x\sqrt{-p},
\end{gathered}\end{equation} 
which span along with $\tilde h_0=1$ a   Lie algebra of functions  isomorphic to the two-photon Lie algebra $\mathfrak{h}_6$~\cite{BCHLS,CLS122} with non-vanishing Poisson brackets given by
\begin{equation}
\begin{aligned}
 & \{\tilde h_1,\tilde h_3\}_\omega =-\frac  12 \tilde h_1 , &&
 \{\tilde h_1,\tilde h_4\}_\omega =-\tilde h_5,& &
 \{\tilde h_1,\tilde h_5\}_\omega =2\tilde h_0,
\\ &
\{\tilde h_2,\tilde h_3\}_\omega =-\tilde h_2, &&
  \{\tilde h_2,\tilde h_4\}_\omega =-2\tilde h_3,&&
 \{\tilde h_2,\tilde h_5\}_\omega =-\tilde h_1, \\&
 \{\tilde h_3,\tilde h_4\}_\omega =-\tilde h_4, &&
\{\tilde h_3,\tilde h_5\}_\omega =-\frac 12 \tilde h_5  .&&
\end{aligned}
\label{ComRel2}
\end{equation} 
Under the identification

\begin{equation}\begin{gathered}
h_1= -\tfrac 1{\sqrt{2} }\tilde h_5,\qquad  h_2= \tfrac 1{\sqrt{2} } \tilde h_1,\qquad  h_3= -2 \tilde h_3,\qquad  h_4= - \tilde h_4,\\ h_5= \tilde h_2,\qquad  h_0= \tilde h_0 ,\end{gathered}\end{equation} 

we recover the Poisson brackets  (\ref{ug}). Furthermore, the   expressions for the vector fields and their Hamiltonian functions written in Table~\ref{table1} in Appendix 1 for the class  P$_5$, in variables $\{u,v\}$,  can be obtained from $X_1,\ldots,X_5$ by applying the change of variables defined by
\bea
&&u= \sqrt{-2p},\quad\  v=x  \sqrt{-2p},\quad\ x=\frac{v}{u} ,\qquad   p=-\frac 12\, u^2 ,\eea
such that $\omega=\dd x\wedge\dd p :=  \dd u\wedge \dd v$.
Notice that $p/\sqrt{-p}=-\sqrt{-p}$  since $p<0$.

The main results of this section are then summarized as follows.

\begin{proposition} The dissipative harmonic oscillator  (\ref{disharoscsys})
and the Hamilton equations  (\ref{Hamil12}) corresponding to the  second-order Riccati equation (\ref{NLe2}) are {\rm P}$_5$-Lie--Hamilton systems with Lie--Hamilton algebra isomorphic to the two-photon one $\mathfrak{h}_6$. Consequently, all are locally diffeomorphic.
\end{proposition}
\end{example}

\subsection{$\mathfrak{h}_2$-Lie--Hamilton systems}

Let us study class I$_{14A}\simeq \mathbb{R} \ltimes \mathbb{R}^r$ with $r=1$ of Table~\ref{table1} in Appendix 1. It admits a basis of vector fields
 $X_1= {\partial_x}$ and $X_2= {\eta_1(x)\partial_y} $. If we require that these two vector fields close a non-abelian Lie algebra, we choose $[X_1,X_2]=X_2$, with no loss of generality, then
 $\eta_1(x)={\rm e}^x $, that is
 \be
 X_1= \frac{\partial}{\partial x},\qquad X_2={\rm e}^x \frac{\partial}{\partial y},\qquad [X_1,X_2]=X_2,
 \ee
that we denote   by ${\mathfrak h}_2\simeq \mathbb{R} \ltimes \mathbb{R}$. This is a Vessiot--Guldberg Lie algebra of Hamiltonian vector fields when endowed with the canonical symplectic form
$\omega=\dd x\wedge \dd y$. Hence, we can choose
\be
h_1= y,\qquad h_2=-{\rm e}^x,\qquad \{h_1,h_2\}_\omega=-h_2.
\ee

In the following, we prove that the complex Bernoulli equation with $t$-dependent real coefficients has an underlying ${\mathfrak h}_2$-Lie algebra.
Furthermore,  we relate this fact with the generalized Buchdahl equations and $t$-dependent Lotka--Volterra systems. It is remarkable that all particular cases of Cayley--Klein Riccati equations (\ref{CKRE2})  with $a_2(t)=0$ are $\mathfrak{h}_2$-Lie--Hamilton systems. Finally, we prove that all of these systems are locally diffeomorphic and belong to the same class  I$_{14A}^{r=1}\simeq {\mathfrak h}_2$.

\begin{example}\normalfont

The {\bf complex Bernoulli equation} \cite{Ma11} with  $t$-dependent real coefficients  takes the form
\begin{equation}\label{complexBernoulli}
 \frac{\dd z}{\dd t}=a_1(t)z+a_2(t)z^n,\qquad n\notin\{0,1\},
\end{equation}
where $z\in \mathbb{C}$ and $a_1(t),a_2(t)$ are arbitrary t-dependent real functions. This is a generalization to the complex numbers
of the usual Bernoulli equation.
By writing $z:=x+iy$, equation (\ref{complexBernoulli}) reads
\begin{align}
 &\frac{\dd x}{\dd t}=a_1(t)x+a_2(t)\sum_{\stackrel{k=0}{k\,{\rm even}}}^n\left( \begin{array}{c}n\\k\end{array}\right) x^{n-k}i^ky^k,\\
&\frac{\dd y}{\dd t}=a_1(t)y+a_2(t)\sum_{\stackrel{k=1}{k\,{\rm odd}}}^n\left( \begin{array}{c}n\\k\end{array}\right) x^{n-k}i^{k-1}y^k.
\label{va}
\end{align}
This can be studied in terms of the $t$-dependent vector field $X=a_1(t)X_1+a_2(t)X_2$,
where
\begin{equation}
\begin{gathered} \label{vb}
 X_1=x\frac{\partial}{\partial x}+y\frac{\partial}{\partial y},\\
 X_2=\sum_{\stackrel{k=0}{k\,{\rm even}}}^n\left( \begin{array}{c}n\\k\end{array}\right)x^{n-k}i^ky^k\frac{\partial}{\partial x}+\sum_{\stackrel{k=1}{k\,{\rm odd}}}^n\left(\begin{array}{c}n\\k\end{array}\right)x^{n-k}i^{k-1}y^k\frac{\partial}{\partial y}
\end{gathered}
\end{equation}
and their commutation relation is
\begin{equation}
[X_1,X_2]=(n-1)X_2,
\end{equation}
which  is isomorphic to   $\mathfrak{h}_2$. In the GKO classification~\cite{BBHLS,GKP92} there is just one Lie algebra isomorphic to $\mathfrak{h}_2$ whose vector fields are not proportional at each point: I$_{14A}$ with $r=1$. So,  $\langle X_1,X_2\rangle$ is a Lie algebra of Hamiltonian vector fields in view of the results of Table \ref{table3} in Appendix 1.

Let us inspect the case $n=2$. So, the vector fields in (\ref{vb}) read $X_1=x\partial_x+y\partial_y$ and
$X_2= (x^2-y^2)\partial_x+ 2xy \partial_y$, and become Hamiltonian with respect to the symplectic form and with Hamiltonian functions
\begin{equation}
\omega=\frac{\dd x\wedge \dd y}{y^2} ,\qquad h_1=-\frac xy,\qquad  h_2=-\frac{x^2+y^2}{y}.
\end{equation}
Additionally, $\{h_1,h_2\}_\omega=-h_2.$
Notice that, in this very particular case,  the Hamiltonian functions  become in the form given in Table~\ref{table3} for the {\it positive Borel subalgebra} of P$_2\simeq \mathfrak{sl}(2) $. In fact, for $n=2$  the equation (\ref{complexBernoulli}) is a particular case of the  complex Riccati equation (\ref{Riccati21}) with $a_0=0$.

Let us now take a closer look at $n=3$. So, the vector field $X_2$ read
\begin{equation}
X_2= \left( x^3 - 3 x y^2\right)\frac{\partial}{\partial x}+ \left( 3  x^2 y -y^3 \right)\frac{\partial}{\partial y}
\end{equation}
 and it can be shown that $X_1$ and $X_2$ become Hamiltonian vector fields with respect to the symplectic form
\begin{equation}
\omega=\frac{x^2+y^2}{x^2y^2}\, \dd x\wedge\dd y.
\end{equation}
Moreover, $X_1$ and $X_2$ admit Hamiltonian functions
\begin{equation}
h_1=\frac{y^2-x^2}{xy},\qquad h_2=-\frac{(x^2+y^2)^2}{xy} ,
\end{equation}
satisfying $\{ h_1,h_2\}_\omega=-2 h_2$. Other new results come out by setting $n>3$.
\end{example}

\begin{example}\normalfont

The {\bf generalized Buchdahl equations} ~\cite{Bu64, CSL05, CN10} are the  second-order differential equations given by \eqref{Bucheq}
which can be rewritten in terms of a first-order system \eqref{Buchdahl1},

\begin{equation}\label{Buchdahl12}
\left\{ \begin{aligned}
 \frac{\dd x}{\dd t}&=y,\\
 \frac{\dd y}{\dd t}&=a(x)y^2+b(t)y .
 \end{aligned}\right.
 \end{equation}
 such that $X=X_1+b(t)X_2$, where the vector
fields corresponded with \eqref{VectBuch}.

These vector fields span a Lie algebra diffeomorphic to  I$_{14A}^{r=1}\simeq {\mathfrak h}_2$ so that (\ref{Buchdahl12}) is a Lie--Hamilton system. The corresponding symplectic form and Hamiltonian functions can be found in~\cite{BBHLS}.
\end{example}

\begin{example}\normalfont
Finally, consider the {\bf particular Lotka--Volterra systems}~\cite{JHL05,Tr96}  given by
  \begin{equation}\label{LV}
 \frac{\dd x}{\dd t}=ax-g(t)(x-ay)x,\qquad
 \frac{\dd y}{\dd t}=ay-g(t)(bx-y)y,\qquad a\ne 0,
\end{equation}
 where $g(t)$ determines the variation of the seasons, while  $a$ and $b$ are constants describing the interactions among the species.
System (\ref{LV}) is associated with  the $t$-dependent vector field $
X=X_1+g(t)X_2$
 where
\begin{equation}
 X_1=ax\frac{\partial}{\partial x}+ay\frac{\partial}{\partial y},\qquad X_2=-(x-ay)x\frac{\partial}{\partial x}-(bx-y)y\frac{\partial}{\partial y} ,
\end{equation}
satisfy
\begin{equation}
[X_1,X_2]=a X_2,\qquad a\ne 0.
\end{equation}
Hence, (\ref{LV}) is a Lie system. Moreover, it has been proven in~\cite{BBHLS} that, except for the case with $a=b=1$, this is also a LH system belonging to the family  I$_{14A}^{r=1}\simeq {\mathfrak h}_2$.

Hence, we conclude this section with the following statement.

\begin{proposition} The complex Bernoulli equation (\ref{va}), the generalized Buchdahl equations (\ref{Buchdahl12}) and the
$t$-dependent Lotka--Volterra systems (\ref{LV}) (with the exception of $a=b=1$) are   LH systems with a Vessiot--Guldberg Lie algebra diffeomorphic to
 I$_{14A}^{r=1}\simeq {\mathfrak h}_2$   in  Table~\ref{table1} in Appendix 1. Thus all of these systems	 are locally diffeomorphic.
\end{proposition}

\end{example}

\section{Dirac--Lie systems}
\setcounter{equation}{0}
\setcounter{theorem}{0}
\setcounter{example}{0}

A step further on the study of {\it structure-Lie systems} is the consideration of Dirac--Lie systems. The introduction of this type of system presents some
advantages over different approaches to certain physical systems. In the following paragraph, the importance of Dirac--Lie systems will be clearly stated.

\subsection{Motivation for Dirac--Lie systems}

Dirac--Lie systems arise due to the existence of certain Lie systems which do not have a Vessiot--Guldberg Lie algebra of Hamiltonian vector
fields with respect to a Poisson structure, but they do with respect to a presymplectic one.
A possible structure incorporating the presymplectic and Poisson structure is the Dirac structure. We will illustrate this fact
with the following example.
\begin{example}\normalfont
Let us reconsider the {\bf third-order Kummer--Schwarz equation} \cite{Be88,GL12},
\begin{equation}\label{ks3nose}
\left\{\begin{aligned}
\frac{dx}{dt}&=v,\\
\frac{dv}{dt}&=a,\\
\frac{da}{dt}&=\frac 32 \frac{a^2}v-2c_0(x)v^3+2b_1(t)v,
\end{aligned}\right.
\end{equation}
on the open submanifold $\mathcal{O}_2=\{(x,v,a)\in {\rm T}^2\mathbb{R}\mid
v\neq 0\}$ of ${\rm T}^2\mathbb{R}\simeq \mathbb{R}^3.$
\noindent
Its associated $t$-dependent vector field is $X^{3KS}=N_3+b_1(t)N_1$. 

If $b_1(t)$ is not a constant, the Vessiot--Guldberg Lie algebra is isomorphic to $\mathfrak{sl}(2,\mathbb{R})$ on $\mathcal{O}_2=\{(x,v,a)\in{T}^2\mathbb{R}\mid v\neq 0\}$
made up with vector fields \eqref{VFKS1}.
Hence, $X^{3KS}$ is a Lie system. 

However, $X^{3KS}$ is not a Lie--Hamilton system when $b_1(t)$ is not a constant. Indeed, in this case $\mathcal{D}^{X^{3KS}}$ coincides with $T\mathcal{O}_2$ on $\mathcal{O}_2$.
 If $X^{3KS}$ were also a Lie--Hamilton system with respect to $(N,\Lambda)$, then $V^{X^{3KS}}$ would consist of Hamiltonian vector fields and the characteristic distribution associated with $\Lambda$ would have odd-dimensional rank on $\mathcal{O}_2$. This is impossible, as the local Hamiltonian vector fields of a Poisson manifold span a generalized distribution of even rank at each point. Our previous argument can easily be generalized to formulate the following `no-go' theorem.

\begin{theorem}\label{The:NoGo} {\bf (Lie--Hamilton no-go Theorem)} If $X$ is a Lie system on an odd-dimensional manifold $N$ satisfying that $\mathcal{D}^X_{x_0}=T_{x_0}N$ for a point $x_0$ in $N$, then $X$ is not a Lie--Hamilton system on $N$.
\end{theorem}

Note that from the properties of ${\rm r}^X$ it follows that, if $\mathcal{D}^X_{x_0}=T_{x_0}N$ for a point $x_0$, then $\mathcal{D}^X_{x}=T_{x}N$ for $x$ in an open neighborhood $U_{x_0}\ni x_0$. Hence, we can merely consider whether $X$ is a Lie--Hamilton system on $N\backslash U_{x_0}$.

Despite the previous negative results, system (\ref{ks3nose}) admits another interesting property:  we can endow the
manifold $\mathcal{O}_2$ with a presymplectic form $\omega_{3KS}$ in such a way that $V^{X^{3KS}}$ consists of Hamiltonian vector fields with respect to it. Indeed, by considering the equations $
\mathcal{L}_{N_1}\omega_{3KS}=\mathcal{L}_{N_2}\omega_{3KS}=\mathcal{L}_{N_3}\omega_{3KS}=0$ and $d\omega_{3KS}=0$, we can readily find the presymplectic form
\begin{equation}
\omega_{3KS}=\frac{dv\wedge da}{v^3}
\end{equation}
on $\mathcal{O}_2$. Additionally, we see that
\begin{equation}\label{3KSHamFun}
\iota_{N_1}\omega_{3KS}=d\left(\frac{2}{v}\right),\qquad \iota_{N_2}\omega_{3KS}=d\left(\frac{a}{v^2}\right), \qquad \iota_{N_3}\omega_{3KS}=d\left(\frac{a^2}{2v^3}+2c_0v\right).
\end{equation}
So, the system $X^{3KS}$ becomes a Lie system with a Vessiot--Guldberg Lie algebra of Hamiltonian vector fields
with respect to $\omega_{3KS}$. As seen later on, systems of this type can be studied through
appropriate generalizations of the methods employed to investigate Lie--Hamilton systems.
\end{example}

\begin{example}\normalfont

Another example of a Lie system which is not a Lie--Hamilton system but admits a Vessiot--Guldberg Lie algebra of Hamiltonian vector fields with respect to a presymplectic form is the {\bf Riccati system}
\begin{equation}\label{Partial1}
\left\{
\begin{aligned}
\frac{ds}{dt}&=-4a(t)u s-2d(t)s, &\frac{dx}{dt}&=(c(t)+4a(t)u)x+f(t)-2u g(t), \\
\frac{du}{dt}&=-b(t)+2c(t) u+4a(t)u^2, &\frac{dy}{dt}&=(2a(t)x-g(t))v,\\
\frac{dv}{dt}&=(c(t)+4a(t)u)v, &\frac{dz}{dt}&=a(t)x^2-g(t)x, \\
\frac{dw}{dt}&=a(t)v^2, &\\
\end{aligned}\right.
\end{equation}
where $a(t)$, $b(t),$  $c(t)$, $d(t)$, $f(t)$ and $g(t)$ are arbitrary $t$-dependent functions. The interest of this system is due to its use in solving diffusion-type equations, Burger's equations, and other PDEs \cite{SSVessiot--Guldberg11}.

Since every particular solution $(s(t),u(t),v(t),w(t),x(t)$, $y(t),z(t))$ of (\ref{Partial1}), with $v(t_0)=0$ ($s(t_0)=0$) for a certain $t_0\in\mathbb{R}$, satisfies
$v(t)=0$ ($s(t)=0$) for every $t$, we can restrict ourselves to analyzing system (\ref{Partial1}) on the submanifold $M=\{(s,u,v,w,x,y,z)\in\mathbb{R}^7\,|\,v\neq 0, s\neq 0\}$. This will simplify the application of our  techniques without omitting any relevant detail.

System (\ref{Partial1}) describes integral curves of the $t$-dependent vector field
\begin{equation}
X^{RS}_t=a(t)X_1-b(t)X_2+c(t)X_3-2d(t)X_4+f(t)X_5+g(t)X_6,
\end{equation}
where
\begin{equation}
\begin{gathered}
 X_1=-4us\frac{\partial}{\partial s}+4u^2\frac{\partial}{\partial u}+4uv\frac{\partial}{\partial v}+v^2\frac{\partial}{\partial w}+4ux\frac{\partial}{\partial x}+2xv\frac{\partial}{\partial y}+x^2\frac{\partial}{\partial z},\\  X_2=\frac{\partial}{\partial u},\quad X_3=2u\frac{\partial}{\partial u}+v\frac{\partial }{\partial  v}+x\frac{\partial}{\partial x}, \quad X_4=s\frac{\partial}{\partial s},\quad  X_5=\frac{\partial}{\partial x},\\ X_6=-2u\frac{\partial}{\partial x}-v\frac{\partial}{\partial y}-x\frac{\partial}{\partial z},\quad X_7=\frac{\partial}{\partial z}.
\end{gathered}
\end{equation}
Their commutation relations are
\begin{equation}
\begin{gathered}
\hspace*{-1.5em}\left[X_1,X_2\right]=4(X_4 -X_3), \quad [X_1,X_3]=-2X_1,\quad [X_1,X_5]=2X_6, \quad [X_1.X_6]=0,\\
[X_2,X_3]=2X_2, \quad [X_2,X_5]=0,\quad [X_2,X_6]=-2X_5, \\
 [X_3,X_5]=-X_5, \quad [X_3,X_6]=X_6,\\
[X_5,X_6]=-X_7, \\
\end{gathered}\end{equation} 
and $X_4$ and $X_7$ commute with all the vector fields.
Hence, system (\ref{Partial1}) is a Lie system associated with a Vessiot--Guldberg Lie algebra $V$ isomorphic to $(\mathfrak{sl}(2,\mathbb{R})\ltimes \mathfrak{h_2})\oplus \mathbb{R}$, where $\mathfrak{sl}(2,\mathbb{R})\simeq \langle X_1,X_2,X_4-X_3\rangle$, $\mathfrak{h_2}\simeq \langle X_5,X_6,X_7\rangle$ and $\mathbb{R}\simeq \langle X_4\rangle$.
It is worth noting that this new example of Lie system is one of the few Lie systems related to remarkable PDEs until now \cite{CGM07}.

Observe that (\ref{Partial1}) is not a Lie--Hamilton system when $V^{X^{RS}}=V$. In this case $\mathcal{D}^{X^{RS}}_p=T_pM$ for any $p\in M$ and, in view of Theorem \ref{The:NoGo} and the fact  that $\dim T_pM=7$, the system $X^{RS}$ is not a Lie--Hamilton system on $M$.

Nevertheless, we can look for a presymplectic form turning $X^{RS}$ into a Lie system with a Vessiot--Guldberg Lie algebra of Hamiltonian vector fields. Looking for a non-trivial solution of the system of equations $\mathcal{L}_{X_k}\omega_{RS}=0$, with $k=1,\ldots,7$, and $d\omega_{RS}=0$, one can find the presymplectic two-form
\begin{equation}
\omega_{RS}=-\frac{4wdu\wedge dw}{v^2}+\frac{dv\wedge dw}{v}+\frac{4w^2du\wedge dv}{v^3}.
\end{equation}
In addition, we can readily see that $d\omega_{RS}=0$ and $X_1,\ldots,X_r$ are Hamiltonian vector fields
\begin{equation}\begin{gathered}\label{HamPar}
\iota_{X_1}\omega_{RS}=d\left(4uw-\frac{8u^2w^2}{v^2}-\frac{v^2}{2}\right),\qquad \iota_{X_2}\omega_{RS}=-d\left(\frac{2w^2}{v^2}\right),
\\  \iota_{X_3}\omega_{RS}=d\left(w-\frac{4w^2u}{v^2}\right),\end{gathered}
\end{equation}
and $\iota_{X_k}\omega_{RS}=0$ for $k=4,\ldots,7$.
\end{example}

Apart from the above examples, other non Lie--Hamilton systems that admit a Vessiot--Guldberg Lie algebra of Hamiltonian vector fields with respect to a presymplectic form can be found in the study of certain reduced Ermakov systems \cite{HG04}, Wei--Norman equations for dissipative quantum oscillators \cite{Dissertationes},  and $\mathfrak{sl}(2,\mathbb{R})$-Lie systems \cite{Pi12}.

A straightforward generalization of the concept of a Lie--Hamilton system to Dirac manifolds would be a Lie system admitting a Vessiot--Guldberg Lie algebra $V$ of vector fields for which there exists a Dirac structure $L$ such that $V$ consists of $L$-Hamiltonian vector fields.
%
%

\subsection{Dirac--Lie Hamiltonians}

In view of Theorem \ref{HamLieSys} in Subsection 3.2, every Lie--Hamilton system admits a Lie--Hamiltonian. Since Dirac--Lie systems are  generalizations of these systems, it is
natural to
investigate whether Dirac--Lie systems admit an ous property.

\begin{definition}
  A {\it Dirac--Lie Hamiltonian structure} is a triple $(N, L, h)$, where $(N, L)$ stands
  for a Dirac manifold and $h$ represents a $t$-parametrized family of
  admissible functions $h_t : N \rightarrow \mathbb{R}$ such that
  Lie$(\{h_t \}_{t \in \mathbb{R}}, \{ \cdot, \cdot \}_L)$ is a
  finite-dimensional real Lie algebra.
  A $t$-dependent vector field $X$ is said to admit, to have or to possess a
  Dirac--Lie Hamiltonian $(N, L, h)$ if $X_t+dh_t \in \Gamma(L)$ for all $t
  \in \mathbb{R}$.
\end{definition}

\begin{note}
For simplicity, we hereafter call Dirac--Lie Hamiltonian structures Dirac--Lie Hamiltonians.
\end{note}

From the above definition, we see that system (\ref{ks3nose}) related to the third-order Kummer--Schwarz equations
possesses a Dirac--Lie Hamiltonian $(N, L^{\omega_{3KS}},h^{3KS})$ and system (\ref{Partial1}), used to analyze diffusion equations,
admits a Dirac--Lie Hamiltonian $(N,L^{\omega_{RS}},h^{RS})$.

Consider the third-order Kummer--Schwarz equation in first-order form (\ref{ks3nose}). Recall that $N_1$, $N_2$, and $N_3$ are Hamiltonian vector fields with respect to
the presymplectic manifold $(\mathcal{O}_2,\omega_{3KS})$.  It follows from relations (\ref{3KSHamFun}) that the vector fields $N_1$, $N_2,$ and $N_3$ have Hamiltonian
functions
\begin{equation}\label{Fun3KS}
h_1=-\frac{2}{v},\qquad h_2=-\frac{a}{v^2},\qquad h_3=-\frac{a^2}{2v^3}-2c_0v,
\end{equation}
respectively. Moreover,
\begin{equation}
\{ h_1, h_3 \}= 2 h_2, \quad \{h_1,
h_2 \} = h_1, \quad \{ h_2, h_3 \}=
h_3,
\end{equation}
where $\{\cdot,\cdot\}$ is the Poisson bracket on ${\rm Adm}(\mathcal{O}_2,\omega_{3KS})$  induced by $\omega_{3KS}$. In consequence, $h_1,h_2, $ and $h_3$ span a finite-dimensional real Lie algebra
isomorphic to $\mathfrak{sl}(2,\mathbb{R})$. Thus,
every $X^{3KS}_t$ is a Hamiltonian vector field with Hamiltonian function $
h^{3KS}_t = h_3 + b_1 (t) h_1$ and the space
${\rm Lie}(\{h^{3KS}_t\}_{t\in\mathbb{R}},\{\cdot,\cdot\})$ becomes a
finite-dimensional real Lie algebra. This enables us to associate $X^{3KS}$ to a curve in ${\rm Lie}(\{h^{3KS}_t\}_{t\in\mathbb{R}},\{\cdot,\cdot\})$. The similarity of  $(\mathcal{O}_2,\omega_{3KS},h^{3KS})$
with Lie--Hamiltonians are immediate.

If we now turn to the Riccati system (\ref{Partial1}), we will see that we can obtain a similar result. More specifically, relations
(\ref{HamPar}) imply that $X_1,\ldots,X_7$ have Hamiltonian functions
\begin{equation}
h_1=\frac{(v^2-4 u w)^2}{2v^2}, \qquad h_2=\frac{2\omega^2}{v^2},\qquad h_3=\frac{4w^2u}{v^2}-w,
\end{equation}
and $h_4=h_5=h_6=h_7=0$. Moreover, given the Poisson bracket on admissible functions induced by $\omega_{3KS}$, we see that
\begin{equation}
\{h_1,h_2\}=-4h_3,\qquad \{h_1,h_3\}=-2h_1,\qquad \{h_2,h_3\}=2h_2.
\end{equation}
Hence, $h_1,\ldots,h_7$ span a real Lie algebra isomorphic to $\mathfrak{sl}(2,\mathbb{R})$ and, as in the previous case, the $t$-dependent vector fields
$X^{RS}_t$ possess Hamiltonian functions $h^{RS}_t=a(t)h_1-b(t)h_2+c(t)h_3$. Again, we can associate $X^{RS}$ to a curve $t\mapsto h^{RS}_t$ in the finite-dimensional real Lie algebra $({\rm Lie}(\{h^{RS}_t\}_{t\in\mathbb{R}}),\{\cdot,\cdot\})$.

Let us analyze the properties of Dirac--Lie structures. Observe first that there may be several systems associated with the same Dirac--Lie Hamiltonian. For instance, the systems $X^{RS}$ and
\begin{multline}
X^{RS}_2=a(t)X_1-b(t)X_2+c(t)X_3-2d(t)X_4
\\
+f(t)z^3X_5+g(t)X_6+h(t) z^2 X_7\hspace*{2em}
\end{multline}
admit the same Dirac--Lie Hamiltonian  $(N, L^{\omega_{RS}},h^{RS})$. It is remarkable that $X_2^{RS}$ is not even a Lie system in general. Indeed, in view of
\begin{equation}
[z^2X_7,z^nX_5]=nz^{n+1}X_5,\qquad n=3,4,\ldots,
\end{equation}
we easily see that the successive Lie brackets of $z^nX_5$ and $z^2X_7$ span an infinite set of vector fields which are linearly independent over $\mathbb{R}$. So, in those cases in which $X_5$ and $X_7$ belong to $V^{X^{RS}_2}$, this Lie algebra becomes infinite-dimensional.

In the case of a Dirac--Lie system, Proposition \ref{CasCon} in Chapter \ref{Chap:GeomFund} shows easily the following.

\begin{corollary} Let $(N,L,X)$ be a Dirac--Lie system admitting a Dirac--Lie
Hamiltonian $(N,L,h)$. Then, we have
the exact sequence of Lie algebras
\begin{equation}
0\hookrightarrow {\rm
Cas}(\{h_t\}_{t\in\mathbb{R}},\{\cdot,\cdot\}_L)\hookrightarrow
{\rm
Lie}(\{h_t\}_{t\in\mathbb{R}},\{\cdot,\cdot\}_L)\stackrel{B_L}{\longrightarrow}
\pi(V^X)\rightarrow 0\,,
\end{equation}
where ${\rm
Cas}(\{h_t\}_{t\in\mathbb{R}},\{\cdot,\cdot\}_L)={\rm Lie}(\{h_t\}_{t\in\mathbb{R}},\{\cdot,\cdot\}_L)\cap {\rm
Cas}(N,L)$.
In other words, we have that ${\rm Lie}(\{h_t\}_{t\in\mathbb{R}},\{\cdot,\cdot\}_L)$ is a Lie
algebra extension of the space $\pi(V^X)$ by ${\rm
Cas}(\{h_t\}_{t\in\mathbb{R}},\{\cdot,\cdot\}_L)$.
\end{corollary}

\begin{theorem}\label{MT}
  Each Dirac--Lie system $(N,L,X)$ admits a Dirac--Lie Hamiltonian $(N,L,h)$.
\end{theorem}

\begin{proof}
  Since $V^X\subset {\rm Ham}(N,L)$ is a finite-dimensional Lie
algebra, we can define
a linear map $T : X_f
  \in V^X \mapsto f \in C^{\infty} (N)$ associating
  each $L$-Hamiltonian vector field in $V^X$ with an associated $L$-Hamiltonian function, e.g., given a basis $X_1,\ldots,X_r$ of $V^X$ we define $T(X_i)=h_i$, with
$i=1,\ldots,r$, and extend $T$ to $V^X$ by linearity. Note that the
  functions $h_1, \ldots, h_r$ need not be linearly independent over $\mathbb{R}$,
  as a function can be Hamiltonian for two different $L$-Hamiltonian vector fields $X_1$ and $X_2$ when $X_1-X_2\in {\rm G}(N,L)$. Given the system $X$,
there exists a smooth curve
  $h_t =T (X_t)$ in $\mathfrak{W}_0\equiv {\rm Im}\,T$ such that $X_t+dh_t \in \Gamma(L)$. To ensure
  that $h_t$ gives rise to a Dirac--Lie Hamiltonian, we need to
  demonstrate that $\dim\, {\rm Lie}(\{h_t \}_{t \in \mathbb{R}}, \{ \cdot,
  \cdot \}_L)<\infty$. This will be done by
constructing a
  finite-dimensional Lie algebra of functions containing the curve $h_t$.

  Consider two elements $Y_1,Y_2\in V^X$. Note that the functions $\{T (Y_1), T (Y_2)\}_L$ and $T ([Y_1, Y_2])$ have the
same
  $L$-Hamiltonian vector field. So, $\{T (Y_1), T (Y_2)\}_L - T ([Y_1,
  Y_2]) \in {\rm Cas}(N,L)$ and, in view of Proposition \ref{CasCon} from Chapter \ref{Chap:GeomFund}, it Poisson
commutes with all other admissible functions. Let us define $\Upsilon : V^X
\times V^X \rightarrow C^{\infty} (N)$
  of the form
  \begin{equation}    \label{formula2}
\Upsilon (X_1, X_2) = \{T(X_1), T(X_2)\}_L - T [X_1, X_2].
  \end{equation}
  The image of $\Upsilon$ is contained in a
  finite-dimensional real Abelian Lie subalgebra of Cas$(N,L)$ of the form
  \begin{equation}
 \mathfrak{W}_\mathcal{C} \equiv \langle \Upsilon (X_i, X_j) \rangle,
     \end{equation} 
 for $ i, j = 1, \ldots, r, $ and being $X_1, \ldots, X_r$ a basis for $V^X$. From here, it follows that
\begin{equation}
   \{ \mathfrak{W}_\mathcal{C}, \mathfrak{W}_\mathcal{C}\}_L = 0,
     \quad \{ \mathfrak{W}_\mathcal{C}, \mathfrak{W}_0 \}_L = 0,
     \quad \{ \mathfrak{W}_0, \mathfrak{W}_0 \}_L \subset
     \mathfrak{W}_\mathcal{C} + \mathfrak{W}_0 . \end{equation} 
  Hence, $(\mathfrak{W} \equiv \mathfrak{W}_0 + \mathfrak{W}_\mathcal{C},\{\cdot,\cdot\}_L)$ is a
  finite-dimensional real Lie algebra containing the curve $h_t$, and $X$ admits a Dirac--Lie Hamiltonian $(N, L, T(X_t))$.
\end{proof}
The following proposition is easy to verify.
\begin{proposition}\label{TanOrb} Let $(N,L,X)$ be a Dirac--Lie system. If $(N,L,h)$ and
$(N,L,\bar h)$ are two Dirac--Lie Hamiltonians for $(N,L,X)$, then
\begin{equation}
h=\bar h+f^X,
\end{equation}
where $f^X\in C^\infty(\mathbb{R}\times N)$ is a $t$-dependent function such
that each $f^X_t:x\in N \mapsto f^X(x,t)\in \mathbb{R}$ is a Casimir function that is constant on every
integral manifold $\mathcal{O}$ of $\mathcal{D}^X$.
\end{proposition}
Note that if we have a Dirac--Lie Hamiltonian $(N,L,h)$ and we define a linear map $\widehat T:  h\in {\rm Lie}(\{h_t\}_{t\in\mathbb{R}},\{\cdot,\cdot\})\mapsto X_h\in {\rm Ham}(N,L)$, the space $\widehat{T}({\rm Lie}(\{h_t\}_{t\in\mathbb{R}},\{\cdot,\cdot\})$ may span an infinite-dimensional Lie algebra of vector fields. For instance, consider again the Lie--Hamiltonian $(\mathcal{O}_2,\omega_{3KS},h^{3KS}_t=h_3+b_1(t)h_1)$ for the system (\ref{ks3nose}). The functions $h_1, h_2,$ and $h_3$ are also Hamiltonian for the vector fields
\begin{equation}\begin{gathered}
N_1=2v\frac{\partial}{\partial a}+e^{v^2}\frac{\partial}{\partial x},\quad N_2=v\frac{\partial}{\partial v}+2a\frac{\partial}{\partial a},\quad 
\\
N_3=a\frac{\partial }{\partial v}+\left(\frac 32 \frac{a^2}v-2c_0v^3\right)\frac{\partial}{\partial a},\end{gathered}
\end{equation}
which satisfy
\begin{equation}
\stackrel{j-{\rm times}}{\overbrace{[N_2,[\ldots,[N_2}},N_1]\ldots]]=f_j(v)\frac{\partial}{\partial x}+2(-1)^jv\frac{\partial}{\partial a}, \quad f_j(v)\equiv\,\, \stackrel{j-{\rm times}}{\overbrace{v\frac{\partial}{\partial v}\ldots v\frac{\partial}{\partial v}}}\!\!(e^{v^2}).
\end{equation}
In consequence, ${\rm Lie}(\widehat{T}({\rm Lie}(\{h_t\}_{t\in\mathbb{R}},\{\cdot,\cdot\})),[\cdot,\cdot])$ contains an infinite-dimensional Lie algebra of vector fields  because the functions $\{f_j\}_{j\in\mathbb{R}}$ form an infinite family of linearly independent functions over $\mathbb{R}$.  So, we need to impose additional conditions to ensure that the image of $\widehat{T}$ is finite-dimensional.

The following theorem yields an alternative definition of a Dirac--Lie system.

\begin{theorem} Given a Dirac manifold $(N,L)$, the triple $(N,L,X)$ is a
Dirac--Lie system if and only if there exists a curve
$\gamma:t\in\mathbb{R}\rightarrow \gamma_t\in\Gamma(L)$ satisfying that $\rho(\gamma_t)=X_t\in {\rm Ham}(N,L)$ for every $t\in\mathbb{R}$ and
${\rm Lie}(\{\gamma_t\}_{t\in\mathbb{R}},[[\cdot,\cdot]]_C)$ is a
finite-dimensional real Lie algebra.
\end{theorem}
\begin{proof} Let us prove the direct part of the theorem.
 Assume that $(N,L,X)$ is a Dirac--Lie system.
In virtue of Theorem \ref{MT}, it admits a Dirac--Lie Hamiltonian
$(N,L,h)$, with $h_t=T(X_t)$ and $T:V^X\rightarrow {\rm Adm}(N,L)$ a linear morphism associating
each element of $V^X$ with one of its $L$-Hamiltonian functions.
We aim to prove that  the curve in $\Gamma(L)$ of the form $\gamma_t=X_t+d(T(X_t))$ satisfies that $\dim {\rm Lie}(\{\gamma_t\}_{t\in\mathbb{R}},[[\cdot,\cdot]]_C)<\infty$.

The sections of $\Gamma(L)$ of the form
\begin{equation}\label{gen}
X_1+dT(X_1)\,\,,\ldots,\,\,X_r+dT(X_r)\,\,,\,\,d\Upsilon(X_i,X_j), 
\end{equation}
with $i,j=1,\ldots,r,$ $X_1,\ldots,X_r$ is a basis of $V^X$ and $\Upsilon:V^X\times V^X\rightarrow {\rm Cas}(N,L)$ is the map
 (\ref{formula2}), span a finite-dimensional Lie algebra $(E,[[\cdot,\cdot]]_C)$. Indeed,
\begin{equation}
[[X_i+dT(X_i),X_j+dT(X_j)]]_C=[X_i,X_j]+d\{T(X_i),T(X_j)\}_L,  
\end{equation}
for $i,j=1,\ldots,r$.
Taking into account that $\{T(X_i),T(X_j)\}_L-T([X_i,X_j])=\Upsilon(X_i,X_j)$, we see that the above is a linear combination of the
generators (\ref{gen}). Additionally, we have that
\begin{equation}
[[X_i+dT(X_i),d\Upsilon(X_j,X_k)]]_C=d\{T(X_i),\Upsilon(X_j,X_k)\}_L=0.
\end{equation}
So, sections (\ref{gen}) span a finite-dimensional subspace $E$ of  $(\Gamma(L),[[\cdot,\cdot]]_C)$. As $\gamma_t\in E$, for all $t\in\mathbb{R}$, we conclude the direct part of the proof.

The converse is straightforward from the fact that $(L,[[\cdot,\cdot]]_C,\rho)$ is a Lie algebroid. Indeed, given the curve $\gamma_t$ within a finite-dimensional real Lie algebra of sections $E$ satisfying that $X_t=\rho(\gamma_t)\in {\rm Ham}(N,L)$, we have that $\{X_t\}_{t\in\mathbb{R}}\subset \rho(E)$ are $L$-Hamiltonian vector fields. As $E$ is a finite-dimensional Lie algebra and $\rho$ is a Lie algebra morphism, $\rho(E)$ is a finite-dimensional Lie algebra of vector fields and $(N,L,X)$ becomes a Dirac--Lie system.
\end{proof}

The above theorem shows the interest of defining a class of Lie systems related to general Lie algebroids.

\subsection{Diagonal prolongations}

We analyze the properties of diagonal prolongations of Dirac--Lie systems. As a result, we discover new features that can be applied to study their superposition rules.
The process will be similar to the one described in Subsection 3.8 for prolongations and the coalgebra method.
Let $\tau:E\to N$ be a vector bundle. Its {\it diagonal prolongation} to $N^m$ is the {\it cartesian product bundle} $E^{[m]}=E\times\cdots\times E$ of $m$ copies of $E$, viewed as a vector bundle over $N^m$ in a natural way
 \begin{equation}E^{[m]}_{(x_{(1)},\dots,x_{(m)})}=E_{x_{(1)}}\oplus\cdots\oplus E_{x_{(m)}}\,.
\end{equation}
Every section $X:N\to E$ of $E$ has a natural {\it diagonal prolongation} to a section $X^{[m]}$ of $E^{[m]}$
\begin{equation}
X^{[m]}(x_{(1)},\dots,x_{(m)})=X(x_{(1)})+\cdots +X(x_{(m)})\,.
\end{equation}
Given a function $f:N\rightarrow \mathbb{R}$, we call {\it diagonal prolongation} of $f$ to $N^m$ the function $\widetilde{f}^{[m]}$ on $N^{m}$ of the form
$\widetilde{f}^{[m]}(x_{(1)},\ldots,x_{(m)})= f(x_{(1)})+\ldots+f(x_{(m)})$.

We can consider also sections $X^{(j)}$ of $E^{[m]}$ given by
\begin{equation}\label{prol1}
X^{(j)}(x_{(1)},\dots,x_{(m)})=0+\cdots +X(x_{(j)})+\cdots+0\,.
\end{equation}
It is clear that, if $\{X_i\mid i=1,\ldots,p\}$ is a basis of local sections of $E$, then $\{X_i^{(j)}\mid i=1,\ldots,p,j=1,\ldots,m\}$	 is a basis of local sections of $E^{[m]}$. Note that all this can be repeated also for {\it generalized vector bundles}, like generalized distributions.

Since there are obvious canonical isomorphisms
\begin{equation}
(TN)^{[m]}\simeq TN^m\quad\text{and}\quad (T^*N)^{[m]}\simeq T^*N^m\,,\end{equation} 
we can interpret the diagonal prolongation $X^{[m]}$ of a vector field on $N$ as a vector field $\widetilde{X}^{[m]}$ on $N^m$, and the diagonal prolongation $\alpha^{[m]}$ of a 1-form on $N$ as a 1-form $\widetilde{\alpha}^{[m]}$ on $N^m$. In the case when $m$ is fixed, we will simply write $\widetilde{X}$ and $\widetilde{\alpha}$. The proof of the following properties of diagonal prolongations is straightforward.

\begin{proposition}\label{p*} The {\it diagonal prolongation} to $N^m$ of a vector field $X$ on
$N$ is the unique vector field $\widetilde X^{[m]}$ on $N^m$, projectable under the map
$\pi:(x_{(1)},\ldots,x_{(m)})\in N^m\mapsto x_{(1)}\in N$ onto $X$ and invariant under the permutation of variables $x_{(i)}\leftrightarrow x_{(j)}$, with $i,j=1,\ldots,m$.
The {\it diagonal prolongation} to $N^m$ of a 1-form $\alpha$ on $N$ is the unique 1-form $\widetilde \alpha^{[m]}$ on $N^m$ such
that $\widetilde \alpha^{[m]}(\widetilde X^{[m]})=\widetilde{\alpha(X)}^{[m]}$ for every vector field $X\in \Gamma(TN)$. We have $d\widetilde\alpha=\widetilde{d\alpha}$ and $\mathcal{L}_{\widetilde X^{[m]}}\widetilde\alpha^{[m]}=\widetilde{\mathcal{L}_X\alpha}^{[m]}$.
In particular, if $\alpha$ is closed (exact), so is its diagonal prolongation $\widetilde \alpha^{[m]}$ to $N^m$.
\end{proposition}

Using local coordinates $(x^a)$ in $N$ and the induced system $(x^a_{(i)})$ of coordinates in $N^m$, we can write, for $X=\sum_aX^a(x)\partial_{x^a}$ and $\alpha=\sum_a\alpha_a(x)dx^a$,
\begin{equation}\label{prol-coord}
\widetilde X^{[m]}=\sum_{a,i}X^a(x_{(i)})\partial_{x^a_{(i)}}\quad\text{and}\quad
\widetilde \alpha^{[m]}=\sum_{a,i}\alpha_a(x_{(i)})dx^a_{(i)}\,.
\end{equation}

Let us fix $m$. Obviously, given two vector fields $X_1$ and $X_2$ on $N$, we have $\widetilde{ [X_1,X_2]}=[\widetilde X_1,\widetilde X_2]$. In consequence, the prolongations to $N^m$ of the elements of a finite-dimensional real Lie algebra $V$ of vector fields on $N$ form a real Lie algebra $\widetilde V$ isomorphic to $V$. Similarly to standard vector fields, we can define the diagonal prolongation of a $t$-dependent vector field $X$ on $N$ to $N^m$ as the only $t$-dependent vector field $\widetilde X$ on $N^m$ satisfying that $\widetilde X_t$ is the prolongation of $X_t$ to $N^m$ for each $t\in\mathbb{R}$.

When $X$ is a Lie--Hamilton system, its diagonal prolongations are also Lie--Hamilton systems in a natural way \cite{BCHLS}.
Let us now focus on proving an ue of this result for Dirac--Lie systems.

\begin{definition} Given two Dirac manifolds $(N,L_N)$ and $(M,L_M)$, we say that $\varphi:N\rightarrow M$ is a {\it forward Dirac map} between them if $(L_M)_{\varphi(x)}\!=\!\mathfrak{P}_\varphi(L_N)_x$, where
\begin{equation}
\mathfrak{P}_\varphi(L_N)_x\!=\!\{\varphi_{*x}X_{x}+\omega_{\varphi(x)}\in T_{\varphi(x)}M\oplus T_{\varphi(x)}^*M\!\mid\! X_x+ (\varphi^*\omega_{\varphi(x)})_x\!\in\! (L_N)_x\},\end{equation} 
 for all $x\in N$.
\end{definition}

\begin{proposition} Given a Dirac structure $(N,L)$ and the natural isomorphism
\begin{equation}
(TN^m\oplus_{N^m}T^*N^m)_{(x_{(1)},\ldots,x_{(m)})}\simeq (T_{x_{(1)}}N\oplus T^*_{x_{(1)}}N)\oplus \cdots\oplus (T_{x_{(m)}}N\oplus T^*_{x_{(m)}}N),
\end{equation}
the diagonal prolongation $L^{[m]}$, viewed as a vector subbundle in $TN^m\oplus_{N^m}T^*N^m=\mathcal{P}N^{[m]}$,
is a Dirac structure on $N^m$.

The forward image of $L^{[m]}$ through each $\pi_{i}:(x_{(1)},\ldots,x_{(m)})\in N^m\rightarrow x_{(i)}\in N$, with $i=1,\ldots,m$, equals $L$. Additionally, $L^{[m]}$ is invariant under the permutations $x_{(i)}\leftrightarrow x_{(j)}$, with $i,j=1,\ldots,m$.
\end{proposition}
\begin{proof}
Being a diagonal prolongation of $L$, the subbundle $L^{[m]}$ is invariant under permutations $x_{(i)}\leftrightarrow x_{(j)}$ and each element of a basis $X_i+\alpha_i$ of $L$, with $i=1,\ldots,n$, can naturally be considered as an element $X^{(j)}_i+\alpha^{(j)}_i$ of the $j$th-copy of $L$ within $L^{[m]}$. This gives rise to a basis of $L^{[m]}$, which naturally becomes a smooth $m n$-dimensional subbundle of $\mathcal{P}N^m$. Considering the natural pairing $\langle\cdot,\cdot\rangle_+$ of $\mathcal{P}N^m$ and using $\langle \alpha^{(i)}_j,X^{(k)}_l\rangle=0$ for $i\neq k$, we have
\begin{eqnarray}
&\left\langle \left(X^{(i)}_{j}+\alpha^{(i)}_{j}\right)(x_{(1)},\ldots,x_{(m)}),\left(X^{(k)}_{l}+
\alpha^{(k)}_{l}\right)(x_{(1)},\ldots,x_{(m)})\right\rangle_+\!\!=\nonumber\\
&\delta^i_k\left\langle \left(X_{j}+\alpha_{j}\right)(x_{(i)}),\left(X_{l}+\alpha_{l}\right)(x_{(i)})\right\rangle_+=0,
\end{eqnarray}
for every $p=(x_{(1)},\ldots,x_{(m)})\in N^m$.
As the pairing is bilinear and vanishes on a basis of $L^{[m]}$, it does so on the whole $L^{[m]}$, which is therefore isotropic. Since $L^{[m]}$ has rank $mn$, it is maximally isotropic.

Using that 
\begin{equation}
[X^{(i)}_j,X^{(k)}_l]=0,\quad \iota_{X^{(i)}_j}d\alpha_{l}^{(k)}=0, 
\quad\text{ and }\quad \mathcal{L}_{X^{(i)}_j}\omega_{l}^{(k)}=0\end{equation} 
 for $i\neq k=1,\ldots,m$ and $j,l=1,\ldots,\dim\, N$, we obtain
\begin{equation}
[[X^{(i)}_j+\alpha^{(i)}_j,X^{(k)}_l+\alpha^{(k)}_l]]_C=\delta^{i}_{k}[[X^{(i)}_j+\alpha^{(i)}_j,X^{(i)}_l+\alpha^{(i)}_l]]_C\in \Gamma(L^{[m]}).
\end{equation}
So, $L^{[m]}$ is involutive. Since it is also maximally isotropic, it is a Dirac structure.

Let us prove that $\mathfrak{P}_{\pi_a}(L^{[m]})=L$ for every $\pi_a$. Note that
$(X^{(a)}_j+\alpha^{(a)}_j)_p\in L^{[m]}_p$ is such that $\pi_{a*}(X^{(a)}_j)_p=(X_j)_{x_{(a)}}$ and $(\alpha_j)_{x_{(a)}}\circ(\pi_{*a})_p =(\alpha^{(a)}_j)_p$ for every $p\in \pi^{-1}_a(x_{(a)})$. So, $(X_j+\alpha_j)_{x_{(a)}}\in (\mathfrak{P}_{\pi_a}(L^{[m]}))_{x_{(a)}}\subset L_{x_{(a)}}$ for $j=1,\ldots,n$ and every $x_{(a)}\in N$. Using that $X_j+\alpha_j$ is a basis for $L$ and the previous results, we obtain $L\subset \mathfrak{P}_{\pi_a}(L^{[m]})$. Conversely, $\mathfrak{P}_{\pi_a}(L^{[m]})\subset L$. Indeed, if $(X+\alpha)_{x_{(a)}}\in \mathfrak{P}_a(L^{[m]})$, then there exists an element $(Y+\beta)_p\in L^{[m]}_p$, with $p\in \pi^{-1}(x_{(a)})$, such that $\pi_{a*}Y_p=X_{x_{(a)}}$ and $(\alpha)_{x_{(a)}}\circ (\pi_{*a})_p=\beta_p$. Using that $(Y+\beta)_p=\sum_{ij}c_{ij}(X^{(i)}_j+\alpha^{(i)}_j)_p$ for a unique set of constants $c_{ij}$, with $i=1,\ldots,m$ and $j=1,\ldots,n$, we have
$\pi_{a*}(\sum_{ij}c_{ij}(X^{(i)}_j)_p)=\sum_{j}c_{aj}(X_j)_{x_{(a)}}=X_{x_{(a)}}$. Meanwhile, $\beta_p=\alpha_{x_{(a)}}\circ (\pi_{*a})_p$ means that
$\sum_{j}c_{aj}(\alpha_j)_{x_{(a)}}=\alpha_{x_{(a)}}$. So, $(X+\alpha)_{x_{(a)}}=\sum_{j}c_{aj}(X_j+\alpha_j)_{x_{(a)}}\in L_{x_{(a)}}$.
\end{proof}

\begin{corollary}\label{COR1} Given a Dirac structure $(N,L)$, we have $\rho_m (L^{[m]})=\rho(L)^{[m]}$, where $\rho_m$ is the projection $\rho_m:\mathcal{P}N^m\rightarrow TN^m$.  Then, if $X$ is an $L$-Hamiltonian vector field
with respect to $L$, its diagonal prolongation $\widetilde X^{[m]}$ to $N^m$ is an $L$-Hamiltonian vector field with respect to $L^{[m]}$. Moreover, $\rho^*_m (L^{[m]})=\rho^*(L)^{[m]}$, where $\rho^*_m$ is the canonical projection $\rho^*_m:\mathcal{P}N^m\rightarrow T^*N^m$.
\end{corollary}

\begin{corollary} If $(N,L,X)$ is a Dirac--Lie system, then
$(N^m,L^{[m]},\widetilde X^{[m]})$ is also a Dirac--Lie system.
\end{corollary}
\begin{proof} If $X$ admits a Vessiot--Guldberg Lie algebra $V$ of Hamiltonian vector fields with respect to $(N,L)$, then $\widetilde X$ possesses a Vessiot--Guldberg Lie algebra $\widetilde V$ given by the diagonal prolongations of the elements of $V$, which are $L^{[m]}$-Hamiltonian vector fields, by construction of $L^{[m]}$ and Corollary \ref{COR1}.
\end{proof}

Similarly to the prolongations of vector fields, one can define prolongations of functions and 1-forms in an obvious way.

\begin{proposition}\label{Prop} Let $X$ be a vector field and $f$ be a function on $N$. Then
\begin{itemize}
\item[(a)] If $f$ is a $L$-Hamiltonian function for $X$, its diagonal prolongation $\widetilde{f}$ to $N^m$ is an $L^{[m]}$-Hamiltonian function of the diagonal prolongation $\widetilde X$ to $N^m$.
\item[(b)] If $f\in {\rm Cas}(N,L)$, then $\widetilde f\in {\rm Cas}(N^m,L^{[m]})$.
\item[(c)] The map $\lambda:({\rm Adm}(N,L),\{\cdot,\cdot\}_L) \ni f \mapsto \widetilde f \in ({\rm Adm}(N^m,L^{[m]}),\{\cdot,\cdot\}_{L^{[m]}})$ is an injective Lie algebra morphism.

\end{itemize}
\end{proposition}
\begin{proof}
Let $f$ be an $L$-Hamiltonian function for $X$. Then, $X+df\in \Gamma(L)$ and $\widetilde X+d\widetilde f=\widetilde X+\widetilde{df}$ is an element of $\Gamma(L^{[m]})$. By a similar argument, if $f\in {\rm Cas}(N,L)$, then $\widetilde f\in {\rm Cas}(N^m,L^{[m]})$.
Given $f,g\in {\rm Adm}(N,L)$, we have $\widetilde{\{f,g\}_L}=\widetilde{X_fg}=\widetilde{X_f}\widetilde{g}=X_{\widetilde{f}}\widetilde{g}=\{\widetilde f,\widetilde g\}_{L^{[m]}}$, i.e., $\lambda(\{f,g\}_L)=\{\lambda(f),\lambda(g)\}_{L^{[m]}}$. Additionally, as $\lambda$ is
linear, it becomes a Lie algebra morphism. Moreover, it is easy to see that $\widetilde f=0$ if and only if $f=0$. Hence, $\lambda$ is injective.
\end{proof}
Note, however, that we cannot ensure that $\lambda$ is a Poisson algebra morphism, as in general $\widetilde {fg}\neq \widetilde f\widetilde g$.

Using the above proposition, we can easily prove the following corollaries.
\begin{corollary} If $h_1,\ldots,h_r:N\rightarrow \mathbb{R}$ is a family of functions on a Dirac manifold $(N,L)$ spanning a finite-dimensional real Lie algebra of functions with respect to the Lie bracket $\{\cdot,\cdot\}_L$, then their diagonal prolongations $\widetilde h_1,\ldots,\widetilde h_r$ to $N^m$ close an isomorphic Lie algebra of functions with respect to the Lie bracket $\{\cdot,\cdot\}_{L^{[m]}}$ induced by the Dirac structure $(N^m,L^{[m]})$.
\end{corollary}

\begin{corollary} If $(N,L,X)$ is  a Dirac--Lie system admitting a Lie--Hamiltonian  $(N,L,h)$, then $(N^m,L^{[m]},\widetilde X^{[m]})$ is a Dirac--Lie system with a Dirac--Lie Hamiltonian $(N^m,L^{[m]},h^{[m]})$, where $h^{[m]}_t=\widetilde h_t^{[m]}$  is the diagonal prolongation of $h_t$ to $N^m$.
\end{corollary}

\subsection{Superposition rules and $t$-independent constants of motion}

Let us give a first straightforward application of Dirac--Lie systems to obtain constants of the motion.

\begin{proposition}
  \label{Cas}Given a Dirac--Lie system $(N, L, X)$, the elements of
  ${\rm Cas}(N, L)$ are constants of the motion for $X$. Moreover, the set $\mathcal{I}^X_L$ of
 its admissible $t$-independent constants of the motion form a Poisson
  algebra $( \mathcal{I}^X_L, \cdot, \{ \cdot, \cdot \}_L)$.
\end{proposition}

\begin{proof}
  Two admissible functions $f$ and $g$ are $t$-independent constants of the motion for $X$ if and
  only if $X_t f = X_t g = 0$ for every $t \in \mathbb{R}$. Using that every
$X_t$ is a derivation of the associative algebra  $C^\infty(N)$, we see that given $f,g\in \mathcal{I}^X_L$, then $f+g$, $\lambda f$,  and $f\cdot g$ are
also constants of the motion for $X$ for every $\lambda \in \mathbb{R}$. Since the sum and product of admissible functions are admissible functions, then $\mathcal{I}_L^X$ is closed under the sum and product of elements and real constants. So $(\mathcal{I}_L^X,\cdot)$ is an associative
subalgebra of $(C^\infty(N),\cdot)$.

As $(N,L,X)$ is a Dirac--Lie system, the vector fields $\{X_t \}_{t \in
\mathbb{R}}$
  are $L$-Hamiltonian. Therefore,
\begin{equation}
 X_t \{f, g\}_L = \{X_tf, g\}_L + \{f, X_tg \}_L.
\end{equation}
 As $f$ and $g$ are constants
  of the motion for $X$, then $\{f, g\}_L$ is so also. Using that $\{f,g\}_L$ is also an admissible function, we
finish the proof.
\end{proof}
The following can easily be proven.
\begin{proposition}\label{Prop:Ham}
  Let $(N,L,X)$ be a Dirac--Lie system possessing a Dirac--Lie Hamiltonian
  $(N, L, h)$. An admissible function $f : N \rightarrow \mathbb{R}$ is
  a constant of the motion for $X$ if and only if it Poisson commutes with all
  the elements of Lie$(\{h_t \}_{t \in \mathbb{R}}, \{ \cdot, \cdot
  \}_{L})$.
\end{proposition}

Consider a Dirac--Lie system $(N,L^\omega,X)$ with $\omega$ being
a symplectic structure and $X$ being an autonomous system. Consequently, ${\rm Adm}(N,L)=C^\infty(N)$ and the above
proposition entails that $f\in C^\infty(N)$ is a constant of the motion for $X$ if and only if it Poisson commutes
with a Hamiltonian function $h$ associated with $X$. This shows that Proposition \ref{Prop:Ham}
recovers as a particular case this well-known result \cite{Abra}.
Additionally, Proposition \ref{Prop:Ham} suggests us that the role played by
autonomous Hamiltonians for autonomous Hamiltonian systems is performed by
finite-dimensional Lie algebras of admissible functions associated with a
Dirac--Lie
Hamiltonian for Dirac--Lie systems. This fact can be employed, for instance, to study $t$-independent
first-integrals of Dirac--Lie systems, e.g., the maximal
number of such first-integrals in involution, which would lead to the
interesting analysis of integrability/superintegrability and action/angle variables for Dirac--Lie
systems \cite{NTZ12}.

Another reason to study $t$-independent constants of motion for
Lie systems is their
use in deriving superposition rules \cite{CGM00}. More
explicitly, a superposition rule for a Lie system can be obtained through the $t$-independent
constants of the motion of one of its diagonal
prolongations \cite{CGM07}. The following proposition provides some ways of obtaining such constants.

\begin{proposition} If $X$ be a system possessing a $t$-independent constant of the motion $f$, then
\begin{enumerate}
 \item The diagonal prolongation $\widetilde f^{[m]}$ is a $t$-independent constant of the motion for $\widetilde X^{[m]}$.
 \item If $Y$ is a $t$-independent Lie symmetry of $X$, then $\widetilde Y^{[m]}$ is a $t$-independent Lie symmetry of $\widetilde X^{[m]}$.
 \item If
$h$ is a $t$-independent constant of the motion for $\widetilde X^{[m]}$, then $\widetilde Y^{[m]}h$ is another $t$-independent constant of the motion for $\widetilde X^{[m]}$.
\end{enumerate}
\end{proposition}
\begin{proof} This result is a straightforward application of Proposition \ref{p*} and the properties of the diagonal prolongations of $t$-dependent vector fields.
\end{proof}

\begin{proposition}\label{Prop:Cas} Given a Dirac--Lie system $(N,L,X)$ that admits a Dirac--Lie Hamiltonian  $(N,L,h)$ such that $\{h_t\}_{t\in\mathbb{R}}$ is contained in a finite-dimensional  Lie algebra of admissible functions $(\mathfrak{W},\{\cdot,\cdot\}_L)$. Given the momentum map $J:N^m	\to\mathfrak{W}^*$ associated with the Lie algebra morphism $\phi:f\in \mathfrak{W}\mapsto \widetilde f\in {\rm Adm}(N^m,L^{[m]})$, the pull-back $J^*(C)$  of any Casimir function $C$ on $\mathfrak{W}^*$ is a constant of the motion for the diagonal prolongation $\widetilde X^{[m]}$. If $\mathfrak{W}\simeq{\rm Lie}(\{\widetilde h_t\}_{t\in\mathbb{R}},\{\cdot,\cdot\}_{L^{[m]}})$, the function $J^*(C)$ Poisson commutes with all $L^{[m]}$-admissible constants of the motion of $\widetilde X^{[m]}$.
\end{proposition}

\begin{example}\normalfont
Let us use the above results to devise a superposition rule for the {\bf third-order Kummer--Schwarz equation} in first-order form (\ref{ks3nose}) with $c_0=0$, the so called {\it Schwarzian equations} \cite{SS12,TTX01}. To simplify the presentation, we will always assume $c_0=0$ in this section. It is known (cf. \cite{CGL11}) that the derivation of a superposition rule for this system can be reduced to obtaining certain three $t$-independent constants of the motion for the diagonal prolongation $\widetilde{X}^{3KS}$ of $X^{3KS}$ to $\mathcal{O}_2^2$. In \cite{CGL11} such
constants were worked out through the method of characteristics which consists on solving a series of systems of ODEs. Nevertheless, we can determine such constants more easily through Dirac--Lie systems.
System (\ref{ks3nose}) is represented by the $t$-dependent vector field
\begin{equation}\label{Exotro}
X^{3KS}_t=v\frac{\partial}{\partial x}+a\frac{\partial}{\partial v}+\left(\frac 32
\frac{a^2}v-2c_0v^3+2b_1(t)v\right)\frac{\partial}{\partial
a}=Y_3+b_1(t)Y_1,
\end{equation}
where the vector fields on $\mathcal{O}_2=\{(x,v,a)\in{T}^2\mathbb{R}\mid v\neq 0\}$ given by

\begin{equation}\label{VFKS1otro}
\begin{gathered}
N_1=2v\frac{\partial}{\partial a},\quad N_2=v\frac{\partial}{\partial v}+2a\frac{\partial}{\partial a},\\
N_3=v\frac{\partial}{\partial x}+a\frac{\partial}{\partial v}+\left(\frac 32\frac{a^2}v-2c_0v^3\right)\frac{\partial}{\partial a},
\end{gathered}
\end{equation}
If so,  $t$-dependent vector field $\widetilde X^{3KS}$ is spanned by a linear combination of the diagonal prolongations of $N_1$, $N_2,$ and $N_3$ to $\mathcal{O}_2^2$. From (\ref{VFKS1otro}), we have
\begin{equation}
\begin{gathered}
\widetilde N_1=\sum_{i=1}^2v_i\frac{\partial}{\partial a_i},\quad \widetilde N_2=
\sum_{i=1}^2\left(v_i\frac{\partial}{\partial v_i}+2a_i\frac{\partial}{\partial a_i}\right),\\ \widetilde N_3=\sum_{i=1}^2\left(v_i\frac{\partial}{\partial x_i}+a_i\frac{\partial}{\partial v_i}+\frac 32
\frac{a_i^2}{v_i}\frac{\partial}{\partial a_i}\right).
\end{gathered}
\end{equation}
From Proposition \ref{Prop} and functions (\ref{Fun3KS}), the vector fields $\widetilde N_1,\widetilde N_2,\widetilde N_3$ are $L^{[2]}$-Hamiltonian with $L^{[2]}$-Hamiltonian functions
\begin{equation}
\widetilde h_1=-\frac 2{v_1}-\frac 2{v_2},\qquad \widetilde {h}_2=-\frac{a_1}{v_1^2}-\frac{a_2}{v_2^2},\qquad \widetilde {h}_3=-\frac {a_1^2}{2v_1^3}-\frac {a_2^2}{2v_2^3}.
\end{equation}
Indeed, these are the diagonal prolongations to $\mathcal{O}_2^2$ of the $L$-Hamiltonian functions of $N_1, N_2$, and $N_3$.
 Moreover, they span a real Lie algebra of functions isomorphic to that one spanned by $h_1,h_2,h_3$ and to $\mathfrak{sl}(2,\mathbb{R})$. We can then define a Lie algebra morphism $\phi:\mathfrak{sl}(2,\mathbb{R})\rightarrow C^\infty(N^2)$ of the form $\phi(e_1)=\widetilde h_1$, $\phi(e_2)=\widetilde h_2$ and $\phi(e_3)=\widetilde h_3$, where $\{ e_1,e_2,e_3\}$ is the standard basis of $\mathfrak{sl}(2,\mathbb{R})$. Using that $\mathfrak{sl}(2,\mathbb{R})$ is a simple Lie algebra, we can compute the Casimir invariant on $\mathfrak{sl}(2,\mathbb{R})^*$ as $e_1e_3-e_2^2$ (where $e_1,e_2,e_3$ can be considered as functions on $\mathfrak{sl}(2,\mathbb{R})$). Proposition 	\ref{Prop:Cas} ensures then that $\widetilde h_1\widetilde h_3-\widetilde h_2^2$ Poisson commutes with $\widetilde h_1,\widetilde h_2$ and $\widetilde h_3$. In this way, we obtain a constant of the motion for $\widetilde{X}^{3KS}$ given by
\begin{equation}
 I=\widetilde h_1\widetilde h_3-\widetilde h_2^2=\frac{(a_2v_1-a_1v_2)^2}{v_1^3v_2^3}.
\end{equation}
Schwarzian equations admit a Lie symmetry $Z=x^2\partial/ \partial x$ \cite{OT09}. Its prolongation to ${T}^2\mathbb{R}$, i.e.,
\begin{equation}\label{LieSym2}
Z_P=x^2\frac{\partial}{\partial x}+2vx\frac{\partial}{\partial v}+2(ax+v^2)\frac{\partial}{\partial a},
\end{equation}
is a Lie symmetry of $X^{3KS}$. From Proposition \ref{Prop}, we get that $\widetilde Z_P$ is a Lie symmetry of $\widetilde{X}^{3KS}$. So, we can construct constants of the motion for $\widetilde{X}^{3KS}$ by applying $\widetilde Z_P$ to any of its $t$-independent constants of the motion. In particular,
\begin{equation}
F_2\equiv -\widetilde Z_P \log |I|=x_1+x_2+\frac{2v_1v_2(v_1-v_2)}{a_2v_1-a_1v_2}
\end{equation}
is constant on particular solutions $(x_{(1)}(t),v_{(1)}(t),a_{(1)}(t),x_{(2)}(t),v_{(2)}(t),a_{(2)}(t))$ of $\widetilde{X}^{3KS}$. If $(x_{(2)}(t),v_{(2)}(t),a_{(2)}(t))$ is a particular solution for $X^{3KS}$, its opposite is also a particular solution. So, the function
\begin{equation}
F_3\equiv x_1-x_2+\frac{2v_1v_2(v_1+v_2)}{a_2v_1-a_1v_2}
\end{equation}
is also constant along solutions of $\widetilde{X}^{3KS}$, i.e., it is a new constant of the motion. In consequence, we get three $t$-independent constants of the motion: $\Upsilon_1=I$ and
\begin{equation}
\Upsilon_2=\frac{F_2+F_3}2=x_1+\frac{2v_1^2v_2}{a_2v_1-a_1v_2},\qquad \Upsilon_3=\frac{F_2-F_3}2=x_2-\frac{2v_1v_2^2}{a_2v_1-a_1v_2}.
\end{equation}
This gives rise to three $t$-independent constants of the motion for ${\widetilde X}^{3KS}$. Taking into account that $\partial (\Upsilon_1,\Upsilon_2,\Upsilon_3)/\partial (x_1,v_1,a_1)\neq 0$, the expressions $\Upsilon_1=\lambda_1$, $\Upsilon_2=\lambda_2$, and $\Upsilon_3=\lambda_3$ allow us to obtain the expressions of $x_1,v_1,a_1$ in terms of the remaining variables and $\lambda_1,\lambda_2,\lambda_3$. More specifically,
\begin{equation}\label{suprulesch}
x_1=\frac{4}{\lambda_1(\lambda_3-x_2)}+\lambda_2,\,\, v_1=\frac{4v_2}{\lambda_1(\lambda_3-x_2)^2},\,\, a_1=\frac{8v_2^2+4 a_2(\lambda_3-x_2)}{\lambda_1(\lambda_3-x_2)^3}.
\end{equation}
According to the theory of Lie systems \cite{CGM07}, the map $\Phi:(x_2,v_2,a_2;\lambda_1,\lambda_2,\lambda_3)\in \mathcal{O}_2^2\times \mathbb{R}^3\mapsto (x_1,v_1,a_1)\in \mathcal{O}_2^2$ enables us to write the general solution of (\ref{ks3nose}) into the form
\begin{equation}
(x(t),v(t),a(t))=\Phi(x_2(t),v_2(t),a_2(t);\lambda_1,\lambda_2,\lambda_3).
\end{equation}
This is the known superposition rule for Schwarzian equations (in first-order form) derived in \cite{LSKummer} by solving a system of PDEs. Meanwhile, our present techniques enable us to obtain the same result without any integration. Note that $x(t)$, the general solution of Schwarzian equations, can be written as $x(t)=\tau\circ \Phi(x_2(t),\lambda_1,\lambda_2,\lambda_3)$, with $\tau$ the projection $\tau:(x_2,v_2,a_2)\in { T}^2\mathbb{R}\mapsto x_2\in \mathbb{R}$, from a unique particular solution of (\ref{ks3nose}), recovering a known feature of these equations \cite{OT09}.
\end{example}

\subsection{Bi--Dirac--Lie systems}

It can happen that a Lie system $X$ on a manifold $N$ possesses Vessiot--Guldberg Lie algebras of vector fields with respect to two different Dirac structures. This results in defining two Dirac--Lie systems. For instance, the system of coupled Riccati equations (\ref{CoupledRic}) admits two Dirac--Lie
structures \cite{BBHLS}: the one previously given, $(\mathcal{O},L^\omega,X)$, where $\omega$ is given by (\ref{wR}), and a second one, $(\mathcal{O},L^{\bar \omega},X)$, with
\begin{equation}
\bar \omega=\sum_{i< j=1}^4\frac{dx_i\wedge x_j}{(x_i-x_j)^2}.
\end{equation}

\begin{definition} A {\it bi--Dirac--Lie system} is a four-tuple $(N,L_1,L_2,X)$, where $(N,L_1)$ and $(N,L_2)$ are two different Dirac manifolds and $X$ is a Lie system on $N$ such that $V^X\subset {\rm Ham}(N,L_1)\cap {\rm Ham}(N,L_2)$.
\end{definition}

Given a bi--Dirac--Lie system $(N,L_1,L_2,X)$, we can apply indistinctly the methods of the previous sections to $(N,L_1,X)$ and $(N,L_2,X)$ to obtain superposition rules, constants of the motion, and other properties of $X$. This motivates studies on constructions of this type of structures.

Let us depict a new procedure to build up bi--Dirac--Lie systems from $(N,L^\omega,X)$ whose $X$ possesses a $t$-independent Lie symmetry $Z$.  This method is a generalization to nonautonomous systems, associated with presymplectic manifolds, of the method  for autonomous Hamiltonian systems devised in \cite{CMR02}.

Consider a Dirac--Lie system $(N,L^{\omega},X)$, where $\omega$ is a presymplectic structure, and a $t$-independent Lie symmetry $Z$  of $X$, i.e. $[Z, X_t] = 0$ for all $t \in
\mathbb{R}$. Under the  above assumptions, $\omega_Z=\mathcal{L}_Z\omega$ satisfies $d\omega_Z=d \mathcal{L}_Z \omega =
\mathcal{L}_Z d
\omega = 0$, so $(N,\omega_Z)$ is a presymplectic manifold. The vector fields of $V^X$ are still
Hamiltonian with respect to $\left( N, \omega_Z \right)$. Indeed, we can see that Theorem \ref{MT} ensures that $X$
admits a Dirac--Lie Hamiltonian $(N,L^\omega,h)$ and
\begin{equation}
\begin{gathered}
\left[Z,X_t\right]=0\Longrightarrow \iota_{X_t}\circ\mathcal{L}_Z=\mathcal{L}_Z\circ\iota_{X_t} \\
\Downarrow
\\\iota_{X_t} \omega_Z = \iota_{X_t}  \mathcal{L}_Z
  \omega = \mathcal{L}_Z \iota_{X_t} \omega = -\mathcal{L}_Z dh_t =- d
(Zh_t),
\end{gathered}
\end{equation}
$ \forall t\in\mathbb{R}.$ So, the vector fields $\{X_t\}_{t\in\mathbb{R}}$ are
$L^{\omega_Z}$-Hamiltonian. Since the successive Lie brackets and linear combinations of
$L$-Hamiltonian vector fields and elements of $V^X$ are $L$-Hamiltonian vector fields, the whole Lie algebra  $V^X$ is Hamiltonian with respect to $\omega_Z$. Consequently, $(N,L^{\omega_Z},X)$ is also a Dirac--Lie system. 
Also, $(N,L^{\omega_Z},Zh_t)$ is also a Lie--Hamiltonian for $X$. Moreover,
\begin{equation}
\{\bar h_t,\bar h_{t'}\}_{L^{\omega_Z}}=X_t(\bar h_{t'})=X_t(Zh_{t'})=Z(X_th_{t'})=Z\{ h_t,h_{t'}\}_{L^{\omega}}, \forall t\in\mathbb{R}\,.
\end{equation}
Summarizing, we have the following proposition.

\begin{proposition} If $(N,L^\omega,X)$ is a Dirac--Lie system for which $X$ admits a $t$-independent Lie symmetry $Z$,
then $(N,L^\omega,L^{\mathcal{L}_Z\omega},X)$ is a bi--Dirac--Lie system. If $(N,L^\omega,h)$ is a Dirac--Lie Hamiltonian for $X$, then $(N,L^{\mathcal{L}_Z\omega},Zh)$ is a Dirac--Lie Hamiltonian for $X$ and there exists an exact sequence of Lie algebras
\begin{equation}
(\{h_t\}_{t\in\mathbb{R}},\{\cdot,\cdot\}_{L^{\omega}})\stackrel{Z}{\longrightarrow} (\{Zh_t\}_{t\in\mathbb{R}},\{\cdot,\cdot\}_{L^{\omega_Z}})\rightarrow 0\,.
\end{equation}
\end{proposition}

Note that, given a Lie--Hamilton system $(N,L^\omega,X)$, the triple $(N,L^{\omega_Z},X)$ need not be a Lie--Hamilton system: $\omega_Z$ may fail to be a symplectic two-form (cf. \cite{CMR02}). This causes that the theory of Lie--Hamilton systems cannot be applied to study $(N,L^{\omega_Z},X)$, while the  methods of our work do.

Let us illustrate the above theory with an example. Recall that Schwarzian equations  admit a Lie symmetry $Z=x^2\partial/\partial x$. As a consequence, system (\ref{firstKS33}), with $c_0=0$, possesses a $t$-independent Lie symmetry $Z_P$ given by \eqref{LieSym2} and
\begin{equation}
\omega_{Z_P}\equiv \mathcal{L}_{Z_P}\omega_{3KS}=-\frac{2}{v^3}(xdv\wedge da+vda\wedge dx+adx\wedge dv).
\end{equation}
Moreover,
\begin{equation}
\begin{gathered}
\iota_{Y_1}\omega_{Z_P}=-d({{Z_P}}h_1)=-d\left(\frac{4x}{v}\right),\quad
\iota_{Y_2}\omega_{Z_P}=-d({{Z_P}}h_2)=d\left(2-\frac{2ax}{v^2}\right),\\
\iota_{Y_3}\omega_{Z_P}=-d({{Z_P}}h_3)=d\left(\frac {2a}v-\frac{a^2x}{v^3}\right).
\end{gathered}
\end{equation}
So, $Y_1,Y_2$, and $Y_3$ are Hamiltonian vector fields with respect to $\omega_{Z_P}$. Moreover, since
\begin{align}\{ {Z_P}h_1, {Z_P}h_2\}_{L^{\omega_{Z_P}}}&= Z_Ph_1\,,\nonumber\\
\{ {Z_P}h_2, {Z_P}h_3\}_{L^{\omega_{Z_P}}}&= {Z_P}h_3\,,\\
\{ {Z_P}h_1,{Z_P}h_3\}_{L^{\omega_{Z_P}}}&=2{Z_P}h_2\,\nonumber,
\end{align}
we see that $Z_P h_1$, $Z_Ph_2,$ and $Z_Ph_3$ span a new finite-dimensional real Lie algebra. So,
if $(\mathcal{O}_2,L^\omega,h)$ is a Lie--Hamiltonian for $X$, then $(\mathcal{O}_2,L^{\omega_{Z_P}},Z_Ph)$ is a Dirac--Lie Hamiltonian for $X$.

Let us devise a more general  method to construct bi--Dirac--Lie systems. Given a Dirac manifold $(N,L)$ and a closed two-form $\omega$ on $N$, the sections on $TN\oplus_N T^*N$ of the form
\begin{equation}
X+\alpha-\iota_X\omega,\end{equation} 
where $X+\alpha\in \Gamma(L)$, span a new Dirac structure $(N,L^{\omega})$ \cite{BR03}. When two Dirac structures are connected by a transformation of this type, it is said that they are {\it gauge equivalent}. Using this, we can prove the following propositions.
\begin{proposition} Let $Z$ be a vector field on $N$. Then, the Dirac structures $L^\omega$ and $L^{\omega_Z}$, with $\omega_Z=\mathcal{L}_Z\omega$, are gauge equivalent.
\end{proposition}
\begin{proof} The Dirac structure $L^\omega$ is spanned by sections of the form $X-\iota_X\omega$, with $X\in \Gamma(N)$, and the Dirac structure $L^{\omega_Z}$ is spanned by sections of the form $X-\iota_X\omega_Z$. Recall that $d\omega=d\omega_Z=0$. So, $L^{\omega_Z}$ is of the form
\begin{equation}
X-\iota_X\omega-\iota_X(\omega_Z-\omega),\qquad X-\iota_X\omega\in \Gamma(L^\omega).
\end{equation}
As $d(\omega_Z-\omega)=0$, then $L^\omega$ and $L^{\omega_Z}$ are connected by a gauge transformation.
\end{proof}




\subsection{Dirac--Lie systems and Schwarzian--KdV equations}
Let us give some final relevant applications of our methods. In particular, we devise a procedure to construct traveling wave solutions for some relevant nonlinear PDEs by means of Dirac--Lie systems. For simplicity, we hereafter denote the partial derivatives of a function $f:(x_1,\ldots,x_n)\in \mathbb{R}^n\mapsto f(x_1,\ldots,x_n)\in\mathbb{R}$ by $\partial_{x_i}f$.
\begin{example}\normalfont
Consider the so called {\bf Schwarzian Korteweg de Vries equation} (SKdV equation)\cite{As10}
\begin{equation}\label{SKdV}
\{\Phi,x\}\partial_x\Phi=\partial_t\Phi,
\end{equation}
where $\Phi:(t,x)\in\mathbb{R}^2\rightarrow \Phi(t,x)\in\mathbb{R}$ and
\begin{equation}
\{\Phi,x\}\equiv \frac{\partial^3_x \Phi}{\partial_x\Phi}-\frac 32\left(\frac{\partial^2_x\Phi}{\partial_x \Phi}\right)^2,
\end{equation}
which is equivalent to that introduced in \eqref{schderi}.
This PDE has  been attracting some attention due to its many interesting properties \cite{As10,GRL98,RBMG03}. For instance, Dorfman established
a bi-symplectic structure for this equation \cite{Do89}, and many others have been studying its solutions and generalizations \cite{As10,RBMG03}. As a relevant result, we can mention that, given a solution $\Phi$ of the SKdV equation, the function $\{\Phi,x\}$ is a particular solution of the Korteweg-de Vries equation (KdV equation) \cite{Mar11}
\begin{equation}\label{KdV}
\partial_tu=\partial_x^3u+3u\partial_xu .
\end{equation}
\end{example}
We now  look for traveling wave solutions of (\ref{SKdV}) of the type $\Phi(t,x)=g(x-f(t))$ for a certain fixed $t$-dependent function $f$ with $df/dt=v_0\in \mathbb{R}$. Substituting $\Phi=g(x-f(t))$ within (\ref{SKdV}), we obtain that $g$ is a particular solution of the Schwarzian equation
\begin{equation}\label{Traveling}
\frac{d^3g}{dz^3}=\frac 32\frac{(d^2g/dz^2)^2}{dg/dz}-v_0\frac{dg}{dz},
\end{equation}
where $z\equiv x-f(t)$. We already know that the Schwarzian equations can be studied through the superposition rule (\ref{suprulesch}), which can better be obtained by using that Schwarz equations can be studied through a Dirac--Lie system, as seen in this section. More specifically, we can generate all their solutions from a known one as
\begin{equation}\label{Sup}
g_2(z)=\frac{\alpha g_1(z)+\beta}{\gamma g_1(z)+\delta},\qquad \alpha\delta-\beta\gamma\neq 0,\qquad \alpha,\beta,\gamma,\delta\in\mathbb{R}.
\end{equation}
In addition, (\ref{Traveling}) is a HODE Lie system, i.e., when written as a first-order system
by adding the variables $v=dx/dz$ and $a=dv/dz$, it becomes a Lie system $X$,
namely one of the form (\ref{Exotro}). It can be proven that (\ref{Traveling}) can be integrated for any $v_0=df/dt$.
For instance, particular solutions of this equation read
\begin{equation}
\bar{g}_1(z)={\rm th}\left[\sqrt{v_0/2}z\right]\,\,\,\, (v_0>0),\qquad g_1(z)=\frac{1}{z+1}\,\,\,\,(v_0=0).
\end{equation}
Note that $g_1(z)$ has the shape of a solitary stationary solution, i.e., $\lim_{x\rightarrow \pm\infty}g_1(x-\lambda_0)=0$  for every $\lambda_0\in\mathbb{R}$. Meanwhile, $\bar{g}_1$ is a traveling wave solution. Moreover, the general solution of (\ref{Traveling}) in both cases can be obtained from (\ref{Sup}).

\section{Jacobi--Lie systems}
\setcounter{equation}{0}
\setcounter{theorem}{0}
\setcounter{example}{0}

Following the research line of this first part of the thesis, we here study Lie systems with Vessiot--Guldberg Lie algebras of Hamiltonian vector fields with respect to Jacobi manifolds. Roughly speaking, a {\it Jacobi manifold} is a manifold $N$ endowed with a local Lie algebra $(C^\infty(N),\{\cdot,\cdot\})$ \cite{Co87,Do87,Kiri,LM87,Ry00}. Using that Poisson manifolds are a particular case of Jacobi manifolds, we can consider Jacobi--Lie systems as a generalization of Lie--Hamilton systems. Although each Jacobi manifold gives rise to an associated Dirac manifold, not all Hamiltonian vector fields with respect to the Jacobi manifold become Hamiltonian with respect to its associated Dirac manifold (cf. \cite{Co87}). Hence, not every Jacobi--Lie system can be straightforwardly understood as a Dirac--Lie system. Even in that case, the Jacobi manifold allows us to construct a Dirac manifold to study the system.
The main difference between Jacobi--Lie systems and Lie--Hamilton systems is that Jacobi manifolds do not naturally give rise to Poisson brackets on a space of smooth functions on the manifold, which makes difficult to prove ues and/or extensions of the results for Lie--Hamilton systems.

In this section, we extend to Jacobi--Lie systems some of the main structures found for Lie--Hamilton systems, e.g., Lie--Hamiltonians \cite{CLS122}, and we classify all Jacobi--Lie systems on $\mathbb{R}$ and $\mathbb{R}^2$ by determining all Vessiot--Guldberg Lie algebras of Hamiltonian vector fields with respect to Jacobi manifolds on $\mathbb{R}$ and $\mathbb{R}^2$. This is achieved by using the local classification of Lie algebras of vector fields on $\mathbb{R}$ and $\mathbb{R}^2$ derived  by Lie \cite{Lie1880} and improven by  Gonz{\'a}lez-L{\'o}pez,  Kamran and   Olver (GKO) \cite{GKP92} (see also~\cite{BBHLS}). Summarizing, we
obtain that every Lie system on the real line is a Jacobi--Lie system and we show that every Jacobi--Lie system on the plane admits a Vessiot--Guldberg Lie algebra diffeomorphic to one of the 14 classes indicated in Table \ref{table9} in Appendix 1.

\subsection{Jacobi--Lie systems}
We now
introduce Jacobi--Lie systems as Lie systems admitting a Vessiot--Guldberg Lie
algebra of Hamiltonian vector fields relative to a Jacobi manifold.

\begin{definition} A {\it Jacobi--Lie system} $(N,\Lambda,R,X)$ consists of the
Jacobi manifold $(N,\Lambda,R)$ and a Lie system $X$  admitting a
Vessiot--Guldberg Lie algebra of Hamiltonian vector fields with respect to
$(N,\Lambda,R)$.
\end{definition}

\begin{example}\normalfont
We reconsider the {\bf continuous Heisenberg group} \cite{We00} given in Chapter \ref{Chap:GeomFund}, which describes the space of matrices
\begin{equation}\label{Hei2}
\mathbb{H}=\left\{\left(\begin{array}{ccc}1&x&z\\0&1&y\\0&0&1\end{array}
\right)\bigg|\,x,y,z\in\mathbb{R}\right\},
\end{equation}
endowed with the standard matrix multiplication,
where $\{x,y,z\}$ is the natural coordinate system on $\mathbb{H}$ induced by (\ref{Hei2}).

A straightforward calculation shows that the Lie
algebra $\mathfrak{h}$ of left-invariant vector fields on $\mathbb{H}$ is
spanned by
\begin{equation}
X^L_1=\frac{\partial}{\partial x},\qquad X^L_2=\frac{\partial}{\partial
y}+x\frac{\partial}{\partial z},\qquad X^L_3=\frac{\partial}{\partial z}.
\end{equation}
Consider now the system on $\mathbb{H}$ given by
\begin{equation}\label{H}
\frac{{\rm d}h}{{\rm d}t}=\sum_{\alpha=1}^3b_\alpha(t)X^L_\alpha(h),\qquad h\in\mathbb{H},
\end{equation}
for arbitrary $t$-dependent functions $b_1(t),b_2(t)$ and $b_3(t)$. Since the associated $t$-dependent vector field $X_t^\mathbb{H}=b_1(t)X_1^L+b_2(t)X_2^L+b_3(t)X_3^L$ takes values in
a finite-dimensional Lie algebra of vector fields, the system $X^\mathbb{H}$ is a Lie system.
The interest of $X^\mathbb{H}$ is due to its appearance in the solution of
the so called {\it quantum Lie systems} as well as Lie systems admitting a Vessiot--Guldberg Lie algebra
isomorphic to $\mathfrak{h}$ \cite{Dissertationes}.

Let us show that system (\ref{H}) gives rise to a Jacobi--Lie system.
Consider the bivector $\Lambda_{\mathbb H}$ given by
\begin{equation}\label{BivectorH2}
\Lambda_\mathbb{H}\equiv -y\frac{\partial}{\partial y}\wedge\frac{\partial}{\partial z}+\frac{\partial}{\partial x}\wedge\frac{\partial}{\partial y}
\end{equation}
and the vector field $R_\mathbb{H}\equiv \partial/\partial z$. 
Then,
\begin{equation}\begin{gathered}
X^L_1=[\Lambda_{\mathbb H},-y]_{SN}-yR_\mathbb{H},\qquad X^L_2=[\Lambda_{\mathbb H},x]_{SN}+xR_\mathbb{H},
\\
 X^L_3=[\Lambda_{\mathbb H},1]_{SN}+R_\mathbb{H}.                                                                                                                                                                 \end{gathered}\end{equation} 
That is, $X^L_1,X^L_2$ and $X^L_3$ are Hamiltonian vector fields with Hamiltonian functions $h_1=-y$, $h_2=x$ and $h_3=1$, respectively.
Hence, we obtain that $(\mathbb{H},\Lambda_\mathbb{H},R_\mathbb{H},X^\mathbb{H})$ is a Jacobi--Lie system. It is remarkable that each Hamiltonian function $h_i$ is a first-integral of $X_i^L$ and $R_\mathbb{H}$ for $i=1,2,3$, respectively.
\end{example}

\begin{lemma}
The space of good Hamiltonian functions $G(N,\Lambda,R)$ with respect to a Jacobi structure $(N,\Lambda,R)$ is a Poisson algebra with respect to the Jacobi bracket $\{\cdot,\cdot\}_{\Lambda,R}$. Additionally, $\star_g:f\in C^\infty(N)\mapsto \{g,f\}_{\Lambda,R}\in C^\infty(N)$, for any $g\in G(N,\Lambda,R)$, is a derivation on $(C^\infty(N),\cdot)$.
 \end{lemma}
\begin{proof}
First, we prove that the Jacobi bracket of two good Hamiltonian functions is a good Hamiltonian function.
For general functions $u_1,u_2\in C^\infty(N)$, we have
\begin{equation}
R\{u_1,u_2\}_{\Lambda,R}=R(\Lambda({\rm d} u_1,{\rm d}u_2)+u_1Ru_2-u_2Ru_1).
\end{equation}
If $u_1,u_2 \in G(N,\Lambda,R)$, then $Ru_1=Ru_2=0$. Using this and $[\Lambda,R]_{SN}=0$, we obtain
\begin{equation}
R\{u_1,u_2\}_{\Lambda,R}=R(\Lambda({\rm d}u_1,{\rm d}u_2))=[R,[[\Lambda,u_1]_{SN},u_2]_{SN}]_{SN}=0.
\end{equation}
Hence, $\{u_1,u_2\}_{\Lambda,R}\in G(N,\Lambda,R)$, which becomes a Lie algebra relative to $\{\cdot,\cdot\}_{\Lambda,R}$. Note also that $R(u_1\cdot u_2)=0$ and $u_1\cdot u_2\in G(N,\Lambda,R)$.

Given an arbitrary $u_1\in G(N,\Lambda,R)$ and any $u_2,u_3\in C^\infty(N)$, we have that
\begin{align}
\star_{u_1}(u_2\cdot u_3)&=\Lambda({\rm d}u_1,{\rm d}(u_2u_3))+u_1R(u_2u_3)-u_2u_3Ru_1\nonumber\\
&=X_{u_1}(u_2\cdot u_3)=u_3X_{u_1}u_2+u_2X_{u_1}u_3=u_3\star_{u_1}u_2+u_2\star_{u_1}u_3.
\end{align}
So, $\star_{u_1}$ is a derivation on $(C^\infty(N),\cdot)$ and also  on $(G(N,\Lambda,R),\{\cdot,\cdot\}_{\Lambda,R})$. From this it trivially follows that
 $(G(N,\Lambda,R),\cdot,\{\cdot,\cdot\}_{\Lambda,R})$ is a Poisson algebra.
\end{proof}

\subsection{Jacobi--Lie Hamiltonians}

\begin{definition} We call {\it Jacobi--Lie Hamiltonian} a quadruple $(N,\Lambda,R,h)$,
where $(N,\Lambda,R)$ is a Jacobi manifold and $h:(t,x)\in\mathbb{R}\times N\mapsto h_t(x)\in
N$ is a $t$-dependent function such that
$(\{h_t\}_{t\in\mathbb{R}},\{\cdot,\cdot\}_{\Lambda,R})$ is a finite-dimensional real Lie
algebra. Given a system $X$ on $N$, we say that $X$ admits a {\it Jacobi--Lie Hamiltonian}  $(N,\Lambda,R,h)$ if
$X_t$ is a Hamiltonian vector field with Hamiltonian function $h_t$ (with respect to $(N,\Lambda,R)$) for each $t\in \mathbb{R}$.
\end{definition}

\begin{example}\normalfont
So, we find that $h_t=b_1(t)h_1+b_2(t)h_2+b_3(t)h_3= -b_1(t) y+b_2(t)x+b_3(t) $ is a Hamiltonian
function for each $X^\mathbb{H}_t$ in \eqref{H}, with $t\in\mathbb{R}$. In addition,
\begin{equation}
\{h_1,h_2\}_{\Lambda_\mathbb{H},R_\mathbb{H}}=h_3,\qquad \{h_1,h_3\}_{\Lambda_\mathbb{H},R_\mathbb{H}}=0,\qquad \{h_2,h_3\}_{\Lambda_\mathbb{H},R_\mathbb{H}}=0.
\end{equation}
In other words, the functions $\{h_t\}_{t\in\mathbb{R}}$ span a finite-dimensional real Lie
algebra of functions with respect to the Poisson bracket (\ref{BivectorH}) induced by the Jacobi
manifold. Thus, $X^\mathbb{H}$ admits a Jacobi--Lie Hamiltonian $(N,\Lambda_\mathbb{H},R_\mathbb{H},h)$.
\end{example}

\begin{theorem} Given a Jacobi--Lie Hamiltonian $(N,\Lambda,R,h)$, the
system $X$ of the form $X_t=X_{h_t}$,  $\forall t\in\mathbb{R}$, is a Jacobi--Lie
system. If $X$ is a Lie system whose $\{X_t\}_{t\in\mathbb{R}}$ are good Hamiltonian vector fields, then it admits a Jacobi--Lie Hamiltonian.
\end{theorem}

\begin{proof}

Let us prove the direct part. By assumption, the Hamiltonian functions $\{h_t\}_{t\in\mathbb{R}}$ are contained in a finite-dimensional Lie algebra 
${\rm Lie}(\{h_t\}_{t\in\mathbb{R}}, \{\cdot,\cdot\}_{\Lambda,R})$. The Lie algebra morphism
$\phi_{\Lambda,R}:f\in C^{\infty}(N)\mapsto X_f\in $ Ham$(N,\Lambda,R)$ maps the curve $h_t$ into a curve $X_t$ within 
$\phi_{\Lambda,R}({\rm Lie}(\{h_t\}_{t\in\mathbb{R}}, \{\cdot,\cdot\}_{\Lambda,R})).$

Since Lie($\{h_t\}_{t},\{\cdot,\cdot\}_{\Lambda,R}$) is finite-dimensional and $\phi_{\Lambda,R}$ is a Lie algebra morphism,
$$\phi_{\Lambda,R}( {\rm Lie}(\{h_t\}_{t},\{\cdot,\cdot\}_{\Lambda,R}))$$
is a finite-dimensional Lie algebra. Hence, $X$
takes values in a finite-dimensional Lie algebra of Hamiltonian vector fields and it is a Jacobi--Lie system.

 Let us prove the converse. Since the elements of $\{X_t\}_{t\in\mathbb{R}}$ are good Hamiltonian vector fields by assumption and Lie$(\{X_t\}_{t\in\mathbb{R}})=V^X$, every element of $V^X$ is a good Hamiltonian vector field and we can choose
a basis $X_1,\ldots, X_r$ of $V^X$ with good Hamiltonian functions $h_1,\ldots, h_r$. The Jacobi bracket $\{h_i,h_j\}_{\Lambda,R}$ is a good Hamiltonian function for $[X_i,X_j]$.

Since $[X_i,X_j]=\sum_{k=1}^rc_{ijk}X_k$ for certain constants $c_{ijk}$, we obtain that each function:
\begin{equation}
s_{ij}=\{h_i,h_j\}_{\Lambda,R}-\sum_{k=1}^rc_{ijk}h_k,\qquad i<j,
\end{equation}
is the difference of two good Hamiltonian functions with the same Hamiltonian vector field. Hence, $\{s_{ij},h\}_{\Lambda,R}=0$ for all $h\in C^\infty(N)$. Using this, we obtain that the linear space generated by $h_1,\ldots,h_r,s_{ij}$, with $1\leq i<j\leq r$, is a finite-dimensional Lie algebra relative to $\{\cdot,\cdot\}_{\Lambda,R}$. If $X=\sum_{\alpha=1}^rb_\alpha(t)X_\alpha$, then $(N,\Lambda,R,h=\sum_{\alpha=1}^rb_\alpha(t)h_\alpha)$ is a Jacobi--Lie Hamiltonian for $X$.
\end{proof}

One of the relevant properties of the Jacobi--Lie Hamiltonians is given by the following proposition, whose proof is straightforward.

\begin{proposition} Let $(N,\Lambda,R,X)$ be a Jacobi--Lie system admitting a Jacobi--Lie Hamiltonian $(N,\Lambda,R,h)$ of good Hamiltonian functions $\{h_t\}_{t\in\mathbb{R}}$. A function $f:N\rightarrow \mathbb{R}$ is a $t$-independent constant of motion for $X$ if and only if $f$ commutes with all the elements of ${\rm Lie}(\{h_t\}_{t\in\mathbb{R}},\{\cdot,\cdot\}_{\Lambda,R})$ relative to $\{\cdot,\cdot\}_{\Lambda,R}$.
\end{proposition}

\subsection{Jacobi--Lie systems on low dimensional manifolds}
Let us prove that every Lie system on the real line gives rise to a Jacobi--Lie system. In the case of two-dimensional manifolds, we display, with the aid of the GKO classification \cite{BBHLS,GKP92} of Lie algebras of vector fields on the plane given in Table \ref{table1} in Appendix 1, all the possible Vessiot--Guldberg Lie algebras related to Jacobi--Lie systems on the plane given in Table \ref{table9} in Appendix 1.

\begin{example}\normalfont
Let us show that the {\bf coupled Riccati equation}
\begin{equation}\label{coupRiceq2}
\frac{{\rm d}x_i}{{\rm d}t}=a_0(t)+a_1(t)x_i+a_2(t)x_i^2,\qquad i=1,\ldots,n,
\end{equation}
with $a_0(t),a_1(t),a_2(t)$ being arbitrary $t$-dependent functions,
%

 can be associated with a Jacobi--Lie system for $n=1$, which proves that every Lie system on the real line can be considered as a Jacobi--Lie system. Recall that (\ref{coupRiceq2}) is a Lie system with a Vessiot--Guldberg Lie algebra $V$ spanned by (\ref{VF1}). Observe that $V$ consists of Hamiltonian vector fields
with respect to a Jacobi manifold on $\mathbb{R}$ given by $
\Lambda=0$ and $R=\frac{\partial}{\partial x_1}.$
Indeed, the vector fields
$X_1,X_2,X_3\in V$ admit Hamiltonian functions
\begin{equation}
h_1=1,\qquad h_2=x_1,\qquad h_3=x_1^2.
\end{equation}
\end{example}

We now classify Jacobi--Lie systems $(\mathbb{R}^2,\Lambda,R,X)$, where we may assume $\Lambda$ and $R$ to be locally equal or different from zero. There exists just one Jacobi--Lie system with $\Lambda=0$ and $R=0$: $(\mathbb{R}^2,\Lambda=0,R=0,X=0)$. Jacobi--Lie systems of the form $(\mathbb{R}^2,\Lambda\neq 0,R=0)$ are Lie--Hamilton systems, whose Vessiot--Guldberg Guldberg Lie algebras were obtained in \cite{BBHLS}. In Table \ref{table9} in Appendix 1, we indicate these cases by writing {\it Poisson}. 
A Jacobi--Lie system $(\mathbb{R}^2,\Lambda=0,R\neq 0,X)$ is such that if $Y\in V^X$, then $Y=fR$ for certain $f\in C^\infty(\mathbb{R}^2)$. All cases of this type can be easily obtained out of the bases given in Table \ref{table1} in Appendix 1. We describe them by writing $(0,R)$ at the last column.

Propositions \ref{Prop1} and \ref{Prop2}, show that the Vessiot--Guldberg Lie algebras of Table \ref{table9} in Appendix 1 that do not fall into the mentioned categories are not Vessiot--Guldberg Lie algebras of Hamiltonian vector fields with respect to  Jacobi manifolds $(\mathbb{R}^2,\Lambda\neq 0,R\neq 0)$. This means that every $(\mathbb{R}^2,\Lambda,R,X)$ admits a Vessiot--Guldberg Lie algebra belonging to one of the previously mentioned classes\footnote{To exclude P$_1$ with $\alpha\neq 0$, we need a trivial modification of Proposition \ref{Prop2} using exactly the same line of thought.}.

\begin{lemma} Every Jacobi manifold on the plane with $R\neq 0$ and $\Lambda\neq 0$ admits a local coordinate system $\{s,t\}$ where $R=\partial_{s}$ and $\Lambda=\partial_s\wedge\partial_t$.
\end{lemma}
\begin{proof} Since it is assumed $R\neq 0$, there exist local coordinates $\{s,t_0\}$ on which $R=\partial_s$. Meanwhile, we have that $\Lambda=\Lambda(s,t_0)\partial_s\wedge\partial_{t_0}$. Since $[R,\Lambda]_{SN}=0$, we get that $\partial_s\Lambda=0$ and $\Lambda=\Lambda(t_0)$. Hence, $\Lambda=\Lambda(t_0)\partial_s\wedge\partial_{t_0}$. As we consider $\Lambda\neq 0$, we can define a new variable $t=t(t_0)$ such that ${\rm d}t/{\rm d}t_0\equiv \Lambda^{-1}(t_0)$. Finally, $\Lambda=\partial_{s}\wedge \partial_t$.
\end{proof}
\begin{definition} We call the local coordinate variables $\{s,t\}$ of the above lemma {\it local rectifying coordinates} of the Jacobi manifold on the plane.
\end{definition}
\begin{lemma} Let  $(\mathbb{R}^2,\Lambda,R)$ be a Jacobi manifold with $R_\xi\neq 0$ and $\Lambda_\xi\neq 0$ at every $\xi\in \mathbb{R}^2$. The Lie algebra morphism $\phi: C^\infty(\mathbb{R}^2)\rightarrow {\rm Ham}(\mathbb{R}^2,\Lambda, R)$ has non-trivial kernel. On local rectifying coordinates, we have $\ker \phi=\langle e^{t}\rangle$.
\end{lemma}
\begin{proof} If $f\in C^\infty(\mathbb{R}^2)$ belongs to $\ker \phi$, then  $\widehat {\Lambda} ({\rm d}f)+fR=0$. In local rectifying coordinates, we get

$
\partial_sf\partial_t-\partial_tf\partial_s+f\partial_s=0\Rightarrow \left\{\begin{array}{c}\partial_sf=0\\\partial_tf=f\end{array}\right.\Rightarrow f=\lambda e^{t},\qquad \lambda\in \mathbb{R}.
$
\end{proof}
\begin{proposition}\label{Prop1} If a Lie system on the plane is related to a Vessiot--Guldberg Lie algebra $V$ of vector fields containing two non-zero vector fields $X_1,X_2$ satisfying
$[X_1,X_2]=X_1$ and $X_1\wedge X_2=0$, then $V$ does not consist of Hamiltonian vector fields relative to any Jacobi manifold with $R\neq 0$ and $\Lambda\neq 0$.
\end{proposition}
\begin{proof}Let us take a rectifying coordinate system for $(\mathbb{R}^2,\Lambda,R)$.  Since $\phi$ is a morphism of Lie algebras, we get that $X_1,X_2$ amount to the existence of non-zero functions $h_1$ and $h_2$ such that $\{h_1,h_2\}_{\Lambda,R}=h_1+g$, where $g\in \ker \phi$. So,
\begin{equation}
\{h_1,h_2\}_{\Lambda,R}=\Lambda({\rm d}h_1,{\rm d} h_2)+h_1Rh_2-h_2Rh_1=h_1+g  .
\end{equation}
Meanwhile, $X_1\wedge X_2=0$ implies that
\begin{equation}
\widehat{\Lambda}({\rm d}h_1)\wedge \widehat{\Lambda}({\rm d}h_2)+R\wedge [h_1\widehat {\Lambda}({\rm d}h_2)-h_2\widehat {\Lambda}({\rm d}h_1)]=0.
\end{equation}
Using local rectifying coordinates, we see that $\widehat{\Lambda}({\rm d}h_i)=(Rh_i)\partial_t-\partial_th_iR$  and $R\wedge \widehat {\Lambda}({\rm d}h_i)=(Rh_i)\Lambda$ for $i=1,2$. Hence,
\begin{equation}
\left[(Rh_1)\partial_t-\partial_th_1R\right]\wedge \left[(Rh_2)\partial_t-\partial_th_2R\right] +[h_1(Rh_2)-h_2(Rh_1)]\Lambda =0
\end{equation}
and

\begin{equation}\begin{gathered}
0=(Rh_1\partial_th_2-Rh_2\partial_th_1)\Lambda +(h_1Rh_2-h_2Rh_1)\Lambda
\\
\Updownarrow
\\
 \Lambda({\rm d}h_1,{\rm d}h_2)+h_1Rh_2-h_2Rh_1=0.\end{gathered}\end{equation} 
This amounts to $\{h_1,h_2\}_{\Lambda,R}=0$, which implies that $0=h_1+g$ and $X_1=0$. This is impossible by assumption and $X_1$ and $X_2$ cannot be Hamiltonian. 
\end{proof}

%
\begin{proposition}\label{Prop2} There exists no Jacobi manifold on the plane with $\Lambda\neq 0$ and $R\neq 0$ turning the elements of a Lie algebra diffeomorphic to $\langle \partial_x,\partial_y,x\partial_x+\alpha y\partial_y\rangle$, with $\alpha\notin \{0,-1\}$, into Hamiltonian vector fields.
\end{proposition}

\begin{proof} Let us proceed by reduction to absurd. Assume $(N,\Lambda,R)$ to be a Jacobi manifold turning the above mentioned vector fields into Hamiltonian ones. Taking local rectifying coordinates for the Jacobi manifold, it turns out that every Lie algebra diffeomorphic to the previous one can be written in terms of three vector fields $X_1,X_2, X_3$
diffeomorphic to $\partial_x,\partial_y,x\partial_x+\alpha y \partial_y$. Then, they satisfy
$
[X_1,X_2]=0, [X_1,X_3]=X_1, [X_2,X_3]=\alpha X_2, X_1\wedge X_2\neq 0
$
and
\begin{equation}\label{mu}
X_3=\mu_2X_1+\alpha\mu_1X_2
\end{equation}
for certain first-integrals $\mu_1$ and $\mu_2$ for $X_1$ and $X_2$, respectively. From \eqref{mu} and using that $X_1,X_2,X_3$ are Hamiltonian for certain Hamiltonian functions $h_1,h_2,h_3$, correspondingly, we obtain
\begin{equation}
\widehat{\Lambda}({\rm d}h_3)+h_3R=\mu_2\widehat{\Lambda}({\rm d}h_1)+\mu_2h_1R+\alpha\mu_1\widehat{\Lambda}({\rm d}h_2)+\alpha\mu_1h_2R
\end{equation}
and, by means of the rectified expression for $\Lambda$ and $R$, we get
\begin{equation}\label{sys1}
\begin{gathered}
\frac{\partial h_3}{\partial s}=\mu_2\frac{\partial h_1}{\partial s}+\alpha\mu_1\frac{\partial h_2}{\partial s},\\
\frac{\partial h_3}{\partial t}=\mu_2\frac{\partial h_1}{\partial t}+\alpha\mu_1\frac{\partial h_2}{\partial t}-\mu_2h_1-\alpha\mu_1h_2+h_3.
\end{gathered}
\end{equation}

Since $[X_1,X_3]=X_1$, then $\{h_1,h_3\}_{\Lambda,R}=h_1+\lambda_1e^t$, where $e^t$ is a function with zero Hamiltonian vector field and $\lambda_1\in\mathbb{R}$. Hence,
\begin{equation}
h_1+\lambda_1e^t=\{h_1,h_3\}_{\Lambda,R}=\Lambda({\rm d}h_1,{\rm d}h_3)+h_1(Rh_3)-h_3(Rh_1).
\end{equation}
Simplifying and using previous expressions (\ref{sys1}), we find that

\begin{equation}
h_1+\lambda_1e^t=\mu_1\left(\frac{\partial h_2}{\partial t}\frac{\partial h_1}{\partial s}-\frac{\partial h_1}{\partial t}\frac{\partial h_1}{\partial s}+h_1\frac{\partial h_2}{\partial s}-h_2\frac{\partial h_1}{\partial s}\right)=\alpha\mu_1\{h_1,h_2\}_{\Lambda,R}.
\end{equation}

As $[X_1,X_2]=0$, then $\{h_1,h_2\}_{\Lambda,R}=\lambda e^t$ for a certain constant $\lambda\in\mathbb{R}$. Hence,
$
h_1=(\alpha\mu_1\lambda-\lambda_1)e^t
$
and $\lambda\neq0 $. Proceeding analogously for $\{h_2,h_3\}_{\Lambda,R}=\alpha(h_2+\lambda_2e^t)$, we obtain
\begin{equation}
h_2+\lambda_2e^t=\mu_2\{h_2,h_1\}_{{\Lambda,R}}\Rightarrow h_2=(-\mu_2\lambda/\alpha-\lambda_2)e^t.
\end{equation}

Writing the compatibility condition for the system (\ref{sys1}), we reach to
\begin{equation}
(\alpha+1)\lambda\left(\frac{\partial \mu_1}{\partial s}\frac{\partial \mu_2}{\partial t}-\frac{\partial \mu_2}{\partial s}\frac{\partial \mu_1}{\partial t}\right)=0.
\end{equation}
This implies that ${\rm d}\mu_1\wedge {\rm d}\mu_2=0$. Since $\mu_1$ is a first-integral for $X_1$ and $\mu_2$ is a first-integral for $X_2$, this means that $X_1\wedge X_2=0$, which is impossible by assumption. This finishes the proof.
\end{proof}

 \chapter{Lie symmetry for differential equations}\markboth{Lie symmetry for differential equations}{Chapter 4}
\label{Chap:LieSymm}
\section{What can a symmetry transformation achieve?}
\setcounter{equation}{0}
\setcounter{theorem}{0}
\setcounter{example}{0}
\bigskip

Invariance of a differential equation under a transformation is synonymous of {\it existence of symmetry} and, consequently, of {\it conserved quantities}.
The invariance of a differential equation under a transformation \cite{Nucci3} helps us to achieve partial or complete integration of such an equation.
 Many of the existing solutions to physical phenomena described by differential equations have been obtained through symmetry arguments \cite{Olver,PS,Stephani}.
 
A conserved quantity for a first-order differential equation can lead to its integration by quadrature, whilst
for higher-order ones, it leads to a reduction of their order \cite{Olver,Stephani}. Nevertheless, finding conserved quantities is a nontrivial task.

The most famous and established method for finding point symmetries is the {\it classical Lie symmetry method} (CLS) developed by Lie in 1881 \cite{Lie81,Lie90,Lie93,Olver,Stephani}.
Although the CLS represents a very powerful tool, it yields cumbersome calculations to be solved by hand. Notwithstanding, the increased number of available software packages for symbolic calculus
has made of generalizations of the CLS analysis very tractable approaches to find conservation laws, explicit solutions, etc.

If we impose that symmetries leave certain submanifold invariant, we find the class of {\it conditional or nonclassical symmetries method} (NSM) introduced in 1969 by
Bluman and Cole \cite{BC1}. The origin of this proposal rooted in the obtention of exact solutions for the linear
{\it heat equation} \cite{Evans} that were not deducible from the former CLS.
The NSM method has successfully been applied by many authors \cite{ArriBeck,ClarkWinter,Nucci1, NucciAmes}
and the author of this manuscript in \cite{EstLejaSar,EstLejaSar2}. A few examples will be displayed in forthcoming sections.
A remarkable difference between the classical and the nonclassical Lie symmetry analysis is that the latter provides us with 
no longer linear systems of differential equations to calculate the symmetries.


During the last decades of the XX century, attention has been focused on Lie symmetry computation methods for
nonlinear evolution equations of hydrodynamic type
in Plasma Physics, Cosmology and other fields \cite{AbloClark,BK,Dissertationes}. These equations have solutions in form of
{solitons}. In order to study and to derive solutions of this kind of equations, the classical and nonclassical Lie symmetry approaches have proven their efficiency.
Given their recurrent appearance in the Physics and Mathematics literature,
it is important to settle an algorithmic method for the calculation of their symmetries, eventual reduction and search of their solutions.
For this reason, we will dedicate this part of this thesis to calculating symmetries and reducing differential equations,
ranging from ODEs to hierarchies of PDEs and
Lax pairs. In this way, the plan of the chapter goes as follows.

{\bf Section 1: The CLS analysis for ODEs and applications}: We explain the CLS for ODEs.
We consider Lie transformations of the most general type and we give theorems which facilitate the calculation of Lie symmetries. 
A number of steps will summarize the quasi-algorithmical symmetry search and it will be applied to some particular ODEs:
the second- and third-order Kummer--Schwarz equations, the first- and second-order Riccati equations and the Milne--Pinney equations, revised in Chapter \ref{Chap:LieSystems}.
 
{\bf Section 2: Lie symmetry for Lie systems}: We inspect certain types of Lie symmetries for Lie systems. We prove that every Lie system admits a Lie algebra of Lie symmetries  whose geometric properties are determined by the Vessiot--Guldberg Lie algebra of the Lie system. 
To illustrate our results, we analyze a particular type of Lie symmetries for $\mathfrak{sl}(2,\mathbb{R})$-Lie systems and we apply our findings to inspect
Riccati equations, the Cayley-Klein Riccati equations (defined for the first time in this thesis), quaternionic Riccati equations with real $t$-dependent coefficients, generalized Darboux-Brioschi-Halphen systems
and second-order Kummer--Schwarz equations.
 Next, we analyze a particular type of Lie symmetries for Aff($\mathbb{R}$)-Lie systems and we apply our results to Buchdahl equations.
To conclude, we generalize the search of this type of Lie symmetries to PDE Lie systems. In a similar fashion, we search for a particular type of Lie symmetries and illustrate
the interest of our theory by studying partial Riccati equations. 

It is remarkable that the application of our methods to Lie systems, leads to obtaining Lie symmetries for all Lie systems sharing isomorphic Vessiot--Guldberg Lie algebras.
In this way, we are concluding results for several different Lie systems with the same Vessiot--Guldberg Lie algebra, simultaneously.

{\bf Section 3: The CLS and NSM for PDEs}: This section contains the theory for Lie point symmetry search applied to PDEs. We explain in a number of steps
how to apply the method and to implement the difference between the CLS and the NSM.
For PDEs, we can reduce the system to lower dimension by getting rid of one of the independent variables through symmetry arguments. We will explain the process of reduction
with the aid of the method of characteristics. We will illustrate the calculation of classical Lie symmetries for PDEs with the example of the inviscid Burgers equation.

{\bf Section 4: Application to PDEs and spectral problems in Physics}:  We start by applying both the CLS and NSM analysis to nonlinear PDEs  of hydrodynamical type, and their corresponding Lax pairs.
 We start with the example of the Bogoyanlevski--Kadomtsev--Petviashvili equation in  $2+1$ dimensions ($2+1$-BKP, henceforth) and its corresponding Lax pair, a two component, nonisospectral problem defined over the complex field, in $2+1$ dimensions.
 The author of this thesis and her collaborators have obtained both the classical and nonclassical symmetries and have compared them.
 Through the obtained symmetries, we have reduced both the equation and Lax pair and we have achieved several reductions. Two of them are
 of special interest.
 The first corresponds with the KdV equation in $1+1$ dimensions, what implies that $2+1$-BKP is a generalization 
 of the $1+1$ KdV equation.
The second interesting reduction can be interpreted from a physical point of view.  We proceed by studying this reduction and its associated Lax pair in $1+1$ dimensions, with a second iteration
through the Lie symmetry calculus (both classical and nonclassical) and later reduction to an ODE
 (considering the nonisospectral parameter as another independent variable). 
Specifically, we prove that the nonclassical symmetries are more global and contain
 the classical case, as a particular. A second list of reductions is shown for the $1+1$ dimensional case.

{\bf Section 5: Application to spectral problems associated with hierarchies of PDEs}: Next, we apply the CLS and NSM to complete hierarchies
 of nonlinear PDEs and their corresponding Lax pairs.
A {\it hierarchy} is a set of differential equations that are related through a {\it recursion operator} \cite{CHJ}. The $n$-th iteration
 of the recursion operator provides us with the $n$-th member of the hierarchy.
 In particular, we take a look at two examples of hierarchies of PDEs of a similar form:
 the so called hierarchy of Camassa--Holm in $2+1$ dimensions \cite{CH1} and the Qiao hierarchy or modified Camassa--Holm hierarchy in $2+1$ dimensions \cite{Qiao2,Qiao3,Qiao1}
  (they are correspondingly denoted as CHH$(2+1)$ and Qiao$2+1$ or mCHH($2+1$)).
We will make an exhaustive description of the nonclassical symmetries attending to different values of the functions 
 and constants of integration present in the symmetries.

A number of reductions arise for both hierarchies. 
We will show the reductions of their corresponding Lax pairs and we check whether the nonisospectral character is inherited within lower dimensions.
 For CHH($2+1$), among its reductions we find the positive and negative Camassa--Holm hierarchies in $1+1$ dimensions
 and their corresponding Lax pairs \cite{EstLejaSar}.
Some of the reductions of the Lax pair will maintain their nonisospectral character, what is quite inusual for $1+1$ dimensions. For the mCHH($2+1$), reductions are physically interpretated too \cite{EstLejaSar1}.

\section{The CLS for ODEs} 
\setcounter{equation}{0}
\setcounter{theorem}{0}
\setcounter{example}{0}
%

Consider a $p$-th order system of $q$ ordinary differential equations, with an independent variable $t$ and $k$ dependent fields $u_1(t),\ldots,u_k(t),$ 
given by
\begin{equation}\label{prODE}
\Psi^{l}(t,u_j^{r)}(t))=0,\qquad l=1,2,\dots,q,
\end{equation}
where 
$$
u_j^{r)}\!\!=\frac{\partial^r u_j}{\partial t^r},\qquad r=1,\ldots,p,\qquad j=1,\dots,k.
$$
and occasionally we will use $u=(u_1,\ldots,u_k).$ 
Eventually, we will also write $(u_j)_{\!\!\!\KeepStyleUnderBrace{t\dots t}_{r-\text{times}}}\!\!$ for $\partial^ru_j/\partial t^r$.
A solution to this system of $p$-th order ODEs is a set of functions $u_j(t)$ such that 
\begin{equation}\label{PS}
\Psi^{l}\left(t,u_j(t),u_j^{1)}(t),\ldots,u_j^{p)}(t)\right)=0,\qquad l=1,\ldots,q,\quad j=1,\dots,k.
\end{equation}

Geometrically, the variables $\{t,u^{r)}_j(t)\,|\,j=1,\ldots,k,r=1,\ldots,p\}$ can be understood as a coordinate system for ${\rm J}^p\pi\simeq \mathbb{R}\times \text{T}^pN$ with $\pi:(t,u_1,\ldots,u_k)\in N_\mathbb{R}\equiv \mathbb{R}\times N\mapsto t\in\mathbb{R}$. Recall that we write 
\begin{equation}
j^p_tu\equiv (t,u_j,u_j^{1)},\ldots,u_j^{p)}),\quad \forall j
\end{equation}
for a general point of ${\rm J}^p\pi$.
Moreover, the differential equation (\ref{prODE}) can be understood as a submanifold $\mathcal{E}=\Psi^{-1}(0)\subset {\rm J}^p\pi$, where $\Psi=(\Psi^1,\ldots,\Psi^q)$. 
Meanwhile, the particular solutions of (\ref{prODE}) are can be understood sections $u:t\in \mathbb{R}\rightarrow (t,u(t))\in N_\mathbb{R}$ of the bundle $(N_\mathbb{R},\mathbb{R}, \pi)$ whose prolongation to ${\rm J}^p\pi$ satisfies (\ref{prODE}), i.e. we have (\ref{PS}).

Let us consider the $\epsilon$-parametric
group of transformations 
\begin{equation}
\left\{
\begin{aligned}\label{pointtransep211}
\bar {t}&\rightarrow t+\epsilon \xi(t,u_j)+O(\epsilon^2),\\
{\bar u}_j&\rightarrow u_j+\epsilon \eta_{j}(t,u_j)+O(\epsilon^2),\\
\end{aligned}\right.\qquad j=1,\dots,k.
\end{equation}

The associated vector field with the trasformation reads
\begin{equation}
 X=\xi(t,u_j)\frac{\partial}{\partial t}+ \sum_{j=1}^k\eta_{j}(t,u_j)\frac{\partial}{\partial u_j}
\end{equation}


\begin{definition}
We call a transformation \eqref{pointtransep211} a {\it Lie point transformation} if the coefficients $\xi(t,u_j)$ and $\eta_{j}(t,u_j)$ 
do not depend on higher-order derivatives $u_j^{r)}$ for $r=1,\dots,p$.
 \end{definition}

Up to $p$-order derivatives, transformation \eqref{pointtransep211} reads
\begin{equation}
\left\{
\begin{aligned}\label{pointtransep21}
t&\rightarrow t+\epsilon \xi+O(\epsilon^2),\\
{u}_j&\rightarrow u_j+\epsilon \eta_{j}(t,u_j)+O(\epsilon^2),\\
{u}^{1)}_j&\rightarrow {u}^{1)}_j+\epsilon \eta_{j}^{1)}(t,u_j)+O(\epsilon^2),\\
{u}^{2)}_j&\rightarrow {u}^{2)}_j+\epsilon {\eta}_{j}^{2)}(t,u_j)+O(\epsilon^2),\\
& 
\dots\\
{u}^{p)}_j\!\!&\rightarrow {u}^{p)}_j\!\!+\epsilon \eta_{j}^{p)}(t,u_j)\!\!+O(\epsilon^2), 
\end{aligned}\right.\qquad i=1,\dots,k.
\end{equation}
on ${\rm J}^p\pi$. 
 We shall drop the dependency $(t,u_j)$ in $\xi(t,u_j), \eta_{j}(t,u_j), \eta_{j}^{1)}(t,u_j), \dots$, etc., for
convenience and express them simply as $\xi$, $\eta_j$, $\eta_{j}^{1)}, \dots$, etc., in the forthcoming expressions.
Geometrically, this amounts to a Lie group action $\Phi:\mathbb{R}\times {\rm J}^p\pi \rightarrow {\rm J}^p\pi.$

The infinitesimal generator of this $\epsilon$-parametric group of transformation reads
%
\begin{equation}
\label{compsymvf1}
{X}^p=\xi\frac{\partial}{\partial t}+\sum_{j=1}^k \left(\eta_{j}\frac{\partial}{\partial u_j}+\eta_{{j}}^{1)}\frac{\partial}{\partial u_j^{1)}}+ \eta_{j}^{2)}\frac{\partial}{\partial u^{2)}_{j}}
 +\dots+  \eta^{p)}_{j}\frac{\partial}{\partial u^{p)}_{j}}\right),
\end{equation}
where we have not specified the dependency of the coefficients of the infinitesimal generator, due to simplicity.

Recall that we gave in \eqref{prolongcomplete} in Chapter \ref{Chap:GeomFund} how to obtain the coefficients $\eta_j^{p)}$.
 From a geometric viewpoint, transformation \eqref{pointtransep211} can be viewed as the infinitesimal part of a Lie group action $\Phi:(\epsilon;t;u)\in \mathbb{R}\times N_\mathbb{R}\mapsto (\bar t,\bar u)\equiv \Phi(\epsilon;t,u)\in N_\mathbb{R}$. Meanwhile, transformation \eqref{pointtransep21} can be understood as a Lie group action 
$\widehat{\Phi}:(\epsilon;j^p_tu)\in \mathbb{R}\times {\rm J}^p\pi\mapsto j^p_{\bar t}\bar u\equiv \widehat{\Phi}(j^p_tu)\equiv \widehat{\Phi}(\epsilon;j^p_tu)\in {\rm J}^ p\pi$.
\begin{definition}\label{Def:LiePoint}
We say that transformation \eqref{pointtransep21} is a {\it Lie point symmetry} of (\ref{prODE}) if it maps points of $\mathcal{E}$ into points of $\mathcal{E}$, namely
\begin{equation}
\Psi^l(t,u_j(t),u_j^{r)}(t)) \Rightarrow \Psi^l(\bar{t},\bar{u}_j(\bar t),\bar{u}_j^{r)}(\bar t)),\qquad \forall j, r, l.
\end{equation} 
\end{definition}

Notice that the invariance under transformations is independent of the choice of the coordinate system in which the system of ODEs and its solutions are given.  Only the explicit form
of the symmetry will change.

\begin{lemma}
Given a Lie point symmetry (\ref{pointtransep21}), we have that 
if $u_j(t)$ is a solution of (\ref{prODE}), then  $\bar u_j(\bar t)$ for all $j=1,\dots,k$, is a new solution of the same equation.
\end{lemma}

\begin{theorem}
We say that the infinitesimal generator ${X}^p$ is a {\it Lie point symmetry} of the system of ODEs of $p$-order  given by (\ref{prODE})
 if
\begin{equation}
{X}^p(\Psi^l)=0,\qquad\qquad  l=1,\dots,q,
\end{equation}
on the submanifold $\mathcal{E}=\Psi^{-1}(0)$, with $\Psi=(\Psi^1,\ldots,\Psi^q)$. 
\end{theorem}
\begin{proof}
According to Definition \ref{Def:LiePoint} of Lie point symmetry, if $\Psi(j^p_tu)=0$, then $\Psi(\widehat{\Phi}(j^p_tu))=0$ for every $\epsilon\in \mathbb{R}$.
 Hence,
%
\begin{equation}
\frac{d}{d\epsilon}\Big|_{\epsilon=0}\Psi^l(\widehat{\Phi}(j_t^pu))=\frac{d}{d\epsilon}\Big|_{\epsilon=0}\Psi^l(j_t^pu)=0, 
\end{equation} 
$\forall j^ p_tu\in \Psi^{-1}(0),$ and $l=1,\ldots,q.$ Meanwhile,
\begin{align}
\frac{d}{d\epsilon}\Big|_{\epsilon=0}\Psi^l(\widehat{\Phi}(j_t^pu))&=d\Psi^l\Big|_{j_t^pu}\left(\frac{d}{d\epsilon}\widehat{\Phi}(j_t^pu)\right)\nonumber\\
&=d\Psi^l\Big|_{j^p_tu}({X^ p}(j_t^pu))={X^p}(\Psi^l)(j_{\bar t}^p\bar u),
\end{align}
for all $j_t^pu\in \Psi^{-1}(0)$ and $l=1,\ldots,q$.  Therefore, ${X^p}\Psi^l=0$ on $\Psi^{-1}(0)$.
\end{proof}

\begin{definition}
We call {\it invariant surface condition} of a system of  ODEs the relation
\begin{equation}\label{invsurfcondODE}
\eta_{{j}}-(u_{j})_t\xi=0,\qquad j=1,\ldots,k,
\end{equation}
\end{definition}

\subsection{Algorithmic computation of Lie point symmetry for systems of ODEs}

It is possible to follow an algorithmic procedure to compute the Lie point symmetries of $p$-order systems of ODEs. The process goes as follows.
\begin{enumerate}
\item Introduce the Lie point transformation \eqref{pointtransep21} in \eqref{prODE}.
\item Select the terms in the zero-order in $\epsilon$, which retrieves the initial equations.
 Isolate the higher-order derivative, $u^{p)}_{j}$, appearing in each of the untransformed equations in system \eqref{prODE}. 
\item Select the terms in the first-order of $\epsilon$. Introduce the prolongations computed according to \eqref{prolongcomplete} given in
Chapter \ref{Chap:GeomFund}, together with the higher-order derivatives $u^{p)}_{j}$ obtained in the previous step.
 Take the coefficients of the remaining derivatives of the dependent
variables in different orders, from the highest-order remaining, $u^{p-1)}_{j}$ to the first-order $u^{1)}_{j}$ for $1\leq j\leq k$.
It is legit to set each of these conditions equal to zero, given the definition
of Lie point  symmetries in which $\xi$ and $\eta_{j}$ do not depend on derivatives of any order.
\item  At this point, we decide for a classical or nonclassical approach. For the nonclassical, we implement the invariant surface condition \eqref{invsurfcondODE}, 
namely $u_{j}^{1)}=\eta_{j}/\xi$, with $j=1,\ldots,k$.
This leads to an overdetermined system of differential equations (which is no longer linear) for the coefficients $\xi$ and $\eta_{j}$ of the infinitesimal generator.

The more possible conditions we impose, the more general the symmetries will be. {\it This implies that nonclassical symmetries
are more general than the classical ones and they must include them as a particular case.}
\item Solving the resulting system is not an easy task, specially in the nonclassical approach, but it leads to the desired symmetries.
\end{enumerate}
We aim to apply the theory exposed for the CLS to some ODEs with physical relevance.  
The equations we inspect have been devised in Chapter \ref{Chap:LieSystems} of this manuscript.

\medskip

From now on, we will use the following notation in the forthcoming examples and sections. We will denote the sucessive derivatives
with respect to $t$ as $u_j^{r)}=(u_j)_{\!\!\!\KeepStyleUnderBrace{t,\dots,t}_{r-\text{times}}}\!\!$.

\begin{example}\normalfont
Let us study the {\bf second-order Kummer--Schwarz equation} \cite{Be88,LSKummer,Ma94,Ta89}
\begin{equation}\label{2KS}
u_{tt}-\frac{3}{2}\frac{u_t^2}{u}+2c_0u^3-2c_1(t)u=0,
\end{equation}
with $c_0$ being a real constant and $c_1(t)$ being any $t$-dependent function.
It is a particular case of the Gambier equations \cite{CGL11,GCG} and appears in cosmological models \cite{NR02}.
We search for a general Lie point symmetry for (\ref{2KS}). As usual, we denote by $\xi$ the coefficient of the infinitesimal generator associated with $t$
and the one associated with $u$ is denoted by $\eta$. Hence, the infinitesimal $\epsilon$-parametric transformation reads
\begin{equation}\label{symtrans1}
\left\{
\begin{aligned}
\bar{t}&\rightarrow t+\epsilon \xi(t,u),\\ \bar{u}&\rightarrow u+\epsilon \eta(t,u).\end{aligned}\right.
\end{equation}
In similar fashion, the functions present in the transformation must be expanded in a Taylor expansion up to first-order in $\epsilon$
\begin{equation}\label{symtrans2}
c_1(\bar t)\rightarrow c_1(t)+\epsilon \frac{dc_1(t)}{dt}\xi(t,u).
\end{equation}
Since \eqref{2KS} is a second-order equation, we prolong this transformation to first and second-order derivatives as well,
\begin{equation}\label{symtrans3}
\left\{\begin{aligned}
\bar{u}_{\bar t}&\rightarrow u_t+\epsilon \eta_{t}(t,u),\\ {\bar u}_{\bar{t} \bar{t}}&\rightarrow u_{tt}+\epsilon \eta_{tt}(t,u)\end{aligned}
\right.
\end{equation}
and we calculate the prolongations $\eta_{t}$ and $\eta_{tt}$ according to \eqref{prolongcomplete} in Chapter \ref{Chap:GeomFund}.
Introducing \eqref{symtrans1} into \eqref{2KS} leads us to an overdetermined system which can
only be solved in case of considering particular cases of symmetry.
Let us for example suppose that $\xi(t,u)$ is not a function of $u$, i.e., $\xi=\xi(t).$ Hence,
\begin{equation}\label{symm2KS}
\xi(t)=-G_{1}(t)-K_1,\quad \eta(t,u)=\frac{dG_1}{dt}u,
\end{equation}
for a certain function $G_1=G_1(t)$.
\end{example}

\begin{example}\normalfont
The {\bf third-order Kummer-Schwarz} equation reads
\begin{equation}\label{3KS}
u_{ttt}-\frac{3}{2}\frac{u_{tt}^2}{u}+2C_0(u)u_t^3-2b_0(t)u_t=0.
\end{equation}
We propose the same Lie transformation as in \eqref{symtrans1}. This time, we need
to expand the functions $C_0(u)$ and $b_0(t)$
\begin{equation}\label{symmtrans2}
 C_0(\bar u)\rightarrow C_0(u)+\epsilon \frac{\partial C_0(u)}{\partial u}\eta(t,u),\quad b_0(\bar u)\rightarrow b_0(u)+\epsilon \frac{\partial b_0(t)}{\partial t}\xi(t,u).
\end{equation}
Since \eqref{3KS} is a third-order equation, transformation \eqref{symtrans1} has to be extended up to the third-order derivative 
apart from considering \eqref{symtrans3},
\begin{equation}\label{symmtrans3}
\bar{u}_{\bar{t} \bar{t} \bar{t}}\rightarrow u_{ttt}+\epsilon \eta_{ttt}(t,u),
\end{equation}
where $\eta_{ttt}$ is calculated according to \eqref{prolongcomplete} in Chapter \ref{Chap:GeomFund}.
Invariability of \eqref{3KS} under such a transformation arises in an overdetermined 
system of differential equations whose unique possibility reads
\begin{equation}\label{symm3KS}
 \xi=K_1,\quad \eta(u)=K_2u,
\end{equation}
with $K_1$ and $K_2$ being two constants and the two following conditions being satisfied
\begin{equation}\label{3KScond}
 \frac{db_0(t)}{dt}=0,\quad C_0(u)=\frac{C_1}{u^2},
\end{equation}
where $C_1$ is another constant of integration. Conditions \eqref{3KScond} had been known previously \cite{LSKummer}, but here we have retrived them
by our Lie symmetry approach which strictly needs \eqref{3KScond} being satisfied in order to \eqref{symtrans3}, \eqref{symmtrans2} and \eqref{symmtrans3} be a Lie symmetry transformation.

\end{example}

\begin{example}\normalfont
Consider now the {\bf first-order Riccati equation}, 
\begin{equation}\label{1ricc}
u_t=a_0(t)+a_1(t)u+a_2(t)u^2.
\end{equation}
This equation is very important in Theoretical Physics, it appeared in the introduction of Witten's Supersymmetric Quantum Mechanics \cite{Witten}. 
A general Lie point symmetry transformation for these equations reads
\begin{equation}\label{symtrans4}\left\{
\begin{aligned}
\bar{t}&\rightarrow t+\epsilon \xi(t,u),\\ \bar{u}&\rightarrow u+\epsilon \eta(t,u).
\end{aligned}\right.
\end{equation}
Applying this transformation to the time-dependent functions of \eqref{1ricc},
\begin{equation}
a_i(\bar t)\rightarrow a_i(t)+\epsilon \frac{\partial a_i(t)}{\partial t} \xi(t,u),\quad i=0,1,2.
\end{equation}
Introducing this transformation into \eqref{symtrans4}, we need to compute the first-order prolongation
\begin{equation}
\bar{u}_{\bar t}\rightarrow u_{t}+\epsilon \eta_{t}(t,u)
\end{equation}
with the computation of $\eta_{t}$ according to \eqref{prolongcomplete} in Chapter \ref{Chap:GeomFund}. A resulting nontrivial system of equations for the symmetries
suggests the consideration of an Ansatz $\xi=\xi(t)$.
This leads to
\begin{equation}
\xi(t)=\frac{1}{\sqrt{a_2(t)}\sqrt{a_0(t)}},\quad \eta(t,u)=\left(-\frac{1}{2}\frac{\frac{da_2(t)}{dt}}{a_2^{3/2}\sqrt{a_0(t)}}+\frac{1}{2}\frac{\frac{da_0(t)}{dt}}{\sqrt{a_2(t)}a_0(t)^{3/2}}\right)u.
\end{equation}
\end{example}
\begin{example}\normalfont
The {\bf second-order Riccati equation} accounts with multiple applications in scientific fields (see Chapter \ref{Chap:LieSystems} and \cite{CC87,GL99}). It takes the form
\begin{equation}\label{2Ricc}
u_{tt}+(f_0(t)+f_1(t)u)u_t+C_0(t)+C_1(t)u+C_2(t)u^2+C_3(t)u^3=0,
\end{equation}
with
\begin{equation}\label{funct2Ricc}
f_1(t)=3\sqrt{C_3(t)},\quad f_0(t)=\frac{C_2(t)}{\sqrt{C_3(t)}}-\frac{1}{2C_3(t)}\frac{dC_3}{dt}(t),
\end{equation}
where $C_1(t),C_2(t)$ are arbitrary $t$-dependent functions and $C_3(t)>0$ \cite{PW}. Conditions \eqref{funct2Ricc} are retrieved from the property
of integrability of the equation in the Painlev\'e sense.
We search for a transformation as in \eqref{symtrans4}, which needs to be prolongated up to the second derivative as
\begin{equation}\label{symtrans7}
{\bar u}_{\bar t}\rightarrow u_{t}+\epsilon \eta_{t}(t,u),\quad \bar{u}_{\bar{t} \bar{t}}\rightarrow u_{tt}+\epsilon \eta_{tt}(t,u),
\end{equation}
with $\eta_{t}$ and $\eta_{tt}$ being calculated with the aid of \eqref{prolongcomplete} in Chapter \ref{Chap:GeomFund}.
The functions appearing in \eqref{2Ricc} are Taylor expanded up to first order as
\begin{equation}\label{symtrans5}
C_j(\bar t)\rightarrow C_j(t)+\epsilon \frac{d C_j(t)}{d t} \xi(t,u),\quad j=1,2,3.
\end{equation}
and
\begin{equation}\label{symtrans6}
f_0(\bar t)\rightarrow f_0(t)+\epsilon \frac{d f_0(t)}{d t} \xi(t,u),\quad f_1(\bar t)\rightarrow f_1(t)+\epsilon \frac{d f_1(t)}{d t} \xi(t,u).
\end{equation}
We consider the Ansatz $\xi=\xi(t)$.
Under such a circumstance, we find
\begin{equation}
\xi(t)=F(t),\quad \eta(t,u)=g_1(t)u+g_2(t),
\end{equation}
where 
\begin{equation}
{F(t)=-\frac{\int{\sqrt{C_3(t)}g_1(t)dt}}{\sqrt{C_3(t)}}}.
\end{equation}
The function $g_2(t)$ is not arbitrary, it satisfies that
\begin{multline}
g_2(t)=-\frac{\frac{dg_1(t)}{dt}}{\sqrt{C_3(t)}}+\frac{1}{3}\frac{\frac{dC_2(t)}{dt}\int{\sqrt{C_3(t)}g_1(t)dt}}{C_3(t)^{3/2}}
\\
+\frac{1}{3}\frac{C_2(t)g_1(t)}{C_3(t)}-\frac{1}{3}\frac{C_2(t)\frac{dC_3(t)}{dt}\int{\sqrt{C_3(t)}g_1(t)dt}}{C_3(t)^{5/2}}.
\end{multline}
Nevertheless, the nonarbitrary function $g_1(t)$ has to satisfy a cumbersome equation whose explicit expression
is omitted but can be straightforwardly obtained after a long but easy calculation. Given this
situation, we try to simplify our result by imposing further additional conditions. For example, let us suppose that $g_2(t)=0$.
Now, the symmetries have a more pleasant expression
\begin{equation}
\xi(t)=F(t),\quad \eta(t,u)=g_1(t)u,
\end{equation}
with the additional relations
\begin{equation}
F(t)=-\frac{\int{\sqrt{C_3(t)}g_1(t)dt}}{\sqrt{C_3(t)}},\quad g_1(t)=\frac{C_2(t)e^{\frac{1}{3}\int{\frac{C_2(t)}{\sqrt{C_3(t)}}dt}}}{C_3(t)}.
\end{equation}
The process of symmetry calculation recovers \eqref{funct2Ricc} in order 
to 
\labelcref{symtrans4,symtrans5,symtrans6,symtrans7} be a symmetry of \eqref{2Ricc}. 
Relations \eqref{funct2Ricc} had previously been devised in other papers \cite{CLS12}.
\end{example}

\begin{example}\normalfont
The {\bf Milne--Pinney equation}, namely
\begin{equation}\label{MPsys}
 u_{tt}+\omega(t)^2u-\frac{K}{u^3}=0.
\end{equation}
where $\omega(t)$ is the frequency of oscillation of the system and $K>0$ is a constant, was studied in Chapter \ref{Chap:LieSystems} \cite{SIGMA,CLS122,Pi50}.
It is defined in $\mathbb{R}-\{0\}$ and it is invariant under parity. That is, for a particular solution $x(t)$, we have that $\bar{x}(t)=-x(t)$ is another solution.

From the point of view of the Lie analysis, we search for a Lie point symmetry  
\begin{equation}\left\{
\begin{aligned}
 \bar {t}&\rightarrow t+\epsilon \xi(t,u),\\ \bar{u}&\rightarrow u+\epsilon \eta(t,u).
\end{aligned}
\right.\end{equation} 
Similarly, we expand in Taylor series up to first-order the function $w(t)$
\begin{equation}
w(\bar t)\rightarrow w(t)+\epsilon \frac{dw(t)}{dt}\xi(t,u).
\end{equation}
Extending the transformation to the second derivative, we have $\bar{u}_{\bar{t} \bar{t}}=u_{tt}+\epsilon \eta_{tt}(t,u),$
where $\eta_{tt}$ is calculated according to \eqref{prolongcomplete} in Chapter \ref{Chap:GeomFund}.
The resulting symmetries read
\begin{equation}
\xi(t,u)=-4\omega(t)\frac{d\omega(t)}{dt}F(t)-4\omega(t)^2\frac{dF(t)}{dt},\quad \eta(t,u)=\frac{dF(t)}{dt}u,
\end{equation}
where $F(t)$ reads
\begin{equation}
F(t)=\exp\left({-2\int{\frac{10\omega_t^3 \omega+\omega_t^2+\omega\omega_{tt}+4\omega_t\omega_{tt}\omega^2-2\omega^3\omega_{ttt}-16\omega^6 \omega_t^2}{16\omega \omega_t-16\omega^3 \omega_{tt}+44\omega_t^2 \omega^2+16\omega^6-32\omega^7 \omega_t+1}dt}}\right).
\end{equation}
\end{example}

\section{Lie symmetries for Lie systems}
\setcounter{equation}{0}
\setcounter{theorem}{0}
\setcounter{example}{0}

Given the interest of Lie symmetries for the study of differential equations, we devote this section to the study of Lie symmetries for Lie systems. 
We pioneer the study and application of certain types of  Lie symmetries for higher-order Lie systems and PDE Lie systems, i.e., the generalizations of Lie systems to the realms of higher-order ODEs and PDEs \cite{CGL11,CGM07,SIGMA,EstLucasSar}.
As a byproduct, we develop one of the few applications of the theory of Lie systems in the investigation of systems of PDEs.
Additionally, only a few particular results about Lie symmetries for Lie systems had appeared before in \cite{CLS122}. We here accomplish a careful and exhaustive study
and calculation of such Lie symmetries \cite{EstLucasSar}.

\subsection{On certain Lie symmetries for Lie systems}

We are now concerned with the study of certain Lie symmetries for Lie systems. We prove that the features of these Lie symmetries are determined by the algebraic structure of a Vessiot--Guldberg Lie algebra of the Lie system. 

Consider a general the Lie system $X$ given by 
\begin{equation}\label{genLiesys12}
X=\sum_{\alpha=1}^r b_{\alpha}(t)X_{\alpha},
\end{equation}
where $b_1,\ldots,b_r$ are some $t$-dependent functions. Then, \eqref{genLiesys12} admits a Vessiot--Guldberg Lie algebra $V$ with basis $X_1,\ldots,X_r$ and structure constants $c_{\alpha\beta\gamma}$. 
Let us study the Lie symmetries of $X$ of the form
\begin{equation}\label{LieSym}
Y=f_0(t)\frac{\partial}{\partial t}+\sum_{\alpha=1}^rf_\alpha(t)X_\alpha,
\end{equation}
where $f_0,\ldots,f_r$ are certain $t$-dependent functions.  We denote by {\it $\mathcal{S}^V_X$ the space of Lie symmetries}.
Recall that $Y\in\mathcal{S}^V_X$ if and only if
\begin{equation}\label{consym}
[Y,\bar X]=h \bar X
\end{equation}
for a function $h\in C^\infty(\mathbb{R}\times N)$, with $\bar X$ being the autonomization of $X$ \cite{Olver}. From this, it immediately follows that $\mathcal{S}^V_X$ is a real Lie algebra of vector fields.

Using the properties of $\mathcal{S}_X^V$, we now characterize the elements of $\mathcal{S}^V_X$ as particular solutions of a family of Lie systems.

\begin{lemma}\label{SLS} The vector field $Y$ of the form (\ref{LieSym}) is a Lie symmetry for the Lie system (\ref{genLiesys12}) if and only if the $t$-dependent functions $f_0,\ldots,f_r$ satisfy the system of differential equations
\begin{equation}\label{SysFun}
\frac {{\rm d}f_0}{{\rm d}t}=b_0(t),\quad \frac{{\rm d}f_\alpha}{{\rm d}t}=f_0\frac{{\rm d}b_\alpha}{{\rm d}t}(t)+b_\alpha(t)b_0(t) +\sum_{\beta,\gamma=1}^rb_\beta(t) f_\gamma c_{\gamma\beta\alpha},
\end{equation}
for a certain $t$-dependent function $b_0$ and $\alpha=1,\ldots,r$.
\end{lemma}
\begin{proof}

From (\ref{genLiesys12}), (\ref{LieSym}) and (\ref{consym}), we have that

\begin{align}
\!\!\left[Y,\bar{X}\right]&=\left[f_0\frac{\partial}{\partial t}\!+\!\sum_{\alpha=1}^
{r}f_\alpha X_\alpha,\frac{\partial}{\partial
t}+\sum_{\beta=1}^rb_\beta X_\beta\right]\!\!\nonumber\\
&=-\frac{{\rm d}f_0}{{\rm d}t}\frac{\partial}{\partial t}+\!\!\sum_{\alpha=1}^r\left(\left(f_0\frac{{\rm d} b_\alpha}{{\rm d} t}-\frac{{\rm d} f_\alpha}{{\rm d}
t}\right)X_\alpha+
\sum_{\beta=1}^rb_\beta f_\alpha[X_\alpha,X_\beta]\right)\nonumber
\\
&=-\frac{{\rm d}f_0}{{\rm d}t}\frac{\partial}{\partial t}+\sum_{\alpha=1}^{r}\left(\left(f_0\frac{{\rm d} b_\alpha}{{\rm d} t}-\frac{{\rm d} f_\alpha}{{\rm d} t}\right)X_\alpha+\sum_{\beta,\gamma=1}^{r}b_{\beta} f_\alpha c_{
\alpha\beta\gamma}X_{\gamma}\right)\nonumber\\
&=-\frac{{\rm d}f_0}{{\rm d}t}\frac{\partial}{\partial t}+\sum_{\alpha=1}^{r}\left(f_0\frac{{\rm d} b_\alpha}{{\rm d} t}-\frac{{\rm d} f_\alpha}{{\rm d}
t}+{\sum_{\beta,\gamma=1}^{r}{b_{\beta}f_\gamma c_{
\gamma\beta\alpha}}}\right)X_{\alpha}\nonumber\\
&=h \left(\frac{\partial }{\partial t}+\sum_{\alpha=1}^{r}b_{\alpha}X_{\alpha}\right).
\end{align}

Thus, $[Y,\bar X]=h\bar X$ is equivalent to
\begin{equation}
\left(-\frac{{\rm d}f_0}{{\rm d}t}-h\right)\frac{\partial}{\partial t}+\sum_{\alpha=1}^{r}\left[\sum_{\beta,\gamma=1}^{r}{{b_{\beta}f_\gamma c_{
\gamma\beta\alpha}}}+f_0\frac{{\rm d} b_\alpha}{{\rm d} t}-\frac{{\rm d} f_\alpha}{{\rm d}
t}-hb_{\alpha}\right]X_{\alpha}=0.\end{equation} 
Since $\partial/\partial t, X_1,\ldots,X_r$, are linearly independent over $\mathbb{R}$, we obtain that $[Y,\bar X]=h\bar X$ if and only if (\ref{SysFun}) is fulfilled with $h$ being an arbitrary $t$-dependent function such that $b_0(t)=-h(t)$ for every $t\in\mathbb{R}$.
 Hence, the elements of $\mathcal{S}_X^V$ are the particular solutions of (\ref{SysFun}) for arbitrary $t$-dependent functions $b_0$. 
\end{proof}

\begin{definition} We call (\ref{SysFun}) the {\it symmetry system} of the Lie system (\ref{genLiesys12}) with respect to its Vessiot--Guldberg Lie algebra $V$. We write $\Gamma_X^V$ for the $t$-dependent vector field associated with (\ref{SysFun}), 
that is, 
\begin{equation}
 \Gamma_X^V=b_0(t)\frac{\partial}{\partial f_0}+\sum_{\alpha=1}^r\left(f_0\frac{{\rm d}b_{\alpha}}{{\rm d} t}(t)+b_0(t)b_{\alpha}(t)+\sum_{\gamma,\beta=1}^r b_{\beta}(t)f_{\gamma}c_{\gamma \beta \alpha}\right)\frac{\partial}{\partial f_{\alpha}}.
\end{equation}
\end{definition}
Note that (\ref{genLiesys12}) may have different Vessiot--Guldberg Lie algebras (see \cite{Dissertationes} for details). 
 Let us prove that (\ref{SysFun}) is a Lie system.

\begin{theorem}\label{The:SLS}
The system $\Gamma_{X}^V$ is a Lie system possessing a Vessiot--Guldberg Lie algebra
\begin{equation}\label{deVS}
(A_1\oplus_S A_2)\oplus_S V_L\simeq (\mathbb{R}^{r+1}\oplus_S\mathbb{R}^r)\oplus_S V/Z(V),
\end{equation}
where
\begin{equation}\begin{gathered}
A_1=\langle Z_0,\ldots, Z_r\rangle \simeq \mathbb{R}^{r+1},\qquad A_2=\langle W_1,\ldots, W_r\rangle \simeq \mathbb{R}^r,
\\ V_L=\langle Y_1,\ldots,Y_r\rangle \simeq V/Z(V),\end{gathered}\end{equation} 
with
\begin{equation}\label{symfields}
Y_\alpha=\sum_{\beta,\gamma=1}^rf_\beta
c_{\beta\alpha \gamma}\frac{\partial}{\partial f_\gamma}, \qquad W_{\alpha}=f_0\frac{\partial}{\partial f_\alpha},\qquad Z_0=\frac{\partial}{\partial f_0},\qquad Z_\alpha=\frac{\partial}{\partial f_\alpha},
\end{equation}
with $\alpha=1,\ldots,r$, we write $A \oplus_S B$ for the semi-direct sum of the ideal $A$ of $A+B$ with $B$, and $Z(V)$ is the center of the Lie algebra $V$. 
\end{theorem}
\begin{proof}
By defining $t$-dependent functions
\begin{equation}
c_0\equiv b_0,\quad \bar b_0\equiv 0, \quad c_\alpha\equiv b_0b_\alpha, \quad \bar b_\alpha\equiv \frac{{\rm d}b_\alpha}{{\rm d}t},\qquad \alpha=1,\ldots,r,
\end{equation}
we can write
\begin{equation}
\Gamma^V_X(t,f)=\sum_{\alpha=0}^r c_{\alpha}(t)Z_{\alpha}(f)+\sum_{\alpha=1}^r\left[\bar{b}_{\alpha}(t)W_{\alpha}(f)+b_{\alpha}(t)Y_{\alpha}(f)\right],\qquad f\in\mathbb{R}^{n+1}.
\end{equation} 
Hence, $\Gamma^V_X$ is a $t$-dependent vector field taking values in the linear space $V^S\equiv A_1+A_2+V_L$. Let us show that $V^S$ is also a Lie algebra of vector fields. To do so, let us first prove that $V_L\simeq V/Z(V)$. Consider $Y_{\alpha}, Y_{\beta}\in V_L$. Recalling that $[X_{\alpha},X_{\beta}]=\sum_{\gamma=1}^rc_{\alpha\beta\gamma}X_{\gamma}$, we obtain
\begin{align}
\left[Y_{\alpha},Y_{\beta}\right]&=\sum_{i,j,m,n=1}^{r}\left[c_{i \alpha j}f_{i}\frac{\partial}{
\partial f_{j}}, c_{m
\beta n}f_{m}\frac{\partial}{\partial f_{n}}\right]
\nonumber\\
&
=\sum_{i,j,m,n=1}^{r}c_{i \alpha j}c_{m
\beta n}\left(f_{i}\delta_{j}^{m}\frac{\partial}{
\partial x_{n}}-f_{m}\delta_{n}^{i}\frac{\partial}{\partial
f_{j}}\right)
\nonumber\\
&=\sum_{i,m,n=1}^{r}c_{i\alpha m}c_{m \beta n}f_{i}
\frac{\partial}{\partial f_{n}}-\sum_{n,j,m=1}^{r}c_{n\alpha j}c_{m \beta n}f_{m
}\frac{\partial}{\partial f_j}
\nonumber\\
&=\sum_{i,m,n=1}^{r}(c_{i\alpha m}c_{m \beta n}+c_{m\alpha n}c_{\beta im})f_{i
}\frac{\partial}{\partial f_n}.
\end{align}
Using the Jacobi identity for the structure constants $c_{\alpha\beta\gamma}$, we see that
\begin{equation}\label{sc}
\sum_{m,n=1}^{r}\left(c_{i\alpha m}c_{m \beta n}+c_{\alpha \beta m}c_{m i n}+c_{\beta i m}c_{m \alpha n}\right)=0,\qquad \forall i,\alpha,\beta=1,\ldots,r.
\end{equation}
From this, 
\begin{align}
\left[Y_{\alpha},Y_{\beta}\right]&=\sum_{i,m,n=1}^{r}(c_{i\alpha m}c_{m \beta n}+c_{m\alpha n}c_{\beta im})f_{i
}\frac{\partial}{\partial f_n}=-\sum_{i,m,n=1}^{r}c_{\alpha \beta m}c_{m i n}f_{i
}\frac{\partial}{\partial f_n}
\nonumber\\
&=\sum_{m=1}^rc_{\alpha \beta m}Y_{m}.
\end{align}
So, $Y_1,\ldots,Y_{r}$ span a Lie algebra. We can define a Lie algebra morphism $\phi:V\rightarrow V_L$ of the form $\phi(X_\alpha)=Y_\alpha$ for 
$\alpha=1,\ldots,r$. 
The vector fields $Y_1,\ldots,Y_r$ do not need to be linearly independent. Let us show this. We can assume with no loss of generality that $X_1,\ldots,X_s$, with $s\leq r$, form a basis for $\ker \phi$. Since $\phi(X_\alpha)=0$ for $\alpha=1,
\ldots,s$, we have $f_\beta c_{\beta\alpha\gamma}=0$ for $\alpha=1,\ldots,s$ and $\beta,\gamma=1,\ldots,r$. Thus, we see that $[X_\alpha,X_\beta]=0$ for $\alpha=1,\ldots,s$ and $\beta=1,\ldots,r$. This means that $X_\alpha\in Z(V)$. Conversely,  we get by similar arguments that if $X\in Z(V)$, then $X\in \ker \phi.$ Hence, $X\in \ker\phi$ if and only if $X\in Z(V)$. In consequence, $\ker \phi=Z(V)$ and $Y_1,\ldots,Y_r$ span a Lie algebra isomorphic to $V/Z(V)$. 

It is obvious that $A_1$ is an ideal of $A_1+A_2$. Moreover, as $[A_1,V_L]\subset A_1$ and $[A_2,V_L]\subset A_2$, then $A_1\oplus_{S} A_2$ is an ideal of $V^S$. Consequently, $V^S$ is a Lie algebra of the form (\ref{deVS}).
\end{proof}

\begin{definition} We say that the {\it Lie systems $X_1$ and $X_2$ are isomorphic} when they take values in two isomorphic Vessiot--Guldberg Lie algebras $V_1, V_2$ and there exists a Lie algebra isomorphism $\phi:V_1\rightarrow V_2$ such that $(X_2)_t=\phi((X_1)_t)$ for each $t\in\mathbb{R}$.

\end{definition}

\begin{proposition}\label{SymTheo} Given two isomorphic Lie systems $X_1$ and $X_2$ related to Vessiot--Guldberg Lie algebras $V_1, V_2$, their symmetry systems relative to such Lie algebras are, up to a change of basis in $V_1$ and/or $V_2$, the same.
\end{proposition}

\subsection{Lie algebras of Lie symmetries for Lie systems}
In this section we study different Lie subalgebras of $\mathcal{S}^V_X$. Their interest resides in the fact that, when finite-dimensional, they can be integrated to form Lie group actions of symmetries for $X$. In turn, they can be employed to simplify the Lie system they are referred to.

\begin{lemma}
The space of functions $C^{\infty}(\mathbb{R})$ can be endowed with a Lie bracket given by
\begin{equation}\label{LSbracket}
 \{f,\bar f\}_{\mathbb{R}}=f\frac{{\rm d} \bar f}{{\rm d}t}-\bar f\,\frac{{\rm d} f}{{\rm d} t},\qquad \forall f,\bar f\in C^\infty(\mathbb{R}).
\end{equation}
\end{lemma}
\begin{proof}
In order to prove that \eqref{LSbracket} is a Lie bracket, we must show that (\ref{LSbracket}) is  bilinear, antisymmetric and satisfies the Jacobi identity.
From its definition (\ref{LSbracket}) is clearly bilinear and antisymmetric. To see that (\ref{LSbracket}) holds the Jacobi identity, we consider the map $\phi:f\in C^{\infty}(\mathbb{R}) \mapsto f\partial/\partial t\in \Gamma({\rm T}\mathbb{R})$. Observe that $\phi$ is a linear isomorphism. 

Moreover, it follows that 
\begin{align}
\phi(\{f,\bar f\}_\mathbb{R})&=\{f,\bar f\}_\mathbb{R}\frac{\partial}{\partial t}=\left(f\frac{{\rm d} \bar f}{{\rm d}t}-\bar f\,\frac{{\rm d}f}{{\rm d}t}\right)\frac{\partial}{\partial t}=\left[f\frac{\partial}{\partial t},\bar f\frac{\partial}{\partial t}\right]
\nonumber\\
&=[\phi(f),\phi(\bar f)],\,\, 
\end{align}
for arbitrary $f,\bar f\!\in\! C^\infty(\mathbb{R})$. By using the Jacobi identity for vector fields on $\mathbb{R}$ with respect to the Lie bracket $[\cdot,\cdot]$, we obtain
\begin{align}
 \phi (\{\{f,\bar f\}_\mathbb{R},\bar{\bar f}\}_\mathbb{R}&+\{\{\bar f,\bar{\bar{f}}\}_\mathbb{R},f\}_\mathbb{R}+\{\{\bar{\bar f},f\}_\mathbb{R},\bar{f}\}_\mathbb{R})=
\nonumber\\
&=[[\phi(f),\phi(\bar{f})],\phi(\bar{\bar f})]+[[\phi(\bar f),\phi(\bar{\bar f})],\phi(f)]+[[\phi(\bar{\bar f}),\phi(f)],\phi(\bar{f})]
\nonumber\\
&=0,
\end{align}
$ \forall f,\bar f,\bar{\bar f}\in C^\infty(\mathbb{R}).$ Since $\phi$ is a linear isomorphism, then
\begin{equation}
\{\{f,\bar f\}_\mathbb{R},\bar{\bar f}\}_\mathbb{R}+\{\{\bar f,\bar{\bar f}\}_\mathbb{R},f\}_\mathbb{R}+\{\{\bar{\bar f},f\}_\mathbb{R},\bar f\}_\mathbb{R}=0
\end{equation}
and (\ref{LSbracket}) satisfies the Jacobi identity giving rise to a Lie bracket on $C^\infty(\mathbb{R})$. Moreover, $\phi$ becomes a Lie algebra isomorphism.
\end{proof}
\begin{definition}
Let $X$ be a Lie system on $N$ with a Vessiot--Guldberg Lie algebra $V$ and let $\mathfrak{W}$ be a nonempty set of
$t$-dependent functions that form a Lie algebra with respect to the Lie bracket defined in \eqref{LSbracket}. We call $\mathcal{S}_{X,\mathfrak{W}}^{V}$ the space
\begin{equation}
 \mathcal{S}_{X,\mathfrak{W}}^{V}=\left\{Y\in \mathcal{S}_X^{V} \mid  Y=f_0\frac{\partial}{\partial t}+\sum_{\alpha=1}^{r}f_\alpha X_{\alpha},\,\, f_0\in \mathfrak{W}\right\},
\end{equation}
where $X_1,\ldots,X_r$ is a basis for $V$.
\end{definition}

\begin{proposition}\label{finite}
The space of symmetries $\mathcal{S}_{X,\mathfrak{W}}^{V}$ is a Lie algebra of symmetries of $X.$
\begin{proof}
Since $\mathfrak{W}$ and $\mathcal{S}_X^V$ are linear spaces, the linear combinations of elements of $\mathcal{S}^V_{X,\mathfrak{W}}$ belong to $\mathcal{S}^V_{X,\mathfrak{W}}$. So, this space becomes a vector space. Moreover, given two elements, $Y,Y^*\in \mathcal{S}_{X,\mathfrak{W}}^{V}$, their Lie bracket reads
\begin{align}
 [Y,Y^*]&=\left[f_0(t)\frac{\partial}{ \partial t}+\sum_{\alpha=1}^{r}f_\alpha(t) X_{\alpha},f_0^*(t)\frac{\partial}{\partial t}+\sum_{\beta=1}^{r}{f}^*_\beta(t) X_{\beta}\right]
\nonumber\\&=
 \{f_0,{f^*_0}\}_{\mathbb{R}}\frac{\partial}{ \partial t}+\sum_{\beta=1}^{r}\left[\left(f_0\frac{{\rm d}{f^*_{\beta}}}{{\rm d}t}-{f^*_0}\frac{df_{\beta}}{d t}\right)X_{\beta}+\sum_{\alpha,\gamma=1}^{r}f_{\alpha}{f^*_{\beta}}c_{\alpha\beta\gamma}X_{\gamma}\right]
\nonumber\\&=
\{f_0,{f^*_0}\}_{\mathbb{R}}\frac{\partial}{\partial t}+\sum_{\gamma=1}^{r}\left[\left(f_0\frac{{\rm d}{f^*_{\gamma}}}{{\rm d} t}-{f^*_0}\frac{{\rm d} f_{\gamma}}{{\rm d} t}\right)+\sum_{\alpha,\beta=1}^{r}f_{\alpha}{f^*_{\beta}}c_{\alpha\beta\gamma}\right]X_{\gamma}.
\end{align}
Since $\mathcal{S}_X^V$ is a Lie algebra and $Y,Y^*\in\mathcal{S}^V_{X}$, then $[Y,Y^*]\in \mathcal{S}_X^V$. As  additionally $\{f_0,f^*_0\}_\mathbb{R}\in \mathfrak{W}$, then $[Y,Y^*]\in \mathcal{S}^V_{X,\mathfrak{W}}$. Hence, the Lie bracket of elements of $\mathcal{S}^V_{X,\mathfrak{W}}$ belongs to $\mathcal{S}^V_{X,\mathfrak{W}}$, which becomes a Lie algebra.
\end{proof}
\end{proposition}

\begin{corollary}
Given a Lie system $X$ on $N$ related to a Vessiot--Guldberg Lie algebra $V,$ the elements of $\mathcal{S}^{V}_{X,\mathfrak{W}}$
with
\begin{enumerate}
 \item $\mathfrak{W}=\{f_0\in C^\infty(\mathbb{R})\,\,|\,\,{\rm d}f_0/{\rm d}t=0\}$,
 \item $\mathfrak{W}=\{f_0\in C^\infty(\mathbb{R})\,\,|\,\,f_0=0\}$,
\end{enumerate}
are finite-dimensional Lie algebras of vector fields. In the second case, $S^V_{X,\mathfrak{W}}$ is isomorphic to $V.$
\end{corollary}

\begin{proof} In both cases, $\mathfrak{W}$ is nonempty. In the first case, the functions with ${\rm d}f_0/{\rm d}t=0$ are constant. These functions form an Abelian Lie algebra with
respect to the Lie bracket $\{\cdot,\cdot\}_\mathbb{R}$. In view of Proposition \ref{finite}, the space $\mathcal{S}^V_{X,\mathfrak{W}}$ is a Lie algebra.

In the second case, the function zero
is also a zero-dimensional Lie algebra relative to $\{\cdot,\cdot\}_\mathbb{R}$. Since $f_0=0$ and using Proposition \ref{finite}, we obtain that $\mathcal{S}^V_{X,\mathfrak{W}}$ can be understood as a Lie algebra of $t$-dependent vector fields taking values in $V$.
To prove that $\mathcal{S}^V_{X,\mathfrak{W}}\simeq V$, let us consider the morphism which maps each $t$-dependent vector field with its value at $t=0$, namely
$$\begin{array}{rccc}
\phi:&\mathcal{S}^V_{X,\mathfrak{W}}&\longrightarrow &V\\
    &Z&\mapsto &Z_0.
\end{array}
$$
Let $X_1,\ldots,X_r$ be a basis for $V$. From Lemma (\ref{SLS}), we have that $(f_1(t),\ldots,f_r(t))$ is a particular solution of the system
\begin{equation}
\frac{ {\rm d}f_\alpha}{{\rm d}t}=\sum_{\delta,\beta=1}^rb_\beta(t) f_\delta
c_{\delta \beta \alpha},\qquad \alpha=1,\ldots,r.
\end{equation}
For each initial condition $f_\alpha(0)=c_\alpha \in \mathbb{R}$, with $\alpha=1,\ldots,r$, i.e., by fixing $Z_0$, there exists a unique solution of the above system. Hence, there exists a unique $t$-dependent vector field $Z$ of $\mathcal{S}^V_{X,\mathfrak{W}}$
with $Z_0=\sum_{\alpha=0}^r c_\alpha X_\alpha$. Thus, $\phi$ is a bijection. Using that for two vector fields $Z_1,Z_2\in \mathcal{S}^V_{X,\mathfrak{W}}$ we have $[Z_1,Z_2]\in \mathcal{S}^V_{X,\mathfrak{W}}$ and $[Z_1,Z_2]_t=[(Z_1)_t,(Z_2)_t]$, we see that
$\phi$ is a Lie algebra morphism and $\mathcal{S}^V_{X,\mathfrak{W}}\simeq V$.
\end{proof}

\subsection{Applications to systems of ODEs and HODEs}

Let us work out the symmetry systems and related Lie symmetries for some Lie systems of interest. In particular, we will illustrate that Proposition
 \ref{SymTheo} enables us to determine simultaneously Lie symmetries for different Lie systems with isomorphic Vessiot--Guldberg Lie algebras.
\subsubsection{Lie symmetries for sl(2,$\mathbb{R}$)-Lie systems}\label{LSSL2}

Let us obtain the symmetry systems and related Lie symmetries for $\mathfrak{sl}(2,\mathbb{R})$-Lie systems. This shall be used in following subsections to obtain simultaneously Lie symmetries of isomorphic $\mathfrak{sl}(2,\mathbb{R})$-Lie systems appearing in the physics and/or mathematical literature.

Let us choose a basis of vector fields $\{X_1,X_2,X_3 \}$ of $V\simeq \mathfrak{sl}(2,\mathbb{R})$ with commutation relations
\begin{equation}\label{ComSl2}
[X_1,X_2]=X_1,\quad [X_1,X_3]=2X_2,\quad [X_2,X_3]=X_3.
\end{equation}
Every Lie system with Vessiot--Guldberg Lie algebra $V$ can be brought into the form
\begin{equation}\label{sl2}
X=b_1(t)X_1+b_2(t)X_2+b_3(t)X_3
\end{equation}
for certain $t$-dependent functions $b_1$, $b_2$ and $b_3$.
The Lie symmetries of $\mathcal{S}_X^V$ take the form
\begin{equation}
Y=f_0(t)\frac{\partial}{\partial t}+f_1(t)X_1+f_2(t)X_2+f_3(t)X_3,\end{equation} 
where $f_0,f_1,f_2,f_3$ are some $t$-dependent functions to be determined. In view of (\ref{SysFun}) and the commutation relations (\ref{ComSl2}), the symmetry system for $X$ relative to $V$ reads
\begin{equation}\label{slsys}
\left\{\begin{aligned}
\frac{{\rm d}f_0}{{\rm d}t}&=b_0(t),\\
 \frac{{\rm d}f_1}{{\rm d}t}&=f_0\frac{{\rm d} b_1}{{\rm d} t}(t)+f_1b_2(t)-f_2b_1(t)+b_0(t)b_1(t),\\
 \frac{{\rm d}f_2}{{\rm d}t}&=f_0\frac{{\rm d} b_2}{{\rm d} t}(t)+2f_1b_3(t)-2f_3b_1(t)+b_0(t)b_2(t),\\
 \frac{{\rm d}f_3}{{\rm d}t}&=f_0\frac{{\rm d} b_3}{{\rm d} t}(t)+f_2b_3(t)-f_3b_2(t)+b_0(t)b_3(t).\\
\end{aligned}\right.
\end{equation}

As stated in Theorem \ref{The:SLS}, this is a Lie system. Indeed, system \eqref{slsys} is related to the $t$-dependent vector field
\begin{align}
\Gamma^{\mathfrak{sl}(2,\mathbb{R})}_X=\frac{{\rm d}b_1(t)}{{\rm d}t}W_1&+\frac{{\rm d}b_2(t)}{{\rm d}t}W_2+\frac{{\rm d}b_3(t)}{{\rm d}t}W_3+b_0(t)Z_0+b_0(t)b_1(t)Z_1+b_0(t)b_2(t)Z_2\nonumber\\
&+b_0(t)b_3(t)Z_3+b_1(t)Y_1+b_2(t)Y_2+b_3(t)Y_3,
\end{align}
where
\begin{equation}\label{com1}
 Z_\alpha=\frac{\partial}{\partial f_\alpha},\qquad \alpha=0,1,2,3,\qquad W_\beta=f_0\frac{\partial}{\partial f_\beta},\qquad \beta=1,2,3
\end{equation}
and
\begin{equation}\label{com3}
 Y_1=-f_2\frac{\partial}{\partial f_1}-2f_3\frac{\partial}{\partial f_2},\quad Y_2=f_1\frac{\partial}{\partial f_1}-f_3\frac{\partial}{\partial f_3},\quad Y_3=2f_1\frac{\partial}{\partial f_2}+f_2\frac{\partial}{\partial f_3}.
\end{equation}
These vector fields hold
\begin{equation}\label{comm0}
 [Y_1,Y_2]=Y_1,\quad [Y_1,Y_3]=2Y_2,\quad [Y_2,Y_3]=Y_3.
\end{equation}
Since $Z(V)=\{0\}$, then $V_L=\langle Y_1,Y_2,Y_3\rangle$ is a Lie algebra isomorphic to $V/Z(V)\simeq\mathfrak{sl}(2,\mathbb{R})$ as stated in Theorem \ref{The:SLS}.
The rest of commutation relations read
\begin{equation}
\begin{array}{lllll}
&[Y_1,Z_0]=0,\quad &[Y_1,Z_1]=0, &[Y_1,Z_2]=Z_1, &[Y_1,Z_3]=2Z_2,\\
&[Y_2,Z_0]=0,\quad &[Y_2,Z_1]=-Z_1, &[Y_2,Z_2]=0, &[Y_2,Z_3]=Z_3,\\
&[Y_3,Z_0]=0,\quad &[Y_3,Z_1]=-2Z_2, &[Y_3,Z_2]=-Z_3, &[Y_3,Z_3]=0.
\end{array}
\end{equation}
Moreover,
\begin{equation}
\begin{array}{llll}
&[Y_1,W_1]=0,\quad &[Y_1,W_2]=W_1,\quad &[Y_1,W_3]=2W_2\\
&[Y_2,W_1]=-W_1,\quad &[Y_2,W_2]=0,\quad &[Y_2,W_3]=W_3,\\
&[Y_3,W_1]=-2W_2,\quad &[Y_3,W_2]=-W_3,\quad &[Y_3,W_3]=0
\end{array}
\end{equation}
and
\begin{equation}\label{comm3}
\begin{gathered}
\left[Z_0,W_j\right]=Z_j,\quad [Z_i,W_j]=0,\quad  [W_i,W_j]=0,\quad i,j=1,2,3,\\ 
[Z_\alpha,Z_\beta]=0,\quad \alpha,\beta=0,\ldots,3.
\end{gathered}
\end{equation}
Hence, $A_1=\langle Z_0,Z_1,Z_2,Z_3 \rangle$ is an ideal of $A_1+A_2$. And $A_1+A_2$ is an ideal
of $A_1+A_2+V_L$, with $A_2=\langle W_1,W_2,W_3 \rangle$.

\begin{example}\normalfont

We study a particular type of {\bf first-order Riccati equation} \cite{Ince,Riccati}
\begin{equation}\label{Ricc}
\frac{{\rm d}x}{{\rm d}t}=\eta(t)+x^2,
\end{equation}
where $\eta$ is an arbitrary $t$-dependent function.

It is well-known that the Riccati equation is a Lie system with a Vessiot--Guldberg Lie algebra isomorphic to $V^{\rm Ric}\simeq \mathfrak{sl}(2,\mathbb{R})$ \cite{CLS122}.
Indeed, equation (\ref{Ricc}) has the associated $t$-dependent vector field $X^{\rm Ric}=X^{\rm Ric}_3+\eta(t)X^{\rm Ric}_1,$ where
\begin{equation}\label{VecRic}
X^{\rm Ric}_1=\frac{\partial}{\partial x},\quad X^{\rm Ric}_2=x\frac{\partial}{\partial x} \quad X^{\rm Ric}_3=x^2\frac{\partial}{\partial x}
\end{equation}
satisfy the same commutation relations as $X_1,X_2,X_3$ in (\ref{ComSl2}). In view of this, (\ref{Ricc}) is related to a $t$-dependent vector field taking values in a finite-dimensional Lie algebra of vector fields isomorphic to $\mathfrak{sl}(2,\mathbb{R})$. Then, (\ref{Ricc}) is an $\mathfrak{sl}(2,\mathbb{R})$-Lie system. Moreover, we can consider $X^{\rm Ric}$ as a particular case of system (\ref{sl2}). By applying the results of the previous sections to generic $\mathfrak{sl}(2,\mathbb{R})$-Lie systems, we find that the symmetry system for $X^{\rm Ric}$ is of the form (\ref{slsys})  with $b_1=\eta, b_2=0$ and $b_3=1$, namely
\begin{equation}\label{ric2}
\left\{\begin{aligned}
\frac{{\rm d}f_0}{{\rm d}t}&=b_0(t),\\
 \frac{{\rm d}f_1}{{\rm d}t}&=f_0\frac{{\rm d}\eta}{{\rm d}t}(t)-\eta(t) f_2+b_0(t)\eta(t),\\
 \frac{{\rm d}f_2}{{\rm d}t}&= 2f_1-2\eta(t) f_3,\\
 \frac{{\rm d}f_3}{{\rm d}t}&=f_2+b_0(t).
\end{aligned}\right.
\end{equation}

We can recover and generalize the results given in \cite{GS10} by means of our approach. From their expressions in (\ref{ric2}), we can differentiate ${\rm d}f_3/{\rm d}t$ twice and ${\rm d}f_2/{\rm d}t$ once. By substituting ${\rm d}^2f_2/{\rm d}t^2$ in ${\rm d}^3f_3/{\rm d}t^3$ and using the remaining equations in (\ref{ric2}), we obtain that	
\begin{equation}
\frac{{\rm d}^3f_3}{{\rm d}t^3}=\frac{{\rm d}^3f_0}{{\rm d}t^3}+2\frac{{\rm d}f_1}{{\rm d}t}-2\frac{{\rm d}\eta}{{\rm d}t}f_3-2\eta(t)\frac{{\rm d}f_3}{{\rm d}t}.                                                                                                                                                                  \end{equation} 
By substituting the value of ${{\rm d}f_1}/{{\rm d}t}$ from (\ref{ric2}) and using that $f_2={\rm d}f_3/{\rm d}t-b_0(t)$, we obtain the following equation for $f_3$ in terms of the coefficients of \eqref{Ricc} and $f_0$
\begin{equation}\label{rCS}
\frac{{\rm d}^3f_3}{{\rm d}t^3}=\frac{{\rm d}^3f_0}{{\rm d}t^3}+4b_0\eta(t)+2\frac{{\rm d}\eta}{{\rm d}t}f_0-2\frac{{\rm d}\eta}{{\rm d}t}f_3-4\eta(t)\frac{{\rm d}f_3}{{\rm d}t}.
\end{equation}
From this, we can retrieve the following corollary given in \cite{GS10}.

\begin{corollary} The Riccati equation (\ref{Ricc}) admits the Lie symmetry
\begin{equation}
Y=f_0\frac{\partial}{\partial t}-\frac 12 \frac{{\rm d}^2f_0}{{\rm d}t^2}\frac{\partial}{\partial x}-\frac{{\rm d}f_0}{{\rm d}t}x\frac{\partial}{\partial x},\end{equation} 
where
\begin{equation}\label{conRicEsp}
\frac{{\rm d}^3f_0}{{\rm d}t^3}+4b_0(t)\eta(t)+2\frac{{\rm d}\eta}{{\rm d}t}f_0=0.
\end{equation}

\end{corollary}

\begin{proof}
Since we are looking for Lie symmetries with $f_3=0$, equation (\ref{rCS}) reduces to (\ref{conRicEsp}). Moreover, by substituting $f_3=0$ in (\ref{ric2}), we obtain that
\begin{equation}
\frac{{\rm d}f_3}{{\rm d}t}=b_0(t)+f_2=0,\qquad \frac{{\rm d}f_2}{{\rm d}t}=2f_1,
\end{equation} 
which yields $f_2=-{{\rm d}f_0}/{{\rm d}t}$ and $2f_1=-{{\rm d}^2f_0}/{{\rm d}t^2}$. Hence, the corollary follows.
\end{proof}

Going back to general symmetries of (\ref{Ricc}), we can obtain some of its Lie symmetries by solving (\ref{ric2}) for certain values of $\eta(t)$. These
possibilities are summarized in Table \ref{table8} in Appendix 1.
\end{example}

\begin{example}\normalfont
Reconsider the {\bf Cayley--Klein  Riccati equation} appeared in \eqref{CKRE} within Chapter \ref{Chap:LieSystems}.
Its associated first-order system (\ref{CKRE2}) was described in Chapter \ref{Chap:LieSystems} by means of the $t$-dependent vector field
\begin{equation}\label{otracK}
X^{\rm CK}=b_1(t)X^{\rm CK}_1+b_2(t)X^{\rm CK}_2+b_3(t)X^{\rm CK}_3,
\end{equation}
where 
\begin{equation}
X^{\rm CK}_1=\frac{\partial}{\partial x},\quad X^{\rm CK}_2=x\frac{\partial}{\partial x}+y\frac{\partial}{\partial y},\quad X_3^{\rm CK}=(x^2+\iota ^2 y^2)\frac{\partial}{\partial x}+2xy\frac{\partial}{\partial y}
\end{equation}
satisfy the same commutation relations  as $X_1,X_2,X_3$ in (\ref{ComSl2}). Since $X_1,X_2,X_3$ span a Lie algebra $V\simeq \mathfrak{sl}(2,\mathbb{R})$, the vector fields $X^{\rm CK}_1,X^{\rm CK}_2$ and $X^{\rm CK}_3$ span a real Lie algebra, $V^{\rm CK}$, which is isomorphic to the Lie algebra $\mathfrak{sl}(2,\mathbb{R})$ independently of the square of $\iota$. Then, \eqref{otracK} is an $\mathfrak{sl}(2,\mathbb{R})$-Lie system. 

Moreover, we can define a the Lie algebra isomorphism $\phi:V^{\rm CK}\rightarrow V^{\rm }$ mapping
\begin{equation}
\phi(X_\alpha^{\rm CK})=X_{\alpha},\qquad \alpha=1,2,3,
\end{equation}
where $X_\alpha$, with $\alpha=1,2,3$, are the generic vector fields in (\ref{ComSl2}).
Hence, we have that $\phi (X_t^{\rm CK})=X_t$ for every $t\in \mathbb{R}$ with $X$ being the generic $\mathfrak{sl}(2,\mathbb{R})$-Lie system (\ref{sl2}). Then, (\ref{otracK}) is isomorphic to  (\ref{sl2}) and, in view of Proposition \ref{SymTheo}, the symmetry system of $X^{\rm CK}$ becomes (\ref{slsys}).
\end{example}

\begin{example}\normalfont
Let us consider the {\bf quaternionic Riccati equation} with $t$-dependent real coefficients.
Quaternions are the elements of the real vector space $\mathbb{H}\simeq\mathbb{R}^4$ of elements of the form $q=q_0+q_1{\rm i}+q_2{\rm j}+q_3{\rm k}$, where $q_0,q_1,q_2,q_3\in \mathbb{R}$ and with the standard sum of elements and multiplication by real numbers. We can define a multiplication of quaternions (see \cite{CS03} for details) by assuming that real numbers commute with all quaternions and
that the following operations are fulfilled
\begin{equation}
\begin{gathered}
{\rm i}^2=-1,\quad {\rm j}^2=-1,\quad {\rm k}^2=-1,\\
{\rm i\cdot j}=-{\rm j\cdot i}={\rm k},\quad {\rm k\cdot i}=-{\rm i\cdot k}={\rm j},\quad {\rm j\cdot k}=-{\rm k\cdot j}={\rm i}.
\end{gathered}
\end{equation}
\noindent
The {quaternionic Riccati equation}  \cite{W714} takes the form
\begin{equation}\label{qric}
\frac{{\rm d}q}{{\rm d}t}=b_1(t)+a_1(t)q+qa_2(t)+qb_3(t)q,
\end{equation}
where $q$ and the $t$-dependent functions $a_1,a_2,b_1,b_3:\mathbb{R}\rightarrow \mathbb{H}$ take values in $\mathbb{H}$ \cite{CS03}. The existence of periodic solutions for particular cases of (\ref{qric}) has been studied in \cite{CM06,W714} and, for real $t$-dependent coefficients, in \cite{W714}. In this work, we focus on the latter case.

Writing $q$ in coordinates, we obtain that \eqref{qric} reads
\begin{equation}
\left\{\begin{aligned}\label{qLS}
\frac{{\rm d}q_0}{{\rm d}t}&=b_1(t)+b_2(t)q_0+b_3(t)(q_0^2-q_1^2-q_2^2-q_3^2),\\
\frac{{\rm d}q_1}{{\rm d}t}&=b_2(t)q_1+2b_3(t)q_0q_1,\\
\frac{{\rm d}q_2}{{\rm d}t}&=b_2(t)q_2+2b_3(t)q_0q_2,\\
\frac{{\rm d}q_3}{{\rm d}t}&=b_2(t)q_3+2b_3(t)q_0q_3,
\end{aligned}\right.
\end{equation}
where $b_2(t)\equiv a_1(t)+a_2(t).$ This system is related to the $t$-dependent vector field
\begin{equation}
X^{\mathbb{H}}=b_1(t)X^{\mathbb{H}}_1+b_2(t)X^{\mathbb{H}}_2+b_3(t)X^{\mathbb{H}}_3,
\end{equation}
with
\begin{equation}
\begin{gathered}
X^{\mathbb{H}}_1=\frac{\partial}{\partial q_0},\qquad X^{\mathbb{H}}_2=q_0\frac{\partial}{\partial q_0}+q_1\frac{\partial}{\partial q_1}+q_2\frac{\partial}{\partial q_2}+q_3\frac{\partial}{\partial q_3},
\\
X^{\mathbb{H}}_3=2q_0\left(q_1\frac{\partial}{\partial q_1}+q_2\frac{\partial}{\partial q_2}+q_3\frac{\partial}{\partial q_3}\right)+(q_0^2-q_1^2-q_2^2-q_3^2)\frac{\partial}{\partial q_0},\\
\end{gathered}\end{equation} 
which satisfy the same commutation relations as $X_1,X_2$ and $X_3$ in (\ref{ComSl2}). So, $X_1^{\mathbb{H}},X_2^{\mathbb{H}},X_3^{\mathbb{H}}$ span a real Lie algebra $V^{\mathbb{H}}$ isomorphic to $\mathfrak{sl}(2,\mathbb{R})$. Hence, the quaternionic Riccati equation \eqref{qric} with $t$-dependent real coefficients is a $\mathfrak{sl}(2,\mathbb{R})$-Lie system isomorphic to (\ref{sl2}) with respect to the Lie algebra isomorphism $\phi:V^{\mathbb{H}}\rightarrow V$ being given by
\begin{equation}
\phi(X^{\mathbb{H}}_\alpha)=X_\alpha,\qquad \alpha=1,2,3.
\end{equation}
Moreover, we have that $\phi(X^\mathbb{H}_t)=X_t$ for every $t\in\mathbb{R}$. Hence, the symmetry system for (\ref{qric}) is (\ref{slsys}). If we assume for instance $f_0=k\in \mathbb{R}$, and $b_1(t)=\eta(t)$, $b_2(t)=0$ and $b_3(t)=1$, we obtain that (\ref{qric}) is isomorphic to the Lie system (\ref{ric2}). Hence, for certain values of $\eta(t)$ given in Table \ref{table8} in Appendix 1, we can derive several Lie symmetries for quaternionic Riccati equations.
\end{example}

\begin{example}\normalfont
We now show that our theory in particular, and the whole theory of Lie systems in general, can be used to study autonomous systems of first-order ODEs.
We consider the {\bf generalized Darboux--Brioschi--Halphen system} (DBH system)\cite{Darboux}
\begin{equation}\label{Partial}
\left\{
\begin{aligned}
\frac{{\rm d}w_1}{{\rm d}t}&=w_3w_2-w_1w_3-w_1w_2+\tau^2,\\
\frac{{\rm d}w_2}{{\rm d}t}&=w_1w_3-w_2w_1-w_2w_3+\tau^2,\\
\frac{{\rm d}w_3}{{\rm d}t}&=w_2w_1-w_3w_2-w_3w_1+\tau^2,\\
\end{aligned}\right.
\end{equation}
where
\begin{multline}
\tau^2\equiv\alpha_1^2(\omega_1-\omega_2)(\omega_3-\omega_1)+\alpha_2^2(\omega_2-\omega_3)(\omega_1-\omega_2)
\\
{+\alpha_3^2(\omega_3-\omega_1)(\omega_2-\omega_3)\hphantom{aaa}}
\end{multline} 
and $\alpha_1,\alpha_2,\alpha_3$ are real constants.

When $\tau=0$, system (\ref{Partial}) retrieves the classical DBH system solved by Halphen \cite{CH03,Darboux,Halphen} which appears in the study of triply orthogonal surfaces and the vacuum Einstein equations for hyper-K\"ahler Bianchi-IX metrics. For $\tau\neq 0$, the generalized DBH system can be considered as a
 reduction of the self-dual Yang--Mills equations corresponding to an infinite-dimensional gauge group of diffeomorphisms of a
three-dimensional sphere \cite{CH03}. 

It can be proven that (\ref{Partial}) is an $\mathfrak{sl}(2,\mathbb{R})$-Lie system. Indeed, it is the associated system to the $t$-dependent vector field
\begin{multline}
    X_t^{\rm DBH}=(w_3w_2-w_1(w_3+w_2)+\tau^2)\frac{\partial}{\partial w_1} + (w_1w_3-w_2(w_1+w_3)+\tau^2)\frac{\partial}{\partial w_2}\\ + (w_2w_1-w_3(w_2+w_1)+\tau^2)\frac{\partial}{\partial w_3}=-X_3^{\rm DBH}\,.
\end{multline}
This vector field span a Lie algebra $V^{\rm DBH}$ of vector fields along with
\begin{equation}
    X_1^{\rm DBH}=\frac{\partial}{\partial w_1} + \frac{\partial}{\partial w_2} + \frac{\partial}{\partial w_3},\quad X_2^{\rm DBH}=w_1\frac{\partial}{\partial w_1} + w_2\frac{\partial}{\partial w_2} + w_3\frac{\partial}{\partial w_3}
\end{equation} 
satisfying the commutation relations (\ref{ComSl2}). In consequence, $X_1^{\rm DBH},X_2^{\rm DBH}$ and $X_3^{\rm DBH}$ span a three-dimensional Lie algebra of vector fields $V^{\rm DBH}$ isomorphic to $\mathfrak{sl}(2,\mathbb{R})$ and then $X^{\rm DBH}$ is an $\mathfrak{sl}(2,\mathbb{R})$-Lie system. Since $X_1,X_2,X_3$ admit the same structure constants as (\ref{ComSl2}), the symmetry system for $X^{\rm DBH}$ becomes (\ref{slsys}) with $b_1(t)=b_2(t)=0$ and $b_3(t)=-1$, namely
\begin{equation}
\frac{{\rm d}f_0}{{\rm d}t}=b_0(t),\qquad \frac{{\rm d}f_1}{{\rm d}t}=0,\qquad \frac{{\rm d}f_2}{{\rm d}t}=-2f_1,\qquad  \frac{{\rm d}f_3}{{\rm d}t}=-f_2-b_0(t).
\end{equation}
Hence, for $b_0(t)=0$, we obtain
 \begin{equation}
 Y=t_0\frac{\partial}{\partial t}+\lambda_1 X_1-(2\lambda_1 t-\lambda_2)X_2+(\lambda_1t^2-\lambda_2t+\lambda_3)X_3,
\end{equation}
with $\lambda_1,\lambda_2,\lambda_3,t_0\in\mathbb{R}$.
 Evidently, these vector fields span a Lie algebra of Lie symmetries isomorphic to $\mathfrak{sl}(2,\mathbb{R})$ for $t_0=0$. For $b_0(t)=c_0$, we obtain
\begin{equation}
 Y=(c_0t+t_0)\frac{\partial}{\partial t}+\lambda_1 X_1-(2\lambda_1 t-\lambda_2)X_2+(\lambda_1t^2-(\lambda_2+c_0)t+\lambda_3)X_3,
\end{equation}
with $\lambda_1,\lambda_2,\lambda_3,t_0,c_0\in\mathbb{R}.$
 Finally, for $b_0(t)=c_0t$, we get
\begin{multline}
 Y=\left(t_0+\frac{c_0t^2}2\right)\frac{\partial}{\partial t}+\lambda_1 X_1-(2\lambda_1 t-\lambda_2)X_2
\\
+\left[\lambda_1t^2-\left(\lambda_2+\frac{c_0t}2\right)t+\lambda_3\right]X_3,\,\, \end{multline} 
with $\lambda_1,\lambda_2,\lambda_3,t_0,c_0\in\mathbb{R}$. Since $\mathfrak{W}=\langle 1,t,t^2\rangle$ is a Lie algebra with respect to the Lie bracket, $\{\cdot,\cdot\}_\mathbb{R}$, we obtain in view of Proposition \ref{finite} that
\begin{equation}
 \mathcal{S}^V_{X,\mathfrak{W}}=\{Y\in \mathcal{S}^V_X|  f_0\in \langle 1,t,t^2\rangle\}
\end{equation}
 is a Lie algebra of Lie symmetries. By choosing appropriately the constant coefficients of the above vector fields and setting $\tau=0$, we recover the Lie algebra of symmetries isomorphic to $\mathfrak{sl}(2,\mathbb{R})$ for classical DBH systems \cite{Nu05}.

The study of systems of HODEs through Lie systems implies the addition of extra variables
in order to express a higher-order system as a first-order one. The introduction of these extra
variables frequently results in the obtainment of non-Lie point symmetries, as we shall exemplify in forthcoming examples.
Certain non-local Lie symmetries can be identified with the prolongations of certain vector fields.

\begin{lemma}
 Given a Lie algebra of Lie point symmetries $V$, its prolongations form a Lie algebra $\widehat{V}$ isomorphic to the former.
\end{lemma}
\begin{proof}
 Consider the map
\begin{equation}
\begin{aligned}
\Phi:\Gamma({\rm T}(\mathbb{R}\times N))&\rightarrow \Gamma({\rm T}[\mathbb{R}\times {\rm T}^{p}N])\\
X&\mapsto \widehat{X},
\end{aligned}
\end{equation}
mapping sections of ${\rm T}(\mathbb{R}\times N)$, i.e., vector fields on $\mathbb{R}\times N$, onto sections of ${\rm T}(\mathbb{R}\times {\rm T}^pN)$, i.e., vector fields on the manifold $\mathbb{R}\times{\rm T}^pN$, where ${\rm T}^pN$ is the 
$p$-order tangent space. Recall that, roughly speaking, ${\rm T}^pN$ is the space of equivalence classes of curves in $N$ with the same Taylor expansion up to order $p$. It admits a differentiable structure induced by the variables $x_1,\ldots,x_n$ of the coordinate systems of $N$ and the induced variables $x_i^{k)}$ describing the derivatives  in terms of $t$ up to order $p$ of the coordinates of a curve within $N$. Given the so called {\it contact one-forms} $\theta^k_{i}={\rm d}x^{k)}_i-x_i^{k+1)}{\rm d}t$ with $i=1,\dots,n$ and $k=0,\ldots,p-1$ on $\mathbb{R}\times {\rm T}^pN$, we say that $\widehat{X}$ is the prolongation of $X\in \Gamma({\rm T}(\mathbb{R}\times N))$ to $\mathbb{R}\times {\rm T}^pN$ if
and only if every $\mathcal{L}_{\widehat X}\theta^k_i$ belongs to the contact ideal spanned by all the contact forms and ${\rm J}^{p}_{\pi_{*}}\widehat{X}=X$, with ${\rm J}^p_{\pi*}$ being the tangent map to the projection ${\rm J}^p_\pi:(t,x,x^{k)})\in \mathbb{R}\times {\rm T}^pN\mapsto (t,x)\in\mathbb{R}\times N$ with $k=1,\ldots,p$. This implies that $\Phi$ is $\mathbb{R}$-linear and injective. Additionally,
\begin{equation}
 \mathcal{L}_{[\widehat{X_1},\widehat{X_2}]}\theta^k_i=\left(\mathcal{L}_{\widehat {X_1}}\mathcal{L}_{\widehat {X_2}}-\mathcal{L}_{\widehat {X_2}}\mathcal{L}_{\widehat {X_1}}\right)\theta^k_i
\end{equation}
belongs to the ideal of the contact forms because  $\mathcal{L}_{\widehat{X_1}}$ and  $\mathcal{L}_{\widehat{X_2}}$ do so.
Moreover,
\begin{equation}
{\rm J}^{p}_{\pi_{*}}[\widehat{X_1},\widehat{X_2}]=[{\rm J}^{p}_{\pi_{*}}\widehat{X_1},{\rm J}^{p}_{\pi_{*}}\widehat{X_2}]=[X_1,X_2].
\end{equation}
Hence, $[\widehat {X_1},\widehat {X_2}]$ must be the prolongation of $[X_1,X_2]$, i.e., $[\widehat {X_1},\widehat {X_2}]=\widehat{[X_1,X_2]}$. In this way, $\Phi$ is a Lie algebra morphism. Obviously, given a Lie algebra of Lie point symmetries $V$, its prolongations form a Lie algebra $\widehat{V}=\Phi(V)$ isomorphic to $V$.
\end{proof}
\begin{definition}
 We call $\widehat{V}$ the Lie algebra whose elements are the prolongations of the elements in ${V}.$
\end{definition}
\end{example}

\begin{example}\normalfont
As an example of a second-order Lie system possessing a Vessiot--Gulberg Lie algebra isomorphic to $\mathfrak{sl}(2,\mathbb{R})$,
we recall the {\bf second-order Kummer--Schwarz equation} given in \eqref{KS22} in Chapter \ref{Chap:LieSystems}.
In form of a first order system, it took the expression
\begin{equation}\label{FirstOrderKummerugh}
\left\{\begin{aligned}
\frac{dx}{dt}&=v,\\
\frac{dv}{dt}&=\frac 32 \frac{v^2}x-2c_0x^3+2b_1(t)x,
\end{aligned}\right.
\end{equation}
on ${\rm T}\mathbb{R}_0$, with
$\mathbb{R}_0=\mathbb{R}-\{0\}$.
Such system is associated with the $t$-dependent vector field $M^{\rm KS}=M_3^{\rm KS}+\eta(t)M_1^{\rm KS}$,
with vector fields 
\begin{equation}
 \begin{gathered}
M_1^{\rm KS}=2x\frac{\partial}{\partial v},\qquad M_2^{\rm KS}=x\frac{\partial}{\partial x}+2v\frac{\partial}{\partial v},
\\ M_3^{\rm KS}=v\frac{\partial}{\partial x}+\left(\frac{3}{2}\frac{v^2}{x}-2c_0x^3\right)\frac{\partial}{\partial v}
\end{gathered}\end{equation} 
satisfy the same commutation relations as the vector fields $X_1,X_2,X_3$ in (\ref{ComSl2}) and they therefore span a Vessiot--Guldberg Lie algebra isomorphic to $\mathfrak{sl}(2,\mathbb{R})$.
Since the basis $M_1^{KS},M_2^{KS},M_3^{KS}$ have the same structure constants as the $\mathfrak{sl}(2,\mathbb{R})$-Lie systems analyzed in Subsection \ref{LSSL2}, we can define, for instance, a Lie algebra morphism $\phi$ mapping $M_1^{KS},M_2^{KS},M_3^{KS}$ to the basis $X^{\rm Ric}_1,X^{\rm Ric}_2,X^{\rm Ric}_3$ for Riccati equations (\ref{Ricc}). In such a case, \eqref{FirstOrderKummerugh} maps to $X^{\rm Ric}$ and in view of Proposition \ref{SymTheo}, the symmetry system for \eqref{FirstOrderKummerugh} is of the form of the symmetry system (\ref{ric2}). As a consequence, the particular solutions $f_0$, $f_1$, $f_2$ and $f_3$ for Riccati equations, detailed in Table \ref{table8}, are valid for \eqref{FirstOrderKummerugh} as well.
\end{example}

\subsection{Lie symmetries for Aff($\mathbb{R}$)-Lie systems}

In this section, we aim to obtain elements of $\mathcal{S}^{V}_X$ for Aff$(\mathbb{R})$-Lie systems. We choose a basis of vector fields $\{X_1,X_2\}$ of $V\simeq$ Aff$(\mathbb{R})$
with $[X_1,X_2]=X_1,$
and express the most general Lie system with Vessiot--Guldberg Lie algebra $V$ as $X=a(t)X_1+b(t)X_2.$
Let us now look for its Lie symmetries of the form
\begin{equation}\label{wzor}
Y=f_0(t)\frac{\partial}{\partial t}+f_1(t)X_1+f_2(t)X_2.
\end{equation}

The symmetry condition gives rise to a symmetry system
\begin{equation}\label{Asys}
\left\{\begin{aligned}
 \frac{{\rm d}f_0}{{\rm d}t}&=b_0(t),\\
\frac{{\rm d}f_1}{{\rm d}t}&=f_0\frac{{\rm d}a}{{\rm d}t}(t)+a(t)b_0(t)+b(t)f_1-a(t)f_2,\\
\frac{{\rm d}f_2}{{\rm d}t}&=f_0\frac{{\rm d}b}{{\rm d}t}(t)+b(t)b_0(t),
\end{aligned}\right.
\end{equation}
associated with the $t$-dependent vector field
\begin{align}
\Gamma^{{\rm Aff}(\mathbb{R})}_X=b_0(t)Z_0+\frac{{\rm d} a}{{\rm d} t}W_1+\frac{{\rm d} b}{{\rm d} t}W_2&+b_0(t)a(t)Z_1
\nonumber\\
&+b_0(t)b(t)Z_2+a(t)Y_1+b(t)Y_2
\end{align}
where 
\begin{equation}
\begin{gathered}\label{com4}
Y_1=-f_2\frac{\partial}{\partial f_1},\quad Y_2=f_1\frac{\partial}{\partial f_1},\quad W_1=f_0\frac{\partial}{\partial f_1},
\\
 W_2=f_0\frac{\partial}{\partial f_2},\quad Z_\alpha=\frac{\partial}{\partial f_\alpha},
\end{gathered}\end{equation} 
with $\alpha=0,1,2.$
Since, $[Y_1,Y_2]=Y_1$, then $V_L=\langle Y_1,Y_2\rangle$ gives rise to a Lie algebra isomorphic to Aff$(\mathbb{R}).$
Moreover, 
%
\begin{equation}\label{comm4}
\begin{aligned}
\left[Z_0,W_1\right]&=Z_1,& [Z_0,W_2]&=Z_2,& [Z_i,W_j]&=0,& [Z_\alpha,Z_\beta]&=0, \\
[Y_1,W_1]&=0,& [Y_1,W_2]&=W_1,& [Y_1,Z_0]&=0,& [Y_1,Z_1]&=0,
\\ [Y_1,Z_2]&=Z_1,
&[Y_2,W_1]&=-W_1,& [Y_2,W_2]&=0,& [Y_2,Z_0]&=0,\\ [Y_2,Z_1]&=-Z_1,& [Y_2,Z_2]&=0.&&&&
\end{aligned}
\end{equation}
with $i,j=1,2$ and $\alpha,\beta=0,1,2.$ 
In this way, $A_1=\langle Z_0,Z_1,Z_2\rangle$ is an ideal of $A_1+A_2$, with $A_2=\langle W_1,W_2\rangle$  and $A_1\oplus_S A_2$ is an ideal of $A_1+A_2+V_L$. Hence, system (\ref{Asys}) possesses a Vessiot--Guldberg Lie algebra 
\begin{equation}\label{VG}
(A_1\oplus_S A_2)\oplus_S V_L\simeq (\mathbb{R}^3\oplus_S\mathbb{R}^2)\oplus_S{\rm Aff}(\mathbb{R}).
\end{equation}
We can solve (\ref{Asys}) when $b_0(t)=0$. In this case, $f_0=k\in \mathbb{R}$ and (\ref{Asys}) reduces to a  trivial equation for $f_2$ and a linear one for $f_1$. The general solution reads

\begin{equation}\label{sol}
\begin{gathered}
f_1(t)=\left[\int_0^t{\left[k\frac{{\rm d}a}{{\rm d}t'}(t')-a(t')(kb(t')+c_1)\right]e^{-\int_0^{t'}{b(t''){\rm d}t''}}{\rm d}t'}+c_2\right]e^{\int_0^{t}{b(t'){\rm d}t'}},\\
f_2(t)=kb(t)+c_1,
\end{gathered}
\end{equation}
where $c_1$ and $c_2$ are integration constants.

\begin{example}\normalfont
In order to illustrate Aff$(\mathbb{R})$-Lie systems through a physical example, we reconsider the {\bf Buchdahl equation} \cite{Bu64,CSL05,CN10}, 
which was analyzed in Chapter \ref{Chap:LieSystems} as a first-order system 
  \begin{equation}\label{Buchdahl12sym}
\left\{ \begin{aligned}
 \frac{\dd x}{\dd t}&=y,\\
 \frac{\dd y}{\dd t}&=a(x)y^2+b(t)y .
 \end{aligned}\right.
 \end{equation}

 Recall that we proven there that (\ref{Buchdahl12sym}) is a Lie system \cite{BBHLS}. Indeed, \eqref{Buchdahl12sym} describes the integral curves of the $t$-dependent vector field
\begin{equation}
X^{BD}=v\frac{\partial}{\partial x}+(f(x)v^2+a_2(t)v)\frac{\partial}{\partial
v}=X_1-a_2(t)X_2,\end{equation} 
where
\begin{equation}
X_1^{BD}=v\frac{\partial}{\partial x}+ f(x)v^2\frac{\partial}{\partial v},\qquad
X_2^{BD}=-v\frac{\partial }{\partial v},\end{equation} 
satisfy $[X_1,X_2]=X_1$. That is, $X$ is an ${\rm Aff}(\mathbb{R})$-Lie system. By applying Theorem \ref{The:SLS}, we see that the Lie symmetries of \eqref{Buchdahl12sym} of the form
\begin{equation}
Y=f_0(t)\frac{\partial}{\partial t}+f_1(t)X_1+f_2(t)X_2
\end{equation} 
are determined by the first-order system (\ref{Asys}) with $a(t)=1$ and $b(t)=-a_2(t)$, i.e.,
\begin{equation}\label{Buchdahl2}
\left\{\begin{aligned}
\frac{{\rm d}f_0}{{\rm d}t}&=b_0(t),\\
\frac{{\rm d}f_1}{{\rm d}t}&=b_0(t)-a_2(t)f_1-f_2,\\
\frac{{\rm d}f_2}{{\rm d}t}&=-f_0\frac{{\rm d}a_2}{{\rm d}t}-b_0(t)a_2(t),
\end{aligned}\right.
\end{equation}
which is associated to the $t$-dependent vector field
\begin{equation}
\Gamma_X^{{\rm BD}}=b_0(t)Z_0-\frac{{\rm d}a_2}{{\rm d}t}W_2+b_0(t)Z_1-a_2(t)b_0(t)Z_2+Y_1-a_2(t)Y_2,
\end{equation}
where the vector fields $Z_0,Z_1, Z_2,Y_1,Y_2,W_2$ are those detailed in \eqref{com4} and have
the commutation relations \eqref{comm4}. Hence, these vector fields span a Lie algebra (\ref{VG}). Therefore, we can obtain the Lie symmetries for this system of the form (\ref{wzor}) by substituting $a(t)=1$ and $b(t)=-a_2(t)$ in (\ref{sol}).
\end{example}

\subsection{Lie symmetries for PDE Lie systems}

Let us consider a system of PDEs 
\begin{equation}\label{pde}
 \frac{\partial x_i}{\partial t_l}=X_{li}(t,x),\qquad i=1,\ldots,n,\qquad l=1,\ldots,s,
\end{equation}
where the $X_{li}:\mathbb{R}^s\times \mathbb{R}^n\mapsto {\rm T}(\mathbb{R}^s\times \mathbb{R}^n) \in X(t,x)$ are arbitrary functions.
Its particular solutions are sections $x:t\equiv (t_1,\ldots,t_s) \in\mathbb{R}^s\mapsto x(t)\in \mathbb{R}^s\times \mathbb{R}^n$ of the bundle $(\mathbb{R}^s\times \mathbb{R}^n,\mathbb{R}^s,\pi_s:(t,x)\in \mathbb{R}^s\times \mathbb{R}^n\mapsto t\in \mathbb{R}^s)$.  In particular, we recover the simple case of ODEs when $s=1$.

Let us assume $t\in \mathbb{R}^s$. We call {\it $t$-dependent vector field} $X$ on $N$ a map $X:\mathbb{R}^s\times N\rightarrow {\rm T}N$ such that $\tau\circ X=\pi_s$. 
 We call $\bar{X}^l$, with $l=1,\ldots,s$, the {\it autonomization} of $X$ with respect to $t_l$, i.e.,
\begin{equation}\label{porx}
 \bar{X}^l(t,x)\equiv\frac{\partial}{\partial t_l}+X(t,x).
\end{equation}

\begin{definition}
 A first-order system of a number $n\times s$ PDEs of the form \eqref{pde}
\begin{equation}\label{pdesys}
 \frac{\partial x_i}{\partial t_l}=\sum_{\alpha=1}^rb_{\alpha l}(t)X_{\alpha}(x),\qquad l=1,\ldots,s,\quad i=1,\dots,n,
\end{equation}
is a {\it PDE Lie system} if
 \begin{itemize}
  \item  There exist vector fields $X_{1},\ldots,X_{r}$ on $N$ spanning an $r$-dimensional real Lie algebra $V$ satisfying that each $t$-dependent vector field $X_{l}(t,x)=\sum_{i=1}^n X_{li}(t,x)\cdot$ $\cdot{\partial}/{\partial x_i}$, where $l=1,\ldots,s$,
 can be written in the form
\begin{equation}
 X_{l}(t,x)=\sum_{\alpha=1}^r{b_{{\alpha l}}(t)X_{\alpha}(x)},\qquad l=1,\dots,s
\end{equation}
for certain $t$-dependent functions $b_{\alpha l}$.
\item Let $c_{\alpha\beta\gamma}$ be the structure constants of $X_1,\ldots,X_r$, i.e., $[X_\alpha,X_\beta]=\sum_{\gamma=1}^rc_{\alpha\beta\gamma}X_\gamma$. Then,
\begin{equation}\label{conint}
 \frac{\partial b_{{\gamma} k}}{\partial t_l}(t)-\frac{\partial b_{{\gamma} l}}{\partial t_k}(t)+\sum_{\alpha,\beta=1}^rb_{{\alpha} l}(t)b_{{\beta} k}(t)c_{\alpha \beta \gamma}=0,
\end{equation}
with $ l\neq k=1,\ldots,s,\quad \gamma=1,\ldots,r.$
\end{itemize}
If these conditions are satisfied, we call $V$ a {\it Vessiot--Guldberg Lie algebra for the PDE Lie system (\ref{pde})}.
\end{definition}

\begin{theorem}\label{the:MainPDE}Let $V$ be a finite-dimensional real Lie algebra of vector fields on $\mathbb{R}^n$ with a basis $X_{1},\ldots,X_r$.
Given a PDE Lie system as \eqref{pdesys},
then
\begin{equation}\label{SymSpe}
Y(t,x)=\sum_{\beta=1}^rf_{\beta}(t)X_{\beta}(x)
\end{equation}
is a Lie symmetry of (\ref{pdesys}) if and only if $[\bar{X}^l,Y]=0$ for every $l$, where $\bar X^l$ is the $l$-autonomization given by (\ref{porx}) of the $t$-dependent vector field
\begin{equation}
X_l(t,x)\equiv \sum_{\alpha =1}^ rb_{\alpha l }(t)X_\alpha(x),\qquad l=1,\ldots,s.
\end{equation} 
\end{theorem}
\begin{proof}
The coordinate systems $\{t_1,\ldots,t_s\}$ on $\mathbb{R}^s$ and $\{x_1,\ldots,x_n\}$ on $\mathbb{R}^n$ induce a coordinate system on the fiber bundle ${\rm J}^1\pi\simeq \mathbb{R}^s\times \mathbb{R}^ n\times\mathbb{R}^{ns}$ with respect to the projection
$\pi:(t_l,x_j)\in\mathbb{R}^s\times\mathbb{R}^n\mapsto t_l\in \mathbb{R}^s$ of the form $t_l,x_j,x_{j,l}$, where $x_{j,l}=\partial x_j/\partial x_l$ with $1\leq j\leq n$ and $1\leq l\leq s$.
A vector field $Y=\sum_{k=1}^n\eta_k(t,x)\partial/\partial x_k,$ $t\in \mathbb{R}^s$ and $x\in \mathbb{R}^n$, is a Lie symmetry of (\ref{pdesys}) if and only if
\begin{equation}
\widehat{Y}F^i_l=0,\qquad\qquad  i=1,\ldots, n,\qquad l=1,\ldots,s\end{equation} 
on the submanifold $\mathcal{S}=\cap_{i=1}^n\cap_{l=1}^s (F^i_l)^{-1}(0)$, with $\widehat{Y}:{\rm J}^1\pi\rightarrow {\rm T}({\rm J}^1\pi)$ being the prolongation of $Y,$ namely
\begin{equation}
\widehat{Y}=\sum_{k=1}^n \left[\eta_k\frac{\partial}{\partial x_k}+\sum_{q=1}^s\left(\frac{\partial \eta_k}{\partial t_q}+\sum_{j=1}^n\frac{\partial \eta_k}{\partial x_j}x_{j,q}\right)\frac{\partial}{\partial x_{k,q}}\right],
\end{equation} 
and
\begin{equation}
\begin{gathered}
F_l^i=x_{i,l}-\sum_{\alpha=1}^rb_{\alpha l}(t)X_{\alpha}(x),\qquad i=1,\ldots,n,\qquad l=1,\ldots,s,\\
X_\alpha=\sum_{i=1}^nX_\alpha^i(x)\frac{\partial}{\partial x_i},\qquad\alpha=1,\ldots,r.
\end{gathered}\end{equation} 
By assumption, $\eta_k(t,x)=\sum_{\beta=1}^rf_{\beta}(t)X_{\beta}^k(x)$ for $k=1,\ldots,n$. So,
\begin{equation}
 \widehat{Y}=\sum_{k=1}^n\sum_{\beta=1}^r\left[ f_{\beta}(t)X_{\beta}^k\frac{\partial}{\partial x_k}+\sum_{q=1}^s \left(\frac{\partial f_{\beta}(t)}{\partial t_q}X_{\beta}^k+\sum_{j=1}^nf_{\beta}(t)\frac{\partial X_{\beta}^k}{\partial x_j}x_{j,q}\right) \frac{\partial}{\partial x_{k,q}}\right].
\end{equation}
Substituting this in $\widehat{Y}F^i_l=0$, we obtain
\begin{align}
\sum_{k=1}^n \sum_{\beta=1}^r&\left[f_{\beta}(t)X_{\beta}^k\left(-\sum_{\alpha=1}^rb_{\alpha l}(t)\frac{\partial X_{\alpha}^i}{\partial x_k}\right)
\right.+\left.
{}
 \sum_{q=1}^s\left(\frac{\partial f_{\beta}(t)}{\partial t_q}X_{\beta}^k
+
\sum_{j=1}^n{f_{\beta}(t)\frac{\partial X_{\beta}^k}{\partial x_j}x_{j,q}}\right)\delta^{l}_q\delta^i_k\right]=
\nonumber\\
&=\sum_{\beta=1}^r\left[ -\sum_{k=1}^n \sum_{\alpha=1}^r f_{\beta}(t)b_{\alpha l}(t)X_{\beta}^k\frac{\partial X_{\alpha}^i}{\partial x_k}+\frac{\partial f_{\beta}(t)}{\partial t_l}X_{\beta}^i+\sum_{j=1}^n f_{\beta}(t)\frac{\partial X_{\beta}^i}{\partial x_j}x_{j,l}\right]=0
\nonumber\\.
\end{align}
Restricting the above expression to the submanifold $\mathcal{S}=\cap_{i=1}^n\cap_{l=1}^s (F^i_l)^{-1}(0)$ and renaming indexes appropriately, we obtain
\begin{equation}
 \sum_{\beta=1}^r \left[\frac{\partial f_{\beta}(t)}{\partial t_l}X_{\beta}^i+ \sum_{k=1}^n \sum_{\alpha=1}^r\left(f_{\beta}(t)\frac{\partial X_{\beta}^i}{\partial x_k}b_{\alpha l}(t)X_{\alpha}^k-f_{\beta}(t)b_{\alpha l}(t)X_{\beta}^k\frac{\partial X_{\alpha}^i}{\partial x_k}\right)\right]=0.
\end{equation}
Hence,
\begin{equation}
\sum_{\beta=1}^r \left[ \frac{\partial f_{\beta}(t)}{\partial t_l}X_{\beta}^i+\sum_{\gamma=1}^r\sum_{k=1}^n f_{\beta}(t)b_{\gamma l}(t)\left(\frac{\partial X_{\beta}^i}{\partial x_k}X_{\gamma}^k-X_{\beta}^k\frac{\partial X_{\gamma}^i}{\partial x_k}\right)\right]=0,
\end{equation}
whose right-hand side becomes, for each fixed $l$, the coefficients in the basis $\partial/\partial x^i$, with $i=1,\ldots,n$, of $[\bar X^l,Y]$. So, the above amounts to
\begin{align}
 \left[\bar{X}^l,Y\right]&=\left[\frac{\partial}{\partial t_l}+\sum_{k=1}^n \sum_{\gamma=1}^r b_{\gamma l}(t)X_{\gamma}^k\frac{\partial}{\partial x_k},\sum_{i=1}^n \sum_{\beta=1}^r f_{\beta}(t)X_{\beta}^i \frac{\partial}{\partial x_i}\right]\nonumber\\
&=\left[\frac{\partial}{\partial t_l}+\sum_{\gamma=1}^r b_{\gamma l}(t)X_\gamma,\sum_{\beta=1}^r f_{\beta}(t)X_\beta\right]=0.
\end{align}
Then, $Y$ is a Lie symmetry of (\ref{pdesys}) if and only if the condition
\begin{equation}\label{symcond2}
 \left[\bar{X}^l,Y\right]=0
\end{equation}
is satisfied for $l=1,\ldots,s.$
\end{proof}

\begin{theorem}\label{MTPDELieSystems}
Given a Lie symmetry of the form (\ref{SymSpe}) for the system (\ref{pdesys}), the
coefficients $f_1(t),\dots,f_r(t)$ satisfy a PDE Lie system admitting a Vessiot--Guldberg Lie algebra $V^{S}\simeq V/Z(V)$, where we recall that $V$ is a Vessiot--Guldberg Lie algebra for (\ref{pdesys}).
\end{theorem}
\begin{proof}
Let $X_1,\ldots,X_r$ be a basis for $V$ with structure constants $c_{\alpha\beta\gamma}$. From Theorem \ref{the:MainPDE}, we have
\begin{align}
\left[\bar{X}^l,Y\right]&=\left[\frac{\partial}{\partial t_l}+\sum_{\alpha=1}^{r}{b_{\alpha l}(t)X_{\alpha}},\sum_{\delta=1}^r f_{\delta}(t)X_{\delta}\right]=\sum_{\delta=1}^r{\left({\frac{\partial f_{\delta}}{\partial t_l}X_{\delta}}+\sum_{\alpha=1}^r{b_{\alpha l}(t)f_{\delta}[X_{\alpha},X_{\delta}]}\right)}\nonumber\\
&=\sum_{\pi=1}^r{\left(\frac{\partial f_\pi}{\partial t_l}+\sum_{\alpha=1}^r \sum_{\delta=1}^r b_{\alpha l}(t)f_{\delta}c_{\alpha \delta \pi}\right)X_\pi}=0.
\end{align}
Since $X_1,\dots,X_r$ are linearly independent over $\mathbb{R}$ and the coefficients of the above expression are only $t$-dependent, we get that the above amounts to
\begin{equation}\label{PDESymSys}
\frac{\partial f_\pi}{\partial t_l}=\sum_{\alpha,\delta=1}^r b_{\alpha l}(t)f_{\delta}c_{\delta\alpha  \pi},\qquad \pi=1,\ldots,r,\quad l=1,\ldots,s.
\end{equation}

To prove that this is a PDE Lie system, we define the vector fields
\begin{equation}
Y_\alpha=\sum_{\delta,\pi=1}^rc_{\delta \alpha \pi}f_\delta\frac{\partial}{\partial f_\pi},\qquad \alpha=1,\ldots,r.
\end{equation}
We have already proven that $[Y_\alpha,Y_\beta]=\sum_{\delta=1}^rc_{\alpha\beta\delta}Y_\delta$ in Theorem \ref{The:SLS}. So, these vector fields span a Lie algebra isomorphic to $V/Z(V)$ (for a proof of this fact follow the same line of reasoning as in Theorem \ref{The:SLS}).
In terms of these vector fields, we see that (\ref{PDESymSys}) is related to the $t$-dependent vector fields $X_l(t,x)=\sum_{\alpha=1}^rb_{\alpha l}(t)X_\alpha(x)$, with $l=1,\ldots,s$.
Additionally, to be a PDE Lie system, the above system (\ref{PDESymSys}) must satisfy the condition
\begin{equation}\label{conint}
\sum_{\alpha,\mu=1}^r\left(\frac{\partial b_{\alpha \pi}}{\partial \kappa}-\frac{\partial b_{\alpha \kappa}}{\partial \pi}+\sum_{\delta,\epsilon=1}^rb_{\delta \kappa}b_{\epsilon \pi}c_{\delta\epsilon\alpha}\right)f_\mu c_{\mu\alpha\sigma}=0,
\end{equation}
with $\kappa\neq \pi=1,\ldots,r,\quad\sigma=1,\ldots,r.$
The expression in brackets vanishes due to the integrability condition for system (\ref{pdesys}). Hence, (\ref{PDESymSys}) is a PDE Lie system. We call \eqref{PDESymSys} the {\it symmetry system} for (\ref{pdesys}) relative to $V$.
\end{proof}

\begin{definition}
 Given a PDE Lie system $X$ with a Vessiot--Guldberg Lie algebra $V$, we call {\it $\mathcal{S}_X^{V}$
the space of Lie symmetries} of $X$ that are also $t$-dependent vector fields taking values in $V$.
\end{definition}

We can straightforwardly prove that the space $\mathcal{S}_X^{V}$ is a Lie algebra.

\subsection{Lie symmetries for $\mathfrak{sl}(2,\mathbb{R})$-PDE Lie systems}

An $\mathfrak{sl}(2,\mathbb{R})$-PDE Lie system is a PDE Lie system admitting a Vessiot--Guldberg Lie algebra isomorphic to $\mathfrak{sl}(2,\mathbb{R}).$
Let us obtain the elements of $\mathcal{S}_X^V$ for this case. 
Let us choose a basis of vector fields $\{X_1,X_2,X_3\}$ for $V$ satisfying the same commutation relations as in \eqref{ComSl2}.
Let us write a general PDE Lie system whose autonomization for a fixed value $l$ is
\begin{equation}
 \bar{X}^l=\frac{\partial}{\partial t_l}+b_{1l}(t)X_1+b_{2l}(t)X_2+b_{3l}(t)X_3,
\end{equation}
with $t=(t_1,\ldots,t_s) \in \mathbb{R}^s,\, 1\leq l\leq s$
and a certain type of possible Lie symmetry $Y=f_1(t)X_1+f_2(t)X_2+f_3(t)X_3$, where $f_1(t),f_2(t),f_3(t)$ are $t$-dependent functions to be determined by the symmetry condition (\ref{symcond2}).
This leads us
to the system of $s$ first-order PDEs
\begin{equation}
\left\{\begin{aligned}
&\frac{\partial f_1}{\partial t_l}=b_{2l}(t)f_1-b_{1l}(t)f_2,\\
 &\frac{\partial f_2}{\partial t_l}=2(b_{3l}(t)f_1-b_{1l}(t)f_3),\\
 & \frac{\partial f_3}{\partial t_l}=b_{3l}(t)f_2-b_{2l}(t)f_3,
\end{aligned}\right.
\end{equation}
with $l=1,\ldots,s$. Expressed in terms of $t$-dependent vector fields, for a fixed value of $l$
\begin{equation}
\Gamma^{\mathfrak{sl}(2,\mathbb{R})}_{l}=b_{1l}(t)Y_{1}+b_{2l}(t)Y_{2}+b_{3l}(t)Y_{3}, \quad l=1,\ldots,s,
\end{equation}
where
\begin{equation}
\begin{gathered}\label{eq2}
 Y_{1}=-f_2\frac{\partial}{\partial f_1}-2f_3\frac{\partial}{\partial f_2},\quad Y_{2}=f_1\frac{\partial}{\partial f_1}-f_3\frac{\partial}{\partial f_3},
\\
 Y_{3}=2f_1\frac{\partial}{\partial f_2}+f_2\frac{\partial}{\partial f_3}
\end{gathered}\end{equation} 
close the commutation relations in (\ref{ComSl2}). So, they span a Lie algebra isomorphic to $\mathfrak{sl}(2,\mathbb{R})$.

\begin{example}\normalfont
Let us consider the {\bf partial Riccati equation}, i.e., the PDE system
\begin{equation}\label{PDERicc}
 \frac{\partial x}{\partial t_1}=b_{11}(t)+b_{21}(t)x+b_{31}(t)x^2,\qquad
\frac{\partial x}{\partial t_2}=b_{12}(t)+b_{22}(t)x+b_{32}(t)x^2,
\end{equation}
with the $t$-dependent coefficients satisfying the appropriate integrability condition (\ref{conint}).
Such systems appear in the study of WZNW equations and multidimensional Toda systems \cite{FGRSZ99}. Observe that the partials $\partial x/\partial t_1$ and $\partial x/\partial t_2$ are related to the $t$-dependent vector fields
${X}^{\rm {pRic}}_{t_1}=b_{11}(t)X_{1}^{\rm {pRic}}+b_{21}(t)X_{2}^{\rm {pRic}}+b_{31}(t)X_{3}^{\rm {pRic}}$ and
${X}^{\rm {pRic}}_{t_2}=b_{12}(t)X_{1}^{\rm {pRic}}+b_{22}(t)X_{2}^{\rm {pRic}}+b_{32}(t)X_{3}^{\rm {pRic}}$,
with 
\begin{equation}
X_{1}^{\rm {pRic}}=\frac{\partial}{\partial x},\quad X_{2}^{\rm {pRic}}=x\frac{\partial}{\partial x},\quad X_{3}^{\rm {pRic}}=x^2\frac{\partial}{\partial x}\end{equation} 
satisfying the commutation relations (\ref{ComSl2}). That is, the vector fields $\langle X_{1}^{\rm {pRic}},X_{2}^{\rm {pRic}},$ $X_{3}^{\rm {pRic}}\rangle$ span a Vessiot--Guldberg Lie algebra $V^{\rm {pRic}}\simeq \mathfrak{sl}(2,\mathbb{R}).$ Since we assume that the functions $b_{ij}(t)$ with $i=1,2,3$ and $j=1,2$ satisfy (\ref{conint}), we get that (\ref{PDERicc}) is a PDE Lie system.

Let us look for Lie symmetries of the form $Y=f_1(t)X^{\rm pRic}_{1}+f_2(t)X^{\rm pRic}_{2}+f_3(t)X^{\rm pRic}_{3}$ for (\ref{PDERicc}). In view of Theorem \ref{MTPDELieSystems}, such Lie symmetries are solutions of the system of PDEs
\begin{equation}
\left\{\begin{aligned}
&\frac{\partial f_1}{\partial t_j}={b_{2j}}(t)f_1-{b_{1j}}(t)f_2,\\
&\frac{\partial f_2}{\partial t_j}=2({b_{3j}}(t)f_1-{b_{1j}}(t)f_3),\\
&\frac{\partial f_3}{\partial t_j}={b_{3j}}(t)f_2-{b_{2j}}(t)f_3,
\end{aligned}\right.
\end{equation}
 with $j=1,2.$
This resulting system can be interpreted in terms of the $t$-dependent vector fields $\Gamma^{\rm{pRic}}_j=b_{1j}(t)Y_{1}+b_{2j}(t)Y_{2}+b_{3j}(t)Y_{3}$, with $j=1,2$ and (\ref{eq2}).
These vector fields have the same structure constants as the $X^{\rm pRic}_1,X^{\rm pRic}_2,X^{\rm pRic}_3$. Therefore, \eqref{PDERicc} is a PDE Lie system with a Vessiot--Guldberg Lie algebra isomorphic to $\mathfrak{sl}(2,\mathbb{R}).$
\end{example}

\section{The CLS and NSM for PDEs}
\setcounter{equation}{0}
\setcounter{theorem}{0}
\setcounter{example}{0}

Let us now consider a manifold ${\rm J}^p\pi$ locally isomorphic to $\mathbb{R}^n\times {\rm T}^p\mathbb{R}^k$ where we now assume the bundle induced by $\pi:(x,u)\in\mathbb{R}^n\times\mathbb{R}^k\simeq N_{\mathbb{R}^n}\mapsto x\in \mathbb{R}^n$,
where $x=(x_1,\dots,x_n)$ and $u=(u_1,\dots,u_k)$. The manifold ${\rm J}^p\pi$ admits a coordinate system $\{x_1,\ldots,x_n,u_1,\ldots, u_k,u_J\}_{|J|=p}$, where $J$  is a multi-index $J\equiv (j_1,\ldots, j_n)$ and $|J|=p$ means that $|J|=j_1+\ldots +j_n$. 
When necessary, we will write 
\begin{equation}
u_{J}=(u_j)_{x_{i_1}^{j_1},\ldots,x_{i_n}^{j_n}}\equiv \frac{\partial^{|J|}u}{\partial x_{i_1}^{j_1}\partial x_{i_2}^{j_2},\ldots,\partial x_{i_n}^{j_n}},\quad j_1+\dots+j_n=p.
\end{equation}
We study systems of PDEs of $p$-order of the form
\begin{equation}\label{genpde}
\Psi^l=\Psi^l\left(x_i,u_j,(u_j)_{x_i},(u_j)_{x_{i_1}^{j_1},x_{i_2}^{j_2}},\dots,(u_j)_{x_{i_1}^{j_1},x_{i_2}^{j_2},x_{i_3}^{j_3},\dots,x_{i_n}^{j_n}}\right),\qquad l=1,\ldots q.
\end{equation}
such that $j_1+\dots+j_n=p$, $1\leq i_1,\dots,i_n\leq n$ and $x_1\leq x_i\leq x_n$, $j=1,\dots,k$.
Geometrically, we understand this expression as a submanifold of $\mathcal{E}\subset {\rm J}^p\pi$.

We search for an $\epsilon$-parametric group of transformations on $N_{\mathbb{R}^n}$, namely a Lie group action $\Phi:\mathbb{R}\times N_{\mathbb{R}^n}\longrightarrow N_{\mathbb{R}^n}.$
Infinitesimally, this $\epsilon$-parametric group reads
\begin{equation}
 \left\{\begin{aligned}
\bar{x}_i&\rightarrow x_i+\epsilon \xi_i(x,u)+O\left(\epsilon^2\right),\\
\bar{u}_j&\rightarrow u_j+\epsilon \eta_{j}(x,u)+O\left(\epsilon^2\right),
\end{aligned}\right.\end{equation} 
for all  $i=1,\ldots,n$ and $ j=1,\ldots,k.$


\begin{definition}
Given a system of PDEs differential equations \eqref{genpde},
we say that $\Phi$ is a {\it Lie point symmetry} for \eqref{genpde} if
for every particular solution $(x,u(x))$ is mapped into a new particular solution $(\bar x,\bar u(\bar x))\equiv \Phi\circ (x,u(x))$.
\end{definition}

\begin{definition}
We call {\it invariant surface conditions} the set of expressions
\begin{equation}\label{invsurfcond}
\eta_{{j}}-\sum_{i=1}^n(u_{ j})_{x_i}\xi_i=0,\quad \quad j=1,\dots,k.
\end{equation}
\end{definition}
\begin{definition}
A {\it nonclassical Lie point  symmetry} is a transformation that does not only leave the PDE problem \eqref{genpde} invariant,
but the set surface conditions \eqref{invsurfcond} as well.
\end{definition}
For systems of PDEs with derivatives that up to order $p$, we propose a general transformation

\begin{equation}
 \left\{\begin{aligned}\label{gentranspde}
&\bar{x}_i\rightarrow x_i+\epsilon \xi_i(x,u)+O(\epsilon^2), \\
&\bar{u}_j\rightarrow u_j+\epsilon \eta_{j}(x,u)+O(\epsilon^2), \\
&(\bar{u}_j)_{\bar{x}_i}\rightarrow (u_j)_{x_i}+\epsilon (\eta_{j})_{x_i}+O(\epsilon^2),\\
&(\bar{u}_j)_{{\bar x}_{i_1}^{j_1},{\bar x}_{i_2}^{j_2}}\rightarrow ({u}_j)_{x_{i_1}^{j_1},x_{i_2}^{j_2}}+\epsilon (\eta_{j})_{x_{i_1}^{j_1},x_{i_2}^{j_2}}+O(\epsilon^2),\\
&\qquad \qquad \dots  \\
&(\bar{u}_j)_{\bar{x}_{i_1}^{j_1},\bar{x}_{i_2}^{j_2},\dots,\bar{x}_{i_n}^{j_n}}\rightarrow (u_j)_{x_{i_1}^{j_1},x_{i_2}^{j_2},\dots,x_{i_n}^{j_n}}+\epsilon (\eta_{j})_{x_{i_1}^{j_1},x_{i_2}^{j_2},\dots,x_{i_n}^{j_n}}+O(\epsilon^2),
\end{aligned}\right.\end{equation} 
\noindent
such that $j_1+\dots+j_n\leq p$, $x_1\leq x_i\leq x_n$ and where $(\eta_{j})_{x_i},\dots,(\eta_{j})_{x_{i_1}^{j_1},x_{i_2}^{j_2},\dots,x_{i_n}^{j_n}}$ are the prolongations of $\eta_{j}$
from first, up to $p$-order. We have not specified the dependency on $(x_i,u_j)$ for a matter of simplicity.

We can write the {\it $p$-order infinitesimal generator $X^p$} on ${\rm T}^pN_{\mathbb{R}^n}$ for a transformation \eqref{gentranspde}
 of the system of $p$-order differential equations \eqref{genpde}, as
%
\begin{align}\label{compsymvf}
{X}^p=\sum_{i=1}^n \xi_i\frac{\partial}{\partial x_i}+\sum_{j=1}^k \eta_{j}\frac{\partial}{\partial u_j}&+\sum_{i=1}^n\sum_{j=1}^k (\eta_{{j}})_{x_i}\frac{\partial}{\partial (u_j)_{x_i}}
\nonumber\\
& \hspace*{-3em}+\sum_{i_1,i_2=1}^n\sum_{j=1}^k (\eta_{j})_{x_{i_1}^{j_1},x_{i_2}^{j_2}}\frac{\partial}{\partial (u_{j})_{x_{i_1}^{j_1},x_{i_2}^{j_2}}}
\nonumber\\
&\hspace*{-3em}+\ldots+\sum_{i_1,i_2,\ldots,i_n=1}^n \!\!\!\!\cdots\,\,\sum_{j=1}^k (\eta_{j})_{x_{i_1}^{j_1}\dots x_{i_n}^{j_n}}\frac{\partial^{|J|}}{\partial (u_{j})_{x_{i_1}^{j_1}\ldots x_{i_n}^{j_n}}}.
\end{align}
such that $j_1+\dots+j_n\leq p$

\subsection{Algorithmic computation of Lie point  symmetries for systems of PDEs}

It is possible to follow an algorithmic procedure to compute Lie point  symmetries of PDEs. We resume the process in
the following number of steps

\begin{enumerate}
\item Introduce the $\epsilon$-parametric group of transformations \eqref{gentranspde} in the PDE problem \eqref{genpde}.
\item Take the zero-order coefficient in $\epsilon$.
We recover the initial, untransformed equations.
 Isolate the higher-order derivative $(u_{j})_{x_{i_1}^{j_1} \dots x_{i_n}^{j_n}}$ in terms of the remaining ones by using the relations \eqref{genpde}.
\item Take the coefficient of first-order in $\epsilon$. Introduce all the prolongations $(\eta_{u_{j}})_{x_{i_1}^{j_1} \dots x_{i_n}^{j_n}}$,
$\dots, (\eta_{u_{j}})_{x_i}$, computed according to \eqref{prolongcomplete} given in Chapter \ref{Chap:GeomFund}. Introduce the correspondent expressions of the higher-order derivatives $(u_{j})_{x_{i_1}^{j_1}\dots x_{i_n}^{j_n}}$ 
of the untransformed equations retrieved in the previous step.

Select the coefficients of the different order derivatives, from the $p-1$ order $(u_{j})_{x_{i_1}^{j_1}\dots x_{i_{n}}^{j_{n}-1}}$, (since the highest-order has already been substituted in the previous step)
down to the first-order $(u_{j})_{x_i}$ and such that $j_1+\dots+(j_{n}-1)=p-1$. It is legit to set such coefficients equal to zero, given the definition
of point symmetries, in which $\xi_i$ and $\eta_{j}$ do not depend on derivatives of any order.
We obtain an overdetermined linear system of differential equations for the coefficients $\xi_{i}$ and $\eta_{j}$ of the infinitesimal
generator \eqref{compsymvf}.

\item At this point, we have to decide whether we search for {\it classical or nonclassical Lie point  symmetries}.
We implement the invariant surface conditions \eqref{invsurfcond} in case of searching for nonclassical ones. 

If $\xi_{\hat i}\neq 0$, where $1\leq \hat{i}\leq n$, for a particular fixed value $\hat i$ of the index $i$,
\begin{equation}
(u_{ j})_{x_{\hat i}}=\frac{\eta_{{ j}}-\sum_{i\neq \hat{i}=1}^n (u_{ j})_{x_i}\xi_i}{\xi_{\hat i}},
\quad \quad 1\leq {j}\leq k.
\end{equation}
In forthcoming examples we shall see that there exists a plethora of cases depending on whether certain $\xi_i$ are nonzero or null.

\end{enumerate}

\subsection{Reduction method of systems of PDEs}

Given a $p$-order system of PDEs  \eqref{genpde} and an $\epsilon$-parametric group of transformations \eqref{gentranspde} for such a system,
there exists an associated infinitesimal generator $X^p$ given in \eqref{compsymvf} for such a transformation which 
is tangent to the submanifold $\mathcal{E}$ given by 
\begin{equation}
0=\Psi^{l}\left(x_i,u_j,(u_j)_{x_i},(u_j)_{x_{i_1}^{j_1},x_{i_2}^{j_2}},\dots,(u_j)_{x_{i_1}^{j_1} \dots x_{i_n}^{j_n}}\right),\qquad l=1,\ldots,q,
\end{equation}
on ${\rm J}^p\pi$, such that $j_1+\dots+j_n=p$, namely $X^p$ satisfies that $X^p\Psi^l=0$ for $l=1,\ldots,q$ on $\mathcal{E}$. 
Observe that 
\begin{multline}
\xi_i\Psi_{x_i}^l+\eta_{j}\Psi_{u_j}^l+(\eta_{j})_{x_i}\Psi^l_{(u_j)_{x_i}}+(\eta_{j})_{x_{i_1}^{j_1},x_{i_2}^{j_2}}\Psi^l_{(u_j)_{x_{i_1}^{j_1}x_{i_2}^{j_2}}}
\\
+\ldots+(\eta_{j})_{x_{i_1}^{j_1}\ldots x_{i_n}^{j_n}}\Psi^l_{(u_j)_{x_{i_1}^{j_1},\dots,x_{i_n}^{j_n}}}=0,
\end{multline}
where we denote by $\Psi^l_y$ the partial derivative $\partial \Psi^l/\partial y$.
Geometrically, this means that a tangent vector to $\mathcal{E}$ has coordinates
\begin{equation}
\left(\xi_i,\eta_{j},(\eta_{j})_{x_i},(\eta_{j})_{x_{i_1}^{j_1},x_{i_2}^{j_2}},\dots,(\eta_{j})_{x_{i_1}^{j_1} \dots x_{i_n}^{j_n}}\right)
\end{equation}
such that $j_1+\dots+j_n=p$, $i_1,\dots,i_n=1,\dots,n,$ $j=1,\dots,k$ and $x_1\leq x_i\leq x_n$ in the natural basis given by
\begin{equation}\label{basis1}
\frac{\partial}{\partial x_i},\quad\,\,\frac{\partial}{\partial u_j},\quad\,\,\frac{\partial}{\partial (u_j)_{x_{i}}},\quad\,\,\frac{\partial}{\partial (u_j)_{x_{i_1}^{j_1}x_{i_2}^{j_2}}},\quad\,\,\ldots\,\,\quad \frac{\partial}{\partial (u_j)_{x_{i_1}^{j_1}\ldots x_{i_n}^{j_n}}}.
\end{equation}
Then, \eqref{prODE} is equivalent to saying that the vector 
\begin{equation}
\left(\Psi_{x_i}^l,\Psi_{u_j}^l,\Psi^l_{(u_j)_{x_i}},\Psi^l_{(u_j)_{x_{i_1}^{j_1},x_{i_2}^{j_2}}},\dots,\Psi^l_{(u_j)_{x_{i_1}^{j_1},\dots,x_{i_n}^{j_n}}}\right)
\end{equation}
in the basis (\ref{basis1}) is normal to the submanifold $\mathcal{E}\subset {J}^p\pi$ with respect to the standard Euclidean product. Observe that this
is so for every vector tangent to $\mathcal{E}$.
In other words, the graph of the solution must be a union of integral curves of this vector field. These integral curves are called
the {\it characteristic curves} of the original PDE.

The integral curves of $X^p$ are given by the particular solutions of 
\begin{equation}\label{LCeqPDE2}
\frac{dx_1}{\xi_{1}}=\dots=\frac{dx_n}{\xi_{n}}=\frac{du_1}{\eta_{1}}=\dots=\frac{du_k}{\eta_{k}},
\end{equation}
the so called {\it Lagrange-Charpit system} \cite{Stephani}. The recommended procedure to solve the above system is to fix one of the variables, for example $x_1$, and integrate by pairs
\begin{equation}
\begin{aligned}\label{intLEeqPDE}
&\frac{dx_1}{\xi_{1}}-\frac{dx_2}{\xi_{2}}=0,\quad \dots \quad \frac{dx_1}{\xi_{1}}-\frac{dx_n}{\xi_{n}}=0,\\
&\frac{dx_1}{\xi_{1}}-\frac{du_1}{\eta_{1}}=0,\quad \dots \quad \frac{dx_1}{\xi_{1}}-\frac{du_k}{\eta_{k}}=0.
\end{aligned}\end{equation} 

Each of these one-form differential equations gives rise to a constant of integration, a total of $n+k-1$, respectively
$z_2,\dots,z_n,U_1,\dots,U_k$.  
\begin{definition}
We call $z_2,\dots,z_n$ {\it reduced variables} and  $U_1,\dots,U_k$ {\it reduced scalar fields}   arising from the integration of the above mentioned system.
\end{definition}
One of variables from the $n+k$-tuple $\{x_1,\dots,x_n,u_1,\dots,u_k\}$
can be written in terms of the remaining $n+k-1$ functions $\{z_2,\dots,z_n,U_1,\dots,U_k\}$. 
\begin{equation}
\left\{\begin{aligned}\label{redd1}
&z_{\hat i}=z_{\hat i}\left(x_1,x_{\hat i},\xi_{1}(x_1,x_{\hat i}),\xi_{\hat i}(x_1,x_{\hat i})\right),&& 2\leq\hat{i}\leq n,\\
&U_{\hat j}=U_{\hat j}\left(x_1,u_{\hat j},\xi_{1}(x_1,u_{\hat j}),\eta_{\hat j}(x_1,u_{\hat j})\right), && 1\leq \hat{j}\leq k.
\end{aligned}\right.
\end{equation}
\noindent
To obtain the reduction, we introduce the new variables by inverting \eqref{redd1} as
\begin{equation}
\left\{\begin{aligned}\label{red1inv}
&x_{\hat i}=F_{\hat i}\left(x_1,z_{\hat i},\xi_1(x_1,z_{\hat i}),\xi_{\hat i}(x_1,z_{\hat i})\right),&&2\leq{\hat i}\leq n, \\
&u_{\hat j}=G_{\hat j}\left(x_1,U_{\hat j},\xi_1(x_1,U_{\hat j}),\eta_{{\hat j}}(x_1,U_{\hat j})\right), && 1\leq {\hat j}\leq k.
\end{aligned}\right.
\end{equation}
The transformation has to be taken to the derivatives as well. To reduce the initial problem, we make use of the change of coordinates in
the tangent space
\begin{equation}
\begin{aligned}\label{yanose}
x_1&=\frac{\partial}{\partial x_1}+\sum_{l=2}^n\frac{\partial z_l}{\partial x_{1}}\frac{\partial}{\partial z_l}+\sum_{m=1}^k\frac{\partial U_m}{\partial x_{1}}\frac{\partial}{\partial U_m},
\\
x_{ i}&=\sum_{l=2}^n\frac{\partial z_l}{\partial x_{ i}}\frac{\partial}{\partial z_l}+\sum_{m=1}^k\frac{\partial U_m}{\partial x_{ i}}\frac{\partial}{\partial U_m},
\\
u_{ j}&=\sum_{l=2}^n\frac{\partial z_l}{\partial u_{ j}}\frac{\partial}{\partial z_l}+\sum_{m=1}^k\frac{\partial U_m}{\partial u_{ j}}\frac{\partial}{\partial U_m},
\end{aligned}
\end{equation}
and extended for every $2\leq{i}\leq n$ and $1\leq{j}\leq k$, where \eqref{redd1} gives us the coefficients for \eqref{yanose}.
For higher-order derivatives we only have to apply these relations recursively.

\begin{example}
\normalfont

Consider the simplest case of a nonlinear equation, that is the particular case of the {\bf Burgers equation} \cite{BTBZ,Hopf}
\begin{equation}\label{burger}
u_t+uu_x=\mu u_{xx}.
\end{equation}
This model contributes to the case of gas dynamics and traffic flow. It is equivalent to the Navier-Stokes equation for
impressible flow with the pressure term removed. The coefficient $\mu$ is known as the viscosity.
For a zero value of $\mu$ we have
\begin{equation}\label{burgermuzero}
u_t+uu_x=0,
\end{equation}

\noindent
which is known as the inviscid Burgers equation, which is the prototype equation developing solutions in form of discontinuities (shock waves) \cite{BTBZ}.
Equation \eqref{burgermuzero} is a first-order partial differential equation with one dependent variable and two independent variables, i.e. a submanifold in ${\rm J}^1(\mathbb{R}^2,\mathbb{R})$.
This equation, although seemingly trivial at a first glance, shows the phenomena of narrowing and distortion of waves.
Let us solve it by the method of the characteristics. We search for curves $x=x(s), t=t(s)$.
such that $u=u(x(s),t(s))$ is constant along such curves.
Then,
\begin{equation}
0=\frac{du}{ds}=u_x\frac{dx}{ds}+u_t\frac{dt}{ds}=u_x\frac{dx}{ds}-uu_x\frac{dt}{ds}=u_x\frac{d}{ds}(x-ut).
\end{equation}

\begin{figure}
  \begin{center}
  \end{center}
\end{figure}
This means that $x-ut=cte.$ Then, the general solution can be written explicitly as $u=f(x-ut)$, with $f$ an arbitrary function $f(x)=u(x,0).$
This implies that every point of the curve $f(x)$ travels with a velocity equal to its height. The higher points travel
at a faster speed than the lower ones.
As a consequence, the curve suffers distortion and gets narrower. 

We propose a general transformation according to the general theory exposed in this section
%
\begin{equation}
 \left\{\begin{aligned}\label{burgtrans}
&{\bar x}\rightarrow x+\epsilon \xi_{1}(x,t,u)+O(\epsilon^2),\\
&{\bar t}\rightarrow t+\epsilon \xi_{2}(x,t,u)+O(\epsilon^2),\\
&{\bar u}\rightarrow u+\epsilon \eta (x,t,u)+O(\epsilon^2)
\end{aligned}\right.
\end{equation}
By direct prolongation of the transformation to the first derivative, we have the prolongated transformation
\begin{equation}
 \left\{\begin{aligned}\label{burgtrans2}
&{\bar u}_x\rightarrow u_x+\epsilon \eta_x(x,t,u)+O(\epsilon^2),\\
&{\bar u}_t\rightarrow u_t+\epsilon \eta_t(x,t,u)+O(\epsilon^2).
\end{aligned}\right.
\end{equation}
Making use of the formula \eqref{prolongcomplete} for first order prolongations and considering the invariant surface condition
\begin{equation}
\eta-u_x\xi_1-u_t\xi_2=0,\end{equation} 
we arrive at
\begin{equation}
 \left\{\begin{aligned}\label{burgtrans2}
\eta_x&=u_x\frac{\partial \eta_{u}}{\partial u}+\frac{\partial \eta}{\partial x}-u_x\frac{\partial \xi_1}{\partial x}-u_t\frac{\partial \xi_2}{\partial x}-u_x^2\frac{\partial \xi_1}{\partial u}-\frac{\partial \xi_2}{\partial u}u_xu_t,\\
\eta_t&=u_t\frac{\partial \eta_{u}}{\partial u}+\frac{\partial \eta}{\partial t}-u_x\frac{\partial \xi_1}{\partial t}-u_t\frac{\partial \xi_2}{\partial t}-u_xu_t\frac{\partial \xi_1}{\partial u}-\frac{\partial \xi_2}{\partial u}u_t^2.
\end{aligned}\right.
\end{equation}

The introduction of the complete transformation \eqref{burgtrans} and \eqref{burgtrans2} into \eqref{burgermuzero} when $\mu=0$ provides us with 
a system in different orders in $\epsilon$.
The first-order in $\epsilon$ is system of differential equations
\begin{align}
\frac{\partial \eta}{\partial t}-\xi_{1t}u_x+\xi_{2t}uu_x+u\frac{\partial \eta_u}{\partial x}-u\xi_{1x}u_x+\xi_{2x}u^2u_x+\eta_{u}u_x=0.
\end{align}
where we have substituted the prolongations $\eta_x$ and $(\eta_{u})_t$ and the relation retrieved from the zero-order term in $\epsilon$ which
is $u_t=-uu_x.$
The remaining coefficients of derivatives, as for $u_x$ can be set equal to zero, because of the definition of Lie point symmetries.
In this way, we obtain the following result
\begin{align}
\xi_1&=a_1+a_2x+a_3t,\\
\xi_2&=b_1+b_2x+b_3t,\\
\eta_{u}&=a_3+(a_2-b_3)u-b_2u^2,
\end{align}
with $a_1,a_2,a_3,b_1,b_2,b_3\in \mathbb{R}$. These symmetries are classical Lie symmetries.

\end{example}

\section{Applications to PDEs in Physics}
\setcounter{equation}{0}
\setcounter{theorem}{0}
\setcounter{example}{0}

In this section we aim to calculate Lie symmetries of certain PDEs of different nature. In particular, we study evolution equations appearing
in the literature of hydrodynamic systems. Some of these equations belong to certain hierarchies of PDEs from which they are one of their
lower-order members. We will study the complete hierarchies from the point of view of Lie point symmetry and we will obtain their reductions.
Associated with such equations we will present their corresponding Lax pairs.
We find really interesting to calculate the Lie point  symmetries of a Lax pair. It is innovative in a way that we can see {\it how
the spectral parameter and the eigenfunction reduce under the symmetry.}
The interest resides in the fact that Lax pairs in higher-order dimensions, when reduced to a lower dimension by the symmetry, they can lose their nonisospectral character
giving rise to an isospectral problem, which could be tractable with the aid of the inverse scattering method, for example.
In this section we aim to review the {\bf Bogoyanlevskii--Kadomtsev--Petviashvili equation} in $1+1$ dimensions ($1+1$-BKP) and the {\bf Bogoyanlevskii--Kadomtsev--Petviashvili equation} in $2+1$ dimensions ($2+1$-BKP) and their corresponding Lax pairs by calculating
their symmetries and reductions. The first is an isospectral Lax pair, while the second is nonisospectral.

Generally, equations with nonisospectral Lax pairs in $2+1$ dimensions may provide more realistic models in the propagation of small amplitude surface waves
in straits or large channels with no dramatic varying depth and width and vorticity \cite{DLW,Xma,TZZ}.

From our experience, it is easier to deal with $2+1$ dimensional Lax pairs rather than with $1+1$ dimensional versions. Therefore, a good way of proceeding
in order to study $1+1$ versions is to start by $2+1$ nonisospectral LPs and obtain their nontrivial reductions to $1+1$ through classical and
non-classical methods.

The nonisospectral Lax pairs are interesting because the compatibility condition $(\psi_{xxt}=\psi_{txx})$ implies that the spectral
parameter $\lambda$ is a function, instead of a constant as in the isospectral cases, and shall be included as an scalar field in the Lie point  transformation.
If we wish to know the $1+1$ reductions of the spectral problem, it is specially important to establish how the spectral parameter reduces,
simultaneously, their compatibility reductions yield the reduction of the nonlinear problem \cite{Est1}.
This means that the spectral parameter $\lambda$ has to satisfy an equation, which is the {\it nonisospectral condition}, and that the symmetries are not only
symmetries of the LP and the nonlinear problem, but of the equation for the spectral parameter as well. We will illustrate these facts in the following subsections.

\subsection{The $2+1$ dimensional BKP equation and corresponding Lax pair}

We consider the $2+1$-BKP \cite{Est1,EH00}, studied for the first 
time here from the point of view of the classical and nonclassical Lie point symmetry symmetry analysis \cite{EstLejaSar2}.
The equation can be expressed as follows
\begin{equation}\label{BKP}
\left(u_{xt}+u_{xxxy}+8u_xu_{xy}+4u_{xx}u_y\right)_x+\sigma u_{yyy}=0,\quad \sigma=\pm 1,
\end{equation}
i.e. a submanifold of ${\rm J}^4(\mathbb{R}^3,\mathbb{R})$.
It has been proven that this equation is the $2+1$ dimensional reduction of a $3+1$ dimensional {\bf Kadomtsev--Petviashili equation} (KP) \cite{Dru,KP}, that it has the Painlev\'e property and admits a Lax pair representation.
It resembles the Calogero-Bogoyanlenvski-Schiff (CBS) equations by similarity in terms \cite{Bogo3,Calo,SE13}.
The solutions of \eqref{BKP} have solitonic behavior. In particular, this equation contains {\it lump solutions}, or solitons that decay polynomialy in all directions and have
non-trivial interactions \cite{Est1,EP08,MS96}.
Expression \eqref{BKP} can be presented in the non-local form as a submanifold of ${\rm J}^4(\mathbb{R}^3,\mathbb{R}^2)$ of the form
\begin{equation}\begin{aligned}\label{cbkp}
&\omega_{yy}=u_{xt}+u_{xxxy}+8u_xu_{xy}+4u_{xx}u_y,\\
&u_y=\omega_x,
\end{aligned}
\end{equation}
where we have chosen $\sigma=-1$ and have introduced the auxiliary scalar field $\omega(x,y,t)$. If $u(x,y,t)$ is a solution for $\sigma=-1$, then $u(x,iy,it)$ is a solution for $\sigma=1$.
This implies that the sign of $\sigma$ makes no real distinction in the problem.
The $2+1$-BKP equation in non-local form \eqref{cbkp}, has an associated spectral problem which is complex and has two component $(\psi,\chi)$ 
giving rise to a
submanifold of  ${\rm J}^2(\mathbb{C}^3,\mathbb{C}^4)$ \cite{EH00}
\begin{equation}\begin{aligned}\label{plst}
        &\psi_{xx}=-i\psi_y-2u_x\psi,\\
       &\psi_t=2i\psi_{yy}-4u_y\psi_x+(2u_{xy}+2i\omega_y)\psi,
       \end{aligned}
       \end{equation}
and its complex conjugate
\begin{equation}\begin{aligned}\label{plstcc}
        &\chi_{xx}=i\chi_y-2u_x\chi,\\
       &\chi_t=-2i\chi_{yy}-4u_y\chi_x+(2u_{xy}-2i\omega_y)\chi,
\end{aligned}
\end{equation}
with $\psi=\psi(x,y,t)$ and $\chi=\chi(x,y,t)$ over $\mathbb{C}^3$. 
The compatibility condition of the cross derivatives $(\psi_{xxt}=\psi_{txx})$ in \eqref{plst} retrives the equations in \eqref{cbkp}.
So does the compatibility condition $(\chi_{xxt}=\chi_{txx})$.
Since \eqref{plst} and \eqref{plstcc} are essentially the same, we shall perform calculations on \eqref{plst} and the results should
be equivalently extrapolated to the eigenfunction $\chi$ in \eqref{plstcc}.
We can denote by 
\begin{equation}\Psi= 
  \left[ {\begin{array}{c}
   \psi  \\
   \chi \\
  \end{array} } \right]
  \end{equation}
   the two component wave function.

We propose the Lie point transformation
\begin{equation}
 \left\{\begin{aligned}\label{bkplietransf}
x&\rightarrow x+\epsilon \xi_1(x,y,t,u,\omega)+O(\epsilon^2),\\
y&\rightarrow y+\epsilon \xi_2(x,y,t,u,\omega)+O(\epsilon^2),\\
t&\rightarrow t+\epsilon \xi_3(x,y,t,u,\omega)+O(\epsilon^2),\\
u&\rightarrow u+\epsilon \eta_u(x,y,t,u,\omega)+O(\epsilon^2),\\
\omega&\rightarrow \omega+\epsilon \eta_{\omega}(x,y,t,u,\omega)+O(\epsilon^2),\\
\psi&\rightarrow\psi+\epsilon\eta_{\psi}(x,y,t,u,\omega,\lambda)+O(\epsilon^2)
\end{aligned}\right.
\end{equation}
where we have dropped the bar notation $\bar{x}_i$ and $\bar{u}_j$ for simplicity.
Associated with this transformation, there exists an infinitesimal generator
\begin{equation}\label{infgenbkp}
X=\xi_1\frac{\partial}{\partial x}+\xi_2\frac{\partial}{\partial y}+\xi_3\frac{\partial}{\partial t}+\eta_u\frac{\partial}{\partial u}+\eta_{\omega}\frac{\partial}{\partial \omega}+\eta_{\psi}\frac{\partial}{\partial \psi}.
\end{equation}
where the subscripts in $\eta$ have been added according to the field to which each $\eta$ is associated.
This transformation must leave \eqref{cbkp} invariant.
We compute the first- and second-order derivatives appearing in the equations
\begin{equation}
 \left\{\begin{aligned}
u_x&\rightarrow u_x+\epsilon(\eta_u)_{x}+O(\epsilon^2),&\omega_x&\rightarrow \omega_x+\epsilon(\eta_{\omega})_{x}+O(\epsilon^2),
\\
\omega_y&\rightarrow \omega_y+\epsilon(\eta_{\omega})_{y}+O(\epsilon^2),& 
u_{xx}&\rightarrow  u_{xx}+\epsilon(\eta_u)_{xx}+O(\epsilon^2),\\ u_{xy}&\rightarrow u_{xy}+\epsilon(\eta_u)_{xy}+O(\epsilon^2),& u_{xt}&\rightarrow u_{xt}+\epsilon(\eta_u)_{xt}+O(\epsilon^2),\\
u_{xxxy}&\rightarrow u_{xxxy}+\epsilon(\eta_u)_{xxxy}+O(\epsilon^2).&&
\end{aligned}\right.
\end{equation}
The Lax pair must be simultaneously invariant \eqref{plst} and \eqref{plstcc}.
\begin{equation}
 \left\{\begin{aligned}
\psi_{xx}&\rightarrow\psi_{xx}+\epsilon (\eta_{\psi})_{xx}+O(\epsilon^2),& \psi_{yy}&\rightarrow \psi_{yy}+\epsilon (\eta_{\psi})_{yy}+O(\epsilon^2),\\
\psi_{x}&\rightarrow\psi_{x}+\epsilon (\eta_{\psi})_{x}+O(\epsilon^2), & \psi_{y}&\rightarrow\psi_{y}+\epsilon (\eta_{\psi})_{y}+O(\epsilon^2),\\ \psi_{t}&\rightarrow\psi_{t}+\epsilon (\eta_{\psi})_{t}+O(\epsilon^2).&&
\end{aligned}\right.
\end{equation}
The prolongations $(\eta_u)_{xxxy}$, $(\eta_u)_{xt}$, $(\eta_u)_{xy}$, $(\eta_u)_{xx}$, $(\eta_u)_{x}$, $(\eta_u)_{y}$, $(\eta_\omega)_{x}$, $(\eta_\omega)_{y}$, $(\eta_{\psi})_{yy}$, $(\eta_{\psi})_{xx}$,
$(\eta_{\psi})_{t}$, $(\eta_{\psi})_{y}$, $(\eta_{\psi})_{x}$ can be calculated according to \eqref{prolongcomplete} given in Chapter \ref{Chap:GeomFund}.
In the case of nonclassical Lie symmetries, we have to include the invariant surface conditions
by extracting $u_t$, $\omega_t$ and $\psi_t$ from
\begin{align}
&\xi_1u_x+\xi_2u_y+\xi_3u_t=\eta_u,\nonumber\\
&\xi_1\omega_x+\xi_2\omega_y+\xi_3\omega_t=\eta_{\omega},\\
&\xi_1\psi_x+\xi_2\psi_y+\xi_3\psi_t=\eta_{\psi}\nonumber
\end{align}
and from the original, untransformed equations, we make use of
\begin{equation}
\begin{aligned}\label{cbkp2}
&\omega_{yy}=u_{xt}+u_{xxxy}+8u_xu_{xy}+4u_{xx}u_y,\\
&u_y=\omega_x,
\end{aligned}
\end{equation}
and the untransformed equations of the Lax pair
\begin{equation}
\begin{aligned}
        &\psi_{xx}=-i\psi_y-2u_x\psi,\\
       &\psi_t=2i\psi_{yy}-4u_y\psi_x+(2u_{xy}+2i\omega_y)\psi.
       \end{aligned}
\end{equation}
Introducing such relations we arrive at the nonclassical symmetries
\begin{align}\label{csymbkp}
&\xi_1=\frac{\dot A_3(t)}{4}x+A_1(t),\\
&\xi_2=\frac{\dot A_3(t)}{2}y+A_2(t),\\
&\xi_3=A_3(t),\\
&\eta_u=-\frac{\dot A_3(t)}{4}u+\frac{\dot A_2(t)}{8}x+\frac{\ddot{A_3}(t)}{16}xy+\frac{\dot{A_1}(t)y}{4}+B_1(t),\\
&\eta_{\omega}=-\frac{\dot A_3(t)}{2}w+\frac{\dot A_1(t)}{4}x+\frac{\ddot{A_3}(t)}{32}x^2+B_3(t)y+\frac{\ddot{A_2}(t)}{16}y^2
\\
&\hspace{2em}+\frac{\dddot{A}_3(t)}{96} y^3+B_2(t),\\
&\eta_{\psi}=\left[-2\lambda-\frac{\dot A_3(t)}{8}+i\left(\frac{\dot A_2(t)}{4}y+\frac{\ddot{A}_3(t)}{16}y^2+2\int{B_3(t)dt}\right)\right]\psi.
\end{align}

These symmetries depend on six arbitrary functions of time, $A_1(t),A_2(t),A_3(t)$ and $B_1(t),$ $B_2(t),$ $B_3(t)$, 
which shall serve us as a way to classify the possible reductions, and a constant $\lambda$.
The classical symmetries are a particular case of the nonclassical and lead to less general reductions. For this matter, we will only specify the
nonclassical ones.

To reduce the problem, we have to solve the Lagrange-Charpit system
\begin{equation}\label{LCBKP21}
\frac{dx}{\xi_1}=\frac{dy}{\xi_2}=\frac{dt}{\xi_3}=\frac{du}{\eta_u}=\frac{d\omega}{\eta_{\omega}}=\frac{d\psi}{\eta_{\psi}}.
\end{equation}
\noindent
The most relevant possible reductions agree with the cases 

\begin{table}[H]
\begin{center}
\begin{tabular}{lllclll}\toprule
\multicolumn{3}{c}{$\text{Case I}: A_3(t)\neq 0$}&&\multicolumn{3}{c}{$\text{Case II}: A_3(t)=0$}
\\ 
\midrule
1.&$A_1(t)\neq 0$& $A_2(t)\neq 0$&&1.& $A_1(t)\neq 0$&$A_2(t)\neq 0$\\
2.& $A_1(t)\neq 0$& $A_2(t)=0$&&2.& $A_1(t)\neq 0$&$A_2(t)=0$\\
3.& $A_1(t)=0$ & $A_2(t)\neq 0$&&3.& $A_1(t)=0$ & $A_2(t)\neq 0$\\
\bottomrule
\end{tabular}
 \end{center}
\caption{Reductions for $2+1$-BKP}
\label{Tab1}
\end{table}

We shall use the next notation for the reduced variables
\begin{equation}
 x,y,t\rightarrow x_1,x_2
\end{equation}
and for the reduced fields
\begin{align}
& \omega(x,y,t)\rightarrow \Omega(x_1,x_2),\nonumber\\
& u(x,y,t)\rightarrow U(x_1,x_2),\\
& \psi(x,y,t)\rightarrow \Phi(x_1,x_2).\nonumber
\end{align}
As a matter of simplification, we shall drop the dependency $x_1,x_2$ of all the fields in the forthcoming expressions.

\begin{itemize}
\item {\it Case I.1. $A_3(t)\neq 0, A_1\neq 0, A_2\neq 0$}
\begin{itemize}
\item Reduced variables 
\begin{equation}
x_1=\frac{x}{A_3(t)^{1/4}}-\int{\frac{A_1(t)}{A_3(t)^{5/4}}dt},\quad x_2=\frac{y}{A_3(t)^{1/2}}-\int{\frac{A_2(t)}{A_3(t)^{3/2}}dt}.
\end{equation}
\item Reduced fields
\begin{align}
u(x,y,t)&=\frac{U(x_1,x_2)}{A_3(t)^{1/4}}+\frac{x_1x_2}{16}\frac{\dot A_3(t)}{A_3(t)^{1/4}}
+\frac{x_1}{16}\left(2\frac{A_2(t)}{A_3(t)^{3/4}}+B(t)\frac{\dot A_3(t)}{A_3(t)^{1/4}}\right)
\nonumber\\
&+\frac{x_2}{16}\left(4\frac{A_1(t)}{A_3(t)^{1/2}}+A(t)\frac{\dot A_3(t)}{A_3(t)^{1/4}}\right),
\end{align}

\begin{align}
\omega(x,y,t)&=\frac{\Omega(x_1,x_2)}{A_3(t)^{1/2}}+\frac{x_2^3}{192}\left(2A_3(t)^{1/2}\ddot{A_3}(t)-\frac{\dot{A}_3(t)^2}{A_3(t)^{1/2}}\right)+\frac{x_1^2}{32}\frac{\dot{A}_3(t)}{A_3(t)^{1/2}}
\nonumber\\
&+\frac{x_1}{16}\left(4\frac{A_1(t)}{A_3(t)^{3/4}}+A(t)\frac{\dot{A}_3(t)}{A_3(t)^{1/2}}\right)
\nonumber\\
&+\frac{x_2^2}{64}\left[4\dot{A_2}(t)-2\frac{A_2(t)\dot{A}_3(t)}{A_3(t)}+\frac{B(t)}{A_3(t)^{1/2}}\left(2A_3(t)\ddot{A_3}(t)-\dot{A}_3(t)^2\right)\right]
\nonumber\\
&+ x_2\left[\frac{\int{B_3(t)dt}}{A_3(t)^{1/2}}-\frac{1}{16}\frac{A_2(t)^2}{A_3(t)^{3/2}}+\frac{B(t)}{16}\left(2\dot{A_2}(t)-\frac{A_2(t)\dot{A}_3(t)}{A_3(t)}\right)
\right.\nonumber
\\
&\left.
+\frac{B(t)^2}{64}\left(2A_3(t)^{1/2}\ddot{A_3}(t)-\frac{\dot{A_3}(t)^2}{A_3(t)^{1/2}}\right)\right],
\end{align}
where we have used the definitions
\begin{equation}
A(t)=\int{\frac{A_1(t)}{A_3(t)^{5/4}}dt},\quad B(t)=\int{\frac{A_2(t)}{A_3(t)^{3/2}}dt}.
\end{equation}
\item Reduced equation
\begin{equation}
\begin{aligned}
       &U_{x_2}=\Omega_{x_1},\\
       &\Omega_{x_2x_2}=U_{x_2x_1x_1x_1}+8U_{x_2x_1}U_{x_1}+4U_{x_1x_1}U_{x_2}.
\end{aligned}\end{equation} 
These two equations can be summarized in
\begin{equation}\label{rode}
U_{x_2x_1x_1x_1x_1}+4\Big((U_{x_2}U_{x_1})_{x_1}+U_{x_1}U_{x_1x_2}\Big)_{x_1}-U_{x_2x_2x_2}=0,
\end{equation}
which appears in \cite{Wazwaz} and has {\it multiple soliton solutions}.

\item Reduced eigenfunction
\begin{align}
\psi(x,y,t)&=\frac{\Phi(x_1,x_2)}{A_3(t)^{1/8}}\exp\left[i\frac{x_2^2}{16}\dot{A}_3(t)+i\frac{x_2}{8}\left(2\frac{A_2(t)}{A_3(t)^{1/2}}+B(t)\dot{A}_3(t)\right)\right.\nonumber\\
&\hspace*{-2em}-2\lambda\int{\frac{dt}{A_3(t)}}\nonumber\\
&\hspace*{-2em}\left.+i\int{\left(2\frac{\int{B_3(t)dt}}{A_3(t)}+\frac{B(t)}{4}\frac{\dot{A}_2(t)}{A_3(t)^{1/2}}+\frac{B(t)^2}{16}\ddot{A}_3(t)\right)dt}\right].
\end{align}
\item Reduced Lax pair
\begin{equation}
\begin{aligned}
&\Phi_{x_1x_1}+i\Phi_{x_2}+2U_{x_1}\Phi=0,\\
 &i\Phi_{x_2x_2}-2U_{x_2}\Phi_{x_1}+\left(U_{x_1x_2}+i\Omega_{x_2}+\lambda\right)\Phi=0.
 \end{aligned}\end{equation} 

\end{itemize}

\item {\it The cases I.2. and I.3. will be omitted for being equivalent to I.1.}

\item{\it Case II.1. $A_3=0$, $A_1\neq 0$, $A_2\neq 0$}

\begin{itemize}

\item Reduced variables
\begin{equation}
x_1=A_2(t) x-A_1(t)  y, \quad x_2= \int A_1(t) A_2(t)^2 dt.
\end{equation}

\item Reduced fields

\begin{align}
u(x,y,t)&=A_2(t) U(x_1,x_2)-{A_1(t)^{3}\over A_2(t)^{3}} {y\over 4} +\left({\dot A_1(t)\over A_2(t)}-2 A_1(t) {\dot A_2(t)\over A_2(t)^2}\right) {y^2\over 8}
\nonumber\\&
- {\dot A_2(t)\over A_2(t)}  {x y\over4},\\
\omega(x,y,t)&=-A_1(t) \Omega(x_1,x_2) +\left({\dot A_1(t)\over A_2(t)}-2 A_1(t) {\dot A_2(t)\over A_2(t)^2}\right)  {x y\over4}- {\dot A_2(t)\over A_2(t)}  {x^2\over 8} \nonumber\\
&- {A_1(t)^{3}\over A_2(t)^{3}} {x\over 4} -\left(4\lambda^2A_2(t)^4+3 i \lambda A_1(t)^2+ {3 A_1(t)^4\over 16A_2(t)^4} -i{\dot A_2(t)\over 4A_2(t)}\right) y 
\nonumber\\&
+ \left(3{\dot A_2(t)^2\over A_2(t)^2}-{\ddot A_2(t)\over A_2(t)}\right)  {y^3\over 24}.
\end{align}

\item Reduced equation
\begin{equation}
\begin{aligned}
&\Omega_{x_1}=U_{x_1},\\
&U_{x_1x_1x_1x_1}=U_{x_1x_2}-12U_{x_1}U_{x_1x_1},
\end{aligned}\end{equation} 
which can be equivalently rewritten as
\begin{equation}
 U_{x_1x_1x_1x_1x_1}+12 U_{x_1x_1}^2+ 12 U_{x_1}U_{x_1x_1x_1}-U_{x_1x_1x_2 }=0.
\end{equation}
This reduced equation corresponds with the Korteweg de Vries equation (KdV) in $1+1$ dimensions \cite{Drazin}. 
{\it Therefore, we can conclude
that the equation BKP is a generalization of KdV to $2+1$ dimensions.}

\item Reduced Eigenfunction
\begin{multline}
 \psi(x,y,t)=\Phi(x_1,x_2)e^{-2 \lambda A_2(t)^2 y}
\times
\\
\times
\exp\left[{i \left({A_1(t)\over A_2(t)}{x\over 2}-{A_1(t)^2\over A_2(t)^2}{y\over 4}-{\dot A_2(t)\over A_2(t)}{y^2\over 4}\right)}\right].
\end{multline}

\item Reduced Lax pair
\begin{equation}
\begin{aligned}
&\Phi_{x_1x_1}=2i\lambda\Phi-2U_{x_1}\Phi,\\
&\Phi_{x_2}=8i\lambda \Phi_{x_1}+4 U_{x_1}\Phi_{x_1}-2\Phi U_{x_1x_1},
\end{aligned}\end{equation} 
which is the Lax pair corresponding with the KdV equation.

\end{itemize}

\newpage

\item {\it Case II.2. $A_3(t)=0$, $A_1(t)\neq 0$, $A_2(t)=0$}
\begin{itemize}
\item Reduced variables
\begin{equation}
x_1=y,\quad x_2=t.
\end{equation}
\item Reduced fields
\begin{align}
u(x,y,t)&=U(x_1,x_2)+\frac{B_2(t)}{A_1(t)}x_1+\frac{B_3(t)}{2A_1(t)}x_1^2+\frac{B_1(t)}{A_1(t)}x
\nonumber\\
&+\frac{A_1'(t)}{4A_1(t)}xy,
\\
\omega(x,y,t)&=\frac{1}{2i}\Omega(x_1,x_2)+\left(\frac{B_1(t)A_1'(t)}{2A_1(t)^2}+\frac{B_1'(t)}{2A_1(t)}\right)x_1^2
\nonumber\\
&+\left(\frac{A_1'(t)^2+A_1(t)A_1''(t)}{24A_1^2}\right)x_1^3+\frac{B_2(t)}{A_1(t)}x+\frac{A_1'(t)}{8A_1(t)}x^2
\nonumber\\+\frac{B_3(t)}{A_1(t)}xy.
\end{align}

\item Reduced equation
\begin{equation}
\begin{aligned}
       &U_{x_1}=0,\\
       &U_{x_1x_1x_1}=0,\\
       &\Omega_{x_1x_1}=0.
\end{aligned}\end{equation} 

\item  Reduced Eigenfunction
\begin{align}
\psi(x,y,t)&=\frac{\Phi(x_1,x_2)}{\sqrt{A_1(t)}}\exp\left[\int{8\frac{B_2(t)}{A_1(t)^2}}\left(\lambda-i\int{B_2(t)dt}\right)
\right.\nonumber\\
&
-i\frac{8}{A_1(t)^2}\left(B_1(t)+2\frac{(\lambda-i\int{B_3(t)dt})^2}{A_1(t)}\right)^2
\nonumber\\
&+\frac{2i}{A_1(t)}\left(B_1(t)+\frac{2}{A_1(t)}\left(\lambda-i\int{B_3(t)dt}\right)^2\right)x_1
\nonumber\\  
&\left.{}
+i\frac{A_1'(t)}{4A_1(t)}x_1^2
-\frac{2}{A_1(t)}\left(\lambda-i\int{B_3(t)dt}\right)x\right].
\end{align}

\item Reduced Lax pair
\begin{equation}
\begin{aligned}
       &\Phi_{x_1}=0,\\
        &\Phi_{x_2}=\Omega_{x_1}\Phi.
\end{aligned}\end{equation} 
\end{itemize}

\newpage

\item{\it Case II.3. $A_3(t)=0$, $A_1(t)=0$, $A_2(t)\neq 0$}

\begin{itemize}
\item Reduced variables
\begin{equation}
x_1=\frac{x}{\sqrt{A_2(t)}}-4\int{\frac{B_1(t)}{A_2(t)^{3/2}}dt},\quad x_2=t.
\end{equation}
\item Reduced fields
\begin{multline}
u(x,y,t)=\frac{1}{2\sqrt{A_2(t)}}\left(U(x_1,x_2)+2x_1\int{B_3(t)dt}\right)
\\
+\frac{B_1(t)}{A_2(t)}y+\frac{A_2'(t)}{8A_2(t)}xy,
\end{multline}
\begin{multline}
\omega(x,y,t)=\Omega(x_1,x_2)+\left(\frac{B_1}{\sqrt{A_2}}+\frac{1}{2}A_2'(t)\int{\frac{B_1(t)}{A_2(t)^{3/2}}dt}\right)x_1\\+\frac{A_2'(t)}{16}x_1^2+\frac{B_2(t)}{A_2(t)}y+\frac{B_3(t)}{2A_2(t)}y^2+\frac{A_2''(t)}{48A_2(t)}y^3.
\end{multline}

\item Reduced equation
\begin{equation}
\begin{aligned}
       &\Omega_{x_1}=0,\\
       & U_{x_1x_2}=0,\\
       &U_{x_1x_1x_2}=0.
\end{aligned}\end{equation}

\item Reduced eigenfunction
\begin{align}
\psi(x,y,t)&=\frac{\Phi(x_1,x_2)}{A_2(t)^{1/4}} \times \nonumber \\
&\times e^{2i\int\Big({\frac{B_2(t)}{A_2(t)}+\frac{4(\lambda-i\int{B_3(t)dt})^2}{A_2(t)^2}\Big)dt}}e^{-\frac{2 (\lambda-i\int{B_3(t)dt})}{A_2(t)}y+i\frac{A_2(t)'}{8A_2(t)}y^2}.
\end{align}

\item Reduced Lax pair
\begin{equation}
\begin{aligned}
     &\Phi_{x_2}=0,\\
     &\Phi_{x_1x_1}=(2i\lambda-U_{x_1})\Phi.
  \end{aligned}\end{equation} 
\end{itemize}
\end{itemize}

\subsection{A reduction of the $2+1$-BKP equation and corresponding Lax pair to $1+1$ dimensions}

Let us now study the nontrivial reduction I.1. obtained in the past Subsection 5.1. of the initial
 $2+1$-BKP equation to $1+1$ dimensions \cite{EstLejaSar2}.
We consider this reduction of interest from a possible physical interpretation viewpoint. Reduction II.1. is another nontrivial
reduction corresponding to the very celebrated Korteweg de Vries equation which has been studied in multiple occasions and already accounts with a 
wide physical literature dedicated to it \cite{Bona,Ismail,Segal}.
This is why in this subsection we shall focus on reduction I.1. Let us consider the differential equation in ${\rm J}^4(\mathbb{R}^2,\mathbb{R}^2)$.
\begin{equation}
\begin{aligned}\label{cbkpr}
&u_{x_2x_2}=\Omega_{x_1x_2},\\
&u_{x_1x_1x_1x_2}+8u_{x_1x_2}u_{x_1}+4u_{x_1x_1}u_{x_2}-\Omega_{x_2x_2}=0.
\end{aligned}
\end{equation}
This equation is also integrable in the Painlev\'e sense and possesses an associated linear spectral problem or Lax pair in ${\rm J}^2(\mathbb{C}^3,\mathbb{C}^3)$
\begin{equation}
\begin{aligned}\label{plnontrred}
&\Phi_{x_1x_1}+i\Phi_{x_2}+2u_{x_1}\Phi=0,\\
&i\Phi_{x_2x_2}-2u_{x_2}\Phi_{x_1}+\left(u_{x_1x_2}+i\Omega_{x_2}+\lambda\right)\Phi=0.
\end{aligned}
\end{equation}
whose compatibility condition recovers \eqref{cbkpr}.
We propose a general transformation for the independent and dependent variables as
\begin{equation}
 \left\{\begin{aligned}\label{rbkpLietransf}
x_1&\rightarrow x_1+\epsilon \xi_1(x_1,x_2,u,\Omega)+O(\epsilon^2),\\
x_2&\rightarrow x_2+\epsilon \xi_2(x_1,x_2,u,\Omega)+O(\epsilon^2), \\
\lambda&\rightarrow \lambda+\epsilon \eta_{\lambda}(x_1,x_2,\lambda,\Phi)+O(\epsilon^2),\\
u&\rightarrow u+\epsilon \eta_{u}(x_1,x_2,u,\Omega)+O(\epsilon^2),\\
\Omega&\rightarrow \Omega+\epsilon \eta_{\Omega}(x_1,x_2,u,\Omega)+O(\epsilon^2),\\
\Phi&\rightarrow \Phi+\epsilon \eta_{\Phi}(x_1,x_2,u,\Omega,\lambda,\Phi)+O(\epsilon^2).
\end{aligned}\right.
\end{equation}

Here we can see that we have considered $\lambda$ as an independent variable in order to make the reductions properly.

By definition of symmetry of the equation and its associated Lax pair, this $\epsilon$-parametric group of transformations must leave the equations in \eqref{cbkpr} and \eqref{plnontrred} invariant.
Equivalently, the associated symmetry vector field has the expression
\begin{equation}\label{vfBKP11}
X=\xi_1\frac{\partial}{\partial x_1}+\xi_2\frac{\partial}{\partial x_2}+\eta_{\lambda}\frac{\partial}{\partial \lambda}+\eta_{u}\frac{\partial}{\partial u}+\eta_{\Omega}\frac{\partial}{\partial \Omega}+\eta_{\Phi}\frac{\partial}{\partial \Phi}.
\end{equation}

From the terms in $\epsilon=0$ we retrieve the original, untransformed equations. We retrive the conditions
\begin{equation}
\begin{aligned}\label{cbkpr2}
&u_{x_2x_2}=\Omega_{x_1x_2},\\
&u_{x_1x_1x_1x_2}=\Omega_{x_2x_2}-8u_{x_1x_2}u_{x_1}-4u_{x_1x_1}u_{x_2},\\
&\Phi_{x_1x_1}=-i\Phi_{x_2}-2u_{x_1}\Phi,\\
&i\Phi_{x_2x_2}=2u_{x_2}\Phi_{x_1}-\left(u_{x_1x_2}-i\Omega_{x_2}-\lambda\right)\Phi
\end{aligned}
\end{equation}
\noindent
that shall be used in the forthcoming steps.
First introduce \eqref{rbkpLietransf} into the system of differential equations in \eqref{cbkpr2} and
set the linear term in $\epsilon$ equal to zero. 
Introduce the prolongations $(\eta_{u})_{x_1x_1x_1x_2}$, $(\eta_{u})_{x_1x_2}$, $(\eta_{u})_{x_1x_1}$, $(\eta_{u})_{x_2x_2}$, $(\eta_{u})_{x_1}$, $(\eta_{u})_{x_2}$, $(\eta_{\Omega})_{x_2x_2}$,
$(\eta_{\Omega})_{x_1x_2}$, $(\eta_{\Omega})_{x_2}$, $(\eta_{\Phi})_{x_1x_1}$, $(\eta_{\Phi})_{x_2x_2}$, $(\eta_{\Phi})_{x_1}$, $(\eta_{\Phi})_{x_2}$, calculated following formula \eqref{prolongcomplete} in Chapter \ref{Chap:GeomFund} and $u_{x_2x_2}$, $u_{x_1x_1x_1x_2}$, $\Phi_{x_1x_1}$, $\Phi_{x_2x_2}$ from \eqref{cbkpr2}.
The difference between classical and nonclassical symmetries in this case is the implementation of the invariant surface conditions. Depending 
on the case we choose, we have the classification 

\begin{table}[H]\centering
\begin{tabular}{ccc}\toprule
\multicolumn{3}{c}{Values of infinitesimal generators}\\
\midrule
Case I.&$\xi_1\neq 0$ & $\xi_2\neq 0$\\
Case II & $\xi_1\neq 0$ &  $\xi_2=0$ \\
Case III. &$\xi_1=0$ & $\xi_2\neq 0$\\
\bottomrule
\end{tabular}
\caption{Iterative reduction of the $1+1$-BKP}
\label{Tab2}
\end{table}

Nontrivial results are led by $\xi_1\neq 0, \xi_2\neq 0$, which is case I., we shall consider.
The invariant surface conditions provide us with
\begin{equation}
\begin{aligned}
&u_{x_2}=\frac{\eta_{u}-\xi_1u_{x_1}}{\xi_2},\\
&\Omega_{x_2}=\frac{\eta_{\Omega}-\xi_1\Omega_{x_1}}{\xi_2},\\
&\Phi_{x_2}=\frac{\eta_{\Phi}-\xi_1\Phi_{x_1}-\eta_{\lambda}\Phi_{\lambda}}{\xi_2},
\end{aligned}\end{equation} 
that we substitute and their multiple higher-order 
derivatives.
We now pick the remaining coefficients in first or higher-order derivatives of the fields $u_{x_1x_1},\Omega_{x_2x_2},\dots$ and set them equal to zero. 
It is an overdetermined system of equations whose solutions are the symmetries.
If we divide by $\eta_{\lambda}$ all the expressions, we resulting reductions will be invariant.
This results in
\begin{align}\label{ncsymbkpr}
&\xi_1(x_1,x_2,u,\Omega)=\frac{1}{\eta_{\lambda}}\left(\frac{1}{2}K_1x_1+K_2\right),\\
&\xi_2(x_1,x_2,u,\Omega)=\frac{1}{\eta_{\lambda}}\left(K_1x_2+K_3\right),\\
&\eta_{\lambda}=1,\\
&\eta_{u}(x_1,x_2,u,\Omega)=\frac{1}{\eta_{\lambda}}\left[-\frac{1}{2}K_1u+K_5+C_0(K_3x_1+K_1x_1x_2+2K_2x_2)\right],\\
&\eta_{\Omega}(x_1,x_2,u,\Omega)=\frac{1}{\eta_{\lambda}}\left[-K_1\Omega+K_4x_2+K_6(x_1)+C_0^2\left(K_3x_2^2+\frac{1}{3}K_1x_2^3\right)\right],\\
&\eta_{\Phi}(x_1,x_2,\lambda,U,W,\Phi,\lambda)=\frac{B(x_1,x_2,\lambda)}{\eta_{\lambda}}\Phi.
\end{align}
These symmetries depend on 6 arbitrary constants of integration $K_1,K_2,K_3,K_4,K_5,C_0$ in $\mathbb{R}$ and two arbitrary functions $K_6(x_1)$
and $B(x_1,x_2)$.

In the classical case, the symmetries read
\begin{align}\label{csymbkpr}
&\xi_1(x_1,x_2,u,\Omega)= \frac{1}{2}k_1x_1+k_2,\\
&\xi_2(x_1,x_2,u,\Omega)=k_1x_2+k_3\,\\
&\eta_{\lambda}(x_1,x_2,u,\Omega,\Phi,\lambda)=-2k_1\lambda-ik_4,\\
&\eta_{u}(x_1,x_2,u,\Omega)=-\frac{1}{2}k_1u+k_5,\\
&\eta_{\Omega}(x_1,x_2,u,\Omega)=-k_1\Omega+k_4x_2,\\
&\eta_{\Phi}(x_1,x_2,U,\Omega,\Phi,\lambda)=G(\lambda)\Phi.
\end{align}
These symmetries depend on 5 arbitrary constants of integration $k_1,k_2,k_3,k_4,k_5\in \mathbb{R}$.
It is obvious that the classical case is included within the nonclassical one, as expected.
{\it From here we see that there exist differences between the classical and nonclassical symmetries for this example.
It is interesting, since for many other examples, the classical and nonclassical symmetries have coincided.}
Nevertheless, the difference between the classical and nonclassical case is not substantial in the case of reductions. Here we show a list of reductions
for the nonclassical case which do not differ from the reduced versions obtained in the classical approach.


For the nonclassical case in which $\eta_{\lambda}=1$, we have the subclassification 

\begin{table}[H]\centering
\begin{tabular}{lllclll}\toprule
\multicolumn{3}{c}{Case  I: $k_1\neq 0$}&&\multicolumn{3}{c}{Case  II: $k_1=0$}
\\ 
\midrule
1.&$A_1(t)\neq 0$& $A_2(t)\neq 0$&&1.& $A_1(t)\neq 0$&$A_2(t)\neq 0$\\
& & &&2.& $A_1(t)\neq 0$&$A_2(t)=0$\\
\bottomrule
\end{tabular}
\caption{Subclassification of reductions for $1+1$-BKP}
\label{Tab3}
\end{table}
\noindent
We introduce the following notation for the reduced variables
\begin{equation}
x_1,x_2,\lambda \rightarrow z,\Lambda
\end{equation}
and the reduced fields and eigenfunctions
\begin{equation}
\begin{aligned}
u(x_1,x_2)&\rightarrow V(z,\Lambda),\\
\Omega(x_1,x_2)&\rightarrow W(z,\Lambda),\\
\Phi(x_1,x_2)&\rightarrow \varphi(z,\Lambda).
\end{aligned}\end{equation} 

\newpage
We find the following reductions

\begin{itemize}

\item{\it Case I.1. $k_1\neq 0$, $k_2\neq 0$, $k_3\neq 0$}

\begin{itemize}

\item Reduced variables
\begin{align}
&z(x_1,x_2,\lambda)={k_1 (k_1x_2+k_3) \over (k_1x_1+2 k_2)^2 },\\
&\Lambda(x_1,x_2,\lambda)=k_1^{-5} (2 k_1\lambda +i k_4) (k_1x_1+2 k_2)^4.
\end{align}

\item Reduced fields
\begin{equation}
U(x_1, x_2,\lambda)=k_1{ V(z,
\Lambda) \over (k_1x_1+2 k_2)},
\end{equation}

\begin{align}
\Omega(x_1, x_2,\lambda)&={k_4\over 2 k_1^3}  (k_1x_1+2 k_2)^2 z(x_1, x_2,\lambda)\nonumber\\
&-{k_1^2\over 2 (k_1x_1+2 k_2)^2} {W\big(z(x_1, x_2,\lambda),\Lambda(x_1, x_2,\lambda)\big)\over z(x_1, x_2,\lambda)}.
\end{align}

\item Reduced Eigenfunction
\begin{equation}
\Phi(x_1, x_2,\lambda)=\varphi (z,\Lambda) e^{i \int {G(\lambda)\over k_4 - 2 i k_1\lambda}d\lambda }.
\end{equation}

\item Reduced Lax pair
\begin{align}
2 z^2 \varphi_{zz} +&8 i z^2 V_z ( 2 \Lambda  \varphi_{\Lambda }- z \varphi_z)\nonumber
\\
&+\Big [ W-z (1-6iz) V_z+iz^2(4zV_{zz}-\Lambda) \Big] \varphi=0,\nonumber\\[-1.6ex]
\\[-1.6ex]
16 \Lambda^2  \varphi_{\Lambda\Lambda}&-16 z \Lambda   \varphi_{z\Lambda}+4 \Lambda(3-8 i z^2 V_z)\varphi_{\Lambda }+(i+6 z +16 i z^3 V_z) \varphi_z\nonumber\\
&\hspace{-0.5em}-2 \Big[W+ V+ z (1+6iz) V_z+iz^2(4zV_{zz}- \Lambda) \Big]\varphi=0.
\nonumber
\end{align}

\item Reduced equations
\begin{equation}
\hspace*{-3em}\begin{gathered}
W_z=U_z, \\
\begin{aligned}
16z^6V_{zzzz}+144&z^5V_{zzz}-(1-300z^2+32z^2V+96z^3V_z)z^2V_{zz}\\
&-176z^4V_z^2+2(1+60 z^2-32 z^2 V)zV_z-2 W=0.
\end{aligned}
\end{gathered}\end{equation} 

\end{itemize}

\item{\it Case II.1. $k_1=0$, $k_2\neq 0$, $k_3=0$}

\begin{itemize}

\item Reduced variables
\begin{align}
&z=-\frac{k_2}{k_3} x_1+\frac{k_2^2}{k_3^2}x_2,\\
&\Lambda=\frac{k_3^4}{k_2^4}\lambda+i\frac{k_2}{k_3}x_1.
\end{align}
\item Reduced fields
\begin{align}
&U(x_1,x_2)=-\frac{k_2}{k_3}V(z,\Lambda)+\frac{k_2^2}{8k_3^2}x_1,\\
&\Omega(x_1,x_2)=\frac{k_2^2}{k_3^2}\left(W(\Lambda)+\frac{1}{2}z^2+\frac{k_2}{k_3}zx_1+\frac{1}{2}\frac{k_2^2}{k_3^2}x_1^2\right).
\end{align}

\item Reduced Eigenfunction
\begin{equation}
\Phi(x_1,x_2,\lambda)=\varphi(z,\Lambda).
\end{equation}

\item Reduced Lax pair
\begin{equation}
\begin{aligned}
&\varphi_{\Lambda \Lambda}-\frac{1}{4}\varphi-2V_z\varphi+2i\varphi_{z \Lambda}-\varphi_{zz}-i\varphi_z=0,\\
&(z-i\Lambda)\varphi+\left(\varphi+2\varphi_{\Lambda}+2i\varphi_z\right)V_z-iV_{zz}\varphi+\varphi_{zz}=0.
\end{aligned}\end{equation}

\item Reduced equation
\begin{equation}
\begin{aligned}
&W_z=V_z,\\
&V_{zzzz}+12V_zV_{zz}-1=0.
\end{aligned}\end{equation} 

\end{itemize}

\item{\it Case II.2. $k_1=0$, $k_2=0$, $k_3\neq 0$}
\begin{itemize}
\item Reduced variables
\begin{align}
&z=z_2\left(-\frac{k_4}{k_2}\right)^{2/5},\\
&\Lambda=\frac{-k_2\lambda-ik_4z_1}{\left(-k_2k_4\right)^{1/5}}+z_2\left(-\frac{k_4}{k_2}\right)^{2/5}.
\end{align}
\item Reduced fields
\begin{align}
&u(z_1,z_2)=i\left(\frac{k_4}{-k_2}\right)^{1/5}\left(U(z)+\frac{i}{2}z^2\right),\\
&\omega(z_1,z_2)=\left(\frac{k_4}{-k_2}\right)^{2/5}\left(\Omega(z)-\frac{1}{4}z\right)+\frac{k_4}{k_2}z_1z_2.
\end{align}
\item Reduced Eigenfunction
\begin{equation}
\psi(z_1,z_2)=\varphi(z,\Lambda).
\end{equation}
\item Reduced Lax pair
\begin{equation}
\begin{aligned}
  &\varphi_{\Lambda \Lambda}-i\varphi_{\Lambda}-i\varphi_z=0,\\
   &(i-2z)\varphi_{\Lambda}+i\varphi_z+\left[-\frac{1}{4}+i(z-\Lambda)+\Omega_z\right]\varphi+2\varphi_{z \Lambda}+\varphi_{zz}=0.
 \end{aligned}\end{equation} 
\item Reduced equation
\begin{equation}
\begin{aligned}
       &V_z=0,\\
       &W_{zz}=0.
       \end{aligned}\end{equation} 
\end{itemize}

\end{itemize}

\section{Explicit calculation for hierarchies of PDEs}
\setcounter{equation}{0}
\setcounter{theorem}{0}
\setcounter{example}{0}
A {\it hierarchy} is a set of differential equations that are related via a {\it recursion operator}. The recursive application of
such operator gives members of different orders of the hierarchy.
For instance, let us consider the well-known {\bf Camassa-Holm equation} \cite{CH1} as a submanifold of ${\rm J}^3(\mathbb{R}^2,\mathbb{R})$ (CH$1+1$, henceforth)

\begin{equation}\label{CH1}
u_t+Cu_x-u_{xxt}=2u_xu_{xx}+uu_{xxx}.
\end{equation}
We can interpret $u$ as the fluid velocity and $t$, $x$ as the temporal and spatial coordinates, respectively.
It is an integrable, nonlinear PDE. Though not integrable in the strict defined Painlev\'e sense, there exists a change of 
variables (action-angle variables) such that the evolution equation in the new variables is equivalent to a linear flow
at constant speed.
This change of variables is achieved by studying its associated spectral problem and it is reminiscent of the
fact that integrable classical hamiltonian systems are equivalent to linear flows on tori.
Indeed, \eqref{CH1} is a bi-Hamiltonian model for shallow water waves propagation. When the parameter $C>0\in \mathbb{R}$ the solutions
are smooth solitary waves or solitons. The case $C=0$ that mostly concerns us, has peakon solutions. That is, solitons
with sharp peak, discontinuity in the slope.
\begin{equation}
u=ce^{-|x-ct|}+O(k\log{k}).
\end{equation}
\begin{example}\normalfont
Equation \eqref{CH1} can be understood as the first member of the  n-dimensional {\bf negative Camassa-Holm hierarchy} (NCHH $1+1$) that reads
\begin{equation}\label{NCHH}
u_t=R^{-n}u_x
\end{equation} 
where $n$ is an integer and $R=JK^{-1}$ is the recursion operator formed by the two operators $K=u\partial_x+\partial_x u$ and
$J=\partial_x-\partial_{xxx}$.
Since calculating the inverse of an operator is a difficult computation, we can introduce auxiliary scalar fields $v_1(x,t),\dots,v_n(x,t)$
such that for the nth-iteration of the operator, we can expand \eqref{NCHH} on ${\rm J}^3(\mathbb{R}^2,\mathbb{R}^{n+1})$ as
\begin{equation}
\begin{aligned}\label{NCHHaf}
&u_t=Jv_{[n]},\\
&Jv_{[j-1]}=Kv_{[j]},\quad j=2,\dots,n, \\
&u_x=Kv_{[1]}.
\end{aligned}\end{equation} 
It is easy to prove that when $n=1$ we recover \eqref{CH1}. Then we can say that \eqref{CH1} is the first member
of the negative Camassa-Holm hierarchy. There exists a positive Camassa-Holm hierarchy (PCHH$1+1$) when $n=-1$ in \eqref{NCHH} and has
been considered in \cite{CGP,Ivanov}.
\end{example}

\subsection{The negative Camassa-Holm hierarchy in $2+1$ dimensions and corresponding Lax pair}

This equation was generalized to $2+1$ dimensions in \cite{EstPr}. It is a simple integrable generalization
\begin{equation}\label{NCHH21}
u_t=R^{-n}u_y,
\end{equation}
where $R$ is the same as defined for the $1+1$ case. This hierarchy has been proven to be equivalent to $n$ copies of the {\bf AKNS equation}
in $2+1$ dimensions \cite{EstPrada1,KP1} by a reciprocal transformation. It is well-known that the AKNS has the Painlev\'e property and a nonisospectral Lax pair.
The NCHH's correspondent LP can be obtained by means of the singular manifold method \cite{EstPrada1} or by
applying the inverse reciprocal transform on the known AKNS associated Lax pair.
Similarly, we can introduce additional scalar fields and express it onto ${\rm J}^3(\mathbb{R}^3,\mathbb{R}^{n+1})$ as
\begin{equation}
\begin{aligned}\label{NCHHaf2}
&u_y=Jv_{[n]},\\
&Jv_{[j-1]}=Kv_{[j]},\quad j=2,\dots,n \\
&u_t=Kv_{[1]}.
\end{aligned}\end{equation} 
where $u=u(x,y,t)$ and $v_{[j]}=v_{[j]}(x,y,t)$ are defined over $\mathbb{R}^3$.
If we write explicitly the form of the operator, the hierarchy reads
\begin{equation}
\begin{aligned}\label{NCHHaf3}
&u_y=(v_{[n]})_x-(v_{[n]})_{xxx},\\
&(v_{[j-1]})_x-(v_{[j-1]})_{xxx}=u_xv_{[j]}+2u(v_{[j]})_x,\quad j=2,\dots,n \\
&u_t=u_xv_{[1]}+2u(v_{[1]})_x.
\end{aligned}\end{equation} 
Indeed, \eqref{NCHHaf} can be recovered if we set $\partial/\partial y=\partial/\partial t.$ The PCHH$1+1$ is recovered by $\partial/\partial y=\partial/\partial_x$.
The first member of \eqref{NCHHaf3} is the \eqref{CH1} equivalent in $2+1$ dimensions.
Lie point  symmetries of the \eqref{NCHHaf3} hierarchy were computed in \cite{EstLejaSar}.

The Lax pair associated with the NCHH$(2+1)$ \eqref{NCHHaf2} is a nonisospectral Lax pair on ${\rm J}^2(\mathbb{C}^3,\mathbb{C}^{n+3}).$
\begin{equation}
\begin{aligned}
&\psi_{xx}-\left(\frac{1}{4}-\frac{\lambda}{2}u\right)\psi=0,\\
&\psi_y-\lambda^n\psi_t+\hat{\mathcal A}\psi_x-\frac{\hat{\mathcal A}_x}{2}\psi=0,
\end{aligned}\end{equation} 
where $\hat{\mathcal A}=\sum_{i=1}^n{\lambda^{(n-j+1)}v_{[j]}}$ such that the compatibility condition 
$(\psi_{xxt}=\psi_{txx})$ recovers \eqref{NCHHaf3} and imposes the additional condition
\begin{equation}\label{nicnchh}
\lambda_y-\lambda^n\lambda_t=0,
\end{equation}
which is the {\it nonisospectrality condition}. 
As a consequence, we can interpret physically the spectrum of eigenvalues as a distorted spectrum whose eigenvalues
are no longer constant and their values get narrower.

In order to obtain symmetries of this Lax pair, together with those of \eqref{NCHHaf3}, we propose the following transformation
\begin{equation}
\left\{\begin{aligned}\label{pointtransep2}
&x\rightarrow x+\epsilon \xi_1(x,y,t,u,v_{[j]})+O(\epsilon^2),\\
&y\rightarrow y+\epsilon \xi_2(x,y,t,u,v_{[j]})+O(\epsilon^2),\\
&t\rightarrow t+\epsilon \xi_3(x,y,t,u,v_{[j]})+O(\epsilon^2),\\
&\lambda \rightarrow \lambda+\epsilon \eta_{\lambda}(x,y,t,u,v_{[j]},\lambda,\psi)+O(\epsilon^2),\\
&u\rightarrow u+\epsilon \eta_{u}(x,y,t,u,v_{[j]})+O(\epsilon^2),\\
&v_{[j]}\rightarrow v_{[j]}+\epsilon \eta_{v_{[j]}}(x,y,t,u,v_{[j]})+O(\epsilon^2),\\
&\psi\rightarrow \psi+\epsilon \eta_{\psi}(x,y,t,u,v_{[j]},\lambda,\psi)+O(\epsilon^2),
\end{aligned}\right.\quad j=1,\dots,n,\end{equation}
where $\lambda$ has to be considered as an scalar field due to the nonisospectrality condition \eqref{nicnchh}.
The associated Lie algebra of infinitesimal symmetries is the set of vector fields of the form
\begin{equation}\label{symvfchh}
X=\xi_1\frac{\partial}{\partial x}+\xi_2\frac{\partial}{\partial y}+\xi_3\frac{\partial}{\partial t}+\eta_{\lambda}\frac{\partial}{\partial \lambda}+\eta_{u}\frac{\partial}{\partial u}+\sum_{j=1}^n \eta_{v_{[j]}}\frac{\partial}{\partial v^{(j)}}+\eta_{\psi}\frac{\partial}{\partial \psi}.
\end{equation}

We need to know how the derivatives of the fields transform under this symmetry transformation. We introduce the prolongations
of the action of the group to the different derivatives that appear in \eqref{NCHHaf3}. To calculate the prolongations 
$(\eta_{v_{[j-1]}})_{xxx}$, $(\eta_{v_{[j-1]}})_{x}$, $(\eta_{v_{[j]}})_{x}$, $(\eta_{v_{[1]}})_{x}$, $(\eta_{v_{[n]}})_{xxx}$, $(\eta_{v_{[n]}})_{x}$, $(\eta_{u})_{x}$, $(\eta_{u})_{y}$, $(\eta_{u})_{t}$, $\psi_{xx}$, $\psi_y$, $\psi_t$, $\psi_x$, we follow the process explained in \eqref{prolongcomplete}
given in Chapter \ref{Chap:GeomFund}.
It is therefore necessary that \eqref{pointtransep2} leaves the hierarchy \eqref{NCHHaf3} invariant. We obtain an overdetermined system
for the infinitesimals $\xi_{x_1},\xi_{x_2},\xi_{x_3},\eta_{\lambda},\eta_{u},\eta_{v_{[j]}},\eta_{\psi}$.
According to the procedure for symmetry calculation step by step explained above, from the zero-order in $\epsilon$ we can isolate the higher-order
derivatives appearing in
\begin{equation}
\begin{aligned}
&(v_{[n]})_{xxx}=(v_{[n]})_x-u_y,\\
&(v_{[j-1]})_{xxx}=(v_{[j-1]})_{x}-u_xv_{[j]}-2u(v_{[j]})_x,\quad j=2,\dots,n,\\
&u_t=u_xv_{[1]}+2u(v_{[1]})_x,
\end{aligned}\end{equation} 
and
\begin{equation}
\begin{aligned}\label{zeroorder}
&\psi_{xx}=\left(\frac{1}{4}+\frac{\lambda}{2}u\right)\psi,\\
&\psi_y=\lambda^n\psi_t-\hat{\mathcal A}\psi_x+\frac{\hat{\mathcal A}_x}{2}\psi.
\end{aligned}\end{equation} 
with $\hat{\mathcal A}=\sum_{i=1}^n{\lambda^{(n-j+1)}v_{[j]}}$,
in order to be introduced in the system of differential equations from which to derive the symmetries. 

The generalization to nonclassical symmetries \cite{BluCole,Olver} imposes the following invariant surface conditions
\begin{equation}
\begin{aligned}\label{iscchh}
&\eta_{u}=\xi_1u_x+\xi_2u_y+\xi_3u_t,\\
&\eta_{v_{[j]}}=\xi_1(v_{[j]})_x+\xi_2(v_{[j]})_y+\xi_3(v_{[j]})_t,\quad j=1,\dots,n.\\
&\eta_{\lambda}=\xi_2\lambda_y+\xi_3\lambda_t,\\
&\eta_{\psi}=\xi_1\psi_x+\xi_2\psi_y+\xi_3\psi_t.
\end{aligned}\end{equation} 
which shall be used as in the addressed method for nonclassical Lie point symmetry calculation.

We distinguish between three different types of nonclassical symmetry according to the values of the infinitesimal generators
%

Owing to \eqref{iscchh} there is no restriction in selecting $\xi_i=1$ when $\xi_i\neq 0$ \cite{Olver}.

\begin{table}[H]\centering
\begin{tabular}{lccc}\toprule
\multicolumn{4}{c}{Values of infinitesimal generators}
\\ 
\midrule
Case A.& $\xi_3=1$ & \text{any}& \text{any}\\
Case B. & $\xi_3=0$ &  $\xi_2=1$ &\text{any}\\
Case C.& $\xi_3=0$ & $\xi_2=0$ & $\xi_1=1$\\
\bottomrule
\end{tabular}
\caption{First classification of reductions of NCHH($2+1$)}
\label{Tab4}
\end{table}
\noindent
In the Case A. of nonclassical symmetries for $\xi_3=1$, we eliminate derivatives with respect to $t$ in
\begin{equation}
\begin{aligned}
&u_t=\eta_{u}-\xi_1u_x-\xi_2u_y,\\
&(v_{[j]})_t=\eta_{v_{[j]}}-\xi_1(v_{[j]})_x-\xi_2(v_{[j]})_y,\quad j=1,\dots,n.\\
&\lambda_t=\eta_{\lambda}-\xi_2\lambda_y,\\
&\psi_t=\eta_{\psi}-\xi_1\psi_x-\xi_2\psi_y.
\end{aligned}\end{equation} 
Under these conditions, we obtain the following symmetries
\begin{align}\label{chhhxi31}
\xi_1&=S_1/S_3,\\
\xi_2&=S_2/S_3,\\
\xi_3&=1,\\
\eta_{\lambda}&=\frac{1}{S_3}\left(\frac{a_3-a_2}{n}\right)\lambda,\\
\eta_{u}&=\frac{1}{S_3}\left(\frac{1}{2}\frac{\partial S_1}{\partial x}+a_0\right)\psi,\\
\eta_{v_{[1]}}&=\frac{1}{S_3}\left[v_{[1]}\left(\frac{\partial S_1}{\partial x}-a_3\right)-\frac{\partial S_1}{\partial t}\right],\\
\eta_{v_{[j]}}&=\frac{1}{S_3}\left(\frac{\partial S_1}{\partial x}-a_2\frac{j-1}{n}-a_3\frac{n-j+1}{n}\right)v_{[j]},\quad j=2,\dots,n.\\
\eta_{\psi}&=\frac{1}{S_3}\left(\frac{1}{2}\frac{\partial S_1}{\partial x}+a_0\right)\psi.
\end{align}
where
\begin{equation}
\begin{aligned}
&S_1=S_1(x,t)=A_1(t)+B_1(t)e^x+C_1(t)e^{-x},\\
&S_2=S_2(y)=a_2y+b_2,\\
&S_3=S_3(t)=a_3t+b_3.\label{lass}
\end{aligned}\end{equation} 
with $A_1(t),A_2(t),A_3(t)$  arbitrary functions of $t$. Furthermore, $a_0,a_2,b_2,a_3,b_3$ are arbitrary constants, such that
$a_3$ and $b_3$ cannot be simultaneously 0.

 For the case B., $\xi_3=0, \xi_2=1$ we can now write
 \begin{align}\label{iscxi21}
u_y&=\eta_{u}-\xi_1u_x,\\
 (v_{[j]})_y&=\eta_{v_{[j]}}-\xi_1(v_{[j]})_x,\\
\lambda_y&=\eta_{\lambda},\\
\psi_y&=\eta_{\psi}-\xi_1\psi_x.
\end{align}
which combined with \eqref{pointtransep2} leads us to
\begin{align}\label{symCHHxi21}
\xi_1&=S_1/S_2,\\
\xi_2&=1,\\
\xi_3&=0,\\
\eta_{\lambda}&=\frac{1}{S_2}\left(-\frac{a_2}{n}\right)\lambda,\\
\eta_{u}&=\frac{1}{S_2}\left(-2\frac{\partial S_1}{\partial x}+\frac{a_2}{n}\right)u,\\
\eta_{v_{[1]}}&=\frac{1}{S_2}\left(\frac{\partial S_1}{\partial x}v_{[1]}-\frac{\partial S_1}{\partial t}\right),\\
\eta_{v_{[j]}}&=\frac{1}{S_2}\left(\frac{\partial S_1}{\partial x}-a_2\frac{j-1}{n}\right)v_{[j]},\quad j=2,\dots,n.\\
\eta_{\psi}&=\frac{1}{S_2}\left(-2\frac{\partial S_1}{\partial x}+\frac{a_2}{n}\right)\psi,
\end{align}
where $S_1$ and $S_2$ are those given in \eqref{lass}. Evidently, $a_2$ and $b_2$ cannot be 0 at the same time.
For the last case $\xi_3=\xi_2=0,\xi_1=1$, the invariant surface conditions turn into
\begin{equation}
\begin{aligned}\label{iscxi11}
u_x&=\eta_{u},\\
v_{[j]}&=\eta_{v_{[j]}},\qquad j=1,\dots,n.\\
\psi_x&=\eta_{\psi}.
\end{aligned}\end{equation} 
And the symmetries take a very trivial form
\begin{align}\label{symchhxi11}
\xi_1&=1,\\
\xi_2&=0,\\
\xi_3&=0,\\
\eta_{\lambda}&=0,\\
\eta_u&=-2u,\\
\eta_{v_{[j]}}&=v_{[j]},\qquad j=1,\dots,n.\\
\eta_{\psi}&=\frac{1}{2}\left(1\pm i\sqrt{2\lambda u}\right)\psi.
\end{align}
For any of the cases, we have to solve the following Lagrange-Charpit system with their corresponding values of infinitesimal generators
\begin{equation}
\frac{dx}{\xi_1}=\frac{dy}{\xi_2}=\frac{dt}{\xi_3}=\frac{d\lambda}{\eta_{\lambda}}=\frac{du}{\eta_u}=\frac{dv_{[j]}}{\eta_{v_{[j]}}}=\frac{d\psi}{\eta_{\psi}}.
\end{equation}
For the first case in which $\xi_3=1$, the classification of reductions can be performed attending 
to \cref{Tab5,Tab6}
\begin{table}[h]\centering
\begin{tabular}{lccc}\toprule
\multicolumn{4}{c}{Possible values of the t-dependent functions}
\\ 
\midrule
Case I.& $A_1(t)\neq 0$ & $B_1(t)=0$& $C_1(t)=0$\\
Case II & $A_1(t)=0$ & $B_1(t)\neq 0$ & $C_1(t)=0$\\
Case III.& $A_1(t)=0$ & $B_1(t)=0$ & $C_1(t)\neq 0$\\
\bottomrule
\end{tabular}
\caption{First subclassification of reductions for NCHH($2+1$)}
\label{Tab5}
\end{table}
\begin{table}[h]\centering
\begin{tabular}{lccc}\toprule
\multicolumn{4}{c}{Values of the constants}
\\ 
\midrule
1.& $a_2=0$ & $a_3=0$& $b_2=0$\\
2.& $a_2=0$ & $a_3=0$ & $b_2\neq 0$\\
3.& $a_2=0$ & $a_3\neq 0$ & $b_2=0$\\
4.& $a_2=0$ & $a_3\neq 0$ & $b_2\neq 0$\\
5.& $a_2\neq 0$& \text{any} & \text{any}\\
\bottomrule
\end{tabular}
\caption{Second subclassification of reductions for NCHH($2+1$).
}
\label{Tab6}
\end{table}

Case II yields the same reduced spectral problems as the obtained for I. Case III is easy to prove that it is equivalent to II owing to
the invariance of the Lax pair and equations under the transformation $x\rightarrow -x$, $y\rightarrow -y$ and $t\rightarrow -t$. 
Below we only consider cases I and II.

For each of these cases we have subcasses depending on the values of the constants $a_j$ and $b_j$, which are the 5 subcasses exposed.
We introduce the notation for the reduced variables as
\begin{equation}
x,y,t\rightarrow x_1,x_2\end{equation} 
and the reduced fields and reduced eigenfunction
\begin{equation}
\begin{aligned}
&u(x,y,t)\rightarrow U(x_1,x_2),\\
&v_{[1]}(x,y,t)\rightarrow V_{[1]}(x_1,x_2),\\
&v_{[j]}(x,y,t)\rightarrow V_{[j]}(x_1,x_2),\\
&\lambda(y,t)\rightarrow \Lambda(x_2),\\
&\psi(x,y,t)\rightarrow \Phi(x_1,x_2).
\end{aligned}\end{equation} 

We obtain 5 different nontrivial reductions. We obtain each of them separately with their correspondent reduced variables, reduced fields and transformations.
In the next subsection, the reduced spectral problem will be obtained.

\begin{itemize}

\item {\it Reduction I.1. $B_1(t)=C_1(t)=0, A_1(t)\neq 0, a_2=0, a_3=0, b_2=0$}

By solving the characteristic equation for these values, we obtain
\begin{itemize}
\item Reduced variables 
\begin{equation}
x_1=x-\frac{1}{b_3}\int{A_1(t)dt},\quad x_2=y.
\end{equation}
\item Reduced spectral parameter
\begin{equation}
\lambda=\lambda_0.
\end{equation}
\item Reduced vector fields
\begin{align}
&u(x,y,t)\rightarrow U(x_1,x_2),\\
&v_{[1]}(x,y,t)\rightarrow V_{[1]}(x_1,x_2)-\frac{A_1}{b_3},\\
&v_{[j]}(x,y,t)=V_{[j]}(x_1,x_2),
\end{align}
\item Reduced hierarchy
\begin{align}
&\frac{\partial^3 V_{[n]}}{\partial x_1^3}-\frac{\partial V_{[n]}}{\partial x_1}+\frac{\partial U}{\partial x_2}=0,\\
&\frac{2U\partial V_{[1]}}{\partial x_1}+V_{[1]}\frac{\partial U}{\partial x_1}=0,\\
&\frac{2U\partial V_{[j+1]}}{\partial x_1}+V_{[j+1]}\frac{\partial U}{\partial x_1}+\frac{\partial^3 V_{[j]}}{\partial x_1^3}-\frac{\partial V_{[j]}}{\partial x_1}=0.
\end{align}
which is {\it the positive Camassa-Holm hierarchy} whose first component $n=1$ is a {\it modified Dym equation} \cite{AntoFordy}.
\item Reduced eigenfunction
\begin{equation}
\psi=e^{\frac{a_0t}{b_3}}e^{\frac{\lambda_0^na_0x_2}{b_3}}\Phi(x_1,x_2).
\end{equation} 
\item Reduced Lax pair
\begin{equation}
\begin{aligned}\label{rlpchhI1}
&\Phi_{x_1 x_1}-\left(\frac{1}{4}-\frac{\lambda_0}{2}U\right)\Phi=0,\\
&\Phi_{x_2}+\hat{\mathcal{B}}\Phi_{x_1}-\frac{\hat{\mathcal{B}}}{2}\Phi=0,
\end{aligned}\end{equation} 
with $\hat{\mathcal{B}}=\sum_{j=1}^n \lambda_0^{n-j+1}V_{[j]}(x_1,x_2).$
\end{itemize}
\item {\it Reduction I.2. $B_1(t)=C_1(t)=0, A_1(t)\neq 0, a_2=0, a_3=0, b_2\neq 0$}
\begin{itemize}
\item Reduced variables 
\begin{equation}
x_1=x-\frac{1}{b_3}\int{A_1(t)dt},\quad x_2=\frac{y}{b_2}-\frac{t}{b_3}.
\end{equation}
\item Reduced spectral parameter
\begin{equation}
\lambda=\left(\frac{b_3}{b_2}\right)^{1/n}\lambda_0.
\end{equation}
\item Reduced fields 
\begin{align}\label{rvfI2}
&u(x,y,t)=\left(\frac{b_2}{b_3}\right)^{1/n}U(x_1,x_2),\\
&v_{[1]}(x,y,t)=\left(\frac{1}{b_3}\right)V_{[1]}(x_1,x_2)-\frac{A_1}{b_3},\\
&v_{[j]}(x,y,t)=\left(\frac{1}{b_3}\right)\left(\frac{b_2}{b_3}\right)^{\frac{1-j}{n}}V_{[j]}(x_1,x_2).
\end{align}
\item Reduced hierarchy 
\begin{align}
&\frac{\partial^3 V_{[n]}}{\partial x_1^3}-\frac{\partial V_{[n]}}{\partial x_1}+\frac{\partial U}{\partial x_2}=0,\\
&2U\frac{\partial V_{[1]}}{\partial x_1}+V_{[1]}\frac{\partial U}{\partial x_1}+\frac{\partial U}{\partial x_2}=0,\\
&2U\frac{\partial V_{[j+1]}}{\partial x_1}+V_{[j+1]}\frac{\partial U}{\partial x_1}+\frac{\partial^3 V_{[j]}}{\partial x_1^3}-\frac{\partial V_{[j]}}{\partial x_1}=0.
\end{align}
\item Reduced eigenfunction
\begin{equation}
\psi=e^{\frac{a_0t}{b_3}}e^{\frac{\lambda_0^na_0x_2}{1+\lambda_0^n}}\Phi(x_1,x_2).
\end{equation}
\item Reduced Lax pair
\begin{equation}
\begin{aligned}\label{rlpchhI2}
&\Phi_{x_1 x_1}-\left(\frac{1}{4}-\frac{\lambda_0}{2}U\right)\Phi=0,\\
&\Phi_{x_2}(1+\lambda_0^n)+\hat{\mathcal{B}}\Phi_{x_1}-\frac{\hat{\mathcal{B}}}{2}\Phi=0,
\end{aligned}\end{equation} 
with $\hat{\mathcal{B}}=\sum_{j=1}^n \lambda_0^{n-j+1}V_{[j]}(x_1,x_2).$
\end{itemize}
\item {\it Reduction I.3. $B_1(t)=C_1(t)=0, A_1(t)\neq 0, a_2=0, a_3\neq 0, b_2=0$}

\begin{itemize}
\item Reduced variables 
\begin{equation}
x_1=x-\int{\frac{A_1(t)}{S_3}dt},\quad x_2=a_3y.
\end{equation}
\item Reduced spectral parameter
\begin{equation}
\lambda(y,t)=S_3^{1/n}\Lambda(x_2).
\end{equation}
\item Reduced fields
\begin{align}
&u(x,y,t)=S_3^{-1/n}U(x_1,x_2),\\
&v_{[1]}=\frac{a_3}{S_3}V_{[1]}(x_1,x_2)-\frac{A_1}{S_3},\\
&v_{[j]}=a_3S_3^{\frac{j-1}{n}-1}V_{[j]}(x_1,x_2).
\end{align}
\item Reduced hierarchy 
\begin{align}
&\frac{\partial^3 V_{[n]}}{\partial x_1^3}-\frac{\partial V_{[n]}}{\partial x_1}+\frac{\partial U}{\partial x_2}=0,\\
&2U\frac{\partial V_{[1]}}{\partial x_1}+V_{[1]}\frac{\partial U}{\partial x_1}+\frac{U}{n}=0,\\
&2U\frac{\partial V_{[j+1]}}{\partial x_1}+V_{[j+1]}\frac{\partial U}{\partial x_1}+\frac{\partial^3 V_{[j]}}{\partial x_1^3}-\frac{\partial V_{[j]}}{\partial x_1}=0.
\end{align}
\item Reduced eigenfunction
\begin{equation}
\psi=\Lambda(x_2)^{\frac{a_0n}{a_3}}S_3^{\frac{a_0}{a_3}}\Phi(x_1,x_2)
\end{equation}
\item Reduced Lax pair
\begin{equation}
\begin{aligned}\label{rlpchhI3}
&\Phi_{x_1 x_1}=\left(\frac{1}{4}+\frac{\Lambda(x_2)}{2}U\right)\Phi,\\
&\Phi_{x_2}=-\hat{\mathcal{B}}\Phi_{x_1}+\frac{\hat{\mathcal{B}}}{2}\Phi,
\end{aligned}\end{equation} 
with $\hat{\mathcal{B}}=\sum_{j=1}^n \lambda_0^{n-j+1}V_{[j]}(x_1,x_2)$.
\end{itemize}
\item {\it Reduction I.4. $B_1(t)=C_1(t)=0, A_1(t)\neq 0, a_2=0, a_3\neq 0, b_2\neq 0$}
\begin{itemize}

\item Reduced variables
\begin{equation}
x_1=x-\int{\frac{A_1(t)}{S_3}dt},\quad x_2=\frac{a_3y}{b_2}-\ln{S_3}.
\end{equation}
\item Reduced spectral parameter
\begin{equation}
\lambda(y,t)=\left(\frac{S_3}{b_2}\right)^{1/n}\Lambda(x_2).
\end{equation}
\item Reduced fields
\begin{align}
&u(x,y,t)=\left(\frac{b_2}{S_3}\right)^{1/n}U(x_1,x_2),\\
&v_{[1]}=\frac{a_3}{S_3}V_{[1]}(x_1,x_2)-\frac{A_1}{S_3},\\
&v_{[j]}=\frac{a_3}{S_3}\left(\frac{S_3}{b_2}\right)^{\frac{j-1}{n}}V_{[j]}(x_1,x_2).
\end{align}

\item The reduced hierarchy
\begin{align}
&\frac{\partial^3 V_{[n]}}{\partial x_1^3}-\frac{\partial V_{[n]}}{\partial x_1}+\frac{\partial U}{\partial x_2}=0,\\
&2U\frac{\partial V_{[1]}}{\partial x_1}+V_{[1]}\frac{\partial U}{\partial x_1}+\frac{\partial U}{\partial x_2}+\frac{U}{n}=0,\\
&2U\frac{\partial V_{[j+1]}}{\partial x_1}+V_{[j+1]}\frac{\partial U}{\partial x_1}+\frac{\partial^3 V_{[j]}}{\partial x_1^3}-\frac{\partial V_{[j]}}{\partial x_1}=0.
\end{align}

\item Reduced eigenfunction
\begin{equation}
\psi=\Lambda(x_2)^{\frac{a_0n}{a_3}}S_3^{\frac{a_0}{a_3}}\Phi(x_1,x_2).
\end{equation}
\item Reduced Lax pair
\begin{equation}
\begin{aligned}\label{rlpchhI4}
&\Phi_{x_1 x_1}-\left(\frac{1}{4}-\frac{\Lambda(x_2)}{2}U\right)\Phi=0,\\
&\Phi_{x_2}(1+\Lambda(x_2)^n)+\hat{\mathcal{B}}\Phi_{x_1}-\frac{\hat{\mathcal{B}}}{2}\Phi=0,
\end{aligned}\end{equation} 
with $\hat{\mathcal{B}}=\sum_{j=1}^n \lambda_0^{n-j+1}V_{[j]}(x_1,x_2)$.
\end{itemize}

\item {\it Reduction I.5. $B_1(t)=C_1(t)=0, A_1(t)\neq 0, a_2\neq 0$}
\begin{itemize}
\item The reduced variables
\begin{equation}
x_1=x-\int{\frac{A_1(t)}{S_3(t)}dt},\quad x_2=S_2S_3^{-\frac{a_2}{a_3}}.
\end{equation}
\item Reduced spectral parameter
\begin{equation}
\lambda(y,t)=S_3^{\frac{a_3-a_2}{a_3n}}\Lambda(x_2).
\end{equation}
\item Reduced fields

\begin{align}
&u(x,y,t)=S_3^{\frac{a_2-a_3}{a_3n}}U(x_1,x_2),\\
&v_{[1]}(x,y,t)=\left(\frac{a_2}{S_3}\right)V_{[1]}(x_1,x_2)-\frac{A_1}{S_3},\\
&v_{[j]}(x,y,t)=\left(\frac{a_2}{S_3}\right)V_{[j]}(x_1,x_2)S_3^{\frac{(a_3-a_2)(j-1)}{a_3n}}.
\end{align}

\item Reduced hierarchy
\begin{align}
&\frac{\partial^3 V_{[n]}}{\partial x_1^3}-\frac{\partial V_{[n]}}{\partial x_1}+\frac{\partial U}{\partial x_2}=0,\\
&2U\frac{\partial V_{[1]}}{\partial x_1}+V_{[1]}\frac{\partial U}{\partial x_1}+x_2\frac{\partial U}{\partial x_2}+\frac{a_3-a_2}{a_2}\frac{U}{n}=0,\\
&2U\frac{\partial V_{[j+1]}}{\partial x_1}+V_{[j+1]}\frac{\partial U}{\partial x_1}+\frac{\partial^3 V_{[j]}}{\partial x_1^3}-\frac{\partial V_{[j]}}{\partial x_1}=0.
\end{align}

\item Reduced eigenfunction
\begin{equation}
\psi=\Lambda(x_2)^{\frac{na_0}{a_3-a_2}}S_3^{\frac{a_0}{a_3}}\Phi(x_1,x_2).
\end{equation}

\item Reduced Lax pair
\begin{equation}
\begin{aligned}
&\Phi_{x_1 x_1}=\left(\frac{1}{4}+\frac{\Lambda(x_2)}{2}U\right)\Phi,\\
&\Phi_{x_2}(1+x_2\Lambda(x_2)^2)+\hat{\mathcal{B}}\Phi_{x_1}-\frac{\hat{\mathcal{B}}_{x_1}}{2}\Phi=0,
\end{aligned}\end{equation} 
with $\hat{\mathcal{B}}=\sum_{j=1}^n \lambda_0^{n-j+1}V_{[j]}(x_1,x_2)$.
\end{itemize}

\end{itemize}

For the type of symmetry in which $\xi_3=0,\xi_2=1$, one of the reduced variables is $t$. This means that the integrals that involve $S_1$ can be performed without any restrictions
for the functions $A_1(t)$, $B_1(t)$, $C_1(t)$. We have four different cases 
%
%
\begin{table}[h]\centering
\begin{tabular}{lcc}\toprule
\multicolumn{3}{c}{Values of the arbitrary functions}
\\ 
\midrule
1.&$a_2=0$ & $E=\sqrt{A_1^2-4B_1C_1}=0$\\
2. &$a_2=0$ & $E=\sqrt{A_1^2-4B_1C_1}\neq 0$ \\
3. &$a_2\neq 0$ & $E=\sqrt{A_1^2-4B_1C_1}=0$ \\
4. &$a_2\neq 0$ & $E=\sqrt{A_1^2-4B_1C_1}\neq 0$ \\
\bottomrule
\end{tabular}
\caption{Second classification of reductions for CHH$2+1$}
\label{Tab7}
\end{table}

\begin{itemize}
\item {\it Reduction 1. $a_2=0$, $E=\sqrt{A_1^2-4B_1C_1}=0$}

\begin{itemize}
 \item Reduced variables 
\begin{equation}
x_1=\int{\frac{dx}{S_1(x,t)}}-\frac{y}{b_2},\quad x_2=\frac{t}{b_2}.
\end{equation}
\item Reduced spectral parameter
\begin{equation}
\lambda(y,t)=\lambda_0.
\end{equation}
\item Reduced fields 
\begin{align}\label{r12type}
&u(x,y,t)=\frac{U(x_1,x_2)}{S_1^2},\\
&v_{[1]}(x,y,t)=\frac{S_1}{b_2}V_{[1]}(x_1,x_2)+S_1\frac{d}{dt}\left(\int{\frac{dx}{S_1(x,t)}}\right),\\
&v_{[j]}(x,y,t)=\frac{S_1}{b_2}V_{[j]}(x_1,x_2).
\end{align}
\item Reduced hierarchy
\begin{align}
&\frac{\partial^3 V_{[n]}}{\partial x_1^3}-\frac{\partial U}{\partial x_1}=0,\\
&2U\frac{\partial V_{[1]}}{\partial x_1}+V_{[1]}\frac{\partial U}{\partial x_1}-\frac{\partial U}{\partial x_2}=0,\\
&2U\frac{\partial V_{[j+1]}}{\partial x_1}+V_{[j+1]}\frac{\partial U}{\partial x_1}+\frac{\partial^3 V_{[j]}}{\partial x_1^3}=0.
\end{align}
\item Reduced eigenfunction
\begin{equation}
\psi(x,y,t)=\sqrt{S_1}e^{\frac{a_0y}{b_2}}e^{\frac{a_0t}{b_2\lambda_0^2}}\Phi(x_1,x_2).
\end{equation}
\item Reduced Lax pair
\begin{equation}
\begin{aligned}\label{rlp1type2}
&\Phi_{x_1 x_1}+\frac{\lambda_0}{2}U\Phi=0,\\
&\lambda_0^n\Phi_{x_2}=(\hat{\mathcal{B}}-1)\Phi_{x_1}-\frac{\hat{\mathcal{B}}}{2}\Phi,
\end{aligned}\end{equation} 
with $\hat{\mathcal{B}}=\sum_{j=1}^n \lambda_0^{n-j+1}V_{[j]}(x_1,x_2).$
\end{itemize}

\item {\it Reduction 2. $a_2=0$, $E=\sqrt{A_1^2-4B_1C_1}\neq 0$}
\begin{itemize}
\item Reduced variables 
\begin{equation}
x_1=E\left(\int{\frac{dx}{S_1(x,t)}}-\frac{y}{b_2}\right),\quad x_2=\frac{1}{b_2}\int{E(t)dt}.
\end{equation}
\item Reduced spectral parameter
\begin{equation}
\lambda(y,t)=\lambda_0.
\end{equation}
\item Reduced fields 
\begin{align}
&u(x,y,t)=\frac{E^2}{S_1^2}U(x_1,x_2),\\
&v_{[1]}(x,y,t)=\frac{S_1}{b_2}V_{[1]}(x_1,x_2)+S_1\frac{d}{dt}\left(\int{\frac{dx}{S_1(x,t)}}\right)+S_1\frac{E_t}{E^2}x_1,\\
&v_{[j]}(x,y,t)=\frac{S_1}{b_2}V_{[j]}(x_1,x_2).
\end{align}
 \item Reduced hierarchy
\begin{align}
&\frac{\partial^3 V_{[n]}}{\partial x_1^3}-\frac{\partial V_{[n]}}{\partial x_1}-\frac{\partial U}{\partial x_1}=0,\\
&2U\frac{\partial V_{[1]}}{\partial x_1}+V_{[1]}\frac{\partial U}{\partial x_1}-\frac{\partial U}{\partial x_2}=0,\\
&2U\frac{\partial V_{[j+1]}}{\partial x_1}+V_{[j+1]}\frac{\partial U}{\partial x_1}+\frac{\partial^3 V_{[j]}}{\partial x_1^3}-\frac{\partial V_{[j]}}{\partial x_1}=0.
\end{align}

\item Reduced eigenfunction
\begin{equation}
\psi=\sqrt{\frac{S_1}{E}}e^{\frac{a_0y}{b_2}}e^{\frac{a_0t}{b_2\lambda_0^n}}\Phi(x_1,x_2).
\end{equation}
\item Reduced Lax pair

\begin{equation}
\begin{aligned}
&\Phi_{x_1x_1}+\left(\frac{\lambda_0}{2}U-\frac{1}{4}\right)\Phi=0,\\
&\lambda_0^n\Phi_{x_2}=(\hat{\mathcal{B}}-1)\Phi_{x_1}-\frac{\hat{\mathcal{B}}_{x_1}}{2}\Phi,
\end{aligned}\end{equation} 
with $\hat{\mathcal{B}}=\sum_{j=1}^n \Lambda(x_2)^{n-j+1}V_{[j]}(x_1,x_2).$

\medskip

This reduced equation and Lax pair correspond with the {\it celebrated negative Camassa--Holm hierarchy in $1+1$ dimensions.}

\end{itemize}
\item{\it Reduction 3. $a_2\neq 0$, $E=\sqrt{A_1^2-4B_1C_1}=0$}
\begin{itemize}
\item Reduced variables 
\begin{equation}
x_1=\int{\frac{dx}{S_1(x,t)}}-\frac{\ln{S_2}}{a_2}, \quad x_2=t.
\end{equation}
\item Reduced spectral parameter
\begin{equation}\lambda(y,t)=S_2^{-1/n}\Lambda(x_2),\end{equation}
 where $\Lambda(x_2)$ satisfies the nonisospectral condition
\begin{equation}
n\frac{d\Lambda(x_2)}{dx_2}+a_2\lambda(x_2)^{1-n}=0.
\end{equation}
\item Reduced fields 
\begin{align}\label{r32type}
&u(x,y,t)=\frac{S_2^{1/n}}{S_1^2}U(x_1,x_2),\\
&v_{[1]}(x,y,t)=S_1V_{[1]}(x_1,x_2)+S_1\frac{d}{dt}\left(\int{\frac{dx}{S_1(x,t)}}\right),\\
&v_{[j]}(x,y,t)=S_1S_2^{\frac{1-j}{n}}V_{[j]}(x_1,x_2).
\end{align}

 \item Reduced hierarchy
\begin{align}
&\frac{\partial^3 V_{[n]}}{\partial x_1^3}-\frac{\partial U}{\partial x_1}+\frac{a_2}{n}U=0,\\
&2U\frac{\partial V_{[1]}}{\partial x_1}+V_{[1]}\frac{\partial U}{\partial x_1}-\frac{\partial U}{\partial x_2}=0,\\
&2U\frac{\partial V_{[j+1]}}{\partial x_1}+V_{[j+1]}\frac{\partial U}{\partial x_1}+\frac{\partial^3 V_{[j]}}{\partial x_1^3}=0.
\end{align}

\item Reduced eigenfunction
\begin{equation}
\psi=\sqrt{S_1}\lambda^{-\frac{a_0n}{a_2}}\Phi(x_1,x_2).
\end{equation}

\item Reduced Lax pair
\begin{equation}
\begin{aligned}
&\Phi_{x_1x_1}+\frac{\Lambda(x_2)}{2}U\Phi=0,\\
&\Lambda(x_2)^n\Phi_{x_2}=(\hat{\mathcal{B}}-1)\Phi_{x_1}-\frac{\hat{\mathcal{B}}_{x_1}}{2}\Phi,
\end{aligned}\end{equation} 
with $\hat{\mathcal{B}}=\sum_{j=1}^n \Lambda(x_2)^{n-j+1}V_{[j]}(x_1,x_2).$
\end{itemize}

\item {\it Reduction 4. $a_2\neq 0$ , $E=\sqrt{A_1^2-4B_1C_1}\neq 0$}
\begin{itemize}
\item Reduced variables 
\begin{equation}
x_1=E\left(\int{\frac{dx}{S_1(x,t)}}-\frac{\ln{S_2}}{a_2}\right), \quad x_2=\int{E(t)dt}.
\end{equation}
\item Reduced spectral parameter
\begin{equation}\lambda(y,t)=S_2^{-1/n}\Lambda(x_2),\end{equation}
where $\Lambda(x_2)$ satisfies the nonisospectral condition
\begin{equation}
n\frac{d\Lambda(x_2)}{dx_2}+\frac{a_2}{E(x_2)}\Lambda(x_2)^{n-1}=0.
\end{equation}
\item Reduced fields 
\begin{align}
&u(x,y,t)=\frac{E^2}{S_1^2}S_2^{1/n}U(x_1,x_2),\\
&v_{[1]}(x,y,t)=S_1V_{[1]}(x_1,x_2)+S_1\frac{d}{dt}\left(\int{\frac{dx}{S_1(x,t)}}\right)+S_1\frac{E_t}{E^2}x_1,\\
&v_{[j]}(x,y,t)=S_1S_2^{\frac{1-j}{n}}V_{[j]}(x_1,x_2).
\end{align}
 \item Reduced hierarchy
\begin{align}
&\frac{\partial^3 V_{[n]}}{\partial x_1^3}-\frac{\partial V_{[n]}}{\partial x_1}-\frac{\partial U}{\partial x_1}+\frac{a_2}{nE(x_2)}U=0,\\
&2U\frac{\partial V_{[1]}}{\partial x_1}+V_{[1]}\frac{\partial U}{\partial x_1}-\frac{\partial U}{\partial x_2}=0,\\
&2U\frac{\partial V_{[j+1]}}{\partial x_1}+V_{[j+1]}\frac{\partial U}{\partial x_1}+\frac{\partial^3 V_{[j]}}{\partial x_1^3}-\frac{\partial V_{[j]}}{\partial x_1}=0.
\end{align}

\item Reduced eigenfunction
\begin{equation}
\psi(x,y,t)=\sqrt{\frac{S_1}{E}}\lambda^{-\frac{-a_0n}{a_2}}\Phi(x_1,x_2).
\end{equation}
\item Reduced Lax pair
\begin{equation}
\begin{aligned}
&\Phi_{x_1x_1}+\left(\frac{\Lambda(x_2)}{2}U-\frac{1}{4}\right)\Phi=0,\\
&\Lambda(x_2)^n\Phi_{x_2}=(\hat{\mathcal{B}}-1)\Phi_{x_1}-\frac{\hat{\mathcal{B}}}{2}\Phi,
\end{aligned}\end{equation} 
with $\hat{\mathcal{B}}=\sum_{j=1}^n \Lambda(x_2)^{n-j+1}V_{[j]}(x_1,x_2).$
\end{itemize}
\end{itemize}

\subsection{The Qiao hierarchy in $2+1$ dimensions and corresponding Lax pair}

Another important hierarchy in relation to the NCHH($2+1$) \eqref{NCHHaf3} is the {\bf Qiao hierarchy} (Qiao$2+1$ or mCHH($2+1$)). It was first introduced
as an integrable generalization of the Qiao hierarchy in $1+1$ \cite{Qiao1} to $2+1$ dimensions. This hierarchy
can be proven to be connected with the NCHH($2+1$) by reciprocal transformations \cite{estevez51}. Henceforth we will
refer to it as mCHH($2+1$).
In the reduced form, it is formally equivalent to \eqref{NCHH}
\begin{equation}\label{mNCHH}
u_t=R^{-n}u_y,
\end{equation}
but the operators are substantially different.
In this case, the operator $K=\partial_{xxx}-\partial_x$ and $J=-\partial_x u \partial_x^{-1}u \partial_x$ that form
the recursion operator $R=JK^{-1}$.
In compact form as a submanifold of ${\rm J}^3(\mathbb{R}^3,\mathbb{R}^{n+1})$ the mCHH$2+1$ reads
\begin{equation}
\begin{aligned}\label{mNCHHaf}
&u_y=Jv_{[n]},\\
&Jv_{[j+1]}=Kv_{[j]},\\
&u_t=Kv_{[n]}.
\end{aligned}\end{equation} 
It is necessary to introduce auxiliary scalar fields $v_{[1]}(x,y,t),\dots,v_{[n]}(x,y,t)$ 
as in \eqref{NCHHaf2}, \eqref{NCHHaf3} for utilizing the inverse of the operator $J$.
Also, due to the difficulty of operating with the term $\delta^{-1}u$ in $J$ it is necessary to include another $n$-tuple of scalar fields  $\omega_{[1]}(x,y,t),\dots,\omega_{[n]}(x,y,t).$
Introducing the explicit form of the operators and making use of the additional scalar fields, we expanded expressions to ${\rm J}^{3}(\mathbb{R}^3,\mathbb{R}^{2n+1})$ as
\begin{equation}
\begin{aligned}\label{mNCHHaf2}
&u_y=-(u\omega_{[1]})_{x},\\
&(v_{[j]})_{xx}-v_{[j]}=-u\omega_{[j+1]},\\
&(\omega_{[j]})_{x}=u(v_{[j]})_{x},\\
&u_t=((v_{[n]})_{xx}-v_{[n]})_x,
\end{aligned}
\qquad j=1,\dots,n-1. 
\end{equation} 
%

The Lax pair for mCHH($2+1$) was first introduced in \cite{estevez51} as an integrable generalization
of the mCHH($1+1$)'s Lax pair to $2+1$ dimensions \cite{Qiao2,Qiao3,Qiao1}.
It is a two component Lax pair given geometrically as a submanifold of ${\rm J}^{2}(\mathbb{C}^3,\mathbb{C}^{2n+4})$ whose temporal and spatial parts read correspondingly
\begin{equation}\label{plqiao1}
{\left(\begin{array}{c}
   \phi  \\
   \psi \\
  \end{array}\right) } _{t}=\lambda^n {\left(\begin{array}{c}
   \phi  \\
   \psi \\
  \end{array}\right) } _y+\lambda p  {\left(\begin{array}{c}
   \phi  \\
   \psi \\
  \end{array}\right) } _x +\frac{i\sqrt{\lambda}}{2}\left[\begin{array}{cc}
  0 & q_x-q\\
  q_x+q & 0
  \end{array}\right]_x
   {\left(\begin{array}{c}
   \phi  \\
   \psi \\
  \end{array}\right) }, 
\end{equation}
where $p=\sum_{j=1}^n \lambda^{n-j}(y,t) \omega_{[j]}(x,y,t)$ and $q=\sum_{j=1}^n \lambda^{n-j}(y,t) v_{[j]}(x,y,t)$
and $i=\sqrt{-1}$,
\noindent
and
\begin{equation}\label{plqiao2}
 {\left(\begin{array}{c}
   \phi  \\
   \psi \\
  \end{array}\right) } _{x}= \frac{1}{2}\left[\begin{array}{cc}
  -1 & i\sqrt{\lambda}u\\
  i\sqrt{\lambda}u & 1
  \end{array}\right]
   {\left(\begin{array}{c}
   \phi  \\
   \psi \\
  \end{array}\right) } .
\end{equation}
%
%

The compatibility condition of $(\phi_{xt}=\phi_{tx})$ or $(\psi_{xt}=\psi_{tx})$ recovers \eqref{mNCHHaf2}.
Additionally, from the compatibility condition we obtain the nonisospectrality condition
\begin{equation}\label{nicqiao}
\lambda_t-\lambda^n\lambda_y=0,
\end{equation} 
which is the same as for the NCHH$2+1$.

In order to calculate the Lie point symmetries of the mCHH($2+1$) and its associated Lax pair, we propose a uniparametric Lie point symmetry transformation of the form
\begin{equation}
\left\{\begin{aligned}\label{mchhtransf}
&x\rightarrow x+\epsilon \xi_1\left(x,y,t,u,\omega_{[j]},v_{[j]},\phi,\psi\right)+O\left(\epsilon^2\right),\\
&y\rightarrow y+\epsilon \xi_2\left(x,y,t,u,\omega_{[j]},v_{[j]},\phi,\psi\right)+O\left(\epsilon^2\right),\\
&t\rightarrow t+\epsilon \xi_1\left(x,y,t,u,\omega_{[j]},v_{[j]},\phi,\psi\right)+O\left(\epsilon^2\right),\\
&\lambda\rightarrow \lambda+\epsilon \eta_{\lambda}\left(x,y,t,\lambda,u,\omega_{[j]},v_{[j]},\phi,\psi\right)+O\left(\epsilon^2\right),\\
&u\rightarrow u+\epsilon \eta_{u}\left(x,y,t,u,\omega_{[j]},v_{[j]},\phi,\psi\right)+O\left(\epsilon^2\right),\\
&\omega_{[j]}\rightarrow \omega_{[j]}+\epsilon \eta_{\omega_{[j]}}\left(x,y,t,u,\omega_{[j]},v_{[j]},\phi,\psi\right)+O\left(\epsilon^2\right),\\
&v_{[j]}\rightarrow v_{[j]}+\epsilon \eta_{v_{[j]}}\left(x,y,t,u,\omega_{[j]},v_{[j]},\phi,\psi\right)+O\left(\epsilon^2\right),\\
&\phi\rightarrow \phi+\epsilon \eta_{\phi}\left(x,y,t,\lambda,u,\omega_{[j]},v_{[j]},\phi,\psi\right)+O\left(\epsilon^2\right),\\
&\psi\rightarrow \psi+\epsilon \eta_{\psi}\left(x,y,t,\lambda,u,\omega_{[j]},v_{[j]},\phi,\psi\right)+O\left(\epsilon^2\right),
\end{aligned}\right.\end{equation} 
with  $j=1,\dots,n.$
The Lie algebra corresponds with the set of vector fields
\begin{align}\label{vfmchh}
X=\xi_1\frac{\partial}{\partial x}+\xi_2\frac{\partial}{\partial y}&+\xi_3\frac{\partial}{\partial t}+\eta_{\lambda}\frac{\partial}{\partial \lambda}+\nonumber \\
&\eta_{u}\frac{\partial}{\partial u}+\sum_{j=1}^n \omega_{[j]}\frac{\partial}{\partial \omega_{[j]}}+\sum_{j=1}^n v_{[j]}\frac{\partial}{\partial v_{[j]}}+\eta_{\psi}\frac{\partial}{\partial \psi}+\eta_{\phi}\frac{\partial}{\partial \phi}.
\end{align}
Symmetries of this hierarchy have been particularly difficult to obtain. For this matter, we will restrict ourselves to the classical realm. 
In general, the resulting system of nonlinear PDEs is very difficult to solve, as for the nonclassical case. The classical case has also been tough, 
due to the number of auxiliary fields that have been introduced.
Following a similar procedure as explained in the algorithmic steps in Subsection 4.1, we need the prolongations corresponding with $(\eta_{v_{[n]}})_{xxx}$, $(\eta_{v_{[j]}})_{xx}$, $(\eta_{v_{[j]}})_{x}$, $(\eta_{v_{[n]}})_{x}$, $(\eta_{\omega_{[j]}})_{x}$, $(\eta_{u})_t$, $(\eta_{u})_y$, $(\eta_{u})_x$, $(\eta_{\omega_{[1]}})_x$,
$(\eta_{\psi})_{t}$, $(\eta_{\psi})_{y}$ ,$(\eta_{\psi})_{x}$, $(\eta_{\phi})_{t}$, $(\eta_{\phi})_{y}$, $(\eta_{\phi})_{x},$ 
according to \eqref{prolongcomplete} in Chapter \ref{Chap:GeomFund}.
Also we make use of the zero-order in $\epsilon$ retrieved initial, untransformed equations
whose higher-order derivative has to be introduced into the system of PDEs for the symmetry search
\begin{equation}
\begin{aligned}
u_y&=-(u\omega_{[1]})_x,\\
(v_{[j]})_{xx}&=v_{[j]}-u\omega_{[j+1]},\\
(\omega_{[j]})_{x}&=u(v_{[j]})_{x},\\
(v_{[n]})_{xxx}&=u_t+(v_{[n]})_x,\\
\lambda_t&=\lambda\lambda_y,\\
\phi_t&=\lambda^n \phi_y+\lambda p \phi_x+\frac{i\sqrt{\lambda}}{2}(q_x-q)\psi,\\
\psi_t&=\lambda^n \psi_y+\lambda p \psi_x+\frac{i\sqrt{\lambda}}{2}(q_x+q)\phi,\\
\phi_x&=-\frac{1}{2}\phi+\frac{i\sqrt{\lambda}u}{2}\psi,\\
\psi_x&=\frac{1}{2}\psi+\frac{i\sqrt{\lambda}u}{2}\phi,
\end{aligned}\end{equation} 
with $j=1,\dots,n-1$.

From here, we obtain the results
\begin{align}\label{symmchh}
&\xi_1=A_1(t),\\
&\xi_2=a_2y+b_2,\\
&\xi_3=a_3t+b_3,\\
&\eta_{\lambda}(y,t,\lambda)=\frac{a_2-a_3}{n}\lambda,\\
&\eta_{u}(x,y,t,u)=\frac{a_3-a_2}{2n}u,\\
&\eta_{\omega_{[j]}}(x,y,t,\omega_{[j]})=\delta^{(j,1)}\frac{dA_1(y)}{dy}-\frac{(n-j+1)a_2+(j-1)a_3}{n}\omega_{[j]},\\
&\eta_{v_{[j]}}(x,y,t,v_{[j]})=\delta^{(j,n)}A_n(y,t)-\frac{(2(n-j)+1)a_2+(2j-1)a_3}{2n}v_{[j]}\\
&\eta_{\phi}(x,y,t,\lambda,\psi,\phi)=\gamma(y,t,\lambda)\phi,\\
&\eta_{\psi}(x,y,t,\lambda,\psi,\phi)=\gamma(y,t,\lambda)\psi,
\end{align}
for all $j=1,\dots,n,$
and where the function $\gamma$ satisfies the following equation
\begin{equation}\label{gammacond}
\frac{\partial \gamma(y,t,\lambda)}{\partial t}=\lambda^n \frac{\partial \gamma(y,t,\lambda)}{\partial t},
\end{equation}
which is equivalent to the equation \eqref{burgermuzero}.
The functions are $A_1(y)$, $A_n(y,t)$ and constants $a_2,a_3,b_2,b_3$ are arbitrary and $\delta^{(j,1)}$ and $\delta^{(j,n)}$ are Kronecker deltas.

The reductions can be archieved by the following characteristic system
\begin{equation}\label{charsysmch}
\frac{dx}{\xi_1}=\frac{dy}{\xi_2}=\frac{dt}{\xi_3}=\frac{d\lambda}{\eta_{\lambda}}=\frac{du}{\eta_{u}}=\frac{d\omega_{[j]}}{\eta_{\omega_{[j]}}}=\frac{dv_{[j]}}{\eta_{v_{[j]}}}=\frac{d\psi}{\eta_{\psi}}=\frac{d\phi}{\eta_{\phi}}
\end{equation}
and attending to the  
 values for the arbitrary constants
%
\begin{table}[H]\centering
\begin{tabular}{lcccccccc}\toprule
\multicolumn{3}{c}{Case I: $a_2\neq 0,b_2=0$}&\multicolumn{3}{c}{Case II: $a_2=0,b_2\neq 0$}&\multicolumn{3}{c}{Case III: $a_2=0,b_2=0$}
\\ 
\midrule
1.& $a_3\neq 0$ & $b_3=0$&& $a_3\neq 0$ & $b_3=0$&& $a_3\neq 0$ & $b_3=0$\\
2. & $a_3=0$ & $b_3\neq 0$&& $a_3=0$ & $b_3\neq 0$&& $a_3=0$ & $b_3\neq 0$\\
3.& $a_3=0$ & $b_3=0$&& $a_3=0$ & $b_3=0$&&&\\
\bottomrule
\end{tabular}
\caption{Classification of reductions for mCHH($2+1$)}
\label{Tab8}
\end{table}

If we introduce the next notation for the reduced variables and vector fields
\begin{equation}
\begin{aligned}
x,y,t,&\rightarrow x_1,x_2,\\
\lambda&\rightarrow \Lambda(x_2),\\
U(x,y,t)&\rightarrow U(z_1,z_2),\\
\omega_{[1]}(x,y,t)&\rightarrow \Omega_{[1]}(x_1,x_2),\\
\omega_{[j]}(x,y,t)&\rightarrow \Omega_{[j]}(x_1,x_2),\qquad j=2,\dots,n,\\
v_{[j]}(x,y,t)&\rightarrow V_{[j]}(x_1,x_2),\qquad j=1,\dots,n-1,\\
v_{[n]}(x,y,t)&\rightarrow V_{[n]}(x_1,x_2),\\
\phi(x,y,t)&\rightarrow \Phi(x_1,x_2),\\
\psi(x,y,t)&\rightarrow \Psi(x_1,x_2),\\
p(x,y,t)&\rightarrow P(x_1,x_2),\\
q(x,y,t)&\rightarrow Q(x_1,x_2).
\end{aligned}\end{equation} 
according to the exposed tables, we have

\begin{itemize}

\item {\it Reduction I.1. $a_2\neq 0, b_2=0,a_3\neq 0, b_3=0$}

Solving the characteristic system \eqref{charsysmch} for the values of constants in I.1., 

\begin{itemize}
\item Reduced variables 
\begin{equation}
x_1=x-\frac{1}{a_2}\int{\frac{A_1(y)}{y}dy},\quad x_2=\frac{t}{y^r},
\end{equation}
where $r=a_3/a_2.$

\item The reduced spectral parameter reads
\begin{equation}
\lambda(y,t)=y^{\frac{1-r}{n}\Lambda(x_2)},
\end{equation}
where $\Lambda(x_2)$ satisfies the {\it nonisospectral condition}
\begin{equation}
\frac{d\Lambda(x_2)}{dx_2}=\left(\frac{1-r}{n}\right)\frac{\Lambda(x_2)^{n+1}}{1+rx_2\Lambda(x_2)^n}.
\end{equation}

\item Reduced fields 
\begin{align}\label{rfmchI1}
&u(x,y,t)=y^{\frac{r-1}{2n}}U(x_1,x_2),\\
&\omega_{[1]}(x,y,t)=\frac{A_1(y)}{a_2y}+\frac{\Omega_{[1]}(x_1,x_2)}{y},\\
&\omega_{[j]}(x,y,t)=y^{\frac{(r-1)(1-j)}{n}}\left(\frac{\Omega_{[j]}}{y}\right),\quad j=2,\dots,n,\\
&v_{[j]}(x,y,t)=y^{\frac{(r-1)(1-2j)}{2n}}\left(\frac{V_{[j]}(x_1,x_2)}{y}\right),\quad j=1,\dots,n-1,\\
&v_{[n]}(x,y,t)=y^{\frac{(r-1)(1-2n)}{2n}}\left(\frac{1}{a_2y}\int{\frac{A_n(y,t)y^{\frac{2rn-r+1}{2n}}}{y}dy}
\right.\nonumber
\\
&\hspace*{5em}\left.
{+}\frac{V_{[n]}(x_1,x_2)}{y}\right).
\end{align}
\item Reduced hierarchy 
\begin{align}
&\left((V_{[n]})_{x_1x_1}-V_{[n]}\right)_{x_1}-U_{x_2}=0,\\
&(V_{[j]})_{x_1x_1}-V_{[j]}+U\Omega_{[j+1]}=0,\qquad j=1,\dots,n-1, \\
&\left(\Omega_{[1]}U\right)_{x_1}+\frac{r-1}{2n}U-rx_2U_{x_2}=0,\\
&(\Omega_{[j]})_{x_1}=U(V_{[j]})_{x_1}.
\end{align}

\item Reduced eigenfunctions
\begin{align}
\phi(x,y,t)&=e^{\int{\frac{\Gamma(y,x_2)}{a_2y}dy}}\Phi(x_1,x_2),\\
\psi(x,y,t)&=e^{\int{\frac{\Gamma(y,x_2)}{a_2y}dy}}\Psi(x_1,x_2),\\
p(x,y,t)&=\frac{y^{\frac{r-1}{n}}}{y^r}\left(\frac{A_1(y)\Lambda(x_2)^{n-1}}{a_2}+P(x_1,x_2)\right),\\
q(x,y,t)&=\frac{y^{\frac{r-1}{2n}}}{y^r}\left(\frac{1}{a_2}\int{\frac{A_n(y,t)y^{\frac{2rn-r+1}{2n}}}{y}dy}+Q(x_1,x_2)\right)\\
P(x_1,x_2)&=\sum_{j=1}^n \Lambda(x_2)^{n-j}\Omega_{[j]}(x_1,x_2),\\
Q(x_1,x_2)&=\sum_{j=1}^n \Lambda(x_2)^{n-j}V_{[j]}(x_1,x_2),
\end{align}

\item The reduced spectral problem is
\begin{multline}
(1+rx_2\Lambda^n) {\left(\begin{array}{c}
   \Phi  \\
   \Psi \\
  \end{array}\right) } _{x_2}=\Lambda P {\left(\begin{array}{c}
   \phi  \\
   \psi \\
  \end{array}\right) } _{x_1}
\\
+\frac{i\sqrt{\Lambda}}{2}\left[\begin{array}{cc}
  0 & Q_{x_1}-Q\\
  Q_{x_1}+Q  & 0
  \end{array}\right]_{x_1}
   {\left(\begin{array}{c}
   \phi  \\
   \psi \\
  \end{array}\right) } 
\end{multline}
and
\begin{equation}
{\left(\begin{array}{c}
   \Phi  \\
   \Psi \\
  \end{array}\right) } _{x_1}=\frac{1}{2}\left[\begin{array}{cc}
  -1 & i\sqrt{\Lambda}U\\
  i\sqrt{\Lambda}U  & 1
  \end{array}\right]
  {\left(\begin{array}{c}
   \phi  \\
   \psi \\
  \end{array}\right) } .
\end{equation}
\end{itemize}

\item {\it Reduction I.2. $a_2\neq 0, b_2=0, a_3=0, b_3\neq 0$}

\begin{itemize}

\item Reduced variables 
\begin{equation}
x_1=x-\frac{1}{a_2}\int{\frac{A_1(y)}{y}dy},\quad x_2=\frac{a_2t}{b_3}-\ln{y}.
\end{equation}

\item Reduced spectral parameter
\begin{equation}
\lambda(y,t)=\left(\frac{a_2y}{b_3}\right)^{\frac{1}{n}}\Lambda(x_2),
\end{equation}
with $\Lambda(x_2)$ satisfying the nonisospectral condition
\begin{equation}
\frac{d\Lambda(x_2)}{dx_2}=\frac{\Lambda(x_2)^{n+1}}{n(1+\Lambda(x_2)^n)}.
\end{equation}

\item  Reduced fields
 \begin{align}
 u(x,y,t)&=\left(\frac{a_2y}{b_3}\right)^{-\frac{1}{2n}}U(x_1,x_2),\\
\omega_{[1]}(x,y,t)&=\frac{A_1(y)}{a_2y}+\frac{\Omega_{[1]}(x_1,x_2)}{y},\\
\omega_{[j]}(x,y,t)&=\left(\frac{a_2y}{b_3}\right)^{\frac{j-1}{n}}\frac{\Omega_{[j]}(x_1,x_2)}{y},\\
v_{[j]}(x,y,t)&=\left(\frac{a_2y}{b_3}\right)^{\frac{2j-1}{2n}}\frac{V_{[j]}(x_1,x_2)}{y},\quad j=1,\dots,n-1,\\
 v_{[n]}(x,y,t)&=y^{-\frac{1}{2n}}\int{\frac{A_n(y,t)y^{\frac{1}{2n}}}{a_2y}dy}+\left(\frac{a_2y}{b_3}\right)^{\frac{2n-1}{2n}}\frac{V_{[n]}}{y}.
\end{align}
\item Reduced hierarchy
\begin{align}
 &((V_{[n]})_{x_1x_1}-V_{[n]})_{x_1}-U_{x_2}=0,\\
 &(V_{[j]})_{x_1x_1}-V_{[j]}+U\Omega_{[j+1]}=0,\qquad j=1,\dots,n-1,\\
 &(\Omega_{[1]}U)_{x_1}-U/2n-U_{x_2}=0,\\
 &(\Omega_{[j]})_{x_1}=U(V_{[j]})_{x_1},\qquad j=1,\dots,n.
 \end{align}

\item The reduced eigenfunctions read
\begin{align}
\phi(x,y,t)&=e^{\int{\frac{\Gamma(y,x_2)}{a_2y}dy}}\Phi,\\
\psi(x,y,t)&=e^{\int{\frac{\Gamma(y,x_2)}{a_2y}dy}}\Psi,\\
p(x,y,t)&=\left(\frac{a_2y}{b_3}\right)^{\frac{n-1}{n}}\left(\frac{A_1(y)\Lambda(x_2)^{n-1}}{a_2y}+\frac{P(x_1,x_2)}{y}\right),\\
q(x,y,t)&=y^{-\frac{1}{2n}}\int{\frac{A_n(y,t)y^{\frac{1-2n}{2n}}}{a_2}dy}+\left(\frac{a_2y}{b_3}\right)^{\frac{2n-1}{2n}}\frac{Q(x_1,x_2)}{y},\\
P(x_1,x_2)&=\sum_{j=1}^n \Lambda(x_2)^{n-j}\Omega_{[j]}(x_1,x_2),\\
Q(x_1,x_2)&=\sum_{j=1}^n \Lambda(x_2)^{n-j}V_{[j]}(x_1,x_2),
\end{align}
where the function $\Gamma(y,x_2)$ is obtained through the following identification
\begin{equation}
\Gamma(y,x_2)=\Gamma\left[y,t=\frac{b_3}{a_2}(x_2+\ln{y})\right]
\end{equation}
that yields
\begin{equation}
\gamma_t=\frac{a_2}{b_3}\Gamma_{x_2},\quad \gamma_y=\Gamma_y-\frac{1}{y}\Gamma_{x_2}.
\end{equation}
Therefore, according to \eqref{gammacond}, $\Gamma(y,x_2)$ satisfies the equation
\begin{equation}
(1+\Lambda^n)\Gamma_{x_2}=y\Lambda^y\Gamma_y.
\end{equation}

\item Reduced Lax pair 

\begin{multline}
(1+\Lambda^n) {\left(\begin{array}{c}
   \Phi  \\
   \Psi \\
  \end{array}\right) } _{x_2}=\Lambda P {\left(\begin{array}{c}
   \phi  \\
   \psi \\
   \end{array}\right) }_{x_1}
\\
+\frac{i\sqrt{\Lambda}}{2}\left[\begin{array}{cc}
  0 & Q_{x_1}-Q\\
  Q_{x_1}+Q  & 0
  \end{array}\right]_{x_1}
   {\left(\begin{array}{c}
   \phi  \\
   \psi \\
  \end{array}\right)  } 
\end{multline}
and
\begin{equation}
 {\left(\begin{array}{c}
   \Phi  \\
   \Psi \\
  \end{array}\right) } _{x_1}=\frac{1}{2}\left[\begin{array}{cc}
  -1 & i\sqrt{\Lambda}U\\
  i\sqrt{\Lambda}U  & 1
  \end{array}\right]
  {\left(\begin{array}{c}
   \phi  \\
   \psi \\
  \end{array}\right) }. 
\end{equation}
\end{itemize}

\item {\it Reduction I.3. $a_2\neq 0, b_2=0, a_3=0, b_3=0$}

\begin{itemize}
\item Reduced variables 
\begin{equation}
x_1=x-\frac{1}{a_2}\int{\frac{A_1(y)}{y}\,dy}, \qquad x_2=t.
\end{equation}

\item Reduced spectral parameter

\begin{equation}
\lambda(y,t)=y^{\frac{1}{n}}\,\Lambda(x_2),
\end{equation}
 where $\Lambda(x_2)$ satisfies the {\it non-isospectral condition}
 \begin{equation}
\frac{d\Lambda(x_2)}{dx_2}=\frac{\Lambda(x_2)^{n+1}}{n}.
\end{equation}

\item Reduced fields 
\begin{align} 
   u(x,y,t)&=y^{-\frac{1}{2n}}\,U(x_1,x_2),\\
    \omega_{[1]}(x,y,t)&=\frac{A_1(y)}{a_2y}+\frac{\Omega_{[1]}(x_1,x_2)}{y},\\
    \omega_{[j]}(x,y,t)&=y^{\frac{j-1}{n}}\,\frac{\Omega_{[j]}(x_1,x_2)}{y},\quad \quad j=2,\dots, n,\\
    v_{[j]}(x,y,t)&=y^{\frac{2j-1}{2n}}\,\frac{V_{[j]}(x_1,x_2)}{y},\quad \quad j=1,\dots, n-1,\\
    v_{[n]}(x,y,t)&=y^{-\frac{1}{2n}}\,\left(\int{\frac{A_n(y,t)\,y^{\frac{1}{2n}}}{a_2\,y}\,dy}+V_{[n]}(x_1,x_2)\right).
 \end{align}

\item Reduced hierarchy 
\begin{align}
&(V_{[n]})_{x_1x_1x_1}- (V_{[n]})_{ x_1}-U_{x_2}=0,\\
& (V_{[j]})_{x_1x_1}-V_{[j]}+U\,\Omega_{[j+1]}=0,\quad \quad j=1,\dots, n-1,\\
&\left(\Omega_{[1]}\,U\right)_{x_1}-\frac{U}{2n}=0,\\
&(\Omega_{[j]})_{x_1}=U(V_{[j]})_{x_1},\quad \quad j=1,\dots, n.
\end{align}

\item The reduced spectral functions
\begin{align}
   \phi(x,y,t)&=e^{\int{\frac{\Gamma(y,x_2)}{a_2y}\,dy}}\,\Phi(x_1,x_2),\\
    \psi(x,y,t)&=e^{\int{\frac{\Gamma(y,x_2)}{a_2y}\,dy}}\,\Psi(x_1,x_2),\\
    p(x,y,t)&=y^{-\frac{1}{n}}\,\left(\frac{A_1(y)\,\Lambda(x_2)^{n-1}}{a_2}+P(x_1,x_2)\right),\\
    q(x,y,t)&=y^{-\frac{1}{2n}}\,\left(\frac{1}{a_2}\int{\frac{A_n(y,t)\,y^{\frac{1}{2n}}}{y}\,dy}+Q(x_1,x_2)\right),\\
    P(x_1,x_2)&=\sum_{j=1}^n{\Lambda(x_2)^{n-j}\,\Omega_{[j]}(x_1,x_2)},\\
    Q(x_1,x_2)&=\sum_{j=1}^{n}{\Lambda(x_2)^{n-j}\,V_{[j]}(x_1,x_2)},
\end{align}
%
\noindent
where the function $\Gamma(y,x_2)$ is obtained through the following identification
\begin{equation}
\Gamma(y,x_2)=\gamma\left(y,t=x_2\right)
\end{equation}
that yields
\begin{equation}
\gamma_t=\Gamma_{x_2},\quad\quad
\gamma_y=\Gamma_y.
\end{equation}
Therefore, according to \eqref{gammacond}, the field $\Gamma(y,x_2)$ satisfies the equation
\begin{equation} \Gamma_{x_2}=y\Lambda^n\,\Gamma_y.\end{equation}

\item Reduced Lax pair
\begin{align}
  \begin{pmatrix}
 \Phi\\
 \Psi
\end{pmatrix}_{x_2}&=\Lambda P \begin{pmatrix}
 \Phi\\
 \Psi
\end{pmatrix}_{x_1}+\frac{i\sqrt{\Lambda}}{2}\begin{bmatrix}
 0&Q_{x_1}-Q\\
 Q_{x_1}+Q& 0
\end{bmatrix}_{x_1}\begin{pmatrix}
 \Phi\\
 \Psi
\end{pmatrix}
\end{align}
and 
\begin{align}
  \begin{pmatrix}
 \Phi\\
 \Psi
\end{pmatrix}_{x_1}&=\frac{1}{2}\begin{bmatrix}
 -1& i\sqrt{\Lambda}\,U\\
 i\sqrt{\Lambda}\,U& 1
\end{bmatrix}\begin{pmatrix}
 \Phi\\
 \Psi
\end{pmatrix}.
\end{align}

\end{itemize}

\newpage
\item {\it Reduction II.1. $a_2=0, b_2 \neq 0, a_3\neq 0, b_3=0$}

\begin{itemize}
\item Reduced variables 
\begin{equation}
x_1=x-\frac{1}{b_2}\int A_1(y)\,dy, \qquad x_2=\frac{a_3\, t}{b_2}\,e^{\frac{-a_3y}{b_2}}.
\end{equation}
\item Reduced fields
\begin{align}
 \hspace*{-3em}u(x,y,t)&=e^{\frac{a_3y}{2n\,b_2}}\,U(x_1,x_2), \\
    \hspace*{-3em} \omega_{[1]}(x,y,t)&=\frac{A_1(y)}{b_2}+\frac{a_3}{b_2}\,\Omega_{[1]}(x_1,x_2), \\
  \hspace*{-3em}   \omega_{[j]}(x,y,t)&=\frac{a_3}{b_2}\,e^{\frac{-a_3(j-1)y}{n\,b_2}}\,\Omega_{[j]}(x_1,x_2),\quad \quad j=2,\dots, n,\\
  \hspace*{-3em}   v_{[j]}(x,y,t)&=\frac{a_3}{b_2}\,e^{\frac{-a_3(2j-1)y}{2n\,b_2}}\,V_{[j]}(x_1,x_2),\quad \quad j=1,\dots, n-1, \\
  \hspace*{-3em}   v_{[n]}(x,y,t)&=e^{\frac{-a_3(2n-1)y}{2n\,b_2}}\left(\int   \frac{A_n(y,t)\,e^{\frac{a_3(2n-1)y}{2n\,b_2}}}{b_2}\,dy+\frac{a_3}{b_2}\,V_{[n]}(x_1,x_2)\right).
 \end{align}
\item Reduced hierarchy 
\begin{align}
& (V_{[n]})_{x_1x_1x_1}- (V_{[n]})_{ x_1}-U_{x_2}=0,\\
& (V_{[j]})_{x_1x_1}-V_{[j]}+U\,\Omega_{[j+1]}=0,\quad \quad j=1,\dots, n-1,\\
& \left(\Omega_{[1]}\,U\right)_{x_1}+\frac{U}{2n}-x_2U_{x_2}=0, \\
& (\Omega_{[j]})_{x_1}=U(V_{[j]})_{x_1},\quad \quad  j=1,\dots, n.\end{align}

\item Reduced spectral functions 
\begin{align}
   \phi(x,y,t)&=e^{\int\frac{\Gamma(y,x_2)}{b_2}\,dy}\,\Phi(x_1,x_2),\\
    \psi(x,y,t)&=e^{\int\frac{\Gamma(y,x_2)}{b_2}\,dy}\,\Psi(x_1,x_2),\\
    p(x,y,t)&=e^{\frac{-a_3(n-1)y}{n\,b_2}}\,\left(\frac{A_1(y)\,\Lambda(x_2)^{n-1}}{b_2}+\frac{a_3}{b_2}\,P(x_1,x_2)\right),\\
    q(x,y,t)&=e^{\frac{-a_3(2n-1)y}{2n\,b_2}}\left[\int \frac{A_n(y,t)\,e^{\frac{a_3(2n-1)y}{2n\,b_2}}}{b_2}\,dy
\right.\nonumber
\\
&\left.{+}\frac{a_3}{b_2}\,Q(x_1,x_2)\right],\\
    P(x_1,x_2)&=\sum_{j=1}^n{\Lambda(x_2)^{n-j}\,\Omega_{[j]}(x_1,x_2)},\\
    Q(x_1,x_2)&=\sum_{j=1}^{n}{\Lambda(x_2)^{n-j}\,V_{[j]}(x_1,x_2)},
\end{align}
where the function $\Gamma(y,x_2)$ is obtained through the following identification
\begin{equation}
\Gamma(y,x_2)=\gamma\left(y,t=\frac{b_2}{a_3}\,e^{\frac{a_3}{b_2}}\right)
\end{equation}
that yields
\begin{equation}
\gamma_t=\frac{a_3}{b_2}\,e^{\frac{-a_3y}{b_2}}\,\Gamma_{x_2},\quad\quad
\gamma_y=\Gamma_y-\frac{a_3}{b_2}x_2\,\Gamma_{x_2}.
\end{equation}
Therefore, according to \eqref{gammacond}, $\Gamma(y,x_2)$ satisfies the equation
\begin{equation} a_3(1+x_2\Lambda^n)\Gamma_{x_2}=b_2\Lambda^n\,\Gamma_y.\end{equation}
\item Reduced spectral parameter 
\begin{equation}
\lambda(y,t)=e^{\frac{-a_3y}{n\,b_2}}\,\Lambda(x_2),
\end{equation} 
where $\Lambda(x_2)$ satisfies the {\it non-isospectral condition}
 \begin{equation}\frac{d\Lambda(x_2)}{dx_2}=-\frac{\Lambda(x_2)^{n+1}}{n\left(1+x_2\Lambda(x_2)^n\right)}.
\end{equation}
\item Reduced Lax pair
\begin{multline}
  (1+x_2\Lambda^n)\left(\begin{array}{c}
 \Phi\\
 \Psi
\end{array}\right)_{x_2}=\Lambda P \left(\begin{array}{c}
 \Phi\\
 \Psi
\end{array}\right)_{x_1}
\\
+\frac{i\sqrt{\Lambda}}{2}\left[\begin{array}{cc}
 0&Q_{x_1}-Q\\
 Q_{x_1}+Q& 0
\end{array}\right]_{x_1}\left(\begin{array}{c}
 \Phi\\
 \Psi
\end{array}\right)
\end{multline}
and
\begin{align}
  \left(\begin{array}{c}
 \Phi\\
 \Psi
\end{array}\right)_{x_1}&=\frac{1}{2}\left[\begin{array}{cc}
 -1& i\sqrt{\Lambda}\,U\\
 i\sqrt{\Lambda}\,U& 1
\end{array}\right]\left(\begin{array}{c}
 \Phi\\
 \Psi
\end{array}\right).
\end{align}

\end{itemize}

\item {\it Reduction II.2. $a_2=0, b_2 \neq 0, a_3=0, b_3\neq 0$.}

\begin{itemize}
\item Reduced variables
\begin{equation}
x_1=x-\int{\frac{A_1(y)}{b_2}\,dy}, \qquad x_2=y-\frac{b_2}{b_3}t.
\end{equation}
\item Reduced spectral parameter  
\begin{equation}
\lambda(y,t)=\left(\frac{b_2}{b_3}\right)^{\frac{1}{n}}\Lambda(x_2),\end{equation}  where $\Lambda(x_2)$ satisfies the {\it isospectral condition}
\begin{equation}
\frac{d\Lambda(x_2)}{dx_2}=0.\end{equation} 

\item Reduced fields 
\begin{align} 
   u(x,y,t)&=\left(\frac{b_3}{b_2}\right)^{\frac{1}{2n}}\,U(x_1,x_2),\\
    \omega_{[1]}(x,y,t)&=\frac{A_1(y)}{b_2}+\Omega_{[1]}(x_1,x_2),\\
    \omega_{[j]}(x,y,t)&=\left(\frac{b_2}{b_3}\right)^{\frac{j-1}{n}}\,\Omega_{[j]}(x_1,x_2),\quad \quad j=2,\dots, n, \\
    v_{[j]}(x,y,t)&=\left(\frac{b_2}{b_3}\right)^{\frac{2j-1}{2n}}\,V_{[j]}(x_1,x_2),\quad \quad j=1,\dots, n-1,\\
    v_{[n]}(x,y,t)&=\frac{1}{b_2}\int A_n(y,t)\,dy+\left(\frac{b_2}{b_3}\right)^{\frac{2n-1}{2n}}\,V_{[n]}(x_1,x_2).
\end{align}

\item  Reduced hierarchy 
\begin{align}
& (V_{[n]})_{x_1x_1x_1}- (V_{[n]})_{x_1}+U_{x_2}=0,\\
& (V_{[j]})_{x_1x_1}-V_{[j]}+U\,\Omega_{[j+1]}=0,\quad \quad j=1,\dots, n-1,\\
& \left(\Omega_{[1]}\,U\right)_{x_1}+U_{x_2}=0,\\
&  (\Omega_{[j]})_{x_1}=U(V_{[j]})_{x_1},\quad \quad j=1,\dots, n.
\end{align}

\item Reduced spectral functions
\begin{align}
    \phi(x,y,t)&=e^{\int\frac{\Gamma(y,x_2)}{b_2}\,dy}\,\Phi(x_1,x_2),\\
    \psi(x,y,t)&=e^{\int\frac{\Gamma(y,x_2)}{b_2}\,dy}\,\Psi(x_1,x_2), \\
    p(x,y,t)&=\left(\frac{b_2}{b_3}\right)^{\frac{n-1}{n}}\,\left(\frac{A_1(y)\Lambda(x_2)^{n-1}}{b_2}+P(x_1,x_2)\right),\\
    q(x,y,t)&=\frac{1}{b_2}\int A_n(y,t)\,dy+\left(\frac{b_2}{b_3}\right)^{\frac{2n-1}{2n}}Q(x_1,x_2),\\
    P(x_1,x_2)&=\sum_{j=1}^n{\Lambda(x_2)^{n-j}\,\Omega_{[j]}(x_1,x_2)},\\
    Q(x_1,x_2)&=\sum_{j=1}^{n}{\Lambda(x_2)^{n-j}\,V_{[j]}(x_1,x_2)},
\end{align}
%
\noindent
where the function $\Gamma(y,x_2)$ is obtained through the following identification
\begin{equation}
\Gamma(y,x_2)=\gamma\left(y,t=\frac{b_3}{b_2}(y-x_2)\right)
\end{equation}
that yields
\begin{equation}
\gamma_t=-\frac{b_2}{b_3}\,\Gamma_{x_2},\quad\quad\gamma_y=\Gamma_y+\Gamma_{x_2}.\end{equation} 
Thus, according to \eqref{gammacond}, $\Gamma(y,x_2)$ satisfies the equation
\begin{equation} 
(1+\Lambda^n)\Gamma_{x_2}=-\Lambda^n\,\Gamma_y.
\end{equation}

\item Reduced Lax pair
\begin{multline}
  (1+\Lambda^n)\left(\begin{array}{c}
 \Phi\\
 \Psi
\end{array}\right)_{x_2}=\Lambda P \left(\begin{array}{c}
 \Phi\\
 \Psi
\end{array}\right)_{x_1}
\\
+\frac{i\sqrt{\Lambda}}{2}\left[\begin{array}{cc}
 0&Q_{x_1}-Q\\
 Q_{x_1}+Q& 0
\end{array}\right]_{x_1}\left(\begin{array}{c}
 \Phi\\
 \Psi
\end{array}\right),
\end{multline}
and
\begin{align}
  \left(\begin{array}{c}
 \Phi\\
 \Psi
\end{array}\right)_{x_1}&=\frac{1}{2}\left[\begin{array}{cc}
 -1& i\sqrt{\Lambda}\,U\\
 i\sqrt{\Lambda}\,U& 1
\end{array}\right]\left(\begin{array}{c}
 \Phi\\
 \Psi
\end{array}\right).
\end{align}
\end{itemize}

\item {\it Reduction II.3. $a_2=0, b_2 \neq 0, a_3=0, b_3=0$}

\begin{itemize}
\item Reduced variables
\begin{equation}
x_1=x-\int{\frac{A_1(y)}{b_2}\,dy}, \qquad x_2=t.\end{equation} 

\item Reduced spectral parameter
\begin{equation}
\lambda(y,t)=\Lambda(x_2),\end{equation}  where $\Lambda(x_2)$ satisfies the {\it isospectral condition}
\begin{equation}
\frac{d\Lambda(x_2)}{dx_2}=0.                             \end{equation} 

\item Reduced fields 
\begin{align} 
  u(x,y,t)&=U(x_1,x_2),\\
    \omega_{[1]}(x,y,t)&=\frac{A_1(y)}{b_2}+\Omega_{[1]}(x_1,x_2),\\
    \omega_{[j]}(x,y,t)&=\Omega_{[j]}(x_1,x_2),\quad \quad j=2,\dots, n,\\
    v_{[j]}(x,y,t)&=V_{[j]}(x_1,x_2),\quad \quad j=1,\dots, n-1,\\
    v_{[n]}(x,y,t)&=\frac{1}{b_2}\int A_n(y,t)\,dy+V_{[n]}(x_1,x_2).
\end{align}

\item Reduced hierarchy 
\begin{align}
& (V_{[n]})_{x_1x_1x_1}- (V_{[n]})_{ x_1}-U_{x_2}=0,\\
& (V_{[j]})_{x_1x_1}-V_{[j]}+U\,\Omega_{[j+1]}=0,\quad \quad j=1,\dots, n-1,\\
& \left(\Omega_{[1]}\,U\right)_{x_1}=0, \\
& (\Omega_{[j]})_{x_1}=U(V_{[j]})_{x_1},\quad \quad  j=1,\dots, n.
\end{align}

\item Reduced spectral functions
\begin{align}
    \phi(x,y,t)&=e^{\int\frac{\Gamma(y,x_2)}{b_2}\,dy}\,\Phi(x_1,x_2), \\
    \psi(x,y,t)&=e^{\int\frac{\Gamma(y,x_2)}{b_2}\,dy}\,\Psi(x_1,x_2),\\
    p(x,y,t)&=\frac{A_1(y)\Lambda(x_2)^{n-1}}{b_2}+P(x_1,x_2),\\
    q(x,y,t)&=\frac{1}{b_2}\int A_n(y,t)\,dy+Q(x_1,x_2),\\
    P(x_1,x_2)&=\sum_{j=1}^n{\Lambda(x_2)^{n-j}\,\Omega_{[j]}(x_1,x_2)},\\
    Q(x_1,x_2)&=\sum_{j=1}^{n}{\Lambda(x_2)^{n-j}\,V_{[j]}(x_1,x_2)},
\end{align}
where the function $\Gamma(y,x_2)$ is obtained through the following identification 
\begin{equation}
\Gamma(y,x_2)=\gamma\left(y,t=x_2\right)
\end{equation}
that yields
\begin{equation}
\gamma_t=\Gamma_{x_2},\quad\quad
\gamma_y=\Gamma_y.\end{equation} 
Therefore, according to (\ref{gammacond}), $\Gamma(y,x_2)$ satisfies the equation
\begin{equation}
\Gamma_{x_2}=\Lambda^n\,\Gamma_y.
\end{equation}

\item Reduced Lax pair
\begin{align}
\hspace{-2em} \left(\begin{array}{c}
 \Phi\\
 \Psi
\end{array}\right)_{x_2}&=\Lambda P \left(\begin{array}{c}
 \Phi\\
 \Psi
\end{array}\right)_{x_1}+\frac{i\sqrt{\Lambda}}{2}\left[\begin{array}{cc}
 0&Q_{x_1}-Q\\
 Q_{x_1}+Q& 0
\end{array}\right]_{x_1}\left(\begin{array}{c}
 \Phi\\
 \Psi
\end{array}\right),
\end{align}
and 
\begin{align}
  \left(\begin{array}{c}
 \Phi\\
 \Psi
\end{array}\right)_{x_1}&=\frac{1}{2}\left[\begin{array}{cc}
 -1& i\sqrt{\Lambda}\,U\\
 i\sqrt{\Lambda}\,U& 1
\end{array}\right]\left(\begin{array}{c}
 \Phi\\
 \Psi
\end{array}\right).
\end{align}

\end{itemize}

\item {\it Reduction III.1. $a_2=0, b_2=0, a_3\neq 0, b_3=0$}

\begin{itemize}
\item Reduced variables
\begin{equation}
x_1=x-\frac{A_1(y)}{a_3}\,\ln t, \qquad x_2=y.\end{equation} 

\item Reduced spectral parameter
\begin{equation}
\lambda(y,t)=t^{\frac{-1}{n}}\,\Lambda(x_2),\end{equation}  where $\Lambda(x_2)$ satisfies the \textit{nonisospectral condition}
\begin{equation}
n\frac{d\Lambda(x_2)}{dx_2}+\Lambda(x_2)^{1-n}=0.\end{equation} 

\item Reduced fields
\begin{align}
   u(x,y,t)&=t^{\frac{1}{2n}}\,U(x_1,x_2), \\
    \omega_{[1]}(x,y,t)&=\frac{\ln(t)}{a_3}\,\frac{d\,A_1(y)}{dy}+\Omega_{[1]}(x_1,x_2),\\
    \omega_{[j]}(x,y,t)&=t^{\frac{1-j}{n}}\,\Omega_{[j]}(x_1,x_2),\quad \quad j=2,\dots, n,\\
    v_{[j]}(x,y,t)&=t^{\frac{1-2j}{2n}}\,V_{[j]}(x_1,x_2),\quad \quad j=1,\dots, n-1,\\
    v_{[n]}(x,y,t)&=t^{\frac{1-2n}{2n}}\,\left(\frac{1}{a_3}\int t^{\frac{-1}{2n}}\,A_n(y,t)\,dt+V_{[n]}(x_1,x_2)\right).
\end{align}

\item Reduced hierarchy
\begin{align}
& (V_{[n]})_{x_1x_1x_1}- (V_{[n]})_{ x_1}-\frac{U}{2n}+\frac{\hat A_1(x_2)}{a_3}U_{x_1}=0,\\
& (V_{[j]})_{x_1x_1}-V_{[j]}+U\,\Omega_{[j+1]}=0,\quad \quad j=1,\dots, n-1,\\
& \left(\Omega_{[1]}\,U\right)_{x_1}+U_{x_2}=0, \\
&  (\Omega_{[j]})_{x_1}=U(V_{[j]})_{x_1},\quad \quad  j=1,\dots, n.
\end{align}

\item Reduced spectral functions
\begin{align}
   \phi(x,y,t)&=e^{\int\frac{\Gamma(t,x_2)}{a_3t}\,dt}\,\Phi(x_1,x_2), \\
    \psi(x,y,t)&=e^{\int\frac{\Gamma(t,x_2)}{a_3t}\,dt}\,\Psi(x_1,x_2), \\
   p(x,y,t)&=t^{\frac{1-n}{n}}\,\left(\frac{\ln(t)\,\Lambda(x_2)^{n-1}}{a_3}\,\frac{d\,A_1(y)}{dy}+P(x_1,x_2)\right), \\
    q(x,y,t)&=t^{\frac{1-2n}{2n}}\,\left(\frac{1}{a_3}\int t^{\frac{-1}{2n}}\,A_n(y,t)\,dt+Q(x_1,x_2)\right),\\
    P(x_1,x_2)&=\sum_{j=1}^n{\Lambda(x_2)^{n-j}\,\Omega_{[j]}(x_1,x_2)},\\
    Q(x_1,x_2)&=\sum_{j=1}^{n}{\Lambda(x_2)^{n-j}\,V_{[j]}(x_1,x_2)},
\end{align}
and the function $\Gamma(y,x_2)$ is obtained through the following identification 
\begin{equation}
\Gamma(t,x_2)=\gamma\left(y=x_2,t\right)
\end{equation}
that yields
\begin{equation}
\gamma_t=\Gamma_{t},\quad\quad
\gamma_y=\Gamma_{x_2}.
\end{equation} 
Therefore, according to \eqref{gammacond}, $\Gamma(y,x_2)$ satisfies the equation
\begin{equation}
\Lambda^n\,\Gamma_{x_2}=t\,\Gamma_t.
\end{equation}

\item Reduced Lax pair
\begin{multline}
-\Lambda^n\left(\begin{array}{c}
 \Phi\\
 \Psi
\end{array}\right)_{x_2}=\left(\Lambda P +\frac{\hat A_1}{a_3}\right)\left(\begin{array}{c}
 \Phi\\
 \Psi
\end{array}\right)_{x_1}
\\
+\frac{i\sqrt{\Lambda}}{2}\left[\begin{array}{cc}
 0&Q_{x_1}-Q\\
 Q_{x_1}+Q& 0
\end{array}\right]_{x_1}\left(\begin{array}{c}
 \Phi\\
 \Psi
\end{array}\right),
\end{multline}
and
\begin{align}
\left(\begin{array}{c}
 \Phi\\
 \Psi
\end{array}\right)_{x_1}&=\frac{1}{2}\left[\begin{array}{cc}
 -1& i\sqrt{\Lambda}\,U\\
 i\sqrt{\Lambda}\,U& 1
\end{array}\right]\left(\begin{array}{c}
 \Phi\\
 \Psi
\end{array}\right).
\end{align}
\end{itemize}
and $\hat A_1=\hat A_1(x_2)=A_1(y=y(x_2)).$

\newpage

\item {\it Reduction III.2. $a_2=0, b_2=0, a_3=0, b_3\neq 0$.}

\begin{itemize}
\item Reduced variables
\begin{equation}
x_1=x-\frac{A_1(y)\,t}{b_3}, \qquad x_2=\frac{1}{b_3}\,\int A_1(y)\,dy.
\end{equation} 

\item Reduced spectral parameter
\begin{equation}
\lambda(y,t)=\Lambda(x_2),\end{equation}  where $\Lambda(x_2)$ satisfies the {\it isospectral condition}
\begin{equation}
\frac{d\Lambda(x_2)}{dx_2}=0.
\end{equation} 

\item Reduced fields
\begin{align}
   u(x,y,t)&=U(x_1,x_2),\\
    \omega_{[1]}(x,y,t)&=\frac{t}{b_3}\frac{dA_1(y)}{dy}+\frac{A_1(y)}{b_3}\,\Omega_{[1]}(x_1,x_2),\\
    \omega_{[j]}(x,y,t)&=\frac{A_1(y)}{b_3}\,\Omega_{[j]}(x_1,x_2),\quad \quad j=2,\dots, n,\\
    v_{[j]}(x,y,t)&=\frac{A_1(y)}{b_3}\,V_{[j]}(x_1,x_2),\quad \quad j=1,\dots, n-1,\\
    v_{[n]}(x,y,t)&=\frac{1}{b_3}\int A_n(y,t)\,dt+\frac{A_1(y)}{b_3}\,V_{[n]}(x_1,x_2),
\end{align}

\item Reduced hierarchy
\begin{align}
& (V_{[n]})_{x_1x_1x_1}- (V_{[n]})_{x_1}+U_{x_1}=0,\\
&  (V_{[j]})_{x_1x_1}-V_{[j]}+U\,\Omega_{[j+1]}=0,\quad \quad  j=1,\dots, n-1,\\
& \left(\Omega_{[1]}\,U\right)_{x_1}+U_{x_2}=0, \\
&  (\Omega_{[j]})_{x_1}=U(V_{[j]})_{x_1},\quad \quad j=1,\dots, n.
\end{align}
\item Reduced spectral functions
\begin{align}
    \phi(x,y,t)&=e^{\int\frac{\Gamma(t,x_2)}{b_3}\,dt}\,\Phi(x_1,x_2),\\
    \psi(x,y,t)&=e^{\int\frac{\Gamma(t,x_2)}{b_3}\,dt}\,\Psi(x_1,x_2), \\
    p(x,y,t)&=\frac{t}{b_3}\,\frac{dA_1(y)}{dy}\,\Lambda^{n-1}+\frac{A_1(y)}{b_3}\,P(x_1,x_2),\\
    q(x,y,t)&=\frac{1}{b_3}\int A_n(y,t)\,dt+\frac{A_1(y)}{b_3}\,Q(x_1,x_2).\\
    P(x_1,x_2)&=\sum_{j=1}^n{\Lambda(x_2)^{n-j}\,\Omega_{[j]}(x_1,x_2)},\\
    Q(x_1,x_2)&=\sum_{j=1}^{n}{\Lambda(x_2)^{n-j}\,V_{[j]}(x_1,x_2)},
\end{align}
where the function $\Gamma(y,x_2)$ is obtained through the following identification
\begin{equation}
\Gamma(t,x_2)=\gamma\left(y=y(x_2),t\right)
\end{equation}
that yields
\begin{equation}
\gamma_t=\Gamma_{t},\quad\quad\gamma_y=\frac{A_1}{a_3}\Gamma_{x_2}.
\end{equation} 
Therefore, according to \eqref{gammacond}, $\Gamma(y,x_2)$ satisfies the equation
\begin{equation}
\Lambda^n\,\Gamma_{x_2}=\frac{b_3}{\hat A_1(x_2)}\Gamma_t,\quad \hat A_1(x_2)=A_1(y=y(x_2)).
\end{equation}

\item Reduced spectral problem
\begin{multline}
  -\Lambda^n\left(\begin{array}{c}
 \Phi\\
 \Psi
\end{array}\right)_{x_2}=\left(\Lambda P +1\right)\left(\begin{array}{c}
 \Phi\\
 \Psi
\end{array}\right)_{x_1}
\\
+\frac{i\sqrt{\Lambda}}{2}\left[\begin{array}{cc}
 0&Q_{x_1}-Q\\
 Q_{x_1}+Q& 0
\end{array}\right]_{x_1}\left(\begin{array}{c}
 \Phi\\
 \Psi
\end{array}\right),
\end{multline}
and
\begin{align}
\left(\begin{array}{c}
 \Phi\\
 \Psi
\end{array}\right)_{x_1}=\frac{1}{2}\left[\begin{array}{cc}
 -1& i\sqrt{\Lambda}\,U\\
 i\sqrt{\Lambda}\,U& 1
\end{array}\right]\left(\begin{array}{c}
 \Phi\\
 \Psi
\end{array}\right).
\end{align}
\end{itemize}

\end{itemize}

\chapter{Reciprocal transformations}\markboth{Reciprocal transformations}{Chapter 5}
\label{Chap:RecipTrans}

\section{What is a reciprocal transformation?}
\setcounter{equation}{0}
\setcounter{theorem}{0}
\setcounter{example}{0}
\bigskip

This chapter is dedicated to the study of reciprocal transformations. Reciprocal transformations are a particular type of hodograph transformations 
in which the use of conserved quantities is fundamental \cite{Rogers11,rogers4,rogers5,RogersCarillo,rogers3}.

On a first approximation, {\it hodograph transformations} are transformations involving the interchange of dependent and independent variables \cite{ClarkFokasAblo,EstPrada2}. 
When the variables are switched, the space of independent variables is called the {\it reciprocal space}. In particular case of two variables, we refer to it as the {\it reciprocal plane}.
As a physical interpretation, whereas the independent variables play the role of positions in the reciprocal space, this number is increased by turning
certain fields or dependent variables into independent variables and viceversa \cite{ConteMusette}.
For example, in the case of evolution equations in Fluid Dynamics, usually fields that represent the height of the wave or its velocity, are turned into a new set of independent variables.

{\it Reciprocal transformations} share this definition but require the employment of conservative forms and their properties, as we shall see in forthcoming paragraphs \cite{estevez09,estevez51,EstSar,EstSar2}.

One of the biggest advantages of dealing with hodograph and reciprocal transformations is that many of the equations reported integrable in the bibliography of differential equations, which are considered seemingly different from one another, happen to be related via reciprocal transformations.
If this were the case, two apparently unrelated equations, even two complete hierarchies of PDEs that are linked via reciprocal transformation, are tantamount versions of an unique
problem. In this way, hodograph and reciprocal transformations give rise to a procedure of relating allegedly new equations
to the rest of their equivalent integrable sisters.
\newpage
Let us consider a manifold $N_{\mathbb{R}^n}$ explained in Chapter \ref{Chap:LieSymm}. We call a $p$-order {\it quasilinear PDE} to a
a submanifold of ${\rm J}^p(\mathbb{R}^n,\mathbb{R}^k)$ as
 \begin{equation}
(u_j)_{x_i}=g(u)(u_j)_{x_1^{j_1},x_2^{j_2},x_3^{j_3},\dots, x_n^{j_n}}+f\left(u_j,(u_j)_{x_{i}},\dots,(u_j)_{x_{i_1}^{j_1},x_{i_2}^{j_2},x_{i_3}^{j_3},\dots, x_{i_n}^{j_n}}\right), 
\end{equation}
with $\frac{\partial g}{\partial u}\neq 0$, the subindices $i_1,\dots,i_n$ can take any value between $1,\dots,n$, the sum  $j_1+j_2+j_3+\dots+j_n\leq p$ and  $i=1,\dots,n$, $j=1,\dots,k$, where $u$ denotes $u=(u_1,\dots,u_k)$ and $x=(x_1,\dots,x_n)$.
In \cite{ClarkFokasAblo}, the most general quasilinear equation above is found to be mapped via hodograph transformation into linearizable equations, that is, equations
solvable in terms of either a linear PDE or a linear integral equation.
Indeed, Gardner associated the solution of the KdV equation with the time-independent Schr\"odinger equation and showed, using ideas of the direct
and inverse scattering, that the Cauchy problem for the KdV equation could be solved in terms of a linear equation \cite{Gard1,GGKM2}.
This novelty is today the well-known IST, leading to numerous solutions in branches of water waves, stratified fluids, Plasma Physics, statistical Mechanics
and Quantum Field theory \cite{AbloClark,AbloKruskalSegur,ARS2,AS}.
In this way, hodograph transformations facilitate the solvability of certain PDEs that can either be treated by the IST or by
transformation to a linear PDE, (those said to be linearizable). In \cite{DDH,h00}, hodograph transformations were proven to be a useful instrument to transform equations
with peakon solutions into equations that pass the Painlev\'e test.

During the past decades, more attention has been laid upon reciprocal transformations due to their manageability and quasi-algorithmical way of approach \cite{estevez09,EstSar,EstSar2}. The increasing number of works
in the Physics literature manifests their importance \cite{Abenda,AbendaGrava2,estevez09,EstSar,EstSar2,Ferapontov1,FerapontovRogersSchief,h00,Rogers11}.
For dealing with reciprocal transformations, the concept of conservation laws is fundamental. Conservation laws are characteristic of real physical processes.

%
By {\it conservation law} we understand an expression on of the form
\begin{equation}
\frac{\partial \psi_1}{\partial x_{i_1}}+\frac{\partial \psi_2}{\partial x_{i_2}}=0, 
\end{equation}
for certain two values of the indices in between $1 \leq i_1,i_2\leq n$ and two scalar functions $\psi_1,\psi_2 \in C^{\infty}({\rm J}^pN_{\mathbb{R}^n})$. One can find a more formal definition
of conservation law in \cite{BCDKKSTVV}.

A second significant role of reciprocal transformations is their utility in the identification of hierarchies
which do not pass the Painlev\'e test and in which the SMM is not applicable \cite{estevez51,EstPrada2}.
Precisely, our motivation for the study of reciprocal transformations roots in the study of the Camassa-Holm hierarchy in $2+1$ dimensions \cite{EstLejaSar}. Such hierarchy, 
has been known to be integrable for some time and has an associated linear problem. Nevertheless, in its most commonly expressed form, given in \eqref{NCHHaf2} in Chapter \ref{Chap:LieSymm}, it is not integrable in the sense of the Painlev\'e property, nor the SMM is constructive.

Our conjecture is that if an equation is integrable,
there must be a transformation that will let us turn the initial equation into a new one in which the Painlev\'e test is successful and the SMM can be applied. 
Hence, a Lax pair could be derived and other methods mentioned in the introduction could be worked out.
Therefore, discarding any pathology, one should be able to write down a transformation that brings the equation into a suitable form, in which the Painlev\'e test is applicable.

From now on, this chapter will be dedicated to reciprocal transformations exclusively. 
Reciprocal transformations have a long story. Alongside with the IST, the two procedures gave rise to the discovery of other integrable nonlinear evolution equations similar to the KdV equations. 
For example, Zakharov and Shabat \cite{ZS}
presented the now famous nonlinear Sch\"odinger (NLS) equation, which presents an infinite number of integrals of motion and possesses n-soliton solutions with
purely elastic interaction. A third important equation that arose from the IST, is the modified Korteweg de Vries (mKdV) equation, found by Wadati \cite{wadati1} in 1972,
which also attracted a lot of attention in fields of Plasma Physics, nonlinear Optics or Superconductivity.
In 1971, the study of the blow-up stability in the theory of solitons lead to the full understanding of the special properties of the KdV \cite{KZ}.

In 1928, the invariance of nonlinear gas dynamics, magnetogas dynamics and general hydrodynamic systems under reciprocal transformations
was extensively studied \cite{Ferapontov1,RogersKingstonShadwick}. Stationary and moving boundary problems in soil mechanics and nonlinear heat conduction
have likewise been subjects of much research \cite{FerapontovRogersSchief,Rogers11}.
  
Applications of the reciprocal transformation in continuum mechanics are to be found in the monographs by Rogers and Shadwick \cite{rogers1}.
These transformations have also played an important role in the soliton theory and between hierarchies of PDEs \cite{DDH,h00}. Indeed, the invariance of certain integrable
hierarchies under reciprocal transformations induces auto-B\"acklund transformations \cite{EstSar,EstSar2,rogers6,RogersCarillo,rogers3}.

The most representative and differing properties of reciprocal transformations from the hodograph ones, are: they
map conservation laws to conservation laws and diagonalizable systems to diagolizable systems, but act nontrivially on metrics and
on Hamiltonian structures. For instance, the flatness property or the locality of the Hamiltonian structure are not preserved, in general \cite{AbendaGrava1,AbendaGrava2}.

Finding a proper reciprocal transformation is usually a very complicated task. Notwithstanding, in the case that concerns us, Fluid Mechanics, etc.,
a change of this type is usually reliable, specially for systems of hydrodynamic type with time evolution defined on manifolds $N_{\mathbb{R}^n}$ 
as a submanifold on ${\rm J}^p(\mathbb{R}^{n+1},\mathbb{R}^{2k})$
\begin{equation}
(u_{j})_t=\sum_{l=1}^k v^{j}_{l}(u)(u_l)_{x_i}, \quad \forall 1\leq i\leq n,\quad j=1,\dots,k
\end{equation}
and $v^{j}_{l}(u)$ are $C^{\infty}({\rm J}^pN_{\mathbb{R}^n})$, appearing Gas Dynamics, Hydrodynamics, Chemical Kinetics, Differential Geometry and Topological
Field Theory \cite{DubroNovi,Tsarev1}.

Lately, we have been focusing on differing versions of reciprocal transformations. Specifically, on the composition of two or more reciprocal transformations or 
the composition of a reciprocal transformation with a transformation of different nature. 
For example, the composition of a {\it Miura transformation} \cite{AbloKruskalSegur,Sakovich} and a reciprocal transformation gives rise to the so called {\it Miura-reciprocal transformations}
that helps us relate two different hierarchies, which is the purpose of this chapter.
In particular, we show the example of the $n$-component Camassa--Holm hierarchy CHH($2+1$) which does not possess the PP in the variables expressed in \eqref{NCHHaf3}), in Chapter \ref{Chap:LieSymm}. Nevertheless, it can be proven \cite{EstSar,EstSar2} that there exists a reciprocal transformation
that turns the CHH($2+1$) into $n$ copies of the Calogero-Bogoyavlenskii-Schiff (CBS) equation \cite{cbs,calogero,KP1}. This latter equation possesses the
PP and the SMM can be applied to obtain its LP and other relevant properties.
On the other hand, the $n$-component Qiao hierarchy mCHH($2+1$) \eqref{mNCHHaf2}), given in Chapter \ref{Chap:LieSymm}, has been reported as an integrable hierarchy
for which the PP is not applicable, nor constructive \cite{estevez51,Qiao2,Qiao3,Qiao1}.
Notwithstanding, one can prove \cite{EstSar,EstSar2} that there exists a reciprocal transformation which turns mCHH($2+1$) into $n$ copies
of the modified Calogero-Bogoyavlenskii-Schiff (mCBS) equation, which is known to have the PP, etc \cite{EstPrada1}.

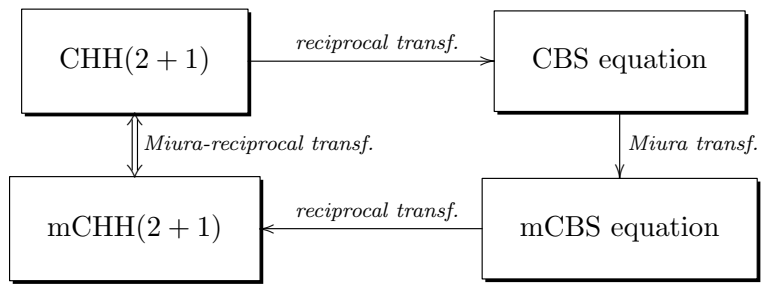
\begin{figure}
\begin{center}
$
\xymatrix{*+<1cm>[F-,]{\text{CHH($2+1$)}} \ar[rrr]^{\textit{reciprocal transf.}} \ar@2{<->}[d]^{\textit{Miura-reciprocal transf.}} &  &  &*+<1cm>[F-,]{\text{CBS equation}}\ar[d]^{\textit{Miura transf.}}\\  *+<1cm>[F-,]{\text{mCHH($2+1$)}} &  &  &*+<1cm>[F-,]{\text{mCBS equation}}\ar[lll]_(0.5){\textit{reciprocal transf.}}}
$
\end{center}
\caption{Miura-reciprocal transformation.}
\label{Fig2}
\end{figure}

Summarizing, CHH($2+1$) and mCHH($2+1$) are related to the CBS and mCBS by reciprocal transformations, correspondingly.
Aside from this property, there exists a Miura transformation \cite{estevez23,estevez25} relating the CBS and the mCBS equations. One then wonders if mCHH($2+1$) is related
to CHH($2+1$) in any way. 
It seems clear that the relationship between mCHH($2+1$) and CHH($2+1$) necessarily includes a composition of a Miura and a reciprocal transformation. 
\Cref{Fig2} clarifies this idea.

\newpage
\noindent
In this way, the plan of this chapter goes as follows

\bigskip

{\bf Section 1: Quasi-algorithmical construction of reciprocal transformations:} We introduce the theoretical framework of a reciprocal transformation
on a general manifold of type $N_{\mathbb{R}^n}$ and give a quasi-algorithmical method for the construction of such a transformation.  
It is our intention to give some (generally common) steps in the construction of the transformation, hence the name ``quasi-algorithmical".
Nevertheless, the study of this theory will become much simpler by applying it to low dimensional cases, that are the cases that concern us.

{\bf Section 2: CHH(2+1) versus CBS:} 
We explicitly construct the reciprocal transformation that turns the $n$-component CHH($2+1$) into $n$ copies of the CBS equation.

{\bf Section 3: mCHH(2+1) versus mCBS:}  We explicitly construct the reciprocal transformation
 that turns the $n$-component mCHH($2+1$) into $n$ copies of the mCBS equation. Here, given the Lax pair for the 
mCBS equation, we will invert the reciprocal transformation and we will achieve a Lax pair for the former $n$-component mCHH($2+1$). Here is where
the role of reciprocal transformations as a tool for deriving Lax pairs will be explicit.

{\bf Section 4: Reciprocal-Miura transformation between CHH(2+1)\\ and mCHH(2+1):} Here we give a Miura transformation between the CBS and mCBS equations. This 
Miura transformation has been well-known in the literature for some years. By means of a composition of a Miura and reciprocal transformations, it is possible
to link the CHH(2+1) and mCHH(2+1) and relate their dependent variables $U$ and $u$, correspondingly, as well as the triple of independent variables in which they are written
$(X,Y,T)$ and $(x,y,t)$, respectively.
We will take a closer look at two particular cases: one is the Qiao equation arosen from the first component of the hierarchy mCHH($2+1$), (that is when $n=1$) and $U$ is independent of $Y$ and
$u$ is independent of $y$. The second particular case is: the Camassa--Holm equation arising from the first-component of the hierarchy CHH($2+1$), ($n=1$) if $T=X$ and $t=x$.

\section{Quasi-algorithmical construction of reciprocal transformations}
\setcounter{equation}{0}
\setcounter{theorem}{0}
\setcounter{example}{0}


%
Let us consider the most general nonlinear system on ${\rm J}^p(\mathbb{R}^n,\mathbb{R}^k)$, 
and we propose a system of $q$ partial differential equations
\begin{equation}\label{genpdert}
\Psi^l=\Psi^l\left(x_{i},u_j,(u_j)_{x_{i_1}},(u_j)_{x_{i_1}^{j_1},x_{i_2}^{j_2}},\dots,(u_j)_{x_{i_1}^{j_1},x_{i_2}^{j_2},x_{i_3}^{j_3},\dots,x_{i_n}^{j_n}}\right),
\end{equation}
for all $l=1,\dots,q$ and $i=1,\dots,n$, $j=1,\dots,k,$ $j_1+\dots+j_n\leq p.$
Suppose that we know a number $n$ of conserved quantities that are expressible in the following manner
\begin{equation}
\begin{gathered}\label{conservedq}
A^{(j)}_{x_{i_1}}\left(x_{i_1},u_j,(u_j)_{x_{i_1}},(u_j)_{x_{i_1}^{j_1},x_{i_2}^{j_2}},\dots\right)=A^{(j')}_{x_{i_2}}\left(x_i,u_j,(u_j)_{x_{i_1}^{j_1}},(u_j)_{x_{i_1}^{j_1},x_{i_2}^{j_2}},\dots\right),\\
 x_{i_1}\neq x_{i_2},\qquad A^{(j)}\neq A^{(j')},\\
1\leq i_1,\dots,i_n\leq n,\qquad j_1+\dots+j_n\leq p, \qquad 1\leq j,j' \leq 2n.
\end{gathered}
\end{equation}
where $A^{(j)}_{x_{i_1}}=\partial A^{(j)}/\partial x_{i_1}$, $\smash{A^{(j')}_{x_{i_2}}=\partial A^{(j')}/\partial x_{i_2}}$ and $A^{(j)},A^{(j')}\in C^{\infty}({\rm J}^pN_{\mathbb{R}^n}),$
which in principle do not need to be equal.

If the number of equations in \eqref{conservedq} is equal to the number of indepent variables, then we propose a transformation for each of the coordinates
$\{x_1,\dots,x_n\}$ to a new set of coordinates $\{z_1,\dots,z_n\}$ as 
\begin{equation}\label{closed}
dz_{i}=A^{(j)}dx_{i_2}+A^{(j')}dx_{i_1}, \qquad \forall 1\leq i_1,i_2\leq n,\quad \forall 1\leq j,j' \leq 2n.
\end{equation}
such that if the property of closeness is satisfied, 
 $d^2z_{i}=0$ for all $z_{i}$, $i=1,\dots,n$ and we recover the conserved quantities \eqref{conservedq}.

Concerning examples of physical nature that we shall deal with, we will focus on the particular case in which only one equation with conserved quantities is used.
This implies that only one independent variable will be transformed.
The transformed independent variable is a particular one. Let us denote it by $x_{\hat i}$. We now search for a function $X(z_1,\dots,z_n)$ such that 
\begin{equation}\label{indintodep}
x_{\hat i}=X(z_{1},\dots,z_{n})
\end{equation}
is turned into a dependent variable.
Simultaneously, we use the conserved quantity to propose the 
 transformation
\begin{equation}
\begin{aligned}\label{rectransf}
&dz_{\hat i}=A^{(j)}dx_{i_2}+A^{(j')}dx_{i_1},\\
&dz_i=dx_i,
\end{aligned}\end{equation} 
$\forall 1\leq i_1\neq i_2\leq n$ and for a fixed value in between $1\leq \hat{i}\leq n,$
where we see that independent variables $x_i, \forall i\neq \hat{i}=1,\dots,n$ have been left untransformed but renamed as $z_i$.

Deriving relation \eqref{indintodep},
\begin{equation}\label{rectransf1}
dx_{\hat i}=\sum_{i=1}^n X_{z_i}dz_i,\quad X_{z_i}=\frac{\partial X}{\partial z_i}
\end{equation}
and by isolating $dz_{\hat i}$ in \eqref{rectransf1}, we have
\begin{equation}\label{rectransf2}
dz_{\hat i}=\frac{dx_{\hat i}}{X_{z_{\hat i}}}-\sum_{i\neq \hat{i}=1}^n \frac{X_{z_{i}}}{X_{z_{\hat i}}}dx_i,
\end{equation}
where we have used that $dz_i=dx_i$ for all $i\neq \hat{i}$ according to \eqref{rectransf}.

Here, by direct comparison of coefficients in \eqref{rectransf} and \eqref{rectransf2} and if we identify $z_{\hat i}$ with $z_{i_1}$, we have that
\begin{equation}
\begin{aligned}\label{rectransfa}
&A^{(j')}=\frac{1}{X_{z_{\hat i}}},\\
&A^{(j)}=-\frac{X_{z_{i}}}{X_{z_{\hat i}}}.
\end{aligned}\end{equation} 
\noindent
It is important that we show the pass to the tangent space. For example, in the case of first-order derivatives, we have
\begin{equation}
\begin{aligned}\label{rectransfts}
u_{x_{\hat i}}&=\frac{\partial u}{\partial z_{\hat i}}\frac{\partial z_{\hat i}}{\partial x_{\hat i}}+\sum_{i\neq \hat{i}=1}^n\frac{\partial u}{\partial z_i}\frac{\partial z_i}{\partial x_{\hat i}}=\frac{u_{z_{\hat i}}}{X_{z_{\hat i}}},\\
u_{x_i}&=\frac{\partial u}{\partial z_{\hat i}}\frac{\partial z_{\hat i}}{\partial x_i}+\sum_{i\neq \hat{i}=1}^n\frac{\partial u}{\partial z_i}\frac{\partial z_i}{\partial x_i}=-\frac{X_{z_i}}{X_{z_{\hat i}}}u_{z_{\hat i}}+u_{z_i}, \quad \qquad \forall i\neq \hat{i}
\end{aligned}
\end{equation} 
This process can be applied recursively for higher-order derivatives in the jet space $\text{J}^pN$.

In this way, using expressions in \eqref{rectransfa}, \eqref{rectransfts}, etc., we transform a system \eqref{genpdert} with initial variables $\{x_1,\dots,x_n\}$ into 
a new system in variables $\{z_{1},\dots,z_{n}\}$ and fields $u_j(z_{1},\dots,z_{n}),\, \forall j=1,\dots,k,$ depending on the new varibles.
From \eqref{rectransfa}, we can extract expressions for
\begin{equation}
z_i,u_j,(u_j)_{z_{i_1}},(u_j)_{z_{i_1}^{j_1},z_{i_2}^{j_2}},\dots,(u_j)_{z_{i_1}^{j_1},z_{i_2}^{j_2},\dots,z_{i_n}^{j_n}}\end{equation} 
for $j_1+\dots+j_n\leq p$, if possible, given the particular form of $A^{(j)}$, $A^{(j')} \in C^{\infty}({\rm J}^pN_{\mathbb{R}^n})$, in each case.
Bearing in mind expression \eqref{indintodep}, the transformation of the initial system \eqref{genpdert} will then read
\begin{equation}
\Psi^l=\Psi^l\left(z_i,X,X_{z_{i}},X_{z_{i_1}^{j_1},z_{i_2}^{j_2}},\dots,X_{z_{i_1}^{j_1},z_{i_2}^{j_2},z_{i_3}^{j_3},\dots,z_{i_n}^{j_n}}\right)
\end{equation}
for all $l=1,\dots,q$ such that $j_1+\dots+j_n\leq p$ and $z_i=z_1,\dots,z_n.$

The forthcoming examples will show the application of this procedure to the particular case in which the number of initial independent variables is 3,
with the identification $x_1=x,x_2=y,x_3=t$.
We deal with hierarchies of PDEs in which a great number of dependent variables is present, as CHH($2+1$) 
and mCHH($2+1$).
{\it Here the necessity of introducing auxiliary independent variables to which such auxiliary fields are reciprocally-transformed is obvious.
Hence, the introduction of intermediate variables $z_4,z_5,\dots$ will account for the dependent variables.}
We depict this fact by showing a clever choice of variables in the next two examples. In order to make things clearer, some slight changes
in the notation of the theory or the dimension of $N_{\mathbb{R}^n}$ shall be altered to fit our concrete examples.

\section{CHH(2+1) versus CBS}
\setcounter{equation}{0}
\setcounter{theorem}{0}
\setcounter{example}{0}
We shall  briefly summarize and improve the results of \cite{estevez51} and \cite{EstPrada2} 
in order to establish the reciprocal transformation that connects the CHH(2+1) and mCHH(2+1)
hierarchies with CBS and mCBS equations, respectively. Many details (specially those referring to the detailed
calculation) are omitted and can be obtained in the above cited references.

It could be useful from the beginning to say that we shall use capital letters for the dependent and
independent variables connected with CHH(2+1), and lower case letters will be used to refer to mCHH(2+1).

The CHH(2+1) hierarchy can be written in a compact form as
\begin{equation}
U_T=R^{-n}U_Y, \label{1}
\end{equation}
where $R$ is the recursion operator defined as
\begin{equation}
R=JK^{-1},\quad K=\partial_{XXX}-\partial_X,\quad J=-\frac{1}{2}\left(\partial_XU+U\partial_X\right).
\label{2}
\end{equation}
Note that the factor $-\frac{1}{2}$ in the definition of $J$ is not essential and it has been introduced to
make the later identification between the time variables easier. 
\noindent
This is the difference between the CHH($2+1$) introduced in \eqref{NCHHaf2} in Chapter \ref{Chap:LieSymm}
and the present.


The $n$ component of this hierarchy can also be rewritten as a set of PDEs by introducing $n$ dependent fields
$\Omega_{[i]}, (i=1,\dots, n)$ in the following way
\begin{equation}
\begin{aligned}
U_Y&=J\Omega_{[1]}\\
J\Omega_{[i+1]}&=K\Omega_{[i]},\\
  U_T&=K\Omega_{[n]}, \label{3}
\end{aligned}
\qquad i=1,\dots, n-1,\end{equation} 
and by introducing two new fields, $P$ and $\Delta$, related to $U$ as
\begin{equation}
U=P^2,\quad\quad P_T=\Delta_X, \label{4}
\end{equation}
we can write the hierarchy in the form of the following set of equations
\begin{equation}
\begin{aligned} &P_Y=-\frac{1}{2}\left(P\Omega_{[1]}\right)_X,\\
&(\Omega_{[i]})_{XXX}-(\Omega_{[i]})_X=-P\left(P\Omega_{[i+1]}\right)_X,  \\
& P_T=\frac{(\Omega_{[n]})_{XXX}-(\Omega_{[n]})_X}{2P}=\Delta_X.\label{5}
\end{aligned}
\qquad i=1,\dots, n-1,\end{equation} 
It was shown in \cite{EstPrada2} that \eqref{5} can be reduced to the the negative Camassa-Holm hierarchy under the reduction $\frac{\partial}{\partial t}=0$. The positive flow can be obtained under the reduction $\frac{\partial}{\partial x}=\frac{\partial}{\partial y}$.

The conservative form of the first two equations allows us to define the exact derivative
\begin{equation}
dz_0= P\,dX-\frac{1}{2}P\Omega_{[1]}\,dY+\Delta\,dT. \label{6}
\end{equation}
A reciprocal transformation \cite{h00,rogers4,rogers5} can be introduced by considering the former independent variable $X$ as a field
depending on  $z_0$, $z_1=Y$ and $z_{n+1}=T$, such that $d^2X=0$. From (\ref{6}) we have
\begin{equation}
\begin{aligned}
dX&= \frac{1}{P}\,dz_0+\frac{\Omega_{[1]}}{2}\,dz_1-\frac{\Delta}{P}\,dz_{n+1},\\
Y&=z_1,\\ 
T&=z_{n+1},\label{7}
\end{aligned}\end{equation} 
If we consider the new field $X=X(z_0,z_1,\dots,z_{n+1})$, by direct comparison
\begin{equation}
\begin{aligned}
&X_{z_0}=\frac{1}{P},\\
&X_{z_1}=\frac{\Omega_{[1]}}{2},\\ \label{8}
&X_{z_{n+1}}=-\frac{\Delta}{P},
\end{aligned}\end{equation} 
where $X_{z_i}=\frac{\partial X}{\partial z_i}$. We can now extend the transformation by introducing a new
independent variable $z_i$ for each field $\Omega_{[i]}$ by generalizing (\ref{8}) as
\begin{equation} X_{z_i}=\frac{\Omega_{[i]}}{2},\qquad i=1,\dots, n.\label{9}\end{equation}
Each  of the former dependent fields $\Omega_{[i]},\,(i=1,\dots, n)$ allows us to define a new dependent variable $z_i$ through  definition (\ref{9}).
It requires some calculation (see \cite{EstPrada2} for  details) but it can be proven that the reciprocal transformation (\ref{7})-(\ref{9}) transforms (\ref{5}) to the following set of $n$ PDEs 
on ${\rm J}^4(\mathbb{R}^{n+2},\mathbb{R})$
\begin{equation}-\left(\frac{X_{z_{i+1}}}{X_{z_0}}\right)_{z_0}=\left[\left(\frac{X_{z_0,z_0}}{X_{z_0}}+X_{z_0}\right)_{z_0}-\frac{1}{2}\left(\frac{X_{z_0,z_0}}{X_{z_0}}+X_{z_0}\right)^2\right]_{z_i}, \label{10}\end{equation}
with $i=1,\dots, n.$ Note that each equation depends on only three  variables $z_0, z_i, z_{i+1}$. This result generalizes the one found in \cite{h00} for the first component of the hierarchy.
The conservative form of (\ref{10}) allows us to define a field $M(z_0,z_1,\dots, z_{n+1})$ such that
\begin{equation}
\begin{aligned}M_{z_i}&=-\frac{1}{4}\left(\frac{X_{z_{i+1}}}{X_{z_0}}\right),\quad\quad i=1,\dots, n,\\
 M_{z_0}&=\frac{1}{4}\left[\left(\frac{X_{z_0,z_0}}{X_{z_0}}+X_{z_0}\right)_{z_0}-\frac{1}{2}\left(\frac{X_{z_0,z_0}}{X_{z_0}}+X_{z_0}\right)^2\right].\label{11}\end{aligned}\end{equation} 
It is easy to prove that each $M_i$ should satisfy the following CBS  equation \cite{cbs,calogero} on ${\rm J}^4(\mathbb{R}^{n+2},\mathbb{R})$
\begin{equation}M_{z_0,z_{i+1}}+M_{z_0,z_0,z_0,z_i}+4M_{z_i}M_{z_0,z_0}+8M_{z_0}M_{z_0,z_i}=0, \label{12}\end{equation}
with $ i=1,\dots, n.$

\section{mCHH(2+1) versus mCBS}
\setcounter{equation}{0}
\setcounter{theorem}{0}
\setcounter{example}{0}
In \cite{estevez51}, the mCHH(2+1) was introduced 
\begin{equation}
u_t=r^{-n}u_y, \label{15}
\end{equation}
where $r$ is the recursion operator, defined as
\begin{equation}
r=jk^{-1},\quad k=\partial_{xxx}-\partial_x,\quad j=-\partial_x\,u\,(\partial_x)^{-1}\,u\,\partial_x
\label{16}
\end{equation}
 This hierarchy generalizes the one introduced by Qiao in \cite{Qiao1},
whose second positive member was studied in \cite{Qiao2,Qiao3,Qiao1}.
We shall  briefly summarize the results of \cite{estevez51} when a  procedure similar to that described for CHH(2+1) is applied to mCHH(2+1).

If we introduce $2n$ auxiliary fields $v_{[i]}$, $\omega_{[i]}$ defined through
\begin{equation}
\begin{aligned}
 u_y&=jv_{[1]},\\
jv_{[i+1]}&=kv_{[i]},\\
(\omega_{[i]})_x&=u(v_{[i]})_x,\\
  u_t&=kv_{[n]}, \label{17}
\end{aligned}
\qquad i=1,\dots, n-1,\end{equation} 
the hierarchy can be written as the system
\begin{equation}
\begin{aligned}
u_y&=-\left(u\omega_{[1]}\right)_x,\\
(v_{[i]})_{xxx}-(v_{[i]})_x&=-\left(u\omega_{[i+1]}\right)_x, \label{18}\\
u_t&=(v_{[n]})_{xxx}-(v_{[n]})_x=\delta_x,
\end{aligned}
\qquad i=1,\dots, n-1,\end{equation} 
which allows to define the exact derivative
\begin{equation}
dz_0= u\,dx-u\omega_{[1]}\,dy+\delta\,dt \label{19}
\end{equation}
and $z_1=y, z_{n+1}=t$.
We can define a reciprocal transformation such that  the former independent variable $x$ is a new field $x=x(z_0,z_1,\dots , z_{n+1})$ depending on $n+2$ variables in the form
\begin{equation}
\begin{aligned}
x_{z_0}&=\frac{1}{u},\\
x_{z_1}&=\omega_{[1]},\label{20}\\
x_{z_{n+1}}&=-\frac{\delta}{u}.
\end{aligned}\end{equation} 
If we introduce the auxiliary variables for the auxiliary fields, $x_{z_i}=\omega_{[i]}$ for $i=2,\dots,n$,
the transformation of the equations (\ref{18}) yields the system of equations on ${\rm J}^4(\mathbb{R}^{n+2},\mathbb{R})$
\begin{equation}\left(\frac{x_{z_{i+1}}}{x_{z_0}}+\frac{x_{z_i,z_0,z_0}}{x_{z_0}}\right)_{z_0}=\left(\frac{x_{z_0}^2}{2}\right)_{z_i},\qquad i=1,\dots, n. \label{21}\end{equation}
Note that each equation depends on only three variables: $z_0, z_i, z_{i+1}$.

The conservative form of (\ref{21}) allows us to define a field $m=m(z_0,z_1,\dots, z_{n+1})$ such that on ${\rm J}^3(\mathbb{R}^{n+2},\mathbb{R}^2)$ we have
\begin{equation} m_{z_0}=\frac{x_{z_0}^2}{2},\quad m_{z_i}=\frac{x_{z_{i+1}}}{x_{z_0}}+\frac{x_{z_i,z_0,z_0}}{x_{z_0}},\qquad i=1,\dots, n.\label{22}\end{equation}
Equation (\ref{21}) has been extensively studied from the point of view of  Painlev\'e analysis \cite{EstPrada1} and it can be considered as the modified version of the CBS equation (mCBS) (\ref{12}). Actually, in \cite{EstPrada1} it was proven that the {\it Miura transformation} that relates (\ref{12}) and (\ref{22}) is
\begin{equation}
4M=x_{z_0}-m. \label{23}
\end{equation}
A non-isospectral Lax pair  was  obtained for (\ref{22}) in \cite{EstPrada1}. By inverting this Lax pair through the reciprocal transformation (\ref{20}) the following spectral problem was obtained for mCHH(2+1). This Lax pair \cite{EstPrada1} reads 

\begin{align} 
\left(\begin{array}{c} \phi \\ \hat\phi
\end{array}\right)
_x &=\frac{1}{2} \left[\begin{array}{cc} -1& I\sqrt{\lambda}u \\ I\sqrt{\lambda}u
& 1
\end{array}\right]
 \left(\begin{array}{c} \phi \\ \hat\phi
\end{array}\right)
,\label{24}
\\
\left(\begin{array}{c} \phi \\ \hat\phi
\end{array}\right)
_t &=\lambda^n 
\left(\begin{array}{c} \phi \\ \hat\phi
\end{array}\right)
_y+\lambda a 
\left(\begin{array}{c} \phi \\ \hat\phi
\end{array}\right)
_x+
I
\frac{\sqrt{\lambda}}{2} \left[\begin{array}{cc} 0& b _{xx}-b_x  \\
b_{xx}+b_x &0
\end{array}\right]
\left(\begin{array}{c} \phi \\ \hat\phi
\end{array}\right)
,
\end{align}
where
 \begin{equation}
a=\sum_{i=1}^n\lambda^{n-i}\omega_{[i]}, \quad b=\sum_{i=1}^n\lambda^{n-i}v_{[i]},\quad I=\sqrt{-1}\quad i=1,\dots,n,
      \end{equation} 
and $\lambda(y,t)$ is a non-isospectral parameter that satisfies
\begin{equation}\lambda_x=0,\quad \lambda_t-\lambda^n\lambda_y=0.\label{25}\end{equation}
Although the Painlev\'e test cannot be applied to mCHH(2+1), reciprocal transformations are a tool that can be used to write the hierarchy as a set of mCBS equations to which  the Painlev\'e analysis (the SMM in particular) can be successfully applied.

\section{Reciprocal-Miura transformations}
\setcounter{equation}{0}
\setcounter{theorem}{0}
\setcounter{example}{0}
As  stated in the previous section, there are two reciprocal transformations (\ref{8}) and  (\ref{20}) that relate CHH(2+1)  and mCHH(2+1)  hierarchies with CBS (\ref{12}) and mCBS (\ref{22}) respectively. Furthermore, it is known that a Miura transformation (\ref{23}) relates CBS and mCBS. The natural question that arises is whether the mCHH(2+1)  hierarchy can be considered as the modified version of CHH(2+1).
Evidently, the relationship between both hierarchies cannot be a simple Miura transformation because they are written in different variables $(X,Y,T)$ and $(x,y,t)$. The answer is provided by the relationship of both sets of variables with the same set $(z_0,z_1,z_{n+1})$. 
By combining \eqref{6} and \eqref{19}, we have
\begin{equation}
\begin{aligned}
 P\,dX-\frac{1}{2}P\Omega_{[1]}\,dY+\Delta\,dT&=u\,dx-u\omega_{[1]}\,dy+\delta\,dt,\\
Y&=y,\\ T&=t, \label{26}
\end{aligned}\end{equation} 
which yields the required relationship between the independent variables of CHH(2+1) and those of mCHH(2+1). 
The Miura transformation (\ref{23}), also provides the following results
\begin{equation}
\begin{aligned} 4M_{z_0}&=x_{z_0,z_0}-m_{z_0}\Longrightarrow \frac{X_{z_0,z_0}}{X_{z_0}}+X_{z_0}=x_{z_0}, \\ 4M_{z_i}&=x_{z_0,z_i}-m_{z_i}\Longrightarrow -\frac{X_{z_{i+1}}}{X_{z_0}}=x_{z_0,z_i}-\frac{x_{z_0,z_0,z_i}}{x_{z_0}}-\frac{x_{z_{i+1}}}{x_{z_0}}, \label{27}\end{aligned}\end{equation} 
with $i=1,\dots, n,$ where (\ref{11}) and (\ref{22}) have been used.
With the aid of (\ref{8}), (\ref{9}) and (\ref{20}), the following results 
 arise from (\ref{27}) (see Appendix 2 for further details)
\begin{equation}
\begin{aligned} \frac{1}{u}&=\left(\frac{1}{P}\right)_X+\frac{1}{P},\\ P\Omega_{[i+1]}&= 2(v_{[i]}-(v_{[i]})_x)\quad\Longrightarrow\quad \omega_{[i+1]}=\frac{(\Omega_{[i+1]})_X+\Omega_{[i+1]}}{2},\\ \Delta &=(v_{[n]})_x-v_{[n]}.\label{28}
\end {aligned}\end{equation} 
with $  i=1,\dots, n-1$. Furthermore, (\ref{26}) can be written as
\begin{equation} dx=\left[1-\frac{P_X}{P}\right]dX+\left[\omega_{[1]}-\frac{\Omega_{[1]}}{2}\left(1-\frac{P_X}{P}\right)\right]dY+\frac{\Delta-\delta}{u}dT.\label{29}\end{equation}
The cross derivatives of (\ref{29}) imply (see Appendix 2) that
\begin{equation}
\begin{aligned} \left[1-\frac{P_X}{P}\right]_Y&=\left[\omega_{[1]}-\frac{\Omega_{[1]}}{2}\left(1-\frac{P_X}{P}\right)\right]_X\Longrightarrow\quad\omega_{[1]}=\frac{(\Omega_{[1]})_X+\Omega_{[1]}}{2},\\
\left[1-\frac{P_X}{P}\right]_T&=\left[\frac{\Delta-\delta}{P}\left(1-\frac{P_X}{P}\right)\right]_X \Longrightarrow \quad\frac{\delta}{u}=\left(\frac{\Delta}{P}\right)_X+\frac{\Delta}{P} \label{30}\end{aligned}\end{equation} 
and therefore with the aid of (\ref{30}), (\ref{29}) reads
\begin{equation} dx=\left[1-\frac{P_X}{P}\right]dX-\frac{P_Y}{P}dY-\frac{P_T}{P}dT.\label{31}\end{equation}
This exact derivative can be integrated as
\begin{equation}x=X-\ln P.\end{equation}
By summarizing the above conclusions, we have proven that the mCHH(2+1)  hierarchy
 \begin{equation}
u_t=r^{-n}u_y,\quad u=u(x,y,t), \end{equation}  can be considered as the modified version of the Camassa- Holm hierarchy
  \begin{equation}
U_T=R^{-n}U_Y,\quad U=U(X,Y,T).  \end{equation} The transformation that connects the two hierarchies  involves the reciprocal transformation
\begin{equation}x=X-\frac{1}{2}\ln U\end{equation}
as well as the following  transformation between the fields
\begin{equation}
\begin{aligned}\frac{1}{u}&=\frac{1}{\sqrt U}\left(1-\frac{U_X}{2U}\right) ,\quad\Longrightarrow\quad \omega_{[i]}=\frac{(\Omega_{[i]})_X+\Omega_{[i]}}{2},\quad i=1,\dots, n,\\
 \frac{\delta}{u}&=\left(\frac{\Delta}{\sqrt U}\right)_X+\frac{\Delta}{\sqrt U}\label{34}.\end{aligned}\end{equation} 
\subsection{Particular case 1: The Qiao equation}
We are now restricted to the first component of the hierarchies $n=1$ in the case in which the field $u$ is independent of $y$ and $U$ is independent of Y.
\begin{itemize}
\item From \eqref{4} and \eqref{5}, for the restriction of CHH(2+1) we have
\begin{equation}
\begin{aligned}U&=P^2, \\U_T&=(\Omega_{[1]})_{XXX}-(\Omega_{[1]})_{X},\\ (P\Omega_{[1]})_X&=0.\label{39}\end{aligned}\end{equation} 
which can be summarized as
\begin{equation}
\begin{aligned}\Omega_{[1]}&=\frac{k_1}{P}=\frac{k_1}{\sqrt U},\\
U_T&=k_1\left[\left(\frac{1}{\sqrt U}\right)_{XXX}-\left(\frac{1}{\sqrt U}\right)_{X}\right],\label{40}\end{aligned}\end{equation} 
that is the {\it Dym equation}.

\item The reduction of mCHH(2+1)  can be achieved from \eqref{18} in the form
\begin{equation}
\begin{aligned}(\omega_{[1]})_x&=u(v_{[1]})_x,\\u_t&=(v_{[1]})_{xxx}-(v_{[1]})_{x},\\ \left(u\omega_{[1]}\right)_x&=0,\end{aligned}\end{equation} 
which can be written as
\begin{equation}\label{41}
\begin{aligned}\omega_{[1]}&=\frac{k_2}{u} \quad\Longrightarrow
\quad v_{[1]}=\frac{k_2}{2u^2},\\
u_t&=k_2\left[\left(\frac{1}{2u^2}\right)_{xx}-\left(\frac{1}{2u^2}\right)\right]_{x}.\end{aligned}\end{equation} 
that is the Qiao equation.

\item From \eqref{28} and \eqref{34} it is easy to see that $k_1=2k_2$. By setting $k_2=1$, we can conclude that the Qiao equation
\begin{equation}
u_t=\left(\frac{1}{2u^2}\right)_{xxx}-\left(\frac{1}{2u^2}\right)_{x}\end{equation} 
is the modified version of the Dym equation
\begin{equation}
U_T=\left(\frac{2}{\sqrt U}\right)_{XXX}-\left(\frac{2}{\sqrt U}\right)_{X}.\end{equation} 
\item From \eqref{8} and \eqref{20}, it is easy to see that the independence from $y$ implies that $X_1=X_0$ and $x_1=x_0$, which means that  the CBS and modified CBS equations \eqref{12} and \eqref{22} reduce to the following potential versions of the KdV and modified KdV equations
\begin{equation}
\begin{gathered} \left(M_{z_2}+M_{z_0,z_0,z_0}+6M_{z_0}^2\right)_{z_0}=0,\\
x_{z_2}+x_{z_0,z_0,z_0}-\frac{1}{2}x_{z_0}^3=0.\end{gathered}\end{equation} 
\end{itemize}

\subsection{Particular case 2: The Camassa-Holm equation}
If we are restricted to the $n=1$ component when $T=X$ and $t=x$, the following results hold
\begin{itemize}
\item From \eqref{4} and \eqref{5}, for the restriction of CHH(2+1) we have
\begin{equation}
\begin{aligned} &\Delta=P=\sqrt U,\\& U=(\Omega_{[1]})_{XX}-\Omega_{[1]},\\ &U_Y+U(\Omega_{[1]})_X+\frac{1}{2}\Omega_{[1]} U_X=0\end{aligned}\end{equation} 
which is the Camassa--Holm equation equation.

\item The reduction of mCHH(2+1)  can be obtained from \eqref{18} in the form
\begin{equation}
\begin{aligned}&\delta=u=(v_{[1]})_{xx}-v_{[1]},\\& u_y+(u\omega_{[1]})_x=0,\\& (\omega_{[1]})_x-u(v_{[1]})_x=0,\end{aligned}\end{equation} 
which can be considered as a modified Camassa-Holm equation.

\item From \eqref{8} and \eqref{20}, it is easy to see that $X_2=x_2=-1$. Therefore, the reductions of \eqref{12} and \eqref{22} are
\begin{equation}
M_{z_0,z_0,z_0,z_1}+4M_{z_1}M_{z_0,z_0}+8M_{z_0}M_{z_0,z_1}=0,\end{equation} 
which is the AKNS equation and
\begin{equation}
\left(\frac{x_{z_1,z_0,z_0}-1}{x_{z_0}}\right)_{z_0}=\left(\frac{x_{z_0}^2}{2}\right)_{z_1},\end{equation} 
that is the modified AKNS equation.

\end{itemize}

\chapter{Conclusions and outlook}\markboth{Conclusions}{Chapter 6}

\section{Lie systems}
\setcounter{equation}{0}
\setcounter{theorem}{0}
\setcounter{example}{0}


\begin{itemize}

\item There remains a lot of work to continue on Lie systems. Apart from the obtained results, e.g., linearizability conditions, constants of motion, Lie symmetries,
superposition rules for such Lie systems, etc., further properties need to be analyzed: the existence of several Lie--Hamiltonian
structures for a Lie system, the study of conditions for the existence of
Lie--Hamilton systems, other methods to derive superposition rules, the analysis of
integrable and superintegrable Lie--Hamilton systems, etc. 

\item Our work on the second-order Riccati equations has introduced many advantages with respect to previous methods.
We directly transform a quite general family of second-order Riccati equations into Lie--Hamilton systems by 
Legendre transforms. This is a much simpler approach than the proposed in the Physics literature, based on more elaborated geometric theories which map second-order Riccati equations into  $\mathfrak{sl}(3,\mathbb{R})$-Lie systems.
Our method reduces the explicit integration of second-order Riccati equations to solving Lie systems related to $\mathfrak{sl}(2,\mathbb{R})$, e.g., Riccati equations.
We have also provided a new approach to Kummer--Schwarz equations (second and third-order) that
retrieves results in a simpler way. 

\item In this thesis we have detailed numerous new applications of Lie systems. This has enlarged the potential applications of these systems.
We have shown for the first time that Lie systems appear in the study of viral models and certain types of Lotka--Volterra systems. We have also proven that Riccati equations over several composition algebras, e.g., quaternions or dual-study complex numbers, are also Lie systems. Other new applications concern  Darboux--Brioschi--Halphen systems, certain systems with quadratic nonlinearities, planar systems with trigonometric nonlinearities appearing in the theory of integrable systems, complex Bernoulli equations, etcetera. Concerning
PDE-Lie systems, we have studied partial Riccati equations.

\item We have classified Lie--Hamilton systems on the plane attending to their underlying Lie algebras.
 We found that only twelve out of the initial 28 classes of finite-dimensional real Lie algebras of vector fields on the plane detailed in the GKO classification, are of Lie--Hamilton type. In turn, these classes give rise to 12 families of Lie algebras of Hamiltonian vector fields. This led to classifying Lie--Hamilton systems on the plane.
In particular, we have given examples belonging to the classes P$_2$,  I$_2$,  I$_4$,  I$_5$,  I$_{14A}$, I$_{14B}$ and I$_{15}$ of our classification of finite-dimensional Lie algebras of Hamiltonian vector fields. 
We have shown that Kummer--Schwarz, Milne--Pinney equations (both with $c>0$) and complex Riccati equations with $t$-dependent coefficients are related to
the same Lie algebra P$_2$, a fact which was used to explain the existence of a local diffeomorphism that maps each of these systems into another. We also showed that
the  $t$-dependent harmonic oscillator, arising from Milne--Pinney equations when $c=0$,   corresponds to class   I$_5$ and   this can  only    be related through   diffeomorphisms  to the 
Kummer--Schwarz equations with $c=0$, but not with complex Riccati equations.
We plan to study the existence and the maximal number of cyclic limits for planar Lie systems so as to investigate the so called second Hilbert's number ${\rm H}(2)$ for these systems. This could be a first step to analyze the XVI Hilbert's problem through our Lie techniques.

\item We have proven the efficiency of the coalgebra method for obtaining superposition rules in the case of $\mathfrak{sl}(2)$- and $\mathfrak{so}(3)$-Lie--Hamilton systems. 
Obtaining superposition rules for all of the two-photon Lie--Hamilton systems remains another open problem. As a byproduct of this further construction, Lie--Hamilton systems defined on subalgebras of P$_5$ would be also obtained as particular cases.
Recall that P$_5$ has relevant Lie--Hamilton subalgebras: $\mathfrak{sl}(2)$,  P$_1\simeq \overline{ \mathfrak{iso}}(2)$ and
 I$_8\simeq \overline{  \mathfrak{iso}  } (1,1)\simeq \mathfrak{h}_4$. Work on these lines is currently in progress.

\item For structure-Lie systems, we have proposed a method for generating bi--Dirac--Lie systems. Superposition rules for Dirac--Lie systems have been achieved for the first time.
We have gone a step further by classifying Lie algebras of Hamiltonian vector fields with respect to a Jacobi structure.
Based on the previous GKO classification, we have established which Lie algebras in the GKO classification are Hamiltonian in the Jacobi sense. Therefore,
we are classifying Jacobi--Lie systems on low dimensional manifolds, since the GKO is a planar classification. Another classification Table
in terms of the Jacobi structure has arisen and has been displayed.
It is not shocking to find a smaller number of Jacobi--Lie systems with physical or mathematical applications, given the restrictive nature of the Jacobi structure.

\item For future research, we aim at finding new types of Lie systems related to other geometric
structures. For instance, it
would be interesting to study the existence of Lie systems admitting a
Vessiot--Guldberg Lie algebra of Hamiltonian vector fields
with respect to almost or twisted Poisson structures. We are also interested in developing a generalization of the theory of Lie systems to the framework of Lie algebroids.
The latter has shown to be very fruitful in Geometric Mechanics and Control Theory.
Moreover, a further analysis of the properties of Dirac--Lie
systems is being performed. 
The theory of partial superposition rules for partial differential equations is still an open question, specifically from the point of view of
the coalgebra method. We would like to develop a solid geometrical framework and expand the applications of Lie systems to partial differential equations.
The author has plans of continuing a similar procedure as the one presented for Lie--Hamilton systems, applied to the realm of $k$-symplectic Lie systems.
Also, she is particularly interested in research in other types of geometries: as K\"ahler or Calabi--Yau with more applications in Theoretical Physics,
given her Physics career background. 
\end{itemize}

\section{Lie symmetries}
\setcounter{equation}{0}
\setcounter{theorem}{0}
\setcounter{example}{0}

\begin{itemize}
\item The wide applicability of Lie symmetries has suggested the inspection of Lie symmetries of Lie systems.
We have shown that these symmetries can be obtained by solving a system of differential equations that is another Lie system possessing a Vessiot--Guldberg Lie algebra related
to the Vessiot-Guldberg Lie algebra of the Lie system. Our results have been generalized to the realm of higher-order differential equations
and partial differential Lie systems. Our procedures enable us to find Lie symmetries of isomorphic partial differential Lie systems simultaneously.

\item We have calculated the Lie point symmetries of (nonlinear) partial differential equations with strong applications in the field of Fluid Dynamics. An important example
is that of the BKP equation and its corresponding two-component Lax pair in $2+1$ dimensions. 
We have reduced both the equation and its Lax pair attending to different choices of the arbitrary functions and constants of integration present in the symmetries. 
Out of 6 reductions, 2 of them happen to be knowledgeable equations in the literature: one is the
Korteweg de Vries equation and another is an equation showing multisoliton solutions. In this way, we can say that BKP in $2+1$ is a generalization of the KdV equation in $1+1$.

\item Next, we perform an iterative process of symmetry search on another nonlinear equation (obtained as one of the reductions of the previous step) and its corresponding two component Lax pair in $1+1$.
The spectral parameter has been conveniently introduced in the reduction and it needs to be considered as an independent variable for further reduction.
%
The nonclassical approach has resulted in 6 constants of integration and 2 arbitrary functions, whilst the
classical approach gives rise to symmetries only containing 5 constants of integration. 
Another 4 interesting reductions have been contemplated out of this second reduction. 

\item We have calculated Lie point symmetries for higher-order hierarchies of differential equations and their corresponding Lax pairs.
It seems natural to expect the same Lie symmetries for both the hierarchy and the Lax pair. Nevertheless, the results obtained do
not match our expected ones, as we have proven for the Camassa--Holm hierarchy and its Lax pair. The difference between 
the symmetries of the hierarchy and the Lax pair is one arbitrary function of time in the Lax pair but a constant appearing for the equation.
It is our future duty to identify the reasons for these differing results.
In the case of the Camassa--Holm hierarchy, we have started with the non-isospectral Lax pair in $2+1$ dimensions. We have searched for non-classical
Lie symmetries of this non-isoespectral problem and each possibility has led us to reduced spectral problems in $1+1$ whose compatibility conditions
give rise to $1+1$ dimensional reduced hierarchies. Each of these possibilities comes from the choice of the 5 arbitrary constants and 3 arbitrary time
dependent functions that appear in the symmetries.
The most outstanding result of this research is that our procedure also provides the reduction of the eigenfunction, as well as that of the spectral parameter.
In many cases, the spectral parameter also proves to be nonisospectral, even in the $1+1$ reductions.
We achieved several reductions that can be summarized in 9 nontrivial cases. 5 of the arosen hierarchies have a nonisopectral Lax pair, 2 of them
are the positive and negative Camassa--Holm hierarchies.
The rest of reductions happen to be isospectral, as most problems found in $1+1$ in the literature.
It is interesting then to remark that we have found nonisospectral problems in $1+1$, which are not abundant.

A similar problem to this one has constituted another section of this second part. We have dealt with the Qiao hierarchy and its associated nonisospectral Lax pair in $2+1$ dimensions.
The interest of this problem resides in the spinorial or two component Lax pair. We have followed a very similar procedure as for the Camassa--Holm hierarchy in $2+1$ dimensions,
with the difference that this time we have only pursued classical symmetries, instead of nonclassical. The reason for contemplating this change is the difficulty
in dealing with the arising nonlinear equations for the obtainaince of the symmetry. Solving the classical case has appeared more reliable to us
in this particular case. The symmetries depend on 4 arbitrary constants and 2 arbitrary functions whereas there exists a function that 
has to satisfy an equation identical to the nonisospectrality condition. Each possibility contemplated leads to a possible reduction, we are counting a
total of 8. Then again, we find nonisospectral reductions in $1+1$ dimensions, which are not frequent. These results confirm the group interpretation of the
spectral parameter proposed by other authors.

\item A future idea of research is identifying the differences between the classical and nonclassical symmetries obtained for our examples. It is noticeable
that several differences occur, in particular, in the character of the arbitrary functions and constants arising out of the integration.
It is our idea to guess where their differences root. It is also interesting to compare these symmetries with contact symmetries and obtain
more through the contact symmetry method.

\end{itemize}

\section{Reciprocal transformations}
\setcounter{equation}{0}
\setcounter{theorem}{0}
\setcounter{example}{0}

\begin{itemize}

\item We have proposed a reciprocal transformation between two hierarchies of interest: the Camassa--Holm and Qiao hierarchies in $2+1$ dimensions.
We have presented both hierarchies in $2+1$ dimensions and we have discussed some general properties of the Camassa--Holm hierarchy for a field $U(X,Y,T)$
and the Qiao hierarchy for $u(x,y,t)$. We have constructed reciprocal transformations that connect both hierarchies with the CBS and mCBS equations, respectively.
The Lax pair for both of the hierarchies can be retrieved through those of the CBS and mCBS. If we consider
the Lax pair of the CBS and mCBS and undo the reciprocal transformation, we achieve the Lax pair of the hierarchies.
There exists a Miura transformation between the CBS and mCBS equations. So, it is possible to achieve a relation
between the fields $U(X,Y,T)$ and $u(x,y,t)$ and the variables $(X,Y,T)$ and $(x,y,t)$. This carefully constructed relation proves that the Qiao hierarchy
is a transformed version of that of Camassa--Holm.
Two particular cases of interest have been contemplated: one is the relation between the two hierarchies when they are independent of $Y$ and $y$. 
This shows that the Qiao equation is a modified Dym equation.
Another is the modified Camassa--Holm equation when the reduction $T=X$ is applied to the first component of the Camassa--Holm hierarchy.

\item We can say that somehow reciprocal transformations help us reduce the number of available nonlinear equations in the literature, as two
seemingly different equations can be turned from one into another by reciprocal transformation, which is the case of the reviewed example.
A question for future work is whether there exists a canonical description for differential equations. Intuitively, we expect that if two equations are essentially the
same, although apparently different in disguised versions, they must share the same singular manifold equations. But this is just an initial guess worth of further
research in the future. 
Also, reciprocal transformations have proven their utility in the derivation of Lax pairs, as one initial equation whose Lax pair is unknown, can be 
interpreted as the reciprocally-transformed equation whose Lax pair is well-known. In this way, undoing the transformation on the latter Lax pair,
we achieve the Lax pair of the former. As we know, obtaining Lax pairs is a nontrivial subject. The common way is to impose ad hoc forms
for such linear problems and make them fit according to the compatibility condition. Accounting for reciprocal transformations, we do not
face the problem of imposing ad hoc Ans\"atze.
To finish, we mention that reciprocal transformations permit us to obtain (sometimes) equations integrable in the Painlev\'e sense if they did
not have this property before the change. This is due to the non invariability of the Painlev\'e test under changes of variables.

\item Some possible future research on this topic would consist of understanding whether the singular manifold equations can constitute a canonical representation
of a partial differential equation and designing techniques to derive Lax pairs in a more unified way.
Also, the trial of composition of reciprocal transformations with transformations of other nature, as performed in the case of Miura-reciprocal transformations,
can lead to more unexpected but desirable results, as in the example exposed.

\end{itemize}

 
\backmatter
\setcounter{chapter}{0}
\chapter{\it Appendix 1: Tables}\markboth{Tables}{Appendix 1}
\label{Chap:appendix1}
\setcounter{equation}{0}
\setcounter{theorem}{0}
\setcounter{example}{0}

\begin{table}[h]\centering
 {\footnotesize
 \noindent
\caption{{\footnotesize The GKO classification of  the $8+20$  finite-dimensional real Lie algebras of vector fields on the plane and their most relevant
characteristics. The vector fields which are written between brackets form a modular generating system.
The functions $\xi_1(x),\ldots,\xi_r(x)$ and $1$ are linearly independent over $\mathbb{R}$ and the functions $\eta_1(x),\ldots,\eta_r(x)$ form a basis of solutions for an $r$-order differential equation with constant coefficients \cite[pp.~470--471]{HA75}. Finally, $\mathfrak{g}=\mathfrak{g}_1\ltimes \mathfrak{g}_2$ stands for the semi-direct sum (as Lie algebras) of $\mathfrak{g}_1$ by $\mathfrak{g}_2$. {\rm Dom} stands for the domain of the Lie algebra of vector fields.}}
\label{table1}
 \hspace*{-0.5cm}\begin{tabular}{ p{.5cm} p{2cm}    p{9.5cm} l}
\hline
&  &\\[-1.9ex]
\#&Primitive & Basis of vector fields $X_i$ &Dom\\[+1.0ex]
\hline
 &  &\\[-1.9ex]
P$_1$&$A_\alpha\simeq \mathbb{R}\ltimes \mathbb{R}^2$ & $ \{ { {\partial_x} ,    {\partial_y} \},   \alpha(x\partial_x + y\partial_y)  +  y\partial_x - x\partial_y},\quad \ \alpha\geq 0$&$\mathbb{R}^2$ \\[+1.0ex]
P$_2$&$\mathfrak{sl}(2)$ & $\{ {\partial_x},   {x\partial_x  +  y\partial_y} \},   (x^2  -  y^2)\partial_x  +  2xy\partial_y$&$\mathbb{R}^2_{y\neq 0}$\\[+1.0ex]
P$_3$&$\mathfrak{so}(3)$ &${     \{{ y\partial_x  -  x\partial _y},     { (1  +  x^2  -  y^2)\partial_x  +  2xy\partial_y} \},   2xy\partial_x  +  (1  +  y^2  -  x^2)\partial_y}$&$\mathbb{R}^2$\\[+1.0ex]
P$_4$&$\mathbb{R}^2\ltimes\mathbb{R}^2$ &$ \{ {\partial_x},   {\partial_y}\},  x\partial_x + y\partial_y,   y\partial_x - x\partial_y$&$\mathbb{R}^2$\\[+1.0ex]
P$_5$&$\mathfrak{sl}(2 )\ltimes\mathbb{R}^2$ &${ \{ {\partial_x},   {\partial_y}\},  x\partial_x - y\partial_y,  y\partial_x,  x\partial_y}$&$\mathbb{R}^2$\\[+1.0ex]
P$_6$&$\mathfrak{gl}(2 )\ltimes\mathbb{R}^2$ &${ \{{\partial_x},    {\partial_y}\},   x\partial_x,   y\partial_x,   x\partial_y,   y\partial_y}$&$\mathbb{R}^2$\\[+1.0ex]
P$_7$&$\mathfrak{so}(3,1)$ &${ \{ {\partial_x},   {\partial_y}\},   x\partial_x\!+\! y\partial_y,   y\partial_x \!-\! x\partial_y,   (x^2 \!-\! y^2)\partial_x \!+\! 2xy\partial_y,  2xy\partial_x \!+\! (y^2\!-\!x^2)\partial_y}$  &$\mathbb{R}^2$\\[+1.0ex]
P$_8$&$\mathfrak{sl}(3 )$ &${ \{ {\partial_x},    {\partial_y}\},   x\partial_x,   y\partial_x,   x\partial_y,   y\partial_y,   x^2\partial_x + xy\partial_y,   xy\partial_x  +  y^2\partial_y}$&$\mathbb{R}^2$\\[+1.5ex]
\hline
&  &\\[-1.5ex]
\#& Imprimitive\!\! & Basis of vector fields $X_i$ &Dom\\[+1.0ex]
\hline
&  &\\[-1.5ex]
I$_1$&$\mathbb{R}$ &$ \{ {\partial_x} \}$ & $\mathbb{R}^2$\\[+1.0ex]
I$_2$&$\mathfrak{h}_2$ & $ \{ {\partial_x} \},  x\partial_x$& $\mathbb{R}^2$\\[+1.0ex]
I$_3$&$\mathfrak{sl}(2 )$ (type I) &$ \{ {\partial_x}\},  x\partial_x,  x^2\partial_x$& $\mathbb{R}^2$\\[+1.0ex]
I$_4$&$\mathfrak{sl}(2 )$ (type II) & ${ \{ {\partial_x  +  \partial_y},    {x\partial _x + y\partial_y}\},   x^2\partial_x  +  y^2\partial_y}$ &$\mathbb{R}^2_{x\neq y}$\\[+1.0ex]
I$_5$&$\mathfrak{sl}(2 )$ (type III) &${\{ {\partial_x},    {2x\partial_x + y\partial_y}\},   x ^2\partial_x  +  xy\partial_y}$&$\mathbb{R}^2_{y\neq 0}$\\[+1.0ex]
I$_6$&$\mathfrak{gl}(2 )$ (type I)& ${\{ {\partial_x},    {\partial_y}\},   x\partial_x,   x^2\partial_x}$&$\mathbb{R}^2$\\[+1.0ex]
I$_7$&$\mathfrak{gl}(2 )$ (type II)& ${ \{ {\partial_x},   {y\partial_y} \},     x\partial_x,    x^2\partial_x +  xy \partial_y}$&$\mathbb{R}^2_{y\neq 0}$ \\[+1.0ex]
I$_8$&$B_\alpha\simeq \mathbb{R}\ltimes\mathbb{R}^2$ &${ \{ {\partial_x},    {\partial_y}\},   x\partial_x  +  \alpha y\partial_y},\quad  0<|\alpha|\leq 1$&$\mathbb{R}^2$\\[+1.0ex]
I$_9$&$\mathfrak{h}_2\oplus\mathfrak{h}_2$ &${\{ {\partial_x},    {\partial_y}\},   x\partial_x,  y\partial_y}$&$\mathbb{R}^2$\\[+1.0ex]
I$_{10}$&$\mathfrak{sl}(2 )\oplus \mathfrak{h}_2$ & ${\{ {\partial_x},    {\partial_y} \},   x\partial_x,  y\partial_y,  x^2\partial_x }$&$\mathbb{R}^2$\\[+1.0ex]
I$_{11}$&$\mathfrak{sl}(2 )\oplus\mathfrak{sl}(2 )$ &$ \{ {\partial_x},    {\partial_y}\},   x\partial_x,   y\partial_y,   x^2\partial_x ,   y^2\partial_y $&$\mathbb{R}^2$\\[+1.0ex]
I$_{12}$&$\mathbb{R}^{r + 1}$ &$\{ {\partial_y} \},   \xi_1(x)\partial_y, \ldots , \xi_r(x)\partial_y,\quad   r\geq 1$&$\mathbb{R}^2$ \\[+1.0ex]
I$_{13}$&$\mathbb{R}\ltimes \mathbb{R}^{r + 1}$ &$ \{ {\partial_y} \},   y\partial_y,    \xi_1(x)\partial_y, \ldots , \xi_r(x)\partial_y,\quad   r\geq 1$ &$\mathbb{R}^2$\\[+1.0ex]
I$_{14}$&$\mathbb{R}\ltimes \mathbb{R}^{r}$ & ${\{ {\partial_x},   {\eta_1(x)\partial_y} \},  {\eta_2(x)\partial_y},\ldots ,\eta_r(x)\partial_y},\quad r\geq 1$&$\mathbb{R}^2$\\[+1.0ex]
I$_{15}$&$\mathbb{R}^2\ltimes \mathbb{R}^{r}$ &  ${\{ {\partial_x},    {y\partial_y}\} ,    {\eta_1(x)\partial_y},\ldots, \eta_r(x)\partial_y},\quad  r\geq 1$&$\mathbb{R}^2$\\[+1.0ex]
I$_{16}$&$C_\alpha^r\simeq \mathfrak{h}_2\ltimes\mathbb{R}^{r + 1}$ & ${ \{ {\partial_x},    {\partial_y} \},   x\partial_x  +  \alpha y\partial y,   x\partial_y, \ldots, x^r\partial_y},\quad   r\geq 1,\qquad \alpha\in\mathbb{R}$&$\mathbb{R}^2$\\[+1.0ex]
I$_{17}$&$\mathbb{R}\ltimes(\mathbb{R}\ltimes \mathbb{R}^{r})$ &$ \{ {\partial_x},    {\partial_y} \},   x\partial_x  +  (ry  +  x^r)\partial_y ,   x\partial_y, \ldots,  x^{r - 1}\partial_y,\quad   r\geq 1$ &$\mathbb{R}^2$\\[+1.0ex]
I$_{18}$&$(\mathfrak{h}_2\oplus \mathbb{R})\ltimes \mathbb{R}^{r + 1}$ & $ \{ {\partial_x},    {\partial_y}\},   x\partial_x,   x\partial_y,   y\partial_y,   x^2\partial_y, \ldots,x^r\partial_y,\quad r\geq 1$ &$\mathbb{R}^2$\\[+1.0ex]
I$_{19}$&$\mathfrak{sl}(2 )\ltimes \mathbb{R}^{r + 1}$ &  $ \{ {\partial_x},    {\partial_y} \},   x\partial_y,    2x\partial _x  +  ry\partial_y,   x^2\partial_x  +  rxy\partial_y,   x^2\partial_y,\ldots, x^r\partial_y ,\quad   r\geq 1$&$\mathbb{R}^2$ \\[+1.0ex]
I$_{20}$&$\mathfrak{gl}(2 )\ltimes \mathbb{R}^{r + 1}$ &  $ \{ {\partial_x},    {\partial_y} \},   x\partial_x,   x\partial_y,   y\partial_y,   x^2\partial_x  +  rxy\partial_y,   x^2\partial_y,\ldots, x^r\partial_y,\quad   r\geq 1$ &$\mathbb{R}^2$\\[+1.5ex]
\hline
 \end{tabular}}
\end{table}

\begin{table}[t] {\footnotesize
 \noindent
\caption{{\small The   classification of the $4+8$ finite-dimensional real  Lie algebras of Hamiltonian vector fields on $\mathbb{R}^2$. For I$_{12}$, I$_{14A}$ and I$_{16}$, we have $j=1,\dots,r$ and $ r\ge1$; in  I$_{14B}$ the index $j=2,\dots, r$.
  }}
\label{table3}
\medskip
\noindent\hfill
\hspace*{-0.6cm}\begin{tabular}{ l l l l l }
\hline
&&&  &\\[-1.5ex]
\#&Primitive  & Hamiltonian functions $h_i$& $\omega$ &  Lie--Hamilton algebra\ \\[+1.0ex]
\hline
&  & &  &\\[-1.5ex]
P$_1$& $A_0\simeq {\mathfrak{iso}}(2)$ & ${y, \ -x, \ \tfrac 12 (x^2+y^2)},\ 1$ & ${\rm d}x\wedge {\rm d}y$ & $ \overline{\mathfrak{iso}}(2)$\\[+1.0ex]
P$_2$& $\mathfrak{sl}(2 )$ & $\displaystyle{- \frac 1y, \ -\frac xy, \  -\frac{x^2+y^2}{y}}$ & $\displaystyle{    \frac{ {\rm d}x\wedge {\rm  d}y }{ y^{2}  } }$ & $\mathfrak{sl}(2)$ or ${\mathfrak{sl}}(2)\oplus\mathbb{R}$\\[+2ex]
P$_3$& $\mathfrak{so}(3)$ &$\displaystyle { \frac{-1}{2 (1+x^2+y^2) },\  \frac{ y}{1+x^2+y^2},  }$ &$\displaystyle \frac{\dd x\wedge \dd y}{(1+x^2+y^2)^{2}}$ & $\mathfrak{so}(3)$ or ${\mathfrak{so}}(3)\oplus\mathbb{R}$\\[+2.5ex]
& &$\displaystyle { - \frac{x}{1+x^2+y^2} }$, \ 1 &  &  \\[+2.5ex]
 P$_5$& $\mathfrak{sl}(2 )\ltimes\mathbb{R}^2$  & ${y,\  -x, \ xy,\  \frac 12 y^2,\  -\frac 12 x^2, \ 1}$&$\dd x\wedge \dd y$  &$\overline{\mathfrak{sl}(2 )\ltimes \mathbb{R}^2}\simeq \mathfrak{h}_6$    \\[+1.5ex]
\hline
 &&&  &\\[-1.5ex]
\# & Imprimitive &  Hamiltonian functions $h_i$& $\omega$  &  Lie--Hamilton algebra \\[+1.0ex]
\hline
 &  &\\[-1.5ex]
I$_1$& $\mathbb{R}$ &$\int^y{f(y')\dd y'}$ & $f(y)\dd x\wedge \dd y$ & $ \mathbb{R}$ or $\mathbb{R}^2$\\[+1.0ex]
I$_4$& $\mathfrak{sl}(2 )$ (type II) & $\displaystyle{ \frac 1 {{x-y}   } ,\ \frac{x+y}{2(x-y)},\ \frac{xy}{x-y} }$ &$\displaystyle   {\frac{\dd x\wedge \dd y} {{(x-y)^{2}}}}$ &$\mathfrak{sl}(2 )$ or $\mathfrak{sl}(2)\oplus\mathbb{R}$\\[+2.0ex]
I$_5$& $\mathfrak{sl}(2 )$ (type III) &$\displaystyle { {-\frac{1}{2y^{2}},\ -\frac{x}{y^{2}},\ -\frac{x^2}{2 y^{2}  }  }} $  &$\displaystyle{\frac{\dd x\wedge \dd y}{y^3}}$ &$\mathfrak{sl}(2 )$ or  ${\mathfrak{sl}}(2)\oplus\mathbb{R}$\\[+2.0ex]
I$_8$& $B_{-1}\simeq  { {\mathfrak{iso}}}(1,1)$ &${y,\ -x,\ xy,\ 1 }$ & $\dd x\wedge \dd y$ & ${ \overline{\mathfrak{iso}}}(1,1) \simeq \mathfrak{h}_4$\\[+2.0ex]
I$_{12}$& $\mathbb{R}^{r+1}$ &$-\int^x \!\! {f(x')\dd x'}, - \int^x \!\! {f(x')\xi_j(x')\dd x'}$ & $f(x)\dd x\wedge \dd y$ & $\mathbb{R}^{r+1}$ or $\mathbb{R}^{r+2}$ \\[+2.0ex]
I$_{14A}$& $\mathbb{R} \ltimes \mathbb{R}^{r}$ (type I) &$y,\  - \int^x {\eta_j(x')\dd x'} $,\quad $1\notin  \langle \eta_j \rangle$ & $ \dd x\wedge \dd y$ & $\mathbb{R}\ltimes\mathbb{R}^{r}$ or $(\mathbb{R}\ltimes \mathbb{R}^{r})\oplus\mathbb{R}$\\[+2.0ex]
I$_{14B}$& $\mathbb{R} \ltimes \mathbb{R}^{r}$ (type II) &$y,\ -x, \ - \int^x{\eta_j(x')\dd x'},\ 1 $ & $ \dd x\wedge \dd y$ & $ \overline{(\mathbb{R} \ltimes \mathbb{R}^{r}) }$\\[+2.0ex]
I$_{16}$& $C_{-1}^r \simeq  {\mathfrak{h}_2\ltimes\mathbb{R}^{r+1}}$ & $\displaystyle{ {y,\  -x, \ xy,  \  -\frac{x^{j+1}}{j+1} ,\ 1} }$ & $\dd x\wedge \dd y$ & $ \overline{\mathfrak{h}_2\ltimes\mathbb{R}^{r+1}} $\\[+2.0ex]
    \hline
 \end{tabular}
\hfill}
\end{table}

\begin{table}[t] {\footnotesize
  \noindent
\caption{{\small Specific Lie--Hamilton systems according to the family given in the classification  of Table~\ref{table1}.   All of these systems have $t$-dependent real coefficients.}}
\label{table2}
\medskip
\noindent\hfill
\begin{tabular}{ l   l l }
\hline
 & &\\[-1.5ex]
LH algebra&\# & LH systems \\[+1.0ex]
\hline
 &    &\\[-1.5ex]
$\mathfrak{sl}(2 )$& P$_2$&  Milne--Pinney  and Kummer--Schwarz   equations with $c>0$   \\[2pt]
 &  &  Complex Riccati equation\\[+1.5ex]
 $\mathfrak{sl}(2 )$& I$_4$&Milne--Pinney  and Kummer--Schwarz   equations with $c<0$ \\[2pt]
  &  &Split-complex Riccati equation\\[2pt]
   &  &Coupled Riccati equations \\[2pt]
   &  &Planar diffusion Riccati system for $c_0=1$\\[+1.5ex]
   
 $\mathfrak{sl}(2 )$& I$_5$& Milne--Pinney  and Kummer--Schwarz   equations with $c=0$ \\[2pt]
 &  & Dual-Study   Riccati equation\\[2pt]
  &  & Harmonic oscillator \\[+2pt]
  &  &Planar diffusion Riccati system for $c_0=0$\\[+1.5ex]
  
${\mathfrak h}_6\simeq \overline{\mathfrak{sl}(2 )\ltimes \mathbb{R}^2}$& P$_5$& Dissipative harmonic oscillator \\[2pt]
  &  & Second-order Riccati equation in Hamiltonian form\\[+1.5ex]
${\mathfrak h}_2\simeq \mathbb{R} \ltimes \mathbb{R}$& I$_{14A}^{r=1}$& Complex Bernoulli equation\\[2pt]
 &  & Generalized Buchdahl equations  \\[2pt]
  &  & Lotka--Volterra systems  \\[+1.5ex]
    \hline
\end{tabular}
\hfill}
\end{table}

\begin{table}[t] {\footnotesize
 \noindent
\caption{{\small
{\footnotesize Functions $1,\xi_1(x),\ldots,\xi_r(x)$ are linearly independent and $\eta_1(x),\ldots,\eta_r(x)$ form a basis of solutions for a system of $r$-order linear differential equations with constant coefficients.
Notice that $\mathfrak{g}_1\ltimes \mathfrak{g}_2$ stands for the semi-direct sum of $\mathfrak{g}_1$ by $\mathfrak{g}_2$, i.e., $\mathfrak{g}_2$ is an ideal of $\mathfrak{g}_1\ltimes\mathfrak{g}_2$.}}}
\label{table9}
\medskip
\noindent\hfill
 \begin{tabular}{ p{.5cm} p{2.1cm}    p{7.0cm} l}
\hline
&  &\\[-1.9ex]
\#&Lie algebra & Basis of vector fields $X_i$ & Jacobi
\\[+1.0ex]
\hline
 &  &\\[-1.9ex]
P$_1$&$A_\alpha\simeq \mathbb{R}\ltimes \mathbb{R}^2$ & $  { {\partial_x} ,    {\partial_y}  ,   \alpha(x\partial_x + y\partial_y)  +  y\partial_x - x\partial_y},\quad \ \alpha\geq 0$&$(\alpha=0)$ Pois.
\\[+1.0ex]
P$_2$&$\mathfrak{sl}(2)$ & $ {\partial_x},   {x\partial_x  +  y\partial_y}  ,   (x^2  -  y^2)\partial_x  +  2xy\partial_y$&Poisson
\\[+1.0ex]
P$_3$&$\mathfrak{so}(3)$ &${     { y\partial_x  -  x\partial _y},     { (1  +  x^2  -  y^2)\partial_x  +  2xy\partial_y}  ,    }$&
\\[+1.0ex]
 &  &${      2xy\partial_x  +  (1  +  y^2  -  x^2)\partial_y}$&Poisson
\\[+1.0ex]
P$_4$&$\mathbb{R}^2\ltimes\mathbb{R}^2$ &$   {\partial_x},   {\partial_y} ,  x\partial_x + y\partial_y,   y\partial_x - x\partial_y$&No
\\[+1.0ex]
P$_5$&$\mathfrak{sl}(2 )\ltimes\mathbb{R}^2$ &${   {\partial_x},   {\partial_y} ,  x\partial_x - y\partial_y,  y\partial_x,  x\partial_y}$&Poisson\\[+1.0ex]
P$_6$&$\mathfrak{gl}(2 )\ltimes\mathbb{R}^2$ &${  {\partial_x},    {\partial_y} ,   x\partial_x,   y\partial_x,   x\partial_y,   y\partial_y}$& No\\[+1.0ex]
P$_7$&$\mathfrak{so}(3,1)$ &${   {\partial_x},   {\partial_y} ,   x\partial_x\!+\! y\partial_y,   y\partial_x \!-\! x\partial_y,   (x^2 \!-\! y^2)\partial_x \!+\! 2xy\partial_y ,}$  &
\\[+1.0ex]
 &  &${    2xy\partial_x \!+\! (y^2\!-\!x^2)\partial_y}$  &No
\\[+1.0ex]
P$_8$&$\mathfrak{sl}(3 )$ &${   {\partial_x},    {\partial_y} ,   x\partial_x,   y\partial_x,   x\partial_y,   y\partial_y,   x^2\partial_x + xy\partial_y,   xy\partial_x  +  y^2\partial_y}$& No
\\[+1.5ex]
I$_1$&$\mathbb{R}$ &$   {\partial_x}  $ & $(0,\partial_x)$, Pois.
\\[+1.0ex]
I$_2$&$\mathfrak{h}_2$ & $   {\partial_x}  ,  x\partial_x$& $(0,\partial_x)$
\\[+1.0ex]
I$_3$&$\mathfrak{sl}(2 )$ (type I) &$   {\partial_x} ,  x\partial_x,  x^2\partial_x$&$(0,\partial_x)$
\\[+1.0ex]
I$_4$&$\mathfrak{sl}(2 )$ (type II) & ${   {\partial_x  +  \partial_y},    {x\partial _x + y\partial_y} ,   x^2\partial_x  +  y^2\partial_y}$ &Poisson\\[+1.0ex]
I$_5$&$\mathfrak{sl}(2 )$ (type III) &${  {\partial_x},    {2x\partial_x + y\partial_y} ,   x ^2\partial_x  +  xy\partial_y}$&Poisson
\\[+1.0ex]
I$_6$&$\mathfrak{gl}(2 )$ (type I)& ${  {\partial_x},    {\partial_y} ,   x\partial_x,   x^2\partial_x}$&No
\\[+1.0ex]
I$_7$&$\mathfrak{gl}(2 )$ (type II)& ${   {\partial_x},   {y\partial_y}  ,     x\partial_x,    x^2\partial_x +  xy \partial_y}$&No
 \\[+1.0ex]
I$_8$&$B_\alpha\simeq \mathbb{R}\ltimes\mathbb{R}^2$ &${   {\partial_x},    {\partial_y} ,   x\partial_x  +  \alpha y\partial_y},\quad  0<|\alpha|\leq 1$&$(\alpha=-1)$ Pois.\\[+1.0ex]
I$_9$&$\mathfrak{h}_2\oplus\mathfrak{h}_2$ &${  {\partial_x},    {\partial_y} ,   x\partial_x,  y\partial_y}$&No
\\[+1.0ex]
I$_{10}$&$\mathfrak{sl}(2 )\oplus \mathfrak{h}_2$ & ${  {\partial_x},    {\partial_y}  ,   x\partial_x,  y\partial_y,  x^2\partial_x }$&No
\\[+1.0ex]
I$_{11}$&$\mathfrak{sl}(2 )\oplus\mathfrak{sl}(2 )$ &$   {\partial_x},    {\partial_y} ,   x\partial_x,   y\partial_y,   x^2\partial_x ,   y^2\partial_y $&No\\[+1.0ex]
I$_{12}$&$\mathbb{R}^{r + 1}$ &$  {\partial_y}  ,   \xi_1(x)\partial_y, \ldots , \xi_r(x)\partial_y $&$(0,\partial_y)$, Pois.
\\[+1.0ex]
I$_{13}$&$\mathbb{R}\ltimes \mathbb{R}^{r + 1}$ &$   {\partial_y}  ,   y\partial_y,    \xi_1(x)\partial_y, \ldots , \xi_r(x)\partial_y $ &$(0,\partial_y)$\\[+1.0ex]
I$_{14}$&$\mathbb{R}\ltimes \mathbb{R}^{r}$ & ${  {\partial_x},   {\eta_1(x)\partial_y}  ,  {\eta_2(x)\partial_y},\ldots ,\eta_r(x)\partial_y} $&Poisson
\\[+1.0ex]
I$_{15}$&$\mathbb{R}^2\ltimes \mathbb{R}^{r}$ &  ${  {\partial_x},    {y\partial_y}  ,    {\eta_1(x)\partial_y},\ldots, \eta_r(x)\partial_y} $&No\\[+1.0ex]
I$_{16}$&$C_\alpha^r\simeq \mathfrak{h}_2\ltimes\mathbb{R}^{r + 1}$ & ${   {\partial_x},    {\partial_y}  ,   x\partial_x  +  \alpha y\partial y,   x\partial_y, \ldots, x^r\partial_y} ,\quad \alpha\in\mathbb{R}$&$(\alpha=-1)$ Pois.
\\[+1.0ex]
I$_{17}$&$\mathbb{R}\ltimes(\mathbb{R}\ltimes \mathbb{R}^{r})$ &$   {\partial_x},    {\partial_y}  ,   x\partial_x  +  (ry  +  x^r)\partial_y ,   x\partial_y, \ldots,  x^{r - 1}\partial_y $ &No
\\[+1.0ex]
I$_{18}$&$(\mathfrak{h}_2\oplus \mathbb{R})\ltimes \mathbb{R}^{r + 1}$ & $   {\partial_x},    {\partial_y} ,   x\partial_x,   x\partial_y,   y\partial_y,   x^2\partial_y, \ldots,x^r\partial_y$ &No
\\[+1.0ex]
I$_{19}$&$\mathfrak{sl}(2 )\ltimes \mathbb{R}^{r + 1}$ &  $   {\partial_x},    {\partial_y}  ,   x\partial_y,    2x\partial _x  +  ry\partial_y,   x^2\partial_x  +  rxy\partial_y,   x^2\partial_y,\ldots, x^r\partial_y  $&No
\\[+1.0ex]
I$_{20}$&$\mathfrak{gl}(2 )\ltimes \mathbb{R}^{r + 1}$ &  $  {\partial_x},    {\partial_y} ,   x\partial_x,   x\partial_y,   y\partial_y,   x^2\partial_x  +  rxy\partial_y,   x^2\partial_y,\ldots, x^r\partial_y $ &No
\\[+1.5ex]
\hline
 \end{tabular}
\hfill}
\end{table}

\begin{table} {\footnotesize
 \noindent
\caption{{\footnotesize Lie symmetries for Riccati equations (\ref{Ricc}) for different $\eta(t)$.} We assume $f_0=k\in\mathbb{R}$.  We have $f_2={\rm d}f_3/{{\rm d}t}$ and  
$f_1=k\eta(t)-\int{\eta(t) f_3{\rm d}t}.$ $Airy_A$ and $Airy_B$ denote the Airy and Bairy functions.  $J,Y$ are the Bessel functions of first and second kind. Function $f_3$ follows from (\ref{conRicEsp}).
}
\label{table8}
\noindent\hfill
\hspace*{-.6cm} \begin{tabular}{  p{1.4cm}    p{11cm} l}

\hline
 &\\[-1.9ex]
$\eta(t)$ & $f_3(t)$ \\[+1.0ex]
\hline
 &\\[-1.9ex]
$\dfrac{k}{at+b}$& $\displaystyle{k+c_1(at+b)J^2\left(1,2\sqrt{(at+b)/a^2}\right)+c_2(at+b)Y^2\left(1,2\sqrt{{at+b}/{a^2}}\right)}$\\
&$\qquad\qquad\qquad \qquad\qquad +\,c_3(at+b)J\left(1,2\sqrt{t+b/{a}}\right)Y\left(1,2\sqrt{t+b/{a}}\right)$ \\
$\dfrac{k}{(at+b)^2}$& $\displaystyle{-\frac{kat}{b}+c_1\left(\frac{at+b}{a}\right)^{\frac{a+\sqrt{a^2-4k}}{a}}+c_2\left(\frac{at+b}{a}\right)^{-\frac{-a+\sqrt{a^2-4k}}{a}}+c_3(at+b),}$\\
$\displaystyle{at+b}$ &$\displaystyle{k+c_1{\rm Airy}_A\left(-\frac{at+b}{a^{2/3}}\right)^2+c_2{\rm Airy}_B\left(-\frac{at+b}{a^{2/3}}\right)^2}$\\
&$\qquad\qquad\qquad \qquad\qquad\qquad \qquad\qquad \displaystyle{+c_3{\rm Airy}_A\left(-\frac{at+b}{a^{2/3}}\right){\rm Airy}_B\left(-\frac{at+b}{a^{2/3}}\right)}$
\\[+1.0ex]
\hline
 \end{tabular}
\hfill}
\end{table}

\chapter{\it Appendix 2: Auxiliar calculations for Chapter 5}\markboth{Auxiliar calculations}{Appendix 2}
\label{Chap:appendix2}
\setcounter{equation}{0}
\setcounter{theorem}{0}
\setcounter{example}{0}

This appendix contains intermediate calculations for the obtainance of some equations displayed in Chapter \ref{Chap:RecipTrans}.
All numerical labels of the equations are in relation with those of Chapter \ref{Chap:RecipTrans}.

\section*{Obtaining \ref{28} eq.1.}

Equation (\ref{27}) provides
\begin{equation*}
x_{z_0}=X_{z_0}+\partial_{z_0}(\ln X_{z_0}).
\end{equation*}
If we use the fact that $X_{z_0}=\frac{1}{P}$ and $x_{z_0}=\frac{1}{u}$, we obtain
\begin{equation*}
\frac{1}{u}=\frac{1}{P}-\partial_{z_0}(\ln P),
\end{equation*}
and by using (\ref{6}) we have
\begin{equation*}
\frac{1}{u}=\frac{1}{P}-\frac{1}{P}(\ln P)_X
\end{equation*}
which yields (\ref{28}) eq.1.

\section*{Obtaining \ref{28} eq.2.}

By taking $i=1,\dots, n-1$ in (\ref{27}), we have
\begin{equation*}
-\frac{X_{z_{i+1}}}{X_{z_0}}=x_{z_0,z_i}-\frac{x_{z_0,z_0,z_i}}{x_{z_0}}-\frac{x_{z_{i+1}}}{x_{z_0}}, \quad i=1,\dots, n-1
\end{equation*}
If we use (\ref{7}) and (\ref{19}) the result is
\begin{equation*}
\frac{P\Omega_{[i+1]}}{2}=\partial_{z_0}(\omega_{[i]})-u\partial_{z_0z_0}(\omega_{[i]})-u\omega_{[i+1]}, \quad i=1,\dots, n-1
\end{equation*}
And now (\ref{19}) gives us
\begin{equation*}
-\frac{P\Omega_{[i+1]}}{2}=\frac{(\omega_{[i]})_x}{u}-\left(\frac{(\omega_{[i]})_x}{u}\right)_x-u\omega_{[i+1]}, \quad i=1,\dots, n-1
\end{equation*}
If we use the following expressions arising from (\ref{18})
\begin{equation*}
(\omega_{[i]})_x=u(v_{[i]})_x,\quad u\omega_{[i+1]}=v_{[i]}-(v_{[i]})_{xx},\quad i=1,\dots, n-1,
\end{equation*}
the result is
\begin{equation*}
-\frac{P\Omega_{[i+1]}}{2}=(v_{[i]})_x-v_{[i]},\quad i=1,\dots, n-1
\end{equation*}
as  is required in (\ref{28}),
and $u\omega_{[i+1]}=v_{[i]}-(v_{[i]})_{xx},\quad  i=1,\dots, n-1$
can be written as
\begin{equation*}
u\omega_{[i+1]}=\left(v_{[i]}-(v_{[i]})_x\right)+\left((v_{[i]})_x-(v_{[i]})_{xx}\right)=\left(\frac{P\Omega_{[i+1]}}{2}\right)+\left(\frac{P\Omega_{[i+1]}}{2}\right)_x,\quad i=1,\dots, n-1
\end{equation*}
We have $\partial_x=\frac{u}{P}\partial_X$. Therefore,
\begin{equation*}
u\omega_{[i+1]}=\left(\frac{P\Omega_{[i+1]}}{2}\right)+\frac{u}{P}\left(\frac{P\Omega_{[i+1]}}{2}\right)_X, \quad i=1,\dots, n-1
\end{equation*}
\begin{equation*}
\omega_{[i+1]}=\left(\frac{P\Omega_{[i+1]}}{2u}\right)+\frac{1}{2P}\left(P_X\Omega_{[i+1]}+P(\Omega_{[i+1]})_X\right), \quad i=1,\dots, n-1
\end{equation*}
We can eliminate $u$ with the aid of the first equation in (\ref{28}). The result is
\begin{equation*}
\omega_{[i+1]}=\frac{(\Omega_{[i+1]})_X+\Omega_{[i+1]}}{2}, \quad i=1,\dots, n-1.
\end{equation*}

\section*{Obtaining \ref{30}}

By taking $i=n$ in (\ref{27}) we have
\begin{equation*}
-\frac{X_{z_{n+1}}}{X_{z_0}}=x_{z_0,z_n}-\frac{x_{z_0,z_0,z_n}}{x_{z_0}}-\frac{x_{z_{n+1}}}{x_{z_0}}.
\end{equation*}
Equations (\ref{8}) and (\ref{20}) allow us to write the above equation as
\begin{equation*}
\Delta=\partial_{z_0}(\omega_{[n]})-u\partial_{z_0z_0}(\omega_{[n]})+\delta
\end{equation*}

With the aid of (\ref{19}), it reads
\begin{equation*}
\Delta=\frac{(\omega_{[n]})_x}{u}-\left(\frac{(\omega_{[n]})_x}{u}\right)_x+\delta.
\end{equation*}
If $(\omega_{[n]})_x=u(v_{[n]})_x$ and $\delta=(v_{[n]})_{xx}-v_{[n]}$ are used,
we obtain (\ref{30}). 
 
%
%
%
%


\end{document}